\newcommand\ThesisContents{%

\pagenumbering{roman}

\begin{titlepage}
        \begin{center}
        \vspace*{1.0cm}

        \Huge
        \textbf{\ThesisTitle}

        \vspace*{1.0cm}

        \normalsize
        by \\

        \vspace*{1.0cm}

        \Large
        \ThesisAuthor \\

        \vspace*{3.0cm}

        \normalsize
        A thesis \\
        presented to the University of Waterloo \\ 
        in fulfillment of the \\
        thesis requirement for the degree of \\
        \ThesisProgramme \\
        in \\
        \ThesisSubject \\

        \vspace*{2.0cm}

        Waterloo, Ontario, Canada, \ThesisYear \\

        \vspace*{1.0cm}

        \copyright\ \ThesisAuthor\ \ThesisYear\\
        \end{center}
\end{titlepage}

\pagestyle{plain}
\setcounter{page}{2}

\ifFinalVersion\ifElectronicVersion
  \noindent
  I hereby declare that I am the sole author of this thesis.  This is a true copy of the thesis, including any required final revisions, as accepted by my examiners.

  \bigskip
  
  \noindent
  I understand that my thesis may be made electronically available to the public.
\else
 \noindent
 I hereby declare that I am the sole author of this thesis.

 \smallskip

 \noindent
 I authorize the University of Waterloo to lend this thesis to other
 institutions or individuals for the purpose of scholarly research.

 \bigskip

 \noindent
 I further authorize the University of Waterloo to reproduce this
 thesis by photocopying or by other means, in total or in part,
 at the request of other institutions or individuals for the purpose
 of scholarly research.
\fi
\fi
\newpage

\begin{center}\textbf{Abstract}\end{center}
In the study of quantum computation, data is represented in terms of linear operators which form a generalized model of probability, and computations are most commonly described as products of unitary transformations, which are the transformations which preserve the quality of the data in a precise sense.
This naturally leads to \emph{unitary circuit models}, which are models of computation in which unitary operators are expressed as a product of ``elementary'' unitary transformations.
However, unitary transformations can also be effected as a composition of operations which are not all unitary themselves: the \emph{one-way measurement model} is one such model of quantum computation.

In this thesis, we examine the relationship between representations of unitary operators and decompositions of those operators in the one-way measurement model.
In particular, we consider different circumstances under which a procedure in the one-way measurement model can be described as simulating a unitary circuit, by considering the combinatorial structures which are common to unitary circuits and two simple constructions of one-way based procedures.
These structures lead to a characterization of the one-way measurement patterns which arise from these constructions, which can then be related to efficiently testable properties of graphs.
We also consider how these characterizations provide automatic techniques for obtaining complete measurement-based decompositions, from unitary transformations which are specified by operator expressions bearing a formal resemblance to path integrals.
These techniques are presented as a possible means to devise new algorithms in the one-way measurement model, independently of algorithms in the unitary circuit model.
\newpage

\ifFinalVersion
\begin{center}\textbf{Acknowledgements}\end{center}

	Many people have helped to contribute to the writing of this thesis.
	The following deserve special recognition:

	\begin{description}
	\item[Elham Kashefi,]
		for introducing me to the one-way measurement model and to the questions about it which she was interested in; for the many stimulating discussions in Waterloo, Oxford, Grenoble, and Edinburgh; and for being a patient hostess.

	\item[Robert Raussendorf,]
		for the handful of discussions in Waterloo that we had about the model which he helped to define.
		His feedback helped me to better understand the differences in perspectives about how to present the problems which Elham and I were interested in, at a time when I was unsure of how to communicate them to the wider community.

	\item[Michele Mosca and Ashwin Nayak,]
		for being very liberal supervisors, and highly supportive of international voyages to conferences and collaborators alike.

	\item[Anne Broadbent and Joe Fitsimons,]
		for insights which helped to enrich my perspective of flows and measurement-based quantum computation.

	\item[Martin Pei]
		for being, quite simply, an excellent collaborator.

	\item[John Watrous and Daniel Gottesman,]
		for being the instructors wose provided me with (more than anyone else) the \emph{conceptual} framework necessary to study and describe the subject of this thesis.

	\item[Marg Feeney,]
		for being very patient in reminding me of deadlines and requirements, and helpful in answering my questions about administrivia; she is the most excellent administrative staff member that I've ever encountered.
	\end{description}

	Finally, I'd like to thank Donny Cheung, Dave Evans, Alma Juarez, Cory Kasper, Ben Korvermaker, Mike LaCroix, Eva M$\upiota \upchi \upalpha \uplambda$, Judy Nowak, Mike Patterson, Carlos P\'erez-Delgado, Joel Reardon, Bill Rosgen, Lana Sheridan, Craig Sloss, Devin Smith, Jason Skomorowski, Stephanus du Toit, Rob Warren, Greg Zaverucha, and Joanna Zimbiecka for helping to make my time as a student in Waterloo as pleasant as it was.

\newpage
\fi

\tableofcontents
\newpage

\addcontentsline{toc}{chapter}{\textbf{List of Figures}}
\listoffigures

\addcontentsline{toc}{chapter}{\textbf{List of Algorithms}}
\listofalgorithms
\newpage

\newpage

	\definecolor{RefColor}{rgb}{0.5,0,0}

\bgroup
\addcontentsline{toc}{chapter}{\textbf{Preamble}}

\makeatletter
\let\@chapter\@old@chapter
\makeatother

\renewcommand\thesection{\arabic{section}}

\chapter*{Preamble \\\Large\it and overview of the thesis}

\section{Introduction}
\bigcapsword

Quantum computation developed as an attempt to respond to two questions:
\begin{enumerate}
\item
	How can we build a computing device to efficiently simulate physical systems?

\item
	What class of problems can be efficiently solved by physical systems, when we regard them as performing computations?
\end{enumerate}
These questions have helped to spur research, continuing in the tradition of Landauer~\cite{Landauer61,Landauer91}, to study information and computations in physical terms.

Richard Feynman~\cite{Feynman82} put forward the thesis that a model of computation founded on quantum mechanics was necessary and fruitful, in order to more efficiently solve problems such as the simulation of quantum mechanical systems.
Because of the interesting interplay between wave mechanics and discrete event detection at work in quantum mechanics, it seems to defy efficient simulation by ``classical'' models of computation, such as Turing machines~\cite{Turing36}.
Feynman proposed a model for a computer which exploited quantum mechanical effects as an antidote to this situation.
In 1985, David Deutsch proposed a programmable model of universal computation~\cite{Deutsch85}, and posited that the difficulty of efficiently simulating quantum mechanics with classical models of computation was a sign of a fundamental computational advantage of models founded on quantum mechanical principles.
This prompted the study of quantum/classical computational separations~\cite{BV93,Simon97}, which eventually led to Peter Shor's discovery of efficient algorithms for the discrete logarithm and integer factoring problems in such a model~\cite{Shor94}.
As these problems are widely regarded as being ``difficult'' to solve by classical models of computation, the discovery of these algorithms sparked significant interest in the question of how a quantum computer might be physically realized.

Due to the link with Schr\"odinger evolution in quantum mechanics, quantum computation is most often (and usually conveniently) described in terms of \emph{unitary transformations}, which are transformations of those configurations of the system which preserve the property of yielding point-mass probability distributions for a suitably chosen measurement operation.
The ``standard'' model of quantum computation is therefore that of \emph{unitary circuits} (described in Section~\ref{sec:unitaryCircuits}), which expresses unitary transformations by products of other unitary transformations, reducing the evolution of quantum states to some set of elementary operations which are considered to be likely to be physically realizable.

In 2001, Robert Rauss\-endorf and Hans Briegel advanced a model~\cite{RB01} for a quantum computer by performing controlled measurement operations performed in a fixed state of a regular spin network.
This gave rise to an alternative model of quantum computation, called the \emph{one-way measurement model} (described in Chapter~\ref{One-way measurement-based computation}).
This model contrasted sharply with the picture of quantum computation via unitary circuits in that measurement operations were used to effectively simulate unitary transformation on a fixed initial state, whereas measurements are often portrayed as being prototypical examples of non-unitary operations.
While it is not exceedingly difficult (having the results of~\cite{RB01} in hand) to describe simple principles by which universal quantum computation may be driven by measurements on a spin network, this contrast of perspectives on quantum computation between transformation by unitaries, and transformation by measurements, makes it difficult to interpret measurement based procedures as ``first class'' descriptions of unitary transformations of quantum states.

In this thesis, we explore the relationship between the one-way measurement model and unitary transformations, with the aim of exploring, on the one hand, techniques to recognize one-way measurement procedures which represent unitary transformations and translate them into the circuit model, and on the other hand to identify when expressions for unitary transformations may be easily translated to a one-way measurement procedure.
A secondary, underlying theme is the subject of combinatorial structures which unify different representations of unitary transformations: throughout the thesis, it is by identifying and recognizing such combinatorial structures that the task of translating between different representations of unitary transformations (such as unitary circuits and one-way measurement-based procedures) becomes tractable.

\section{Structure of the thesis}

This thesis contains five chapters, discounting the present one.

The first, introductory chapter, introduces the mathematical tools for describing quantum computation, including three individual models of importance:
\begin{enumerate}
\item
	\emph{Unitary circuit models} --- more properly speaking, this is a family of computational models, in which elementary unitary operations are composed to produce unitary operations of greater complexity.
	In practise, however, almost all models of unitary circuits considered are essentially equivalent to a ``standard'' choice of elementary gates, which we introduce here.
	We also define a non-standard, but closely related, unitary circuit model which will come to dominate the analysis of unitary circuits throughout the thesis; and a non-standard representation of unitary circuits which also plays an important role in the analysis of the later chapters.

\item
	\emph{Classically controlled unitary circuits} --- an elaboration of unitary circuit models by explicitly representing a classical control, we present these models essentially in order to concretely describe the way in which they are usually translated back into ``fully quantum'' unitary circuits.

\item
	\emph{Clifford circuits} --- a model of computation with restricted power, the limited operations of this model nevertheless provides useful analytical tools for describing transformations of a special class of states by measurements.
\end{enumerate}

The second chapter introduces measurement-based computation, with emphasis on the \emph{one-way measurement-based model} in particular.
This model is the main subject of the thesis.
In particular, we describe how universal quantum computation is effected in both the original \emph{cluster-state} model, and the less restrictive model based on \emph{open graphical encodings}, presenting two standard constructions for measurement-based computation to do so.

The third chapter is concerned with algorithms to recognize a certain class of one-way measurement-based procedures that perform unitary transformations, and to translate these procedures into unitary circuits.
This is done by the graph-theoretical characterization and examination of a combinatorial structure arising out of the non-standard circuit representation introduced in Chapter~\ref{Fundamentals of quantum computation}.
Using this combinatorial structure, we may reduce the problem of recognizing and translating such patterns to previously solved problems in graph theory.
We then consider how the structure can be generalized to extend the scope of these recognition and translation algorithms.
We also consider how combinatorial structures underlying one-way measurement-based computations may allow the certification that they perform unitary transformations, even in the absence of efficient algorithms to translate them to unitary circuits.

The fourth chapter takes advantage of the certifying structures described in the preceding chapter, to suggest techniques to automatically synthesize decompositions of unitary transformations (as unitary circuits, or as measurement-based procedures) from concise matrix expressions for such transformations.
These techniques are presented as a stepping stone towards devising ``native'' idioms for solving problems in the one-way measurement model, which would contribute to the usefulness of the one-way measurement model as a useful model of computation, in addition to being a potentially useful proposal for implementation.
At the same time, we consider the matrix expressions themselves, in particular noting their recurring role in quantum computation, as well as the possibility that they may naturally arise from representations of propagators of for physical systems as path integrals.

In the final chapter, we consider natural research directions, which essentially consist of the ways that the analysis of Chapter~\ref{Measurement Pattern Interpolation} may be most naturally extended.

\section{Contributions presented in this thesis}

The following is a list of the contributions made by myself, as presented in this thesis, to the understanding of one-way measurement-based computation, the context in which these results arose, and the subsequent work in the literature which followed.
The description of where the results have been previously published, in each case, may be found in the pre-amble of the corresponding chapters.

\subsection*{Graph-theoretic descriptions of flows, and flow-finding algorithms}

Sections~\ref{sec:graphth-charn} and~\ref{sec:flowAlgs} present work by myself (in the case of~\ref{sec:extreme}, with Martin Pei) on ``flows'', which may be interpreted as a certificate for the unitarity of a one-way measurement-based procedure~\cite{DK06}, arising from a particular construction of measurement-based procedures (which is described in Section~\ref{sec:DKPstandardization} as the \emph{simplified DKP construction}).

The characterizations of Section~\ref{sec:graphth-charn} led to the first efficient algorithm for determining whether a geometry has a flow, albeit conditioned on the input and output subsystems of the corresponding measurement-based procedure being of the same size (corresponding to the case of a unitary bijection).
These algorithms were made possible by an identification of a uniqueness result in~\ref{sec:unique}.
The result of~\ref{sec:extreme} then helps to further bound the running time.

Subsequently, a faster and more general algorithm for finding flows unconditionally has been presented in~\cite{MP08}.
This algorithm is likely to be optimal; bounds on its running time are also improved by the result of Section~\ref{sec:extremal}.
An examination of the algorithm of~\cite{MP08} imply a second graph-theoretic characterization of flows, which I briefly describe in Section~\ref{sec:improvedFlowAlgs} in relation to other work in Chapter~\ref{One-way measurement-based computation}.

\subsection*{Translation of one-way measurement procedures to unitary circuits}

One application of flows as a certificate of unitarity is to obtain a unitary circuit which is equivalent~\cite{DK06}, both in complexity (in a sense described in Section~\ref{sec:semanticsReduce}) and in the transformation which it performs.
However, an explicit description of such an algorithm has not yet appeared in the literature, and involves a non-trivial amount of combinatorial analysis.

Section~\ref{sec:semanticDKP} presents an explicit, efficient algorithm to translate one-way measurement procedures with flow certificates to unitary circuits in this way.
In particular, Section~\ref{sec:dependencies} characterize the dependency relations of measurement patterns with flows which have been fully standardized: this completes the work begin in~\cite{BK2007} on the dependency relations in measurement patterns, which is pertinent to the topic of depth complexity.
Section~\ref{sec:starDecomposition} then presents a new decomposition result for geometries with flows, which is pertinent to the faster flow-finding algorithm of~\cite{MP08}.
These results allow for the verification of the correctness of the measurement dependencies in the measurement procedure, and then construction an appropriate circuit.

\subsection*{Extension of flows}

In Sections~\ref{sec:eflow}, I present an extension of the ``flow'' certificate, which I call ``extended flows''.
These extended flows may be taken as a certificate of unitarity for a different construction for one-way procedures (which is described in Section~\ref{sec:RBB} as the \emph{simplified RBB construction}).
Exploiting the description given in Section~\ref{sec:improvedFlowAlgs} of the algorithm of~\cite{MP08}, I present an efficient algorithm to detect extended flows, with essentially no change to the running time of~\cite{MP08}.
I also present a sketch of how the efficient mapping to unitary circuits, which was presented for flows, may be similarly extended.

\subsection*{Measurement Pattern Interpolation}

In Chapter~\ref{Measurement Pattern Interpolation}, as part of joint work with Elham Kashefi, Vincent Danos, and Martin Roetteler, I present the definition of a quadratic form expansion.
Similar objects have occurred previously in the literature; ours was the first to explicitly connect them with measurement-based computation, although a link through graph-states~\cite{Schl04} was previously known.
The contribution of Chapter~\ref{Measurement Pattern Interpolation} as a whole is to outline how algorithms may be naturally compiled for the one-way measurement model, and to propose that techniques along these lines may be applicable to the problem of approximating path integrals for physical propagators.
There does not appear to be prior work on compiling for the one-way measurement model, except for the standard constructions via the circuit model.

The algorithms presented in Section~\ref{sec:interpolateGflows} are essentially applications of a principle forwarded by Drs. Kashefi and Danos: that for any certificate that a measurement pattern performs a unitary transformation which is based on the stabilizer formalism, there exists an efficient algorithm to produce such a measurement pattern from a description of how it acts conditioned on one particular sequence of measurement results.
The contribution of Section~\ref{sec:interpolateGflows} consists of the analysis of applying this principle to each ``instance'', including making the necessary connections with the implementation of diagonal operations presented in~\cite{BB06}, and with the concept of ``fractional weight'' flows.

Section~\ref{sec:synthClifford} presents a refinement of the formulae for Clifford group operations presented in~\cite{DM03}, which allows them to be presented as quadratic form expansions.
Having found this connection, I present an algorithm to obtain a measurement pattern, essentially via an application of the stabilizer formalism~\cite{GotPhD}.
This construction of one-way measurement procedures for the Clifford group elements is shown to be minimal with respect to the reduction techniques of~\cite{HEB04}, and is proven to run in time comparable to the algorithm of~\cite{AG04} for producing unitary circuits for Clifford group (and strictly faster than an algorithm using~\cite{AG04} as a subroutine to produce minimal one-way measurement procedures for the Clifford group).

\section{Conventions}

\subsection*{Notation and Pseudocode}

Throughout the thesis, Dirac (``bra-ket'') notation is used, along with conventional notations for Hilbert spaces: see \eg~\cite{NC00} for an introduction to how it is used conventionally.
However, owing to the regular discussion of operations in terms of the measurement-based model, transformations of quantum states are often represented as completely positive trace preserving linear superoperators (or \emph{CPTP maps}, described on page~\pageref{dfn:CPTP}) rather than as unitary matrices acting on column vectors.
Other notations pertaining to linear operations are a minor variation on that of~\cite{Watrous07}.

Special notations for operators, states, and other variables are usually introduced where they are used.
However, it may be useful to note the following conventions, which may help to avoid possible ambiguities of interpretation:
\begin{itemize}
\item
	The constant $i$ in italics always represents a square root of $-1$ in the complex numbers $\C$\,.
	In particular, it is never used as a variable to  iterate over the terms of a sum, or the vertices of a path.
	(For the latter two purposes, variables such as $h, j, \ell, \ldots$ are most often used.) 

\item
	The constant $\e$ in an upright typeface always represents Euler's constant, the base of natural logarithms.
	(This should be contrasted with an italic $e$, with may represent a generic element of a finite ground set $V$.)

\item
	The identity operator (or superoperator) for a physical system is always represented as $\idop$\,.
	(This should be contrasted with the variable $I$, which always represents a subset of some larger ground-set $V$.)

\item
	Superoperators are always presented as capital sans-serif letters such as \,$\Ent{}{}$, or fraktur letters such as $\mathfrak P$.
	Operations in fraktur font in particular always represent procedures in a measurement-based model of computation.
	(This should be contrasted with capital italic letters, which usually represent unitary operations.)

\item
	Vectors over $\Z_2$ (the integers modulo $2$) are represented in a serif bold-face font, usually as a variable $\vec x \in \Z_2^V$ for some set $V$.
	(This should be contrasted with the sans-serif letters $\sfx$, $\sfy$, $\sfz$, which are generic labels common to the notation for certain state vectors, measurements, and unitary transformations.)

\item
	In the presentation of an algorithm, $a = b$ always denotes an equality comparison.
	(This should be contrasted with $a \gets b$, which denotes an assignment.)

\item
	In the presentation of an algorithm, the notation $\INCR{a}$ represents an instruction to increment the integer variable $a$.

\item
	In the presentation of an algorithm, a \emph{procedure} takes as input \emph{only} those variables which are explicitly specified in its invocation: \ie\ it has no ``side effects''.
	(This should be contrasted with a \emph{subroutine}, which may affect the values of data which are presented as being ``global variables'' available to the subroutine.)

\subsection*{Colours and Hyperlinking}

In the figures, I have tried wherever appropriate to emphasize distinct elements using grayscale shading rather than colour.
Where I found it important to provide additional emphasis with colour, I have tried also to supplement this with non-coloured visual cues (\eg\ different styles of dashed lines).

This thesis was written using PDF\LaTeX, and the electronic version is hyperlinked.
Links are coloured, I hope unobtrusively, according to the nature of the reference:
\begin{description}
\item[\textcolor{RefColor}{red}] represents a link to another part of the main text, \eg\ Theorem~\ref{thm:churchHilbertSpace};
\item[\textcolor{CiteColor}{green}] represents a link to the bibliography, \eg\ \cite{HEB04};
\item[\textcolor{UrlColor}{blue}] represents a link to a URL on the internet, \eg\ \arXiv[0806.1972].
\end{description}
The entries of the Table of Contents, List of Figures, and List of Algorithms are also hyperlinked, although they are not coloured.

\end{itemize}

\egroup

	\pagenumbering{arabic}
	\pagestyle{headings}

	\chapter[Fundamentals of quantum computation]{
		Fundamentals (of quantum computation)}

Computation is the act of transforming a piece of data by physical processes, into a form which may be interpreted as representing \eg\ the solution to a mathematical problem.
Data may be stored as any reliably measurable (and manipulable) property of a physical system, in which case computations are most appropriately described in terms of operations on that system.
Observations along these lines made by Landauer~\cite{Landauer61} spurred research into computation as a physical process, culminating in a series of papers by Benioff~\cite{Benioff80,Benioff81,Benioff82} proposing an implementation of Turing machines~\cite{Turing36} along quantum mechanical principles.

In~\cite{Feynman82}, Feynman considered the complementary question of whether or not quantum mechanical phenomena could be efficiently simulated by the existing models of computation, and concluded that a new type of computer was required in order to intrinsically model quantum mechanical phenomena.
A model of a computation operation using quantum mechanical states as data was presented by Deutsch~\cite{Deutsch85}, and the theory of \emph{quantum computation} was subsequently developed by Bernstein and Vazirani~\cite{BV93,BV97} and Yao~\cite{Yao93}; however, the mathematical tools for this description of information were largely developed by von Neumann~\cite{Neumann32}.

Quantum computation is most often described in terms of \emph{unitary circuits}, in which quantum data is represented as unit $\ell_2$-norm \emph{state vectors} over $\C^d$ for some $d \ge 2$, which generalize the notion of a point-mass probability distribution; transformations of the state are given by \emph{unitary matrices}, which are the linear transformations which preserve the set of these vectors.
In this model, measurements are postponed to the end of the computation, or (in the case of a computation which is to be used as a subroutine of a larger quantum computation) not performed at all, and so the description of measurements and post-measurement states is often simplified.
However, it will prove essential for the discussion of the \emph{one-way measurement-based model} to establish conventions and definitions for the more general description of quantum computation in terms of \emph{density operators} which generalize both state vectors and arbitrary probability distributions, and the transformations which preserve this set of objects, which are \emph{completely positive and trace-preserving superoperators} (or CPTP maps).

In this chapter, I will introduce the principal mathematical objects of quantum computation and standard results concerning universal quantum computation, abstracted away from particular applications or implementations.
I also introduce the conventions used in this thesis to describe \emph{unitary circuits}, the principal language for describing quantum computation.

\paragraph{New results.}

Unsurprisingly, the vast majority of the work in this chapter is prior art; however, the section ``Non-standard representations of unitary circuits'' (pages~\pageref{sec:nonstandardRepns}\thru\pageref{sec:nonstandardRepnsEnd}) presents new results of an elementary nature, which fit most naturally in this chapter.

\section{Quantum data, and quantum transformations}
\label{sec:quantumData}

\subsection{Qualitative remarks on the nature of quantum data}

We begin with some remarks on quantum states, and the differences between them and classical probability, as a pre-amble to the technical definitions that follow.

The utility of probabilistic theories of evolution and computation is that they allow the transformation of data when one does not have perfect information about a system.
A probability distribution (usually) represents some amount of information about a physical system, but (usually) not complete information; and transformations of such a distribution represents the information that can be obtained about that system under certain conditions.
In particular, a probability distribution is a \emph{model} of the state of a physical system, or of the distribution of inputs to a problem, and stochastic transformations represent a model of the evolution of a system, without committing to any answer as to whether a more precise model for the same physical system is possible.
However, there is an implicit assumption that there is no penalty for coarseness of resolution of events compared to the possible states of the system in a traditional probabilistic model, except in the coarseness of the corresponding predictions.

Let us say that one set of events $R$ is a \emph{refinement} of another set $S$ if any system which is determined with respect to $R$ (\ie, such that there is an event $r \in R$ which occurs with certainty), then the system is also determined with respect to $S$.
Conceptually, there if a function from each $r \in R$ to some $s \in S$, which determines the events in $S$ which occur conditioned on the events in $R$.
The more refined a set of events is, the more precisely it allows description of a physical system; and a \emph{maximally} refined set of events $R$ for a system is one where any property which can be tested simultaneously with the events of $R$ can in principle be \emph{defined} in terms of events $r \in R$.
\label{discn:pureState}
Then, let us say that a system is in a \emph{pure state} if, for some maximally refined set of events $R$ for the system, the system can be described by the event $r \in R$.
Classical probability theory is then a means of dealing with probabilities of events and stochastic processes on the premise that there may be in principle a unique set of distinguishable pure states, and that analysis of other distributions may be performed with respect to this unique set of states of complete knowledge.

By contrast, quantum mechanics presents us with physical systems in which there seems to be a mismatch between observable events (such as the outcomes of measurements in experiments) and the possible physical states of a system.
It is possible to prepare physical systems such as the polarization of light signals, or the orientation of the spin of ions, in different ways which correspond to point-mass distributions of certain events, but where these events cannot be simultaneously tested.
This results in states which correspond to maximal information regarding certain physical properties, but seem only to be distinguishable from one another statistically, through multiple preparations.

What is more, the way in which such physical systems evolve demonstrate that the inability to resolve between different maximally refined sets of events in such systems is crucial: they may exhibit different evolutions depending on whether they interact with other systems, even when those interactions may only be used to test for some event.
An example is with light filtered with a polarizing lens: vertically polarized light cannot pass through a horizontally polarized lens.
However, there is some probability of the light passing through the horizontal filter if another filter (aligned diagonally to the horizontal axis) is placed between the source of polarized light and the horizontal filter.
The intermediate filter causes an increase in the probability of light being detected after the original blocking filter, suggesting that the light has been transformed \emph{en route}.
However, unless the physical system is sensitive to the context in which the intermediate filter is being used, there is no reason why this filter should have a different effect than it would when used to either let diagonally polarized light pass through, or filter out light with the opposite diagonal polarization, in an attempt to measure in a different experiment which of the two diagonal polarizations it may have.

The example of diagonally versus horizontally or vertically polarized light illustrates that physical systems can be prepared in different pure states which can only be distinguished statistically, and that these states are sensitive to testing for properties of the system, and are generally changed by measurement of a property to a state consistent with having a definite value of that property.
This suggested to Einstein, Podolsky, and Rosen~\cite{EPR35} that quantum mechanics is incomplete, as they posited that there must be a single underlying maximally refined set of events, which contained complete information corresponding to what they referred to as \emph{elements of reality} (\ie\ point-mass functions over any testable set of events).
This corresponds to the some of the current interpretation of quantum mechanics (\eg\ the Bohmian interpretation~\cite{Bohm52}) involving \emph{hidden variables} which we are prevented from testing precisely.
However, the more common view has been that there do not exist properties which we are prevented for fundamental reasons from testing, and that there does not exist a \emph{unique} maximally refined set of events, although there do exist multiple \emph{incompatible} such maximally refined sets, in the sense that (by necessity) they cannot be tested for simultaneously, and that testing for one will disturb point-mass functions over the others.

The apparent absence of a maximally refined set of events for certain physical systems prompted the development of early quantum theory, in which tests for events (and the set of events which is being tested for) must be specified explicitly.
This led to the specialized concept of \emph{measurement} in quantum mechanics (and consequently in quantum computation as well) which will play a central role in this thesis.

\subsection{Density operators}

Quantum states may be described by \emph{density operators}, and property testing on these states may be represented as \emph{projections} on these operators~\cite{Neumann55}.
We will illustrate this formalism of quantum states and measurements as an extension of classical probability theory, as follows.

\subsubsection{Description of classical probability distributions}

We describe probability in terms of an event space $\E(S)$ consisting of the set of functions $f: S \to \R$\,, for some finite set of events $S$.
The set of probability distributions $\Prob(S)$ in this space are the non-negative functions $p \in \E(S)$ of unit $\ell_1$-norm,
\begin{gather}
		\norm{p}_1
	\;\;=\;\;
		\sum_{s \in S} p(s)
	\;\;=\;\;
		1	\;.
\end{gather}
Point-mass functions are given by characteristic functions $\bigchi_{\ens{s}}$ for $s \in S$, where for $E \subset S$, $\bigchi_E$ is the characteristic function of $E$\,:
\begin{gather}
		\bigchi_E(s)
	\;\;=\;\;
		\begin{cases}
			1\,,	&	\text{if $s \in E$}	\\
			0\,,	&	\text{otherwise}
		\end{cases}\,;
\end{gather}
we may then decompose $p \in \Prob(S)$ as a \emph{convex combination}\footnote{%
A \emph{convex combination} of a set $X$ is a linear combination of the form $\smash{\sum\limits_{x \in X} \lambda(x) x}$, for some non-negative function $\lambda: X \to \R$ with unit $\ell_1$-norm.}
of operators $\bigchi_{\ens{s}}$ for $s \in S$,
\begin{gather}
	p \;=\; \sum\limits_{s \in S} p(s) \bigchi_{\ens{s}}	\;\;.
\end{gather}
Events correspond to subsets $E \subset S$, and the probability of $E$ corresponds to the sum of $p(s)$ for $s \in E$\,; equivalently, we can express it in terms of the adjoint functional $\bigchi_E\herm$\,, which maps characteristic functions $\bigchi_A$ to their ``overlap'' $\card{E \inter A}$\,:
\begin{gather}
	\label{eqn:simpleProbEvent}
		\Pr_{p}(E)
	\;=\;
		\bigchi_E\herm \; p
	\;=\;
		\anglebr{\bigchi_E, p}	\;,
\end{gather}
where $\anglebr{p,q}$ is the usual inner product over Euclidean space, and $A\herm$ is the Hermitian adjoint for an operator $A$.
Joint probability mass functions $g \in \Prob(S \x T)$ are maps from pairs of events $(s,t) \in S \x T$ to their probabilities $g(s,t)$\,, and can be described as convex combinations of the point-mass functions $\bigchi_{(s,t)}$\,; for independently distributed variables $p \in \Prob(S)$ and $q \in \Prob(T)$\,, we may describe such a joint distribution as a tensor product of the distributions $p$ and $q$\,, by identifying distributions in $\Prob(S \x T)$ with the unit $\ell_1$-norm functions in the event space $\E(S) \ox \E(T)$.
The probability distributions $\Prob(S \x T)$ then correspond to the non-negative, unit trace distributions of $\E(S) \ox \E(T) \,\cong\, \E(S \x T)$.

Transformations $U$ of probability distributions are given by the linear transformations which preserve $\ell_1$-norm on $\E(S)$, \ie\ by \emph{stochastic transformations}.
By virtue of their linearity, they may be characterized by conditional probabilities, \ie\ how $U$ transforms point-mass functions $\bigchi_{\ens{s}}$\,.
Independent transformations $U$ on an event space $\E(R)$ and $V$ on an event space $\E(T)$ then given by the tensor product $U \ox V$; in particular, the marginal distribution over $S$ of a probability distribution over $S \x T$ can be recovered by setting the map $V$ to be the unique ``stochastic'' transformation $V: \E(T) \to \R$, using the equivalence $\E(S) \ox \R \,\cong\, \E(S \x \ens{1}) \,\cong\, \E(S)$\,.

An isomorphic choice of mathematical structure would be to substitute real-valued diagonal matrices, and the Euclidean inner product with the Hilbert--Schmidt inner product,
\begin{gather}
		\anglebr{A,B}_{\Tr{}}
	\;\;=\;\;
		\Tr(A\herm B)	\;.
\end{gather}
For a maximally refined set of measurable events $R$ on a system, let $\ens{\ket{s}\big.}_{s \in S}$ represent the standard basis for the Hilbert space $\cH[S]$\,.
Then, we may use the following constructions for $\E(S)$ and $\Prob(S)$:
\begin{subequations}
\label{eqn:embedHilbertSchmidt}
\begin{align}
		\Pi_{\ens{s}}
	\;\;=&\;\;
		\ket{s}\bra{s}
		\;\;\text{is a rank-$1$ projector $\big($\cf\ point-mass functions $\bigchi_{\ens{s}}\big)$;}
	\\[1ex]
		\E(S)
	\;\;=&\;\;
		\Span_{\R}\ens{\ket{s}\bra{s} \,\big|\, s \in S}
		\;\;\big(\text{\cf\ span of the functions $\bigchi_{\ens{s}}$}\big);
	\\[1ex]
	\begin{split}
			\Prob(S)
		\;\;=&\;\;
			\ens{p \in \E(S) \,\big|\, \text{$p$ non-negative, with unit trace}}
		\\[-0.1ex]&
			\;\;\;\text{(\cf\ non-negative functions with unit $\ell_1$-norm);}
	\end{split}
	\\[1ex]
	\begin{split}
			\smash{\Pr_{p}(E)}
		\;\;=&\;\;
			\Tr(\Pi_E \; p)
 		\;\;\text{for $p \in \Prob(S)$ and $E \subset S$,\; where~} \smash{\Pi_E \;=\; \sum_{s \in E} \ket{s}\bra{s}}
 		\\[-0.3ex]&\;\;(\text{\cf\ Equation~\eqref{eqn:simpleProbEvent} above}).
	\end{split}
\end{align}
\end{subequations}
In particular, distinct point-mass distributions are orthogonal with respect to the Hilbert--Schmidt inner product, and probabilities of events $E \subset R$ are given by the \emph{operator functionals} $\Pi_E\conj$ characterized by mapping other projectors $\Pi_S$ to the overlap $\Pi_E\conj \Pi_A \,=\, \anglebr{\Pi_E, \Pi_A}_{\Tr{}} \,=\, \card{E \inter A}$.

\subsubsection{Generalization to arbitrary quantum distributions}

As we noted above, a single maximally refined set of events does not seem to suffice to describe the pure states accessible to quantum systems.
For a quantum system which admits an event space $\E(S)$ as constructed above, we will generalize the set of pure states to all rank-$1$ projectors $\ket{\psi}\bra{\psi}$\,: a maximally refined set of events is then a maximal collection
\begin{gather}
	\label{eqn:rankOneOrthProjectors}
	B = \ens{\ket{\psi_j}\bra{\psi_j}\Big.}_{j = 1}^{\card{S}}
\end{gather}
of rank-$1$ projections on $\cH[S]$\,, subject to these projectors being orthogonal with respect to the Hilbert--Schmidt inner product.
We may then consider events of which this set is a refinement, which correspond to sets of projectors which all commute with the projections in $S$; and we may consider states of the system which are convex combinations of elements of $S$.

Considering all such sets of orthogonal rank-$1$ projectors prompts us to generalize the event space of the system from $\E(S)$ to the set of all matrices which can be decomposed as a real-valued linear combination of orthogonal rank-$1$ projectors: that is, the Hermitian operators (normal operators with real eigenvalues) $\Herm(S)$ over $\cH[S]$.
Distributions describing states of the system are then as follows:

\begin{definition}
	The set of \emph{density operators} over $S$ is the set of positive semidefinite Hermitian operators over $\cH[S]$ with unit trace,
	\begin{align}
			\D(S)
		\;\;=&\;\;
			\ens{\rho \in \Herm(S) \,\Big|\, \rho \ge 0 \text{~and~} \Tr(\rho) = 1}\,.
	\end{align}
\end{definition}
As before, joint distributions are formed by tensor products of the event space: we have
\begin{gather}
	\Herm(S \x T) \;=\; \Herm(S) \ox \Herm(T) \,,
\end{gather}
which can be easily proven by considering a basis for $\Herm(S \x T)$ over $\R$ consisting of the tensor product of two such bases for $\Herm(S)$ and $\Herm(T)$.
For joint distributions $\tau \in \D(S \x T)$, we then take the non-negative, unit-trace elements of $\Herm(S \x T)$.

\subsubsection{Transformations of density operators}

The natural extension of a transformation between probability distributions is a \emph{linear superoperator} --- that is, a linear operation that itself acts on linear operations $\rho \in \Herm(R)$ --- which in particular maps density operators to density operators.
It follows that valid transformations of quantum data consist of \emph{trace-preserving, positive} superoperators $\Upphi: \D(\cH[S]) \to \D(\cH[S'])$\,, \ie\ that $\Upphi$ maps operators of $\L(\cH[S])$ to operators with the same trace, and positive semidefinite operators to other positive semidefinite operators.
In order for such a superoperator $\Upphi$ to be valid when applied to part of a joint distribution of two systems, we also require that $\Upphi \ox\, \id_{\D(T)}$ be a valid transformation (and therefore a positive operator) on $\D(S \ox T)$ for any $T$\,: we say that $\Upphi$ is \emph{completely positive}.
\begin{definition}
	\label{dfn:CPTP}
	A \emph{transformation} of a set of density operators $\D(S)$ is a completely positive and trace preserving mapping (or \emph{CPTP map}) $\Upphi: \D(S) \to \D(S')$ for some state-space $\D(S')$.
\end{definition}
It is possible to show (see \eg~\cite{Watrous07}) that for $\Upphi: \L(\cH[S]) \to \L(\cH[S'])$ completely positive and trace preserving, there exist linear operators $\ens{A_j}_{j = 1}^{n}$, for some $n \in \N$ and $A_j: \cH[S] \to \cH[S']$ non-zero, such that
\begin{gather}
	\label{eqn:KrausCondition}
		\Upphi(\rho)
	\;\;=\;\;
		\sum_{j = 1}^n A_j \,\rho\, A_j\herm	\;,
	\quad\text{~where~}\quad
		\sum_{j = 1}^n
		A_j\herm A_j
	\;\;=\;\;
		\idop_{\cH[S]}	\;.
\end{gather}
Such operators $A_j$ are called \emph{Kraus} operators, and are non-unique. 
For another set $T$, if $\Uppsi: \L(\cH[T]) \to \L(\cH[T'])$ is another CPTP map given by
\begin{align}
		\Uppsi(\sigma)
	\;\;=&\;\;
		\sum_{k = 1}^m B_k \,\sigma\, B_k\herm	\;,
\end{align}
we may form the transformation performed by independent transformations $\Upphi: \D(S) \to \D(S')$ and $\Uppsi : \D(T) \to \D(T')$ on a \emph{joint} distribution over $\D(S \x T)$ by taking tensor products of the Kraus operators: we write
\begin{align}
		(\Upphi \ox \Uppsi)(\tau)
	\;\;=&\;\;
		\sum_{j = 1}^n \sum_{k = 1}^m \,
			(A_j \ox B_k) \, \tau \, (A_j \ox B_k)\herm
\end{align}
for $\tau \in \L(\cH[S \x T]) \to \L(\cH[S \x T])$.

We may use this to determine the marginal distributions of joint density operators $\tau \in \D(S \x T)$\,.
Consider the map $\tr[T] : \Herm(T) \to \R$ which maps every Hermitian operator over $\cH[T]$ to its trace, given \eg~by
\begin{gather}
	\label{eqn:traceOut}
		\tr[T](\sigma)
	\;\;=\;\;
		\sum_{t \in T} \bra{t} \sigma \ket{t}	\;:
\end{gather}
note that $\tr[T]$ is the unique CPTP map from $\Herm(T)$ to $\R$.
Then, we may obtain the marginal distribution $\tau_S$ over $\D(S)$ of a density operator $\tau \in \D(S \x T)$ by the \emph{partial trace} over $T$, defined by
\begin{gather}
		\Big(\id_{\D(S)} \,\ox\, \tr[T]\Big)(\tau)
	\;\;=\;\;
		\sum_{t \in T} \Big( \idop_{\cH[S]} \ox \bra{t} \Big) \,\tau\, \Big( \idop_{\cH[S]} \ox \ket{t} \Big)	\;.
\end{gather}
We refer to the process of applying such a map as \emph{tracing out} the system represented by the state space $\D(T)$; and we commonly write such a map by $\tr[T]$, leaving the identity on any other subsystems implicit in the description of the map.

\subsubsection{Quantum and classical registers}

Based on the preceding, we will make the following definitions:
\begin{definition}
	\label{def:register}
	A \emph{quantum register} $\sfQ$ over a finite set $Q$ is an abstract physical system with state-space $\D(Q) \le \Herm(Q)$, consisting of unit trace operators $\rho \ge 0$ supported on a Hilbert space $\cH[Q]$.
	Similarly, a \emph{classical register} $\sfC$ over a finite set $C$ is an abstract physical system with state-space $\Prob(C) \le \D(C)$; and where furthermore, we require that for any joint state $\rho \in \D(C \x R_1 \x \cdots \x R_n)$ of $\sfC$ together with any collection of registers $\ens{\sfR_1, \ldots, \sfR_n}$, $\rho$ is a convex combination of operators $\vec{p}_j \ox \sigma_j$, for $\vec{p}_j \in \Prob(C)$ and $\sigma_j \in \D(R_1 \x \cdots \x R_n)$.
	For either a classical register with state-space $\P(R)$ or quantum register with state-space $\D(R)$, the \emph{dimension} of the register is $\card{R}$.
\end{definition}
The motivation for defining classical registers in addition to quantum registers is to abstractly represent physical systems which we cannot (either for fundamental reasons or practical limitations) put into general quantum states: examples include read-out devices in laboratory settings, and hand-written notes on paper, whose states do not exhibit obvious quantum behavior but which still may not be determinable in advance.

\begin{notation}
	Let $X$ and $Y$ be convex sets in complex Hilbert spaces.
	In analogy to the linear extension $V \x W \incl V \ox W$ for vector spaces, we will let $X \ox Y$ represent the \emph{convex} extension of the set $X \x Y$, which consists of convex combinations of tensor products $\rho \ox \sigma$ for $\rho \in X$ and $\sigma \in Y$.
\end{notation}
In particular, for $\rho$ a joint state of a classical register $\sfC$ together with any collection of registers $\ens{\sfR_1, \ldots, \sfR_n}$ as in Definition~\ref{def:register} above, we have $\rho \in \Prob(C) \ox \D(R_1 \x \cdots \x R_n)$.

Given the definition of a classical register above, the following Lemma allows us to interpret aggregates of classical registers as a single, compound classical register:
\begin{lemma}
	Let $\rho$ be a joint state of a collection of quantum registers $\ens{\sfQ_1,\ldots,\sfQ_n}$ and classical registers $\ens{\sfC_1,\ldots,\sfC_m}$.
	Then we have
	\begin{gather}
		\rho \in \D(Q_1 \x \cdots \x Q_n) \ox \Prob(C_1 \x \cdots \x C_m),
	\end{gather}
	where $Q_j$ (respectively $C_j$) are the finite sets over which the registers $\sfQ_j$ (respectively $\sfC_j$) are defined.
\end{lemma}
\begin{proof}
	We prove this by induction on $m$.
	If $m = 0$, there is nothing to prove.
	Otherwise, suppose the statement holds for all $m < M$ for some integer $M \ge 1$.
	As $\sfC_m$ is a classical register, $\rho$ is a convex combination of the form
	\begin{align}
			\rho
		\;\;=\;\;
			\sum_{c_m \in C_m} \lambda_m(c_m) \;\sigma_{c_m} \ox \ket{c_m}\bra{c_m}	\,,
	\end{align}
	where $\lambda_m \in \Prob(C_m)$ is the marginal distribution of $\rho$ on $\sfC_m$\,. 
	For $\lambda_m(t) > 0$, the density operators $\sigma_t$ are the states of the other registers conditioned on $\sfC_m$ being in the state $\ket{t}\bra{t}$: we may verify this for events $E \in \Herm(Q_1 \x \cdots \x Q_n \x C_1 \x \cdots \x C_{m-1})$ by
	\begin{align}
			\Pr_{\rho}(E|\sfC_m = t)
		\;\;=&\;\;
			\frac{\Pr_{\rho}(E,t)}{\Pr_{\rho}(t)}
		\;\;=\;\;
			\frac{\Tr\Big(\Big[E \ox \ket{t}\bra{t}\Big] \rho\Big)}{\Tr\Big(\ket{t}\bra{t} \,\tr[R](\rho)\Big)}
		\notag\\[2ex]=&\;\;
			\frac{%
				\Tr\paren{\sum\limits_{c_m} \, \lambda_m(c_m) \; E \, \sigma_{c_m} \ox \ket{t}\!\bracket{t}{c_m}\!\bra{c_m} \;}
			}{%
				\Tr\paren{\sum\limits_{\vec r} \sum\limits_{c_m} \, \lambda_m(c_m) \; \bra{\vec r} \sigma_{c_m} \ket{\vec r} \,\ox\, \ket{t}\!\bracket{t}{c_m}\!\bra{c_m} }
			}
		\notag\\[2ex]=&\;\;
			\frac{\lambda_m(t) \Tr(E \, \sigma_t)}{\lambda_m(t)}
		\;\;=\;\;
			\Tr(E \, \sigma_t)	\;;
	\end{align}
	we may let $\sigma_t$ be arbitrary for $\lambda_m(t) = 0$.
	Let $\tilde C = C_1 \x \cdots \x C_{m-1}$\,: by hypothesis, we can decompose each $\sigma_{c_m}$ as a combination
	\begin{gather}
			\sigma_{c_m}
		\;\;=\;\;
			\sum_{\tilde{\vec c} \in \tilde C} \tilde\lambda_{c_m}(\tilde{\vec c}) \; \tau_{(\tilde{\vec c},c_m)} \ox \ket{\tilde{\vec c}}\bra{\tilde{\vec c}}
	\end{gather}
	again for some marginal distribution $\tilde \lambda_{c_m} \in \Prob(\tilde{C})$ depending on $c_m \in C_m$, and for conditional states $\tau_{(\tilde{\vec c},c_m)} \in \D(Q_1 \x \cdots Q_n)$.
	As before, we may identify tuples and standard basis vectors as follows:
	\begin{subequations}
	\begin{gather}
			\vec{c} \;=\; (\tilde{\vec c}, c_m) \,\in\, C_1 \x \cdots \x C_{m-1} \x C_m \;=\; C \,,
		\\[1ex]
			\ket{\vec c} \;=\; \ket{\;\! \big.(\tilde{\vec c}, c_m) \,} = \ket{\tilde{\vec c}} \ox \ket{c_m} \;\in\; \cH[C]	\;;
	\end{gather}
	\end{subequations}
	then, if we let $\lambda(\vec c) = \lambda_m(c_m) \tilde{\lambda}_{c_m}(c_1, \ldots, c_{m-1})$, we have
	\begin{align}
			\rho
		\;\;=&
			\sum_{\substack{\tilde{\vec c} \in \tilde C \\[0.2ex] c_m \in C_m}}  \lambda_m(c_m) \tilde\lambda_{c_m}(\tilde{\vec c}) \;\;\tau_{(\tilde{\vec c},c_m)} \ox \ket{\tilde{\vec c}}\bra{\tilde{\vec c}} \ox \ket{c_m}\bra{c_m}
		\notag\\[1ex]=&\;\;
			\sum_{\vec c \in C} \;\lambda(\vec c) \;\tau_{\vec c} \ox \ket{\vec c}\bra{\vec c}	\,,
	\end{align}
	where for each $\vec c \in C_1 \x \cdots \x C_m$, we have $\tau_{\vec c} \in \D(Q_1 \x \cdots \x Q_n)$\,.
	Then $\rho$ is a convex combination of the form described by the Lemma.
\end{proof}
This allows us to ``factor'' a composite system, consisting of classical and quantum registers, into classical and quantum \emph{subsystems} which we may treat as a single (composite) quantum register, together with a single (composite) classical register; and the states of these subsystems as consisting of ``classically probabilistic'' and ``irreducibly quantum'' data.

\subsection{Measurement of quantum data}

As we noted before, quantum systems are sensitive to testing for events, which requires that we describe how they evolve under event testing.
We will define three senses of measurement, starting with the following:
\begin{definition}
	\label{def:projectiveMeas}
	For a finite set $Q$, a set of operators $\ens{\Pi_r}_{r \in R} \subset \Herm(Q)$ is a \emph{complete set of orthogonal projectors} if
	\begin{gather}
			\sum_{r \in R} \;\! \Pi_j
		\;\;=\;\;
			\idop_{Q}
		\qquad\text{~and~}\qquad
			\Pi_j \Pi_k
		\;\;=\;\;
			\Kdelta[j][k]\, \Pi_j\,,
	\end{gather}
	where
	\begin{gather}
			\Kdelta[j][k]
		\;\;=\;\;
			\begin{cases}
				1	\,,	&	\text{if $j = k$}	\\
				0	\,,	&	\text{otherwise}	
			\end{cases}
	\end{gather}
	is the Kronecker delta.
	A \emph{projective measurement} with respect to $\ens{\Pi_r}_{r \in R}$ on a quantum register $\sfQ$ over $Q$ is a mapping $\sfP: \D(Q) \to \D(Q)$ given by
	\begin{gather}
			\sfP(\rho)
		\;\;=\;\;
			\sum_{r \in R} \Pi_r\;\! \rho \,\Pi_r	\;.
	\end{gather}
	We call $\sfP$ a \emph{complete} projective measurement if each $\Pi_r$ has rank $1$.
\end{definition}
Note that for any state $\rho$ with which the projectors $\Pi_j$ all commute, we have $\sfP(\rho) = \rho$\,: that is, if there is a set of orthogonal point-mass distributions (pure states) on the system which can be used to decompose both $\rho$ and the projectors $\Pi_s$, the measurement does not affect the state of the system.
However, for any other $\rho \in \D(R)$, a projective measurement will disturb the state.

We usually suppose that measurement corresponds to a process of actually measuring a physical parameter, the result of which is recorded in the state of another system used for read-out (represented by a \emph{classical} register).
This corresponds to a sense of measurement defined by von Neumann~\cite{Neumann55}:
\begin{definition}
	\label{def:vonNeumannMeas}
	A \emph{von Neumann measurement} with respect to a complete set of orthonormal projectors $\ens{\Pi_r}_{r \in R} \subset \Herm(Q)$, on a quantum register $\sfQ$ over a finite set $Q$, is a mapping $\sfN: \D(Q) \,\to\; \D(Q) \ox \Prob(R)$ given by
	\begin{align}
			\sfN(\rho)
		\;\;=&\;\;
			\sum_{r \in R}
				\Big(\Pi_r \ox \ket{r}\Big) \rho \Big(\Pi_r \ox \bra{r}\Big)
		\;\;=\;\;
			\sum_{r \in R}
				\, \Pi_r \:\!\rho\, \Pi_r \,\ox\, \ket{r}\bra{r}	\;.
	\end{align}
	We call $\sfN$ a \emph{complete} von Neumann measurement if each $\Pi_r$ has rank $1$.
\end{definition}
The output space consists of joint distributions across the original quantum system $\sfQ$ and a classical \emph{result register} $\sfR$, whose marginal distribution may be interpreted as simply a probability distribution.

A special case of a von Neumann measurement which it will be often useful to discuss is a measurement \emph{with respect to an observable}.
In quantum mechanics, Hermitian operators $A$ may correspond to physical properties of a system (such as energy or angular momentum), where the property described by $A$ has the expected value $\lambda$ for any system in a state $\rho$ such that $\Tr(A \rho) = \lambda$.
We may describing measurement of an observable $A$ with a von Neumann measurement, as follows:
\begin{definition}
	\label{def:measObservableVN}
	Let $A \in \Herm(Q)$ be an operator with spectrum given by
	\begin{gather}
		\Lambda = \ens{\lambda \in \R \,\big|\, \det(A - \lambda \idop_Q) = 0}\,.
	\end{gather}
	A von Neumann measurement \emph{of the observable $A$} on a quantum register $\sfQ$ over $Q$ is a mapping $\sfN: \D(Q) \,\to\; \D(Q) \ox \Prob(\Lambda)$ given by
	\begin{align}
			\sfN(\rho)
		\;\;=&\;\;
			\sum_{\lambda \in \Lambda}
				\, \Pi_\lambda \:\!\rho\, \Pi_\lambda \,\ox\, \ket{\lambda}\bra{\lambda}	\;,
	\end{align}
	where each operator $\Pi_\lambda$ is the projector onto the $\lambda$-eigenspace of $A$.
\end{definition}
As a mathematical concept, this can be trivially extended to measurement of any normal operator on a quantum register, but this will not be necessary for this thesis.

We will regard the state of a classical subsystem (such as a result register) to be known in practise --- that is, without an infinite regress of measurements --- even if it is not determined in advance; then, we may consider the state of the quantum registers conditioned on a particular state of the classical subsystem.
For a post-measurement state $\bar\rho$ resulting from a von Neumann measurement, we may consider post-measurement distributions $\bar\rho_R \in \D(R)$ conditioned on a particular value of the result register after measurement $t$, which for an arbitrary projector $E$ over $\cH[R]$ describing (by a conventional abuse of notation) an event $E$, is given by
\begin{align}
		\Tr(E \, \bar\rho_{\text{\raisebox{-0.2ex}{$R$}}}) 
	\;\;=\;\;
		\Pr_{\bar\rho}(E|t)
	\;\;=&\;\;
		\frac{\Pr_{\bar\rho}(E,t)}{\Pr_{\bar\rho}(t)}
	\;\;=\;\;
		\frac{\Tr\Big(\big[E \ox \ket{t}\bra{t}\big] \bar\rho\Big)}{\Tr\Big(\ket{t}\bra{t} \, \tr[R](\bar\rho)\Big)}
	\notag\\[2ex]=&\;\;
		\frac{%
			\Tr\Big(E\, \Pi_t \rho\, \Pi_t \,\ox\, \ket{t}\bra{t}\Big) 
		}{%
			\Tr\paren{\sum\limits_{r \in R} \bra{r}\Pi_t \rho \Pi_t\ket{r} \cdot \ket{t}\bra{t}}
		}
	\notag\\[2ex]=&\;\;
		\Tr\paren{ E \sqparen{\frac{\Pi_t \rho\, \Pi_t}{\Tr\paren{\Pi_t \rho\, \Pi_t}}}}	\;,
\end{align}
which entails by linearity that
\begin{gather}
		\bar\rho_{\text{\raisebox{-0.2ex}{$R$}}}
	\;\;=\;\;
		\frac{\Pi_t \rho\, \Pi_t}{\Tr\paren{\Pi_t \rho\, \Pi_t}}
\end{gather}
conditioned on the result register being in the state $t$.

A final variety of measurement which we consider is one in which the system being measured is in a sense destroyed by the measurement, as in the case of photons: this occurs \eg\ in the detection of photons.
In this case, a measurement is performed and a result obtained, but no conditional state of the original system meaningfully exists.
We may describe such a measurement by:
\begin{definition}
	\label{def:destructiveMeas}
	Let $\ens{\big.\ket{\psi_j}\big.}_{j = 1}^n \subset \cH[Q]$ be a collection of (not necessarily normalized) vectors such that
	\begin{gather}
			\sum_{j = 1}^n \ket{\psi_j}\bra{\psi_j} \;=\; \idop_Q	\,,
	\end{gather}
	and consider a mapping $r: \ens{1, \ldots, n} \to R$.
	Then a \emph{destructive measurement}\footnote{%
	Destructive measurements as defined here correspond to \emph{positive operator-valued measurements}, or POVMs (see \eg~\cite{NC00}): we may define
	\begin{gather*}
			E_t
		\;\;=
			\sum\limits_{j\;\!:\;\!r(j) = t}	\ket{\psi_j}\bra{\psi_j}	\,,
	\end{gather*}
	from which we obtain
	\begin{gather*}
			\Tr(E_t \, \rho)
		\;=\;
			\Tr\paren{\sum\limits_{j = 1}^n \ket{\big.t} \bracket{\big.t}{\big.r(j)}\!\bra{\big.\psi_j} \rho \ket{\big.\psi_j}\bra{\big.r(j)}}
		\;=\;
			\Tr\paren{\ket{t}\bra{t} \Big.\Meas{}{}(\rho)}	\;; 
	\end{gather*}
	the primary difference between POVMs and destructive measurements as presented here is in the explicit description of the classical register retaining the measurement result.}
	on a state space $\D(Q)$ with respect to the states $\ens{\big.\ket{\psi_j}\big.}_{j = 1}^n$ and mapping $r$ is a transformation $\Meas{}{} \,:\, \D(Q) \to \E(R)$ of the form
	\begin{align}
			\Meas{}{}(\rho)
		\;=\;
			\sum_{j = 1}^n \,\ket{r(j)\big.}\bra{\psi_j\big.} \,\rho\, \ket{\big.\psi_j}\bra{\big.r(j)}	\,.
	\end{align}
	We say that $\Meas{}{}$ is an \emph{orthonormal} (destructive) measurement if the states $\ket{\psi_j}$ are orthonormal, and that it is also \emph{complete} if $r$ is injective.
\end{definition}%
We will generally be interested in the special case of complete orthonormal (destructive) measurements, in which case the vectors $\ket{\psi_j}$ form an orthonormal basis.
In particular, we may define a sense of the destructive measurement of an observable:
\begin{definition}
	\label{def:measObservableDestructive}
	Let $A \in \Herm(Q)$ be an operator with spectrum given by
	\begin{gather}
		\Lambda = \ens{\lambda \in \R \,\big|\, \det(A - \lambda \idop_Q) = 0}\,,
	\end{gather}
	and for each $\lambda \in \Lambda$, let $\ens{\big.\ket{\psi_{\lambda, j}}\big.}_{j = 1}^{m_\lambda}$ be an orthogonal basis of the $\lambda$-eigenspace of $A$.
	Then a destructive measurement \emph{of the observable $A$} on a quantum register $\sfQ$ over $Q$ is a mapping $\Meas{}{}: \D(Q) \,\to\; \Prob(\Lambda)$ given by
	\begin{align}
			\Meas{}{}(\rho)
		\;\;=&\;\;
			\sum_{\lambda \in \Lambda} \sum_{j = 1}^{m_\lambda}
				\, \ket{\lambda}\bra{\psi_{\lambda,j}} \:\!\rho\, \ket{\psi_{\lambda,j}} \bra{\lambda}
		\;\;=\;\;
			\big(\tr[Q] \,\circ\, \sfN\,\big)(\rho)	\;,
	\end{align}
	where $\sfN: \D(Q) \to\, \D(Q) \ox \Prob(\Lambda)$ is a von Neumann measurement of the observable $A$, and $\tr[Q]$ is a trace-out operation on $Q$.
\end{definition}

Orthonormal destructive measurements and projective measurements can both can be recovered from von Neumann measurements: projective measurements represent the marginal distribution of the system which is measured, and projective destructive measurements represent the marginal distribution of the result register.

\subsection{Quantum state vectors and qubits}

Although it will often be important to describe quantum computation in terms of the more general mathematical apparatus described so far, we now provide some specialized definitions and notation which will help to streamline discussion in many cases.

\subsubsection{State vectors for pure states}

Recall the definition of \emph{pure state} on page~\pageref{discn:pureState} as a point-mass distribution over a maximally refined set of events.
As we described on page~\pageref{eqn:rankOneOrthProjectors}, a point-mass function corresponds to a choice of one projector $\ket{\psi_j}\bra{\psi_j}$, which is characterized by vector $\ket{\psi_j}$ out of an orthonormal basis.
In those cases when we restrict our attention to pure states, we may simplify our analysis by considering the following in place of density operators:
\begin{definition}
	For a quantum register $\sfR$ system in a pure state $\rho \in \D(R)$, the \emph{(pure) state vector} of $\sfR$ is a unit $\ell_2$-norm vector $\ket{\psi} \in \cH[R]$ such that $\rho = \ket{\psi}\bra{\psi}$.
\end{definition}
Note that state vectors are defined only up to scalar factors, as we have $\big[\;\! \omega \ket{\psi} \:\!\big] \big[\;\! \omega\conj \!\bra{\psi} \:\!\big] = \ket{\psi}\bra{\psi}$ for any vector $\ket{\psi}$ and scalar factor $\omega \in \C$ with $\abs{\omega} = 1$.
As well, we may more generally represent some pure state $\rho$ by any non-zero vector $\ket{\psi}$ by first ``normalizing'' $\ket{\psi}$ to obtain a unit (pure state) vector for $\rho$, as follows:
\begin{gather}
		\rho
	\;\;=\;\;
		\sqparen{\sfrac{\ket{\psi}}{\,\norm{\,\ket{\psi}\,}_2}}
		\sqparen{\sfrac{\bra{\psi}}{\,\norm{\,\ket{\psi}\,}_2}}
	\;\;=\;\;
		\sfrac{\ket{\psi}\bra{\psi}}{\bracket{\psi}{\psi}}	\,.
\end{gather}
Thus, we will primarily be interested in proportionality rather than equality when treating pure-state vectors.

\subsubsection{Qubits and qubit state vectors}

We will be primarily interested in a specific type of quantum register, which defines the units of quantum data that we will manipulate:
\begin{definition}
	A \emph{qubit} is a quantum register over the set $\ens{0,1}$\,: we may write $\D(2)$ for the state-space of a qubit, and $\cH = \Span_\C\big\{ \ket{0}, \ket{1} \big\}$ for the Hilbert space on which the operators of $\D(2)$ are supported.
	(We will occasionally write $\D(2^n) = \D(2 \x \cdots \x 2)$ for the state space of $n$ generic qubits.)
	By analogy, we consider a \emph{classical bit} (considered as a physical system) to be a classical register over $\ens{0,1}$.
\end{definition}

The emphasis that will be placed on bits and qubits in this thesis motivate some simplifying notations and conventions:
\begin{notation}
	We will tend to represent qubits by lower-case roman letters.
	We will denote the state space of a qubit $v$ by $\D(v)$\,, the Hilbert space on which it is supported by $\cH[v]$\,, and the identity operator on $\cH[v]$ by $\idop_v$\,, to distinguish these from spaces/operators corresponding to different qubits.
\end{notation}
\begin{notation}
	In a mild abuse of notation, the joint state spaces of multiple qubits $u,v,\ldots$ will be denoted $\D(u, v, \ldots)$, the Hilbert space which supports it by $\cH^{\ox \ens{u,v,\ldots}}$, and the identity operator of that Hilbert space variously as $\idop_2^{\ox \ens{u,v,\ldots}}$ and $\idop_{u,v,\ldots}$\,.
	(We will also tend to represent the identity superoperator $\id_{\D(u,v,\ldots)}$ acting on $\D(u,v,\ldots)$ by the latter two symbols.)
\end{notation}
We will also extend the above conventions to state-spaces and operations on bits, although these will be less common in practise.

\begin{notation}
	We will define the following alternative and short-hand notations for various qubit state vectors:
	\begin{align}
		\begin{split}
				\ket{\splus \sfz} \;=&\; \ket{0}\Big.,
			\\[1ex]
				\ket{\sminus \sfz} \;=&\; \ket{1}\Big.,
		\end{split}
			&\quad\;
		\begin{split}
				\ket{\splus \sfx} \;=&\; \tfrac{1}{\sqrt 2}\Big[ \ket{0} + \ket{1} \Big],
			\\[1ex]
				\ket{\sminus \sfx} \;=&\; \tfrac{1}{\sqrt 2}\Big[ \ket{0} - \ket{1} \Big],
		\end{split}
			&\quad\;
		\begin{split}
				\ket{\splus \sfy} \;=&\; \tfrac{1}{\sqrt 2}\Big[ \ket{0} + i \ket{1} \Big],
			\\[1ex]
				\ket{\sminus \sfy} \;=&\; \tfrac{1}{\sqrt 2}\Big[ \ket{0} - i \ket{1} \Big].	
		\end{split}
	\end{align}
	We often further abbreviate $\ket{\splus} = \ket{\splus \sfx}$ and $\ket{\sminus} = \ket{\sminus \sfx}$.
	We may also write $\ket{\pm}$ or $\ket{\spm \sfx}$ to denote one or both of the state vectors $\ket{\splus \sfx}$ or $\ket{\sminus \sfx}$, and similarly for $\ket{\spm \sfy}$ and $\ket{\spm \sfz}$.
\end{notation}

\section{Unitary evolution and elementary gate sets}
\label{sec:unitaryEvolution}

We have already partially described transformations that can be performed on quantum data: by CPTP maps in general, and by measurement operations (projective, von Neumann, and destructive) as a special case.
We now describe the theory by which general CPTP maps may be decomposed into (or approximated by) the product of discrete elementary operations, which are supposed to be physically implementable in principle.
At the same time, we present the theoretical importance of \emph{unitary transformation} of quantum data, which will motivate the questions explored in this thesis.

\subsection{Unitary evolution of quantum states}

Evolution of pure states (described by state-vectors $\ket{\psi_t} \in \cH[Q]$) in time are governed by the Schr\"odinger equation~\cite{Schrodinger26,Neumann55},
\begin{gather}
		i \hbar \sfrac{\d}{\d t}\ket{\psi_t}
	\;\;=\;\;
		H_t \ket{\psi_t}	\,,
\end{gather}
where $H_t$ is the \emph{Hamiltonian} of the system, a linear operator on $\cH[Q]$ which may depend on time, whose eigenvalues are possible energy levels of pure states of the system.
Energy is represented here by real numbers: abstracting away the physical content of this equation, $H_t$ is therefore a Hermitian operator.
Then, we have
\begin{align}
		\sfrac{\d}{\d t} \bracket{\psi_t}{\psi_t}
	\;\;=&\;\;
		\sqparen{\sfrac{\d}{\d t} \bra{\psi_t}} \ket{\psi_t}
		\;+\;
		\bra{\psi_t} \sqparen{\sfrac{\d}{\d t} \ket{\psi_t}}
	\notag\\[1ex]=&\;\;
		\sqparen{ \sfrac{i}{\hbar} \bra{\psi_t} H_t\herm } \ket{\psi_t}
		\;+\;
		\bra{\psi_t} \sqparen{ \sfrac{-i}{\hbar} H_t \ket{\psi_t}}
	\;\;=\;\;
		0\;;
\end{align}
that is, Schr\"odinger evolution preserves the $\ell_2$-norm of vectors (\ie\ it is well-defined as an evolution of pure state vectors).
The evolution of an input state $\ket{\psi_0}$ from $t = 0$ to some $t = \tau$ can then be described by some operator $U_\tau$ on the pure states of the system which preserves the $\ell_2$-norm.
For a Hilbert space $\cH[Q]$ on a finite set $Q$, we may consider the action of $U_\tau$ on the standard basis of $\cH[Q]$\,: 
we have $\bra{q} U_\tau\herm U_\tau \ket{q} = 1$ by the preservation of norm of the states $\ket{q}$ for $q \in Q$; and for any combination $\ket{\phi} = \frac{1}{\sqrt 2} \ket{p} + \frac{1}{\sqrt 2} \e^{i \theta} \ket{q}$ for distinct $p,q \in Q$ and arbitrary angles $\theta \in \R$, we have
\begin{align}
		1
	\;\;=&\;\;
		\bra{\phi} U_\tau\herm U_\tau \ket{\phi}
	\notag\\=&\;\;
		\tfrac{1}{2}\sqparen{\,\bra{p}U_\tau\herm U_\tau \ket{p} \,+\, \e^{i \theta} \bra{p} U_\tau\herm U_\tau \ket{q}
				\,+\, \e^{-i \theta} \bra{q}U_\tau\herm U_\tau \ket{p} \,+\, \bra{q} U_\tau\herm U_\tau \ket{q}\,}
	\notag\\=&\;\;
		1 \;+\; \Re\Big(\e^{i \theta} \bra{p}U_\tau\herm U_\tau \ket{q}\Big)	\;.
\end{align}
Then $\e^{i \theta} \bra{p}U_\tau\herm U_\tau \ket{q}$ has no real component for any angle $\theta$, which implies $\bra{p}U_\tau\herm U_\tau\ket{q} = 0$ for distinct $p,q \in Q$\,.
We therefore have $U_\tau\herm U_\tau = \idop_Q$\,: that is, $U_\tau$ is a \emph{unitary operation}.
Applied to the evolution of density operators, the time-evolution $\Upphi_\tau$ of density operators corresponding to pure states from time $t = 0$ to $t = \tau$ is then
\begin{subequations}
\begin{align}
		\Upphi_\tau\big(\ket{\psi_0}\bra{\psi_0}\big)
	\;\;=\;\;
		\ket{\psi_\tau} \bra{\psi_\tau}
	\;\;=&\;\;
		U_\tau \ket{\psi_0}\bra{\psi_0} U_\tau\herm\,,
\end{align}
which by linearity implies that
\begin{gather}
	\label{eqn:unitaryCPTP}
		\Upphi_\tau(\rho)
	\;\;=\;\;
		U_\tau \,\rho\: U_\tau\herm
\end{gather}
for all $\rho \in \D(Q)$, which is a valid CPTP map.
We say that a CPTP map is unitary if it can be expressed as in~\eqref{eqn:unitaryCPTP}; and we denote the set of unitary operators on $\cH[S]$ by $\U(S)$.
\end{subequations}

Precisely how \emph{measurement processes} (as described by CPTP maps of the kind described in Definitions~\ref{def:projectiveMeas} through~\ref{def:measObservableDestructive}, on pages~\pageref{def:projectiveMeas}\thru\pageref{def:measObservableDestructive}) might arise from unitary evolution, represent an explicit departure from unitary evolution, or represent something entirely subjective, is a subject of ongoing research and debate (see \eg~\cite{BCFFS00,GRW86,Popper82,Wallace02,Zurek03}).
Such considerations are beyond the scope of this thesis: in practise, we adopt the approach of von Neumann~\cite{Neumann55}, and treat unitary evolution and measurements as complementary ways in which states may be transformed; and we supplement this also with maps of the form $\rho \mapsto \rho \ox \sigma$ for density operators $\rho$ and $\sigma$, corresponding to augmenting the system by introducing an auxiliary system in the state $\sigma$.

\paragraph{A remark on continuous time evolution.}

As we have noted above, Schr\"odinger evolution of physical systems can be described as being continuous in time.
Despite this, we will generally describe computation as being performed by \emph{discrete} transformations, \eg\ by applying unitary transition operators, rather than Schr\"odinger evolution.
The reasons for this are practical in nature: because we can only perform operations with finite precision in any particular setting, and because arbitrarily high precision can only be achieved with a commensurate decrease in the speed of computation, we limit ourselves to models of quantum computation where (like measurement and the introduction of auxiliary systems) the evolution of a quantum system is performed in discrete steps.
As well, the same unitary operation can be realized in different ways by Schr\"odinger evolution: for instance, considering time-independent Hamiltonians alone, a unitary $U_\tau$ can be achieved in time $\tau$ by Schr\"odinger evolution by any Hamiltonian $H$ satisfying $U_\tau = \exp(- i \hbar\, H \tau)$, where
\begin{gather}
		\exp(A)
	\;\;=\;\;
		\sum_{n \in \N} \sfrac{A^n}{n!}
\end{gather}
for any Hermitian matrix $A$; the set of such Hamiltonians is an equivalence class of operators with common eigenvectors, and whose eigenvalues differ by integer multiples of $2\pi$.
By considering discrete-time unitary evolution, we abstract away details such as the particular Hamiltonian, which in an implementation may be determined later.

\subsection{Elementary sets of gates}

To perform computations, we suppose that we are provided a set of elementary operations (or \emph{elementary gates}) as follows:
\begin{definition}
	\label{def:elemGatesTypes}
	A set $\Gates$ of elementary gates on a set of classical registers $B$ and a set of quantum registers $V$ is a set of operators $\Op{\sigma:\alpha/\delta}{}$\,, each of which consists of some CPTP map
	\begin{gather}
		\Op{}{} \;:\; \D\big(\ens{0,1}^{n+m}\big) \;\to\; \D\big(\ens{0,1}^{n+\ell}\big)	\;,
	\end{gather}
	which operates on finite sequences of registers given by $\sigma$, $\alpha$, and $\delta$ (where the input and output distributions of $\Op{}{}$ are also consistent with the additional constraints imposed by the state space for bits as classical registers).
	In particular, $\alpha = (a_j)_{j = 1}^\ell$ is a sequence of \emph{allocated} registers which are produced as output but which were not present as input; $\delta = (d_j)_{j = 1}^m$ is a sequence of \emph{discarded} registers which are taken as input but not produced as output; and $\sigma = (s_j)_{j = 1}^n$ is a sequence \emph{stable} registers taken as input and produced as output, yielding an operation
	\begin{gather}
		\Op{\sigma:\alpha/\delta}{} \;:\; \D\big(\sigma;\delta\big) \;\to\; \D\big(\sigma;\alpha\big)	\;.
	\end{gather}
	We say that the \emph{type} of the operation is $[\sigma:\alpha/\delta]$.
\end{definition}
The sequence of registers for an elementary operation, once fixed, is significant.
This may be illustrated with a map $\Meas{a,b:r/\vide}{}$ performing a von Neumann measurement on one of two qubits $a$ and $b$, producing a single result bit $r$, defined by
\begin{gather}
		\Meas{a,b:r/\vide}{}(\rho)
	\;\;=\;\;
		\sum_{\beta \in \ens{0,1}}
			\Big(\idop_a \ox \ket{\beta}\bra{\beta}_b\Big) \,\rho\, \Big(\idop_a \ox \ket{\beta}\bra{\beta}_b\Big)  \ox \ket{\beta}\bra{\beta}_r \,;
\end{gather}
then, for \eg\ two qubits $u$ and $v$ in independent pure states, $\Meas{u,v:r/\vide}{}$ performs a measurement on $v$ (producing the result in $r$) while leaving the state of $u$ undisturbed, while $\Meas{v,u:r/\vide}{}$ does the opposite.

In order for operations on bits and qubits --- or more generally, any register --- to be unambiguous, there must be exactly one ``copy'' of a given register which is to be operated upon; and any register to be operated on by some operation cannot have been explicitly discarded by the preceding operations without being re-allocated.
We formalize these constraints through the concept of the types defined for each gate, as follows:
\begin{figure}[t]
	\begin{center}
		\includegraphics[width=50mm,height=30mm]{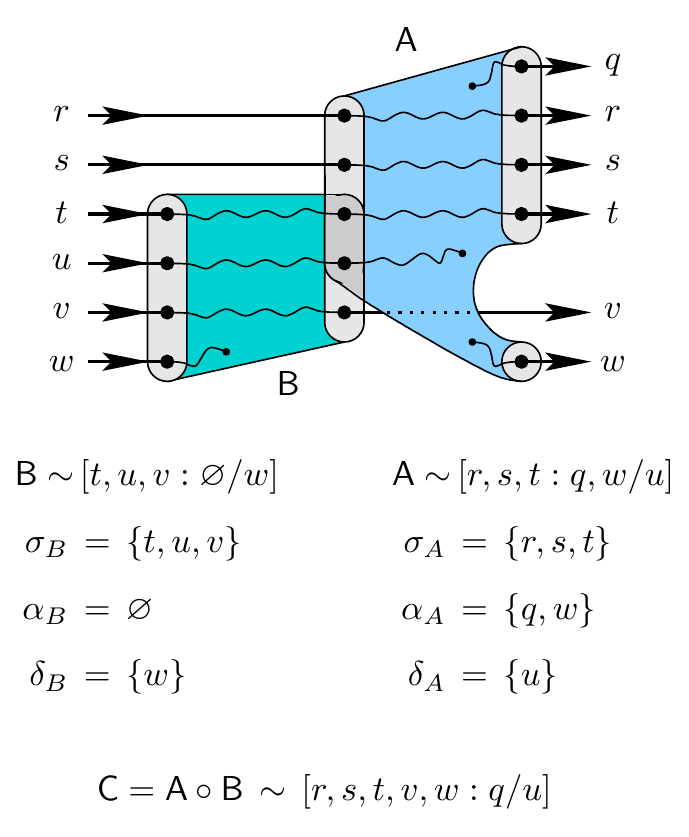}
	\end{center}
	\caption[Illustration of the type of the composition of two operations]{\label{fig:typeCompose}\colouronline%
		Illustration of the type of the composition of two operations $\mathsf A$ and $\mathsf B$ acting on named registers (represented as ``wires'' progressing in time from left to right).
		Registers not explicitly required by the input or output of $\mathsf B$ are assumed to be available to be either allocated, discarded, or transformed by $\mathsf A$ as necessary, and vice-versa.
		Otherwise, the types of $\mathsf A$ and $\mathsf B$ must agree on what bits or qubits are available to be operated on after the application of $\mathsf B$ and before the application of $\mathsf A$.
	}
\end{figure}
\begin{lemma}
	\label{lemma:typeCompose}
	For two operations $\mathsf{A}_{\sigma_A:\alpha_A/\delta_A}$ and $\mathsf{B}_{\sigma_B:\alpha_B/\delta_B}$ with types $[\sigma_A:\alpha_A/\delta_A]$ and $[\sigma_B:\alpha_B/\delta_B]$ respectively, the composition $\mathsf{A}_{\sigma_A:\alpha_A/\delta_A} \,\circ\, \mathsf{B}_{\sigma_B:\alpha_B/\delta_B}$ is well-formed if and only if
	\begin{romanum}
	\item
		there is no intersection between the registers in $\alpha_A$, and the registers in $\alpha_B$ or $\sigma_B$ which are explicitly specified in the output of $\mathsf{B}_{\sigma_B:\alpha_B/\delta_B}$\,;

	\item
		there is no intersection between the registers of $\delta_B$, and the registers in $\delta_A$ or $\sigma_A$ which are explicitly specified in the input of $\mathsf{A}_{\sigma_A:\alpha_A/\delta_A}$\,.
	\end{romanum}
	If the above conditions hold, the composition $\mathsf{C}_{\sigma_C:\alpha_C/\delta_C} = \mathsf{A}_{\sigma_A:\alpha_A/\delta_A} \,\circ\, \mathsf{B}_{\sigma_B:\alpha_B/\delta_B}$ has type $[\sigma_C:\alpha_C/\delta_C]$, where the sequences $\sigma_C$, $\alpha_C$, and $\delta_C$ are arbitrary orderings of the registers in the following ``set'' constructions:
	\begin{subequations}
	\label{eqn:typeCompose}
	\begin{align}
 			\sigma_C
		\;\;=&\;\;
			(\sigma_A \inter \sigma_B) \union (\alpha_A \inter \delta_B)	\,,
		\\[1ex]
			\alpha_C
		\;\;=&\;\;
			(\alpha_A \setminus \delta_B) \union (\alpha_B \setminus \delta_A)	\,,
		\\[1ex]
			\delta_C
		\;\;=&\;\;
			(\delta_B \setminus \alpha_A) \union (\delta_A \setminus \alpha_B)	\,.
	\end{align}
	\end{subequations}
\end{lemma}
The type composition relations of~\eqref{eqn:typeCompose} are illustrated in Figure~\ref{fig:typeCompose}.\footnote{%
The emphasis given here on the composability of CPTP maps is meant largely to prefigure similar discussions for particular, ``non-standard'' representations of quantum computation.}

A particular set $\Gates$ of elementary operations defines a specific \emph{circuit model}, consisting of the well-formed compositions of those gates:
\begin{definition}
	Given a set $\Gates$ of elementary gates on a set $B$ of bits and $V$ of qubits,
	a \emph{circuit in the model defined by $\Gates$} (or \emph{in the $\Gates$ model}) is the composition $C$ of a finite sequence of operations of $\Gates$.
	The \emph{size} of such a circuit is the number of operations it uses, and the \emph{depth} of the circuit is the minimum integer $D$ such that the circuit may be decomposed in the form,
	\begin{gather}
			C \;\;=\;\; \ordprod[>]_{t = 1}^D \big(\Op{t,1}{} \ox \Op{t,2}{} \ox \cdots \ox \Op{t,m_t}{}\big)
	\end{gather}
	\ie\ as an ordered product of $D$ parallel (tensor) products of gates on independent sets of qubits.
\end{definition}

From this point onwards, we will consider only models of quantum computation acting on bits and qubits (\ie\ where the set of elementary gates only acts on aggregates of registers over $\ens{0,1}$).
We also assume that the number of available bits and qubits is infinite, and that the only limitation on the number of qubits which a circuit may operate upon is imposed by the set of elementary gates.

\begin{notation}
	Operators $U$ written in italics represent linear operators (\eg\ unitary transformations), and superscripts denote exponents of these operators.
	Operators $\Upupsilon$ written in upright sans-serif represent CPTP maps.
	Superscripts of CPTP maps denote either real-valued parameters or parameters provided by an interactive classical control, rather than exponents.
	For two CPTP maps $\mathsf A$ and $\mathsf B$, we will often write $\mathsf A \mathsf B$ for $\mathsf A \circ \mathsf B$ when this composition is well-defined (although we will continue to write in terms of compositions for much of this section).
\end{notation}

\subsection{Universality for quantum computation/unitary evolution}
\label{sec:universal}

In order to have a robust model of computation, it is helpful not only to have a set of discrete operations with which to define a sense of computational complexity, but also to have a well-defined notion of the range of operations which that set generates.
This motivates a discussion of senses of \emph{universality} of quantum operations.

\subsubsection{Universality for generating/approximating distributions}

For quantum computation on aggregates of qubits, the natural candidate is the the ability to generate all CPTP maps $\Upphi: \D(2^n) \to \D(2^n)$ for any $n \ge 1$.
Alternatively, if we wish to consider models of computation with only a finite set of elementary operations, we cannot generate the uncountably\footnote{%
``Uncountable'' is used here in the set-theoretic sense, meaning that the cardinality of the set of such maps is greater than that of the natural numbers.}
many CPTP maps from finite compositions from a finite set of operations; however, we may define a sense of \emph{approximate} universality by defining a sense in which a CPTP map may be approximated with some precision $\epsilon > 0$.
We will present a formal definition of approximating CPTP maps below.
However, defining universality in terms of generating or approximating arbitrary CPTP maps misses important models of computation (such as some models of measurement-based quantum computation, which we present in Section~\ref{sec:oneWayModel}) which do not take quantum states as input, but which achieve what is essentially the main practical motivation of quantum computation: a means of efficiently preparing a larger class of probability distributions via measurement than is apparently possible with, say, randomized Turing machines.

The complementary alternative --- considering only probability distributions that may be prepared by a set of gates --- presents no distinction between quantum and classical computation.
While the set of functions computable by classical and by quantum computers from explicitly specified primitives are the same, it is commonly thought that the functions which can be \emph{efficiently} computed by each are not the same.
We might wish to investigate descriptions of universality for quantum computation which strictly subsumes what classical computers can achieve, if we are interested in possible indications of the differences in what can be efficiently done by classical computation and by quantum computation.
One approach could be to explicitly consider computational resources in any discussion of universality: to describe \eg\ pertinent bounds (or lack of bounds) on time, space, multiparty communication, \etc\ when describing senses of universality.
A discussion of different senses of universality is presented in~\cite{NDMB2007}, which raises the issue of addressing universality and computational complexity at the same time.
However, for the sake of simplicity, it is more convenient to completely separate concerns of efficiency and universality.\footnote{%
The rarity of such senses of universality in the literature is likely due, no doubt, to the success of the Strong Church-Turing thesis, which posits that any ``physically realizable'' model of computation is polynomial-time equivalent to the Turing machine model of computation.
Even in light of the apparent failure of the Strong Church-Turing thesis with the advent of quantum computation, there seems to be remarkably few natural senses of universal computation arising out of ``physically realistic'' models of computation which are not provably poly-time equivalent, except for those which are equivalent to constraining another ``physically realistic'' model in some way.}

A reasonable compromise position is to define a sense of universality which strictly extends the sense of generating probability distributions, but which does not presume that arbitrary quantum inputs are possible inputs to the operations of the model --- that is, simply the generation or approximation of arbitrary density operators.
The natural distance on probability distributions is that induced by the $\ell_1$-norm, which --- using the correspondence of~\eqref{eqn:embedHilbertSchmidt} --- is the sum of the absolute values of the entries of a diagonal matrix.
By~\cite[Lemma~10.2]{KSV2002}, the \emph{operator $\ell_1$-norm} (also called the trace norm) on linear operators $M$ between finite-dimensional Hilbert spaces,
\begin{subequations}
\begin{align}
		\norm{M}_1
	\;\;=&\;\;
		\sum_{j} \,\sigma_j	\;,
		&&\text{\small where $\sigma_1 \ge \sigma_2 \ge \ldots \ge 0$ are the singular values of $M$},
	\\=&\;\;
		\Tr\paren{\sqrt{M\herm M\,}}	\;,
		&&\text{\small for $\sqrt{M\herm M\,} = S$ such that $S \ge 0$ and $S^2 = M\herm M$};
\end{align}
and the corresponding metric on density operators,
\begin{gather}
		d(\rho,\sigma)
	\;\;=\;\;
		\norm{\rho - \sigma}_1
\end{gather}
\end{subequations}
are the natural generalization to arbitrary density operators.
Then, we define a sense of universality for quantum computation as follows:
\begin{definition}
	\label{def:universalQC}
	For two density operators $\rho, \tilde\rho$, we say that \emph{$\tilde\rho$ approximates $\rho$ with precision~$\epsilon$} if $0 \le \norm{\tilde\rho - \rho}_1 < \epsilon$.
	A collection of elementary gates $\Gates$ is \emph{universal for quantum computation} if any density operator $\rho \in \D(2^n)$ on $n$ qubits can be generated as a circuit in the $\Gates$ model, for every $n \ge 1$; alternatively, $\Gates$ is \emph{approximately universal for quantum computation} if any such operator $\rho$ can be approximated to arbitrary precision $\epsilon > 0$ by some circuit in the $\Gates$ model.
\end{definition}
This definition of universality for quantum computation corresponds to the definition of ``\textbf{CQ}-universality'' presented in~\cite{NDMB2007}.

\subsubsection{Universality for generating/approximating CPTP maps}

Although generation or approximation of density operators is what we have chosen to define universality, it is certainly sufficient to be able to generate or approximate all CPTP maps on $n$ qubits (corresponding to ``\textbf{QQ}-universality'' as described in~\cite{NDMB2007}).
The ultimate significance of a CPTP map from the point of view of quantum computation is in the probability distributions arising from measurement.
A natural definition of an approximation $\tilde\Upphi$ of a CPTP map $\Upphi$ is then when the result of a measurement (\eg\ a destructive measurement as in Definition~\ref{def:destructiveMeas}) on either $\tilde\Upphi(\rho)$ or $\Upphi(\rho)$ yields similar probability distributions for any input state $\rho$; more generally, we may consider measurements performed on $\big[ \tilde\Upphi \ox \idop \big](\rho)$ and $\big[ \Upphi \ox \idop \big](\rho)$, where $\rho$ is a state of some larger system.
Then, considering upper bounds on the difference of the effects of two maps $\tilde\Upphi : \D(Q) \to \D(Q')$ and $\Upphi : \D(Q) \to \D(Q')$ on arbitrary density operators, we are led to the following metric on superoperators:
\begin{align}
		\Delta\big(\tilde\Upphi, \Upphi\big)
	\;\;=&\;\;
		\sup_{\card{A} \ge 1}\;\;\;
		\sup_{\rho \in \D(Q \x A)}
		\quad
		d\paren{\Big(\tilde\Upphi \ox \idop_A\Big)(\rho)\,,\; \Big(\Upphi \ox \idop_A\Big)(\rho)}
	\notag\\[1.5ex]=&\;\;
		\sup_{\card{A} \ge 1}\;\;
		\sup_{\substack{M \in \Herm(Q \x A) \\ M \ge 0,\, M \ne 0}}
		\frac{\norm{\Big(\tilde\Upphi \ox \idop_A\Big)(M) - \Big(\Upphi \ox \idop_A\Big)(M)}_1}{\norm{M}_1}	\;.
\end{align}
We may simplify this definition slightly by removing the emphasis on positive operators, and defining the \emph{superoperator norm} (introduced as the \emph{diamond norm} in~\cite{AKN98}), in the fashion of~\cite{KSV2002}:\footnote{%
This definition actually corresponds to the description of $\norm{\Uppsi}_\lozenge$ on page~109 and to the statement of Theorem~11.1 on page~110, in~\cite{KSV2002}.
The definition presented here also differs in that the descriptions in~\cite{KSV2002} do not involve a supremum over $\card{A}$.}
\begin{gather}
		\norm{\Uppsi}_\lozenge
 	\;\;=\;\;
 		\sup_{\card{A} \ge 1}\;\;
 		\sup_{\substack{M \in \L(\cH[Q] \ox \cH[A]) \\ M \ne 0}}
 		\frac{\norm{\Big(\Uppsi \ox \idop_A\Big)(M)}_1}{\norm{M}_1}	\;.		
\end{gather}
It is easy to see that the value of the inner supremum is non-decreasing with the size of $\card A$, as we may isometrically embed $\cH[Q] \ox \cH[A]$ into $\cH[Q] \ox \cH[A']$ for any $\card{A'} \ge \card{A}$.
For the outer supremum, it suffices to take $\card{A} = \card{Q}$, as shown in~\cite[Theorem~11.1]{KSV2002}; then, we may equivalently define
\begin{gather}
		\norm{\Uppsi}_\lozenge
 	\;\;=\;\;
 		\sup_{\substack{M \in \L(\cH[Q] \ox \cH[Q]) \\ M \ne 0}}
 		\frac{\norm{\Big(\Uppsi \ox \idop_Q\Big)(M)}_1}{\norm{M}_1}	\;.		
\end{gather}
Therefore, we present the following extension of Definition~\ref{def:universalQC} for the purposes of describing the generation or approximation of CPTP maps:
\begin{definition}
	\label{def:universalCPTP}
	For two CPTP maps $\Upphi, \tilde\Upphi$ with the same domain and range, we say that \emph{$\tilde\Upphi$ performs $\Upphi$} if $\tilde\Upphi = \Upphi$\,, and that \emph{$\tilde\Upphi$ approximates $\Upphi$ with precision~$\epsilon$} if \mbox{$0 \le \norm{\big.\smash{\tilde\Upphi - \Upphi}\big.}_\lozenge < \epsilon$}.
	A collection of elementary gates $\Gates$ is \emph{universal for CPTP transformation} if any CPTP map $\Upphi: \D(2^n) \to \D(2^m)$ from $n$ qubits to $m$ qubits can be performed by a circuit in the $\Gates$ model, for every $n,m \ge 1$; alternatively, $\Gates$ is \emph{approximately universal for CPTP transformation} if any such map $\Upphi$ can be approximated to arbitrary precision $\epsilon > 0$ by some circuit in the $\Gates$ model.
\end{definition}
Because a density operator $\rho \in \D(Q)$ is equivalent to a CPTP map $\D(1) \to \D(Q)$, (approximate) universality for CPTP transformation entails (approximate) universality for quantum computation.

\paragraph{Remark on approximation of composite CPTP maps.}
We can approximate any composition of CPTP maps $\Uppsi_N \circ \cdots \circ \Uppsi_2 \circ \Uppsi_1$ by using approximations to each of the maps $\Uppsi$, with an imprecision which is at worst the sum of the imprecisions of each of the components: that is, imprecision in approximations compose \emph{additively}, as shown in~\cite{AKN98}.
We may show this by first noting that for superoperators $\Uppsi: \L(\cH[Q]) \to \L(\cH[Q'])$ and $\Upphi: \L(\cH[Q']) \to \L(\cH[Q''])$, we have
\begin{align}
	\label{eqn:superoperatorNormSubMultiplic}
		\smaller{\norm{\Upphi \circ \Uppsi}_\lozenge}
		\;\;\smaller=&\quad
		\smaller{
	 		\sup_{\substack{M \in \L(\cH[Q] \ox \cH[Q]) \\ M \ne 0}}
 			\frac{\norm{\Big((\Upphi \circ \Uppsi) \ox \idop_Q\Big)(M)}_1}{\norm{M}_1}
		}
	\notag\\[1.8ex]\smaller=&\quad
		\smaller{
	 		\sup_{\substack{M \in \L(\cH[Q] \ox \cH[Q]) \\ M \ne 0}}
			\;\;
 			\sqparen{
				\frac{\Bigg.\norm{\bigg(\Upphi \ox \idop_Q\bigg)\Big[\Big(\Uppsi \ox \idop_Q\Big)(M)\Big]}_1}{\norm{\Big(\Uppsi \ox \idop_Q\Big)(M)}_1}
			}	
 			\sqparen{
				\frac{\Bigg.\norm{\Big(\Uppsi \ox \idop_Q\Big)(M)}_1}{\norm{M}_1}
			}
		}
	\notag\\[1.8ex]\smaller\le&\;\;
		\smaller{
	 		\sqparen{
				\sup_{\substack{M' \in \L(\cH[Q'] \ox \cH[Q']) \\ M \ne 0}}
 				\frac{\bigg.\norm{\Big(\Upphi \ox \idop_{Q'}\Big)(M')}_1}{\norm{M'}_1}
			}	
 			\sqparen{
				\sup_{\substack{M \in \L(\cH[Q] \ox \cH[Q]) \\ M \ne 0}}
				\frac{\bigg.\norm{\Big(\Uppsi \ox \idop_Q\Big)(M)}_1}{\norm{M}_1}
			}
		}
	\notag\\[1.8ex]\smaller=&\quad
		\smaller{
			\norm{\Upphi}_\lozenge \norm{\Uppsi}_\lozenge	\;,
		}
\end{align}
where we implicitly take the supremum over all auxiliary spaces for the second-last inequality.
It is possible to show that any CPTP map has unit superoperator norm:\footnote{%
Using the definition on page of $\norm{\Uppsi}_\lozenge$ in~\cite[page~110]{KSV2002}, and Theorem~11.1 which follows shortly after, this follows from Theorem~\ref{thm:churchHilbertSpace} below.
The fact that CPTP maps have \emph{at most} unit superoperator norm may alternatively be obtained from Theorem~\ref{thm:churchHilbertSpace} and the sub-multiplicative property shown above, together with the fact that $\norm{\Upphi \ox \Uppsi}_\lozenge = \norm{\Upphi}_\lozenge \norm{\Uppsi}_\lozenge$\,, and from the fact that the trace operation and isometries each have unit superoperator norm.}
then, if $\tilde\Uppsi: \L(Q) \to \L(Q')$ is a CPTP map approximating $\Uppsi$ with precision $\epsilon$ and $\tilde\Upphi: L(Q') \to \L(Q'')$ is a CPTP map approximating $\Upphi$ with precision $\epsilon'$, we have
\begin{align}
	\label{eqn:errorAdditiveCPTP}
		\norm{\tilde\Upphi \circ \tilde\Uppsi - \Upphi \circ \Uppsi}_\lozenge
	\;\;\le&\;\;
		\norm{\tilde\Upphi \circ \big(\tilde\Uppsi - \Uppsi\big)}_\lozenge \;+\;
		\norm{\big(\tilde\Upphi - \Upphi\big) \circ \Uppsi}_\lozenge
	\notag\\[1ex]\le&\;\;
		\norm{\tilde\Upphi}_\lozenge \norm{\tilde\Uppsi - \Uppsi}_\lozenge \;+\;
		\norm{\tilde\Upphi - \Upphi}_\lozenge \norm{\Uppsi\Big.}_\lozenge
	\;\;<\;\;
		\epsilon + \epsilon'\;.
\end{align}
Thus, to approximate a composite CPTP map within a small margin of error, it suffices to approximate some set of gates which is universal for CPTP transformation, with sufficiently good precision to bound the cumulative error.

\subsubsection{Universality for generating/approximating unitary transformations}

One of the primary tools of quantum computation is to reduce this sense of universality for quantum computation to the ability to perform/approximate unitary transformations in particular.
This is possible due to the following corollary (see \eg~\cite[Lecture~5]{Watrous07} and~\cite{KSW2008} for explicit treatments) to a theorem by Stinespring~\cite{Stinespring1955}:
\begin{theorem}
	\label{thm:churchHilbertSpace}
	For Hilbert spaces $\cH[]$ and $\cK$ of finite dimension, a superoperator $\Upphi: L(\cH[]) \to \L(\cK)$ is a CPTP map if and only if, for some auxiliary Hilbert space $\cE$ and some isometry $T: \cH[] \to \cK \ox \cE$, we have
	\begin{gather}
	\label{eqn:churchHilbertSpace}
			\Upphi(\rho)
	 	\;\;=\;\;
			\tr[\cE](T \rho \:T\herm)\;.
	\end{gather}
\end{theorem}
If we define $\sfT(\rho) = T \rho \,T\herm$, this is a decomposition of $\Upphi$ into an isometry and a trace-out operation.
When $\cH[]$, $\cE$, and $\cK$ as above are restricted to tensor products of $\cH$, the isometry $T: \cH[] \to \cK \ox \cE$ can be decomposed into the preparation of an arbitrary state $\ket{\phi} \in \cA \cong \cH^{\ox m}$ (for some number $m$ of auxiliary qubits such that $\cH[] \ox \cA \cong \cK \ox \cE$), and a unitary transformation $U: \cH[] \ox \cA \to \cK \ox \cE$\,:
\begin{gather}
		T
	\;\;=\;\;
		U \big[\idop_{\cH[]} \ox \ket{\phi}_{\cA}\big]	\;.
\end{gather}
Then, the decomposition of a CPTP map $\Upphi$ on qubits as in~\eqref{eqn:churchHilbertSpace} can be further elaborated as being of the form
\begin{subequations}
\begin{gather}
			\Upphi(\rho)
	 	\;\;=\;\;
			\tr[\cE]\Big(U \big[\rho \ox \ket{\phi}\bra{\phi} \big] U\herm\Big)
		\;\;=\;\;
			\Big(\tr[\cE] \circ \Upupsilon \circ \mathsf A\,\Big)(\rho)	\;,
	\\[1ex]
		\text{where}\quad	\Upupsilon(\rho) \;=\; U \rho U\herm,	\qquad \mathsf A(\rho) = \rho \ox \ket{\phi}\bra{\phi}	.
\end{gather}
\end{subequations}
The characterization of CPTP maps by such decompositions, is an additional motivation (apart from the physical motivation of Schr\"odinger evolution) to analyze quantum computation in terms of unitary transformations, even if we are interested in CPTP maps in general.

In order to obtain a similar sense of universality as for CPTP maps, it will be useful to discuss a norm on operators analogous to the superoperator norm described above.
We define the \emph{operator sup-norm} of an operator $M: \cH[Q] \to \cH[Q']$ by
\begin{gather}
		\norm{M}_\infty
	\;\;=\;\;
		\sup_{\substack{\vec v \in \cH[Q] \\ \vec v \ne \vec 0}}
		\frac{\norm{M \vec v}_2}{\norm{\vec v}_2}	\;;
\end{gather}
it is easy to verify that $\norm{M}_\infty$ is the largest singular value of $M$.
By a similar derivation as in~\eqref{eqn:superoperatorNormSubMultiplic}, it is easy to prove that
\begin{subequations}
\begin{align}
		\norm{L M}_\infty
	\;\;\le&\;\;
		\norm{L}_\infty \norm{M}_\infty	\;;
\intertext{it is also easy to verify that the following properties hold:}
		\norm{M}_\infty
	\;\;=&\;\;
		1
		\;\;\;\text{for $U$ a unitary embedding},
	\\[0.3ex]
		\norm{M}_\infty
	\;\;=&\;\;
		\norm{M}_1
		\;\;\text{for $M$ with rank $1$},		
	\\[0.5ex]
		\norm{\big.\smash{M\herm}\big.}_\infty
	\;\;=&\;\;
		\norm{\big.M}_\infty	\;,
	\\[0.5ex]
		\norm{L \ox M}_\infty
	\;\;=&\;\;
		\norm{L}_\infty \norm{M}_\infty	\;.
\end{align}
\end{subequations}
Then for operators $U, \tilde U \in \U(N)$ and $\ket{\psi} \in \cH[N] \ox \cH[M]$ for positive integers $N$ and $M$, we have
\begin{align}
	\lefteqn{
		\smaller{
			\norm{
				(\tilde U \ox \idop_M) \ket{\psi}\bra{\psi} (\tilde U\herm \ox \idop_M)
				\,-\,
				(U \ox \idop_M) \ket{\psi}\bra{\psi} (U\herm \ox \idop_M)
			}_1
		}
	}\qquad\qquad
	\notag\\\smaller\le&\quad
 		\smaller{
			\norm{	(\tilde U \ox \idop_M) \ket{\psi}\bra{\psi} \Big(\big[\tilde U\herm - U\herm\big] \ox \idop_M\Big)	}_1
		\;\;+\;\;\;
			\norm{	\Big(\big[\tilde U - U\big] \ox \idop_M \Big) \ket{\psi}\bra{\psi} (U\herm \ox \idop_M)	}_1
 		}
	\notag\\\smaller=&\quad
 		\smaller{
			\norm{	(\tilde U \ox \idop_M) \ket{\psi}\bra{\psi} \Big(\big[\tilde U\herm - U\herm\big] \ox \idop_M\Big)	}_\infty
		\;\;+\;\;\;
			\norm{	\Big(\big[\tilde U - U\big] \ox \idop_M \Big) \ket{\psi}\bra{\psi} (U\herm \ox \idop_M)	}_\infty
 		}
	\notag\\\smaller\le&\quad
		\smaller{
			\norm{\Big.\, \tilde U \,\Big.}_\infty \; \norm{\Big.\ket{\psi}\bra{\psi}\Big.}_\infty \; \norm{\Big.\tilde U\herm - U\herm \Big.}_\infty
		\;\;+\;\;\;
		\norm{\Big.	\tilde U - U \Big.}_\infty \; \norm{\Big.\ket{\psi}\bra{\psi}\Big.}_\infty \; \norm{\Big. U\herm \Big.}_\infty
		}
	\notag\\\smaller=&\quad
		\smaller{	2 \norm{\tilde U - U}_\infty	\;.	}
\end{align}
By convexity, the same inequality holds if we replace $\ket{\psi}\bra{\psi}$ with any operator $\rho \in \D(N \x M)$.
As a result, for $\Upphi(\rho) = U \rho U\herm$ and $\tilde\Upphi(\rho) = \tilde U \rho \tilde U\herm$ we then have
\begin{gather}
	\label{eqn:unitaryApprox}
		\norm{\tilde\Upphi - \Upphi}_\lozenge
	\;\;\le\;\;
		2 \norm{\tilde U - U}_\infty	\;.
\end{gather}
Finally, note that replacing a unitary $U$ with some operator $U' = \e^{i \theta} U$ (for any real $\theta$) in the definition of the map $\Upphi$ as above yields the same CPTP map $\Upphi$.
Because our interest in unitary operators arises from the CPTP maps which they induce as above, Theorem~\ref{thm:churchHilbertSpace} together with~\eqref{eqn:unitaryApprox} motivates the following definition:%
\begin{definition}
	\label{def:universalUnitary-1}
	For two unitary operations $U, \tilde U \in \U(N)$ for some $N \ge 1$, we say that \emph{$\tilde U$ performs $U$} if $\tilde U$ is proportional to $U$, and that \emph{$\tilde U$ approximates $U$ with precision~$\epsilon$} if there is an angle $\theta \in \R$ such that $\norm{\big.\smash{\tilde U - \e^{i\theta} U}}_\infty \le \frac{\epsilon}{2}$\,. 
	A set $\sS$ of unitary operations is \emph{universal for unitary transformation} if any unitary $U$ on $n$ qubits can be performed by a composition of operators from $\sS$, for every $n \ge 1$; alternatively, $\sS$ is \emph{approximately universal for unitary transformation} if any such map $U$ can be approximated to arbitrary precision $\epsilon > 0$ by some product of operators from $\sS$.
\end{definition}
This definition differs from the standard treatments of approximate universality (see \eg~\cite{KSV2002,NC00}) by the multiplication of $\epsilon$ by $\frac{1}{2}$\,, which has been inserted for consistency with Definition~\ref{def:universalCPTP}, and by the insertion of the scalar factor $\e^{i\theta}$.
The former variation is inconsequential because of the Solovay-Kitaev Theorem, which we discuss below; the latter variation is made to facilitate description of unitary circuits, starting in Section~\ref{sec:unitaryCircuits}.

It is useful to note that a similar derivation as in~\eqref{eqn:errorAdditiveCPTP} allows us to show that errors in approximating unitaries are also additive: if $\norm{\big.\smash{\tilde U - U}}_\infty < \epsilon$ and $\norm{\big.\smash{\tilde V - V}}_\infty < \epsilon'$ for $U,V,\tilde U, \tilde V \in \U(N)$ for some $N \ge 1$, we then have
\begin{align}
	\label{eqn:errorAdditiveUnitary}
		\norm{\tilde U \tilde V - U V}_\infty
	\;\;\le&\;\;
		\norm{\tilde U \big(\tilde V - V\big)}_\infty \;+\;
		\norm{(\tilde U - U) V}_\infty
	\;\;<\;\;
		\epsilon + \epsilon'\;.
\end{align}

Because we will be interested in describing unitary CPTP maps with non-unitary CPTP maps, we may more generally define the following:
\begin{definition}
	\label{def:universalUnitary-2}
	A set $\Gates$ of elementary operations is \emph{universal for unitary transformation} if any unitary CPTP map $\Upphi$ from $n$ qubits to $n$ qubits, for any $n \ge 1$, can be performed by a composition of operations from $\Gates$, and is \emph{approximately universal for unitary transformation} if any such map $\Upphi$ can be approximated to arbitrary precision $\epsilon > 0$ by some composition of operations from $\Gates$.
\end{definition}
Note that for approximate universality for unitary transformation of a set $\Gates$, the approximating compositions may not themselves be unitary; they may only approach arbitrarily close to being unitary.
By Theorem~\ref{thm:churchHilbertSpace}, a set of elementary gates which is (approximately) universal for unitary transformation, allows for the preparation of states on arbitrarily many qubits, and which allows the tracing-out of qubits, is also (approximately) universal for CPTP transformation.

These results and concepts allow us to reduce the notion of universal quantum computation to (approximate) universality for unitary transformation: but there remains a question of how sensitive quantum computational complexity is to the particular choice of elementary operations.
This question was addressed by Kitaev~\cite[Lemma~4.7]{Kitaev97}, who notes that the result was independently discovered by Robert Solovay: detailed treatment of the result can be found in~\cite{DN2006,KSV2002}.
The following is a paraphrasing of the result as described by~\cite{DN2006}:
\begin{theorem*}[Solovay-Kitaev Theorem]
	Let $U \in \U(N)$ be a unitary operator for some $N \ge 1$, and $\sS$ a set of unitary operations which is closed under inversion and generates a dense subgroup of $\U(N)$.
	Then there is a $\polylog(1/\epsilon)$-time deterministic algorithm which produces a product $\tilde U \,=\, \tilde U_L \cdots \tilde U_2 \, \tilde U_1$\,, where $L \in \polylog(1/\epsilon)$ and $\tilde U_j \in \sS$ for each $1 \le j \le L$, such that $\tilde U$ approximates $U$ with precision $\epsilon$.
\end{theorem*}
There are several variations on the algorithm, with different complexities for the run-time and the size of the approximating product; the complexity for both described in~\cite{KSV2002} is $O(\log(1/\epsilon)^{3 + \delta})$ for arbitrary $\delta > 0$.
This result holds for any fixed $N$\,; however, the running-time of the algorithm for producing an approximating unitary $\tilde U \in \U(N)$ (\eg~as described by~\cite[Theorem~8.5]{KSV2002}) is exponential in $N^2$.
Nevertheless, from a theoretical point of view, the particular set of unitaries by which we describe unitary circuits is essentially a matter of convenience, provided that we compare only sets of elementary operations which act on a small, and fixed, number of dimensions.

\section{Unitary circuit models}
\label{sec:unitaryCircuits}

While quantum computation is likely in practise to involve some aspect of classical control, it is often presented in terms of decompositions of CPTP maps
\begin{gather}
	\label{eqn:unitaryCircuitDecomp}
		\Upphi \;\;=\;\; \Meas{}{} \circ \Upupsilon \circ \mathsf A	\;,
\end{gather}
where $\mathsf A$ prepares some number of qubits in a fixed joint state, $\Upupsilon$ is a unitary operation on the set of qubits (and in particular does not involve any explicit interaction with any classical data registers), and $\Meas{}{}$ is a composition of complete orthonormal (destructive) measurements on some of the qubits and trace-outs.
By the discussion following Theorem~\ref{thm:churchHilbertSpace} on page~\pageref{thm:churchHilbertSpace}, every CPTP map can be decomposed in such a form.

\vspace{1ex}\noindent
In this section, we consider models of computation of this type:
\begin{definition}
	\label{def:unitaryCircuitModel}
	A model of quantum computation is a \emph{unitary circuit model} if its elementary gates consist only of preparation of some number of qubits in a fixed initial state, unitary transformations of some number of qubits, complete (orthogonal \& destructive) measurements on some number of qubits, and trace-out operations.
\end{definition}
Because unitary quantum circuits do not have any operations which can act on classical bits, we may assume without loss of generality that all measurements are performed at the end; similarly, we may assume that any auxiliary qubits are introduced at the beginning, leading to a decomposition as in~\eqref{eqn:unitaryCircuitDecomp}.

\subsection{Particular unitary operations of interest}
\label{sec:particularUnitaries}

Throughout the rest of this thesis, we will be interested in referring to the particular unitary operations defined as follows (in matrix form):
\begin{definition}
	\label{def:Pauli}
	The \emph{Pauli operators} are the operators $\ens{\idop_2, X, Y, Z}$, where the latter three operations are defined by
	\begin{align}
			X
		\;\;=&\;\;
			\left[\begin{matrix} \,0 & \,1 \\ \,1 & \,0 \end{matrix}\,\right],
		&
			Y
		\;\;=&\;\;
			\left[\begin{matrix} \,0 & \!-i \\ \,i & \,0 \end{matrix}\,\right],
		&
			Z
		\;\;=&\;\;			
			\left[\begin{matrix} \,1 & \,0 \\ \,0 & \!-1 \end{matrix}\,\right];
	\end{align}
	and we define the $X$-rotation, $Y$-rotation, and $Z$-rotation operators for angles $\alpha \in \R$ by
	\begin{subequations}
	\begin{align}
			R_\sfx(\alpha)
		\;\;=&\;\;
			\exp(-iX\alpha/2)
		\;\;=\;\;
			\left[\begin{matrix}	\;\cos(\tfrac{\alpha}{2}) & -i\sin(\tfrac{\alpha}{2}) \\[1ex]
													 -i\sin(\tfrac{\alpha}{2}) & \;\cos(\tfrac{\alpha}{2}) \end{matrix}\,\right]	,
		\intertext{}
		\notag\\[-2em]
			R_\sfy(\alpha)
		\;\;=&\;\;
			\exp(-iY\alpha/2)
		\;\;=\;\;
			\left[\begin{matrix}	\;\cos(\tfrac{\alpha}{2}) & -\sin(\tfrac{\alpha}{2}) \\[1ex]
													 \;\sin(\tfrac{\alpha}{2}) & \;\cos(\tfrac{\alpha}{2}) \end{matrix}\,\right]	,			
		\intertext{}
			R_\sfz(\alpha)
		\;\;=&\;\;
			\exp(-iZ\alpha/2)
		\;\;=\;\;
			\left[\begin{matrix} \;\e^{\minus i \alpha/2} & \,0 \\[1ex] \,0 & \e^{i \alpha/2} \end{matrix}\,\right].
	\end{align}
	\end{subequations}
\end{definition}
\begin{definition}
	We define the \emph{controlled-$Z$ transform} $\cZ$, the \emph{controlled-$X$ transform} $\cX$, the \emph{Hadamard transform} $H$, and the \emph{$J(\alpha)$ transform} as follows:
	\begin{subequations}
	\label{eqn:basicGates}
	\begin{align}
			\cZ
		\;\;=&\;\;
			\left[\begin{matrix} \,1 & \,0 & \,0 & \,0 \\ \,0 & \,1 & \,0 & \,0 \\ \,0 & \,0 & \,1 & \,0 \\ \,0 & \,0 & \,0 & \!-1 \end{matrix}\right],
		&
			\cX
		\;\;=&\;\;
			\left[\begin{matrix} \,1 & \,0 & \,0 & \,0 \\ \,0 & \,1 & \,0 & \,0 \\ \,0 & \,0 & \,0 & \,1 \\ \,0 & \,0 & \,1 & \,0 \end{matrix}\,\right],
		\\[2ex]
			H
		\;\;=&\;\;
			\sfrac{1}{\sqrt 2} \left[\begin{matrix} \,1 & \,1 \\[0.5ex] \,1 & \!-1 \end{matrix}\right]	,
		&
			J(\alpha)
		\;\;=&\;\;
			\sfrac{1}{\sqrt 2}\left[\begin{matrix} \,\e^{\minus i\alpha/2} & \,\e^{i \alpha/2} \\[0.5ex]
																 \,\e^{\minus i\alpha/2} & \!-\e^{i \alpha/2} \end{matrix}\right] .
	\end{align}
	\end{subequations}
\end{definition}

\subsection{Two related unitary circuit models}
\label{sec:standardCircuitModels}

We may use the unitaries defined in the previous section to describe two useful families of elementary gates, which are universal or approximately universal for unitary transformation for appropriate choices of angle parameters.
We describe how this may be done by summarizing the development presented in~\cite[Chapter~4]{NC00}.
We start with single-qubit unitaries.

First, note that an arbitrary single-qubit unitary $U \in \SU(2)$ can be decomposed as a product of $Y$ and $Z$ transformations,
\begin{gather}
		U
	\;\;=\;\;
		R_\sfz(\alpha) R_\sfy(\beta) R_\sfz(\gamma),
\end{gather}
for some angles $\alpha, \beta, \gamma \in (-\pi, \pi]$.
Note that we may decompose 
\begin{gather}
		R_\sfy(\beta)
	\;\;=\;\;
		R_\sfz(\pion 2) H R_\sfz(\beta) H R_\sfz(\mpion 2)\,,
\end{gather}
which implies that
\begin{gather}
	U \;\;=\;\; R_\sfz(\alpha + \tfrac{\pi}{2}) \,H\, R_\sfz(\beta) \,H\, R_\sfz(\gamma - \tfrac{\pi}{2}),
\end{gather}
so that $\SU(2)$ can be generated with $Z$-rotations and Hadamard transforms alone.
For approximation of single-qubit unitaries, following the development in \cite{BMPRV99}, the operations
\begin{align}
		J(\pion 4)
	\;=&\;\;
		H R_\sfz(\pion 4),
	&
		J(\mpion 4)
	\;=&\;\;
		H R_\sfz(\mpion 4)
	\;=\;
		H J(\pion 4)\herm H,
\end{align}
have eigenvalues $\e^{\pm i \theta}$ for $\theta = \arccos(\frac{2 + \sqrt 2}{4})$, which is an irrational multiple of $\pi$\,.
Such an angle then generates a dense subgroup of $\R / 2\pi\Z$, the group of real angles modulo complete cycles of $2\pi$.\footnote{%
We may effectively prove this by the Pigeon-Hole principle. For any $\epsilon > 0$, let $N$ be the smallest positive integer such that $\frac{\pi}{N} < \epsilon$.
We may divide the interval $[0,2\pi)$ into $N$ partitions $I_m = [\frac{2m\pi}{N}, \frac{2(m+1)\pi}{N})$, which we use to represent a connected subset $I_m + 2\pi\Z$ of the compact group $\R/2\pi\Z$.
Because $\theta$ is an irrational multiple of $\pi$, $(\theta + 2\pi\Z)$ is an element of $\R/2\pi\Z$ with infinite order.
As a result, there is at least one interval $I_{\widetilde m} + 2\pi\Z$ which contains at least two distinct elements $t, t'$ of the group $\gen{\theta + 2\pi\Z} \le \R/2\pi\Z$\,.
Then, there are integers $K$ and $L$ such that $(K\theta - L\theta) = (t - t') \in I_0 + 2\pi\Z$ --- that is, for some $M \in \Z$, there exists an angle $\theta' = K\theta - L\theta + 2M\pi \in [0, \frac{2\pi}{N})$.
Note that any element of $\R$ differs from a multiple of $\theta'$ by at most $\frac{\pi}{N} < \epsilon$.
Thus, for any $\epsilon$ and for any $x \in \R/2\pi\Z$, there exists a multiple of $\theta + 2\pi\Z$ which is within distance $\epsilon$ of $x$.}
Let $\ket{\spm \hat{\sfn}}$ represent the unit $\e^{\pm i \theta}$-eigenvectors of $J(\pion 4)$; then, if we define operators
\begin{gather}
		R_{\hat n}(\alpha)
	 \;\;=\;\;
		\e^{i \alpha} \ket{\splus \hat{\sfn}}\bra{\splus \hat{\sfn}}
		\;\;+\;\;
		\e^{-i \alpha} \ket{\sminus \hat{\sfn}}\bra{\sminus \hat{\sfn}}	\,,
\end{gather}
we may take powers of $J(\pion 4) = R_{\hat \sfn}(\theta)$ to obtain arbitrarily good approximations to $R_{\hat n}(\alpha)$ for any angle $\alpha \in (-\pi, \pi]$.
Similarly, if $\ket{\spm \hat{\sfm}}$ represent the unit $\e^{\pm i \theta}$-eigenvectors of $J(\mpion 4)$, and we define the operators
\begin{gather}
		R_{\hat \sfm}(\alpha)
	 \;\;=\;\;
		\e^{i \alpha} \ket{\splus \hat{\sfm}}\bra{\splus \hat{\sfm}}
		\;\;+\;\;
		\e^{-i \alpha} \ket{\sminus \hat{\sfm}}\bra{\sminus \hat{\sfm}}	\,,
\end{gather}
we may take powers of $J(\mpion 4) = R_{\hat \sfm}(\theta)$ to obtain arbitrarily good approximations to $R_{\hat \sfm}(\alpha)$ for any angle $\alpha \in (-\pi, \pi]$.
By~\cite[Section~4.2]{NC00} (but see~\cite{NC00errata} for an erratum), because $J(\pion 4)$ and $J(\mpion 4)$ do not commute, we may then approximate any unitary $U \in \SU(2)$ to arbitrary precision using the group generated by $\ens{\big.J(\pion 4),\, J(\mpion 4)\big.}$.

The set of operations $\ens{\cX, U}_{U \in \U(2)}$ is universal for unitary transformation~\cite[Section~4.5.2]{NC00}.
By the preceding remarks about single-qubit unitaries, we may perform arbitrary $n$-qubit unitaries using controlled-not transformations, Hadamard transformations, and arbitrary $Z$-rotations; and using the remarks on approximation of unitary operations in Section~\ref{sec:universal}, we approximate arbitrary unitaries (up to scalar factors) using controlled-not transformations and $J(\spm\pion 4)$ transformations.
Finally, using the following equivalencies,
\begin{align}
		\cX
	\;=&\;\;
		(\idop_2 \ox H) \,\cZ\, (\idop_2 \ox H)	\,,
	&
		J(0)
	\;=&\;\;
		H	,
	&
		J(\alpha)
	\;=&\;\;
		H R_\sfz(\alpha)	\,,
\end{align}
we obtain the following:
\begin{lemma}
	\label{lemma:universalUnitaries}
	For a set of angles $\A \subset \R$, the two parameterized sets of unitary transformations $\ens{H, R_\sfz(\alpha),\cZ}_{\alpha \in \A}$ and $\ens{J(\alpha), \cZ}_{\alpha \in \A}$ are both universal for unitary transformation when $\A = \R$, and approximately universal for unitary transformation when $\A = \frac{\pi}{4}\Z$.
\end{lemma}
Because $R_\sfz(\mpion 4) = R_\sfz(\frac{7\pi}{4})$ and because we may approximate $H$ to arbitrary precision using $J(\spm \pion 4)$, we have actually shown that the sets $\ens{H, R_\sfz(\pion 4),\cZ}$\,, $\ens{J(0), J(\pion 4),\cZ}$, and $\ens{J(\mpion 4), J(\pion 4),\cZ}$ are approximately universal for unitary transformation; however, the somewhat weaker statement presented in Lemma~\ref{lemma:universalUnitaries} will be more useful in further discussion, as each set of unitaries described can be described by the same family of angles, and can be used to exactly generate the other.

We define the following unitary CPTP maps corresponding to the unitaries described in Lemma~\ref{lemma:universalUnitaries}, acting on density operators $\rho \in \D(2)$ and $\varrho \in \D(2 \x 2)$:
\begin{subequations}
\label{eqn:cptpElemUnitaries}
\begin{align}
		\sfH(\rho)
	\;\;=&\;\;
		H \rho H	\,,
	&
		\sfR^\alpha(\rho)
	\;\;=&\;\;
		R_\sfz(\alpha) \,\rho\, R_\sfz(\alpha)\herm	\,,
	\\
		\sfJ^\alpha(\rho)
	\;\;=&\;\;
		J(\alpha) \,\rho\, J(\alpha)\herm\,,
	&
		\Ent{}(\varrho)
	\;\;=&\;\;
		\cZ\,\varrho\,\cZ	\,.
\end{align}
\end{subequations}
The map $\Ent{}{}$ we will tend to call an \emph{entangling} or \emph{entangler} operation, referring to the fact that it maps states of two independent qubits to \emph{entangled} states (\ie\ joint states which are not independent), provided that neither of the qubits are initially in a state which is a convex combination of the standard basis.\footnote{%
The map $\rho \mapsto \Ent{}{}(\rho)$ is not the only map with this property, of course, but it is the map which will arise most often in this thesis.
In particular, $\Ent{}{}$ is the only entangling operation which is allowed in the one-way measurement-based model of quantum computing described in Section~\ref{sec:oneWayModel}.}
We may define three more non-unitary CPTP maps: preparation maps $\New{}{\sfx}$ and $\New{}{\sfz}$ for the pure states $\ket{\splus} = \ket{\splus \sfx}$ and $\ket{0} = \ket{\splus \sfz}$ respectively,
\begin{align}
	\begin{split}
		\New{v}{\sfx} : \ens{1}& \to \D(v)	\,,
	\\[1ex]
		\New{v}{\sfx}(1) \;\;=&\;\; \ket{\splus \sfx} \bra{\splus \sfx}_v	\,,
	\end{split}
	&
	\begin{split}
		\New{v}{\sfz} : \ens{1}& \to \D(v)	\,,
	\\[1ex]
		\New{v}{\sfz}(1) \;\;=&\;\; \ket{\splus \sfz} \bra{\splus \sfz}_v	\,;
	\end{split}
\end{align}
and complete measurements $\Meas{}{\sfz}$ in the standard basis (discarding the qubit $v$ on which it operates and allocating a bit $\s[v]$ to store the result),
\begin{align}
	\begin{split}
	\begin{gathered}
			\Meas{v}{\sfz}
		\,:\,
			\D(v) \to \Prob(\s[v])
		\\[1ex]
			\Meas{v}{\sfz}(\rho)
		\;\;=\;\;
			\sum_{r \in \ens{0,1}} \bra{r}_v \rho \,\ket{r}_v \,\ox\, \ket{r}\bra{r}_{\s[v]}	\;.
	\end{gathered}
	\end{split}
\end{align}
Then, we define the following sets of gates:
\begin{definition}
	\label{def:elemGateSets}
	For a set $\A \subset \R$, an infinite set $V$ of qubits, and an infinite set $B$ of bits, we define the following parameterized sets of elementary gates,
	\begin{subequations}
	\label{eqn:elemGateSets}
	\begin{align}
			\mspace{-50mu}\Phases(\A)
		\;\;=&\;\;
			\ens{\sfH_v\,,\; \sfR_v^\alpha\,,\; \Ent{u,v}\,,\; \New{v}{\sfz}\,,\; \New{v}{\sfx}\,,\; \Meas{v}{\sfz}\,,\;  \tr[v]
			\,\Big|\, \alpha \in \A \text{~and~} v,w \in V}\,,
		\\[1ex]
			\Jacz(\A)
		\;\;=&\;\;
			\ens{\sfJ_v^\alpha\,,\; \Ent{u,v}\,,\; \New{v}{\sfz}\,,\; \New{v}{\sfx}\,,\; \Meas{v}{\sfz}\,,\; \tr[v]
				\,\Big|\, \alpha \in \A \text{~and~} v,w \in V}\,.
	\end{align}
	\end{subequations}
	where the map $v \mapsto \s[v]$ is an injective map from $V$ to $B$.
\end{definition}	
The circuit models arising from these two sets of elementary operations are both unitary circuit models; by Lemma~\ref{lemma:universalUnitaries}, both are (approximately) universal for quantum computation for an appropriate choice of angles $\A \subset \R$.
As well, circuits over each gate set may be easily transformed to the other using the equivalences $\sfJ^\alpha = \sfH \circ \sfR^\alpha$ and $\sfR^\alpha = \sfH \circ \sfJ^\alpha$.

\subsection{Representations of unitary circuits}
\label{sec:repnsUnitaryCircuits}

One of the topics of this thesis is a description of transformations between unitary circuits, and the \emph{one-way measurement-based model} of quantum computation, which we describe in Chapter~\ref{One-way measurement-based computation}.
As well, we will often be interested in presenting unitary circuits diagrammatically.
We now present different ways in which unitary circuits may be represented.

\subsubsection{Standard representations of unitary circuits}

The two conventional ways of representing unitary circuits are as products of unitary operators (as we have been doing so far), and by \emph{circuit diagrams}.

When writing unitary circuits as products of operators, we conventionally fix representation of unitary operators with respect to the standard basis.
For instance, for an $n$-qubit unitary, we would write something of the form
\begin{gather}
	\label{eqn:genericTensor}
		U
	\;\;=\;\;
		\Big[\;{U_{\vec x}}^{\vec y}\;\Big]_{\vec x, \vec y \in \ens{0,1}^n}
\end{gather}
where superscripts denote row indices and subscripts denote column indices.
We identify the bit-positions in the strings $\vec x, \vec y \in \ens{0,1}^n$ with particular qubits:\footnote{%
This identification is an injection from the set $\ens{1,\ldots,n}$ to some subset $L$ of $n$ distinct qubits among the set of all qubits in the model of computation being considered; this induces a particular isometry between $\U\big(\cH^{\!\!\ox n}\big)$ and the unitary group $\U\big(\cH^{\!\!\ox V}\big)$, where $V$ is the set of \emph{all} qubits admitted by the model of computation.}
for a sequence $(v_j)_{j = 1}^n$ of such qubits, we may then write a particular unitary operation acting on these qubits as $U_{v_1,\ldots,v_n}$.
This describes a transformation of pure state vectors, which is how quantum computation is conventionally described for unitary circuit models.
Products of such operators are then defined as performing the appropriate operation on the qubits listed as operands, and performing the identity otherwise: for example, we may write
\begin{gather}
		W_{b,c} V_a U_{a,b,c} T_b \ket{\psi}_{b,c}
	\;\;=\;\;
		(\idop_2 \ox W) (V \ox \idop_2 \ox \idop_2) U (\idop_2 \ox \ket{\psi})
\end{gather}
for generic unitary operators $W \in \U(8)$, $V,T \in \U(2)$, $U \in \U(8)$, and $\ket{\psi} \in \cH^{\ox 2}$, labelling the tensor factors as $a$, $b$, and $c$ in order.
The operator specified by this notation clearly depends on the order of the composition, but only up to permutations of commuting operators.

\begin{figure}[t]
	\begin{center}
		\includegraphics{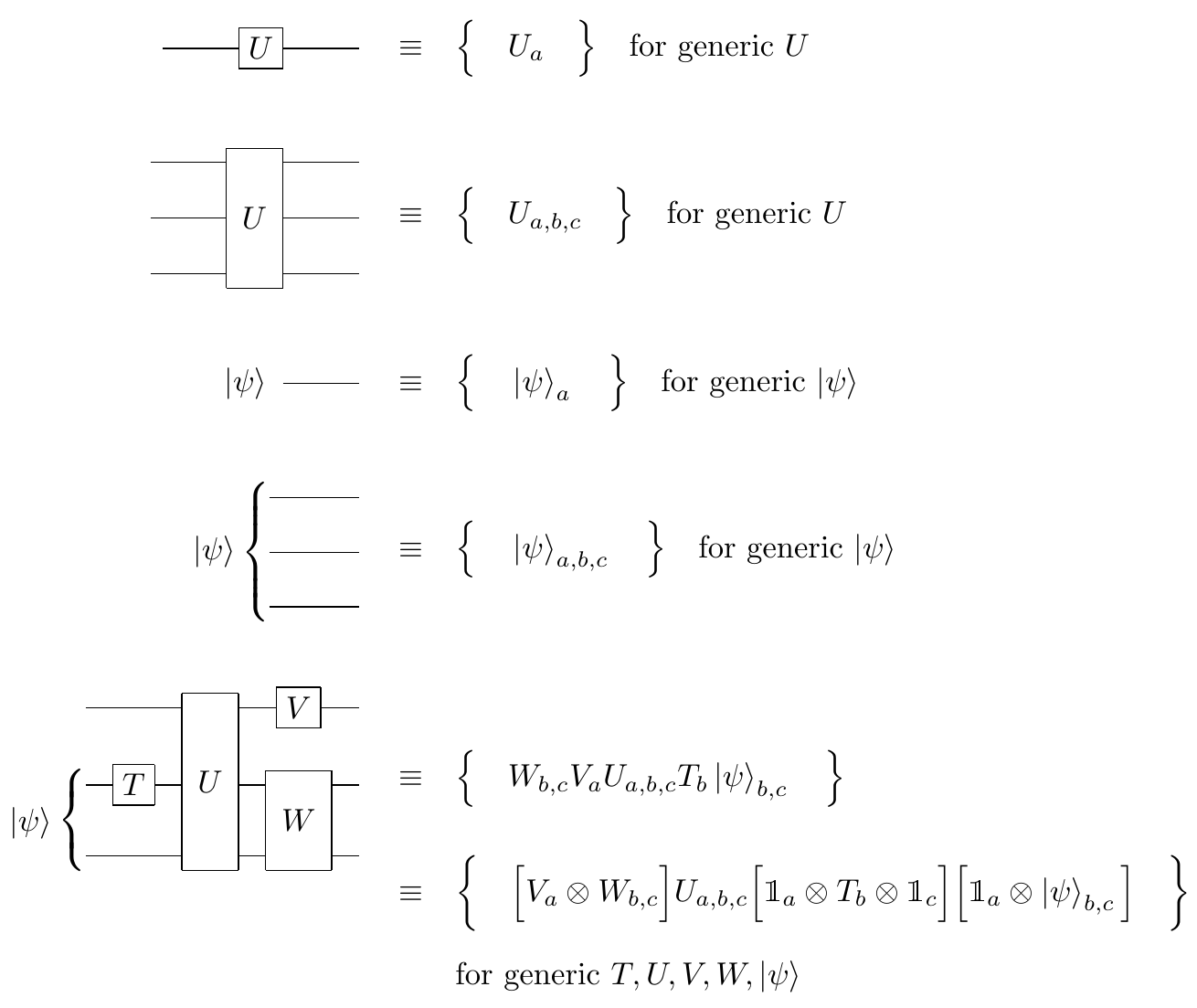}
	\end{center}
	\caption[Examples of simple quantum circuit diagrams.]{\label{fig:qcircuit-1}%
		Examples of simple quantum circuit diagrams.
		In the operator expressions for these examples, we arbitrarily label the qubits of each circuit from top to bottom in alphabetical order ($a, b, c, \ldots$).
		Diagrams compose from left to right, in contrast to the right-to-left composition in written notation; this represents an arrow of time in the diagrams from left to right.}
\end{figure}
\begin{figure}[tp]
	\begin{center}
		\includegraphics{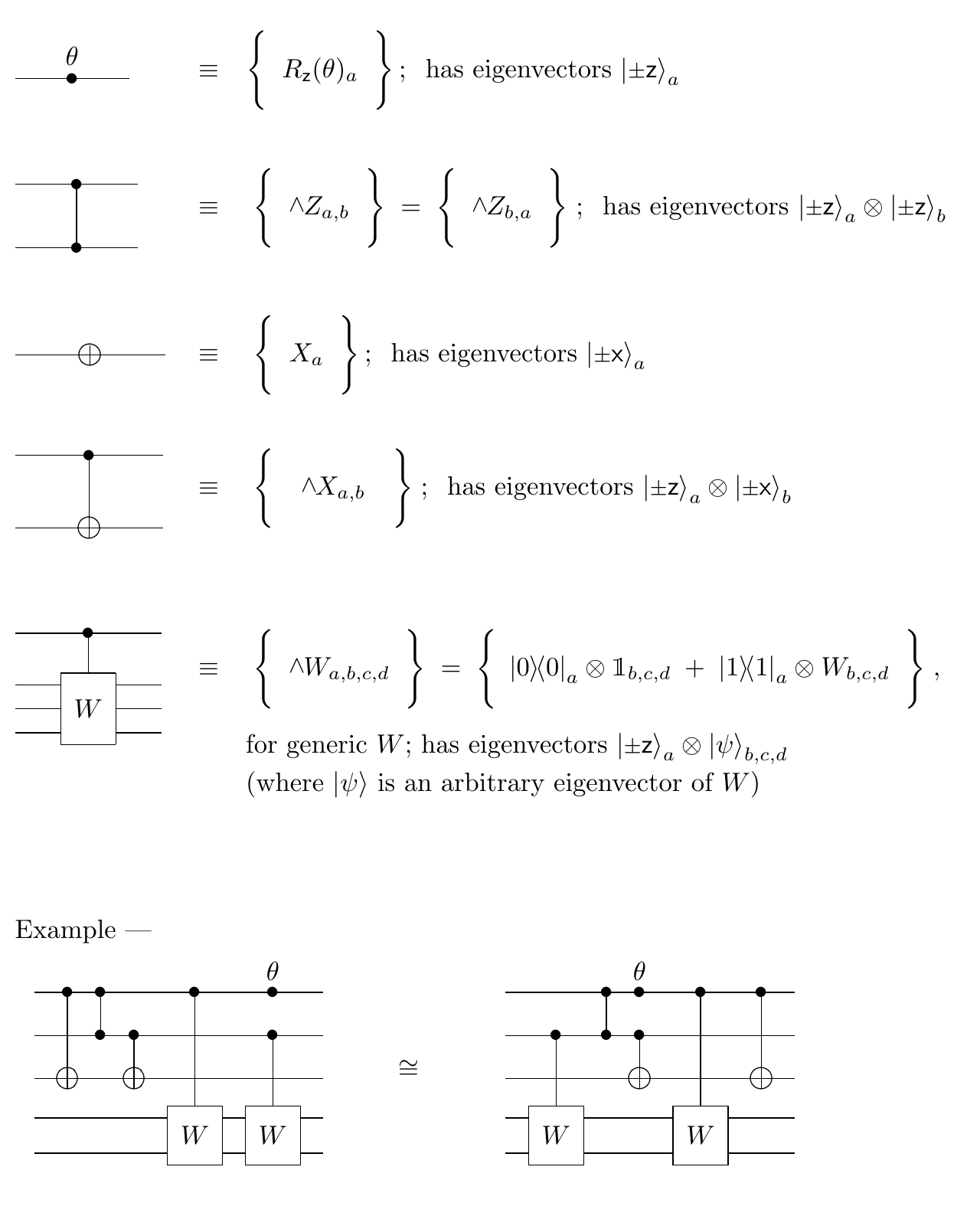}
	\end{center}
	\caption[Examples of special notations for quantum circuit diagrams.]{\label{fig:qcircuit-2}%
		Examples of special notations for quantum circuit diagrams.
		In the operator expressions for these examples, we again arbitrarily label the qubits of each circuit from top to bottom in alphabetical order ($a, b, c, \ldots$).
		The bottom-most circuits illustrate two equivalent circuits under commutation relations, where these relations are made visually apparent by common symbols on collections of wires.}
\end{figure}
Circuit diagrams are an alternative method of representing quantum circuits, as in Figures~\ref{fig:qcircuit-1} and~\ref{fig:qcircuit-2}.
Distinct qubits are represented by distinct \emph{wires} (often drawn as progressing from left to right, as already illustrated in Figure~\ref{fig:typeCompose}), schematically representing the trajectory of a qubit.
Operations on qubits are then drawn with boxes or other symbols representing multi-qubit interactions.
Figure~\ref{fig:qcircuit-1} illustrates simple examples of quantum circuit diagrams for some generic unitary operators and state vectors in terms of products of operators on specified qubits.
This notation allows operators which can be taken in tensor product to be represented more clearly as being simultaneously applicable, but is still strongly dependent on the sequence of operations specified --- and has difficulty expressing operations on qubits which are not adjacent in the linear arrangement of qubits in the diagram.

Circuit diagrams readily admit an improvement over the operator notation, by making commutation relations more apparent through the following notational devices.
Given an operator $U$ which operates on some collection of $n$ qubits, if the eigenvectors of $U$ can be expressed by vectors $\ket{\phi_1} \ox \cdots \ox \ket{\phi_m}$ where each $\ket{\phi_j}$ ranges over (possibly different) orthonormal bases for $\cH^{\!\!\ox n_j}$, where $n_1 + \cdots + n_m = n$ is a partition of $n$, we may represent $U$ using special symbols on each of the partition classes of qubits corresponding to the tensor decomposition of the eigenvectors.
Then, for different operators represented in this way, the operators commute whenever they act on common sets qubits and are represented by common symbols.
Standard notational devices are illustrated in Figure~\ref{fig:qcircuit-2}, along with an example which exploits these devices to show the equivalence of two circuits.
Such devices can also partially overcome the problem of specifying operations which act on non-adjacent qubits.

\subsubsection{Non-standard representations of unitary circuits}
\label{sec:nonstandardRepns}

One of the main topics of this thesis is translations between different representations of quantum computations.
Therefore, it will be useful to have a notation for unitary circuits in which irrelevant details such as the order of commuting gates are essentially absent, while being at the same time easy to produce by an algorithm.
This motivates the definition of an alternative representation for unitary circuits. 

One alternative notation (albeit one which is rare in the quantum computing literature) is Einstein summation notation~\cite[Section~B5]{Einstein1916}, where we provide different indices for the domain and range of each operator (\eg\ as in~\eqref{eqn:genericTensor} above, for the rows and columns), and identify the output indices of operators with the input indices of any subsequent operator acting on the same qubit.
We implicitly sum over any indices which are repeated (representing matrix multiplication of the matrix representation of the operators).
Thus, for example, for a composition of operators
\begin{subequations}
\begin{align}
	\label{eqn:arbitraryOperatorProduct}
		C
	\;\;=\;\;
		W_{b,c} V_a U_{a,b,c} T_b \ket{\psi}_{b,c}
\end{align}
for generic unitary operators $W \in \U(8)$, $V,T \in \U(2)$, $U \in \U(8)$, and $\ket{\psi} \in \cH^{\ox 2}$, we may equivalently write
\begin{align}
	{C_{a_0}}^{\!\!\!\!a_2,b_3,c_2}
	\;\;=\;\;
	\Big[{W_{b_2,c_1}}^{\!\!\!\!b_3,c_2}\Big]
	\Big[{V_{a_1}}^{\!\!\!\!a_2}\Big]
	\Big[{U_{a_0,b_1,c_0}}^{\!\!\!\!a_1,b_2,c_1}\Big]
	\Big[{T_{b_0}}^{\!\!\!\!b_1}\Big]
	\Big[\ket{\psi}^{b_0,c_0}\Big] .
\end{align}
Indices which are repeated are \emph{bound} variables over which we sum, and indices which are not repeated correspond to \emph{free} variables corresponding to rows (for subscripted indices) or columns (for superscripted indices) of a matrix representation of an operator.
(The brackets here are provided for additional emphasis.)
Because the sense of composition of tensors here is provided by the indices, we may freely rearrange the factors of such an expression without changing the meaning of the expression, which implies that the following composition of operators
\begin{align}
		\Big(\bra{a_2} \ox& \bra{b_3} \ox \bra{c_2}\Big) C \ket{a_0}
	\notag\\=&\;\;
		\sum_{a_1} \sqparen{\sum_{b_0,b_1,b_2} \paren{\sum_{c_0,c_1} 
			{W_{b_2,c_1}}^{\!\!\!\!b_3,c_2} \;\;
			{V_{a_1}}^{\!\!\!\!a_2} \;\;
			{U_{a_0,b_1,c_0}}^{\!\!\!\!a_1,b_2,c_1} \;\;
			{T_{b_0}}^{\!\!\!\!b_1} \;\;
			\ket{\psi}^{b_0,c_0}
		}} 
\end{align}
\end{subequations}
(where each index $a_h, b_j, c_k$ ranges over $\ens{0,1}$) can be expressed as a sum over a commuting product of scalars.
Thus, this notation has added flexibility over the conventional operator notation, in that the order of the factors is not significant.
However, because the meaning is preserved by indicating the order of the multiplication in the tensor indices, circuits which are equivalent up to rearrangements of commuting operations still give rise to inequivalent expressions: for instance, although it is easy to verify that the operations $\cZ_{a,b}$, $\cZ_{a,c}$, and $\cZ_{b,c}$ commute, the two products 
\begin{subequations}
\label{eqn:tripleCZproduct}
\begin{gather}
		\cZ_{b,c} \; \cZ_{a,c} \; \cZ_{a,b}
	\;\;\equiv\;\;
		{\cZ_{b_1,c_1}}^{\!\!\!\!b_2,c_2} \;\; {\cZ_{a_1,c_0}}^{\!\!\!\!a_2,c_1} \;\; {\cZ_{a_0,b_0}}^{\!\!\!\!a_1,b_1}	\,,
	\\[1ex]
		\cZ_{a,c} \; \cZ_{b,c} \; \cZ_{a,b}
	\;\;\equiv\;\;
		{\cZ_{a_1,c_1}}^{\!\!\!\!a_2,c_2} \;\; {\cZ_{b_1,c_0}}^{\!\!\!\!b_2,c_1} \;\; {\cZ_{a_0,b_0}}^{\!\!\!\!a_1,b_1}	\,,	
\end{gather}
\end{subequations}
still give rise to expressions which are not equivalent in Einstein notation by a relabeling of the indices and re-ordering of the terms.

A compromise between Einstein summation notation, and the approach of using special symbols in circuit diagrams, would be an Einstein-like tensor notation in which an index can be repeated multiple times when it corresponds to elements of an orthonormal basis which are preserved by several tensors --- that is, where the tensor indices may be interpreted as indicating elements of an orthonormal basis $\ens{\ket{\phi_j}}$ of some subset of the qubits, such that for several successive unitary operators $U$ in succession in the circuit, there exist a collection of unitaries $V_j$ for which $U \big( \ket{\phi_j} \ox \ket{\psi} \big) = \ket{\phi_j} \ox V_j \ket{\psi}$ holds for all states $\ket{\psi}$ of the remaining qubits.

\vspace{1ex}
\noindent To this end, we may consider a variant of Einstein summation notation, as follows:
\vspace{-0.8ex}
\begin{notation*}[Stable index tensor notation]
	\label{def:stableIndex}
	For a unitary operator $U$ acting on qubits $v_1, \ldots, v_n$, let $\sigma$ and $\delta$ be sequences of $n_1$ and $n_2$ qubits respectively (where $n = n_1 + n_2$) such that the eigenvectors of $U$ consist entirely of vectors $\ket{\vec{s}}_\sigma \ox \ket{\psi}_\delta$, for standard basis vectors $\ket{\vec s}$ on the qubits $\sigma$ and vectors $\ket{\psi}$ which are \emph{not known} to be decomposable as tensor products over the standard basis for qubits in $\delta$.
	We then write the matrix coefficients of $U$ in the standard basis by
	\begin{subequations}
	\label{eqn:stableIndex}
	\begin{gather}
			{U_{\vec x}}^{\vec y}
		\;\;=\;\;
				U\pseu[\vec s:\vec a/\vec d]	\;,
	\end{gather}
	where $\vec{x} = (\vec s ; \vec d)$ and $\vec{y} = (\vec s; \vec a)$ are decompositions of $\vec x$ and $\vec y$ into the appropriate substings.
	That is, we define $U\pseu[\vec s: \vec a / \vec d]$ by the operator equality
	\begin{align}
			U
		\;\;=&
			\sum_{\substack{\vec s \in \ens{0,1}^{n_1} \\ \vec{a},\vec{d} \in \ens{0,1}^{n_2}}}		\!\!
				U\pseu[\vec s:\vec a/\vec d]
				\;\; \Big( \ket{\vec s}\bra{\vec s}_\sigma \ox \ket{\vec a}\bra{\vec d}_\delta \Big)	\;.
	\end{align}
	\end{subequations}
	We call the indices of $\vec s$ \emph{stable}, the indices of $\vec d$ \emph{deprecated}, and the indices of $\vec a$ \emph{advanced}.
	The sequence of the advanced indices correspond to the same qubits being operated on as the sequence of deprecated qubits, in order.
	A product of such expressions is well-formed if each index is advanced at most once and deprecated at most once; and we sum implicitly over indices which are both advanced and deprecated in a product expression.
\end{notation*}

\begin{notation*}[A further shorthand notation]
	For a generic unitary operator $U \in \U(2^n)$ which is not known to have eigenvectors of the form $\ket{s}_{v_1} \ox \ket{\psi}_{v_2, \ldots, v_n}$ for arbitrary standard basis states $\ket{s} = \ket{\spm \sfz}$, for any qubit $v_1$ that $U$ acts upon, we may represent $U$ by the symbol $U\pseu[:\vec a/\vec d]$ (\ie\ without stable indices).
	Conversely, if the eigenvectors of such an operator $U$ are all standard basis vectors, we may represent $U$ by the symbol $U\pseu[\vec s]$ (\ie\ without advanced or deprecated indices).
	Finally, for a unitary embedding, there may be advanced indices which do not correspond to any deprecated indices; we may pad out the sequence of deprecated indices by a space, dot, or similar placeholder in this case.
\end{notation*}

\begin{example}
	For a product of generic operators $C = W_{b,c} V_a U_{a,b,c} T_b \ket{\psi}_{b,c}$ as described in~\eqref{eqn:arbitraryOperatorProduct}, we may write
	\begin{gather}
			C\pseu[:a_2,b_3,c_2/a_0,\;\cdot\;,\;\cdot\;]
		\;\;=\;\;
			W\pseu[:b_3,c_2/b_2,c_1]V\pseu[:a_2/a_1]U\pseu[:a_1,b_2,c_1/a_0,b_1,c_0]T\pseu[:b_1/b_0]\ket{\psi}\pseu[:b_0, c_0/] .
	\end{gather}
	Because the operators are not specified as preserving standard basis vectors over any of their operands, there are no stable indices; then, the meaning of this expression essentially reduces to that of Einstein notation, where we just sum over repeated indices (\ie\ indices which are advanced in some factor and deprecated in some factor).
\end{example}

\begin{example}
	\label{eg:commutingControls}
	For some generic unitary $W \in \U(4)$, we may define the operator 
	\begin{gather}
		\ctrl W \;=\; \ket{0}\bra{0} \ox \idop_2^{\,\ox 2} \;+\; \ket{1}\bra{1} \ox W
	\end{gather}
	analogously as in the example of Figure~\ref{fig:qcircuit-2}; then, the stable index representations of the circuits $\cX_{a,d} \,  \ctrl W_{a,b,c}$ and $\ctrl W_{a,b,c} \, \cX_{a,d}$ on four qubits may be given by
	\begin{align}
		\cX\pseu[a:d_1/d_0] \ctrl W\pseu[a:b_1,c_1/b_0,c_0]
		&&
		\text{and}
		&&
		\ctrl W\pseu[a:b_1,c_1/b_0,c_0] \cX\pseu[a:d_1/d_0] \;,
	\end{align}
	respectively.
	Note that these are equivalent up to permutations of the factors.
\end{example}
The informal semantics of stable index tensor notation is to provide a description of circuits in terms of computational paths.%
\label{discn:stableIndexPathIntegral}
In particular, stable index notation for circuits was partially inspired by the \emph{path-labelling} of circuits in~\cite{DHHMNO04}, which is explicitly concerned with descriptions of circuits in terms of computational paths: the tensor indices here correspond to labels of ``wire segments'' in that article, which are separated by \eg\ Hadamard gates, or other operations which do not preserve the standard basis.
For a given operator, the stable indices then represent qubits for which each vector of the standard basis is preserved, with coefficients being assigned (conditioned on a particular value of the stable indices) to transitions from certain values of the deprecated indices to various values of the advanced indices.
For products of operators which preserve standard basis vectors on a common set of qubits, it is not necessary (as is done in Einstein notation) to introduce distinct indices for the domain and the range of the operator, given that the coefficients for all of the cross-terms will be zero.
Stable index tensor notation then consists simply of elaborating Einstein notation by so-called ``stable'' indices.

By construction, there is a close analogy between \emph{stable, deprecated, and advanced indices} in the stable index notation, and \emph{stable, discarded, and allocated qubits} as described in Definition~\ref{def:elemGatesTypes}.\label{discn:typeCorrespondance}
This correspondence could be made exact if we imposed the constraint for any CPTP map $\Upphi$, we discard any qubit for which $\Upphi$ does not preserve the marginal distributions $\ket{0}\bra{0}$ and $\ket{1}\bra{1}$, substituting it if desired by a differently distributed qubit, and never re-allocating any qubit which have already been discarded.
This correspondence will play a useful role in our discussion of the one-way measurement-based model in Chapter~\ref{One-way measurement-based computation}.

\paragraph{Constructing stable index representations.}
We may automatically generate a stable index representation $S$ for a unitary circuit from a representation as a composition of operators $C$ on specified qubits, as follows.
Let $L$ be the set of qubit labels in $C$.
We may then transliterate the terms in $C$, defining a set of tensor indices incrementally as we do so, as follows:
\begin{enumerate}
\item
	For the first operator in $C$ to act on a given qubit $v$, we define a tensor index $v_0$\,, and designate $v_0$ as \emph{the current tensor index} for $v$.

\item
	For any operator $U_{a,b,\ldots}$ in $C$, we translate $U_{a,b,\ldots}$ into a stable index term $U\pseu[\vec s_U:\vec a_U/\vec d_U]$ with tensor indices as follows.
	For any qubit $v$ upon which $U$ acts, let $v_j$ be the current tensor index for $v$ (for some $j \in \N$).
	\begin{itemize}
	\item
		If $U_{a,b,\ldots}$ is a unitary embedding which allocates $v$, then $v_0$ is an advanced index in $\vec{a}_U$, without a corresponding deprecated index.

	\item
		If $U_{a,b,\ldots}$ is a unitary embedding which takes a state on $v$ as an input, and $U_{a,b,\ldots}$ does not preserve standard basis states on $v$, then we define a \emph{new} current tensor index $v_{j+1}$ for $v$; then $v_{j+1}$ is an advanced index in $\vec{a}_U$, corresponding to the deprecated index $v_j$ in $\vec{d}_U$.

	\item
		If $U_{a,b,\ldots}$ is a unitary embedding which takes a state on $v$ as an input, and $U_{a,b,\ldots}$ preserves standard basis states on $v$, then $v_j$ remains the current tensor index for $v$, and is a stable index in $\vec{s}_U$.
	\end{itemize}
	In particular, every new index which is advanced by a gate $U\pseu[\vec s:\vec a/\vec d]$ must be distinct from any other index being advanced, as well as from all of the indices which have occurred to that point.
	Performing this transliteration of operators, sequentially for all terms in $C$, produces the corresponding stable index tensor expression $S$.
\end{enumerate}
For a given circuit decomposition $C$ of a unitary operation, we will refer to the above construction (or one which differs from it only by a relabelling of indices) as the stable index tensor expression $S$ corresponding to $C$.
This construction leads to a natural (partial) mapping $f$ on the set $V(S)$ of indices of a stable index tensor expression $S$; for each index $v_j$, we define $f(v_j) = v_{j+1}$ if the latter is well-defined.
More generally, for each index $v$ which is deprecated in some term of $S$, we define $f(v)$ to be the corresponding index which is advanced.
If $I$ is the set of ``input indices'' (ones which are not advanced by any term in $S$) and $O$ the set of ``output indices'' (which are not deprecated by any term in $S$), $f$ is then an injective map from $V(S) \setminus O$ to $V(S) \setminus I$ by construction.

Example~\ref{eg:commutingControls} on page~\pageref{eg:commutingControls} illustrates the intuitive purpose of stable index notation: by allowing stable indices to be repeated multiple times while remaining ``free'' variables, we also discard some redundant combinatorial information about the order of the product of operators.
If we allow the terms of the product to be permuted arbitrarily, information about order of commuting operators is then lost completely.
Circuits which are congruent up to the re-ordering of commuting gates may then give rise to ``equivalent'' stable index tensor expressions, in the following sense:
\begin{definition}
	A \emph{homomorphism} of stable index tensor expressions is a bijection $\lambda$ of the \emph{index labels} of one stable index expression $S_1$ to those of another such expressions $S_2$, such that for any vector $(\vec s; \vec a; \vec d)$ of index labels for which there is an operator $U\pseu[\vec s:\vec a/\vec d]$ in $S_1$, there is an operator $W\pseu[\lambda(\vec s):\lambda(\vec a)/\lambda(\vec d)]$ in $S_2$, where we write $\lambda(\vec s) = \lambda(s_j)_{j = 1}^n = \big(\lambda(s_1), \ldots, \lambda(s_n)\big)$ , and similarly for $\lambda(\vec a)$ and $\lambda(\vec d)$.
	An \emph{isomorphism} of stable index tensor expressions is such a homomorphism where there is also a bijective mapping $\tau$ of \emph{terms} of $S_1$ to those of $S_2$, where we require
	\begin{gather}
			\tau\paren{\bigg.U\pseu[\vec s:\vec a/\vec d]}
		\;\;=\;\;
			U\pseu[\lambda(\vec s):\lambda(\vec a)/\lambda(\vec d)]
	\end{gather}
	for each term $U\pseu[\vec s:\vec a/\vec d]$ in $S_1$.
\end{definition}
\paragraph{Remark.}
Homomorphisms of stable index representations are \emph{combinatorial} homomorphisms, not \emph{algebraic} ones.
In particular, for two isomorphic stable index representations $S = U_n\pseu[\ast] \cdots U_2\pseu[\ast] U_1\pseu[\ast]$ and $S' = U'_n\pseu[\ast] \cdots U'_2\pseu[\ast] U'_1\pseu[\ast]$, whose isomorphism is witnessed by an appropriate bijection $\tau$ of the terms, it need not be the case that $\tau\big( U_j\pseu[\ast] \big) = U'_j\pseu[\ast]$.

Isomorphic stable index expressions represent not only equivalent unitary operations (which follows from the implicit summation convention), but equivalent circuit decompositions as well:
\begin{lemma}
	Let $C_1$ and $C_2$ be two sequences of unitary operators on a common set of qubits, composed from a common set of unitaries.
	If the stable index tensor representations of $C_1$ and $C_2$ are isomorphic, then $C_1$ and $C_2$ are congruent up to re-ordering of commuting operators.	
\end{lemma}
\begin{proof}
	Consider the coarsest equivalence relation $\cong$ on circuits which consist of permutations of the gates of $C_1$, such that $C \cong C'$ if $C$ differs from $C'$ by a transposition of commuting gates.
	By definition, the equivalence classes of $\cong$ are sets of circuits which are congruent up to re-ordering of commuting operators.
	Consider stable index tensor expressions $S$ and $S'$ arising from two unitary circuits $C$ and $C'$, where $S$ acts on the same index set as $S'$, and $S$ differs from $S'$ only by a transposition of two terms $U_1\pseu[\vec s_1: \vec a_1/\vec d_1]$ and $U_2\pseu[\vec s_2:\vec a_2/\vec d_2]$, where without loss of generality the former precedes the latter in $C$, and vice-versa in $C'$.
	Then, the corresponding gates $U_1$ and $U_2$ in $C$ occur with $U_1$ preceding $U_2$, and occur in $C'$ in the opposite order.
	Consider the index sets for these two operations in the stable index tensor expression:
	\begin{itemize}
	\item
		If the sequences of indices $(\vec s_1; \vec a_1; \vec d_1)$ and $(\vec s_2; \vec a_2; \vec d_2)$ do not overlap, the operations $U_1$	and $U_2$ act on disjoint sets of qubits, and so they commute.
	
	\item
		Suppose the sequences of indices $(\vec s_1; \vec a_1; \vec d_1)$ and $(\vec s_2; \vec a_2; \vec d_2)$ overlap, and let $v$ be an index occurring in both sequences.
		In the construction of $S$, because $v$ occurs in $U_2\pseu[\vec s_2:\vec a_2/\vec d_2]$, it cannot be a deprecated index in $U_1\pseu[\vec s_1:\vec a_1/\vec d_1]$\,; and because it occurs in $U_1\pseu[\vec s_1:\vec a_1/\vec d_1]$, it cannot be an advanced index in $U_2\pseu[\vec s_2:\vec a_2/\vec d_2]$.
		By the construction of $S'$, we similarly have that $v$ cannot be a deprecated index in $U_2\pseu[\vec s_2:\vec a_2/\vec d_2]$ or an advanced index in $U_1\pseu[\vec s_1:\vec a_1/\vec d_1]$.
		Then, $v$ is a stable index of both terms.
	\end{itemize}
	Therefore, the only qubits $v$ on which $U_1$ and $U_2$ both act are those whose standard basis states are preserved by both $U_1$ and $U_2$.
	Then, $U_1$ and $U_2$ commute, in which case $C \cong C'$.

	Let $S_1$ and $S_2$ be the stable index tensor representations of $C_1$ and $C_2$, where the isomorphism is given by an index relabelling map $\lambda$ and mapping of the terms $\tau$.
	Let $S'_2$ be the representation obtained by applying $\lambda$ to all of the indices in $S_2$\,: then $S_1$ and $S'_2$ consist of a product of the same terms in different orders.
	Without loss of generality, we will then suppose that $S_2$ and $S_1$ differ only by a reordering of terms.
	Because the terms of $S_1$ differ from those of $S_2$ by a permutation, by induction on the decomposition of this permutation into transpositions, we therefore have $C_1 \cong C_2$.
\end{proof}
Rearranging the terms of a stable index tensor expression for a circuit not only preserves the meaning (by summation over indices which are both advanced and deprecated) as a unitary operator, but also preserves structural information about the circuit itself, while discarding information about the ordering of commuting operators.
As a consequence, stable index notation facilitates identification of congruent circuits; and we may freely re-arrange the terms of such an expression while preserving a congruence class of unitary circuits.

However, not all congruent circuits have isomorphic tensor index expansions: congruent circuits may still yield non-isomorphic stable index expressions when the common operands of two commuting gates do not have their standard basis states preserved.
\begin{example}
	The two circuits $\cX_{c,b} \; \cX_{a,b} \; H_a$ and $\cX_{a,b} \; \cX_{c,b} \; H_a$ on three qubits $a,b,c$ have the following respective stable index expressions:
	\begin{align}
			\cX\pseu[c:b_3/b_2] \cX\pseu[a_1:b_2/b_1] H\pseu[:a_1/a_0]
		&&
			\text{and}
		&&
			\cX\pseu[a_1:b_3/b_2] \cX\pseu[c:b_2/b_1] H\pseu[:a_1/a_0]	\;.
	\end{align}
	These two expressions are non-isomorphic: in the first expression, the leftmost factor contains the indices $c$ (which is neither advanced nor deprecated in the entire expression) and $b_3$ (which is not deprecated in the entire expression); the second expression does not contain any factor with these properties.
	The inequivalence of the two expressions above is due to the fact that the two operations $\cX_{a,b}$ and $\cX_{b,c}$ preserve the basis $\ket{\spm \sfx}$ on $b$, rather than the basis $\ket{\spm \sfz}$.
\end{example}

One solution to this would be to diagonalize every element of $U$ of a circuit, decomposing them into diagonal operations flanked by ``virtual'' unitaries performing a some change of basis from the standard basis to the eigenbasis of $U$, essentially representing changes in reference frame rather than ``real'' unitary transformations performed by the circuit.\footnote{%
It is plausible that the ``virtual'' gates may represent in some sense \emph{actual} work that would be required to change the Hamiltonian of a physical system, in order to perform time-dependent unitary evolution in some physical implementations.
However, such issues lie outside the scope of this thesis.}
Interpreting the ``virtual'' unitaries strictly as a change of reference frame rather than as part of the evolution of the system, the diagonal representations of each gate $U$ would then represent preservation of the eigenbasis of $U$ in place of the standard basis.
For sequences of operators which commute, a judicious choice of change-of-basis operators would then allow the diagonalized gates to share stable indices in common.
However, we can avoid such decompositions into ``real'' and ``virtual'' unitary operations if we consider elementary gate sets for which the commutation relations\footnote{%
In this context, by ``commutation relations'' we mean the relation of \emph{whether} or not two unitaries commute, for all pairs of unitaries in the set, without any further comment on \eg\ the values of commutators of these operators.} 
are sufficiently restricted:
\begin{definition}
	\label{def:parsimonious}
	A set $S$ of unitary operations is \emph{parsimonious} if the eigenvectors of every multi-qubit gate in $S$ are standard basis states, and if distinct gates in $S$ which act on the same qubits commute if and only if their eigenvectors are standard basis states.\footnotemark
\end{definition}\footnotetext{%
The title of~\cite{DKP06} predates this definition of ``parsimonious'' by four years, and does not refer to the concept we have defined here; however, as we show immediately below, the set of unitary operations described there is parsimonious in this sense.}
\label{discn:elemParsimonious}
For example, for any set $\A \subset \R$, the two universal sets of unitaries $\ens{H, R_\sfz(\alpha),\cZ}_{\alpha \in \A}$ and $\ens{J(\alpha), \cZ}_{\alpha \in \A}$ described in Section~\ref{sec:particularUnitaries} are parsimonious.
This is evident for the former set of unitaries; for the latter, it is sufficient to note that
\begin{align}
		J(\alpha)\herm J(\beta)\herm J(\alpha)J(\beta)
	\;\;=&\;\;
		R_\sfz(-\alpha) \,H R_\sfz(-\beta) \,H \;\! H \;\! R_\sfz(\alpha) \,H R_\sfz(\beta)
	\notag\\=&\;\;
		R_\sfz(-\alpha) R_\sfx(\alpha-\beta) R_\sfz(\beta)	\;,
\end{align}
which is equal to $\idop_2$ if and only if $\alpha = \beta$.

The restrictions imposed on parsimonious sets of unitary operations make it simple to determine whether two operations on some collection of qubits commute.
This allows us to prove the following:
\begin{lemma}
	Let $C_1$ and $C_2$ be two unitary circuits composed from a common parsimonious set of unitaries.
	If $C_1$ is congruent to $C_2$ up to re-ordering of commuting operators, then the stable index representations of $C_1$ and $C_2$ will be isomorphic.
\end{lemma}
\begin{proof}
	Let $K_0 \cong K_1 \cong \cdots \cong K_\ell$ be a sequence of circuits which differ only by a transposition of two commuting gates, where $C_1 = K_0$ and $C_2 = K_\ell$.
	We may show by induction on $\ell$ that the stable index tensor expressions for $C_1$ and $C_2$ are isomorphic by induction on $\ell$ by proving the case $\ell = 1$.

	Let $S_0$ and $S_1$ be the stable index representations of $K_0$ and $K_1$.
	Up to a relabeling, we may assume that the indices of $S_0$ and $S_1$ are all of the form $v_0, v_1, \ldots$ for different qubits $v$ acted on by $K_0$ and $K_1$: we set $v_0$ to be the first index corresponding to a given qubit in either $S_0$ or $S_1$, and $v_{j+1}$ to be an advanced index corresponding to each deprecated index $v_j$ for $j \in \N$.
	Let $U_1$ and $U_2$ be the pair of gates whose order differs between the two circuits; and consider the truncation $S'_0$ of $S_0$ just prior to the terms corresponding to $U_1$ and $U_2$, and similarly for $S'_1$.
	Because $K_0$ and $K_1$ differ only by a transposition of commuting gates $U_1$ and $U_2$, $S'_0 = S'_1$.
	Let $S''_0$ and $S''_1$ be the truncations of $S_0$ and $S_1$ just after the terms corresponding to $U_1$ and $U_2$\,:
	\begin{itemize}
	\item
		If $U_1$ and $U_2$ act on disjoint sets of qubits, the terms corresponding to $U_1$ and $U_2$ in $S''_0$ may differ from those in $S''_1$ only by the indices which are advanced in each; and as the indexing scheme we have fixed is the same for each, and the indices which will be advanced for $U_1$ and $U_2$ are independent of each other in both expressions, $S''_0$ differs from $S''_1$ only by a transposition of those two terms.

	\item
		If $U_1$ and $U_2$ both have the standard basis as their eigenbases, their corresponding terms in $S_0$ and $S_1$ only have stable indices, and therefore do not advance any new indices.
		Then the term in $S_0$ and in $S_1$ corresponding to $U_1$ are the same, and similarly for $U_2$\,; the difference between $S''_0$ and $S''_1$ is a transposition of those two terms, and so they are isomorphic.

	\item
		If $U_1$ and $U_2$ perform the same single-qubit operation, there will be some common tensor index $u_j$ which is current for that qubit prior to the application of $U_1$ and $U_2$ in both $S'_0$ and $S'_1$. 
		Then, the terms corresponding to $U_1$ and $U_2$ in $S_0$ will be $U_2\pseu[:u_{j+2}/u_{j+1}]U_1\pseu[:u_{j+1}/u_j]$, and the corresponding terms in $S_1$ will be $U_1\pseu[:u_{j+2}/u_{j+1}]U_2\pseu[:u_{j+1}/u_j]$ for some indices $v''$ and $w''$.
		Given that $U_1$ and $U_2$ perform the same operation, we then have $S''_0 = S''_1$.
	\end{itemize}
	Because the operations of $K_0$ and $K_1$ are identical after $U_1$ and $U_2$, the differences between $S_0$ and $S_1$ after the terms corresponding to $U_1$ and $U_2$ can only arise from the indices advanced by each term; because we have fixed the indexing scheme, the only difference between $S_0$ and $S_1$ is then the (possibly trivial) transposition of those two terms.
	Thus, $S_0$ and $S_1$ are isomorphic.
\end{proof}
Thus, for unitary circuits constructed from a parsimonious set of gates, isomorphism of stable index tensor expressions is equivalent to congruence up to rearrangements of commuting gates.
We will take advantage of this feature of this tensor notation in later chapters of the thesis.

\noindent We conclude our discussion of stable tensor index expressions with two remarks.
\begin{enumerate}
\item
	\label{item:nonstandardRepsOrderRecovery}%
	\textbf{Recovering the order of unitary operations.}~~
	As we have noted, isomorphism classes of stable tensor index expressions (for a parsimonious set of operations) correspond to congruency classes of unitary circuits under permutations of commuting gates.
	In any such circuit, the order of \emph{non-commuting} gates can be determined from the indexing: for a tensor index $v_j$ which is deprecated in a given stable index expression and for $v_{j+1} = f(v_j)$, any operation acting on $v_{j+1}$ must occur after any (other) operation in the circuit acting on $v_j$\,, with the only operator acting on both being the operator which deprecates $v_j$ and correspondingly advances $v_{j+1}$\,.
	The complement of the relation ``$U$ must occur after $V$'' is then a pre-order\footnote{%
		\label{fn:preorder}%
		A pre-order on a class $S$ is a binary relation which is reflexive ($x \preceq x$ for all $x \in S$) and transitive ($x \preceq y$ and $y \preceq z$ implies $x \preceq z$ for all $x,y,z \in S$).
 		An instructive example from quantum information processing is the binary relation ``$\to*$'' on bipartite pure states $\ket{\Psi} \in \cH[A] \ox \cH[B]$, where $\ket{\Phi} \to* \ket{\Psi}$ if and only if $\ket{\Psi}$ can be obtained by LOCC on $\cH[A]$ and $\cH[B]$ from $\ket{\Phi}$. 
	 	Every state can be obtained via LOCC from itself, and if $\ket{\Phi} \to* \ket{\Psi}$ and $\ket{\Psi} \to* \ket{\Upsilon}$, then $\ket{\Phi} \to* \ket{\Upsilon}$ by composing protocols.
	 	There are also pairs of states with $\ket{\Phi} \to* \ket{\Psi}$ and $\ket{\Psi} \not\to* \ket{\Phi}$, \eg\ for $\ket{\Phi}$ an entangled state and $\ket{\Psi}$ a product state; and there are distinct states with $\ket{\Phi} \to* \ket{\Psi} \to* \ket{\Phi}$, \ie\ if $\ket{\Phi}$ and $\ket{\Psi}$ only differ by local unitaries.
	 	(A result of Nielsen~\cite{Nielsen99} shows that that there also exist pairs of states such that $\ket{\Phi} \not\to* \ket{\Psi}$ and $\ket{\Psi} \not\to* \ket{\Phi}$, so this pre-order exhibits the complete spectrum of possible relations between pairs of elements.)}
	which characterizes the \emph{possible} orderings of the operations in a stable-index expression, up to commuting operations.

\item
	\textbf{Induced combinatorial structures on the indices.}
	The operations of a stable-index tensor expression also induce structures on the tensor indices.
	From any stable-index tensor expression $S$, we may construct a graph $G$ whose vertices are the indices of $S$, and where $v w \in E(G)$ for two indices $v$ and $w$ if and only if there exists an operator in $S$ involving both $v$ and $w$\,: we may call such a graph an \emph{interaction graph} for a stable index expression.
	If $I$ is the set of non-advanced indices of $S$, and $O$ is the set of non-deprecated indices, the function $f$ which maps each deprecated index to an advanced index then describes vertex-disjoint $I$\thru$O$ directed paths in $G$.

	The interaction graph $G$ encodes much of the information of the original expression $S$: in particular, from $G$ and the function $f$, we may recover the order in which tensor indices are deprecated (if they are deprecated at all) in any ordering of the terms of of $S$ consistent with a conventional circuit notation.
	Expressed as a partial order $\preceq$, we clearly have $v \preceq f(v)$, as $f(v)$ is advanced in the same term which deprecates $v$; and for any $w$ which is adjacent to $f(v)$, we have $v \preceq w$, as $f(v)$ and $w$ are acted on by an operator $U$ in common and therefore must be current at the same time at the stage where $U$ is performed.
	These same graph and ordering structures arise in the one-way model, and form the core of the topic of Chapter~\ref{Semantics for unitary one-way patterns}.
\end{enumerate}
\label{sec:nonstandardRepnsEnd}

\section{Classically controlled unitary circuits}

We may extend unitary circuits to obtain a description of quantum computation which involves \emph{classical control}: that is, dependency of operations on classical bits, which in particular may be the results of measurements.
In this section, we briefly consider such models of quantum computation, and their relationship to unitary circuits without classical control.
This will also allow us to define operations which will be necessary to describe the \emph{one-way measurement model} in Chapter~\ref{One-way measurement-based computation}.

We may define a ``classically controlled'' CPTP operation using the following construction:
\begin{definition}
	For a sequence of bit-registers $\vec c = (c_j)_{j = 1}^n$ and a map $f: \ens{0,1}^n \to* S$ for some set $S$, define the completely positive (but \emph{not} trace-preserving) operators $\Cond^{f,s} : \Prob(\vec c) \to \Prob(\vec c)$ for each $s \in S$, given by
	\begin{gather}
			\Cond^{f:s}(\rho)
		\;\;=\;\;
			\Pi_s \,\rho\, \Pi_s \;,
		\qquad\text{where}\quad
			\Pi_s
		\;\;=
			\sum_{\vec x \in f\inv(s)} \ket{\vec x}\bra{\vec x}	.
	\end{gather}
	For sequence of generic registers $\sigma$, $\alpha$, and $\delta$, a \emph{classically controlled} CPTP map is a sum of superoperators, of the form
	\begin{subequations}
	\begin{gather}
			\Upphi_{\sigma:\alpha/\delta}^{f(\vec c)} : \Prob(\vec c) \ox \D(\sigma; \delta) \to \Prob(\vec c) \ox \D(\sigma; \alpha)
		\\[1ex]
			\Upphi^{f(\vec c)}_{\sigma:\alpha/\delta}(\rho)
		\;\;\;=
			\sum_{\vec x \in \ens{0,1}^n} \Cond^{f:\vec x}_{\vec c} \ox \Uppsi_{\sigma:\alpha/\delta}^{(\vec x)}	\;,
	\end{gather}
	\end{subequations}
	for some family of \emph{conditionally applied CPTP maps} $\ens{\Uppsi\supp{s}}_{s \in S}$ which all have the same domain and range.
\end{definition}
A common example of a classically controlled operation is a \emph{classically-controlled not} operation, where we perform an $X$ operation on a qubit depending on the value of a classical bit:
\begin{align}
		\Upphi_{c,q}(\rho)
	\;\;=&\;\;
		\Big[ \ket{0}\bra{0}_c \ox \idop_q \Big] \rho \Big[ \ket{0}\bra{0}_c \ox \idop_q \Big]
		\;+\;
		\Big[ \ket{1}\bra{1}_c \ox X_q \Big] \rho \Big[ \ket{1}\bra{1}_c \ox X_q \Big]
	\notag\\=&\;\;
		\bigg[\Big( \Cond^{\id:0}_c \ox \idop_q \Big) + \Big( \Cond^{\id:1}_c \ox \Xcorr{q}{}\Big) \bigg](\rho)	,
\end{align}
where $\Xcorr{}{}: \D(2) \to \D(2)$ is given by $\Xcorr{}{}(\rho) = X \rho X$.
It is easy to see that classically controlled CPTP maps $\Upphi$ are indeed CPTP maps, by forming the Kraus operators of $\Upphi$ from the tensor products of the $\Cond^{f:s}$ operations and the conditionally applied maps $\Uppsi\supp{s}$ which define $\Upphi$.

We may distinguish two classes of conditionally controlled CPTP maps, which will play an important role in this thesis:
\begin{definition}
	A \emph{classically controlled unitary} operation is a conditionally controlled CPTP map $\Upphi$ whose conditionally applied maps $\ens{\Uppsi\supp{s}}_{s \in S}$ are all unitary operations.
	Similarly, a \emph{classically controlled measurement} operation is a conditionally controlled CPTP map $\Upphi$ whose conditionally applied maps $\ens{\Uppsi\supp{s}}_{s \in S}$ are all measurement operations.
\end{definition}
Our interest in unitary circuit models (as opposed to models with classical control) is largely due to the convenience of omitting explicit reference to classical control in the analysis of quantum algorithms.
However, as instances of computational problems are usually described as being represented with classical information, a robust experimental set-up which is capable of solving different instances of a problem will very probably control which particular unitary evolution occurs by classical control of the unitary evolution.

\begin{definition}
	A model of quantum computation is a \emph{classically controlled unitary circuit model} if its elementary gates consist only of operations which
	\begin{alphanum}
	\item
		prepare some number of qubits in a fixed initial state;

	\item
		perform a (possibly classically-controlled) unitary operation on some number of qubits; or

	\item
		perform a (possibly classically-controlled) complete (orthogonal \& destructive) measurement on some number of qubits.
	\end{alphanum}
\end{definition}

\subsection{Classically controlled extensions, and deferred measurement}

For a given unitary circuit model defined by some gate-set \Gates, it is common to extend to a classically controlled unitary circuit model by including classically controlled versions $\Upupsilon^{c}$ of the unitary gates $\Upupsilon \in \Gates$, which perform either $\Upupsilon$ or the identity superoperator depending on the value of a single classical bit $c$.
We may also consider classically controlled versions $\mathsf K^c$ of measurement gates, where $\mathsf K^{\ssstyle 0}, \mathsf K^{\ssstyle 1} \in \Gates$ are different measurement gates which are performed on a given set of qubits depending on the value of a classical bit $c$.\label{discn:classicalControl}
Such an extended gate set, including controlled versions of unitaries and measurements in $\Gates$, we will call $\textsc{cc}(\Gates)$.\footnote{\label{fn:classicalControlFn}%
We may easily extend this definition to $\textsc{cc}(\Gates)$ when $\Gates$ does not define a unitary circuit model, by considering any pair of conditionally applied CPTP maps $\Uppsi^0, \Uppsi^1 \in \Gates$\,; however, when $\Uppsi^0$ and $\Uppsi^1$ are unitaries, it is conventional to require that one of them be the identity superoperator.}
In such a circuit model, classically controlled CPTP maps may either perform the identity or perform a non-trivial operation on a set of qubits, depending on the results of measurements performed in the middle of the circuit.
This is the primary qualitative difference between unitary circuit models and classically controlled versions of those models.

It is also common to transform classically-controlled circuits into unitary circuits without classical control, following a folklore result reported in~\cite{BV97} and attributed to \cite{Bernstein97}.
For a gate-set $\Gates$, we define a gate-set $\textsc{qc}(\Gates)$ as follows:
\begin{itemize}
\item
	If $\Gates$ doesn't include the measurement operator $\Meas{}{\sfz}$, we include it in $\textsc{qc}(\Gates)$\,;

\item
	For any unitary $\Upupsilon \in \Gates$ performing $\rho \mapsto U \rho U$ for $\rho \in \D(2^n)$, we include a (\emph{coherently}) \emph{controlled} version $\ctrl\Upupsilon \in \textsc{qc}(\Gates)$, which performs the mapping
	\begin{gather}
		\label{eqn:controlledU}
			\ctrl\Upupsilon(\rho)
		\;\;=\;\;
			\Big( \ket{0}\bra{0} \ox \idop_2^{\ox n} \;+\; \ket{1}\bra{1} \ox U \Big)
			\rho
			\Big( \ket{0}\bra{0} \ox \idop_2^{\ox n} \;+\; \ket{1}\bra{1} \ox U\herm \Big);
	\end{gather}

\item
	For any complete measurement $\mathsf K$ given by
	\begin{gather}
			\mathsf K(\rho)
		\;\;=\;\;
			\sum_{\vec x \in \ens{0,1}^n} \ket{\vec x}\bra{\psi_{\vec x}} \rho \ket{\psi_{\vec x}}\bra{\vec x}
	\end{gather}
	for some orthonormal basis $\ens{\big.\ket{\psi_{\vec x}}\big.}_{\vec x \in \ens{0,1}^n}$ of $\cH^{\ox n}$, we include the unitary change of basis operation $\mathsf B^{\mathsf K} \in \textsc{qc}(\Gates)$, defined by
	\begin{gather}
			\mathsf B^{\mathsf K{}{}}(\rho)
		\;\;=\;\;
			B \rho B\herm\,,
		\qquad\text{where}\quad
			B
		\;\;=
			\sum_{\vec x \in \ens{0,1}^n} \ket{\vec x}\bra{\psi_{\vec x}}	\,.
	\end{gather}
	We also include a coherently controlled version $\ctrl \mathsf B^{\mathsf K} \in \textsc{qc}(\Gates)$, defined in terms of $B^{\mathsf K}$ by~\eqref{eqn:controlledU}.

\item
	Every state-preparation, unitary operation, or trace-out operation in $\Gates$ is included in $\textsc{qc}(\Gates)$, but not any measurement operation except for $\Meas{}{\sfz}$.
\end{itemize}
Having defined the gate-set $\textsc{qc}(\Gates)$, we may translate any circuit composed from the gate set $\textsc{cc}(\Gates)$ into a ``coherent'' version in the gate-set $\textsc{qc}(\Gates)$ by replacing classical input registers with quantum input registers (thereby extending the domain of the input), and commuting all of the measurement operations to the end of the circuit, as follows:
\begin{enumerate}
\item
	Decompose each measurement operation $\mathsf K$ on $n$ qubits (other than $\Meas{}{\sfz}$) into a change of basis, from the measurement basis of $\mathsf K$ to the standard basis, followed by a measurement in the standard basis:
	\begin{align}
			\mathsf K_{\alpha/\delta}
		\;\;=\;\;
			\sqparen{\bigotimes_{v \in \delta} \Meas{v}{\sfz}} \circ \mathsf B^{\mathsf K}_{\delta}	\;,
	\end{align}
	where we attribute classical bits $\s[v] \in \alpha$ to each $v \in \delta$, in sequence.
	We similarly replace controlled measurement operators with \emph{classically controlled} changes of basis followed by measurement in the standard basis:
	\begin{align}
			\mathsf K^{c}_{\alpha/\delta}
		\;\;=\;\;
			\sqparen{\bigotimes_{v \in \delta} \Meas{v}{\sfz}} \circ \Big(\mathsf B^{\mathsf K^1}\Big)^{\!c}_{\!\delta}	\,
																				\circ \Big(\mathsf B^{\mathsf K^0}\Big)^{\!1-c}_{\!\delta}	\;.
	\end{align}

\item
	We substitute any classically-controlled operation $\Upupsilon^c$ depending on a classical input bit $c$, including instances of such operations arising from measurements in the previous step, by a (coherently) controlled unitary $\ctrl\Upupsilon$ acting on the input qubit which has replaced the input bit;

\item
	For each qubit $v$ and any unitary $\Upupsilon \in \Gates$, we perform the following substitution repeatedly until there are no further classically controlled unitaries remaining:
	\begin{gather}
			\Upupsilon^{\s[v]}_{\sigma} \; \Meas{v}{\sfz} 
		\;\;\mapsto\;\;
			\Meas{v}{\sfz} \; \ctrl\Upupsilon_{v, \sigma}
	\end{gather}
\end{enumerate}
Because each classically controlled operation in a $\textsc{cc}(\Gates)$ circuit acts either on an input bit or a bit allocated by a measurement, the above procedure will eventually remove every classically controlled unitary (either from the original circuit or from the decomposition of classically controlled measurements), and replace it with a coherently controlled unitary, thereby producing a unitary circuit.
Replacing the classical registers at the input with equivalent quantum registers also leaves the effect of the circuit as a superoperator unchanged on the original input space: we refer to this as a \emph{coherent extension} of the original circuit.
In summary:
\begin{lemma*}[Principle of Deferred Measurement \cite{Bernstein97,BV97}]
	\label{principleDeferredMeas}
	For any gate set $\Gates$ giving rise to a unitary circuit model, every classically controlled circuit $C$ in a gate set $\textsc{cc}(\Gates)$ can be transformed to a unitary circuit $Q$ in the gate set $\textsc{qc}(\Gates)$, where $Q$ is a coherent extension of $C$, and where the difference in the complexity of $Q$ and $C$ is bounded above by the number of measurements in $C$.
\end{lemma*}
The principle of deferred measurement is what allows us in practise to consider quantum computation in terms of unitary circuits.
Further, if $\Gates$ is universal for quantum computation, we may further decompose the operations of $\textsc{qc}(\Gates)$ into operations of $\Gates$; the Solovay-Kitaev theorem then places an upper bound on the factor of increase of the resulting $\Gates$ circuit over the original $\textsc{cc}(\Gates)$ circuit.
Therefore, we do not incur very large computational costs by restricting ourselves to unitary circuit models, rather than always explicitly considering classically controlled unitary circuits.

\section{Clifford circuits and the stabilizer formalism}
\label{sec:cliffordCircuits}

A final model of quantum computation which we will consider in this Chapter is one known \emph{not} to be universal for quantum computation; and furthermore, one which is known to be efficiently simulatable on classical computers.
This model is a controlled-unitary circuit model whose unitary operations are restricted to what is called the \emph{Clifford group}, and whose preparation/measurement operations are similarly restricted.
In this section, we present the results of~\cite{GotPhD} about this model of computation, as it provides an illustrative pre-amble to measurement-based computation.

\subsection{The Pauli and Clifford groups}

\begin{definition}
	The \emph{Pauli group} on $n$ qubits is the group of operators $\cP^{\ox n}$, consisting of $n$-fold tensor products of the group $\cP = \gen{\idop_2\,, X, Y, Z} = \gen{i\idop_2\,, X, Z}$ generated by the Pauli operators (see Definition~\ref{def:Pauli}).
	The \emph{Clifford group} on $n$ qubits is the normalizer of the Pauli group in $\U(2^n)$.
\end{definition}

The following Lemma is proven for the set $\ens{H,R_\sfz(\pion 2), \cX}$ by~\cite{Gottesman98}, and therefore also holds as a corollary for the sets $\smash{\ens{H, R_\sfz(\alpha), \cZ}_{\alpha \in \frac{\pi}{2}\Z}}\,$, and $\smash{\ens{J(\alpha), \cZ}_{\alpha \in \frac{\pi}{2} \Z}}\,$ by the remarks preceding Lemma~\ref{lemma:universalUnitaries}:

\begin{lemma}
	For any $n \ge 1$, the Clifford group on $n$ qubits can be generated by the following sets of unitaries together with tensor products with scalar multiples of the identity: $\ens{H,R_\sfz(\pion 2), \cX}$, $\smash{\ens{H, R_\sfz(\alpha), \cZ}_{\alpha \in \frac{\pi}{2}\Z}}\,$, and $\smash{\ens{J(\alpha), \cZ}_{\alpha \in \frac{\pi}{2} \Z}}\,$.
\end{lemma}

A Clifford group operation $U$ can be characterized (up to an unimportant scalar factor) by the action induced on the Pauli group by conjugation by $U$.
Because $U Y_v U\herm = U (iX_v Z_v) U\herm = i (UX_vU\herm)(UZ_vU\herm)$ for a $Y$ operation on any single qubit $v$, and $U \idop U\herm = \idop$, we may further reduce this to how $U$ transforms tensor products of the operators $X$ and $Z$ on single qubits by conjugation.
This then leads to a natural representation of Clifford group operations as CPTP maps, albeit acting as a transformation of the zero-trace, non-positive tensor products of $X$ and $Z$ operators with tensor powers of $\idop_2$\,.
For instance, it is easy to verify by calculation using the CPTP maps defined by~\eqref{eqn:cptpElemUnitaries} that
\begin{subequations}
\label{eqn:cliffordMapping}
\begin{align}
		\sfH(X)	\;=&\;	Z	\,,								&	\sfH(Y)	\;=&\;	-Y	\,,					&	\sfH(Z)	\;=&\;	X	\,;
	\\[2ex]
		\sfR^{\pi/2}(X)	\;=&\;	Y \,,		&	\sfR^{\pi/2}(Y) \;=&\; -X	\,,			&	\sfR^{\pi/2}(Z)	\;=&\;	Z\,;
	\\[2ex]
		\begin{split}
			\Ent{}(X \ox \idop_2)\;=&\;	X \ox Z\,,
		\\
			\Ent{}(\idop_2 \ox X) \;=&\;	Z \ox X\,,
		\end{split}
	&
		\begin{split}
			\Ent{}(Y \ox \idop_2)\;=&\;	Y \ox Z\,,
		\\
			\Ent{}(\idop_2 \ox Y) \;=&\;	Z \ox Y\,,
		\end{split}
	&
		\begin{split}
			\Ent{}{}(Z \ox \idop_2)	\;=&\;	Z \ox \idop_2\,,
		\\
			\Ent{}{}(\idop_2 \ox Z) \;=&\;	\idop_2 \ox Z\,.
		\end{split}
\end{align}
\end{subequations}
(Note also that the middle column of these equations can be easily computed from the left and right columns by $Y = iXZ = -iZX$.)
The way in which compositions of these CPTP maps transform $\cP^{\ox n}$ can then be easily computed.

The Clifford group acts \emph{transitively} on elements of the Pauli group with eigenvalues $\pm 1$: that is, for any two Hermitian Pauli group elements $P_1, P_2 \in \cP^{\ox n} \setminus \ens{\pm \idop_2^{\ox n}}$\,, there is a Clifford group operation $U$ such that $P_2 = U P_1 U\herm$.
This can easily be seen from the fact that for any Pauli operator $P$, we may apply $\sfR^{\pi/2}$ operations to obtain an operator $P'$ for which at least one qubit $a$ is acted on by $X$, and the others with $Z$\,; and then perform $\Ent{}$ operations on pairs of qubits $(a,v)$ for which $v$ is acted on by $P'$ by a $Z$ operation, which will yield the operator $X_a$ acting only on $a$.
By defining the mapping
\begin{subequations}
\begin{gather}
		\mathsf{HSWAP}_{a,b} \;=\; \Ent{a,b} \sfH_a \sfH_b \Ent{a,b} \sfH_a \sfH_b \Ent{a,b}	\;,
\intertext{which for product operators $\rho \ox \sigma \in \L(\cH[2] \ox \cH[2])$ performs the mapping}
		\mathsf{HSWAP}(\rho \ox \sigma) \;=\; \sfH(\sigma) \ox \sfH(\rho)	\;,
\end{gather}
\end{subequations}
we can see that performing an $\mathsf{HSWAP}_{a,n}$ on $a$ and the $n\th$ qubit, we obtain operation $Z_n$ acting only on the $n\th$ qubit, up to a scalar factor of $\pm 1$ (not directly attributable to an operation on any given qubit).
Finally, we may transform $-Z_n$ to $Z_n$ by conjugation by $X_n \,=\, H_n R_\sfz(\pion 2)_n^2 H_n$\,; and so, any Pauli operator $P \in \cP^{\ox n}$ may be mapped by a Clifford group operation to $Z_n = \idop_2^{\ox n-1} \ox Z$.
We can do this for any Pauli operator; furthermore, by performing such a Clifford transformation in reverse, we may also obtain an arbitrary Pauli operator from a $Z$ on the $n\th$ qubit.
This proves our claim of transitivity.

\subsection{Stabilizer groups and stabilizer codes}

It is easy to verify that any two Pauli operators $P \in \cP = \gen{\idop_2, X, Y, Z}$ either commute or anticommute with each other;  and that therefore the same holds for two elements of $\cP^{\ox n}$.
We may then consider subgroups of $\cP^{\ox n}$ generated by commuting sets of operators.
Any such group may be characterized by a set of independent generators $\ens{S_1,\ldots,S_{n-k}}$ for some $k \le n$\,.
As $\ens{X,Y,Z}$ are all self-inverse, such a group will be of size $2^{n-k}$, and in particular isomorphic to a subgroup of $\Z_2^n$.

We may also show that for any $0 \le k \le n$, the Clifford group acts transitively on abelian subgroups of the Pauli group of size $2^{n-k}$ whose operators all have $+1$ eigenspaces (which we call \emph{Pauli stabilizer groups}, often simply \emph{stabilizer groups}).
We may show this by induction on $n - k$, as follows.
The case of $n = k$ is trivial, as there is only one such group, $\ens{\idop_2^{\ox n}}$.
For $n > k$, let $\ens{S_1,\ldots,S_{n-k}}$ be a set of generators for the group: we may reduce the problem by considering a Clifford group operation $U$ such that $U S_{n-k} U\herm = Z_n$, using the transitivity of the action of the Clifford group on the non-trivial Hermitian elements of $\cP^{\ox n}$
Because two Pauli operators $P_1$ and $P_2$ commute if and only if $U P_1 U\herm$ and $U P_2 U\herm$ commute, it is easy to show that the operators $U S_j U\herm$ must act on the $n\th$ qubit with either the identity or a $Z$ operation.
For those generators $S_j$ such that the latter holds, define $S'_j = S_j S_{n-k}$\,; then, $U S'_j U\herm$ acts on the $n\th$ qubit with the identity.
If we define $S'_j = S_j$ for the other stabilizers, $\{ S'_j \}_{j = 1}^{n-k}$ generates the same abelian group, and by construction is transformed via conjugation by $U$ to a set of operators $\ens{T_1 \ox \idop_2\,,\; \ldots\,,\; T_{n-k-1} \ox \idop_2\,,\; Z_n}$.
The first $n-k-1$ of these generate an abelian subgroup on the first $n-1$ qubits, which by induction we may transform by a Clifford group operation to the group generated by $\ens{Z_{k+1}, \ldots, Z_{n-1}}$.
Again, this transformation may be reversed to yield an arbitrary stabilizer code of size $2^{n-k}$; then any such abelian group of size $2^{n-k}$ to be mapped to any other.

A set of Pauli operators which commutes, by that very fact, shares a common set of eigenvectors.
The group $\gen{Z_{k+1}, Z_{k+2}, \ldots, Z_n}$ in particular admits a $2^k$-dimensional subspace of $\cH^{\ox n}$ which are $+1$-eigenvectors of the entire group (specifically, the set of states of the form $\ket{\psi} \ox \ket{0}^{\ox n-k}$ for arbitrary $\ket{\psi} \in \cH^{\ox k}$); by transitivity of the Clifford group on stabilizer groups of similar size, every stabilizer group of size $2^{n-k}$ which does not include $-\idop_2^{\ox n}$ has a $2^k$-dimensional space of joint $+1$-eigenvectors.
We say that a vector $\ket{\psi}$ is \emph{stabilized} by an operator $M$ if $\ket{\psi} = M\ket{\psi}$\,, \ie\ if it is a $+1$-eigenvector of $M$\,, which motivates the following terminology:

\begin{definition}
	A \emph{stabilizer code} $\cE$ on $n$ qubits is the space of joint $+1$-eigenvectors of a stabilizer group on $n$ qubits.
	In particular, a \emph{stabilizer state} on $n$ qubits is a state $\rho = \ket{\psi}\bra{\psi}$, where $\ket{\psi}$ is a joint $+1$-eigenvector of a commuting group of $2^n$ Pauli operators on $n$ qubits.
	The \emph{stabilizer group of the code} is the group for which the code is the joint $+1$-eigenspace.
\end{definition}

\begin{example}
 	Let $S \in \ens{X,Y,Z}$, and let $\ket{\splus \sfs}$ represent the corresponding state vector in $\ens{\ket{\splus \sfx}, \ket{\splus \sfy}, \ket{\splus \sfz}}$.
 	It is easy to verify that $\ket{\splus \sfs}$ is stabilized by $S$\,; then, for any sequence of non-trivial Pauli operators $\big(S_j\big)_{j = 1}^n$, where each operator $S_j$ also acts on qubit $j$ and stabilizes the state vector $\ket{\splus \sfs_j}$, the product state
 	\begin{gather}
 			\ket{\psi}
 		\;\;=\;\;
 			\bigotimes_{j = 1}^n \;\ket{\splus \sfs_j}
 	\end{gather}
 	is stabilized by the abelian group $\gen{S_1,\,\ldots,\,S_n}$.
 	As a special case, the state $\ket{0}^{\ox n} = \ket{0} \ox \cdots \ox \ket{0}$ is stabilized by the group $\gen{Z_1, \ldots, Z_n}$.
 \end{example}

\subsection{The stabilizer formalism}
\label{sec:stabilizerFormalism}

As stabilizer codes can be characterized by a short list of generators for its stabilizer group, the state space of a system which is initially prepared in a pure state in some stabilizer code and then transformed by Clifford group operations can be efficiently computed.
We may extend this result further: consider the circuits which can be composed of the unitary CPTP maps $\sfH$, $\sfR^{\pi/2}$, and $\Ent{}$; the state preparation maps
\begin{subequations}
\begin{align}
		\New{v}{\sfx}(\rho)
	\;\;=&\;\;
		\rho  \;\ox\; \ket{\splus \sfx}\bra{\splus \sfx}_{v}	\;,
	\\[2ex]
		\New{v}{\sfy}(\rho)
	\;\;=&\;\;
		\rho  \;\ox\; \ket{\splus \sfy}\bra{\splus \sfy}_{v}	\;,
	\\[2ex]
		\New{v}{\sfz}(\rho)
	\;\;=&\;\;
		\rho  \;\ox\; \ket{\splus \sfz}\bra{\splus \sfz}_{v}	\;;
\end{align}
\end{subequations}
the von Neumann measurement operations
\begin{subequations}
\begin{align}
		\sfP_{v:r}^\sfx(\rho)
	\;\;=&\;\;\;
	\text{\small$
		\ket{\splus \sfx}\bra{\splus \sfx}_v \rho \ket{\splus \sfx}\bra{\splus \sfx}_v \;\ox\; \ket{0}\bra{0}_r
		\;\;+\;\;
		\ket{\sminus \sfx}\bra{\sminus \sfx}_v \rho \ket{\sminus \sfx}\bra{\sminus \sfx}_v	\;\ox\; \ket{1}\bra{1}_r \;,
	$}\\[2ex]
		\sfP_{v:r}^\sfy(\rho)
	\;\;=&\;\;\;
	\text{\small$
		\ket{\splus \sfy}\bra{\splus \sfy}_v \rho \ket{\splus \sfy}\bra{\splus \sfy}_v \;\ox\; \ket{0}\bra{0}_r 
		\;\;+\;\;
		\ket{\sminus \sfy}\bra{\sminus \sfy}_v \rho \ket{\sminus \sfy}\bra{\sminus \sfy}_v	\;\ox\; \ket{1}\bra{1}_r  \;,
	$}\\[2ex]
		\sfP_{v:r}^\sfz(\rho)
	\;\;=&\;\;\;
	\text{\small$
		\ket{\splus \sfz}\bra{\splus \sfz}_v \rho \ket{\splus \sfz}\bra{\splus \sfz}_v \;\ox\; \ket{0}\bra{0}_r 
		\;\;+\;\;
		\ket{\sminus \sfz}\bra{\sminus \sfz}_v \rho \ket{\sminus \sfz}\bra{\sminus \sfz}_v	\;\ox\; \ket{1}\bra{1}_r \;;
	$}
\end{align}
\end{subequations}
and the corresponding destructive measurement operations,
\begin{subequations}
\label{eqn:pauliMeas}
\begin{align}
		\Meas{v}\sfx(\rho)
	\;\;=&\;\;\;
	\text{\small$
		\bra{\splus \sfx}_v \rho \ket{\splus \sfx}_v \;\ox\; \ket{0}\bra{0}_{\s[v]}
		\;\;+\;\;
		\bra{\sminus \sfx}_v \rho \ket{\sminus \sfx}_v	\;\ox\; \ket{1}\bra{1}_{\s[v]} \;,
	$}\\[2ex]
		\Meas{v}\sfy(\rho)
	\;\;=&\;\;\;
	\text{\small$
		\bra{\splus \sfy}_v \rho \ket{\splus \sfy}_v \;\ox\; \ket{0}\bra{0}_{\s[v]}
		\;\;+\;\;
		\bra{\sminus \sfy}_v \rho \ket{\sminus \sfy}_v	\;\ox\; \ket{1}\bra{1}_{\s[v]} \;,
	$}\\[2ex]
		\Meas{v}\sfz(\rho)
	\;\;=&\;\;\;
	\text{\small$
		\bra{\splus \sfz}_v \rho \ket{\splus \sfz}_v \;\ox\; \ket{0}\bra{0}_{\s[v]}
		\;\;+\;\;
		\bra{\sminus \sfz}_v \rho \ket{\sminus \sfz}_v	\;\ox\; \ket{1}\bra{1}_{\s[v]} \;.
	$}
\end{align}
\end{subequations}

\refstepcounter{footnote}\footnotetext{\addtocounter{footnote}{-1}%
To be precise, we are using here the slightly extended sense of the function $\Gates \mapsto \textsc{cc}(\Gates)$, described in the footnote (\ref{fn:classicalControlFn}) on page~\pageref{fn:classicalControlFn}, for when $\Gates$ does not define a unitary circuit model.}%
\begin{definition}
	For an infinite set $V$ of qubits and $B$ of bits, the \emph{Clifford circuit model} is the circuit model generated by $\Clifford = \textsc{cc}(\cG)$,\footnotemark\ where
	\begin{gather}
		\cG \;=\;
		\text{\small$
			\ens{ \sfH_v\,,\; \sfR_v^{\pi/2}\,,\; \Ent{u,v}\,,\; \New{v}{\sigma}\,,\; \sfP^\sigma_{v:r}\,,\; \Meas{v}{\sigma}
			\,\Big|\; \sigma \in \ens{\sfx, \sfy, \sfz}, ~r \in B, \text{~and~} v,w \in V}\,.
		$}
	\end{gather}
\end{definition}
Note that circuits in the $\Phases(\frac{\pi}{2}\Z)$ model which do not use trace-out operations are also circuits in the \Clifford\ model.

Assuming an arbitrary input state, the state-space of the output to a Clifford circuit conditioned on particular values of each bit in the circuit (and in particular, conditioned on particulars outcomes of each von Neumann measurement performed) may be efficiently computed by transforming generators of the stabilizer group of the system.
The techniques for doing this, described below, were presented by~\cite{GotPhD} and are referred to collectively as \emph{the stabilizer formalism}.
\label{discn:stabilizerFormalism}

The state preparation operators $\New{v}{\sigma}$ simply adjoin another qubit $v$ on which the existing generators act with the identity, and also add another generator $S_v$ stabilizing the state $\ket{\splus \sigma}_v$.
As noted before, the effect of the unitary operators $\sfH$, $\sfR^{\pi/2}$, and $\Ent{}{}$ may be efficiently computed on each generator.
This leaves the von Neumann measurements $\sfP_{v:r}^\sigma$ and destructive measurement $\Meas{v}\sigma$\,, which we describe following the treatment of~\cite{GotPhD}.
Consider a Pauli operator $P_v$ on a single qubit $v$, for $P \in \ens{X, Y, Z}$.
For the corresponding label $p \in \ens{\sfx, \sfy, \sfz}$, a $\sfP_{v:r}^{p}$ measurement is equivalent to a measurement of the observable $P_v$ (recall Definitions~\ref{def:measObservableVN} and~\ref{def:measObservableDestructive}), and relabelling the measurement results $(-1)^r \mapsto* r$ to obtain measurement results $r \in \ens{0,1}$.
Abusing our earlier terminology, we will momentarily speak of the measurements in the Clifford circuit model as being measurements of the corresponding Pauli operator as an observable in this fashion.

Consider a state $\rho = \ket{\psi}\bra{\psi}$ for $\ket{\psi} \in \cE$, where we let $\cE$ be the space stabilized by a stabilizer group $\sS = \gen{S_1, \ldots, S_{n-k}}$; and define the projectors
\begin{align}
		\Pi_0
	\;\;=\;\;
		\sfrac{1}{2}\paren{\idop_2^{\ox n} + P_v}
	&&
		\text{and}
	&&
		\Pi_1
	\;\;=\;\;
		\sfrac{1}{2}\paren{\idop_2^{\ox n} - P_v}\,.
\end{align}
\begin{itemize}
\item
	If $P_v \in \sS$, then $\ket{\psi}$ is a $+1$-eigenvector of $P_v$, in which case a von Neumann $P_v$ measurement will yield the result $\s[v] = 0$ and leave the state of the quantum part of the system unchanged.

\item
	If $P_v \notin \sS$ but $P_v$ commutes with all of the generators $S_j$, then we may decompose the code $\cE$ stabilized by $\sS$ into two orthogonal subspaces $\cE_0$ and $\cE_1$, stabilized by $\sS_0 = \sS \oplus \gen{P_v}$ and $\sS_1 = \sS \oplus \gen{-P_v}$ respectively.
	(That these subspaces are orthogonal follows from the orthogonality of the $\pm 1$-eigenspaces of $P_v$.)
	Let $\ket{\psi_0} = \Pi_0 \ket{\psi}$ and $\ket{\psi_1} = \Pi_1 \ket{\psi}$\,: then a von Neumann $P_v$ measurement yields the state $\ket{\psi_0}\bra{\psi_0} \ox \ket{0}\bra{0}_r \;+\; \ket{\psi_1}\bra{\psi_1} \ox \ket{1}\bra{1}_r$\,, where $r$ is the bit containing the measurement result.
	For each $r \in \ens{0,1}$, we have
	\begin{gather}
		S_j \ket{\psi_r} \;=\; S_j \Pi_r \ket{\psi} \;=\; \Pi_r S_j \ket{\psi} = \ket{\psi_r}
	\end{gather}
	for every $1 \le j \le n-k$\,; then, the state after measurement conditioned on the result bit having the value $r$ is is in the space $\cE_r$ stabilized by $\sS_r = \sS \oplus \gen{\big.\text{\small$(-1)^r P_v$}}$.
	Because the initial state $\ket{\psi}$ may be an \emph{arbitrary} state in $\cE_0$ or $\cE_1$ (which would then be unaffected by measurement), this fully characterizes the post-measurement state space.
	Because the number of generators of the stabilizer group increases in this case, $\cE_0$ and $\cE_1$ each have half the dimension of $\cE$, and so this operation is non-unitary.

\item
	If $P_v$ anticommutes with a generator $S_j$, without loss of generality we may suppose that $P_v$ anticommutes with $S_1$ in particular.
	Then, define a new set of generators $\{ S'_j \}_{j = 1}^{n-k}$ for $\sS$, such that
	\begin{gather}
			S'_j
		\;\;=\;\;
			\begin{cases}
				S_j		\;,	&	\text{if $j = 1$ or $S_j$ commutes with $P_v$}	\\[0.5ex]
				S_1 S_j	\;,	&	\text{otherwise}
			\end{cases}	\;;
	\end{gather}
	then $S'_1$ anticommutes with $P_v$\,, and $S'_j$ commutes with $P_v$ for $j \ge 2$.
	Let $\sS_0 = \ens{P_v, S'_2, \ldots, S'_{n-k}}$, and $\sS_1 = \ens{-P_v, S'_2, \ldots, S'_{n-k}}$, and let $\cE_0$ and $\cE_1$ be the codes stabilized by these groups.
	Once more, let $\ket{\psi_0} = \Pi_0 \ket{\psi}$ and $\ket{\psi_1} = \Pi_1 \ket{\psi}$\,: then a von Neumann $P_v$ measurement yields the state $\ket{\psi_0}\bra{\psi_0} \ox \ket{0}\bra{0}_r \;+\; \ket{\psi_1}\bra{\psi_1} \ox \ket{1}\bra{1}_r$\,, where $r$ is the bit containing the measurement result.
	For each $r \in \ens{0,1}$, we again have
	\begin{gather}
		S'_j \ket{\psi_r} \;=\; S'_j \Pi_r \ket{\psi} \;=\; \Pi_r S'_j \ket{\psi} = \ket{\psi_r}	\,;
	\end{gather}
	for every $2 \le j \le n-k$\,; however, we have
	\begin{gather}
			S'_1 \ket{\psi_0}
		\;=\;
			\sfrac{1}{2} S'_1 \paren{\idop_2^{\ox n} + P_v} \ket{\psi}
		\;=\;
			\sfrac{1}{2} \paren{\idop_2^{\ox n} - P_v} S'_1 \ket{\psi}
		\;=\;
			\ket{\psi_1}	\;,
	\end{gather}
	and similarly $S'_1 \ket{\psi_1} = \ket{\psi_0}$.
	Then, the state after measurement conditioned on the result bit having the value $r$ is in the space $\cE_r$ stabilized by the group $\sS_r$\,, and the two possible post-measurement results may be mapped to one another by the operation $S'_1$.
	(In particular, if we wish to select for the post-measurement state that would arise for the $0$ state, it suffices to apply the operation $S_1$ if instead the measurement result $1$ arises.)

	Note that the probabilities of the two outcomes in this case are both $\frac{1}{2}$\,: we have
	\begin{align}
			\bracket{\psi_0}{\psi_0}
		\;\;=\;\;
		 	\sfrac{1}{2} \bra{\psi} (\idop_2^{\ox n} + P_v) \ket{\psi}
		\;\;=&\;\;
		 	\sfrac{1}{2} \bra{\psi} S'_1 (\idop_2^{\ox n} - P_v) S'_1 \ket{\psi}
		\notag\\=&\;\;
		 	\bracket{\psi_1}{\psi_1}	\;,
	\end{align}
	which implies that applying a $P_v$ measurement on $\ket{\psi}$ yields $+1$-eigenstates and $-1$-eigenstates of $P_v$ each with equal probability.
	Because $S'_1$ maps the spaces $\cE_0$ and $\cE_1$ into one another, we may consider arbitrary vectors $\ket{\psi_0} \in \cE_0$ and $\ket{\psi_1} \in \cE_1$ such that $\ket{\psi_0} = S'_1 \ket{\psi_1}$\,.
	If $\bracket{\psi_0}{\psi_0} = \bracket{\psi_1}{\psi_1} = \frac{1}{2}$, then $\ket{\psi} = \ket{\psi_0} + \ket{\psi_1}$ is a state vector stabilized by $S'_1$, and also by $S'_j$ for $j \ge 2$.
	Then, for any state vector $\ket{\psi'}$ in either $\cE_0$ or $\cE_1$\,, we may construct a state $\ket{\psi}$ stabilized by $\sS$ for which $\ket{\psi'}$ is a possible post-measurement state.
	This implies that the codes $\cE_0$ and $\cE_1$ characterize the possible post-measurement state-spaces.

	Finally, for any two states $\ket{\psi}, \ket{\phi}$ stabilized by $\sS$, we have
	\begin{gather}
			\bra{\phi} P_v \ket{\psi}
		\;\;=\;\;
			\bra{\phi} S'_1 P_v \ket{\psi}
		\;\;=\;\;
			- \bra{\phi} P_v S'_1 \ket{\psi}
		\;\;=\;\;
			- \bra{\phi} P_v \ket{\psi}	\,,
	\end{gather}
	so that $\bra{\phi} P_v \ket{\psi} = 0$.
	Define the state-vectors
	\begin{gather}
			\ket{\bar\psi_r}
		\;\;=\;\;
			\frac{\Pi_r \ket{\psi}}{\sqrt{\bra{\psi}\Pi_r{\ket{\psi}}\,}\,}
		\;\;=\;\;
			\sqrt{2} \; \Pi_r \ket{\psi}	\;,
		\\[1ex]
			\ket{\bar\phi_r}
		\;\;=\;\;
			\frac{\Pi_r \ket{\phi}}{\sqrt{\bra{\phi}\Pi_r{\ket{\phi}}\,}\,}
		\;\;=\;\;
			\sqrt{2} \;\Pi_r \ket{\phi}
	\end{gather}
	for the states arising after a $P_v$ measurement on $\ket{\psi}$ and $\ket{\phi}$ respectively, conditioned on obtaining the result $r \in \ens{0,1}$.
	Then, the inner product of $\ket{\bar\phi_r}$ with $\ket{\bar\psi_r}$ can be given by
	\begin{align}
			\bracket{\bar\phi_r}{\bar\psi_r}
		\;\;=\;\;
			2 \bra{\phi} \Pi_r \ket{\psi}
		\;\;=&\;\;
			\bra{\phi} (\idop_2^{\ox n} + (-1)^r P_v) \ket{\psi}
		\notag\\=&\;\;
			\bracket{\phi}{\psi}	\;;
	\end{align}
	thus the transformation performed by the $P_v$ measurement (conditioned on either result) preserves inner products, and is therefore a unitary transformation.

\item
	Finally, for any destructive $P_v$ measurement, note that the state of $v$ after a similar von Neumann measurement is characterized by being a $\pm 1$-eigenvector of $P_v$ determined by the measurement result; and in particular is independently distributed from the other qubits.
	Then no information about the rest of the qubits is lost by discarding the qubit.
	If the measurement result is $r \in \ens{0,1}$, let $R_v = (-1)^r P_v$ be the operator stabilizing the state of $v$ in the post-measurement stabilizer group $\sS_r = \{S'_j\}_{j = 1}^{n-k'}$ --- where we may easily verify that $k' = k-1$ in the case where $P_v$ commuted with (but was not generated by) the pre-measurement stabilizer group, and $k' = k$ otherwise.
	Because all the generators $S'_j$ commute, any other operator $S'_j \ne R_v$ must acts on $v$ with either the identity or a $P_v$ operation.
	We may then form a new set of generators
	\begin{gather}
			S''_j
		\;\;=\;\;
			\begin{cases}
				S'_j		\;,	&	\text{if $S'_j = R_v$ or if $S'_j$ acts on $v$ with $\idop_2$}	\\[0.5ex]
				S'_j R_v	\;,	&	\text{otherwise}
			\end{cases}
	\end{gather}
	for $\sS_r$\,, where $R_v \in \{S''_j\}_{j = 1}^{n- k'}$ is the only operator which acts non-trivially on $v$.
	We may then trace $v$ out by removing the generator $R_v$, and tracing out $v$ from the domain of the other generators $S''_j$\,.
\end{itemize}
Thus, the state-space of a system produced by a Clifford circuit, assuming a arbitrary pure input state, can be efficiently computed for any possible value of the measurement results, and even whether the evolution will be unconditionally unitary.
\label{discn:endStabilizerFormalism}

Applied to the special case when there are \emph{no} quantum inputs (\ie\ where all qubits are allocated by the circuit itself, and prepared in one of the states $\ket{\splus \sfx}$, $\ket{\splus \sfy}$, or $\ket{\splus \sfz}$), this yields the following result, reported in~\cite{Gottesman98} and attributed to Emmanuel Knill:\footnote{%
The result stated in~\cite{Gottesman98} is actually an equivalent result, in which the only preparation and measurement operations are $\New{}{\sfz}$ and $\sfP_{v:r}^{\sfz}$ (and the unitary gate $\cX$ is used in place of $\cZ$).}
\begin{theorem*}[Gottesman-Knill Theorem]%
	\label{thm:GottesmanKnill}
	A circuit in the Clifford circuit model which has no quantum inputs can be simulated efficiently on a classical computer.
\end{theorem*}
Note that the \Clifford\ model is not even approximately universal for quantum computation: because there are only finitely many elements of $\cP^{\ox n}$, there are also only finitely many stabilizer groups of size $2^n$ on $n$ qubits, and thus finitely many stabilizer states.
Thus, there exist CPTP maps $\Upphi: \ens{1} \to \D(2^n)$ which cannot be approximated to arbitrary precision with a Clifford circuit.

As we remarked above, circuits in the $\Phases(\frac{\pi}{2}\Z)$ model without trace-out operations are also Clifford circuits; and therefore circuits in the $\Jacz(\frac{\pi}{2}\Z)$ models without trace-out operations can be easily simulated by Clifford circuits; then these models are also not approximately universal for quantum computation.\footnote{\label{fn:parityL}%
A further result of Aaronson and Gottesman~\cite{AG04} show that simulating Clifford circuits is a complete problem for $\ParityL$~\cite{BHMD91}, the class of problems which can be solved by a nondeterministic Turing machine (NTM) which uses only logarithmic workspace, and where in particular the ``\emph{yes}'' instances of the problem have an odd number of accepting paths for that NTM.
Solving linear equations over $\Z_2$ is another problem which is complete for $\ParityL$~\cite{Damm90}, and which is conjectured to be insufficient even to perform universal polynomial-time \emph{classical} computation.}

\section{Conclusion}
\label{sec:introConclusion}

In this Chapter, we have described quantum states as being represented by density operators, and transformations of them as being by CPTP maps, which may be performed or approximated by composing discrete operations.
We have also noted the important role that unitary evolution has in quantum mechanics, and described the role of unitary transformations in defining the most commonly used models of quantum computation.

Other significantly different models of quantum computation exist which are not obviously special cases of either unitary circuit or classically-controlled unitary circuit models, but which can be shown to be polynomial-time equivalent:
\begin{description}
\item[Adiabatic quantum computation]\cite{FGGS00}
	rests on the adiabatic theorem first described by Born and Fock~\cite{BF28} (see~\cite{AR2004} for a modern treatment), and which is a model of computation relying on the evolution of states under a slowly and continuously varying Hamiltonian.
	This model lends itself naturally to optimization problems: an analysis of the strengths and limitations of this model, and its relation to unitary circuits, can be found in~\cite{DMV2001}.

\item[Topological quantum computation]\cite{Kitaev2003}
	is a model which exploits properties of particle statistics to approximate unitaries to arbitrary precision by braiding the trajectories of ``virtual particles'' which can be simulated via the quantum Hall effect~\cite{ASW1984}.
	This is a natural model for describing algorithms for such problems as approximating knot invariants~\cite{AJL2006,LK2006}; and by exploiting the topological properties of the particle trajectories, it is possible to naturally achieve fault-tolerant computation.
	A review of toplogical quantum computation can be found in~\cite{NSSFS2008}.

\item[Continuous-time quantum walks]\cite{FG98}
	represent a generalization of diffusion processes in graphs by random walks by replacing the diffusion equation with Schr\"odinger evolution.
	(Discrete-time quantum walks can also be formulated naturally by generalizing random walks~\cite{Watrous98}, but these may be easily subsumed by classically controlled unitary circuits.)
	This model led to the discovery of a quantum speed-up for evaluating \textsc{nand}-trees~\cite{FGG07}; the universality of this model is shown in~\cite{Childs2008}.
\end{description}

In this thesis, we are primarily interested in one further model of quantum computation, which may be subsumed into classically controlled unitary circuit models, but in which measurements depending on many classical controls play a significant role, resulting in a model with a significantly different character than unitary circuit models.
This model is the \emph{one-way measurement-based model}, which we introduce in the next chapter.
	\chapter[One-way measurement-based computation]{
		One-way measurement-based computation}

In the absence of efficient algorithms for simulating quantum mechanics or for solving difficult number-theoretic problems such as factoring or discrete logarithms, quantum computers represent a potential breakthrough in scientific computing.
However, the technical challenges involved in controlling the evolution of a quantum mechanical system for the purposes of computation have prompted the exploration of different ``computational primitives'' for quantum computation --- meaning essentially different sets of operations which
\begin{inlinum}
\item
	form a model which is universal for quantum computation, and
\item
	may be easier to implement physically.
\end{inlinum}

In order to achieve universal quantum computation via unitary circuits, multi-qubit entangling gates are required: otherwise, only states which are tensor-products of single-qubit states can be produced.
As scalable quantum control of multi-qubit systems is an open research problem, reducing the number of rounds involving multi-qubit interactions is one plausible route to achieving practical scalable quantum computation.
However, to perform or approximate any unitary evolution which cannot be efficiently simulated on a classical computer, many rounds of multi-qubit operations such as $\cX$ or $\cZ$ (as defined in \eqref{eqn:basicGates}) are required in a unitary circuit model.\footnote{%
To be precise, although it is widely believed that the measurement results arising out of quantum mechanics (and arbitrary unitary circuits in particular) cannot be efficiently approximated by a classical model of computation such as Turing machines, any unitary circuit on $n$ qubits in which each qubit is operated on fewer than $O(\log(N))$ times by a two-qubit gate can be simulated in $n \cdot\poly(N)$-time by a classical computer~\cite{Jozsa03}.}

If we are interested in describing unitary transformations of states (or \emph{approximately} unitary transformations, in the sense of approximating a unitary CPTP map with precision $\epsilon$), while allowing classically controlled measurements to play a significant role in the computation, a single round of multi-qubit operations --- or alternatively, access to appropriate multi-qubit resource states --- will suffice to achieve universal quantum computation.
The models which make this possible are the \emph{measurement based models} of quantum computation.

In~\cite{GC99}, Gottesman and Chuang showed that the preparation of a special set of multi-qubit states, together with teleportation~\cite{BBCJPW93}, was universal for quantum computation.
Teleportation is essentially a means of implementing the identity operation by means of measurement and classical control: if we define the \emph{Bell basis} by the states
\begin{small}
\begin{align}
	\begin{split}
			\ket{\beta_{00}}
		\;=&\;
			\sfrac{1}{\sqrt 2} \ket{0} \ox \ket{0} \;+\; \sfrac{1}{\sqrt 2} \ket{1} \ox \ket{1}	\;,
		\\
			\ket{\beta_{01}}
		\;=&\;
			\sfrac{1}{\sqrt 2} \ket{0} \ox \ket{0} \;-\; \sfrac{1}{\sqrt 2} \ket{1} \ox \ket{1}	\;,
	\end{split}
	&
	\begin{split}
			\ket{\beta_{10}}
		\;=&\;
			\sfrac{1}{\sqrt 2} \ket{0} \ox \ket{1} \;+\; \sfrac{1}{\sqrt 2} \ket{1} \ox \ket{0}	\;,
		\\
			\ket{\beta_{11}}
		\;=&\;
			\sfrac{1}{\sqrt 2} \ket{0} \ox \ket{1} \;-\; \sfrac{1}{\sqrt 2} \ket{1} \ox \ket{0}	\;,
	\end{split}
\end{align}
\end{small}
and a \emph{Bell basis measurement} operation by the map
\begin{gather}
		\mathsf B_{v,w}(\rho_{v,w})
	\;\;=\;
		\sum_{x,z \in \ens{0,1}}	\!
			\bra{\beta_{xz}} \rho \ket{\beta_{xz}}_{\!v,w} \; \Big[ \ket{x}\bra{x}_{\s[v]} \ox \ket{z}\bra{z}_{\s[w]} \Big] \;,
\end{gather}
then teleportation (as described in~\cite{GC99}) is the map $\mathsf T_{q/v} : \D(v) \to \D(q)$ given by
\begin{gather}
	\label{eqn:teleportation}
		\mathsf T_{q/v}(\rho_v)
	\;\;=\;\;
		\Big( \Xcorr{q}{\s[v]} \circ \Zcorr{q}{\s[w]} \circ \,\mathsf B_{v,w} \circ \Big[ \idop_v \ox \:\! \ket{\beta_{00}}\bra{\beta_{00}}_{w,q} \Big]\Big)(\rho_v),
\end{gather}
where $\Zcorr{q}{s}$ and $\Xcorr{q}{s}$ are classically-controlled $Z$ and $X$ operations depending on the bit $s$.
The result of~\cite{BBCJPW93} is that $\mathsf T_{q/v}$ as defined in~\eqref{eqn:teleportation} performs the map $\idop_{q/v}$\,: that is, the state of $v$ is ``teleported'' unchanged into $q$.\footnote{%
The description of the output neglects the classical measurement results $\s[v]$ and $\s[w]$, which are also produced as output and which will be uncorrelated and uniformly randomized bits; it is conventional to implicitly discard classical measurement results when they are no longer required.}
This remains true even though the qubit $q$ may never have interacted with the qubit $v$, and may be separated from $v$ by a large distance: if the qubits $w$ and $q$ are controlled by different parties (conventionally named ``Alice'' and ``Bob'') after their joint preparation in the state $\ket{\beta_{00}}$, it suffices for Alice to communicate the results $\s[v]$ and $\s[w]$ to Bob, and for Bob to perform the appropriate operations depending on those bits.
This is illustrated in Figure~\ref{fig:teleportation}.

\begin{figure}[t]
	\begin{center}
		\includegraphics{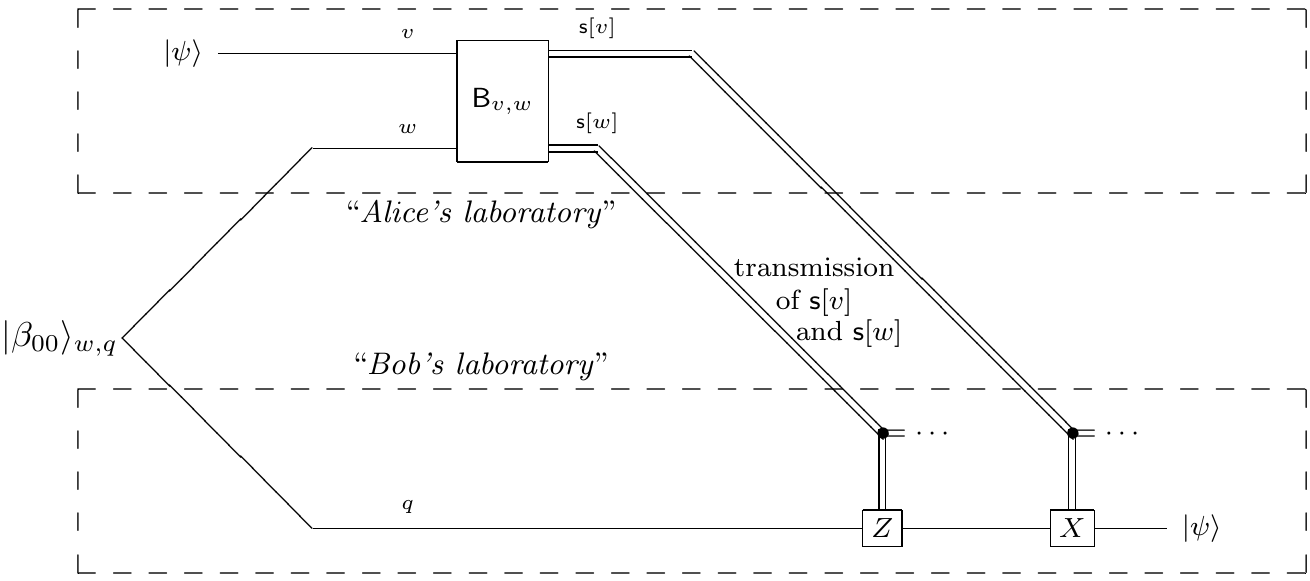}	
	\end{center}
	\caption[An illustration of the ``teleportation'' protocol.]{\label{fig:teleportation}%
		An illustration of the ``teleportation'' protocol defined in~\eqref{eqn:teleportation}.
		We use similar conventions as in Figures~\ref{fig:qcircuit-1} and~\ref{fig:qcircuit-2} to illustrate circuits with measurements and classical control; double-lines represent classical bits produced by measurement.
		The system is divided into two separate ``laboratories'' controlled by separate parties (``Alice'' and ``Bob''), between which we suppose that coherently controlled operations cannot occur.
		Diagonal lines represent the sharing of information resources: the preparation of the joint initial state $\ket{\beta_{00}}$ of $w$ and $q$, and the transmission of the bits $\s[v]$ and $\s[w]$.}
\end{figure}

Teleportation is thus an important primitive for transmitting quantum information, using the correlations of the Bell state $\ket{\beta_{00}}\bra{\beta_{00}}$ and classical communication; \cite{GC99} showed that by generalizing the initial states used as a resource, teleportation (and in particular Bell measurements) could also be used as a \emph{computational} primitive.
In the three years following Gottesman and Chuang's result, three different proposals for quantum computation based on measurement operations arose.
Nielsen~\cite{Nielsen2003} described a model of quantum computation based on four-qubit measurement operations; this together with the result of~\cite{GC99} which was subsequently refined by Leung~\cite{Leung2001} into what is called the \emph{teleportation-based model of quantum computation} in~\cite{CLM2005}.
The third proposal, the \emph{one-way measurement-based model} presented by Robert Raussendorf and Hans Briegel in~\cite{RB01}, forms the starting point for the main topic of this thesis.

In this chapter, we will describe different variants of the one-way measurement model in terms of the \emph{stabilizer formalism} presented in Section~\ref{sec:stabilizerFormalism}, and provide an overview of applications and potential schemes for implementing one-way measurement based quantum computing.

\paragraph{New results.}

Strictly speaking, none of the results of this Chapter are new.
However, we present explicit definitions for three distinct (but closely related) constructions of measurement-based computation, to which we will regularly refer in later chapters.
These constructions are the \emph{simplified} and \emph{complete DKP constructions} (after Danos, Kashefi, and Panangaden~\cite{DKP07}), defined in Section~\ref{sec:DKPstandardization}, and the \emph{simplified RBB construction} (after Raussendorf, Browne, and Briegel~\cite{RBB03}), defined in Section~\ref{sec:RBB}.

\section{Overview and proposals for implementation}

As the one-way measurement model was originally proposed as a scheme for implementations of quantum computation, we will provide a quick overview of the model in order to hilight the aspects which make it seem reasonable as a candidate for scalable quantum computation.
Before describing the model in full detail, we will give an overview of different proposals for fulfilling the requirements of the one-way measurement model.

\subsection{A brief description of the one-way model} 

The one-way measurement model of quantum computing proposed in~\cite{RB01} is a model of computation which involves no explicit multi-qubit operations: instead, it allows the preparation of arbitrary states from a uniform family of states,\footnote{%
A uniform family of states is one which can be efficiently specified by a classical, deterministic Turing machine in time polynomial in the size of the desired state.}
called \emph{cluster states}.

\begin{definition}
	\label{def:clusterState}
	The \emph{$n \x d$ grid} is a graph $G_{n,d}$ whose vertex set consists of \emph{grid sites} $v \in \ens{1,\ldots,n} \x \ens{1,\ldots,d}$, and whose edges are given by adjacent grid sites, \ie\ pairs $vw$ of vertices $v, w \in V(G)$ for which $\norm{v - w}_1 = 1$\,, where $\norm{x}_1$ is the $\ell_1$-norm.
	The \emph{$n \x d$ cluster state} is then a stabilizer state whose stabilizer group is given by $\cC_{n,d} = \gen{K_v}_{v \in V(G)}$, where
	\begin{gather}
			K_v
		\;\;=\;\;
			X_v \prod_{w \in \ngbh{v}} Z_w	\;,
	\end{gather}
	and where $\ngbh{v}$ denotes the vertices adjacent to $v$ in the graph $G_{n,d}$.
\end{definition}

The $n \x d$ cluster state can be produced by evolution by an Ising interaction operating for an appropriate length of time, where each qubit in the grid is initially prepared in the $\ket{+}$ state~\cite{RB01}: several schemes for how this may be done are described in the original article.

Having a cluster state, computation is then performed using single qubit measurements.
Qubits are destructively measured either using $Z$ observables, or observables of the form $M(\theta) = \cos(\theta) X + \sin(\theta) Y$ which anticommute with $Z$; the qubits measured with a $Z$ observable are effectively removed from the computation, while those measured with an observable $M(\theta)$ are used to control what computation is performed.
It is possible to show that each qubit in the cluster is maximally entangled with its' nearest neighbors; then, each measurement result is maximally random.
The different measurement outcomes corresponds to different transformations of the data in the computation: in order to simulate \eg\ a unitary circuit, it becomes necessary to retain the results of the measurements long enough to counteract the randomness, \eg\ by changing the measurement observables of later measurements.
Thus, the one-way measurement model is a \emph{classically controlled} model of computation.
Furthermore, the number of measurement results that an operation may depend upon may be arbitrarily large, in general scaling with the length of the computation.

We will describe the one-way measurement model more fully in Section~\ref{sec:oneWayComputation}: for the time being, we will remark that the necessary elements of the one-way model as a scheme for for implementations of quantum computation are the possibility of creating a family of large entangled states; the ability to perform classically controlled measurements; and the ability to compute and transmit measurement results quickly enough to act as the classical control of further measurements.
We can broaden the criteria for implementation somewhat by observing the following:
\begin{itemize}
\item
	The large entangled states required do not necessarily have to be cluster states.
	We will show how the model of~\cite{RB01} may be generalized to allow quantum computation with general \emph{graph states}, which we may define in analogy to cluster states by replacing the $n \x d$ grid with an arbitrary graph.

\item
	Rather than adapting the measurements, we may allow classically controlled unitary operations just prior to each measurement, which performs a change of reference frame which maps some ``logical'' measurement observable to a particular ``default'' measurement observable.
	Generalizing still further, we may provide for the preparation of arbitrary density operators if we leave some qubits unmeasured to support a final state; in this case, classically controlled unitaries are necessary to enable the preparation of pure states.
\end{itemize}

These are the primary features which make the one-way measurement model an attractive prospect for implementation: it provides a strict divide between a stage involving the creation of entanglement, and a stage involving only single-qubit operations.
The requirement that these operations depend on the results of previous measurements may present a challenge, but this challenge is of a significantly different character than the control of a many-body Hamiltonian for the purpose of performing multi-qubit transformations.
As well, once a cluster state has been prepared, it can in principle be safely stored until ready for use, provided a sufficiently reliable quantum memory.

\subsection{Proposals for implementation} 

One of the promises of the one-way measurement model is that the state involved is quite simple (in particular, it is a stabilizer state) and can in principle be easily prepared.
The original proposal~\cite{RB01} itself suggested that the state may be possible to produce by controlled Ising interactions on qubits prepared in the $\ket{\splus}$ state.
We now consider the proposals for schemes for producing cluster states and graph states, and for performing one-way measurement-based computation generally.

\subsubsection{Optical proposals}

The first proposal for efficient quantum computing by linear optics (albeit not implementing the one-way measurement model) was presented in~\cite{KLM2001}, which introduced an entangling gate using number counting photo-detectors (thus introducing an effective non-linearity) which failed with finite probability using techniques similar to~\cite{GC99}.
However, in order to perform scalable quantum computation, a significant amount of overhead in error correction is required even in the case of ideal apparatus as there is a significant probability of photon loss due to destructive measurement.

The approach of~\cite{KLM2001} was improved upon in~\cite{Nielsen2004}, which creates a cluster state as an entangled resource by successive ``fusion'' operators, used to append qubits onto an existing graph state to make a larger graph state.
The overhead due to the possibility of failure of the fusion operators is then realized as repeated attempts to create a cluster state: in the case of ideal apparatus, no error correction is required, resulting in a reduction of overhead.
A further improvement is made by~\cite{BR2005}, who employ fusion operators realized via a ``simplified'' application of the Hong-Ou-Mandel effect~\cite{HongOuMandel87}\footnote{%
The simplification entailed in the results of~\cite{BR2005} is that the Hong-Ou-Mandel effect does not occur as a part of a Mach-Zehnder interferometer, which is sensitive to perturbations in path length on the order of one wavelength.} to reduce the sensitivity of the interferometer on path perturbations.

Another approach~\cite{LBK2005} known as ``repeat until success'' uses a simple error correcting code in conjunction with linear optics and measurement, to implement entangling gates in such a way that the quantum information of the system is not destroyed when the logical gate fails: this allows cluster states to be built with a much better expected time, as in the ideal case no losses arise from a failed operation.

In all of the above models, as with all optical set-ups, the primary challenges to be overcome are inefficiencies associated with photon creation and detection: see \eg~\cite{LBBKK2006} for a discussion of these issues; more information on linear optical proposals can be found in~\cite{Kok2007}.

\subsubsection{Matter qubits and ``hybrid'' matter-optics approaches}

In~\cite{BK2005}, a scheme for constructing cluster states by interacting ``stationary'' qubits (specifically, defects in diamonds) is proposed which probabilistically entangles pairs of excited qubits by performing interferometry on the photons emitted in decay processes.
Combining this with the ``repeat until success'' techniques from linear optics led to the proposal of~\cite{LBBKK2006}, which promises to provide the best features of both proposals with regards to resilience against noise: \ie\ stable quantum memory provided by the spatially separated ``stationary'' qubits, whose interactions are mediated by deterministic photon gates provided by the optical elements.

An alternative approach to the previous, explicitly hybrid model between matter and optics is to employ a \emph{broker/client model} as described in~\cite{BBFM2006}, in which fragments of graph states are generated using fast-reacting but unstable subsystems (such as electron spin), and then transferred to more stable sub-systems (such as nuclear spins) to reduce the probability of loss..

Optical lattices form another setting for proposals of one-way measurement-based computation, wherein cold atoms are localized in a periodic electromagnetic field.
One such proposal~\cite{CAJ2005} proposes to represent qubits by the local modes of a linear array of non-interacting fermions, which become entangled simply as a consequence of Fermi statistics as the state of the fermions undergo mirror inversion.
Another proposal~\cite{KPA2006} proposes to mediate interactions by using global control to transport a ``virtual'' mediating qubit, represented by a pair of reserved excited states for each lattice site arising from the tuning of the lattice.

Almost all of the models mentioned in this and the preceding section address only the creation of the cluster state (which is admittedly likely to be the more difficult task in performing one-way measurement based computation, due to the need to control multi-qubit operations).
A departure from this is a proposal for producing cluster states with charge and flux qubits with quantum dots is described in~\cite{TLFHN2006}, which is robust to the nonuniformity of the dot formation process.
Measurement adaptations are explicitly performed by controlled rotations, which are performed by moderating the same continuous time Hamiltonian which gives rise to the cluster state itself.

A result of~\cite{Nielsen2006} is that neither the cluster state itself, (nor any other universal resource for the one-way model, can be the unique ground state of any ``physically realistic'' Hamiltonian; despite this, \cite{BR2006} describes a nearest-neighbor two-local Hamiltonian with a unique ground state on the grid, which may be interpreted as an \emph{encoded} cluster state. 
This indicates that a cluster state (or equivalent) is likely to be physically attainable by cooling an appropriate system to its ground state, and performing measurements to obtain the cluster state itself (equivalently, performing single-qubit measurements in such a way as to \emph{simulate} the a one-way measurement-based computation using a cluster).

Finally, a different approach to the creation of a cluster state is to encode the qubits in different energy modes of a single physical system.
One proposal of this type~\cite{MFP2008} is to implement cluster states in the modes of an optical cavity, using highly peaked (or \emph{squeezed}) distributions over continuous variables to represent computational basis states.
The high degree of coherent control over optical cavities is touted as an advantage for this scheme; however, the degree of ``squeezing'' required to perform any given computation is not yet known.

\subsection{Approaches to fault-tolerance in the one-way model}

A pre-requisite of scalable quantum computation is the ability to protect the state of a system against noise.
An alternative to achieving the ability to directly control the interactions of the system versus the environment is to apply generic techniques for protecting quantum information against uncontrolled operations: this approach is what is generally referred to as \emph{fault tolerance} of quantum computation \cite{KLABVZ2002,LW2003}.

Most of the attention to fault-tolerant approaches to one-way measurement-based computation have focused on the linear-optical proposals for implementation, where the noise operations (aside from those induced by imperfect realizations of logical gates) can be described in terms of loss errors and decoherence errors.
A scheme for loss-tolerance is proposed in~\cite{VBR2006}, which (provided perfect logical gates) can cope with photon losses of up to 50\%, by using stabilizer techniques to allow would-be measurement results of lost photons to be inferred.
However, \cite{RRM2007} describe that this yields an exponential blow-up of error rate for other reasons.
A modification of the scheme of~\cite{VBR2006} is also presented in~\cite{RRM2007}, for which they suggest a trade-off exists between the error rates of different types of error.
An optical protocol protecting against both polarization and loss errors is presented in~\cite{DHN2006}, requiring Bell pairs and employing the fusion operator techniques of~\cite{BR2005}.

An analysis applicable to linear optical set-ups, but also to optical lattices, is presented in~\cite{KRE2007}.
Using renormalization theory, they present a means of treating losses incurred due to non-deterministic entangling gates which eliminates the need to ``route'' states (\ie\ conditionally operate in the case of other, failed, operations).
This analysis yields scaling of physical requirements similar to hypothetical schemes where the entangling gates are deterministic, which may substantially improve the prospects of those physical implementations which rely on lossy, probabilistic operations.

Analyses of one-way based computation in decoherence free subspaces (\ie\ subspaces of the state-space of a system which are naturally protected from noise by symmetries of the system) are presented in~\cite{TPK2007} for phase-damping errors in an optical lattice setting; the authors propose that the construction should be extensible to more general errors.
An experimental demonstration for a corresponding proposal in linear optics is presented in~\cite{PTSPKZ2007}.

Finally, a generic scheme for performing topological quantum computation natively in a three-dimensional cluster state (\ie\ in which two-dimensional grid graphs are replaced by three-dimensional grids) is presented in~\cite{RHG2007}.
The computation is performed in this proposal by simulating the evolutions of ``defects'' in a two-dimensional lattice over discrete time-steps; the trajectories of the defects trace out paths in the three-dimensional lattice, which may be braided to effect quantum computation (as briefly described for topological quantum computation in general, in Section~\ref{sec:introConclusion}).

\section{Computing in the one-way measurement model}
\label{sec:oneWayComputation}

The one-way measurement model can be used to achieve universal quantum computation by performing \emph{adaptive single-qubit measurements} --- classically controlled measurement operations, which may depend on the results of prior measurements --- and classically controlled Pauli operations on the output qubits, which compensate for the randomness of the prior measurement results.
(In the original presentation~\cite{RB01}, the correction or \emph{byproduct} operation is never explicitly performed, and is instead used to adapt the final measurement used to produce the read-out of the computation: in order to describe this model of computation as being universal in the sense of Definition~\ref{def:universalQC}, we include the conditional corrections in the description.\footnote{%
The one-way model, as presented in~\cite{RB01}, is a model of ``\textbf{CC}-universal'' quantum computation in the terminology of~\cite{NDMB2007}: however, the way in which we extend it by including classically controlled unitaries to produce a ``\textbf{CQ}-universal'' model of computation is already implicit in the analysis of~\cite{RB01}.})

Every operation after the preparation of the cluster state acts only on individual qubits --- albeit possibly with classical control --- and (except for the classically controlled unitaries by which we extend the model) discards the qubits it acts on.
The cluster state is therefore consumed as a computational resource via measurement; hence the name of the model, which is irreversible, in contrast with the reversibility of closed-system Schr\"odinger dynamics.

We will spend most of this thesis considering aspects of measurement-based models of quantum computation which have been abstracted from the model of~\cite{RB01}, with emphasis on one such model in particular (Definition~\ref{def:oneWayModel}) which retains what can be regarded as the essence of the original proposal in~\cite{RB01}.
In doing so, we will implicitly describe schemes of translating unitary circuits into various models of quantum computation based on adaptive single-qubit measurements.

\paragraph{A remark on terminology.}

Despite the existence of other models of measurement-based computation, such as teleportation-based computation~\cite{Leung2001} or further relaxations of the one-way model as in~\cite{GE2007}, we will generally refer only to the models presented in this Chapter when we use the term \emph{measurement-based quantum computation}.
Because of the emphasis on \emph{unitary} circuits associated with most research in quantum computation, we will use the terms \emph{pattern} or \emph{procedure} in place of the word ``circuit'' when describing well-defined compositions of operations in measurement-based models of computation, following conventions present in the literature (\eg~\cite{DK06,DKP07,BKMP07}), in order to reinforce the prominence of non-unitary operations in the model.

\subsection{The cluster state model}
\label{sec:clusterStateModel}

We consider a slightly extended version of the model of~\cite{RB01} (which we call the \emph{cluster state model}), as follows.
For any $n,d \ge 1$, we include the map $\New{G_{n,d}}{}$ which prepares the $n \x d$ cluster state on fresh qubits.
The measurement operations used in~\cite{RB01} are $\Meas{}{\sfz}$ measurements as in~\eqref{eqn:pauliMeas} and, \emph{\XY-plane measurements} $\Meas{}{\alpha}$, defined for angles $\alpha \in (-\pi, \pi]$ by
\begin{gather}
	\label{eqn:xyPlaneMeas}
		\Meas{v}{\alpha}(\rho)
	\;\;=\;\
		\bra{\psub\alpha}_v \,\rho\, \ket{\psub\alpha}_v		\;\ox\; \ket{0}\bra{0}_{\s[v]}
		\;\;+\;\;
		\bra{\msub\alpha}_v \,\rho\, \ket{\msub\alpha}_v	\;\ox\; \ket{1}\bra{1}_{\s[v]} \;,
\end{gather}
where the states $\ket{\pmsub\alpha}$ are the $\pm 1$-eigenstates of the operator $\cos(\alpha) X + \sin(\alpha) Y$\,,
\begin{gather}
	\label{eqn:pmsubAlpha}
		\ket{\pmsub\alpha}
	\;\;=\;\;
		\sfrac{1}{\sqrt 2} \ket{0} \;\pm\; \sfrac{\e^{i \alpha}}{\sqrt 2} \ket{1}	\;.
\end{gather}
We refer to $\alpha$ as the \emph{measurement angle} of the measurement operator.\footnote{%
Representing the states $\ket{\pmsub\alpha}$ in the Bloch sphere, $\alpha$ is the angle between of the state $\ket{\psub\alpha}$ and the $\ket{\splus} = \ket{\splus \sfx}$ state, \ie\ the \axisX-axis.} 
The \XY-plane measurements are usually classically controlled, in that the sign of the measurement angle may depend on the parity of some sequence of bits --- in particular, some sequence of prior measurement results.\footnote{%
Note that this is a slight departure from the description of classically controlled measurements on page~\pageref{discn:classicalControl}, where the measurement performed depends explicitly on the value of a \emph{bit}, rather than abstractly on the parity of a set of many bits.
While it is likely that the measurement basis would depend explicitly on some bit which stores the value of such a boolean expression in practical implementations, we will not consider explicitly how this parity is evaluated.
However, in a serious consideration of \eg\ the depth-complexity of quantum operations in the one-way model, how the parities are evaluated is an important detail.}
Therefore, we introduce the abbreviation
\begin{gather}
		\Meas{v}{\alpha; \beta}
	\;\;=\;\;
		\Meas{v}{(\minus 1)^\beta \!\!\;\cdot \alpha}
\end{gather}
for a classically controlled measurement performing either $\Meas{}{\alpha}$ or $\Meas{}{\minus \alpha}$, depending on a boolean expression $\beta = \s[a] + \s[b] + \cdots$ representing the sum (modulo $2$) of some set of measurement results.
\label{discn:defaultMeasAngle}
We refer to $\alpha$ as the \emph{default angle} of the measurement (or the \emph{default measurement angle}).
It will be convenient to extend this further to accommodate changes in angle by a possible addition by $\pi$, which is not a part of the model of~\cite{RB01}, but which will facilitate analysis later: therefore, we will write
\begin{gather}
		\Meas{v}{\alpha; \beta, \gamma}
	\;\;=\;\;
		\Meas{v}{(\minus 1)^\beta \!\!\;\cdot \alpha \,+\, \gamma \pi}
\end{gather}
for boolean expressions $\beta$ and $\gamma$\,; we call $\beta$ a \emph{sign dependency} $\gamma$ a \emph{$\pi$-dependency}.
Similarly, as previously mentioned, we also extend the model of~\cite{RB01} by allowing operations $\Xcorr{v}{\beta}$ and $\Zcorr{v}{\beta}$ on arbitrary qubits $v$, which perform the identity if $\beta = 0$ and an $X$ or $Z$ operation (respectively) on $v$ if $\beta = 1$.
Thus, the extended version of the model of~\cite{RB01} can be described as follows:
\begin{definition}
	For any set of angles $\A \subset \R$ the \emph{cluster state model} is the model of computation with the elementary gate set
	\begin{align}
		\mspace{-50mu}
			\Cluster(\A)
		\;\;=\;\;
			\bigg\{& \New{G_{n,d}}{}\,,\; \Meas{v}{\sfz}\,,\; \Meas{v}{\alpha;\beta,\gamma}\,,\; \Xcorr{v}{\beta}\,,\; \Zcorr{v}{\beta}\,,\; \tr[v]
 				\,\bigg|\, n,d \ge 1, ~\alpha \in \A\,,
		\notag\\[-1.5ex]&\quad\qquad
			~\beta,\gamma \;= {\textstyle\sum\limits_{\s[v] \in S} \s[v]} ~~\text{for any $S \subset B$}, ~\text{and}~ v \in \N \x \N \bigg\}\,,
	\end{align}
	where $B$ is an infinite set of bits and $\s: \N \x \N \to B$ is injective.
	We will say that a $\mathfrak P$ procedure in the $\Cluster(\A)$ model is in \emph{standard form} if it can be decomposed as $\mathfrak P \,=\, \mathsf C \circ \mathsf M \circ \smash{\New{G_{n,d}}{}}$\,, where $\mathsf M$ is a composition of \emph{adaptive measurements} $\Meas{v}{\alpha;\beta}$ without $\pi$-dependencies, and where $\mathsf C$ is a composition of \emph{correction operations} $\Xcorr{v}{\beta}$ and $\Zcorr{v}{\beta}$.
	We then call $\New{G_{n,d}}{}$ the \emph{preparation stage} of the procedure, $\mathsf M$ the \emph{measurement stage} of the procedure, and $\mathsf C$ the \emph{correction stage} of the procedure.
\end{definition}
Because cluster states are stabilizer states, procedures in the $\Cluster(\frac{\pi}{2}\Z)$ model can be efficiently simulated by classical circuits by the Gottesman-Knill Theorem~(page~\pageref{thm:GottesmanKnill}).
However, procedures in $\Cluster(\frac{\pi}{4}\Z)$ fall outside of the domain of the stabilizer formalism; no corresponding result is known in that case.

In order to perform universal quantum computation, the one-way model must be able in particular to prepare (or approximate) arbitrary pure states for the qubits on which it act.
The measurement process, however, produces random outcomes.
In order to describe how the cluster state model can produce pure states by adaptive measurements and single-qubit unitary corrections, we give an account of this model in terms of unitary circuits.
This can be facilitated by considering further extensions beyond the cluster state model.

\subsection{The graph state model}

Consider the varieties of states which may be obtained after we perform all $\axisZ$ measurements in a cluster state measurement pattern, and before we apply any other operation.
Because $\axisZ$ measurements are not subject to adaptations based on prior measurements, we may commute the $\axisZ$ measurements to the beginning of the measurement phase of the procedure (whether or not it is in standard form) by applying the relations
\begin{align}
	\label{eqn:ZmeasCommute}
		\Meas{v}{\sfz} \, \Zcorr{v}{\beta} \;=&\; \Meas{v}{\sfz}	\,,
	&
		\Meas{v}{\sfz} \, \Xcorr{v}{\beta} \;=&\; \Shift{v}{\beta} \, \Meas{v}{\sfz}	\,;
\end{align}
for the latter relation, we define the \emph{shift} operator\footnote{%
The shift operator is usually not defined in accounts of measurement-based computation (with~\cite{DKP07} being an exception), probably because it is a very simple ``classical'' operation.
However, in any implementation of quantum computers based on single-qubit measurements, shift operators are likely to play an important role in describing how the parity expressions $\beta$ are computed in the midst of a computation.
Thus, while much of the analysis later in the thesis follows the convention of eliminating shift operations from measurement-based procedures, I explicitly allow shift operators as elementary gates, and extend the sense of ``standard forms'' found in the literature to allow shift operators to occur between measurements.}
$\Shift{v}{\beta}$,
\begin{subequations}
\label{eqn:shiftOptor}
\begin{gather}
		\Shift{v}{\beta}: \Prob(\s[v]) \to \Prob(\s[v])
	\\[1.5ex]
		\mathsf S^{\beta}(\rho)
	\;\;=
		\sum_{x \in \ens{0,1}} \ket{x \oplus \beta}\bra{x}\, \rho \,\ket{x}\bra{x \oplus \beta}
\end{gather}
\end{subequations}
acting on the bit $\s[v]$, where $a \oplus b$ is the sum of bits $a$ and $b$ modulo $2$.

Because the cluster state is a stabilizer state, we may determine the state of the system after the $\axisZ$ measurements.
For a measurement on a qubit $v$, $K_v$ is the only generator of the stabilizer group $\cC_{n,d}$ (as described in Definition~\ref{def:clusterState}) which does not commute with the observable $Z_v$.
By the stabilizer formalism (Section~\ref{sec:stabilizerFormalism}), the state of the other qubits after a $Z_v$ measurement where we obtain $\s[v] = r$ is then given by
\begin{gather}
		\tilde\cC\supp{r}
	\;\;=\;\;
		\gen{\tilde K\supp{r}_u}_{u \in V(G) \setminus v}	\;,
\end{gather}
where we have
\begin{gather}
		\tilde K\supp{r}_u
	\;\;=\;\;
		\begin{cases}[r@{}l]
				K_u				\;=&\; X_u \prod\limits_{w \sim u} Z_w
														\,,	&	\text{if $u \not\sim v$}	\\[3ex]
				(-1)^r Z_v K_u	\;=&\; (-1)^r \,X_u \!\prod\limits_{\substack{w \sim  u \\ w \ne v}} \!Z_w
														\,,	&	\text{if $u \sim v$}
		\end{cases}\,,
 \end{gather}
and where $\sim$ is the adjacency relation in the grid.
In the case where $\s[v] = 0$, the state of the system is stabilized by a group generated by $\tilde K_u\supp{0}$, which like the generators $K_u$ are tensor products of an $X$ operator and some number of $Z$ operators.
Otherwise, we may convert the state of the system to that stabilized by $\tilde\cC\supp{0}$ by performing the operations $\Zcorr{u}{}$ on every qubit $u$ adjacent to $v$ in $G_{n,d}$\,: because $Z_u$ anticommutes with $X_u$ and commutes with $Z_w$ for any qubit $w$, we have
\begin{gather}
		\Zcorr{u}{}\Big(\tilde K\supp{1}_u\Big)
	\;\;=\;\;
		Z_u \sqparen{ - X_u \prod_{w} Z_w } Z_u
	\;\;=\;\;
		X_u \prod_{w} Z_w 
	\;\;=\;\;
		\tilde K\supp{0}_u	\;.
\end{gather}
Then, we can select for the post-measurement state corresponding to $\s[v] = 0$ by performing the operation
\begin{gather}
	\mathsf B_v \;\;=\;\; \prod_{u \sim v} \Zcorr{u}{\s[v]}	\;.
\end{gather}
We call such an operation a \emph{byproduct} operation.
To perform pure-state computation, there is no loss in generality in selecting for the post-measurement state corresponding to $\s[v] = 0$\,, rather than $\s[v] = 1$\,, as they are equivalent up to conditional unitaries which may be performed in the cluster-state model in any case.

As we noted, the post-measurement state produced in this way is stabilized by operators $\tilde K\supp{0}_u$ for $u \in G_{n,d} \setminus v$, which are each tensor products of $X$ and $Z$ operators, but which are distinct from the stabilizers of $\cC_{n,d}$.
Furthermore, by construction, each $\tilde K\supp{0}_u$ acts on $u$ with an $X$ operation, and acts on a qubit $w \in V(G_{n,d})$ with a $Z$ operation if and only if $w \ne v$ and $w \sim u$: that is, if $w$ is adjacent to $u$ in the graph $G_{n,d} \setminus v$.
Thus, a $\axisZ$ measurement on $v$ (followed by the byproduct $\mathsf B_v$ if required) corresponds to removing a vertex represented by $v$ from the $n \x d$ grid graph, producing a state similar to the cluster state, but defined on a different graph.
This motivates the following definition:
\begin{definition}
	For an arbitrary graph $G$, the \emph{graph state} corresponding to $G$ is a stabilizer state $\rho_{\ssstyle G} = \ket{G}\bra{G}$, where $\ket{G}$ is a unit-vector stabilized by the group $\sS_G = \gen{K_{G,u}}_{u \in V(G)}$ for generators given by 
	\begin{gather}
		\label{eqn:graphStabilizers}
			K_{G,v}
		\;\;=\;\;
			X_v \prod_{w \in \ngbh[G]{v}} Z_w	\;,
	\end{gather}
	and where $\ngbh[G]{v}$ denotes the set of vertices adjacent to $v$ in the graph $G$.
\end{definition}
Applying the same analysis as for performing a $\axisZ$ measurement on a qubit $v$ of the $n \x d$ cluster state, we may show that for any graph $G$ and vertex $v \in V(G)$, we have
\begin{gather}
		\sqparen{\prod_{w \in \ngbh[G]{v}} \Zcorr{w}{\s[v]}} \Meas{v}{\sfz} (\rho_{\ssstyle G})
	\;\;=\;\;
		\rho_{\ssstyle [G \setminus v]}	\;.
\end{gather}
Any $\axisZ$ corrections on qubits $w$ which are also to be measured with a $\axisZ$ measurement can be absorbed into the $\Meas{w}{\sfz}$ operation as in~\eqref{eqn:ZmeasCommute}.
Let $S$ be the set of vertices measured with a $\Meas{}{\sfz}$ operation: $\axisZ$ corrections then accumulate on the neighbors of vertices in $S$, yielding the transformation
\begin{gather}
		\sqparen{\prod_{w \in \ngbh{S}} \Zcorr{w}{\,\big(\!\sum_{v \in S_w} \s[v]\big)}}
		\sqparen{\prod_{v \in S} \Meas{v}{\sfz}}	,
\end{gather}
where $\ngbh{S}$ is the set of qubits in $G_{n,d}$ adjacent to (but not contained in) $S$ in the $n \x d$ grid, and $S_w = \ngbh[G_{n,d}]{w} \inter S$.
This operation serves to transform $\rho_{\ssstyle G_{n,d}}$ to a graph state corresponding to some induced subgraph, $\rho_{\ssstyle [G_{n,d} \setminus S]}$\,.

We may abstract this process of producing graph states from cluster states via vertex removal, to describe a more general measurement-based model.
For an \emph{arbitrary} graph $G$, let $\New{G}{}$ be the map preparing the graph state $\ket{G}\bra{G}$.
Then we may consider a \emph{graph state model}  of computation, generalizing the cluster state model, by allowing preparation maps $\New{G}{}$ for arbitrary $G$ (or, for instance, for $G$ from some broader family of graphs) rather than just for grids $G_{n,d}$\,.

\subsection{Open graph states, geometries, and the general ``one-way'' model}
\label{sec:geometriesOneWay}

As with other models of quantum computing, in order to describe how to achieve universal quantum computation in measurement-based models of quantum computing, it is convenient to reduce to universality for unitary transformations.
This motivates another extension beyond graph-state models --- which can only prepare independent graph states and perform classically controlled single-qubit operations on them --- to a model which permits multi-qubit operations on previously allocated qubits, just as we generalized universality for quantum computation to universality for CPTP transformation.

A natural approach (and one also suggested in~\cite{RB01}) is to extend beyond graph states to maps which embed quantum states $\rho \in \D(I)$ into a stabilizer code similar to a graph state.
Here, $I$ is a set representing an \emph{input system} of qubits, which we include in the vertex-set $I \subset V(G)$ of a graph $G$\,: we then consider the stabilizer code described by a group $\sS_{G,I}$ generated by the operators $K_{G,v}$ as in~\eqref{eqn:graphStabilizers}, but ranging only over qubits $v \in V(G) \setminus I$.
In order to describe an explicit map for this embedding, consider how the operators $K_{G,v}$ may be generated by Clifford group operations from simpler operations.
From~\eqref{eqn:cliffordMapping}, we may observe that
\begin{gather}
		K_{G,u}
	\;\;=\;\;
		\sqparen{\prod_{v \in \ngbh[G]{u}} \!\!\Ent{u,v}} (X_u)	\,,
\end{gather}
where this product may be taken in any order; then, the code which is stabilized by $\sS_{G,I} = \gen{K_{G,u}}_{u \in V(G) \setminus I}$ can be obtained from the code stabilized by $\sS_{T,I} = \gen{X_u}_{u \in V(G) \setminus I}$ by performing the operation
\begin{gather}
		\Ent{G}
	\;\;=
		\prod_{uv \in E(G)}
			\!\!\Ent{u,v}	\;.
\end{gather}
The operators $X_u$ also generate the stabilizer group of a graphical code, for the ``trivial'' graph $T$ on the vertices $V(G)$ but with no edges.
This code consists of states $\ket{\splus}_u$ on each $u \in V(G) \setminus I$, with no constraints imposed on the qubits of $I$.
Thus, the map $\cE_{G,I} : \D(I) \to \D(V(G))$ encoding the qubits of $I$ into the code stabilized by $\sS_{G,I}$ can be given by
\begin{gather}%
	\label{eqn:openGraphEncoding}
		\cE_{G,I}
	\;\;=\;\;
		\sqparen{\prod_{uv \in E(G)} \!\! \Ent{u,v}} \sqparen{\prod_{u \in V(G) \setminus I} \!\!\New{u}{\sfx}}
	\;\;=\;\;
		\Ent{G} \, \New{I\comp}{}	\;,
\end{gather}
where we abbreviate $I\comp = V(G) \setminus I$.
We may refer to $\cE_{G,I}$ as an \emph{open graph state} encoding.\footnote{%
This terminology is similar to, but different from, that used in \eg~\cite{BKMP07,MP08}, 
who use the term ``open graph state'' to refer to what we call a \emph{geometry} in Definition~\ref{def:geometry} (in agreement with~\cite{DK05u}).} 
However, in contrast to the $\Cluster(\A)$ model, we do not make the embedding $\cE_{G,I}$ a primitive operation of the corresponding model of computation, and instead define the following model:
\begin{definition}
	\label{def:oneWayModel}
	For any set of angles $\A \subset \R$, infinite set $V$ of qubits, and infinite set $B$ of bits, the \emph{one-way model} is the model of computation with the elementary gate set
	\begin{align}
			\OneWay(\A)
		\;\;=\;\;
			\bigg\{& \New{v}{\sfx}\,,\; \Ent{u,v}\,,\; \Meas{v}{\sfz}\,,\; \Meas{v}{\alpha;\beta,\gamma}\,,\; \Xcorr{v}{\beta}\,,\; \Zcorr{v}{\beta}\,,\; \Shift{v}{\beta}\,,\; \tr[v]
 				\,\bigg|\, u,v \in V, ~\alpha \in \A, 
		\notag\\[-1ex]&\qquad\qquad
			  ~\text{and}~ ~\beta,\gamma \;= {\textstyle\sum\limits_{\s[v] \in S} \s[v]} ~~\text{for any $S \subset B$} \bigg\}\,.
	\end{align}
	We will say that a procedure $\mathfrak P$ in the $\OneWay(\A)$ model is in \emph{standard form} if it can be decomposed as $\mathfrak P \,=\, \mathsf C \,\circ\, \mathsf M \,\circ\, \Ent{G} \circ \New{S}{}$ for $\mathsf C$ consisting of corrections, $\mathsf M$ consisting of shift operations and measurement operations with no $\pi$-dependencies, $\Ent{G}$ consisting of a product of \emph{entangler} operations $\Ent{u,v}$ for distinct pairs $\ens{u,v} \subset V(G)$ for some graph $G$, and $\New{S}{}$ consisting of preparation operations $\New{v}{\sfx}$ for $v \in S \subset V(G)$.
	We refer to these as the \emph{correction}, \emph{measurement}, \emph{entangling}, and \emph{preparation} stages respectively.
\end{definition}
The model of measurement-based quantum computing which later chapters will be primarily concerned with is the model $\OneWay(\A)$ for sets of angles $\A \subset \R$.
We will refer to models of this form generically as \emph{the one-way model}, and refer to ``one-way procedures'', ``one-way patterns'', \etc\ when describing algorithms in the one-way model.

The purpose of extending to open graph encodings is to make possible the \emph{composition} of one-way patterns.
Therefore, in addition to the extension to admit an input subsystem, we also explicitly identify \emph{output} subsystems of measurement-based procedures, which simply consists of those qubits which are not measured (and the classical results of the measurements).
We may then achieve universality for unitary transformations by performing an appropriate set of elementary unitary transformations --- effectively simulating a unitary circuit in doing so.

A specification of an open graph state encoding, together with the set of qubits to be left unmeasured, effectively captures the \emph{structural} information (that aspect which is independent of measurement observables and measurement results) of a one-way measurement procedure, which we may describe as follows:

\begin{definition}
	\label{def:geometry}
	For a measurement based procedure $\mathfrak P$ in the one-way model, the \emph{geometry underlying $\mathfrak P$} is a triple $(G,I,O)$ where
	\begin{romanum}
	\item
		$G$ is the \emph{entanglement graph}, whose vertices are labelled by the qubits on which $\mathfrak P$ acts, and whose edges are given by the operations of the entangling phase of $\mathfrak P$\,;

	\item
		$I \subset V(G)$ is the \emph{input subsystem} of $\mathfrak P$, consisting of the qubits which are not allocated by the preparation phase of $\mathfrak P$\,;

	\item
		$O \subset V(G)$ is the \emph{output subsystem} of $\mathfrak P$, consisting of the qubits which are not discarded by the measurement phase of $\mathfrak P$.
	\end{romanum}
\end{definition}
We will often illustrate geometries in the style shown in Figure~\ref{fig:genericGeom}.
\begin{figure}[t]
	\begin{center}
		\includegraphics{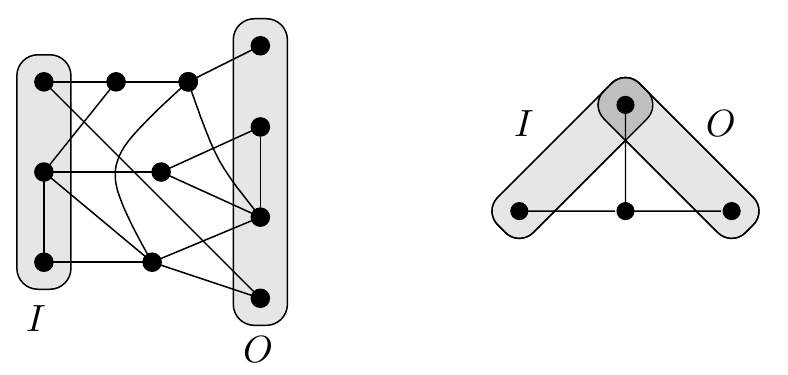}
	\end{center}
	\caption[Illustration of two geometries for one-way procedures.]{\label{fig:genericGeom}%
		Illustration of two geometries for one-way procedures.
		Labels for the vertices have been omitted.
		Note in particular that the sets $I$ and $O$ may overlap.}
\end{figure}

When composing one-way computations, we will often be interested in converting the composition into standard form, in which case no qubit can be allocated after measurement operations have occurred.
Therefore, when composing one-way procedures $\mathfrak P = \mathfrak P_2 \circ \mathfrak P_1$, we will be interested in restricting the qubits which are allocated in $\mathfrak P_2$ to be distinct from any qubits involved in $\mathfrak P_1$.
We can express this in terms of a sense of composing the \emph{geometries} of these patterns, using the constraints described in Lemma~\ref{lemma:typeCompose} (page~\pageref{lemma:typeCompose}):
\begin{definition}
	For two geometries $\cG_1 = (G_1, I_1, O_1)$ and $\cG_2 = (G_2, I_2, O_2)$, we say that $\cG_1$ and $\cG_2$ are \emph{composable} if $\big[V(G_1) \inter V(G_2)\big] \subset \big[I_2 \inter O_1\big]$\,.
	The \emph{composition} $\bar\cG = \cG_2 \circ \cG_1$ is then defined by $\bar\cG = (\bar G, \bar I, \bar O)$, where
	\begin{subequations}
	\begin{align}
			V(\bar G)
		\,\;=&\;\;\;
			V(G_1) \union V(G_2)	\,,
		\\
			E(\bar G)
		\,\;=&\;\;\;
			E(G_1) \symdiff E(G_2)	\,,
		\\
			\bar I
		\;\;=&\;\;\;
			I_1 \,\,\union (I_2 \setminus O_1) \,,
		\\
			\bar O
		\,\;=&\;\;\;
			O_2 \union (O_1 \setminus I_2) \;,
	\end{align}
	\end{subequations}
	where $S \symdiff T = [S \setminus T] \union [T \setminus S]$ is the symmetric difference of $S$ and $T$.
	Two one-way procedures $\mathfrak P_1$ and $\mathfrak P_2$ are said to be \emph{properly composable} if their underlying geometries  $\cG_1$ and $\cG_2$ are composable; the underlying geometry of the procedure $\mathfrak P_2 \circ \mathfrak P_1$ is then $\cG_2 \circ \cG_1$.
\end{definition}
Figure~\ref{fig:geomCompose} illustrates the composition of two geometries.
\begin{figure}[t]
	\begin{center}
		\includegraphics{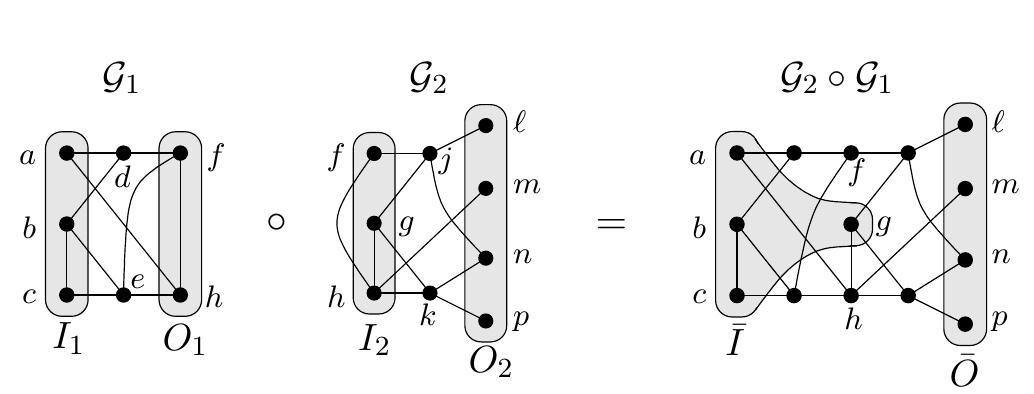}
		\vspace{-2ex}
	\end{center}
	\caption[Illustration of the composition of two generic geometries for one-way procedures.]{\label{fig:geomCompose}%
		Illustration of the composition of two generic geometries for one-way procedures.
		Notice that composition in the diagrams is left-to-right, while the composition in written notation is right-to-left.}
		\vspace{2ex}
\end{figure}
From this point onwards, we will suppose that two composable patterns $\mathfrak P_1$ and $\mathfrak P_2$ are properly composable: we lose no generality in doing this, as we may ``relabel'' the qubits acted on by the patterns $\mathfrak P_1$ and $\mathfrak P_2$ to achieve this if necessary.

\subsection{Universality of the one-way and graph-state models}

We may now illustrate how to perform universal quantum computation in the one-way model, using the construction illustrated in~\cite{DKP07}, which constructs measurement patterns which ``simulate'' unitary circuits.
Consider the one-way patterns
\begin{align}
	\label{eqn:elemOpsDKP}
		\mathfrak J_{w/v}^\alpha
	\;\;=&\;\;
		\Xcorr{w}{\s[v]} \Meas{v}{-\alpha} \Ent{v,w} \New{w}{\sfx}	\;,
	&
		\Ent{u,v}	\;,
\end{align}
whose underlying geometries are illustrated in Figure~\ref{fig:primitiveGeom}.
\begin{figure}[t]
	\begin{center}
		\includegraphics{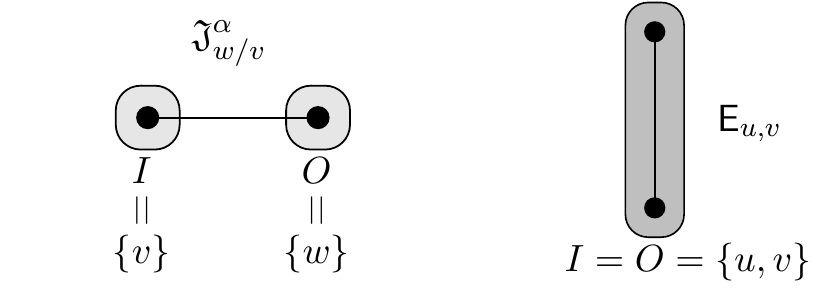}
		\vspace{-2ex}
	\end{center}
	\caption[Illustration of the geometries of two primitive one-way patterns.]{\label{fig:primitiveGeom}%
		Illustration of the geometries of two primitive one-way patterns.
		(Note that the input and output subsystems of the pattern on the right coincide.)}
		\vspace{2ex}
\end{figure}
The single entangler operation $\Ent{u,v}$ in particular is a simple one-way procedure in its own right, and can be interpreted as representing the same operation between two wires in a unitary circuit model such as $\Phases(\A)$.
To describe the $\mathfrak J_{w/v}^\alpha$ operation, note that the first two operations put the system into a stabilizer code which is stabilized by the two operator group $\sS \,=\, \ens{\idop_v \ox \idop_w\,,\; Z_v \ox X_w}$.
We can decompose $\Meas{v}{-\alpha}$ into a $\axisZ$-rotation and an $\axisX$ measurement:
\begin{align}
		\Meas{v}{-\alpha}(\rho)
	\;\;=&\;\;
		\bra{\psub{(\minus\alpha)}}_v\, \rho \,\ket{\psub{(\minus\alpha)}}_v	\;\ox\; \ket{0}\bra{0}_{\s[v]}
		\;\;+\;\;
		\bra{\msub{(\minus\alpha)}}_v\, \rho \,\ket{\msub{(\minus\alpha)}}_v	\;\ox\; \ket{1}\bra{1}_{\s[v]}
	\notag\\[1ex]=&\;\;
		\bra{\splus}_v \Big[ R_\sfz(\alpha)_v \;\rho\; R_\sfz(-\alpha)_v \Big] \ket{\splus}_v	\ox \ket{0}\bra{0}_{\s[v]} \;\;+\;\;
	\notag\\&\qquad\qquad
		\bra{\sminus}_v \Big[ R_\sfz(\alpha)_v \;\rho\; R_\sfz(-\alpha)_v \Big] \ket{\sminus}_v	\ox \ket{1}\bra{1}_{\s[v]}
	\notag\\[1ex]=&\;\;
		\Meas{v}{\sfx} \, \sfR_v^{\alpha} (\rho)	\;.
\end{align}
Because the operator $R_\sfz(\alpha)$ is diagonal, $\sfR^\alpha_v$ commutes with the operation $\Ent{v,w}$ as well as $\New{w}{\sfx}$\,.
Then, we have
\begin{align}
	\label{eqn:factorRotationJ}
		\mathfrak J_{w/v}^\alpha
	\;\;=\;\;
		\Xcorr{w}{\s[v]} \Meas{v}{-\alpha} \Ent{v,w} \New{w}{\sfx}
	\;\;=&\;\;
		\Xcorr{w}{\s[v]} \Meas{v}{\sfx} \sfR_v^{\alpha} \Ent{v,w} \New{w}{\sfx}
	\notag\\=&\;\;
		\Xcorr{w}{\s[v]} \Meas{v}{0} \Ent{v,w} \New{w}{\sfx} \sfR_v^{\alpha} 
	\;\;=\;\;
		\mathfrak J_{w/v}^{\,0} \, \sfR_v^{\alpha} \;.
\end{align}
Then, it suffices to consider the effect of $\mathfrak J_{w/v}^{\,0}$, which by linearity we may characterize by the effect on any four linearly independent $2 \x 2$ operators.
We will consider the effect of $\mathfrak J_{w/v}^{\,0}$ on the Pauli operators via the stabilizer formalism, as follows.
As we noted above, the maps $\Ent{v,w} \New{w}{\sfx}$ encodes a single-qubit state into the code stabilized by $\ens{\idop_v \ox \idop_w\,,\; Z_v \ox X_w}$\,.
An $X_v$ observable measurement then performs an isometry, and performing a $\Xcorr{w}{\s[v]}$ correction selects the post-measurement state result corresponding to $\s[v] = 0$.
Thus, we may reduce how $\mathfrak J_{w/v}^{\,0}$ transforms the Pauli operators on $v$ to how it transforms $X_v$ and $Z_v$ in particular.
We may then compute
\begin{subequations}
\label{eqn:stabilizerJ}
\begin{align}
			\gen{\, X_v \,}
		\;\;\mapstoname{\New{w}{\sfx}}\;\;
			\gen{\begin{array}{c@{\,\ox\,}c@{}l}
					X_v		&	\idop_w	&	\;,	\\
					\idop_v	&	X_w	
			\end{array}}
		\;\;\mapstoname{\Ent{v,w}}\;\;
		&
			\gen{\begin{array}{c@{\,\ox\,}c@{}l}
					X_v	&	Z_w	&	\;,	\\
					Z_v	&	X_w	
			\end{array}}
		\;\;\mapstoname{\Xcorr{w}{\s[v]} \Meas{v}{\sfx}}\;\;
			\gen{\, Z_w	\,}
	\;,
	\\[3ex]
			\gen{\, Z_v \,}
		\;\;\mapstoname{\New{w}{\sfx}}\;\;
			\gen{\begin{array}{c@{\,\ox\,}c@{}l}
					Z_v		&	\idop_w	&	\;,	\\
					\idop_v	&	X_w	
			\end{array}}
		\;\;\mapstoname{\Ent{v,w}}\;\;
		&
			\gen{\begin{array}{c@{\,\ox\,}c@{}l}
					Z_v	&	\idop_w	&	\;,	\\
					Z_v	&	X_w	
			\end{array}}
	\notag\\=\;\;&
			\gen{\begin{array}{c@{\,\ox\,}c@{}l}
					Z_v		&	\idop_w	&	\;,	\\
					\idop_v	&	X_w	
			\end{array}}
		\;\;\mapstoname{\Xcorr{w}{\s[v]} \Meas{v}{\sfx}}\;\;
			\gen{\, X_w	\,}
	\;.
\end{align}
\end{subequations}
In both cases, we have $\mathfrak J_{w/v}^{0}(\sigma_v) \,=\, \sfH_w (\sigma_w)$\,, which then holds for all $\sigma \in \L(\cH[v])$.
Therefore, we have
\begin{align}
	\label{eqn:semanticsJ}
		\mathfrak J_{w/v}^{\alpha}(\rho_v)
	\;\;=\;\;
		\sqparen{\sfH \circ \sfR^{\alpha} (\rho)}_w
	\;\;=\;\;
		\sqparen{\sfJ^{\alpha}(\rho)}_w	\;.
\end{align}
Thus, the operations performed by the patterns $\mathfrak J_{w/v}^{\alpha}$ and $\Ent{v,w}$ correspond to the unitary operations of the $\Jacz(\A)$ unitary circuit model defined in Definition~\ref{def:elemGateSets}.
Using stable index notation, we may then define an operator homomorphism $\Phi$ mapping unitary circuits in the $\Jacz(\A)$ model to (non-standard form) measurement procedures in the $\OneWay(\A)$ model as follows.
For a stable index tensor expression $C$, we may require without loss of generality that the terms of $C$ are in an order corresponding to the order in which the operations are performed in the circuit (\eg~as in an ordinary operator expression for $C$).
Identifying advanced/deprecated tensor indices in $C$ with allocated/discarded qubits in the one way pattern (using the correspondence described on page~\pageref{discn:typeCorrespondance}), we define
\begin{align}
	\label{eqn:circuitPatternHomphm}
		\Phi\Big(J(\alpha)\pseu[:w/v]\Big)
 	\;\;=&\;\;
		\mathfrak J_{w/v}^{\alpha}	\;;
	&
		\Phi\Big(\cZ\pseu[v,w]\Big)
	\;\;=&\;\;
		\Ent{v,w}\;:
\end{align}
the unitary transformation described by a unitary circuit $C$ in stable index notation is then also performed by the one-way procedure $\mathfrak P = \Phi(C)$\,.\footnote{%
	We have deliberately neglected the allocation of the result register $\s[v]$ in this account of how unitaries are performed, in part because we are primarily interested in how the quantum registers are transformed rather than the classical information produced as a result.
	We conventionally ignore classical bits in discussions of the ``transformations performed'' by measurement patterns: this can be formalized by adopting a convention of implicitly tracing out all result registers at the end of a quantum computation.}

\subsection{Standardization of one-way patterns, and the DKP constructions}
\label{sec:DKPstandardization}

The homomorphism $\Phi$ from circuits to one-way patterns is the main instrument in the simplest constructions of standardized measurement patterns, presented by Danos, Kashefi, and Panangaden in~\cite{DKP07}.
In this section, we describe these constructions, which we will refer to as the DKP constructions for one-way patterns.

For any one-way procedure --- and for the one-way procedures arising from the homomorphism $\Phi$ of~\eqref{eqn:circuitPatternHomphm} in particular --- we may obtain a standardized one-way procedure by accumulating correction operations from the beginning of the pattern and moving them to the end.
To do so, we must commute them past preparation operations (which we may easily do) and entangler operations (which induces further corrections), and absorbing them into the measurement operations to describe how measurements are adapted.
We describe how this is done in this section, essentially via the stabilizer formalism, representing the correction operations as a product of Pauli operators, together with boolean expressions representing their classical dependencies.

Starting from the beginning of the pattern, we scan towards the end of the pattern, maintaining a set of correction operations that we encounter.
We attempt to commute each correction operator past the other operators in the pattern, as follows:
\begin{enumerate}
\item
	As the preparation maps $\New{v}{\sfx}$ are always the first operations on any qubit $v \in I\comp$ (in particular because, as we have restricted ourselves to properly composable patterns, no qubit is discarded and subsequently re-allocated), any operations which precede a preparation map also commutes with it.
	Thus, we may trivially commute any correction operations past $\New{v}{\sfx}$ operations.

\item
	Similarly, we may commute correction operations past any entangling operator $\Ent{u,v}$.
	However, operations of the form $\Ent{u,w}$ do not commute with correction operations $\Xcorr{v}{\beta}$\,: in particular, we have
	\begin{align}
			\Ent{u,v} \Xcorr{v}{\beta}
		\;\;=\;\;
			\Zcorr{u}{\beta} \Xcorr{v}{\beta} \Ent{u,v}	\;.
	\end{align}
	Thus, in commuting an $\Xcorr{v}{\beta}$ operation past an entangler acting on $w$ and another qubit $u$, we induce an additional correction $\Zcorr{u}{\beta}$, increasing the complexity of the correction operation.

\item
	Finally, for a measurement on any qubit $v$ that we encounter, we absorb the corrections accumulated on $v$ into the bases of measurement, according to the equations
	\begin{subequations}
	\label{eqn:measAbsorbCorr}
	\begin{gather}
	\begin{alignat}{4}
			\Meas{w}{\alpha;\beta,\gamma} \Xcorr{w}{\beta'}
		\;\;=&\;\;\;
			\Meas{w}{\big[(\minus 1)^{\beta + \beta'} \cdot\, \alpha \,+\, \gamma\pi\big]}
		&\;\;=&\;\;\;
			\Meas{w}{\alpha;(\beta + \beta'),\gamma}	\;,
		\\
			\Meas{w}{\alpha;\beta,\gamma} \Zcorr{w}{\gamma'}
		\;\;=&\;\;\;
			\Meas{w}{\big[(\minus 1)^\beta \cdot\, \alpha \,+\, (\gamma + \gamma')\pi\big]}
		&\;\;=&\;\;\;
			\Meas{w}{\alpha;\beta,(\gamma + \gamma')} \;.
	\end{alignat}
	\end{gather}
	\end{subequations}
	This decreases the complexity of the corrections which we keep track of, transferring the information of the corrections on $v$ to the measurement on $v$.
\end{enumerate}
The result is a measurement procedure with a phase of classically controlled correction operations on the outputs, preceded by a mixed sequence of preparation maps, entanglers, and corrections with both sign- and $\pi$-dependencies.
Again, as the preparations are always the first operation performed on a qubit, and the measurements always the last, the operations in this mixed sequence all commute; we may then separate them into preparation, entangling, and measurement phases.

The complexity of the operations described above may be performed in time polynomial in the number of qubits $n$ operated on in the pattern.
For an arbitrary one-way pattern, at each point in time, we must track at most one $\Xcorr{v}{\beta_v}$ operation and one $\Zcorr{v}{\gamma_v}$ operation per qubit, with both $\beta$ and $\gamma$ being boolean expressions of complexity at most $O(n)$.
Each entangler operation acting on a qubit $v$ then induces a change in the expression $\Zcorr{v}{\gamma_v}$ of complexity $O(n)$\,, and there are $O(\deg_G(v))$ entanglers which may act on $v$, where $G$ is the graph of the geometry underlying the pattern.
Then, the complexity of updating the corrections on each qubit throughout the pattern is at most $O(n \deg_G(v))$\,; accumulated across all qubits, the work required is then $O\big(n \,\smash{\sum\limits_v \deg_G(v)}\big) = O(n m)$, where $m = \card{E(G)}$.

For one-way patterns arising out of the homomorphism $\Phi$, we may achieve a much better bound, arising from the fact that the corrections are not fully general: in particular, the correction on each qubit $w$ is initially just an operation $\Xcorr{w}{\s[v]}$ for some qubit $v$.
Following the same analysis as above, the work required to standardize such a pattern is then at most $O\big(\smash{\sum\limits_v \deg_G(v)}\big) = O(m)$.

In the case of the one-way patterns arising from $\Phi$, the resulting (non-standard form) procedure is one of the constructions for one-way patterns described in~\cite{DKP07}, which we will refer to in later chapters as follows:
\begin{definition}
	\label{def:simplifiedDKP}
	The \emph{simplified DKP construction for one-way patterns} is the procedure for producing $\OneWay(\A)$ patterns from unitary circuits $C$ in the $\Jacz(\A)$ model by
	\begin{enumerate}
	\item
		constructing the measurement pattern $\Phi(C)$, where $\Phi$ is the mapping defined in~\eqref{eqn:circuitPatternHomphm};\footnotemark

	\item
		commuting the preparations and entangler maps of the resulting concatenated procedure to the beginning of the procedure;

	\item
		absorbing the resulting classically controlled corrections $\Xcorr{v}{\beta}$ and $\Zcorr{v}{\gamma}$ to the end of the measurement procedure, absorbing them into the measurements on the qubits in $V(G) \setminus O$ when required by the relations of~\eqref{eqn:measAbsorbCorr}.
	\end{enumerate}
\end{definition}%
\footnotetext{\label{fn:DKPformalSemantics}%
The construction of~\cite{DKP07} does not make use of stable index tensor notation for circuits, but rather through an equivalent consideration of formal semantics of symbols in a measurement calculus corresponding to the CPTP maps in the gate set $\OneWay(\A)$ which we defined above.}
A standardized procedure can be easily obtained from the simplified DKP construction by eliminating $\pi$-dependencies from measurements by observing that $\ket{\pmsub{(\alpha + \pi)}} = \ket{\mpsub\alpha}$.
Then, $\pi$-dependencies may be abstracted away from measurements via shift operators, using the equation
\begin{gather}
	\label{eqn:piElim}
		\Meas{v}{\alpha;\beta,\gamma}
	\;\;=\;\;
		\Meas{v}{(\minus 1)^\beta \cdot\, \alpha \,+\, \gamma \pi}
	\;\;=\;\;
		\Shift{v}{\gamma} \Meas{v}{(\minus 1)^\beta \cdot\, \alpha}
	\;\;=\;\;
		\Shift{v}{\gamma} \Meas{v}{\alpha;\beta}	\;.
\end{gather}
That is, rather than potentially ``rotating the apparatus by 180 degrees'' to exchange the states $\ket{\pmsub\alpha}$ in a given measurement, we conditionally change the measurement result $\s[v]$ by adding (modulo 2) the expression $\gamma$.

There are two further techniques for simplifying a measurement procedure arising from the simplified DKP construction, which are also described in~\cite{DKP07}:

\paragraph{Pauli simplifications.}

It is easy to verify that $\ket{\pmsub0} \;=\; \ket{\spm \sfx}$, and $\ket{\pmsub{\pi\!/\:\!\!2}} = \ket{\spm \sfy}$\,; we may use this to further reduce measurement dependencies for measurements $\Meas{}{\alpha;\beta}$ for $\alpha$ a multiple of $\pion 2$, as follows.
For $\alpha$ any multiple of $\pi$, the measurement operators $\Meas{v}{\alpha}$ and $\Meas{v}{-\alpha}$ on any qubit $v$ are equivalent, as they differ by an integer multiple of $2\pi$\,: changes of the sign of the angle of measurement have no impact, and as a result such a measurement does not have any classical dependencies.
For $\alpha$ an odd multiple of $\pion2$, we have $-\alpha \equiv \alpha + \pi \pmod{2\pi}$, in which case we may also extract the classical dependency into a shift operator
\begin{align}
		\Meas{v}{(\minus 1)^\beta \cdot \,\pion 2}
	\;\;=\;\;
		\Meas{v}{\pion 2 \,+\, \beta \pi}
	\;\;=\;\;
		\Shift{v}{\beta} \Meas{v}{\pion 2}	\;.
\end{align}
These measurements (together with any \axisZ\ measurements $\Meas{v}{\sfz}$) are described as \emph{Pauli measurements}, and the procedure of eliminating dependencies of these measurements on prior measurement results as \emph{Pauli simplifications}.
By performing Pauli simplifications, the Pauli measurements in a one-way pattern may be performed independently of any other measurement result, and in particular they may be commuted to the beginning of the measurement phase of any one-way pattern.

\paragraph{Signal shifting.}

After bringing a one-way pattern from the simplified DKP construction into standard form and performing Pauli simplifications, we may eliminate any shift operators by commuting them to the end of a measurement procedure and discarding them.
For any one-way model operation $\Op{w}{\exprbr{\s[v]}}$ on a qubit $w$ which depends on a measurement result $\s[v]$, we have the relation
\begin{align}
			\Op{u}{\exprbr{\s[w]}} \Shift{w}{\beta}
	\;\;=&\;\;
			\Shift{w}{B} \Op{u}{\exprbr{ \s[w] + \beta }}
\end{align}
where $\exprbr{ \s[v] + \beta }$ is the result of substituting $\s[v]$ in the boolean expression with the value of $\s[v] + \beta$\,.
We may use such relations to commute all shift operators to the end of a one-way procedure.
Because shift operators are purely classical operations, shift operations at the end of a measurement procedure have no effect on the quantum output, and so they may be eliminated.
We call this procedure of eliminating shift operations \emph{signal shifting}.
(If this is performed simultaneously with the standardization procedure described above, we may accumulate shift operators along with the correction operations, and ultimately commute the shift operators past the corrections: an identical analysis for the run-time of this procedure holds as for standardization without signal shifting.)

The measurement pattern that arises from performing these transformations is then a standard form one-way pattern, where the measurement phase begins with a collection of commuting Pauli measurements; and where the operational dependencies of every measurement are represented explicitly in the classical controls of each operation, as a result of eliminating the shift operations.
We will refer to this construction as follows:
\begin{definition}
	\label{def:completeDKP}
	The (\emph{complete}) \emph{DKP construction for one-way patterns} is the procedure for producing $\OneWay(\A)$ patterns from unitary circuits in the $\Jacz(\A)$ model by
	\begin{enumerate}
	\item
		obtaining a measurement pattern from $C$, via the simplified DKP construction of Definition~\ref{def:simplifiedDKP}\,;

	\item
		eliminating $\pi$-dependencies, via the relation~\eqref{eqn:piElim}\,; 

	\item
		performing Pauli simplifications and signal shifting; and

	\item
		commuting Pauli measurements to the beginning of the measurement phase.
	\end{enumerate}
\end{definition}
The analysis of these constructions prove the following result, by the discussion following Definition~\ref{def:elemGateSets} on page~\pageref{def:elemGateSets}:
\begin{theorem}
	The $\OneWay(\A)$ model is
	\begin{inlinum}
	\item
		universal for quantum computation and for unitary transformations when $\A = \R$\,; and

	\item
		approximately universal for quantum computation and for unitary transformations when $\A = \frac{\pi}{4}\Z$\,.
	\end{inlinum}
	Furthermore, we may without loss of generality require that the measurement pattern is standardized, contains no shift operations, and that Pauli measurements are performed before any non-Pauli measurement operation.
\end{theorem}

\subsection{The simplified RBB construction}
\label{sec:RBB}

We may apply the techniques of the preceding sections to show the universality of the cluster-state model, by adapting the construction to the constrained case where the graph is an $n \x d$ grid.
We will do this by showing how the techniques of the preceding sections may be used to describe (a modest extension of) the elementary constructions of Raussendorf, Browne, and Briegel~\cite{RBB03} for quantum computation in the cluster state model.\footnote{%
	The constructions of~\cite[page~5]{RBB03} assume arbitrary precision of measurement angles, allowing the exact implementation of arbitrary single-qubit using a fixed ``horizontal width'' when embedded in the grid.
	For an account of approximate universality, a trivial but useful extension is to describe computations in terms of ``chains'' of $J(\alpha)$ operations on distinct qubits of arbitrary length, and then describe how to compensate for differences in horizontal positions in the grid.
}
We will refer to the resulting construction as ``the simplified RBB construction''.

The main difference between cluster-state based computation and what we have described in Definition~\ref{def:oneWayModel} as general one-way patterns are the topological constraints on the interactions between qubits imposed by the grid graph.
These parallel a commonly imposed constraint in unitary circuit models: a \emph{linear nearest neighbor} unitary circuit model is one where we a impose a linear ordering on the qubits by assigning them to points on a line (possibly corresponding \eg~to physical locations of the qubits as physical systems), and operations are restricted to operate only on adjacent qubits (or more generally, on blocks consecutive qubits, for operations acting on two or more qubits).
Linear nearest neighbor models can simulate more general models by performing swap operations on adjacent qubits, effecting permutations of the qubits in the linear ordering, in order to allow arbitrary collections of qubits to interact.
It is easy to show that the operation
\begin{gather}
	\label{eqn:simpleSwap}
		\mathsf{SWAP}_{u,v}
	\;\;=\;\;
		\sfH_u \sfH_v \Ent{u,v} \sfH_u \sfH_v \Ent{u,v} \sfH_u \sfH_v \Ent{u,v}
\end{gather}
interchanges independent pure states of two neighboring qubits $u$ and $v$: thus, for universality, it suffices to map linear nearest-neighbor unitary circuits into the cluster-state model.

To represent single-qubit gates, we may compose patterns of the form $\mathfrak J_{w/v}^\alpha$ as before.
However, in order to represent a chain of patterns
\begin{gather}
	\mathfrak J_{v_N/v_{N \minus 1}}^{\alpha_{N \minus 1}} \cdots\;\, \mathfrak J_{v_2/v_1}^{\alpha_1} \, \mathfrak J_{v_1/v_0}^{\alpha_0}	\;,
\end{gather}
which we may refer to as a \emph{chain pattern}, we require that the qubits $v_j$ for $0 < j < N$ all have degree $2$ in the corresponding entanglement graph (as any neighbors aside from $v_{j-1}$ and $v_{j+1}$ represents an operation not in the pattern described above).
To embed the corresponding graph state in the grid without additional neighborhood relationships, we must remove every neighbor of these qubits $v_j$ except for those specified by the pattern, which we do using $\axisZ$ measurements.
An illustration of such an embedding is illustrated in Figure~\ref{fig:isolatedChain}.
\begin{figure}[t]
	\begin{center}
		\includegraphics{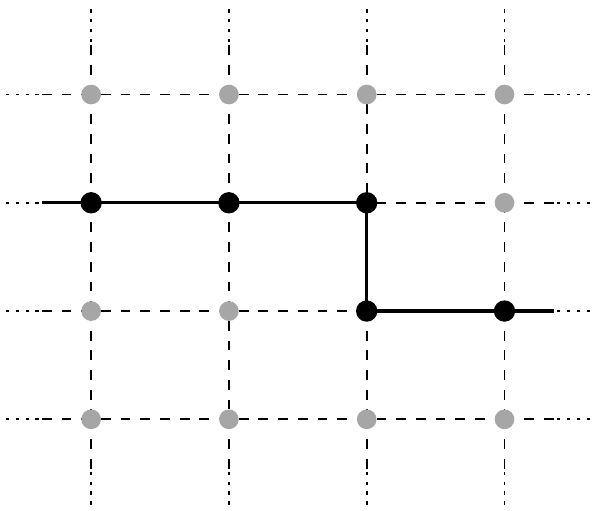}
	\end{center}
	\caption[Illustration of an embedding of a chain in the grid.]{\label{fig:isolatedChain}%
		Illustration of an embedding of a chain in the grid.
		Grey dots and broken lines represent qubits and entanglement relations which are effectively removed from the cluster state by $\axisZ$ measurements, to produce a graph state whose underlying graph contains a sequence of vertices with degree $2$.
	}
\end{figure}
Typically, such chain patterns would be embedded as a horizontal path through the grid, but as Figure~\ref{fig:isolatedChain} also shows, we may also employ more general paths in the grid.

Independent single-qubit unitaries acting on independent qubits may then be embedded in the grid as a collection of vertex-disjoint paths, where in particular the distance between two vertices of distinct paths is at least $2$ (in order to maintain a buffer of at least one grid site between two chain patterns).
The paths so traced out in the grid then correspond to \eg~the wires of a unitary circuit diagram.
In the construction, we may then represent the logical qubits of the quantum circuit by parallel horizontal paths in the grid whose vertical separation is precisely $2$.
Note that the paths which in general may be of different lengths, as each horizontal step corresponds to a single-qubit $J(\alpha)$ transformation for some given $\alpha$.
However, we may extend any path in the grid by a path of length $2$ or more in a way which represents the identity operation, using chain patterns of the form
\begin{align}
	\label{eqn:identityPatterns}
		\mathfrak {Id}^2_{v_2/v_0}
	\;\;=&\;\;
		\mathfrak J_{v_2/v_1}^{\,0} \, \mathfrak J_{v_1/v_0}^{\, 0}		\;,
	&
		\mathfrak {Id}^3_{v_3/v_0}
	\;\;=&\;\;
		\mathfrak J_{v_3/v_2}^{\pion 2} \, \mathfrak J_{v_2/v_1}^{\pion 2} \, \mathfrak J_{v_1/v_0}^{\pion 2}	\;;
\end{align}
by~\eqref{eqn:circuitPatternHomphm}, described in stable index notation, these operations are equivalent to unitary circuits $H \pseu[:v_2/v_1] H \pseu[:v_1/v_0]$ and $J(\pion 2) \pseu[:v_3/v_2] J(\pion 2) \pseu[:v_2/v_1] J(\pion 2) \pseu[:v_1/v_0]\Big.$ respectively.
The former obviously implements the single-qubit identity operation $\idop \pseu[:v_2/v_0]$\,, as $H$ is self-inverse; for the latter, recalling that $J(\pion 2) = H R_\sfz(\pion 2)\Big.$, we may characterize its effect by conjugation on linear operators via the stabilizer formalism:
\begin{subequations}
\begin{align}
		X
	&\;\mapstoname{\; H R_\sfz(\pion 2) \;}\;
		-Y
	\;\mapstoname{\; H R_\sfz(\pion 2) \;}\;
		Z
	\;\mapstoname{\; H R_\sfz(\pion 2) \;}\;
		X	\,,
	\\[1ex]
		Z
	&\;\mapstoname{\; H R_\sfz(\pion 2) \;}\;
		X
	\;\mapstoname{\; H R_\sfz(\pion 2) \;}\;
		-Y
	\;\mapstoname{\; H R_\sfz(\pion 2) \;}\;
		Z	\,;
\end{align}
\end{subequations}
then $J(\pion 2) J(\pion 2) J(\pion 2)$ performs $\idop_2$ as well.
Thus, both $\mathfrak {Id}^2_{v_2/v_0}$ and $\mathfrak {Id}^3_{v_3/v_0}$ map states $\rho_{v_0}$ to $\rho_{v_j}$ for some $j \ge 2$\,; the first using a path of length $2$ in the grid, and the second using a path of length $3$.
Using combinations of these, we can then implement a path of any length $\ell \ge 2$ in the grid, representing a sequence of operations which performs the identity $\idop_2$ on a given qubit.
Thus, whenever required, we may suppose that the paths in the grid corresponding to any two qubits are of the same length, regardless of the number of (non-trivial) unitary transformations performed on them.

In order to achieve universality for quantum computation, it then suffices to implement a logical $\cZ$ operation (or $\cX$ operation) in the cluster-state, between two qubits which are at the same horizontal position, and are vertically separated in the grid by a distance of $2$ (equivalently, by one grid site).
One approach to doing this may be described using a technique described in~\cite[Section~3.8]{BB06}.
Consider the pattern
\begin{gather}
	\label{eqn:simpleZzPattern}
		\mathfrak {Zz}_{u,v}
	\;\;=\;\;
		\Zcorr{v}{\s[a]}
		\Zcorr{u}{\s[a]}
		\Meas{a}{\pion 2}
		\Ent{a,v}
		\Ent{a,u}
		\New{a}{\sfx}
\end{gather}
with input and output subsystems $I = O = \ens{u,v}$\,.
The preparation and entangling phases encode the input state into the code stabilized by $\ens{ \idop_u \ox \idop_a \ox \idop_v \,, Z_u \ox X_a \ox Z_v }$\,: the operation $\Meas{a}{\pion 2}$ then represents a $\axisY$ measurement which performs an isometry on this code transformation, with the subsequent corrections effectively selecting for the result $\s[a] = 0$.
Using the stabilizer formalism, we may then determine
\begin{subequations}
\begin{align}
		\gen{ X_u \ox \idop_v }
	\;\mapstoname{\; \Ent{a,v} \Ent{a,u} \New{a}{\sfx} \;}&\;
		\gen{\begin{array}{c@{\,\ox\,}c@{\,\ox\,}c@{}l}
				X_u	&	Z_a	&	\idop_w	&	\;,	\\
				Z_v	&	X_w	&	Z_w		&
		\end{array}}
	\notag\\[-0.2ex]=&\;
		\gen{\begin{array}{c@{\,\ox\,}c@{\,\ox\,}c@{}l}
				Y_u	&	Y_a	&	Z_w		&	\;,	\\
				Z_v	&	X_w	&	Z_w		&
		\end{array}}
	\;\mapstoname{\; \Zcorr{v}{\s[a]}\Zcorr{u}{\s[a]} \Meas{a}{\pi/2} \;}\;
		\gen{ Y_u \ox Z_w }	\;,		
 	\\[2ex]
		\gen{ \idop_u \ox X_v }
	\;\mapstoname{\; \Ent{a,v} \Ent{a,u} \New{a}{\sfx} \;}&\;
		\gen{\begin{array}{c@{\,\ox\,}c@{\,\ox\,}c@{}l}
				\idop_u	&	Z_a	&	X_w		&	\;,	\\
				Z_v		&	X_w	&	Z_w		&
		\end{array}}
	\notag\\[-0.2ex]=&\;
		\gen{\begin{array}{c@{\,\ox\,}c@{\,\ox\,}c@{}l}
				Z_u	&	Y_a	&	Y_w		&	\;,	\\
				Z_v	&	X_w	&	Z_w		&
		\end{array}}
	\;\mapstoname{\; \Zcorr{v}{\s[a]}\Zcorr{u}{\s[a]} \Meas{a}{\pi/2} \;}\;
		\gen{ Z_u \ox Y_w }	\;,		
	\intertext{}
		\gen{ Z_u \ox \idop_v }
	\;\mapstoname{\; \Ent{a,v} \Ent{a,u} \New{a}{\sfx} \;}&\;
		\gen{\begin{array}{c@{\,\ox\,}c@{\,\ox\,}c@{}l}
				Z_u	&	\idop_a	&	\idop_w	&	\;,	\\
				Z_v	&	X_w		&	Z_w		&
		\end{array}}
	\;\mapstoname{\; \Zcorr{v}{\s[a]}\Zcorr{u}{\s[a]} \Meas{a}{\pi/2} \;}\;
		\gen{ Z_u \ox \idop_w }	\;,		
	\\[2ex]
		\gen{ \idop_u \ox Z_v }
	\;\mapstoname{\; \Ent{a,v} \Ent{a,u} \New{a}{\sfx} \;}&\;
		\gen{\begin{array}{c@{\,\ox\,}c@{\,\ox\,}c@{}l}
				\idop_u	&	\idop_a	&	Z_w	&	\;,	\\
				Z_v		&	X_w		&	Z_w	&
		\end{array}}
	\;\mapstoname{\; \Zcorr{v}{\s[a]}\Zcorr{u}{\s[a]} \Meas{a}{\pi/2} \;}\;
		\gen{ \idop_u \ox Z_w }	\;;		
\end{align}
\end{subequations}
in particular, it is a symmetric operation on $u$ and $v$ which leaves $Z_u$ and $Z_v$ invariant, so it applies a two qubit diagonal operation.
It is then easy to verify that the CPTP map which $\mathfrak {Zz}$ performs is 
\begin{gather}
	\label{eqn:zz}
		\mathfrak {Zz}_{u,v}(\rho)
	\;\;=\;\;
		\sfR^{\pion 2}_u \sfR^{\pion 2}_v \Ent{u,v} (\rho)
	\;\;=\;\;
		\e^{-i\pi Z_u \ox Z_v / 4} \,\rho\,\, \e^{i\pi Z_u \ox Z_v / 4}	\;;
\end{gather}
equivalently, we may substitute any entangler operation $\Ent{u,v}$ in the DKP construction with the composite CPTP map
\begin{gather}
	\label{eqn:clusterEntanglerSubst}
		\Ent{u,v}
	\;\;=\;\;
		\sfR^{\mpion 2}_v \; \sfR^{\mpion 2}_u \; \mathfrak {Zz}_{u,v}	\;.
\end{gather}
To obtain an equivalent one-way pattern, we may decompose $\sfR^{\mpion 2} = \sfJ(0)\sfJ(\mpion 2)$, and then substitute the rotations in~\eqref{eqn:clusterEntanglerSubst} with chain patterns $\mathfrak J_{u''/u'}^{\,0} \, \mathfrak J_{u'/u}^{\mpion 2}$ (and similarly for $v$); this construction is illustrated in Figure~\ref{fig:zzConstr}.
\begin{figure}[t]
	\begin{center}
		\includegraphics[width=146mm]{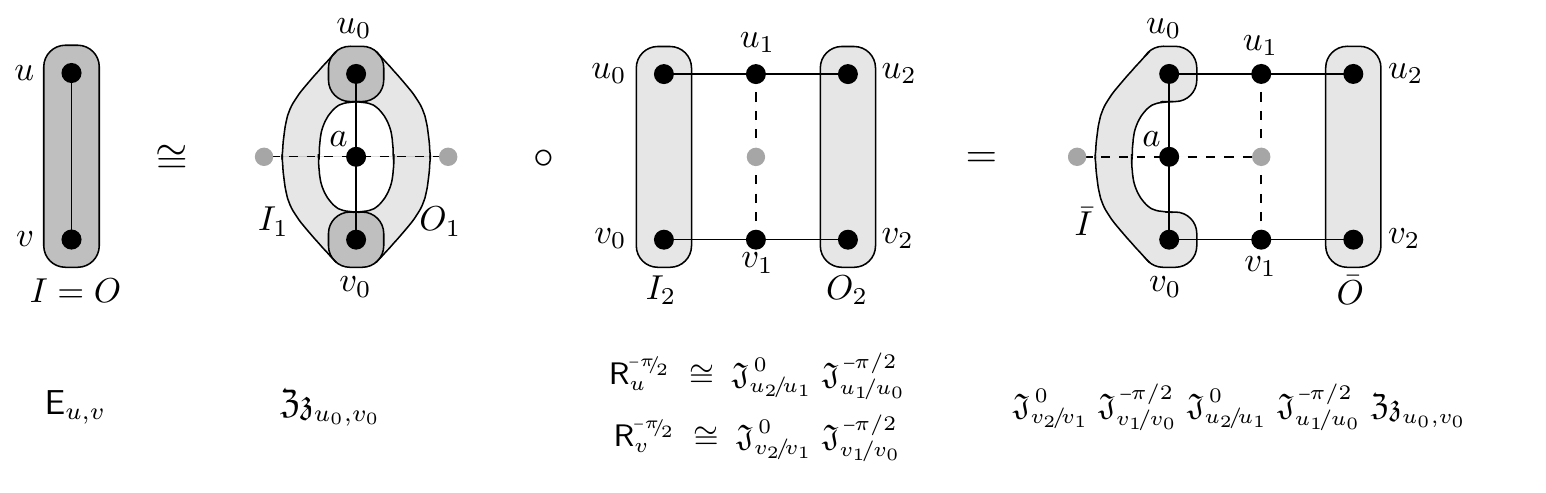}
	\end{center}
	\caption[Illustration of an embedding of \protect{$\protect\ctrl Z_{u,v}$} in the grid in terms of geometries, using the decomposition of~\eqref{eqn:clusterEntanglerSubst}.]{\label{fig:zzConstr}%
		Illustration of an embedding of $\protect\cZ_{u,v}$ in the grid in terms of geometries, using the decomposition of~\eqref{eqn:clusterEntanglerSubst}.
		Congruency relations represent equivalence up to changes of the input/output system.
		Grey dots and broken lines indicate qubits and entanglement relations of the grid state which must be removed via $\axisZ$ measurements in order not to interfere with the corresponding measurement pattern (displayed below each geometry).}
\end{figure}
Alternatively, if the logical entangler operation $\Ent{u,v}$ precedes $\mathfrak J^{\alpha_u}_{u'/u}$ and $\mathfrak J^{\alpha_v}_{v'/v}$  operations in a measurement pattern, we may employ the relations
\begin{align}
		\mathfrak J^{\alpha_v}_{v'/v} \; \mathfrak J^{\alpha_u}_{u'/u} \; \Ent{u,v}
	\;\;=&\;\;
		\mathfrak J^{\,0}_{v'/v} \; \sfR_v^{\alpha_v} \; \sfR_v^{-\pion 2} \; \mathfrak J^{\,0}_{u'/u} \; \sfR_u^{\alpha_u} \sfR_u^{-\pion 2} \; \mathfrak{Zz}_{u,v}
	\notag\\=&\;\;
		\mathfrak J^{\alpha_v - \pi/2}_{v'/v} \; \mathfrak J^{\alpha_u - \pi/2}_{u'/u} \; \mathfrak{Zz}_{u,v}	\;,
\end{align}
which follow from~\eqref{eqn:factorRotationJ}.

\paragraph{A remark on terminology.}

We have taken some license in referring to the above construction as ``the simplified RBB construction'', in that some of the constructions above (\eg~the construction of the pattern $\mathfrak {Zz}$) do not appear explicitly in~\cite{RBB03}.
However, all of these constructions are clearly implicit in that work.
The construction above also completely omits the cleverer constructions of~\cite{RBB03}, including compact patterns for reversing an entire consecutive block of qubits, non-nearest neighbor $\cX$ operations, the quantum Fourier transform over $\Z_{2^k}$ for arbitrary $k$, and addition circuits for $\Z_{2^k}$.

\subsection{Universality of the cluster state model}

We may use the constructions above, and the pattern transformations described for the DKP constructions in Section~\ref{sec:DKPstandardization}, to prove the (approximate) universality of the cluster-state model pattern in standard form, as follows.

By composing the elementary patterns of the simplified RBB construction above, we may construct a one-way pattern in grid graphs $G_{n,d}$ for sufficiently large $n,d \in \N$ which implement arbitrary unitary circuits in a nearest-neighbor version of the $\Jacz(\A)$ model.
Restricting to one-way patterns which initialize each input qubit in the $\ket{+}$ state and performing the appropriate trace-out operations, we may then prepare arbitrary density operators by Theorem~\ref{thm:churchHilbertSpace}.
This yields (approximate) universality for the model $\Cluster(\A)$, for any $\A \subset \R$ for which $\Jacz(\A)$ is (approximately) universal.

Using the standardization procedures of Section~\ref{sec:DKPstandardization}, we may obtain a one-way pattern in which the first stage is to apply an open graph state encoding on the input subsystem: the preparation and entanglement phases of the resulting pattern can then be replaced with the preparation of a sufficiently large cluster state, and removing the qubits which do not play a role in the pattern via $\axisZ$ measurements.
Thus, we may prepare arbitrary density operators in the cluster-state model; and by the approximate universality of $\Jacz(\frac{\pi}{4}\Z)$, the cluster-state model is approximately universal for quantum computing for measurement angles which are multiples of $\pion 4$.
Furthermore, by performing signal shifting and Pauli simplifications then yield a measurement pattern involving no $\pi$-dependencies or shift operations, thus yielding a pattern in the cluster-state model in standard form.

\vspace{1ex}
\noindent The above then reproduces the following result of~\cite{RB01, RBB03}:
\vspace{-0.5ex}
\begin{theorem}
	The $\Cluster(\A)$ model is
	\begin{inlinum}
	\item
		universal for quantum computation when $\A = \R$\,; and

	\item
		approximately universal for quantum computation when $\A = \frac{\pi}{4}\Z$\,.
	\end{inlinum}
	Furthermore, we may without loss of generality require that the measurement pattern is standardized, and that Pauli measurements are performed before any non-Pauli measurement operation.
\end{theorem}

\section{Other constructions in the one-way model}
\label{sec:otherConstrs}

Before closing this chapter, we remark upon two other constructions in the one-way measurement model which we will refer to in later chapters, which lie beyond the DKP and the simplified RBB construction schemes described above.

\subsection{Qubit reversal pattern in the grid.}
\label{sec:qubitReversal}

	Figure~\ref{fig:swapRBB} illustrates a geometry for a one-way pattern for reversing a sequence of consecutive logical qubits in a linear nearest neighbor model, using only $\axisX$ measurements, illustrated in~\cite[Fig.~10]{RBB03}.
	\begin{figure}
		\begin{center}
			\includegraphics[width=58mm]{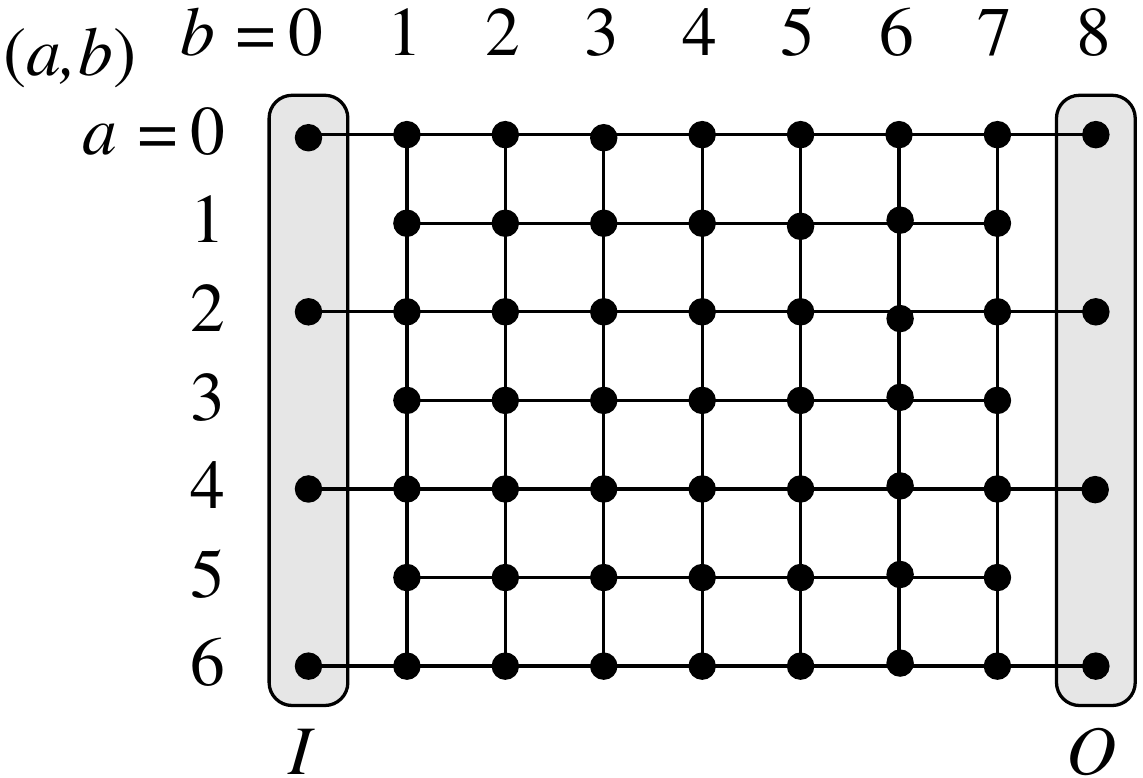}
		\end{center}
		\caption[Geometry for the measurement pattern of~\protect{\cite[Fig.~10]{RBB03}} for reversing a sequence of qubits in the cluster-state model.]{\label{fig:swapRBB}%
		Geometry for the measurement pattern of~\protect{\cite[Fig.~10]{RBB03}} for reversing a sequence of qubits in the cluster-state model.
		Every qubit (except for the elements of $O$) is measured with an $\axisX$ measurement, selecting for the state $\ket{\splus}$: the corresponding correction operations may be determined by the stabilizer formalism.
		Numbering the rows and columns of grid-sites from $0$ at the upper-left, the resulting transformation maps states $\rho_{(0,0),(0,2),(0,4),(0,6)}$ to $\rho_{(8,6),(8,4),(8,2),(8,0)}$\,, \ie\ exchanging the logical qubits of rows $0$ and $6$ as well as rows $2$ and $4$. 
	}
	\end{figure}
	(This pattern is a special case of a more general transformation, which performs a unitary $\e^{i \varphi Z \ox \cdots \ox Z}$ transformation in addition to the qubit reversal by measuring a particular qubit in a different basis.)
	Numbering grid sites $(a,b)$ according to the rows and columns of the grid (starting from $0$ at the upper-left), it is possible to show that the resulting transformation is unitary, and in particular, it reverses order of the the ``logical'' qubits of rows $\ens{0,2,4,6}$.

	This may be shown using \eg~the stabilizer formalism: to do so, however, it is useful to note the presence of correlations in the measurement results of the qubits in the interior block of columns 1\thru7.
	Consider the twelve-qubit set $S$ given by
	\begin{gather}
			S
		\;\;=\;\;
			\ens{ (a,b) ~\text{adjacent to}~ (d,d) \,\big|\, d \in \ens{2,3,4,5,6}\, }	\;:
	\end{gather}
	it is easy to verify that the product of $K_{(a,b)} = X_{(a,b)} Z_{(a\sminus1,b)} Z_{(a\splus1,b)} Z_{(a,b\sminus1)} Z_{(a,b\splus1)}$ for $(a,b) \in S$ is a tensor product of $X_{(a,b)}$ operators, also for $(a,b)$ ranging over $S$.
	If we measure the qubits in this set column-by-column with $X$ observable measurements, this implies that the unique qubit in $S$ in the seventh column will yield the $\pm 1$ result with certainty if the product of the preceding eleven measurements is $\pm 1$.
	\label{discn:qubitReversalCertainty}%
	We may similarly show that every qubit in the seventh column (except for $(0,7)$ and $(6,7)$) will yield measurement results correlated in a similar way with a subset of qubits in the preceding six columns; the measurements on those qubits then perform the identity, while the other measurements will anticommute with some generators of the stabilizer group $\gen{K_v}_{v \notin O}$.

	Note that these correlations arise from cancelling out $Z$ operators in the interior block; by a similar technique of cancelling such $Z$ operators, it is possible to show that $X$ and $Z$ observables on each of the input qubits get mapped by measurements (and selecting for the $\ket{\splus}$ measurement result) to the same observable on the ``reversed'' output qubit.
	By scaling this measurement pattern in the obvious way, we may obtain a qubit-reversal pattern for an arbitrary number of qubits.

	This measurement pattern is noteworthy in that it falls outside of the domain of automated approaches to inferring the semantics of a measurement-based pattern (one of which we describe in Chapter~\ref{Semantics for unitary one-way patterns}), precisely because of the correlations which produce measurement results with certainty.
	In the special case of swapping two consecutive qubits, it is also more compact than the construction of $\mathsf{SWAP}_{u,v}$ described in~\eqref{eqn:simpleSwap} using the construction for $\Ent{u,v}$ in the simplified RBB construction above.

\subsection{Concise patterns for $Z \ox \cdots \ox Z$ Hamiltonians.}
\label{sec:zzMany}

	The construction of the $\mathfrak{Zz}$ pattern in~\eqref{eqn:simpleZzPattern} is a special case of a more general construction in~\cite{BB06} in the more general one-way model for applying unitaries arising from $Z^{\ox k}$ Hamiltonians for $k$ arbitrarily large.
	We may define the states $\ket{\spm \sfy\sfz_{\theta}}$ as follows:
	\begin{gather}
		\label{eqn:defYZtheta}
			\ket{\spm \sfy \sfz_\theta}
		\;\;=\;\;
			\sfrac{1}{\sqrt 2} \ket{\splus \sfx} \;\pm\; \sfrac{\e^{-i\theta}}{\sqrt 2} \ket{\sminus \sfx}	\;;			
	\end{gather}
	it is possible to verify that these are $\pm 1$ eigenvectors of the observable $\cos(\theta) Z + \sin(\theta) Y$.
	Define the measurement operator
	\begin{gather}
			\mspace{-20mu}
			\Meas[\YZ]{a}{\theta}(\rho)
		\;\;=\;\;
			\bra{\splus\sfy\sfz_\theta}_a \rho\, \ket{\splus\sfy\sfz_\theta}_{\!a} \ox \ket{0}\bra{0}_{\s[a]}
			\;+\;\;
			\bra{\sminus\sfy\sfz_\theta}_a \rho\, \ket{\sminus\sfy\sfz_\theta}_{\!a} \ox \ket{1}\bra{1}_{\s[a]}
	\end{gather}
	which performs a $\YZ$-plane measurement similar to the $\XY$-plane measurements defined by~\eqref{eqn:xyPlaneMeas}.
	Then, the measurement pattern
	\begin{subequations}
	\begin{gather}
			\label{eqn:zzMany}
			\smash{\underbrace{\mathfrak{Zz\cdots z}}_{\text{$k$ times}}}_{v_1,\cdots,v_k}^{\,\theta}
		\;\;=\;\;
			\sqparen{\prod_{j = 1}^k \Zcorr{u_j}{\s[a]}}
			\Meas[\YZ]{a}{\theta}
			\sqparen{\prod_{j = 1}^k \Ent{a, u_j}}
			\New{a}{\sfx}
	\end{gather}
	performs a unitary rotation
	\begin{gather}
			\smash{\underbrace{\mathfrak{Zz\cdots z}}_{\text{$k$ times}}}_{v_1,\cdots,v_k}^{\,\theta}(\rho)
		\;\;=\;\;
			\e^{\minus i \theta Z_{v_1} \ox \cdots \ox Z_{v_k}/2} \,\rho \,\;\e^{i \theta Z_{v_1} \ox \cdots \ox Z_{v_k}/2}	\;.
		\\[-1ex]\notag
	\end{gather}
	\end{subequations}%
	\label{discn:reduceZz}
	The operation $\mathfrak{Zz}_{u,v}$ is a special case for $k = 2$ and $\theta = \pion 2$; the geometry for the more general measurement pattern is illustrated in Figure~\ref{fig:zzMany}.

	\begin{figure}[t]
		\begin{center}
			\includegraphics{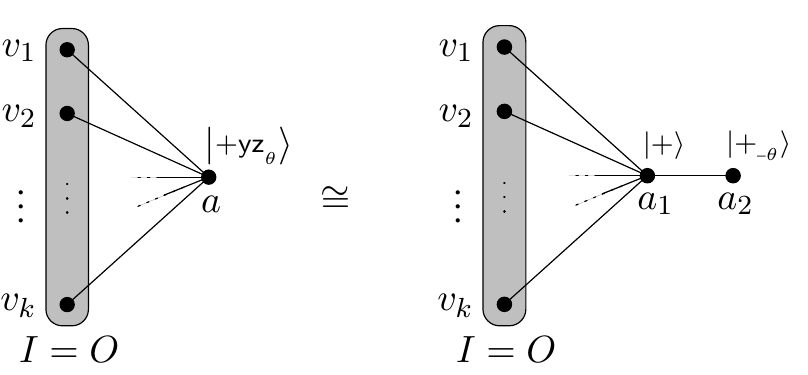}
		\end{center}
		\caption[Illustration of the geometry for the pattern $\protect\mathfrak {Zz\cdots z}^{\theta}$ applied to $k$ qubits.]{\label{fig:zzMany}%
			Illustration of the geometry for the pattern $\mathfrak {Zz\cdots z}^{\theta}$ applied to $k$ qubits, $v_1, \ldots, v_k$\,.
			In the left-most figure, the auxiliary qubit is to be measured in the basis \protect{$\ket{\spm \sfy\sfz_{\theta}}$}, with corrections to be performed to effectively select the positive result.
			The right-most figure illustrates an equivalent pattern using only $\XY$-plane measurements, in which the two auxiliary qubits are to be measured, performing corrections if necessary to select for the measurement results given.
			}	
	\end{figure}
	It is easy to verify the behavior of $\mathfrak{Zz\cdots z}$ on standard basis state vectors, as follows.
	Using the fact that $\cZ \big( \ket{\splus} \ox \ket{0} \big) \,=\, \ket{\splus} \ox \ket{0}$ and $\cZ \big( \ket{\splus} \ox \ket{1} \big) \,=\, \ket{\sminus} \ox \ket{1}$\,, the state vector produced by the preparation and entanglement procedure of $\mathfrak{Zz\cdots z}^{\theta}_{v_1,\ldots,v_k}$ for an input state $\ket{\vec x}_{v_1,\ldots,v_k}$ (for $\vec x \in \ens{0,1}^k$ arbitrary) is
	\begin{align}
			\ket{\psi_{\vec x}}
		\;\;=\;\;
			\sqparen{\prod_{j = 1}^k \cZ_{a, u_j}}
			\mspace{-40mu}&\mspace{40mu}
			\ket{\splus}_a \ox \ket{\vec x}_{v_1,\ldots,v_k}
		\notag\\[1ex]=&\;\;
			\begin{cases}
				\ket{\splus}_a \ox \ket{\vec x}_{v_1,\ldots,v_k}	\,,	&	\text{if $x_1 \oplus \cdots \oplus x_k \equiv 0 \pmod{2}$}	\\[1ex]
				\ket{\sminus}_a \ox \ket{\vec x}_{v_1,\ldots,v_k}	\,,	&	\text{if $x_1 \oplus \cdots \oplus x_k \equiv 1 \pmod{2}$}
			\end{cases}
		\notag\\[2ex]=&\;\;
			\ket{\spm}_a \ox \ket{\vec x}_{v_1,\ldots,v_k}	\quad\text{for $\ket{\vec x}$ a $\pm 1$-eigenvector of $Z^{\ox k}$}.
	\end{align}
	Note that $\bracket{\splus \sfy\sfz_{2\theta}}{\splus} = \frac{1}{\sqrt 2}$ and $\bracket{\splus \sfy\sfz_{\theta}}{\sminus} = \frac{\e^{i\theta}}{\sqrt 2}$\,; then, if we apply the projection $\ket{\splus \sfy\sfz_{\theta}}\bra{\splus \sfy\sfz_{\theta}}$ to $\ket{\psi}$ as described above and renormalize, we obtain
	\begin{align}
			\ket{\splus \sfy\sfz_{2\theta}}\bra{\splus \sfy\sfz_{\theta}}_a \ket{\psi}_{a,v_1,\ldots,v_k}
		\;=&\;\;
			\begin{cases}
				\frac{1}{\sqrt 2} \ket{\vec x}_{v_1,\ldots,v_k}	\,,					&	\text{if $\ket{\vec x}$ a $+1$-eigenvector of $Z^{\ox k}$}	\\[1.5ex]
				\frac{\e^{i\theta}}{\sqrt 2} \ket{\vec x}_{v_1,\ldots,v_k}	\,,	&	\text{if $\ket{\vec x}$ a $-1$-eigenvector of $Z^{\ox k}$}
			\end{cases}
		\notag\\[2ex]\propto&\;\;
			\bigg\{ \e^{\mp i\theta/2} \ket{\vec x}_{v_1,\ldots,v_k}	\,,
				\quad\text{for $\ket{\vec x}$ a $\pm 1$-eigenvector of $Z^{\ox k}$} \bigg\}
		\notag\\[2ex]=&\;\;\;
			\Big[ \e^{-i \theta Z^{\ox k}/2} \, \ket{\vec x} \Big]_{v_1,\ldots,v_k}	\;;
	\end{align}
	a similar analysis for the projection $\ket{\splus \sfy\sfz_{\theta}}\bra{\splus \sfy\sfz_{\theta}}$ yields the result
	\begin{gather}
			\ket{\sminus \sfy\sfz_{\theta}}\bra{\sminus \sfy\sfz_{\theta}}_a \ket{\psi}_{a,v_1,\ldots,v_k}
		\;\;\propto\;\;
			\Big[ \e^{-i \theta Z^{\ox k}/2} \, Z^{\ox k} \ket{\vec x} \Big]_{v_1,\ldots,v_k}	\;,
	\end{gather}
	so that the pattern $\mathfrak{Zz\cdots z}_{v_1,\cdots,v_k}^{\,\theta}$ performs the rotation described, by linearity.

	In order to perform such a measurement-based procedure in the one-way model as described above, we must represent the measurement $\Meas[\YZ]{a}{\theta}$ in terms of $\XY$-plane measurements.
	We may simulate a $\YZ$-plane measurement using $\XY$-plane measurements by noting that
	\begin{gather}
			\sfH\big(\cos(\theta)Z + \sin(\theta)Y\big)
		\;\;=\;\;
			\cos(-\theta) X + \sin(-\theta) Y	\;;
	\end{gather}
	that is, a Hadamard transformation effects a change of reference frame, transforming $\YZ$-plane measurement observables to some corresponding $\XY$-plane measurement observables.
	Then, we have the following congruency,
	\begin{gather}
		\label{eqn:simulateYZwXY}
			\Meas[\YZ]{a}{\theta}
 		\;\;\cong\;\;
			\Meas[\YZ]{\s[a_2]/a_1}{\theta}
		\;\;=\;\;
			\Meas{a_2}{-\theta}
			\Xcorr{a_2}{\s[a_1]}
			\Meas{a_1}{\,0}			
			\Ent{a_1,a_2}
			\New{a_2}{\sfx}	\;,
	\end{gather}
	where congruency is up to changes in the input and output subsystems; the measurement on $a_1$ performs a Hadamard transform to effect the necessary change of reference frame to perform the $\YZ$-plane measurement via $\XY$-plane measurements.
	Absorbing the correction on $a_2$ into the measurement, we may then transform $\mathfrak {Zz\cdots z}^{\theta}$ into a pattern in the one-way model as described in this section, by the congruence
	\begin{gather}
		\label{eqn:zzManyXY}
			\smash{\underbrace{\mathfrak{Zz\cdots z}}_{\text{$k$ times}}}_{v_1,\cdots,v_k}^{\,\theta}
		\;\;\cong\;\;
			\sqparen{\prod_{j = 1}^k \Zcorr{u_j}{\s[a_2]}}
			\Meas{a_2}{-\theta;\s[a_1]}
			\Meas{a_1}{\,0}
			\sqparen{\prod_{j = 1}^k \Ent{a_1, u_j}}
			\Ent{a_1,a_2} \New{a_2}{\sfx} \New{a_1}{\sfx}	\;.
	\end{gather}
	The geometry for this measurement pattern is also illustrated in Figure~\ref{fig:zzMany}.
	Nevertheless, the pattern $\mathfrak{Zz\cdots z}^{\theta}$ described in~\eqref{eqn:zzMany} illustrates the potential usefulness of extending beyond $\XY$-plane measurements in the one-way model; and we will also refer to this construction in later Chapters.

\section{Conclusion}

In this section, we have described the one-way measurement model in detail, by way of defining constructions for measurement procedures which simulate unitary circuit models; and we have also given an overview of approaches to robustly implementing quantum computers via the one-way model. 

Simulation of unitary circuits is the standard approach to obtaining one-way measurement procedures to perform quantum computation: however, it is possible to construct measurement patterns which do not clearly correspond to the simulation of a unitary circuit.
An example of one technique for doing so are the simplification techniques of~\cite{HEB}, which allows the transformation of any one-way measurement procedure involving measurements of Pauli observables into an equivalent procedure without Pauli observable measurements, on fewer qubits.
The way in which this is done takes advantage of a correspondence between \emph{local Clifford operations} (\ie\ products of single-qubit unitaries from the Clifford group) on graph state, and \emph{local complementation} of the graphs which describe the graph states: we describe this correspondence in Lemma~\ref{lemma:HEB04}.
This gives rise to transformations of graph states, and subsequently of measurement-based procedures, where the resulting procedures no longer correspond to a simulation of a unitary circuit in any clear way (see for example \cite[Figure~16]{HEB04}).

This raises the question of the conditions under which we may identify when a measurement pattern may be meaningfully considered to simulate some unitary circuit: this problem is the one which we consider in Chapter~\ref{Semantics for unitary one-way patterns}.
	
	\chapter[Semantics for unitary one-way patterns]{
 		Semantics (for unitary one-way patterns)
}

One-way measurement based quantum computation describes unitary transformations of quantum states as a composition of CPTP maps, many of which are not themselves unitary.
The fact that the resulting transformation is unitary is due to an appropriate combination of measurement observables and entanglement operations which, as in the analysis of the pattern $\mathfrak J^\alpha$ in~\eqref{eqn:stabilizerJ}, may be treated with the stabilizer formalism (Section~\ref{sec:stabilizerFormalism}).
The post-measurement residual state in either case will be an isometric image of the pre-measurement state; and the possible post-measurement states corresponding to the different measurement results may be mapped to one another by single-qubit unitary transformations.
As a result, we may perform a correction after each measurement to simulate postselection of the $+1$ measurement result, yielding a mixed one-way measurement pattern; equivalently, we may absorb the byproduct operations into future measurements, interpreting them as changes of reference frame, in order to obtain a one-way measurement based computation in standard form.

As we are interested in unitary transformations, we are presented with natural decision problems when considering quantum computation in the one-way measurement model. 
Measurements do not transform quantum states unitarily --- is it possible to tell whether a one-way measurement based computation performs a unitary transformation between its input and output subsystems?
Is it possible to describe precisely \emph{how} such computations transform quantum states, by translation to \eg\ a quantum circuit of comparable complexity, using a reasonable set of elementary gates?
If not, is it at least possible to efficiently verify whether it performs some given transformation?

The problem of determining when a measurement problem performs an isometry seems similar to to the more general question of when a classically controlled circuit in general performs an isometry.
By~\cite{MCSB98}, this may be rephrased as asking whether each measurement performed has the same bias towards one measurement result or another, across all possible input states.
This problem is \NP-hard if we require exponential precision, as we can encode the evaluation of arbitrary boolean functions $f: \ens{0,1}^n \to \ens{0,1}$ to produce the state
\begin{gather}
		\ket{\textsc{sat}_f}
	\;\;=\;\;
		\sfrac{1}{\sqrt {2^n}} \sum_{\vec x \in \ens{0,1}^n} (-1)^{f(x)}	\ket{x}
\end{gather}
and perform measurements in the $\ket{\spm}$ basis.
If any such measurement has non-zero probability of yielding the $\ket{\sminus}$ state, this would indicate that $f$ is satisfiable but not a tautology.
However, it is easy to verify that the natural approach to producing a state such as $\ket{\textsc{sat}_f}$ requires the use of non-Clifford group operations, whereas the open graphical encoding procedures described in~\eqref{eqn:openGraphEncoding} consist of Clifford group operations, and prepare states in a known stabilizer code.
As well, we require that the post-measurement residual states be related to each other by local unitary operations in order to be able to adaptively perform the (single-qubit) measurements of a one-way procedure.
It may be hoped that the additional structure imposed by these constraints yields a tractable problem.

These questions arise because simple unitary circuit models such as $\Phases(\A)$ defined in Section~\ref{sec:unitaryCircuits} are the \emph{de facto} standard models of quantum computation: they seem to provide the most natural set of idioms for uniformly expressing algorithms independently of architecture.
(We will commonly refer ``the'' unitary circuit model, by which we will mean the $\Phases(\A)$ or $\Jacz(\A)$ models for some $\A \subset \R$.)
While alternative models are promoted as abstractions of proposals for physically implementing a quantum computer (\eg\ measurement-based models~\cite{GC99,RB01} and the adiabatic model~\cite{FGGS00}), and ``purely physical'' idioms seemingly divorced from uniform circuit families have led to new quantum algorithms (as with the \textsc{nand} tree algorithm of~\cite{FGG07}, inspired by transmission and reflection rates associated with energy barriers), the unitary circuit model provides what seems to be the simplest mathematical model for describing quantum computation.
Because of the present uncertainty as to \emph{which} implementation proposal will prove successful, the unitary circuit model represents a natural candidate for an \emph{intermediate language}\footnote{%
\emph{Intermediate language} refers here to a language used to facilitate the design of a suite of compilers, which translate programs from one or many source language (or from many source languages) to one of several different target architectures.}
for representing algorithms independently of the target architecture, even if alternative models suggest new algorithms.

Therefore, I argue that how effectively one can translate decompositions of unitaries to and from the unitary circuit model is currently an important topic for any proposed alternative model of quantum computation.
Unless some one proposal for implementation comes to dominate over the others, or an unlikely breakthrough (\eg\ the discovery of a proof that $\BQP = \BPP$) is made, translation to a ``simple'' model of quantum computation such as the unitary circuit model is the best that can be reasonably hoped for as a \emph{uniform} way of obtaining low-level semantics of a quantum computation.
In other words: the most reasonable means of understanding the behavior of a quantum algorithm is currently by reduction to the unitary circuit model.
This motivates the problem of effective translations from those one-way pattern which perform unitary embeddings, into the unitary circuit model.

The topic of this chapter is one such reduction, from those one-way measurement patterns which arise out of the DKP constructions to unitary circuits.
We consider this topic by examining a \emph{flow} property which emerges from that construction, which gives rise to efficiently detectable combinatorial structures and combinatorial decompositions of the corresponding measurement pattern.

\vspace{-1ex}
\paragraph{Previous appearances of this work.}

Earlier versions of many of the results of this chapter appear in~\cite{Beaudrap06,Beaudrap08,BP08}.
The work presented in Section~\ref{sec:graphth-charn} (except for~\ref{sec:extreme}) and Section~\ref{sec:flowAlgs} first appeared in a preliminary form in~\cite{Beaudrap06}; the results of the former was published as~\cite{Beaudrap08} in essentially the form presented here.
The results of Section~\ref{sec:extreme} are joint work with Martin Pei, and were published in~\cite{BP08}.
The rest of the work in this Chapter, except where indicated, are developments original to this thesis.

\vspace{-1ex}
\paragraph{Graph theoretic notations.}

We will assume basic familiarity with graph theory: an introduction to the subject and basic definitions can be found in~\cite{Diestel}.
Whenever a graph $G$ is clear from context, $\sim$ will denote the adjacency relation of $G$.
In graphs, we will denote an edge between $v$ and $w$ by $vw$, and in directed graphs, we will let $v \arc w$ represent an arc between $v$ and $w$\,; and in graphs or directed graphs, we will represent (directed) paths by concatenated sequences of edges/arcs, $v_1 v_2 \cdots v_n$ or $v_1 \arc v_2 \arc \cdots \arc v_n$.
We adopt the convention that graphs do not have self-loops on vertices.

\section{Defining semantics by reduction to unitary circuits}
\label{sec:semanticsReduce}

Before proceeding, we will put forward a precise definition of how the semantics of a measurement-based computation may be reduced to another model, such as the unitary circuit model.

We are generally not interested in \emph{arbitrary} translations to and from the unitary circuit models: in this instance, the particular circuit model becomes significant.
For instance, in the case of the one-way measurement model, we may easily obtain a circuit in the $\Phases(\A)$ or $\Jacz(\A)$ models by describing the measurement pattern as a circuit in a classically controlled unitary circuit model, and applying the principle of deferred measurement~(page~\pageref{principleDeferredMeas}) to obtain a unitary circuit in some unitary circuit model $\CoherCtrl$ which includes (coherently) controlled unitary transformations performing conditional change of basis operations to simulate adaptive measurements.
Such a circuit can then be decomposed in \eg\ the $\Jacz(\A)$ model by decomposing the individual transformations of the model $\CoherCtrl$\:\!.
However, such a direct translation scheme may require many more qubits of workspace, and many more multi-qubit operations, than is actually necessary.
For instance, if we translate a circuit of the form $J(\alpha_n) \cdots J(\alpha_2) J(\alpha_1)$ on a single qubit into the one-way model using the DKP construction, and then obtain a unitary circuit from that pattern via the principle of deferred measurement as above, we will obtain a circuit involving $n + 1$ qubits and $O(n)$ two-qubit operations, whereas the original circuit required only operations on a single qubit.

In contrast with the above, we may hope to obtain a translation scheme to unitary circuit models which in some sense preserves the complexity of the computation.
In order to make this sense of complexity-conservation well defined despite the differences in how resources are used in different models of computation, we may make reference (as we have done above for the DKP construction) to another function which translates from unitary circuits to the computational model of interest.
In particular, suppose we have a map $\cR$ which maps unitary circuits in some gate model $\Gates$ to some target model $\M$ of quantum computation.
We would like to consider a mapping $\cS$ such that
\begin{romanum}
\item
	\label{condition:superset}
	$\cS$ is a map from a superset of $\img(\cR)$ to unitary circuits using the gate-set \Gates\ (where in particular, $\dom(\cS)$ includes any procedures in $\M$ which may be in some sense ``congruent'' to the image of a circuit in $\cR$);

\item
	\label{condition:noninflation}
	$\cS \circ \cR$ maps each circuit in $\dom(\cR)$ to an equivalent circuit of lesser or equal gate complexity;

\item
	\label{condition:retraction}
	$\cR \circ \cS$ maps procedure in $\img(\cR)$ to a ``congruent'' procedure in $\cR$.
\end{romanum}
We will not formally define here when two computational procedures are ``congruent'' in an arbitrary model of computation: however, for both the unitary circuit model and for the one-way measurement based model, we will adopt the convention of Section~\ref{sec:unitaryCircuits} that computations are congruent in either model if they only differ by transpositions of commuting operations (and by a possible relabelling of the qubits).

It may seem unproductive to consider a map $\cS$ from a model $\M$ as above to unitary circuits, which is explicitly defined with reference to a \emph{existing} translation procedure $\cR$ from unitary circuits to $\M$, as the most obvious source of a computation in $\img(\cR)$ is to take an existing unitary circuit $C$ and to produce $\cR(C)$.
However, given the current prominence of the unitary circuit model, I would argue that a mature understanding of a model $\M$ of quantum computation currently requires the ability to perform meaningful translations \emph{in both directions} between $\M$ and the unitary circuit model.
Should new insights arise for performing algorithms in the model $\M$, these insights can then also be translated back to the unitary circuit model.

In practise we will be largely interested in mappings between congruency classes of unitary circuits and of other models of computation.
In the case where $\cR$ is an injective map from a unitary circuit model to some model $\M$, we may simply wish for $\cS$ to be an inverse for the action of $\cR$ on congruency classes of unitary circuits.
However, for more general transformations $\cR$ which may map incongruent circuits (but which perform equivalent CPTP maps) to congruent procedures in some other model of computation, the conditions described above may be of interest in the absence of a left inverse for $\cR$.

In this thesis, we are primarily interested in the one-way measurement model, so we consider the case of $\M = \OneWay(\A)$ for some $\A \subset \R$.
\label{discn:representationMap}
A function $\cR$ as above provides an image of unitary circuit ``idioms'' --- structural formulas for circuits, which are used to perform certain tasks --- in the one-way model: a one-way measurement based computation which is in the image of such a function $\cR$ may be said to be a representation of some unitary circuit in the one-way model.
We may therefore call $\cR$ a \emph{representation} map, and the one-way measurement based routines in $\img(\cR)$ \emph{representations} of circuits.

The map $\cS$ is then a means of recognizing one-way measurement patterns which simulate circuits in this sense, revealing semantics for these one-way measurement patterns in terms of unitary circuits, in a way which
\begin{inlinum}
\item
	preserves the structure of some representative of each class of circuits which are translated by $\cR$ into a given one-way measurement based procedure, and

\item
	such that the one-way measurement based algorithm can be recovered from the circuit through the ``representation'' map.
\end{inlinum}
We may call a map $\cS$ of this sort a \emph{semantic} map, and say that a circuit in $\img(\cS)$ provides the \emph{semantics} of a one-way pattern in terms of some unitary circuit model.\label{discn:semanticMap}

It is not difficult to imagine more properties which we would like the map $\cS$ to have beyond the ones described above, for instance:
\begin{itemize}
\item
	It is obviously desirable for $\cS$, and the membership predicate for $\dom(\cS)$, to be efficiently computable.

\item
	It would be convenient for a semantic map $\cS$ to also be able to recognize \emph{parts} of one-way measurement based computations which resemble ``idioms'' from the circuit model, and be able to translate them into components of the corresponding circuit.
	This would allow us to describe the semantics of a one-way measurement pattern by reduction to smaller pieces.

\item
	We might also like $\cS$ to be well defined on the set of \emph{all} one-way measurement based computations performing unitary transformations, which may be a proper superset of the range of $\cR$\,: this may suggest which extensions to the unitary circuit model can be easily translated into a one-way measurement pattern, and which may therefore be useful tools to adopt for designing quantum algorithms.
\end{itemize}
We are of course primarily interested in efficient algorithms, and will restrict our attention without further comment to semantic maps which can be performed in polynomial time in the size of the specification for a one-way measurement based computation.
The second of the extensions above, while beyond the scope of this thesis to formalize, is easy to achieve in an intuitive sense in some instances which will be described in this chapter.
The third of these extensions is more difficult, in part because it is not yet known how to determine whether a one-way measurement based algorithm realizes a unitary transformation or not: for any such ``extended semantics'' map $\cS$, a polynomial-time algorithm for the membership predicate of $\dom(\cS)$ is not yet known.

We have defined semantic maps $\cS$ above relative to a map $\cR$ for representing unitary circuits in the one-way measurement based model, and it is conceivable that for some interesting maps $\cR$, it may be difficult or impossible to find a map $\cS$ which satisfies all of the properties~\ref{condition:superset}\thru\ref{condition:retraction} on pages~\pageref{condition:superset}\thru\pageref{condition:retraction}.
In this case, we may be satisfied by an $\cS$ which \emph{approximately} achieves these properties for a given $\cR$.
For instance, we may require that for a unitary circuit $c$, the number of single-qubit and two-qubit gates in the circuit $(\cS \circ \cR)(C)$ are increased beyond those of $C$ only by small scalar factors (\eg\ less than $1 + \epsilon$ for a suitably small $\epsilon$), or that even if $\cR$ is not a left-inverse of $\cS$, that there is an efficiently computable reduction map $f$ on $\dom(\cS)$ such that $f \circ \cR \circ \cS$ is the identity on $\dom(\cR)$.
However, the less closely we adhere to the conditions that $\cR \circ \cS$ is equivalent to the identity and that $\cS \circ \cR$ does not inflate the complexity of unitary circuits, the less useful the mapping $\cS$ is as a means of providing semantics for a one-way measurement based computation in terms of the unitary circuit model.

The main result of this chapter is to describe a mapping $\cS$ which acts as a semantic map with respect to the DKP construction described in Section~\ref{sec:oneWayComputation}, where we further restrict to those patterns produced from circuits without terminal measurements or trace-out operations.
This allows us to identify a class of one-way patterns which perform unitary transformations, and provide semantics for them in terms of unitary circuits.
The semantic map $\cS$ in this case is defined in terms of a \emph{flow} property of the geometry underlying the one-way pattern, which will be the main subject of this chapter.
Towards the end of the Chapter, we sketch similar results for a semantic map for the simplified RBB construction by using a slight extension of flows, building on the analysis for flows and on improved flow-finding algorithms to do so.

\subsection*{A remark on terminal measurement and trace-out operations}

As we have described it in Section~\ref{sec:oneWayComputation}, the model $\OneWay(\A)$ includes trace-out operations.
When considering how to transform a pattern $\mathfrak P$ in the $\OneWay(\A)$ model to a unitary circuit model, given that the pattern is congruent to a pattern in the image of the DKP construction, we may augment the output subsystem $O$.
This results in a measurement pattern $\mathfrak P'$ which also performs a unitary embedding, and is also congruent to a pattern from the DKP construction; the unitary circuit $C$ corresponding to the original pattern $\mathfrak P$ may be obtained by adding trace-out operations to the circuit $C'$ corresponding to $\mathfrak P'$.
However, patterns produced by the DKP construction from circuits with terminal measurements pose a more difficult problem, as it seems difficult to distinguish between qubits which are measured in order to drive the transformation of data and qubits which are measured in order to yield a classical probabilistic output of interest (speaking of the measurements in their roles of representing operations of the unitary circuit).

Given that we are primarily interested in how unitary transformations may be described by measurement-based computation, we set aside the issue of how to provide semantics for arbitrary patterns which may be produced by the DKP construction, and focus on those which arise from circuits which do not perform measurements or trace-out operations.
Fortunately, by the comments made above, semantics for patterns in $\img(\cR)$ which perform trace-outs may still be efficiently recovered; but for patterns arising from circuits with measurements, it seems that more general tools will be required.
We will pass over this subject for the remainder of the thesis.

\section{Flows: structure underlying the DKP construction}
\label{sec:flowDefn}

As we defined it in Definition~\ref{def:simplifiedDKP}, the simplified DKP construction produces a measurement pattern from a circuit in the $\Jacz(\A)$ model (for some $\A \subset \R$) by transforming each gate $J(\alpha)$ and $\cZ$ into a simple measurement pattern via the mapping $\Phi$ defined in~\eqref{eqn:circuitPatternHomphm}, composing those patterns in the appropriate manner.
The resulting measurement dependencies exhibits a simple structure which may be described in terms of local properties of the entanglement graph, as follows.

For a unitary circuit $C$ in the $\Jacz(\A)$ model, consider the last gate $J(\alpha)$ performed.
(If there is more than one such gate which may be performed in parallel, we may choose an arbitrary one.)
We may decompose $C$ into circuits $C_v C'$, where $C_v$ consists of the final $J(\alpha)$ gate on $v$, and any controlled-$Z$ operations on $v$ which follow it, and $C'$ consists of the rest of the operations of $C$.
We can recursively apply this procedure to decompose $C$ into layers of $\cZ_{u,v}$ gates separated by $J(\alpha)_v$ gates for various qubits $u$ and $v$, where the $\cZ$ gates of each layer following a $J(\alpha)_v$ gate all act on the qubit $v$.
We will refer to this as a \emph{star decomposition} of a $\Jacz(\A)$ circuit.
Consider the patterns produced by the DKP construction, for each sub-circuit $C_v$ consistsing of a $J(\alpha)_v$ gate and the subsequent layer of $\cZ$ gates: writing $C_v$ in stable-index tensor notation (pages~\pageref{sec:nonstandardRepns}\thru\pageref{sec:nonstandardRepnsEnd}), we have
\begin{align}
		\Phi(C_v)
	\;\;=&\;\;
		\Phi\paren{\paren{\prod_u \cZ\pseu[u,v']} J(\alpha)\pseu[:v'/v]}
	\notag\\[1ex]=&\;\;
		\paren{\prod_u \Ent{u,v'}{}} \mathfrak J^\alpha_{v'/v}
	\notag\\[1ex]=&\;\;
		\paren{\prod_u \Ent{u,v'}{}} \Xcorr{v'}{\s[v]} \Meas{v}{-\alpha} \Ent{v,v'} \New{v'}{\sfx}	\;,
	\notag\\[1ex]\cong&\;\;
		\paren{\prod_u \Zcorr{u}{\s[v]}} \Xcorr{v'}{\s[v]} \paren{\prod_u \Ent{u,v'}{}} \Meas{v}{-\alpha} \Ent{v,v'} \New{v'}{\sfx}	\;.
\end{align}
where in the last step we commute the correction operation $\Xcorr{v'}{\s[v]}$ to the right.
The $J(\alpha)$ gate in the original circuit $C$ gives rise to a qubit $v'$ which is a ``successor'' of the qubit $v$\,, in the sense that the pattern $\mathfrak J^\alpha_{v'/v}$ transfers the state of $v$ to $v'$ (up to the given unitary transformation); and every other neighbor of $v'$ in the entanglement graph will be subject to a correction operation which depends on the result $\s[v]$ of the measurement on $v$.

If we compose measurement patterns of this form together, this structure of correction is (almost) preserved: because the $\Zcorr{u}{\s[v]}$ operations commute with any entangling operations acting on the qubits $u$, the only additional correction which can be induced is a correction $\Zcorr{v''}{\s[v]}$ on a possible successor of $v'$ in cases where the same qubit is later acted on by another $J(\alpha')$ gate in $C$.
Absorbing these correction into measurements where appropriate, these give rise to sign- and $\pi$-dependencies on $v$ of the measurements on qubits adjacent to $v'$ in the entanglement graph (other than $v$ itself).

In the case of qubits measured with $\axisX$ measurements (\ie\ whose measurement angle is zero), the ``effect'' of $X$ corrections are trivial; in every other case, however, there is either a non-trivial effect on the basis of measurement, or (speaking somewhat counterfactually) on the result obtained upon measurement in the case of a $\pi$-dependency.
In particular, while $\pi$-dependencies could be removed by performing signal-shifting, \ie\ by introducing an operator which conditionally toggles the measurement result, we could also speak informally of the ``measurement result that would occur'' being affected by the change in the measurement by the addition of $\pi$ to the measurement angle.\footnote{%
	This intuition of ``counterfactual'' change of measurement results could perhaps be given an ontological foundation in terms of hidden variables, thereby removing its counterfactual nature; a formal discussion of this is beyond the scope of this thesis.}
Thus, we interpret the angle adaptations given by the simplified DKP construction as potential ``influences'' on the measurement outcomes, neglecting the imperviousness of $\axisX$ measurements to sign dependencies for the sake of uniformity.

Signal shifting and Pauli measurement simplifications can be applied to reduce the logical depth of a measurement-based computation, and so produce a pattern of the \emph{complete} DKP construction.
However, the dependencies described by the \emph{simplified} DKP construction are local in nature, and can be described simply by the geometry $(G,I,O)$ of the measurement pattern (Definition~\ref{def:geometry}), together with a function $f: O\comp \to I\comp$ mapping each qubit $v$ in the pattern to its ``successor'' (where $O\comp = V(G) \setminus O$, and $I\comp = V(G) \setminus I$), arising from the $\mathfrak J^{\alpha}$ patterns.
In particular, the measurement of each qubit $v \in O\comp$ influences the measurement of its successor $f(v)$ by a sign dependency, and the measurements of the other neighbors of $f(v)$ by a $\pi$-dependency.
Furthermore, these dependencies entail that both $f(v)$ and every qubit adjacent to $f(v)$ --- except for $v$ itself --- would be measured strictly after $v$ in the measurement order arising from the simplified DKP construction.

``Flows'' were defined by Danos and Kashefi~\cite{DK06} in an approximate\footnote{%
The presentations of the one-way model in~\cite{RB01,RBB03} already contained several constructions which do not have ``flows'', which were formulated later.
However, at least in the case of the the simplified RBB construction presented in the previous chapter, there is a simple correspondence to the DKP construction which allows these to be related to geometries with flows.
It was thought that a modest extension of the definition of flows would also capture those one-way measurement based algorithms~\cite{DK05priv}.}
early attempt to characterize those one-way measurement patterns which perform unitary transformations, based on this structure of dependencies arising from this construction.
Without reference to a pattern known to be obtained from the DKP construction, a flow is defined as follows:
\begin{definition}
	\label{def:flow}
	For a geometry $(G,I,O)$, a \emph{flow} is an ordered pair $(f, \preceq)$ consisting of a function $f: O\comp \to I\comp$, and a partial order\footnote{%
		A partial order $\preceq$ is a pre-order (see footnote~(\ref{fn:preorder}) on page~\pageref{fn:preorder}) which is anti-symmetric: that is, for which $x \preceq y$ and $y \preceq x$ imply $x = y$.
		Examples include the divisibility relation $x \,|\, y \,\iff\,  (y \div x \in \N)$ for complex numbers $x$ and $y$\,, and the subset relation $X \subset Y \,\iff\; \forall z: \paren{z \in X \implies* z \in Y}$ on sets $X$ and $Y$.}
	$\preceq$ on $V(G)$, such that the conditions
	\begin{subequations}
	\begin{gather}
			\label{flow:a}			v \sim f(v)	\;,
		\\
			\label{flow:b}		\textbalance{v \preceq f(v)\;,}[\;\;\text{and}]
		\\
			\label{flow:c}		w \sim f(v)	\;\implies\; v \preceq w
	\end{gather}
	\end{subequations}
	hold for all $v \in O\comp$ and $w \in V(G)$.
\end{definition}
Examples of geometries with and without flows are illustrated in Figures~\ref{fig:examplesFlow} and~\ref{fig:exampleNoFlow}.
\begin{figure}[t]
	\begin{center}
			\includegraphics{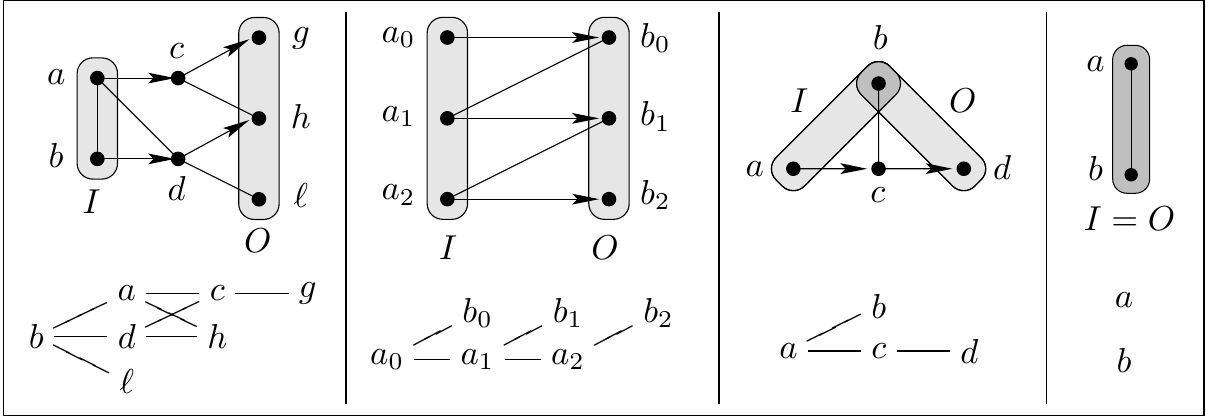}
			\vspace{-2ex}
	\end{center}
	\caption[Examples of geometries with flows.]{\label{fig:examplesFlow}
		Examples of geometries with flows.
		Arrows indicate the action of a function $f: O\comp \to I\comp$ along otherwise undirected edges.
		Corresponding partial orders $\preceq$ for each example are given by Hasse diagrams below the graphs (with minimal elements on the left and maximal elements on the right).
		In the right-most example, the two vertices $a$ and $b$ are incomparable, \ie\ there is no order relation between them.}
		\vspace{1em}
\end{figure}
\begin{figure}[t]
	\begin{center}
			\includegraphics{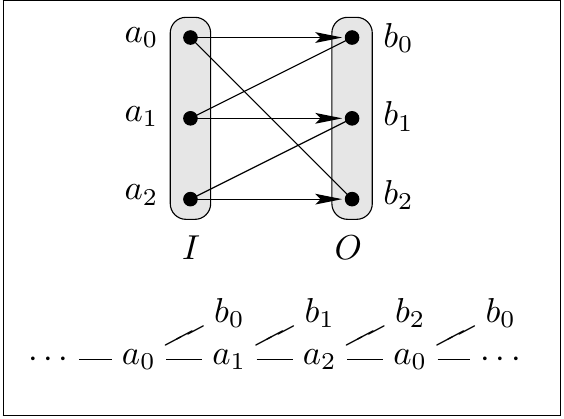}
			\vspace{-2ex}
	\end{center}
	\caption[Example of a geometry with no flow.]{\label{fig:exampleNoFlow}
		Example of a geometry with no flow (\cf\ the second geometry from the left in Figure~\ref{fig:examplesFlow}).
		Arrows indicate the action of an injective function $f: O\comp \to I\comp$ along otherwise undirected edges.
		Also given is a reflexive and transitive binary relation $\preceq$ which satisfies conditions~\eqref{flow:b} and \eqref{flow:c} for this function $f$, but which is not antisymmetric.}
\end{figure}

The function $f$ and the partial order $\preceq$ capture the essential structure of the dependencies in the simplified DKP construction: $f$ represents the mapping of qubits to their successors, implementing logical ``wires'' via single-qubit state transfer, and the partial order $\preceq$ represents \emph{one possible} order (not necessarily one of minimum depth) in which the qubits may be measured to perform a unitary computation.

By~\cite{DK06}, any geometry which has a flow underlies some one-way measurement pattern which can be obtained by the DKP construction, and which therefore performs a unitary transformation: moreover, the proof in~\cite{DK06} is by reduction to unitary circuits.
We may then use flows to obtain a semantic map $\cS$ (as described on page~\pageref{discn:semanticMap}), corresponding to the choice of the \emph{complete} DKP construction as a choice of representation map $\cR$ from unitary circuits to the one-way measurement model, if
\begin{inlinum}
\item
	we can find a flow for a geometry $(G,I,O)$; and
\item
	we can verify whether the \emph{dependencies and corrections} of a measurement pattern are consistent with one arising from the DKP construction.
\end{inlinum}
The main subject of this Chapter is essentially how to perform these tasks efficiently.

\subsection{Direct interpretation of flows in terms of unitary circuit structure}
\label{sec:flowsNonstandardRepn}

On pages~\pageref{sec:nonstandardRepns}\thru\pageref{sec:nonstandardRepnsEnd}, we presented the stable-index representation of unitary circuits, and in particular described the combinatorial structure of the ``interaction graph'', whose vertices consist of tensor indices (corresponding to ``wire segments'') of the unitary circuit.
The edges of this graph consist of pairs of indices which are involved in a single unitary gate; and in particular, there is a function $f$ mapping each deprecated index to a corresponding advanced index, imposing an arrow of time on the indices.

The map $\Upphi$ of~\eqref{eqn:circuitPatternHomphm} from the $\Jacz(\A)$ model to the $\OneWay(\A)$ model, which forms the first part of the simplified DKP construction of Section~\ref{sec:DKPstandaridzation}, identifies the interaction graph of a circuit in the $\Jacz(\A)$ model with the geometry $(G,I,O)$ of the corresponding one-way pattern.
In particular, advanced indices correspond to prepared (or non-input) qubits, and deprecated indices to measured (or non-output) qubits; and the function $f$ on the indices of the stable index expression may then be identified with a function $f: O\comp \to I\comp$ on the geometry $(G,I,O)$.

A flow consists of a function $f$ and a partial order $\preceq$, where $f$ may be considered to correspond to the mapping of deprecated indices to advanced indices in a stable index representation of a unitary circuit, and the partial order $\preceq$ to correspond to the order described on page~\pageref{item:nonstandardRepsOrderRecovery} in the discussion on ``Recovering the order of unitary operations''.
Flows may then also be interpreted as describing the structure of a unitary circuit, expressed as a stable index tensor product, from the geometry $(G,I,O)$ of a measurement pattern.

\section{Graph constructions and characterizations for flows}
\label{sec:graphth-charn}

The main result of this chapter is to characterize flows in graph-theoretic terms, which allows us to reduce the problem of deciding whether a geometry has a flow to problems in graph theory with efficient solutions, and to examine (and solve) an extremal problem which further bounds the running time of those algorithms.
This will ultimately enable us to obtain an efficiently computable semantic map $\cS$ (as described on page~\pageref{discn:semanticMap}) for the DKP construction described in Section~\ref{sec:DKPstandarization}, and provide a strong upper bound on its running time.

In this section, we consider useful constructions of directed graphs, culminating in the aforementioned graph-theoretic characterizations.
These constructions presented in this section will also allow us to more easily describe the generality of the geometries which have flows.
We also make use of these constructions to obtain uniqueness results, and bounds on the number of edges of graphs with flows, which ultimately lead to the first efficient algorithm for determining when a geometry $(G,I,O)$ has a flow.

\subsection{Generalizing to path covers/successor functions}

We will begin by exploring how to relax the notion of a flow to admit a larger class of objects, starting by noting the most important properties of a flow.
\begin{lemma}
	\label{lemma:injectiveFlow}
	If $(f, \preceq)$ is a flow for a geometry $(G,I,O)$, then $f$ is injective.
\end{lemma}
\begin{proof}
	Suppose $x,y \in O\comp$ are such that $f(x) = f(y)$.
	Then $y \sim f(y) = f(x)$, so that $x \preceq y$, and $x \sim f(x) = f(y)$, so that $y \preceq x$\,.
	It follows that $x = y$.
\end{proof}
The limitation to injective functions $f$ motivates a description of flows in terms of vertex-disjoint collections of paths (or more precisely, an adaptation of the concept of a ``path cover''~\cite{Diestel}):
\begin{definition}
	\label{def:pathCover}
	Let $(G,I,O)$ be a geometry. A collection $\cC$ of (possibly trivial\footnote{%
	A (directed) path or walk is \emph{trivial} if it is of length zero, \ie\ it starts and ends at a single vertex without traversing any edges (respectively, arcs).})
	directed paths in $G$ is a \emph{path cover} of $(G,I,O)$ if the following conditions hold:
	\begin{romanum}
	\item
		each $v \in V(G)$ is contained in exactly one path (i.e. the paths cover $G$ and are vertex-disjoint);
	\item
		each path in $\cC$ is either disjoint from $I$, or intersects $I$ only at its initial point;
	\item
		each path in $\cC$ intersects $O$ only at its final point.
	\end{romanum}
\end{definition}
In the case $\card{I} = \card{O}$, a path cover of $(G,I,O)$ is a collection of vertex-disjoint paths from $I$ to $O$ which covers all the vertices of $G$.
For a flow $(f,\preceq)$, there is a natural connection between the function $f$ and path covers for the geometry $(G,I,O)$:
\begin{lemma}
	\label{lemma:orbitPathCover}
	Let $(f, \preceq)$ be a flow on a geometry $(G,I,O)$.
	Then there is a path cover $\cP_f$ of $(G,I,O)$ such that, for all vertices $v,w \in V(G)$, $v \arc w$ is an arc in some path of $\cP_f$ if and only if $w = f(v)$.
\end{lemma}
\begin{proof}
	Let $(f,\preceq)$ be a flow on $(G,I,O)$.
	Define a digraph $P$ on the vertices of $G$, and with arcs $v \arc f(v)$ for $v \in O\comp$.
	Because $f$ is both a function and injective, every vertex in $P$ has maximal out-degree and maximal in-degree $1$.
	Thus, $P$ is a collection of vertex-disjoint dipaths and closed walks.\footnote{%
	A closed walk is one which begins and ends at the same vertex.}
	Furthermore, for every arc $(v \arc w) \in A(P)$, we have $v \preceq w$; by induction, $v \preceq z$ whenever there is a dipath from $v$ to $z$ in $P$.
	Then if $v$ and $z$ are such that there are dipaths from $v$ to $z$ and from $z$ to $v$, then $v \preceq z$ and $z \preceq v$, in which case $x = z$ and the dipaths are trivial.
	Thus, $P$ is acyclic, so $P$ consists entirely of vertex-disjoint dipaths.

	Let $\cP_f$ be the collection of maximal dipaths in $P$.
	We show that $\cP_f$ satisfies each of the criteria of Definition~\ref{def:pathCover}:
	\begin{romanum}
	\item
		Any vertex $v$ which is neither in $\dom(f)$ nor $\img(f)$ will be isolated in $P$: then, the trivial path on $v$ is an element of $\cP_f$.
		All other vertices are in either $\dom(f)$ or $\img(f)$, and so are contained in a non-trivial path of $\cP_f$.
		As these paths are vertex-disjoint, each vertex is contained in exactly one path.

	\item
		Each vertex in $I$ has in-degree $0$, and so may only occur at the beginning of any path in $\cP_f$.

	\item
		The vertices in $P$ which have out-degree $0$ are precisely the output vertices $O$: therefore one occurs at the end of every path, and they may only occur at the end of paths in $\cP_f$.
	\end{romanum}
	Then $\cP_f$ is a path cover, whose paths contain only arcs $v \arc f(v)$, as required.
\end{proof}
It will be occasionally be convenient to alternate between descriptions of problems in terms of path covers for a given geometry, and functions $f$ such that $\cP_f$ is a path cover.
In particular, we may define:
\begin{definition}
	\label{def:successorFn}
	A \emph{successor function} for a geometry $(G,I,O)$ is an injective function $f: O\comp \to I\comp$ such that the maximal orbits of $f$ form a path cover for $(G,I,O)$.
	A successor function $f$ for $(G,I,O)$ is a \emph{flow function} if there exists a partial order $\preceq$ on $V(G)$ such that $(f, \preceq)$ is a flow.
\end{definition}
The terminology of ``successor function'' is meant to be suggestive of the mapping $v_j \mapsto* v_{j+1}$ between successive indices in a stable index tensor expression for a single qubit, which in turn correspond (by the comments made on page~\pageref{discn:stableIndexPathIntegral}) to successive segments of a wire (separated by single-qubit gates which do not preserve standard basis states) in a quantum circuit diagram.
The generalization from flow functions to successor functions will prove very helpful in the analysis below.

\subsection{Generalizing from partial orders to pre-orders}

Figure~\ref{fig:exampleNoFlow} (on page~\pageref{fig:exampleNoFlow}) illustrates a geometry with successor functions, but which has no flow functions.
There are only two possible successor functions $f$ and $f'$ for that geometry, given by
\begin{align}
		f(a_j) \;=&\; b_j	\;\;,
	&
		f'(a_j) \;=&\; b_{(j \minus 1) \bmod{3}}
\end{align}
(\ie\ taking the appropriate representative modulo $3$ for the index in the case of $f'$).
The successor function $f$ is the function actually illustrated in Figure~\ref{fig:exampleNoFlow}, and $f': O\comp \to I\comp$ corresponds to instead directing the diagonal edges of the graph illustrated there.
To show that neither of these are flow functions, we may show that any reflexive and transitive binary relation satisfying the conditions~\eqref{flow:b} and~\eqref{flow:c} will fail to be anti-symmetric; in which case there can be no partial order $\le$ such that either $(f, \le)$ or $(f', \le)$ is a flow for that geometry.
Specifically, to satisfy the flow conditions for $f$, such a binary relation would satisfy the constraints $a_0 \le a_1 \le a_2 \le a_0$\,, while for $f'$ it would have to satisfy the constraints $a_0 \ge a_1 \ge a_2 \ge a_0$\,.
In either case, the vertices $a_j \in I$ are distinct but mutually related vertices: so any such relation fails to be a partial order.
Because no partial order can satisfy the conditions~\eqref{flow:b} and~\eqref{flow:c} for either $f$ or $f'$, that geometry has no flow.

Analysis of the form above will be generally useful, and motivates a generalization from partial orders to pre-orders
	similar to our generalization from flow functions to successor functions.
\begin{definition}
	\label{def:inflRel}
	Let $(G,I,O)$ be a geometry and $f: O\comp \to I\comp$ be a successor function for $(G,I,O)$.
	The \emph{influence relation} $\inflRel$ for $f$ is the binary relation such that $v \inflRel w$ for $v \in O\comp$ and $w \in V(G)$ if and only if $v \ne w$ and either $w = f(v)$ or $w \sim f(v)$.
	An \emph{influencing pre-order} $\preceq$ for $f$ is a pre-order on $V(G)$ which extends $\inflRel$\,, \ie\ a reflexive and transitive relation such that
	\begin{subequations}
	\begin{gather}
 			\label{natural:a}
			v \preceq f(v)
		\\
 			\label{natural:b}
			w \sim f(v) \;\;\implies\;\; v \preceq w
	\end{gather}
	\end{subequations}
	holds	for all $v \in O\comp$ and $w \in V(G)$.
	A \emph{causal order} for $f$ is a influencing pre-order for $f$ which is also a partial order, \ie\ a binary relation such that $(f,\preceq)$ is a flow.
\end{definition}
Note that the influence relation $v \inflRel w$ describes exactly the conditions~\eqref{flow:b} and~\eqref{flow:c} on a causal order: the definition of an influencing pre-order then relaxes nothing more of the flow conditions than the condition that $\preceq$ be anti-symmetric.
The proof given at the beginning of this section that the geometry of Figure~\ref{fig:exampleNoFlow} has no flows is essentially that every influencing pre-order of $f$ or $f'$ must somehow fail to be a full-fledged partial order.

\subsection{Deciding if a successor function is a flow function}
\label{sec:influencing}

Every successor function has an influencing pre-order, and for each $f$, we identify one in particular:
\begin{definition}
	\label{def:naturalPreorder}
	Let $(G,I,O)$ be a geometry and $f: O\comp \to I\comp$ be a successor function for $(G,I,O)$.
	The \emph{natural pre-order} $\preceq$ for $f$ is the transitive and reflexive closure\footnote{%
	That is, $\preceq$ is the coarsest reflexive and transitive relation which extends $\inflRel$, and so may be described as the logical conjunction of all influencing pre-orders.
	As there exists at least one influencing pre-order, namely the relation $R$ such that $v R w$ for all $v, w \in V(G)$, this closure is guaranteed to exist.}
	of the relation $\inflRel$\,. 
\end{definition}
The natural pre-order $\preceq$ for a successor function $f$ is by definition an influencing pre-order, and in particular the coarsest influencing pre-order for $f$.
As a result, it can be used to determine whether or not $f$ is a flow function:
\begin{lemma}
	\label{lemma:causalOrderFlow}
	A successor function $f$ is a flow function if and only if its natural pre-order $\preceq$ is a partial order.
\end{lemma}
\begin{proof}
	If $\preceq$ is a partial order, then $(f, \preceq)$ is a flow, and so $f$ is a flow function.
	For the converse, suppose that there is \emph{some} partial order $\le$ such that $(f, \le)$ is a flow: then $\le$ is an influencing pre-order for $f$.
	Being a partial order, for each distinct pair of vertices $v,w \in V(G)$, at least one of $v \nleqslant w$ or $w \nleqslant v$ holds; then, for such pairs of vertices, at least one of $v \not\preceq w$ or $w \not\preceq v$ holds as well, as $\preceq$ is coarser than $\le$\,.
	Thus $\preceq$ is a partial order.
\end{proof}
We can use this to efficiently decide when a successor function $f$ is a flow function by reduction to the \emph{transitive closure} problem on directed graphs: specifically, by constructing $\preceq$ from the influence relation $\inflRel$ for $f$.
While the algorithms will be simpler to describe in terms of the binary relations $\inflRel$ and $\preceq$ themselves, we will describe the graph-theoretic presentation of this problem: this is the presentation used in the relevant literature, and the graph constructions we describe here will also prove convenient for analysis in the following sections.

\begin{definition}
	\label{def:inflDigraph}
	For a successor function $f$ on a geometry $(G,I,O)$, the \emph{influence digraph} is a directed graph with vertex-set $V(G)$, and with an arc $v \arc w$ between $v, w \in V(G)$ if and only if $v \inflRel w$.
\end{definition}
\begin{figure}[t]
	\begin{center}
		\includegraphics{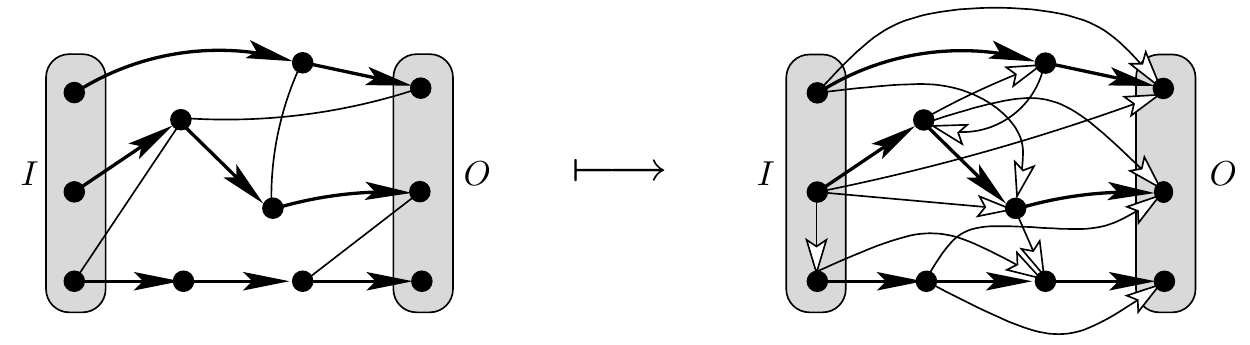}
	\end{center}
	\caption[Illustration of the influence digraph construction.]{\label{fig:inflDigraph}%
		Illustration of the influence digraph construction.
		On the left: a geometry $(G,I,O)$ with a successor function $f$, represented by the corresponding path cover $\cP_f$\,.
 		On the right: the corresponding influence digraph $\sI_f$\,.
 		Solid arrows represent arcs of the form $v \arc f(v)$\,, and hollow arrows represent arcs $v \arc w$ for $w \sim f(v)$ in the graph $G$\,.}
\end{figure}
An example of this construction is illustrated in Figure~\ref{fig:inflDigraph}.
We may then characterize the natural pre-order $\preceq$ of $f$ in terms of the \emph{transitive closure} of $\sI_f$\,:
\begin{definition}
	The \emph{transitive closure of a digraph $D$} is a directed graph $T(D)$ with $V(T(D)) = V(D)$, and such that $(v \arc w) \in A(T(D))$ if and only if there is a non-trivial directed path from $v$ to $w$ in the digraph $D$.
\end{definition}
If we let $\sT_f$ be the transitive closure of $\sI_f$, the natural pre-order of a successor function $f$ is then the relation $\preceq$ such that $v \preceq w$ if and only if either $v = w$ or $(v \arc w) \in A(\sT_f)$ for two vertices $v,w \in V(\sI_f)$.
This allows us to reduce the construction of $\preceq$ to the following problem applied to $\sI_f$:

\begin{problem}[Transitive Closure]
 	For a digraph $D$, construct its transitive closure $T(D)$.
\end{problem}
The transitive closure problem is solvable in polynomial time by the Floyd--Warshall algorithm~\cite{Floyd62,CLRS}.
We will present a faster algorithm using techniques presented in~\cite[Chapters 3\thru4]{Nuutila95},
\label{disc:stronglyConnected}
based on Tarjan's algorithm~\cite{Tarjan72, CLRS} for determining the \emph{strongly connected components} of a digraph (equivalence classes of vertices in a digraph which are mutually reachable by directed paths).
We will describe this algorithm in detail in Section~\ref{sec:flowAlgs}, where we will prove the following:
\begin{theorem}
	\label{thm:findPreorder}
	Let $f$ be a successor function of a geometry $(G,I,O)$, and let $n = \card{V(G)}$ and $k = \card{O}$.
	Then there is an algorithm constructing the natural pre-order $\preceq$ for $f$, or which determines that $\preceq$ is not a partial order, in time $O(k^2 n)$.
\end{theorem}

\subsection{Influencing walks and causal path covers}

While a detailed analysis of the algorithms described above will wait until Section~\ref{sec:flowAlgs}, it will be useful to consider the graph-theoretic structures which are involved in those algorithms.
These structures also present a means of proving uniqueness results which led to the first algorithms for determining whether a geometry has a flow, and extremal results which allow refinements in their running times.

In the previous section, we observed that not only does the geometry of Figure~\ref{fig:exampleNoFlow} have a cyclic underlying graph, but also a complete cycle of relationships between some of its vertices in any influencing pre-order $\le$ for a given successor function for that geometry.
These cycles of relationships will manifest themselves as directed cycles in the corresponding influence digraphs $\sI_f$, but can be traced also to circuit\footnote{%
	In this context, we are using the graph-theoretic term \emph{circuit}, which usually refers to closed walks of length $3$ or more.
}
in the original graph $G$, of the following sort:
\begin{definition}
	\label{def:viciousCircuit}
	Let $G$ be a graph, and $\cP$ a family of vertex-disjoint directed paths in $G$.
	A walk $W = u_0 u_1 \cdots u_\ell$ is an \emph{influencing walk} for $\cP$ if it is a concatenation of zero or more paths (\emph{segments} of the influencing walk) of the following two types:
	\begin{romanum}
	\item
		$v \arc w$, where this is an arc in some path of $\cP$;
	\item
		$v \arc z \arc w$, where $v \arc z$ is an arc in some path of $\cP$ and $w z \in E(G)$ is an edge not covered by $\cP$.
	\end{romanum}
	A \emph{vicious circuit} for $\cP$ is a closed influencing walk for $\cP$ with at least one segment.
\end{definition}
It is easy to show (by induction) that we may also characterize an influencing walk $W$, for a family of vertex-disjoint paths $\cP$, as one whose arcs 
\begin{inlinum}
\item
	never traverse an edge of $G$ in a direction opposite that of an arc of $\cP$, and

\item
	do not traverse two edges in a row which are not covered by $\cP$.
\end{inlinum}
Influencing walks were first identified as objects of interest by Broadbent and Kashefi~\cite{BK2007}, who examine their role in the depth complexity of unitaries in the one-way measurement model.

We will occasionally refer to the two types of segments as \emph{type \iti} and \emph{type \itii} respectively.
The motivation for the two types of segment is in the two ways in which distinct vertices $v, w \in V(G)$ may be related by the influence relation $\inflRel$ for a successor function:
\begin{lemma}
	\label{lemma:influencingNatlPreorder}
	Let $\cP_f$ be a path cover for $(G,I,O)$ with successor function $f$, and let $\inflRel$ and $\preceq$ be the influence relation and the natural pre-order of $f$ respectively.
	Then $v \preceq w$ if and only if there is an influencing walk $W = v \diwalk w$ for $\cP_f$ from $v$ to $w$, and in particular, $v \inflRel w$ if and only if there is an influencing walk $W$ for $\cP_f$ from $v$ to $w$ which consists of one segment.
\end{lemma}
\begin{proof}
	Suppose $W$ is an influencing walk for $\cP_f$ from $v$ to $w$.
	A segment of type \iti\ is a dipath $v \arc w$ between vertices $w = f(v)$, and a segment of type \itii\ is a dipath $v \arc z \arc w$ for vertices $v$, $z$, and $w$ such that $w \sim z$ and $z = f(v)$, and where $w \ne v$ (because the edge $v z$ is covered by the path cover $\cP_f$).
	In both cases, we have $v \inflRel w$\,.
	Similarly, if $v \inflRel w$, then there is either an influencing walk $W = v \arc w$ or an influencing walk $W' = v \arc z \arc w$ for $z = f(v)$.
	The Lemma then holds for $\inflRel$ and influencing walks of one segment.

	An influencing walk in general (of \emph{zero} or more segments) then corresponds to the reflexive and transitive closure of the influence relation, \ie\ to the natural pre-order $\preceq$ of $f$.
	In particular, an influencing walk $v \diwalk w$ of \emph{zero} segments is the trivial walk on $v = w$, in which case $v \preceq w$.
	Influencing walks $W$ of $n \ge 1$ segments from $v$ to $w$ we may decompose into an influencing walk $W'$ of $n-1$ segments from $v$ to some vertex $u$, and a single segment from $u$ to $w$.
	By induction, we have $v \preceq u \inflRel w$, so $v \preceq w$.
	For the converse, suppose $v \preceq w$ for some vertices $v,w \in V(G)$.
	By definition, there is then a sequence of vertices $\paren{u_j}_{j = 0}^\ell$ for some $\ell \in \N$, such that $v = u_0 \inflRel u_1 \inflRel \cdots \inflRel u_\ell = w$.
	Then, for each $0 \le j < \ell$\,, we have $u_j \ne u_{j+1}$, and either $u_{j+1} = f(u_j)$, or $u_{j+1} \sim f(u_j)$, and we may define an dipath $W_j = u_j \arc u_{j+1}$ or $W_j = u_j \arc f(u_j) \arc u_{j+1}$, respectively.
	In the latter case, if $u_{j+1} = f(f(u_j))$, the path $W_j$ is a concatenation of two influencing walk segments of type~\iti; otherwise, $W_j$ is itself a single influencing walk segment of either type~\iti\ or type~\itii.
	In any case, the concatenation $W = W_0 \cdots W_{\ell-1}$ of these paths is an influencing walk for $\cP_f$ from $v$ to $w$.
\end{proof}
Because of the correspondence between segments of an influencing walk and the influence relation $\inflRel$, a single segment influencing walk essentially corresponds to a single arc of the influence digraph $\sI_f$, and an influencing walk in general to a directed walk in $\sI_f$.

The close correspondence between influencing walks for $\cP_f$ and the natural pre-order for $f$ motivates the following definition:
\begin{definition}
	A \emph{causal path cover} $\cC$ for a geometry $(G,I,O)$ is a path cover which has no vicious circuits.
\end{definition}
Using this further definition, we may then easily characterize flows in terms of path covers:
\begin{theorem}
	\label{thm:graphth-charn}
	Let $f$ be a successor function of a geometry $(G,I,O)$ and let $\cP_f$ be the path cover induced by $f$.
	Then $f$ is a flow function if and only if $\cP_f$ is a causal path cover.
\end{theorem}
\begin{proof}
	Let $\preceq$ be the natural pre-order of $f$.
	By Lemma~\ref{lemma:influencingNatlPreorder}, we have $v \preceq w$ if and only if there is an influencing walk for $\cP_f$ from $v$ to $w$\,: then we have $v \preceq w$ and $w \preceq v$ for distinct $v, w \in V(G)$ if and only if there are influencing walks $W' = v \diwalk w$ and $W'' = w \diwalk v$ for $\cP_f$\,.
	If there are such walks, then $C = W' W''$ is a vicious circuit for $\cP_f$.
	Conversely, a vicious circuit $C = \sigma_1 \cdots \sigma_\ell$ for $\cP_f$ (concatenated from some non-zero number of segments $\sigma_j$) can be decomposed into an influencing walk $W' = \sigma_1 \cdots \sigma_{\ell - 1}$ between two distinct vertices $v$ and $w$, and another influencing walk $W'' = \sigma_\ell$ from $w$ to $v$.
	Then if there is such a circuit $C$, we have $v \preceq w$ and $w \preceq v$ for some two distinct vertices $v$ and $w$, so that $\preceq$ is not a partial order.
	Thus, $\preceq$ is a partial order if and only if $\cP_f$ lacks vicious circuits.
	By Lemma~\ref{lemma:causalOrderFlow}, $f$ is a flow function if and only if its natural pre-order $\preceq$ is a partial order; the theorem then holds.
\end{proof}
This completes our graph-theoretic characterization of flows.
The main utility of these constructions and characterizations is to reduce the properties of the natural pre-order $\preceq$ of a function $f$ to influencing walks for $\cP_f$\,, or to walks in the influence digraph $\sI_f$\,.
This change in emphasis will serve to clarify the graph-theoretic results which follow.

\subsection{Bounding the number of edges of a geometry with a flow}
\label{sec:extreme}

For a geometry $(G,I,O)$ with a flow-function $f$, let us call an edge $vw \in E(G)$ a \emph{flow edge} if either $w = f(v)$ or $v = f(w)$, \ie\ if $vw$ is an edge covered by the path-cover $\cP_f$\,; every other edge of $G$ is a \emph{non-flow} edge.
We have seen how influencing walks for path covers $\cP_f$ correspond to the natural pre-order $\preceq$ for $f$\,: and in this terminology, from the remarks made just after Definition~\ref{def:viciousCircuit}, an influencing walk $W$ is essentially a walk which follows the paths of $\cP_f$, but which occasionally jumps across paths by traversing a non-flow edge (under the constraint that it does not traverse two such edges in a row).
Thus, the more non-flow edges there are, the more freedom there is in how an influencing walk may be constructed.

It is reasonable to suppose that the more non-flow edges there are, the more likely there are to be influencing walks $W' = v \diwalk w$ and and $W'' = w \diwalk v$ for some distinct vertices $v, w \in V(G)$\,, producing a vicious circuit for $\cP_f$.
We might then ask if there is an upper bound on the number of non-flow edges that a geometry with a flow may have.
Note that, whether or not a geometry $(G,I,O)$ has a flow, a path cover $\cC$ for $(G,I,O)$ must have exactly $k = \card{O}$ paths, as all the paths of $\cC$ must terminate at an output vertex, and all output vertices terminate such a path.
These paths will collectively cover exactly $n - k$ edges of $G$, where $n = \card{V(G)}$, as there is a one-to-one correspondence between elements of $O\comp$ and the arcs of $\cC$ which leave those vertices; all other edges of $G$ are non-flow edges relative to this path cover $\cC$.
Then, we may describe this question without reference to any particular path cover in terms of the following:

\begin{definition}
	For $n \ge k \ge 1$, define $\Gamma(n,k)$ to be the maximum number of edges in a geometry $(G,I,O)$ which has a flow, subject to $\card{V(G)} = n$ and $\card{O} = k$.
\end{definition}

For the question of whether \emph{there exists} a geometry on $n$ vertices and with $m$ edges which admits a flow, we can consider any graph $G$ with $n$ vertices, under the constraint that it admits some family of vertex-disjoint (directed) paths $\cP = \ens{P_1,\ldots,P_k}$ covering the entire graph.
Without loss of generality, we may then let $O$ be the final points of those paths, and $I$ be (an arbitrary subset of) the initial points of those paths.
Because we may consider the outputs and inputs to be given by a family of paths in this way, we will often refer to the successor function of a family of paths $\cP$ which covers the vertices of a graph, whether or not it is explicitly a path cover for a geometry.

As we noted on page~\pageref{disc:stronglyConnected}, we may reduce this question to the question of whether the strongly-connected components of the influence digraph $\sI_f$ (Definition~\ref{def:inflDigraph}, page~\pageref{def:inflDigraph}) all consist of exactly one vertex.
This holds if and only if $\sI_f$ lacks directed circuits, \ie\ if $\sI_f$ is \emph{acyclic}.
We may then consider whether the successor function $f$ for $\cP$ has a causal order, and consider the problem in terms of adding non-flow edges to a graph $G$, which initially contains only the paths of $\cP$, under the constraint of keeping $\sI_f$ acyclic.
This leads to a simple upper bound on $\Gamma(n,k)$\,:

\begin{theorem}
	\label{thm:extremalUpperBound}
	$\Gamma(n,k) \le kn - \binom{k+1}{2}$ for all integers $n \ge k \ge 1$.
\end{theorem}
\begin{proof}
	We may base this proof on two simple observations about a path cover $\cP_f = \ens{P_1, \ldots, P_k}$ such that $\sI_f$ is acyclic.
	For each $1 \le j \le k$, let $n_j$ denote the number of vertices in the path $P_j$.
	\begin{observation}
		\label{obs:extr1}
		Consider a path $P_j = v_1 \arc v_2 \diwalk v_{n_j}$\,.
		If $v_a v_b \in E(G)$ for $b > a+1$, then $\sI_f$ will contain the directed cycle $v_a \arc v_{a+1} \diwalk v_{b-1} \arc v_a$\,.
		Then, if $\sI_f$ is acyclic, we have $v_a v_b \in E(G)$ for $a < b$ if and only if $b = a+1$, that is if $v_a v_b$ is a flow edge.
	\end{observation}
	\begin{observation}
		\label{obs:extr2}
		Consider any two distinct paths $P_h = v_1 \arc v_2 \diwalk v_{n_h}$ and $P_j = w_1 \arc w_2 \diwalk w_{n_j}$\,.
		If there are edges $v_a w_b\,,\, v_c w_d \in E(G)$ for some $a < c$ and $b > d$, then $\sI_f$ will contain the directed cycle $v_a \diwalk v_{c-1} \arc w_d \diwalk w_{b-1} \arc v_a$.
		Then, if $\sI_f$ is acyclic, we have non-flow edges $v_a w_b\,,\, v_c w_d \in E(G)$ only if the relation
		\begin{gather}
				a < c \;\implies\; b \le d
		\end{gather}
		holds. (The condition $b > d \;\implies\; a \ge c$ is equivalent, up to relabelling).
	\end{observation}
	The first observation implies that the only non-flow edges that $(G,I,O)$ may contain if it has a flow are edges between distinct paths $P_h$ and $P_j$\,; and the second imposes a useful constraint on the endpoints of pairs of non-flow edges.
	We may then assume that these constraints hold for all non-flow edges without loss of generality.

	We define a function $\lambda$ from the non-flow edges to $\N$ as follows.
	For a non-flow edge $v_a w_b$ between the $a\th$ vertex of a path $P_h = v_1 \diwalk v_{n_h}$ and the $b\th$ vertex of a path $P_j = w_1 \diwalk w_{n_j}$, we let
	\begin{gather}
		\lambda(v_a w_b) = a+b	\;.
	\end{gather}
	Suppose $\sI_f$ is acyclic.
	For two distinct edges $v_a w_b$ and $v_c w_d$ are non-flow edges between the two paths $P_h$ and $P_j$.
	Observation~\ref{obs:extr2} then implies that $\lambda(v_a w_b)$ and $\lambda(v_c w_d)$ differ:
	\begin{itemize}
	\item
		If $a < c$, we have $b \le d$, in which case $\lambda(v_a w_b) = a + b < c + d = \lambda(v_c w_d)$\,;

	\item
		If $c < a$, we have $d \le b$, in which case $\lambda(v_c w_d) = c + d < a + b = \lambda(v_a w_b)$\,.
	\end{itemize}
	Then, restricted to the non-flow edges between $P_h$ and $P_j$\,, $\lambda$ is injective.
	Because $2 \le \lambda(e) \le n_h + n_j$ for all non-flow edges $e$ between these paths, there are at most $n_h + n_j - 1$ non-flow edges between those two paths of $\cP$.

	Applying this to all pairs of paths $P_h$ and $P_j$, the number of non-flow edges in $G$ is then bounded above by
	\begin{align}
			\sum_{1\le h < j \le k} (n_h+n_j-1)
		\;\;=&\;\;
			\sfrac{1}{2}\left[ \sum_{h = 1}^k \sum_{j = 1}^k (n_h + n_j - 1) \;\;-\;\; \sum_{h = 1}^k (2n_h - 1) \right]
		\notag\intertext{}
			=&\;\;
			\sfrac{1}{2}\left[ \sum_{h = 1}^k (k n_h + n - k) \;\;-\;\; \sum_{h = 1}^k (2n_h - 1) \right]
 		\notag\\[1ex]=&\;\;
			\sfrac{1}{2}\left[2 k n - k^2 - 2 n + k \right]
		\notag\\[1ex]=&\;\;
 			kn - n - \tfrac{1}{2}(k^2 - k)	\;.
	\end{align}
	As the number of edges in the paths $P_j$ themselves is collectively $n - k$, the total number of edges $G$ may have if $\sI_f$ is acyclic is at most $kn - k - \tfrac{1}{2}(k^2 - k) \;=\; kn - \binom{k+1}{2}$\,.
\end{proof}
This result in itself will prove sufficient to strongly bound the running time of the algorithms presented in this chapter.
For completeness, however, we will consider a construction which saturates this upper bound.
Consider the following construction for any $n \ge k \ge 1$:

\begin{definition}
	\label{def:extremalGraph}
	Let $n_1,n_2,\ldots,n_k$ be an integer partition of $n$ such that $n_1 \le n_2 \le \cdots \le n_k$\,.
	For each $1 \le j \le k$\,, let $P_j \,=\, v_{j,1}\, v_{j,2}\,\cdots\,v_{j,n_j}$\,.
	Define $\sG(n_1,\ldots,n_k)$ to be the graph containing these paths, as well as the following non-flow edges for each $1 \le h < j \le k$\,:
	\begin{romanum}
	\item	
		If $n_h > 1$, then for each $1 \le a < n_h$\,, we include the edge $v_{h,a} v_{j,a}$;
	
	\item
		If $n_j > 1$, then for each $1 \le a < n_h$\,, we include the edge $v_{h,a+1} v_{j,a}$;

	\item
		For each $n_h \le a \le n_j$\,, we include the edge $v_{h,n_h} v_{j,a}$.
	\end{romanum}
	We will describe \iti\thru\itiii\ as the \emph{types} of non-flow edges in $\sG(n_1,\ldots,n_k)$\,.
	We also define $\sD(n_1, \ldots, n_k)$ to be the influence digraph $\sI_f$ for the successor function $f$ of the family of paths $\ens{P_1, \ldots, P_k}$.
\end{definition}
Figure~\ref{fig:extremalEg} illustrates an example of this graph construction for $\sG(6,8,9)$. 
(A diagram of the corresponding digraph $\sD(6,8,9)$ has a rather large number of arcs, and is not shown.)

\begin{figure}[tb]
	\begin{center}
		\qquad\includegraphics[width=120mm]{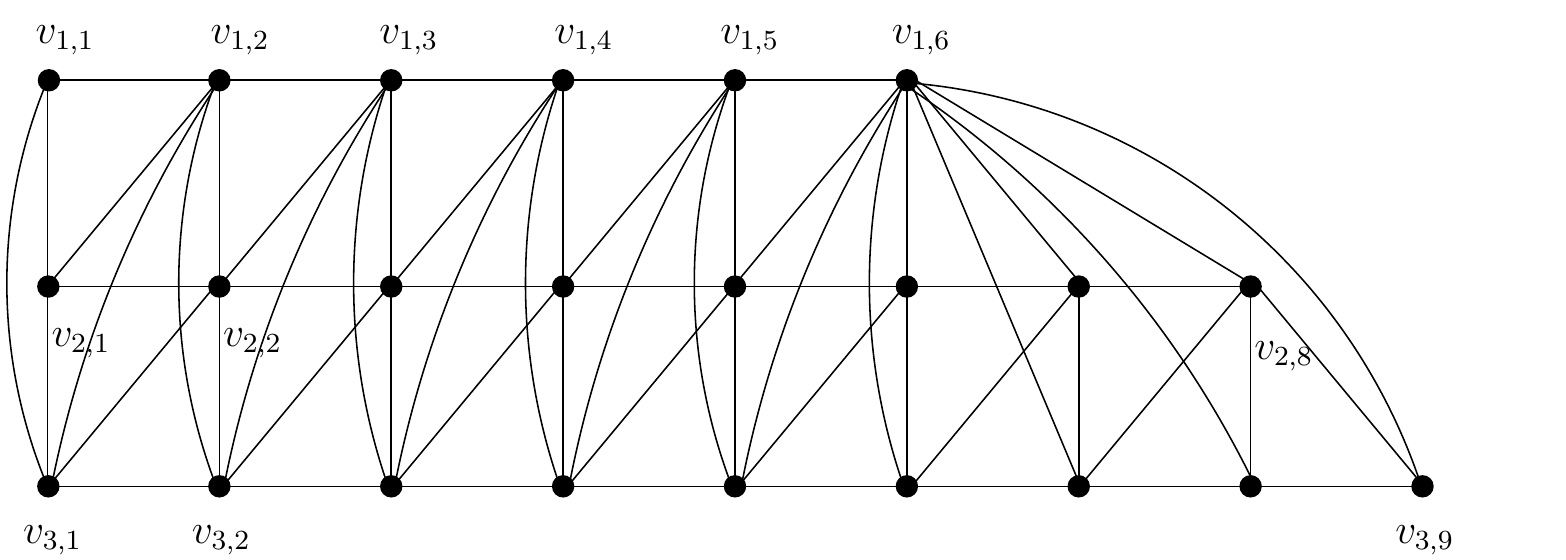}
	\end{center}
	\caption{\label{fig:extremalEg}%
		The graph $\sG(n_1,n_2,n_3)$ for $n_1=6$, $n_2=8$, $n_3=9$.}
\end{figure}

For the sake of brevity, let $G = \sG(n_1, \ldots, n_k)$.
Define $\cP = \ens{P_1, \ldots, P_k}$, and let $f$ be the successor function for this family of paths.
Then each non-flow edge in $G$ induces up to two arcs in the associated digraph $\sI_f$\,.
For each $1 \le h < j \le k$\,, the arcs of $\sI_f$ induced by non-flow edges between the paths $P_h$ and $P_j$ are of six different types, labelled \rma\thru\rmf, which we group together by the type of the non-flow edge which induces them:
\begin{subequations}
\label{eqn:extremalDigraphArcTypes}
\begin{align}
	\iti \;\mapsto
	\;&\;
	\begin{brace}
		\;\rma\;\;	v_{h,a-1} \arc v_{j,a}	&\;\;\;\:\text{for $1 < a \le n_h$ (if $n_h > 1$), and}	\\
		\;\rmb\;\;	v_{j,a-1} \arc v_{h,a}	&\;\;\;\:\text{for $1 < a \le n_h$ (if $n_j > 1$);}
	\end{brace}
	\qquad\\[2ex]
	\itii	\;\mapsto
	\;&\;
	\begin{brace}
		\;\rmc\;\;	v_{h,a} \arc v_{j,a}		&\text{for $1 \le a < n_h - 1$ (if $n_h > 1$), and}	\\
		\;\rmd\;\;	v_{j,a-1} \arc v_{h,a+1}	&\text{for $1 < a \le n_h - 1$ (if $n_h > 1$);}
	\end{brace}
	\qquad\\[2ex]
	\itiii	\;\mapsto
	\;&\;
	\begin{brace}
		\;\rme\;\;	v_{h,n_h-1} \arc v_{j,a}		&\;\:\,\text{for $n_h \le a \le n_j$, and}	\\
		\;\rmf\;\;	v_{j,a-1} \arc v_{h,n_h}		&\;\:\,\text{for $\max \{ n_h \,, 2 \} \le a \le n_j$ (if $n_j > 1$).}
	\end{brace}
	\qquad
\end{align}
\end{subequations}
We may refer to \rma\thru\rmf\ as \emph{rules} for inclusion of arcs in $\sD(n_1, \ldots, n_k)$.
In addition to these arcs, $\sD(n_1, \ldots, n_k)$ also contains arcs $v_{j,a} \arc v_{j,a+1}$ arising from orienting the paths $P_j$ themselves, and arcs $v_{j,a} \arc v_{j,a+2}$ from traversing two path edges in a row (corresponding to the adjacency relation $v_{j, a+2} \sim f(v_{j,a})$, which yields $v_{j,a} \inflRel v_{j,a+2}$).

\begin{lemma}
	\label{lemma:acyclicConstr}
	For any $n \ge k \ge 1$ and any integer partition $n_1 \le \cdots \le n_k$ of $n$, the digraph $\sD(n_1, \ldots, n_k)$ is acyclic.
\end{lemma}
\begin{proof}
	Let $D = \sD(n_1, \ldots, n_k)$ for the sake of brevity.
	All of the arcs in $D$ produced by the rules \rma\thru\rme\ are either of the form $v_{h,a} \arc v_{j,b}$ with $a < b$ and no
	constraints on $h$ and $j$\,, or $v_{h,a} \arc v_{j,a}$ with $h < j$\,.
	In either case, if an arc $v_{h,a} \arc v_{j,b}$ is of one of the types \rma\thru\rme, we have $(a,h) < (b,j)$ in the lexicographic ordering on ordered pairs of integers.
	The same also holds for the arcs $v_{h,a} \arc v_{h,a+1}$ and $v_{h,a} \arc v_{h,a+2}$ yielded by the paths $P_h$.
	Then, if there are arcs in $D$ of the form $v_{j,b} \arc v_{h,a}$ where $(b,j) > (a,h)$\,, they must arise from the rule~\rmf, in which case $a = n_h$\,.

	Note that none of the rules \rma\thru\rmf\ produce arcs which leave the final vertex $v_{h,n_h}$ of any path $P_h$\,, so there are no non-trivial walks in $D$ which leave such a vertex.
	Then, it is easy then to show by induction that if there is a directed walk in $D$ between distinct vertices $v_{h,a}$ and $v_{j,b}$\,, either $(a,h) < (b,j)$ in the lexicographic order, or $b = n_j$\,.

	Let $v_{h,a}$ and $v_{j,b}$ be two vertices, with a directed walk $W$ from $v_{h,a}$ to $v_{j,b}$\,.
	Because of the existence of $W$, we know that $a \ne n_h$\,; then, there is a directed walk from $v_{j,b}$ to $v_{h,a}$ only if $(b,j) < (a,h)$\,.
	We would then have $b = n_j$\,, in which case there are no directed walks from $v_{j,b}$ to \emph{any} other vertices in $D$\,.
	So, for any two distinct vertices $v_{h,a}$ and $v_{j,b}$\,, there cannot both be a directed walk $v_{h,a} \diwalk v_{j,b}$ and also a directed walk $v_{j,b} \diwalk v_{h,a}$\,.
	Thus, $D$ is acyclic.
\end{proof}
As well as giving rise to an acyclic digraph $D(G,P_1,\ldots,P_k)$\,, we also have:
\begin{lemma}
	\label{lemma:constrMaximal}
	$\big\lvert E\big(\sG(n_1,\ldots,n_k)\big) \big\rvert \,=\, kn - \binom{k+1}{2}$\,,\, for any $n \ge k \ge 1$ and integer
	partition $n_1 \le \cdots \le n_k$ of $n$.
\end{lemma}
\begin{proof}
	Between any pair of paths $P_h$ and $P_j$ in $\sG(n_1,\ldots,n_k)$\,, there are $n_h - 1$ edges of type
	\iti, $n_h - 1$ edges of type \itii, and $n_j - n_h + 1$ edges of type \itiii.
	There are then $n_h + n_j - 1$ non-flow edges between $P_h$ and $P_j$\,.
	This saturates the upper bound for non-flow edges between pairs of paths in Theorem~\ref{thm:extremalUpperBound}: summed over all pairs of paths, and including the edges in the paths $P_h$, the total number of edges in $\sG(n_1,\ldots,n_k)$ is then $kn - \binom{k+1}{2}$.
\end{proof}
Lemmas~\ref{lemma:acyclicConstr} and~\ref{lemma:constrMaximal}, together with Theorem~\ref{thm:extremalUpperBound}, prove:
\begin{theorem}
	\label{thm:extremalBound}
	$\Gamma(n,k) = kn - \binom{k+1}{2}$ for all integers $n \ge k \ge 1$.
	That is, for every $n \ge k \ge 1$ and any geometry with $\card{V(G)} = n$ and $\card{O} = k$, the geometry $(G,I,O)$ has a flow only if $m \le nk - \binom{k+1}{2}$, where $m = \card{E(G)}$; and there exist geometries $(G,I,O)$, \eg\ with $G = \sG(n_1, \ldots, n_k)$, for any integer partition $n_1 \le \cdots \le n_k$ of $n$, which saturate this bound.
\end{theorem}
In later sections, we will only need the result of Theorem~\ref{thm:extremalUpperBound}: the matching lower bound only serves to prove that no improvement in the upper bound is possible.

\subsection{Uniqueness in the case \protect{$\card I = \card O$}}
\label{sec:unique}

Influencing walks for path-covers are closely related to \emph{walks which alternate with respect to $\cP$}, as defined by Diestel~\cite{Diestel} in his presentation of a clever proof of Menger's Theorem (a forerunner of the Min-Flow/Max-Cut theorem) by~\cite{BGH2001}.
\begin{definition}[\protect{\cite[page~64]{Diestel}}]
	Let $\cP = \ens{P_1, \ldots, P_k}$ be a collection of vertex-disjoint $I$\thru$O$ paths $P_j = v_{j,1} \cdots v_{j,n_j}$.
	A walk $W = u_0 u_1 \cdots u_\ell$ \emph{alternates with respect to $\cP$} if it starts at a vertex in $I$ not covered by $\cP$\,, and if the following three conditions hold for all $0 \le j \le \ell$\,:
	\begin{romanum}
	\item
		if $u_j u_{j+1}$ is covered by $\cP$, then $u_j = v_{h,a+1}$ and $u_{j+1} = v_{h,a}$ for some $1 \le h \le k$ and $1 \le a < n_h$\,;
	\item
		if $u_j = u_h$ for some $h \ne j$, then $u_j$ is covered by $\cP$\,;
	\item
		if $u_j$ is covered by $\cP$ for some $0 < j < \ell$, then at least one of $u_{j-1} u_j$ or $u_j u_{j+1}$ is covered by $\cP$.
	\end{romanum}
\end{definition}
Such walks share with influencing walks the property that, for any vertex covered by $\cP$, at least one of the two edges incident to the vertex are covered by $\cP$ as well.
By relaxing the requirement that $W$ must start at a vertex not covered by $\cP$, we may also consider ``alternating walks'' with respect to path covers: it is easy to see that closed ``alternating walks'' of this sort correspond to vicious circuits whose arcs have been reversed.

The role that alternating walks play in the proof of Menger's Theorem in~\cite{BGH2001}, which we present as Lemma~\ref{thm:BGH2001} on page~\pageref{thm:BGH2001}, provides some intuition of the importance of the link with alternating walks.
For completeness, the statement of the Theorem is as follows (see \eg~\cite{Diestel}):
\begin{theorem*}[Menger's Theorem]
	Let $G$ be a graph and $I, O \subset V(G)$.
	Then the minimum size of a set $S$ such that $V \setminus S$ is disconnected, with $I$ and $O$ in different components, is equal to the maxmum number of vertex-disjoint $I$\thru$O$ paths in $G$.
\end{theorem*}
Alternating walks arise as a sort of ``difference'' of distinct collections of vertex-disjoint $I$\thru$O$ paths.\footnote{%
The role of ``alternating walks'' and ``augmenting walks'' for vertex-disjoint $I$\thru$O$ paths, phrased in terms of differences of sets of paths, is similar to the analogous notions for \eg\ matchings in graphs, and have a basis (so to speak) in matroid theory.
The analogy here is used here only for illustrative purposes: a thorough discussion of augmenting walks in the context of matroid theory is beyond the scope of this thesis.}
Specifically, an $I$\thru$O$ walk $W$ which alternates with respect to a collection of vertex-disjoint paths $\cP$ can be used to obtain a new collection $\cP'$ with strictly more paths by using $W$ as an \emph{augmenting walk}, as in a network flow:\footnote{%
For a discussion of augmenting walks in the context of network flows, see \eg~\cite{CLRS}.}
interpreting $W$ as a ``difference'' of the two families $\cP'$ and $\cP$, we can obtain $\cP'$ by ``adding'' $W$ to $\cP$.
These addition and subtraction analogies can be made formal by considering the skew-symmetric directed adjacency matrices $A$ and $A'$ of the families of paths $\cP$ and $\cP'$ as directed subgraphs of $G$, given by
\begin{gather}
	\label{eqn:adjDigraphForSubtraction}
		A_{uv}	\;=\;	\begin{cases}
								\;\;1,		&	\text{if $u \arc v$ is an arc of $\cP$}	\\
									-1,		&	\text{if $v \arc u$ is an arc of $\cP$}	\\
								\;\;0,		&	\text{otherwise}
		      	     	\end{cases}
\end{gather}
(and similarly for $A'$ and $\cP'$), and considering the digraphs yielded by taking sums and differences of these adjacency matrices when those sums/differences are $\ens{-1,0,1}$-matrices.

Maximal collections of vertex-disjoint paths do not have such ``augmenting'' walks, and path covers for geometries $(G,I,O)$ are by necessity a maximal collection of such paths.
A difference of two path covers would then not contain any augmenting walks.
However, it is not difficult to see that the symmetric difference will contain walks which are very much like alternating walks, except that they must begin and end at the same point, which is necessarily covered by the path cover.
This intuition suggests that path covers which lack vicious circuits should be unique.
In fact, we may prove something somewhat stronger:

\begin{theorem}
	\label{thm:uniqueFamilyPaths}
	Let $(G,I,O)$ be a geometry such that $\card{I} = \card{O}$.
	If $(G,I,O)$ has a causal path cover $\cC$, then $\cC$ is also the only maximum-size collection of vertex-disjoint dipaths from $I$ to $O$.
\end{theorem}
\begin{proof}
	Suppose that $\cC$ is a path cover for $(G,I,O)$ with successor function $f: O\comp	\to I\comp$, and suppose there is a maximum-size collection $\cF$ of vertex-disjoint	$I$ -- $O$ dipaths which differs from $\cC$.
	Let $S \subset V(G)$ be the set of vertices not covered by $\cF$: because $\card{\cF} = \card{\cC} = \card{I} = \card{O}$, we have $S \inter I = \vide$ and $S \inter O = \vide$, in which case $\cF$ is a path cover for the geometry $(G \setminus S, I, O)$.
	Let $\tilde f$ be the successor function of $\cF$ on that geometry: then $\tilde f$ is injective.
	Because $\card{I} = \card{O}$, $\tilde f$ will also be a bijection from $V(G) \setminus (O \union S)$ to $V(G) \setminus (I \union S)$.
	Then, let $\tilde g$ be the inverse function of $\tilde f$.
	
	Because $\cC$ and $\cF$ differ, there must exist a vertex $v \in O\comp$ such that $v \arc f(v)$ is not an arc of any path of $\cF$. Note also that for $z \in \dom(f)$, $f(z) \notin \dom(f)$ holds only if $f(z) \in O \setminus I \subset \dom(\tilde g)$; that is, $f(z) \in \dom(f) \union \dom(\tilde g)$.
	Then, define a vertex sequence $\paren{u_j}_{j \in \N}$ in $G$ by setting $u_0 = v$, $u_1 = f(v)$, and 
	\begin{align}
			\label{eqn:inflWalkRecurrance}
			u_{j+1}
		\;=&\;
			\begin{cases}
				\; f(u_j)	\;,			&\!\!	\text{$u_j \in S$ or $u_j \ne f(u_{j-1})$}	\\[.5ex]
				\; \tilde{g}(u_j)	\;,	&\!\!	\text{$u_j \notin S$ and $u_j = f(u_{j-1})$}
			\end{cases}
	\end{align}
	for all $j \ge 1$.
	We then have $u_j \sim u_{j+1}$ for all $j \in \N$, so this defines a directed walk $u_0 \arc u_1 \arc \cdots$ in $G$.
	Figure~\ref{fig:exampleInflConstr} illustrates this construction.

\begin{figure}[t]
	\begin{center}
		\includegraphics{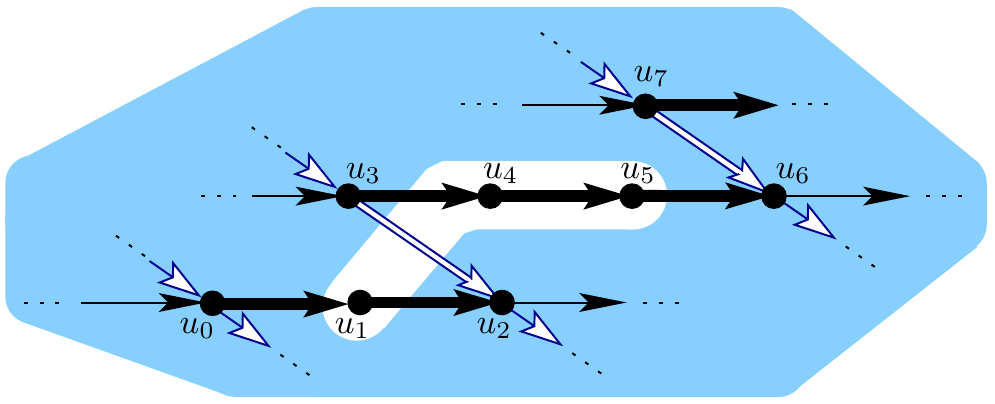}
	\end{center}
	\caption[Illustration of an unbounded influencing walk.]{\label{fig:exampleInflConstr}\colouronline
 		Illustration of an unbounded influencing walk, for a path cover $\cC$ (solid arrows), induced by the presence of another maximum-size collection $\cF$ of paths from $I$ to $O$ (hollow arrows).
		The shaded area represents the set of vertices covered by $\cF$, which in this illustration does not include all of the vertices.
		The arcs of $\cC$ and of $\cF$ which are thicker are those involved in the influencing walk.}
\end{figure}

	\begin{observation}
		\label{obs:consecutiveDistinct}
		Any three consecutive vertices $u_j, u_{j+1}, u_{j+2}$ are distinct.
		That $u_{j+1}$ differs from $u_j$ and $u_{j+2}$ follows from $u_j \sim u_{j+1}$ and $u_{j+1} \sim u_{j+2}$, as we do not permit $G$ to have self-loops.
		We have either $u_2 = f(u_1)$ or $u_2 = \tilde{g}(u_1)$, with the latter being equivalent to $\tilde{f}(u_2) = u_1$.
		In the former case, $u_0$ and $u_2$ lie along on a common path, so $u_0 \ne u_2$; and because $\tilde{f}(u_0) \ne f(u_0) = u_1$ by the choice of $u_0$, we have $u_2 \ne u_1$ in the latter case.
		We may induct similarly, if the proposition holds up to some particular $j \ge 0$:\
		\begin{itemize}
		\item
			Suppose $u_{j+3} = f(u_{j+2})$.
			If we also have $u_{j+2} = f(u_{j+1})$, then $u_{j+3}$ and $u_{j+1}$ lie along a common path of $\cC$.
			Otherwise, $u_{j+1} u_{j+2}$ is an edge not covered by $\cC$ while $u_{j+2} u_{j+3}$ is.
			In either case, $u_{j+3} \ne u_{j+1}$.

		\item
			Suppose $u_{j+3} = \tilde{g}(u_{j+2})$.
			Then $u_{j+2} \notin S$ and $u_{j+2} = f(u_{j+1})$; the latter of which implies that either $u_{j+1} \in S$ or $u_{j+1} = \tilde{g}(u_j)$.
			In the former case, $u_{j+1} \notin \img(\tilde{g})$, so that $u_{j+1} \ne u_{j+1}$\,; and if $u_j = \tilde{f}(u_{j+1})$, we have
			\begin{gather}
				u_{j+3} = \tilde{g}(u_{j+2}) \ne \tilde{g}(u_j) = u_{j+1}	\;.
			\end{gather}
		\end{itemize}
		In any case, $u_{j+1}$, $u_{j+2}$, and $u_{j+3}$ all differ, so the claim holds by induction.
		\end{observation}

	\begin{observation}
		For any $\ell \ge 0$, the walk $W_\ell = u_0 \diwalk u_\ell$ is an influencing walk.
		By definition, $u_0 \arc u_1$ is an arc in some path of $\cC$, so this is an influencing walk, as is the trivial walk on $u_0$.
		Otherwise, suppose that for some particular $\ell \ge 0$ we have both $W_\ell = u_0 \diwalk u_\ell$ and $W_{\ell+1} = u_0 \diwalk u_{\ell+1}$ are influencing walks.
		\begin{itemize}
		\item
			If $u_\ell \arc u_{\ell+1}$ is not an arc of $\cC$, then we have $u_{\ell+1} \ne f(u_\ell)$, in which case we have $u_{\ell + 2} = f(u_{\ell+1})$.
			Similarly, if $u_{\ell+1} \in S$, we also have $u_{\ell 2 1} = f(u_\ell)$.
			In either case, $u_{\ell+1} \arc u_{\ell+2}$ is a segment of type~\iti.

		\item
			If $u_\ell \arc u_{\ell+1}$ is an arc of $\cC$ and $u_{\ell+1} \notin S$, then we have $u_{\ell + 2} = \tilde{g}(u_{\ell+1})$.
			Because $u_{\ell + 2} \ne u_\ell$ by Observation~\ref{obs:consecutiveDistinct}, the arc $u_{\ell+1} u_{\ell+2}$ is not an arc covered by $\cC$, in which case $u_\ell \arc u_{\ell+1} \arc u_{\ell + 2}$ is a segment of type~\itii.
		\end{itemize}
		Then, $W_{\ell + 2} = u_0 \diwalk u_{\ell + 2}$ is either a concatenation of $W_{\ell + 1}$ with a segment of type~\iti, or a concatenation of $W_\ell$ with a segment of type~\itii, and is in either case also an influencing walk.
		The claim then holds by induction.	
	\end{observation}
	Because $G$ is a finite graph, the Pigeon Hole Principle implies that there must be integers	$\ell, \ell' \in \N$ with $\ell < \ell'$, $u_\ell = u_{\ell'}$, and $u_{\ell+1} = u_{\ell'+1}$.
	Because $W_{\ell'+1}$ is an influencing walk, either $u_\ell$ is at the beginning of a segment, or $u_{\ell + 1}$ is (or both).
	if $u_\ell$ is at the beginning of a segment, the closed walk $u_\ell \diwalk u_{\ell'}$ is a concatenation of segments from $W$, and so is a closed influencing walk for $\cC$.
	Otherwise, $u_{\ell+1} \diwalk u_{\ell'+1}$ is a concatenation of segments from $W$, and is in this case a closed influencing walk for $\cC$.
	In either case, there exists a vicious circuit for $\cC$, in which case $\cC$ is not a causal path cover.

	Thus, if $\cC$ is a causal path cover, there can be no such vertex sequence $\paren{u_j}_{j \in \N}$ as defined above, and so there can be no maximum family of vertex-disjoint $I$ -- $O$ paths $\cF$ which differs from $\cC$.
	$\cC$ is then the unique such family of paths.
\end{proof}

In the case where $\card{I} = \card{O}$, \ie\ for geometries of one-way patterns corresponding to unitary circuits without measurements or trace-out, we may then reduce the problem of finding a flow to that of finding an arbitrary collection $\cC$ of vertex-disjoint $I$\thru$O$ paths of maximum size.
If this collection of paths is not a path cover, we know that any path cover for $(G,I,O)$ cannot be causal, and so the geometry has no flow; otherwise, we may test the successor function $f$ of $\cC$ to see if it is a flow function, which by Lemma~\ref{lemma:causalOrderFlow} and Theorem~\ref{thm:findPreorder} (page~\pageref{lemma:causalOrderFlow} and following) can be determined in polynomial time.
This will allow us to prove the following in Section~\ref{sec:flowAlgs}:
\begin{theorem}
	Let $(G,I,O)$ be a geometry with $\card{I} = \card{O} = k$ and $\card{V(G)} = n$.
	Then there is an algorithm which determines that $(G,I,O)$ does not have a flow, or constructs a flow $(f, \preceq)$ for $(G,I,O)$ such that $\preceq$ is the coarsest causal order for $f$, in time $O(k^2 n)$.
\end{theorem}

\section{Flow-finding algorithms}
\label{sec:flowAlgs}

Throughout this section, we will suppose that we are given as input some particular geometry $(G,I,O)$, and define $n = \card{V(G)}$, $m = \card{E(G)}$, and $k = \card{O}$.
In some parts of our analysis, we will further suppose that $\card{I} = \card{O}$.

\paragraph{Partial function data types.}

A mutable variable for a partial function $f$ between two sets $A$ and $B$ (which we will denote $f: A  \partialto B$) will be represented as arrays $f: A \to B \union \ens{\nil}$, where $\nil$ is a reserved value not contained in any set of vertices or indices.
The domain of $f$ is then the set of $x \in A$ for which $f(x) \ne \nil$.
We will also adopt the convention that a partial function $f$ is injective when $\; f(x) = f(y) \ne \nil \;\iff\; x = y \;$ holds for all $x, y \in A$.

\paragraph{Set data types.}

For a mutable set variable $S$, we will consider to be implemented using a mixed array/linked list structure, where an array entry $S(v)$ either contains a value $\nil$ when $v \notin S$, or a node with links to at most two other nodes, representing arbitrarily designated ``previous'' and ``next'' elements of the set (using $\nil$ if one or both of these are not defined) when $v \in S$.
A specially designated master node is kept to point to the ``first'' element of the set, or $\nil$ if the set is empty: newly added elements of the set may be inserted to the beginning of this list of nodes.
This will allow for constant-time elementhood testing, element addition and removal, and comparison with the empty set, and iteration across the set can be done in $O(\card{S})$ steps; it also provides a total ordering of $S$ if required.

\subsection{Eliminating geometries with too many edges}
\label{sec:eliminatingAbundant}

As we proved in Theorem~\ref{thm:extremalUpperBound}, if $G$ has more than $kn - \binom{k+1}{2}$ edges, then $(G,I,O)$ cannot have a flow.
Then, as a preliminary step, we may reject any geometry which has more than this number of edges, so we may assume that $m \in O(kn)$for the purposes of run-time analysis.
(As we proved in Lemma~\ref{lemma:constrMaximal}, the bound of $m \le kn - \binom{k+1}{2}$ is tight for geometries with flows.)

\subsection{Finding a potential flow-function $f$ when $\card I = \card O$}

In the case where $\card I = \card O$, the uniqueness result of Theorem~\ref{thm:uniqueFamilyPaths} allows us to reduce finding a causal path cover to finding an arbitrary maximum-size vertex-disjoint family of $I$\thru$O$ paths.
This problem may be reduced to an instance of network flow, by finding a maximum integer flow in a digraph $N$ with unit edge capacities, constructed from $G$ so that edge-disjoint paths in $N$ correspond to vertex-disjoint paths in $G$~\cite{CH91}.
Here, we describe a different algorithm, based on the proof of the lemma in~\cite{BGH2001} (reproduced here with minor changes in order to facilitate description of the algorithm) used to prove Menger's Theorem :
\begin{lemma}[\protect{\cite[Theorem 3]{BGH2001}}]
	\label{thm:BGH2001}
	Let $G$ be a digraph, $I, O \subset V(G)$ such that $I$ cannot be separated from $O$ by a set of $\kappa$ or fewer vertices, for some $\kappa \in \N$.
	If $\cW \subset G$ is a subgraph consisting of $\kappa$ vertex-disjoint $I$\thru$O$ paths in $G$, there is a subgraph $\cU \subset G$ consisting of $\kappa+1$ disjoint $I$\thru$O$ paths, such that the terminal points of the paths of $\cU$ contain those of $\cW$.
\end{lemma}
\begin{proof}
	We will induct on supersets of $O$ contained in the vertex-set $V(G)$.
	If $O = V(G)$, then $I \subset O$, in which case $\cU = G[I]$ contains at least $\kappa+1$ (trivial) paths from $I$ to $O$.
	Otherwise, we suppose that the Lemma holds for all proper supersets $O' \prsupset O$.

	Let $R = O \inter V(\cW)$: as $\card R = \kappa$\,, removing $R$ from $G$ does not suffice to separate $I$ from $O$, in which case the graph $G \setminus R$ contains a path $P$ from $I$ to $O$.
	If $P$ is disjoint from $\cW$, we may let $\cU = \cW \union P$.
	Otherwise, let $x$ be the final vertex on $P$ which intersects $\cW$, and let $W$ be the path of $\cW$ containing $x$.

	Let $P_1$ be the subpath of $P$ from $x$ to $O$, $W'$ be the subpath of $W$ from $I$ to $x$, and $P_2$ be the subpath of $W$ from $x$ to $O$.
	We may augment the set $O$ by defining $O' = O \union V(P_1) \union V(P_2)$, and correspondingly reduce the path $W$ by defining $\cW' = (\cW \setminus W) \union W'$\,; then $\cW'$ consists of $\kappa$ paths from $I$ to $O'$.
	As $O \prsubset O' \subset V(G)$, by hypothesis there exists a subgraph $\cU' \subset G$ consisting of $\kappa + 1$ vertex-disjoint paths, such that the terminal points of $\cU'$ contain those of $\cW'$.

	Let $\omega \in O'$ be the terminal point in $\cU'$ which is not covered by $\cW'$, and let $z$ be the endpoint of $W$ in $O$.
	If $\omega$ is a vertex of $O \setminus \ens{z}$, then $P_2$ extends a path of $\cU'$ from $x$ to $z$; then $\cU = \cU' \union P_2$ covers $\omega$, as well as all of the vertices of $O$ covered by $\cW$.
	Otherwise, we have $\omega \in V(P_1 \union P_2)$ and $\omega \ne x$, and there is an $\omega$\thru$O$ subpath $Q \subset P_j$ (for one of $j = 1$ or $j = 2$).
	Either $P_1$ or $P_2$ will extend a path of $\cU'$ from $x$ to $O$, and $Q$ will extend a path of $\cU'$ from $\omega$ to $O$.
	Then, $\cU = \cU' \union P_{3-j} \union Q$ covers all of the vertices of $O$ covered by $\cW$, as well as the endpoint of $P$ in $O$.
\end{proof}

The following few sections will concern a translation of this proof to a depth-first search along alternating walks.

\subsubsection{Translation to depth-first search}

The connection between the induction proof above of Theorem~\ref{thm:BGH2001} and walks which alternate with respect to $\cW$ can be made as follows:
\begin{itemize}
\item
	If $P$ is disjoint from $\cW$, it (trivially) alternates with respect to $\cW$.

\item
	If $P$ intersects $\cW$ at a vertex not in $O$, but we have $\omega \in O \setminus \ens{z}$, this corresponds to a case where an $I$\thru$O$ path is found which is totally independent (\ie\ disjoint) from $P$, which may be found \eg~after a search from the terminal point of $P$ has been exhausted.
	Similarly, if $\omega \in V(P_1) \setminus \ens{x}$, this corresponds to a case where an $I$\thru$O$ path is found which intersects $P_1$ at some internal vertex, which may be found \eg~after a search through $x$ has been exhausted.

\item
	If $P_1$ intersects $\cW$ and $\omega \in V(P_2) \setminus \ens{x}$, this corresponds to a case where $\omega$ is an intermediary 
	point of an alternating walk which backtracks (by at least one edge) along $W$ from $\omega$ to $x$, and then progresses along $P_1$ to meet $O$.
	In this case, the induction hypothesis essentially reduces the problem to finding a vertex $\omega$ along $P_2$ such that there is an alternating walk with respect to $\cW'$ from $I$ to $\omega$.
\end{itemize}
In any case, the alternating walk may be used to construct the new family of paths $\cU$ incrementally, by including and excluding path segments as in the induction steps of the proof of Theorem~\ref{thm:BGH2001} above.
This process is illustrated in Figure~\ref{fig:entryPoint}.

\begin{figure}[p]
	\vspace{-2.5ex}
	\begin{center}
	\includegraphics[height=0.82\textheight]{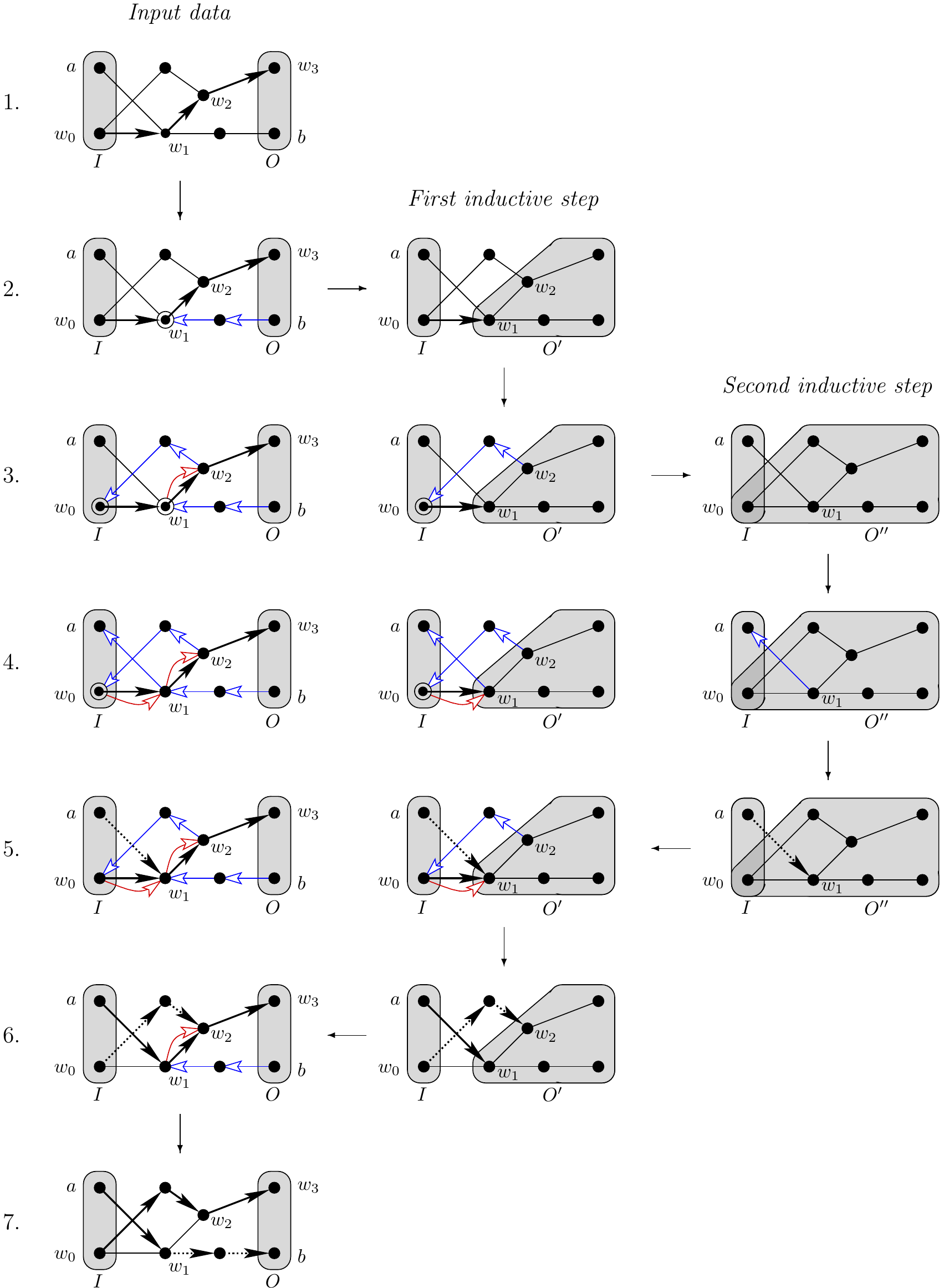}
	\end{center}
	\vspace{-2ex}
	\caption[Illustration of a depth-first search along alternating walks, based on the proof of Theorem~\ref{thm:BGH2001}.]{\label{fig:entryPoint}\colouronline
		Illustration of a depth-first search along alternating walks, based on the proof of Theorem~\ref{thm:BGH2001}.
		Each column corresponds to a successive reduction of the problem by augmenting the set $O$, as in the inductive step: arrows between the diagrams represent the problem reduction suggested by the proof.
		Steps~1\thru3 consist of a search for a path $P$, which stops when some $a \in I$ or a pre-existing path $W$ is encountered. 
		This search in the reduced problems corresponds to a search along alternating walks (indicated here with hollow arrows), illustrated in the left-most column.
		A solution for a reduced problem is extended in the parent problem, corresponding to backtracking along the walk: this amounts to ``subtracting'' the alternating walk from the existing family of paths, as described in the paragraph leading up to the equation~\eqref{eqn:adjDigraphForSubtraction}.
	}
\end{figure}

The fact that this walk should be an alternating walk is not the main concern, but rather that $\cU$ can be obtained at all by searching along a walk with simple structure.
We will proceed by describing the reductions that will make it easy to prove the correctness of a depth-first search.

We may refine the proof of Theorem~\ref{thm:BGH2001} by extending the set $O \mapsto O'$ regardless of whether $P$ intersects $\cW$.
In particular: the question of whether there is an $I$\thru$O$ path $P = p_0 \cdots p_{\ell - 1} p_\ell$, which is either disjoint from $\cW$ or such that $p_j \notin V(\cW)$ for some range of consecutive vertices $r < j \le \ell$, can be reduced to whether there is a similar $I$\thru$O'$ path $P' = p_0 \cdots p_{\ell-1}$, for $O' = O \union \ens{p_{\ell-1}}$.
Similarly, the question of whether there is an $I$\thru$O$ path $P = p_0 \cdots x p_\ell$ such that $x \in V(\cW)$ can be reduced (as it is in the induction step of Theorem~\ref{thm:BGH2001}) to the question of whether there exists a path $P'$ from $I$ to some vertex of the set $O' = O \union \ens{x, x', x'', \ldots}$, where $x x' x'' \cdots$ is the tail of the path $W$ in $\cW$ which covers $x$; we also require that $P'$ does not end at $x$ itself, as this is covered by the truncated path $W' \subset W$ ending at $x$.
The depth-first search may then be described in terms of the valid ways in which we may perform an augmentation $O\supp{r} \mapsto O\supp{r+1}$, in order to reduce the search for a path $P\supp{r}$ from $I$ to $O\supp{r}$ to a search for a path $P\supp{r+1}$ from $I$ to $O\supp{r+1}$.
\label{discn:reductionRefinement}%
To re-iterate, the valid extensions would be:
\begin{enumerate}
\item
	For any $v \in O\supp{r}$, extend to $O\supp{r+1} = O\supp{r} \union \ens{w}$ for any $w \sim v$ such that $w \notin V(\cW\supp{r})$ and $w \notin O\supp{r}$.
	We then attempt to find $\cU\supp{r+1}$ which covers $w$, and the terminal points of $\cW\supp{r+1} = \cW\supp{r}$.

\item
	For any $v \in O\supp{r}$, extend to $O\supp{r+1} = O\supp{r} \union \ens{x, f(x), f^2(x), \ldots}$, for any $x \in V(\cW\supp{r}) \setminus O\supp{r}$, where $f$ is the ``successor'' function mapping each vertex covered by $\cW$ to the one which follows it (orienting the paths towards $O$).
	If $W$ is the path of $\cW\supp{r}$ on which $x$ lies, let $W$ be the segment of that path from $I$ to $x$; we then attempt to find $\cU\supp{r+1}$ which covers the terminal points of $\cW\supp{r+1} = \cW\supp{r} \symdiff \ens{W, W'}$, and one of the vertices $f^\ell(x)$ for $\ell \ge 1$.
\end{enumerate}
The path families $\cW\supp{r}$ we can keep track of by simply ``ignoring'' any arcs of $\cW$ contained within $O\supp{r}$, as such edges are guaranteed not to be in $\cW\supp{r}$ by construction.
The extended set $O\supp{r}$ will be the original set of output vertices $O$ together with the vertices traversed by the depth-first search to that point.

We will not keep track of of $O\supp{r}$ explicitly, but instead by ``marking'' most of the vertices traversed in the walk (which we may also use to avoid non-terminating recursive loops).
The exceptions, those vertices which are traversed but not marked, are what we will call \emph{entry points}.
Let us call a vertex $x$ an \emph{entry point} of the search into a path of $\cW$ if the edge $v x$ is not covered by $\cW$, where $v$ is the previous vertex in the search.
(Examples where a depth-first search encounters entry points are illustrated in Figure~\ref{fig:entryPoint}, where entry points are circled at each stage of the algorithm where they have been traversed only once: see \eg\ the vertex $w_1$ in the left column, in steps $2$\thru$3$.)
When such a vertex is discovered in the $r\th$ problem reduction for some $r \ge 0$, the set $O\supp{r}$ is augmented to a set $O\supp{r+1} = O\supp{r} \union \ens{x, f(x), \ldots}$; however, $x$ is then an element of $O\supp{r+1}$ covered by $W\supp{r+1}$ (specifically, by the path $W$ truncated after $x$), and is not a valid vertex from which to search for an augmented family of paths $\cU\supp{r+1}$.
Nevertheless, a later problem reduction may truncate the path $W'$ on which $x$ lies, in which case it becomes uncovered in that problem reduction, and a valid vertex from which to later perform a depth-first search --- in particular, it may be necessary to traverse $x$ a second time.
(Figure~\ref{fig:entryPoint} illustrates an example where visiting an entry point twice, $w_1$ in that case, is necessary to find the augmented family of paths.)
Because of this special behavior of entry points, we do not mark them when they are first traversed, in order to allow a second traversal if necessary.
However, we may still identify vertices of $O\supp{r}$ as vertices which are in $O$, vertices which are marked, or vertices covered by $\cW$ which immediately precede marked vertices (the latter case representing entry points).

\label{discn:depthFirst}
To implement a depth-first search, we will be interested in describing the problem reductions described above in terms of local transitions in the graph $G$.
We may define
\begin{subequations}
\begin{align}
		f(v)
	\;\;=&\;\;
		\begin{cases}
			\nil\,,	&	v \in O \text{~or $v$ not covered by $\cW$}	\\
			w\,,		&	v \arc w \text{~an arc of $\cF$}
		\end{cases}	\;,
	\\[1ex]
		g(w)
	\;\;=&\;\;
		\begin{cases}
			\nil\,,	&	w \in I \text{~or $v$ not covered by $\cW$}	\\
			v\,,		&	v \arc w \text{~an arc of $\cF$}
		\end{cases}.
\end{align}
\end{subequations}
In a slight abuse of the original terminology, we call $f$ the \emph{successor function} of $\cW$ (which will be a family of vertex-disjoint paths, albeit perhaps not a path cover); we similarly call $g$ the \emph{predecessor function} of $\cW$.
A transition which will be permitted for the depth-first search, in the $r\th$ problem reduction, will be one from a vertex $v \in O\supp{r}$ not covered by $\cW\supp{r}$ to a vertex $w$ satisfying one of the following conditions:
\begin{romanum}
\item
	$v \sim w$, $w$ is not covered by $\cW$, and $w \notin O\supp{r}$: this corresponds to the first reduction in the discussion above.
	The latter two conditions are equivalent to $w$ not covered by $\cW\supp{r}$, and $w$ not a marked vertex.

\item
	$w = f(x) \ne \nil$ for some $x \sim v$ on some path of $\cW$ with $x \notin O\supp{r}$, and where $v x$ is not covered by a path of $\cW\supp{r}$: this corresponds to the first two vertices in the second reduction in the discussion above.
	The conditions that $x$ be covered by $\cW\supp{r}$ and $x \notin O\supp{r}$ is equivalent to $w = f(x)$ being well defined (as we already require), and neither $x$ nor $w$ being marked.

\item
	$w = f(v) \ne \nil$, and $w$ on some path of $\cW\supp{r}$ --- in particular, where either $w \notin O\supp{r}$, or $w \in O\supp{r}$ is the vertex at the end of some path $W \in \cW\supp{r}$: this corresponds to all but the first two vertices in the second reduction of the discussion above.
	In the latter case, $w$ was an entry point of some ancestral problem reduction, in which case $w$ is unmarked, but $f(w)$ is well defined and marked.
	Then, this case is equivalent to $w = f(v) \ne \nil$ and $w$ not a marked vertex.

	In order to perform the second reduction properly, \ie\ all at once as described, this transition must be performed preferentially when it is valid; the reduction $O\supp{r} \mapsto* O\supp{r+1}$ is only complete when the endpoint of the path $W \in \cW\supp{r}$ has been reached, and other transitions explored from that vertex.
\end{romanum}
We mark a vertex when it plays the role of $w$ in such a local transition (in particular, leaving any entry point $x$ initially unmarked).
In each case, we need only consider the neighbors $z \sim v$ which have not yet been marked, such that either $z \notin O$ or $z = f(v)$; in the case where $z$ is an entry point, we further require that $w = f(z)$ has not been marked.
Then, in the depth-first search, it will suffice at each step to explore the unmarked vertices $z$ such that either $z = f(v) \ne \nil$, or such that $z \sim v$ and $z \notin O$.

Algorithms~\ref{alg:iterFindPaths} and~\ref{alg:augmentPaths} together describe an iterative procedure to build a maximum-size family of vertex-disjoint paths, based on this reduction to a depth-first search; where sets of disjoint paths $\cW$ are represented by successor/predecessor functions and set consisting of those vertices not covered by $\cW$.
In the following pages, we use the reduction just described to prove their correctness.

\subsubsection{Overview of Algorithm~\ref{alg:iterFindPaths}}

\begin{algorithm}[t]

	\PROCEDURE \ObtainMaxFamilyPaths(G,I,O):

	\Input{%
		$G$, a graph with subsets $I, O$.
	}

	\Output{%
		successor/predecessor functions $f,g: V(G) \partialto V(G)$ for a maximum-size family $\cP$ of vertex-disjoint $I$\thru$O$ paths
	}

	\BEGIN
		\LET $f,g: V(G) \partialto V(G)$\;
		\LET $\marked: V(G) \to \N$\;
		\LET $\iter \gets 0$\;
		\FORALL $v \in V(G)$ \DO																\nllabel{line:initPathsLoop}
			$\marked(v) \gets 0$\;
			$f(v) \gets \nil$;~
			$g(v) \gets \nil$\;
		\ENDFOR																						\nllabel{line:initPathsLoopEnd}
		\REPEAT																						\nllabel{line:buildPathsLoop}
			$\INCR \iter$\;
			\LET $\foundpath \gets \false$\;
			\FORALL $v \in O$ \SUCHTHAT $v \notin I$ \ANDALSO $g(v) = \nil$ \DO	\nllabel{line:iterSearchLoop}
				$\AlternPathSearch(v)$\;
				\oneline
				\IF $\foundpath$ \THEN \ESCAPE\;
			\ENDFOR																					\nllabel{line:iterSearchLoopEnd}
		\UNTIL {$\foundpath = \false$}														\nllabel{line:buildPathsLoopEnd}
		\RETURN $(f,g)$
	\END

	\caption[An iterative algorithm to obtain a maximum-size family of vertex-disjoint $I$\thru$O$ paths.]{\label{alg:iterFindPaths}%
		An iterative algorithm to obtain a maximum-size family of vertex-disjoint $I$\thru$O$ paths.
		The subroutine $\AlternPathSearch$ is described in Algorithm~\ref{alg:augmentPaths} on page~\pageref{alg:augmentPaths}.}
\end{algorithm}

The procedure $\ObtainMaxFamilyPaths$ described in Algorithm~\ref{alg:iterFindPaths} above uses a subroutine $\AlternPathSearch$ (described in Algorithm~\ref{alg:augmentPaths}, on page~\pageref{alg:augmentPaths}) to obtain larger and larger families of vertex disjoint paths, until a maximum size family is found.
Before turning to the description of the more detailed subroutine $\AlternPathSearch$, we will describe Algorithm~\ref{alg:iterFindPaths} as a whole.

As noted before, we will represent a family of vertex-disjoint paths $\cW$ using partial functions $f$ and $g$, which will store the successor and predecessor functions of $\cW$.
In the loop on lines~\ref{line:initPathsLoop}\thru\ref{line:initPathsLoopEnd}, the partial functions are initialized to represent a family of paths consisting only of trivial paths $v \in I \inter O$ (without loss of generality, we may assume that $\cW$ contains the trivial paths on whatever elements there may be in $I \inter O$): we will consider a vertex $v$ to be covered by $\cW$ if and only if at least one of $v \in I \inter O$, $f(v) \ne \nil$, or $g(v) \ne \nil$ holds.
Also defined and initialized are an array $\marked$ and a counter $\iter$: as we construct larger families of paths, $\iter$ will represent the iteration of the construction, and $\marked$ will keep track of the most recent iteration that each vertex was marked in the depth-first search along alternating walks.
These variables will all be used in the subroutine $\AlternPathSearch$ as global variables, all of which (except for $\iter$) may be modified.

The iterative component of $\ObtainMaxFamilyPaths$ increments $\iter$, and defines a flag $\foundpath$ which will indicate whether the pre-existing family of paths $\cW$ represented by $(f,g)$ has been successfully augmented; this is also a global parameter which may be modified by $\AlternPathSearch$.
The loop on lines~\ref{line:iterSearchLoop}\thru\ref{line:iterSearchLoopEnd} iterates through all of the elements of $O$ which are not covered by $\cW$.
We will maintain the constraint that $f(v) = \nil$ for all $v \in O$: then, $v \in O$ is covered by $\cW$ if and only if $g(v) \ne \nil$ or $g(v) \in I$.

We will later show that so long as $\foundpath = \false$, the elements of $O$ which have been marked are those for which $\AlternPathSearch$ has been invoked, but has failed to find a path $P$ similar to that in the proof of Theorem~\ref{thm:BGH2001}.
We will also show that, once an augmenting walk is found, $\AlternPathSearch$ updates $(f,g)$ to represent a collection of vertex-disjoint paths $\cU$ which covers the terminal points of $\cW$ and one additional vertex in $O$, and that $\foundpath$ is set to $\true$\,; we may then progress to the next iteration of the construction.
If $\foundpath$ retains the value $\false$, the variables $(f,g)$ will remain unchanged since the last iteration.
Finally, if all uncovered elements of $O$ are visited in an iteration without the value of $\foundpath$ changing, the pre-existing path family $\cW$ is of maximum size, and the variables $(f,g)$ representing it are returned.

Having described the data provided to, and the requirements made of, the subroutine $\AlternPathSearch$, we will turn to the description of it in Algorithm~\ref{alg:augmentPaths}.

\subsubsection{Analysis of Algorithm~\ref{alg:augmentPaths}}

\begin{algorithm}[p]

	\SUBROUTINE \AlternPathSearch(v):

	\Data {%
		$G$, a graph with subsets $I, O$;
		\\
		$f$ and $g$, successor/predecessor \fns\ for a family $\cW$ of disjoint $I$\thru$O$ paths;
		\\
		$\iter$, the current iteration number of path-family augmentations;
		\\
		$\foundpath$, a flag for indicating if an augmenting path was found.
	}

	\BEGIN
		$\marked(v) \gets \iter$\;

		\medskip
		\IF $v \in I$ \THEN {\nllabel{line:trivialPath}%
			$\foundpath \gets \true$\;
			\RETURN}{}
		\algline

		\medskip
		\LET $w \gets f(v)$\;																		\nllabel{line:testPathGlob-1}
		\IF $w \ne \nil$ \ANDALSO $\marked(w) < \iter$ \THEN {							\nllabel{line:testPathGlob-2}
				$\AlternPathSearch(w)$\;
				\IF \foundpath \THEN {\nllabel{line:testRemoveVertexPath}
					$f(v) \gets \nil$;~																
					$g(v) \gets \nil$\;																\nllabel{line:removeVertexPath}
					\RETURN
			}{}
		}

		\medskip

		\FOREACH $w \sim v$ \SUCHTHAT $w \notin O$ \ANDALSO $\marked(w) < \iter$ \DO						\nllabel{line:searchLoop}
			\algline
			\IF $f(w) = \nil$ \THEN {\nllabel{line:simpleRedux}%
				$\AlternPathSearch(w)$\;
				\IF \foundpath \THEN {%
					$f(w) \gets v$;~
					$g(v) \gets w$\;
					\RETURN
				}{}
			}
			\ELSE {\nllabel{line:entryPoint}%
				\algline
				\LET $x \gets w$;~
				$w \gets f(x)$\;
				\IF $\marked(w) < \iter$ \THEN {%
					$\AlternPathSearch(w)$\;
					\IF \foundpath \THEN {%
						$f(x) \gets v$;~															
						$g(v) \gets x$\;
						\RETURN
					}{}
				}{}
			}
		\ENDFOR																							\nllabel{line:searchLoopEnd}

	\medskip
	\RETURN
	\END

	\caption[A depth-first search subroutine to find an alternating walk from $I$ to $O$.]{\label{alg:augmentPaths}
		A depth-first search subroutine to find an alternating walk (with respect to a family of disjoint paths $\cW$) from $I$ to $O$, traversed in reverse.}
\end{algorithm}

The subroutine $\AlternPathSearch$ is described in Algorithm~\ref{alg:augmentPaths} (page~\pageref{alg:augmentPaths}).
We will analyze it in terms of the problem reductions which we derived on page~\pageref{discn:depthFirst} from the proof of Theorem~\ref{thm:BGH2001}.
In particular, we will consider an invocation of $\AlternPathSearch$ on a vertex $v$, representing the $r\th$ problem reduction (or part of it) for some $r \ge 0$.
(Note that even if $v \in O$ is not covered by $\cW$, we may have $r > 0$, if $v$ is not the first such vertex for which $\AlternPathSearch$ has been invoked in a given iteration.)

Recall that a vertex $z$ is marked iff $\marked(z) = \iter$: we will assume as in the analysis on page~\pageref{discn:depthFirst} that $O\supp{r}$ consists of vertices $z$ which are marked, vertices $x$ for which $f(x)$ is non-$\nil$ and marked, and the elements of $O$.
A truncated family of paths $\cW\supp{r}$ then corresponds to those vertices covered by $\cW$ which are not marked (and the edges covered by $\cW$ incident to those vertices).
Equivalently, we will say that the variables $(f,g)$ represent a family of vertex disjoint paths $\cW\supp{r}$ from $I$ to $O\supp{r}$ if the following hold:
\begin{romanum}
\item
	the vertices covered by $\cW\supp{r}$ are those which satisfy at least one of
	$v \in I \inter O\supp{r}$\,,
	$\big[ g(v) \ne \nil$ and $g(v) \notin O\supp{r} \big]$\,, or 
	$\big[ f(v) \ne \nil$ and $v \notin O\supp{r} \big]$\,;
	\vspace{0.5ex}

\item
	$\cW$ covers every edge $vw$ for which $v \notin O\supp{r}$, $w = f(v)$, and $v = g(w)$.
\end{romanum}
(The same variables $(f,g)$ may represent multiple families of paths from $I$ to $O\supp{r}$, corresponding to different levels $r$ of problem reduction.)

The first action performed by $\AlternPathSearch$ is to set $\marked(v) \gets \iter$, corresponding to the inclusion of $v$ into $O\supp{r}$ (and if $g(v) \ne \nil$, the inclusion of that vertex into $O\supp{r}$ as well).
If it happens that $v \in I$, then the trivial path on $v$ is a path from $I$ to $O\supp{r}$ which is disjoint from the paths of $\cW\supp{r}$\,; we may then define $\cU\supp{r} = \cW\supp{r} \union \ens{v}$, which covers $v$ and the terminal points of $\cW\supp{r}$, and which will already be represented by $(f,g)$.
We then set $\foundpath \gets \true$, and return to the previous recursive call.

If $v \notin I$, we then test on lines~\ref{line:testPathGlob-1}\thru\ref{line:testPathGlob-2} whether if $v$ is on a path and has an unmarked successor $w = f(v)$.
If so, we are in the midst of a problem reduction of the form $O\supp{r-1} \mapsto* O\supp{r-1} \union \ens{x, f(x), f^2(x), \ldots}$ for some preceding vertex $x$ on the same path as $v$.
To complete the reduction, we must accumulate any unmarked vertices $w, f(w), \ldots$ that there may be, and perform the depth-first search through them as we mark them for inclusion.
To do this, we recur the depth-first search at $w = f(v)$.

We will consider the effects of the conditional block starting on line~\ref{line:testRemoveVertexPath} further on: in any case, we regard all the operations from line~\ref{line:testRemoveVertexPath} onwards as being part of an $r\th$-level problem reduction for some $r \ge 0$, and the state of $\marked$ at this point to define $O\supp{r}$, any further changes made to $\marked$ by further recursive calls notwithstanding.

We require that at line~\ref{line:testRemoveVertexPath}, the variables $(f,g)$ represent a family of vertex disjoint paths $\cW\supp{r}$, and that $v \in O\supp{r}$ is not covered by $\cW\supp{r}$.
(In particular, this will be true when $v \in O$ is uncovered by $\cW$ and is the only vertex in $V(G)$ which is marked, \ie\ for $r = 0$).
We will then prove by induction that, after an invocation of $\AlternPathSearch(v)$ for $v$ not covered by $\cW\supp{r}$, the following hold if $\foundpath$ evaluates to $\true$:
\begin{romanum}
\item
	The variables $(f,g)$ represent a family of vertex-disjoint paths $\cU\supp{r}$ from $I$ to $O\supp{r}$, which covers the terminal points of $\cW\supp{r}$ and one other vertex;

\item
	If $v$ is not covered by $\cW$, then the additional vertex covered by $\cU\supp{r}$ is $v$, and $f(v)$ will remain unchanged;

\item
	The variables $f$ and $g$ may only change for vertices $z \in V(G) \setminus O\supp{r-1}$ (or for $z \in V(G)$ when $r=0$); 
	and they will remain unchanged for any vertex for which a recursive call $\AlternPathSearch(z)$ returns with $\foundpath = \false$.
\end{romanum}
Note that if $(f,g)$ represents a family of vertex disjoint paths $\cW\supp{r}$ which does not cover $v \in I$ at line~\ref{line:trivialPath}, the above conditions are satisfied; this forms the base case of the induction.

As we noted on page~\pageref{discn:depthFirst}, for variables $(f,g)$ representing a family of paths $\cW\supp{r}$ from $I$ to $O\supp{r}$, and for a vertex $v \in O\supp{r}$ which is not covered by $\cW\supp{r}$, the possible problem reductions can be obtained by considering the vertices which are iterated through on line~\ref{line:searchLoop} (of which there are at most $\deg(v)$ in number, and which can easily be iterated over provided an adjacency list representation of $G$).
We will now consider the reductions performed in the loop on lines~\ref{line:searchLoop}\thru\ref{line:searchLoopEnd}, together with the conditional block at line~\ref{line:testRemoveVertexPath}:
\begin{itemize}
\item
	For a vertex $w$ for which $f(w) = \nil$ (on line~\ref{line:simpleRedux}): we may attempt a problem reduction by defining $O\supp{r+1} = O\supp{r} \union \ens{w}$: to do this, we recur the depth-first search at $w$.
	If $\foundpath = \true$ after this reduction attempt, $(f,g)$ represents a new family of paths $\cU\supp{r+1}$ from $I$ to $O\supp{r+1}$ covering $w$ and the terminal points of $\cW\supp{r+1}$.
	Let $W' \in \cU\supp{r+1}$ be the path covering $w$\,: we may obtain a path $W$ covering $v$ by extending $W'$ by the edge $wv$.
	As $v$ is not covered by $\cW\supp{r+1}$, it is not covered by $\cU\supp{r+1}$, and so $\cU\supp{r} = \cU\supp{r+1} \symdiff \ens{W', W}$ is a vertex-disjoint family of paths covering $v$ and the terminal points of $\cW\supp{r}$.
	To represent $\cU\supp{r}$, we set $f(w) \gets v$ and $g(v) \gets w$.
	The only vertices for which $(f,g)$ are changed are then $v$ and those changed by the recursive call, which consist of $w$ and
	some vertices $z \in V(G) \setminus O\supp{r+1} \subset V(G) \setminus O\supp{r-1}$.

\item
	For a vertex $w = f(x)$ (on line~\ref{line:testRemoveVertexPath}): as noted before, if $w$ is well-defined, then we must perform a recursive call on $w$ to complete a reduction of the form $O\supp{r+1} = O\supp{r} \union \ens{x,f(x),\ldots}$ for some $x$ preceding $v$ and $w$ on a common path $P \in \cW\supp{r}$.
	In particular, such an $x$ is by definition an entry point, and as we will see in the next case, $\AlternPathSearch$ is not invoked on vertices which occur as entry points; then $v$ is distinct from $x$, and is not the end point of any path of $\cW\supp{r+1}$.

	If $\foundpath = \true$ after this reduction attempt, $(f,g)$ represents a new family of paths $\cU\supp{r+1}$ from $I$ to $O\supp{r+1}$ covering the terminal points of $\cW\supp{r+1}$, and one more vertex $\omega$ on the segment of $P$ starting at $w$ (in particular, we have $\omega \ne v$).
	Let $W' \in \cU\supp{r+1}$ be the path ending at $\omega$, and let $P'$, $P''$, and $P'''$ be the segment of $W$ ending at $x$, the segment of $W$ strictly bounded between $x$ and $\omega$ (and not including those endpoints\footnote{%
	Note that the path $P''$ may be an ``empty path'', consisting of zero vertices.}),
	and the segment of $W$ starting at $\omega$ respectively.
	By definition, $\omega$ is the first vertex from the end of $W$ for which the recursive call to $\AlternPathSearch$ returned with $\foundpath = \true$\,; in particular, for every vertex $z \in V(P''' \setminus \omega)$, the recursive call returned with $\foundpath = \false$.
	Then, the status of such vertices $z$ remain unchanged with respect to $(f,g,S)$\,: for each such vertex, we have $f(g(z)) = z$, and $g(f(z)) = z$ if $f(z) \ne \nil$.
	As well, by the analysis for the preceding case, the value of $g(\omega)$ will have changed to the vertex $\tilde{w}$ which precedes $\omega$ on $W'$, and $f(\omega)$ will remain unchanged.

	The reduction described in the proof of Theorem~\ref{thm:BGH2001} involves constructing a family of paths $\cU\supp{r}$ by extending the path $W'$ from $I$ to $O\supp{r+1}$ to a path $W = W' P'''$ from $I$ to $O\supp{r}$, and extending the path $P'$ from $I$ to $O\supp{r+1}$ by a single edge (as $x$ must in this case be adjacent to a vertex $\tilde{v} \in O\supp{r}$ which is not covered by $\cW\supp{r}$, from which $x$ was reached in the depth-first search).
	In particular, the vertices of $P''$ are not covered by $\cU\supp{r+1}$, and are not elements of $O\supp{r}$.
	In order to represent $\cU\supp{r}$, it is then necessary to remove the vertices $z \in V(P'')$ from the domains of $f$ and $g$.
	Therefore, we set $f(v) \gets \nil$ and $g(v) \gets \nil$\,; by induction on the length of $P''$, this will then be performed for all vertices in $V(P'')$.
	The only vertices for which $(f,g)$ have changed are then $\omega$, the vertices of $P''$ from $v$ onwards, and some vertices $z \in V(G) \setminus O\supp{r} \subset V(G) \setminus O\supp{r-1}$.

\item
	For a vertex $w$ for which $f(x) \ne \nil$ (on line~\ref{line:entryPoint}): because of the recursive call in the conditional block on line~\ref{line:testRemoveVertexPath}, the vertex $f(v)$ has already been marked, if it is defined; because $w$ is unmarked, we know that $w \ne f(v)$.
	Then either $w$ is an entry point, or $w = g(v)$.
	We relabel the possible entry point $w$ as $x$, overwrite $w \gets f(x)$, and test whether $w = f(x)$ has been marked.
	If not, then we have $x \ne g(v)$, as $f(g(v)) = v$ is a marked vertex; in this case, $x$ is indeed an entry point.
	We may then attempt a problem reduction by defining $O\supp{r+1} = O\supp{r} \union \ens{x, f(x), f^2(x), \ldots}$: to do this, we recur the depth-first search at $w$.
	(Marking $w$ will implicitly include both $x,w \in O\supp{r+1}$\,; the test at line~\ref{line:testRemoveVertexPath} ensures that all further vertices along the same path, up to its endpoint in $O\supp{r}$, will be immediately included as well.)

	If $\foundpath = \true$ after this reduction attempt, $(f,g)$ represents a new family of paths $\cU\supp{r+1}$ from $I$ to $O\supp{r+1}$ covering the endpoints of $\cW\supp{r+1}$, which include $x$, and one more vertex $f^\ell(x)$ for some $\ell \ge 1$.
	By the analysis of the preceding case, the variables $(f,g)$ also represent a family of paths $\cU'$ from $I$ to $O\supp{r} \union \ens{x}$, covering the terminal points of $\cW\supp{r}$ as well as the additional vertex $x$.
	Let $W' \in \cU'$ be the path ending at $x$\,: we may obtain a path $W$ covering $v$ by extending $W'$ by the edge $xv$.
	As $v$ is not covered by either $\cU\supp{r+1}$ or $\cW\supp{r}$ (and in particular not by the path $P \in \cW\supp{r}$ which covers $x$), $\cU\supp{r} = \cU' \symdiff \ens{W, W'}$ is a vertex-disjoint collection of paths covering $v$ and the terminal points of $\cW\supp{r}$.
	To represent $\cU\supp{r}$, we set $f(x) \gets v$ and $g(v) \gets x$.
	The only vertices for which $(f,g)$ are changed are then $v$, the vertices $O\supp{r+1} \setminus O\supp{r} = \ens{x, f(x), f^2(x), \ldots}$ for which the recursive call to $\AlternPathSearch$ returned with $\foundpath = \true$, and some vertices $z \in V(G) \setminus O\supp{r+1} \subset V(G) \setminus O\supp{r-1}$.
\end{itemize}
It is easy to verify that if the preconditions are satisfied that $(f,g)$ represent a family of vertex disjoint paths $\cW\supp{r}$ from $I$ to $O\supp{r}$, and that $v \in O\supp{r}$ is not the endpoint of such a path, that they are also satisfied for each recursive call.
Then, by induction, if initially the variables $(f,g)$ represents a family $\cW\supp{r}$ of vertex disjoint $I$\thru$O\supp{r}$ paths, $v \in O$ is not covered by $\cW\supp{r}$, and $\AlternPathSearch(v)$ returns with $\foundpath = \true$, then afterwards $(f,g)$ represents a family $\cU$ of vertex-disjoint $I$\thru$O\supp{r}$ paths covering $v$ and the terminal points of $\cU$.
If such a collection of paths $\cU$ exists, then it is the only family of vertex disjoint $I$\thru$O$ paths larger than $\cW$ in the a graph $G \setminus S$ where we define
\begin{align}
	S \;=\; \ens{z \in O\supp{r} \,\Big|\, \Big(\, g(z) = \nil \text{~or~} g(z) \in O\supp{r} \,\Big) \text{~and~} z \ne x}	\;,
\end{align}
\ie\ where we remove every element of $O\supp{r}$ not covered by $\cW$ \emph{except} for $x$.
Because Algorithm~\ref{alg:augmentPaths} reduces this problem to finding an appropriate family of paths ending in at least one of its neighbors, Theorem~\ref{thm:BGH2001} guarantees that $\AlternPathSearch(v)$ will terminate with $\foundpath = \true$ in the case that such a family of paths $\cU$ does exist.

\subsubsection{Analysis of Algorithm~\ref{alg:iterFindPaths}}

Because a call to $\AlternPathSearch(v)$ returns with $\foundpath = \true$ if and only if either $v \in I$ or a recursive call returns with $\foundpath = \true$, and because $(f,g)$ are only changed when such a recursive call returns with $\foundpath = \true$, we know that they will be unchanged if $\AlternPathSearch(v)$ returns with $\foundpath = \false$.
If this occurs for an invocation $\AlternPathSearch(v)$ where initially no vertices were marked, the loop on lines~\ref{line:iterSearchLoop}\thru\ref{line:iterSearchLoopEnd} of Algorithm~\ref{alg:iterFindPaths} will select another element $v' \in O$ such that $g(v') = \nil$, and invoke $\AlternPathSearch(v')$.
In this case, the data $(f,g)$ represent $\cW$ as a family of paths from $I$ to $O'$, for some superset $O' \supset O$ such that, for all $w \in O' \setminus O$, $w$ is not itself the endpoint of any family of $I$\thru$O'$ vertex-disjoint paths $\cU'$ which also covers the terminal points of $\cW$.
Then, any attempt to reduce the problem to finding such a family of paths for $w \in O' \setminus O$ will fail, and is therefore unnecessary.
Therefore, by induction, subsequent invocations of $\AlternPathSearch(v')$ in the loop on  lines~\ref{line:iterSearchLoop}\thru\ref{line:iterSearchLoopEnd} of Algorithm~\ref{alg:iterFindPaths} will return with $\foundpath = \true$ if and only if there exists a family $\cU'$ of vertex disjoint $I$\thru$O$ paths which covers $v'$ as well as the terminal points of $\cW$; and $(f,g)$ represents this family of paths.

In each repetition of this loop, $\AlternPathSearch$ is invoked at most once on each vertex $v \in V(G)$ in the depth-first search.
Because the assignments and conditional tests for each vertex can be performed in time $O(1)$, the time consumed by each invocation (not including the time consumed by recursive invocations) is dominated by the time required to iterate through the neighbors of $v$ on lines~\ref{line:searchLoop}\thru\ref{line:searchLoopEnd} of Algorithm~\ref{alg:augmentPaths}, of which there are $\deg(v)$.
Summing over all vertices, the time required to perform one iteration of the ``while'' loop on lines~\ref{line:buildPathsLoop}\thru\ref{line:buildPathsLoopEnd} of Algorithm~\ref{alg:iterFindPaths} is then
\begin{gather}
		O\paren{\sum_{v \in V(G)} \deg(v)	\;}
	\;\;=\;\;
		O\Big(2 \card{E(G)}\Big)
	\;\;=\;\;
		O(m)	\;.
\end{gather}

Because there are at most $k$ vertices in $O$, there can be at most $k$ vertex disjoint paths from $I$ to $O$\,; then, at most $k$ iterations of the ``while'' loop on lines~\ref{line:buildPathsLoop}\thru\ref{line:buildPathsLoopEnd} of Algorithm~\ref{alg:iterFindPaths} can terminate with $\foundpath = \true$; then this loop will terminate in time $O(km)$.
By induction, regardless of the value of $\foundpath$, the variables $(f,g)$ represent a family of vertex disjoint paths $\cW$ from $I$ to $O$ after each such iteration; and by the analysis above, we have $\foundpath = \false$ only if there does not exist a family of vertex disjoint $I$\thru$O$ paths in $G$ which is larger than $\cW$.
The run-time of Algorithm~\ref{alg:iterFindPaths} is dominated by the ``while'' loop, as the initialization loop can be performed in time $O(n)$.
Therefore:

\begin{lemma}
	\label{lemma:obtainMaxFamilyPaths}
	Let $(G,I,O)$ be a geometry with $\card{E(G)} = m$ and $\card{O} = k$.
	Then the procedure $\ObtainMaxFamilyPaths(G,I,O)$ terminates in time $O(km)$, and returns data $(f,g)$ such that there exists a maximum-size family $\cW$ of vertex disjoint $I$\thru$O$ paths with the following properties:
	\begin{romanum}
	\item
		$\dom(f) = V(\cW) \setminus O$, and $\dom(g) = V(\cW) \setminus I$\,;

	\item
		$vw \in E(G)$ is covered by $\cW$ if and only if $v \in \dom(f)$, $w \in \dom(g)$, $w = f(v)$, and $v = g(w)$.
	\end{romanum}
\end{lemma}

\paragraph{Remark on this algorithm and its run-time.}

It should be noted that the algorithms presented above are essentially an adaptation of the Ford-Fulkerson algorithm for Max-Flow,\footnote{%
For more information about the Max-Flow problem and the Ford-Fulkerson algorithm to solve it, the interested reader may consult any standard text on polynomial time graph algorithms, \eg~\cite{CLRS}.
The term ``flow'' in this context is not directly related to the definition used throughout this chapter.}
for a flow network with multiple indistinguishable sources and sinks, and unit edge capacities.
For such a flow-network, a solution to Max Flow is a maximum-size collection of \emph{edge}-disjoint $I$\thru$O$ paths: the adaptation made here is the special treatment of entry points treated as an added special case, to ensure \emph{vertex}-disjointness of the families of paths.
Thus, the run-time of $O(km)$ should be unsurprising, as this is also an upper found for the Ford-Fulkerson algorithm for a flow-network whose maximum integer flow has value $k$.

\subsubsection{Determining if we have found a successor function}

The final step towards finding a potential flow-function (\ie\ a successor function in the sense of Definition~\ref{def:successorFn}, to which we will again restrict ourselves to for the remainder of the chapter) is to check if the family of paths $\cW$ described by the variables $(f,g)$ constructed by $\ObtainMaxFamilyPaths(G,I,O)$ is in fact a path cover.
We may do this by simply verifying that
\begin{gather}
	\dom(f) \union \dom(g) \union (I \inter O) = V(G)	\,,
\end{gather}
that is, that there are no vertices $v \in I\comp \union O\comp$ for which $f(v) = g(v) = \nil$, as the set of vertices which would lie outside of both the domain and the image of a successor function is precisely $I \inter O$.
We can test whether there are any such vertices $v$ in time $O(n)$.

By Theorem~\ref{thm:uniqueFamilyPaths}, if $(G,I,O)$ is a geometry where $\card{I} = \card{O}$ and which has a causal path cover $\cC$, then that path cover is the unique maximum family of vertex-disjoint $I$\thru$O$ paths.
Then, if the family of paths $\cW$ represented by the variables $(f,g)$ is not a path cover, the geometry $(G,I,O)$ does not have a flow.
We may then reject any geometries for which this occurs.

\subsection{Constructing the natural pre-order for a successor function}

As we remarked in Section~\ref{sec:influencing}, the problem of determining if a successor function is a flow function can be reduced to the problem of constructing the natural pre-order $\preceq$ for $f$.
The natural pre-order is the transitive and reflexive closure of the influence relation $\inflRel$ for $f$ defined by Definition~\ref{def:inflRel}: thus, we may reduce finding $\preceq$ to the Transitive Closure problem.
In this section, we will describe algorithms for doing this which take advantage of the addition structure provided by the presence of the successor function $f: O\comp \to I\comp$ (and its inverse function $g: I\comp \to O\comp$) based on techniques presented in Nuutila~\cite{Nuutila95}.

\subsubsection{Adapting Tarjan's Algorithm to find the natural pre-order}

Tarjan's algorithm~\cite{Tarjan72, CLRS} is a depth-first search algorithm for determining the strongly connected components of a digraph $D$, which are the equivalence classes of vertices which are mutually reachable from each other by directed walks.
A straightforward adaptation of Tarjan's algorithm due to~\cite{SS-S88} can be used to solve the Transitive Closure problem by storing the set of vertices which can be reached from each vertex (which we refer to as the \emph{descendants} of a vertex), thus determining the transitive closure of the digraph $D$.
A clear presentation of such an algorithm can be found in~\cite[page~49]{Nuutila95}.

In order to efficiently store the sets of descendants of a given vertex, we may make use of a \emph{chain decomposition} of the digraph $D$, a technique proposed in~\cite{Simon88}.
A chain decomposition of a digraph $D$ is a decomposition of $V(D)$ into a family of vertex-disjoint directed paths $\cP$ in the digraph $D$.
Then, for each vertex $v$ and each path $P \in \cP$, we may store the minimal vertex $w \in V(P)$ such that $w$ is a descendant of $v$.
If it is easy to determine when one vertex within a path $P$ is a descendant of another, we may then use this to obtain an efficient method to make reachability comparisons between arbitrary vertices.

We may adopt these techniques to efficiently obtain a representation of the natural pre-order $\preceq$ of the successor function $f$.
Described in terms of directed graphs, the relevant problem is to find the (reflexive and) transitive closure of the influence digraph $\sI_f$ (Definition~\ref{def:inflDigraph}, page~\pageref{def:inflDigraph}).
Note that for each $v \in O\comp$ and $w = f(v)$, the arc $v \arc w$ is an arc of $\sI_f$\,: then, the path cover $\cP_f$ is a chain decomposition of $\sI_f$.
Thus, if functions $f$ and $g$ have been computed, we have a ready-made chain decomposition of $\sI_f$ which we may exploit.
As well, if $\sI_f$ contains directed circuits (\ie\ if it has strongly connected components consisting of more than one vertex), then the pre-order $\preceq$ is not antisymmetric: rather than storing strongly connected components, as in Tarjan's algorithm, we may simply determine whether we have found a directed circuit in $\sI_f$ and terminate once one is found.

The way in which the natural pre-order $\preceq$ will be constructed is as follows.
We will construct and array $\Path: V(G) \to \N$ which assigns a ``path label'' to each vertex which is constant along paths, and distinct between distinct paths, of $\cP_f$.
For each vertex $v \in V(G)$, we will also store a distance $\Dist(v)$ which measures the distance of each vertex from the initial point of the path on which it lies: more precisely, to the largest integer $\ell \in \N$ such that $g^\ell(v) \ne \nil$.
Thus, for vertices $v,w$ on a common path, we have $v \preceq w$ if and only if $\Dist(v) \le \Dist(w)$.
We will call a pair of such arrays $(\Dist,\Path)$ a \emph{path-labelling} of the orbits of $f$.

To exploit the chain decomposition to efficiently store the partial order $\preceq$\,, we will construct an array of lower bounds for each vertex, which we will use to store the ``ancestors'' rather than the descendants of a vertex (\ie\ for a vertex $w$, the vertices $v$ from which $w$ can be reached by directed walks).
The \emph{infimum} $\inf_P(w)$ of a vertex $w$ in a path $P \in \cP_f$ will simply be the maximal vertex $v$ in $\preceq$ such that $v \in V(P)$ and $v \preceq w$.
Rather than explicitly storing the vertices $\inf_P(w)$, we may instead store the distance of $\inf_P(w)$ from the initial point on its path.
Then, if we label the paths of $\cP_f$ arbitrarily by integers $p \in \ens{1, \ldots, k}$,\footnote{%
Recall that each path of $\cP_f$ ends in $O$ and each vertex of $O$ ends such a path: thus $\card{\cP_f} = \card{O} = k$.}
we may record the infima of vertices $w \in V(G)$ in an array $\Inf: V(G) \x \ens{1, \ldots, k} \to \Z$ such that $\Inf(w,p)$ is the maximum distance of a vertex $v$ such that $\Path(v) = p$ from the initial point of its path, such that $v \preceq w$.
(If there is no such vertex $v$, we may set $\Inf(w,p)$ to some negative integer.)
This suggests the following definition:

\begin{definition}
	For a geometry $(G,I,O)$ with a successor function $f: O\comp \to I\comp$, a \emph{system of infima} for a pre-order $\preceq$ on $V(G)$ is a triple $(\Dist, \Path, \Inf)$ of arrays $\Path, \Dist: V(G) \to \N$ and $\Inf: V(G) \x \N \to \Z$ such that 
	\begin{gather}
		\label{eqn:infimaCriterion}
			v \preceq w
		\;\;\iff\;\;
			\Dist(v) \le \Inf(w, \Path(v))
	\end{gather}
	holds for all $v, w \in V(G)$.
\end{definition}

Algorithms~\ref{alg:findInfima} and~\ref{alg:testNaturalPreorder} together describe an iterative procedure for constructing a system of infima for $\preceq$ as described above, and in doing so, test whether the natural pre-order $\preceq$ is antisymmetric.
In the following pages, we will prove their correctness.

\subsubsection{Analysis of Algorithm~\ref{alg:findInfima}}

Algorithm~\ref{alg:findInfima} describes a recursive subroutine $\FindInfima$, with implicit inputs defined by an embedding procedure shown in Algorithm~\ref{alg:testNaturalPreorder}~(page~\pageref{alg:testNaturalPreorder}).
Before describing the more general algorithm there, we will first describe behavior of $\FindInfima$.

\begin{algorithm}[t]

	\SUBROUTINE \FindInfima(w):

		\Data{%
			$(G,I,O)$, a geometry with $\card{O} = k$
		\\
			$f: V(G) \partialto V(G)$, a successor function for $(G,I,O)$
		\\
			$g: V(G) \partialto V(G)$, the inverse function of $f$
		\\
			an array $\status: V(G) \to \ens{\untouched, \incomplete, \complete}$
		\\
			$\Inf: V(G) \x \ens{1,\ldots,k} \to \Z$, an array to store path-infima for vertices
		}

	\BEGIN
		\algline

		\oneline
		\IF $\status(w) = \incomplete$ \THEN $\foundcircuit = \true$\;					\nllabel{line:testCircuit}
		\oneline
		\IF $\status(w) \ne \untouched$ \THEN \RETURN\;										\nllabel{line:testStatus}
		$\status(w) \gets \incomplete$\;															\nllabel{line:setIncomplete}

		\medskip

		\LET $v \in V(G)$\;
		\FOREACH $z \sim w$ \SUCHTHAT $z \ne f(w)$ \ANDALSO $\big(z = g(w) ~\ORELSE~ g(z) \ne \nil\big)$ \DO		\nllabel{line:predLoop}
			\short
			\IF $z = g(w)$ \THEN {\nllabel{line:directPred}%
				$v \gets z$
			}
			\ELSE {\nllabel{line:indirectPred}
				$v \gets g(z)$
			}	

			\smallskip
			$\FindInfima(v)$\;																								\nllabel{line:recursiveFindInfima}
			\oneline
			\IF $\foundcircuit$ \THEN \RETURN\;																			\nllabel{line:recursiveTestCircuit}

			\smallskip
			\short 
			{\FORALL $p \in \ens{1,\ldots,k}$ \DO																		\nllabel{line:augmentInfLoop}
				\oneline
				\IF $\Inf(w,p) < \Inf(v,p)$ \THEN $\Inf(w,p) \gets \Inf(v,p)$\;
			\ENDFOR}

			\smallskip
	\ENDFOR																																	\nllabel{line:predLoopEnd}

		\medskip
		$\status(w) \gets \complete$\;
		\RETURN
	\END

	\caption[An algorithm to obtain the ``ancestors'' of a vertex $w$.]{\label{alg:findInfima}%
		An algorithm to obtain the ``ancestors'' of a vertex $w$ (and by extension of all vertices $v \prec w$), and to determine in the process whether $\prec$ is antisymmetric.}
\end{algorithm}

Suppose that $(\Dist,\Path)$ is a path-labelling of the orbits of $f$.
We will say that $w$ is \emph{untouched} with respect to a function $\Inf: V(G) \x \ens{1,\ldots,k} \to \N$ if we have
$\status(w) = \untouched$ and 
\begin{gather}
	\Inf(w,p) = \begin{cases}
						\Dist(w)	\,,	&	\text{if $p = \Path(w)$}	\\
						-1	\,,			&	\text{otherwise}
	            \end{cases}
\end{gather}
for all $p \in \ens{1,\ldots,k}$\,; and that $w$ is \emph{complete} with respect to $\Inf$ if for all $v \preceq w$, the following holds:
\begin{romanum}
\item
	$\status(v) = \complete$\,,

\item
	$\Inf(v,p) \ge -1$ for all $p \in \ens{1,\ldots,k}$\,, and
	
\item
	$u \preceq v \;\iff\; \Dist(u) \le \Inf(v, \Path(u))$ for all $u \in V(G)$.
\end{romanum}
For any vertex $w$ which is complete with respect to $\Inf$, we also have $v$ complete with respect to $\Inf$ whenever $v \preceq w$: this represents a subset of $(v,w) \in V(G) \x V(G)$ where the relation $v \preceq w$ is faithfully represented by a comparison as in~\ref{eqn:infimaCriterion}. 
We will be interested in situations where $\FindInfima$ is called for an untouched vertex, conditioned on subsets of $V(G)$ being either untouched or complete with respect to $\Inf$.

Note that the instructions of the loop on lines~\ref{line:predLoop}\thru\ref{line:predLoopEnd} are executed for precisely those vertices $z = f(v) \ne w$ and $z = v = g(w)$, where $v \inflRel w$; and on lines~\ref{line:directPred} and~\ref{line:indirectPred}, we assign to the \emph{variable} $v$ that corresponding vertex $v \inflRel w$.
Then, the iteration over $z$ on line~\ref{line:predLoop} essentially stands in for an iteration over all vertices $v \inflRel w$.
We will use this fact without further comment throughout our analysis of $\FindInfima$.

Next, note that in any invocation of $\FindInfima(w)$ terminates with $\foundcircuit = \false$ only if either $\foundcircuit = \false$ and $\status(w) = \complete$ before the invocation, or if the program control proceeds beyond for loop on lines~\ref{line:predLoop}\thru\ref{line:predLoopEnd}.
The value of $\Inf(w,p)$ can only change for some value of $p \in \ens{1, \ldots, k}$ in the latter case, in which case the value of $\status(w)$ changes from $\untouched$ to $\complete$ in the course of the invocation.
Furthermore, this will only happen if every recursive call to $\FindInfima(v)$ made also returns with $\foundcircuit = \false$, in which case the same is true of all $v \inflRel w$.
Thus, an invocation of $\FindInfima(w)$ terminates with $\foundcircuit = \false$ only if:
\begin{romanum}
\item
	$\foundcircuit = \false$ before the invocation;

\item
	for any $v \in V(G)$ the value of $\Inf(v,p)$ is only changed for some $p \in \ens{1, \ldots, k}$ if $\status(v)$ is also changed; and

\item
	$\status$ is changed by the invocation only in that $\status(v)$ is set to $\complete$ for some set of vertices $v \preceq w$, including $w$ in particular.
\end{romanum}
Using this, we will consider certain conditions under which an invocation $\FindInfima(w)$ will return with $\foundcircuit = \true$.

\begin{definition}
	We call a set of vertices $S$ \emph{noxious} if, when $\status(v) = \untouched$ or $\status(v) = \complete$ for every vertex $v \in V(G) \setminus S$, an invocation of $\FindInfima(s)$ will terminate with $\foundcircuit = \true$ for any $s \in S$.
\end{definition}

\begin{observation*}
	Suppose that $\status(v) = \incomplete \iff v \in S$ for some set $S \subset V(G)$.
	Then $S$ is a noxious set, as an invocation of $\FindInfima(v)$ for any $v \in S$ will terminate with $\foundcircuit = \true$.
\end{observation*}

The concept of a vertex in a noxious set is meant as a generalization of the case above, of a vertex $v \in V(G)$ for which $\status(v) = \incomplete$.
We will apply this concept ultimately to demonstrate simple conditions on $\status$ under which a circuit in $\sI_f$ will cause an invocation of $\FindInfima$ to return with $\foundcircuit = \true$.
The main tool in this approach is the following Lemma:

\begin{lemma}
	\label{lemma:noxiousChain}
	Let $S \subset V(G)$ be a noxious set defined solely in terms of the properties of vertices with respect to adjacency in $G$, and the arrays $\status$, $f$, and $g$.
	Suppose that $w \in V(G)$ is such that for all $v \preceq w$, either $\status(v) = \untouched$, $\status(v) = \complete$, or $v \in S$; and that there is some $s \in S$ such that $s \preceq w$.
	Then an invocation of $\FindInfima(w)$ will terminate with $\foundcircuit = \true$.
\end{lemma}
\begin{proof}
	For any such $w$, let $S'$ be the subset of vertices $s \in S$ such that there is a chain
	\begin{gather}
		s \;=\; u_0 \inflRel u_1 \inflRel \cdots \inflRel u_d = w
	\end{gather}
	for some $d \ge 0$, such that $u_j \notin S$ for all $j > 0$; and let $\ell(w)$ be the length of the longest such chain.
	(We may extend this to vertices $w$ for which there are no vertices $s \preceq w$, by adopting the convention $\ell(w) = \infty \ge N$ for any $N \in \N$.)
	If $\ell(w) = 0$, we have $w \in S$, in which case an invocation of $\FindInfima(w)$ will terminate with $\foundcircuit = \true$ by definition.
	Otherwise, suppose the claim holds for all such vertices $v \in V(G)$ for which $\ell(v) \le L$, and suppose that $\ell(w) = L+1$.
	Let $N$ be the set of vertices $u \inflRel w$ such that $s \preceq u$ for some $s \in S'$, and consider the operations performed in an invocation of $\FindInfima(w)$ under the stated initial conditions, and with $\foundcircuit = \false$.

	If the invocation $\FindInfima(w)$ terminates without setting $v$ to an element of $N$ in the loop from lines~\ref{line:predLoop}\thru\ref{line:predLoopEnd}, this is because the condition on line~\ref{line:recursiveTestCircuit} was met, in which case the invocation terminates with $\foundcircuit = \true$.
	Otherwise, all of the preceding recursive invocations $\FindInfima(v)$ for $v \inflRel w$ terminated with $\status(v) = \complete$ and $\foundcircuit = \false$, in which case the value of $\status(u)$ is unchanged for any vertex $u$ such that $u \not\preceq v$ for all of the preceding values of $v$.
	Also, by the induction hypothesis, because these preceding recursive invocations $\FindInfima(v)$ did not terminate with $\foundcircuit = \true$, we have $\ell(v) > L$.

	Let $u_0 \inflRel w$ be the first vertex of $S'$ to occur as the value of $v$ in the loop on lines~\ref{line:predLoop}\thru\ref{line:predLoopEnd} of the invocation $\FindInfima(w)$.
	Then, any chain $s = u_d \inflRel u_{d-1} \inflRel \cdots \inflRel u_1 \inflRel u_0$, such that $s \in S$ and $u_j \notin S$ for $j < d$, has length $d \le L$ by the definition of $\ell(w)$.
	Because $\ell(v) > L$ for any value of $v \inflRel w$ from an earlier iteration, there cannot be a chain $s \inflRel v_{d-1} \inflRel \cdots \inflRel v_1 \inflRel v$, where $d \le \ell$, for any such $v$; and because there is no such chain for $d > \ell$ by the definition of $\ell(w)$, there cannot be a chain $u_j \inflRel v_{N-1} \inflRel \cdots \inflRel v_1 \inflRel v$ for any length $N$, and for any $0 \le j \le d$.

	Thus, for all $0 \le j \le d$ and for all $v \inflRel w$ from earlier iterations of the for loop on lines~\ref{line:predLoop}\thru\ref{line:predLoopEnd}, we have $u_h \not\prec v$.
	In particular, values of $\status(u_j)$ are unchanged from their initial values before the invocation of $\FindInfima(w)$, and so for each $0 \le j \le d$, $u_j \in S$ before the invocation of $\FindInfima(u_0)$ if and only if $u_j \in S$ before the invocation of $\FindInfima(w)$.
	Therefore, just prior to the invocation of $\FindInfima(u_0)$, there is a chain $s \inflRel u_{d-1} \inflRel \cdots \inflRel u_1 \inflRel u_0$ for $d \le L$, such that $s \in S$ and $u_j \notin S$ for $j < d$.
	By the induction hypothesis, the invocation of $\FindInfima(u_0)$ terminates with $\foundcircuit = \true$: then the invocation of $\FindInfima(w)$ also terminates with $\foundcircuit = \true$.
	By induction on $\ell(w)$, the claim then holds.
\end{proof}

\begin{definition}
	A vertex $w \in V(G)$ is \emph{circuitous} if there is a vertex $v \prec w$ for which $w \prec v$\,, \ie\ if there is a circuit in $\sI_f$ containing $w$.
\end{definition}
The order $\preceq$ is a partial order if and only if there are no such vertices.
We would like to describe simple conditions under which such a vertex causes an invocation of $\FindInfima$ to return with $\foundcircuit = \true$.
We will show:

\begin{lemma}
	\label{lemma:circuitNoxious}
	Suppose that every vertex $v \in V(G)$ is either untouched or complete with respect to $\Inf$, and in particular that every circuitous vertex is untouched with respect to $\Inf$\,.
	Then the set of circuitous vertices is noxious.
\end{lemma}
\begin{proof}
	Let $w$ be a circuitous vertex, and let $N$ be the set of vertices $v \inflRel w$ such that $w \prec v$\,.
	Because $\status(w) = \untouched$ initially, it follows that any vertex $u \in V(G)$ for which $w \preceq u \preceq w$ holds is not complete with respect to $\Inf$\,; therefore each such $u$ is untouched with respect to $\Inf$.
	In particular, for each vertex $u_j$ in any chain $w \inflRel u_{d-1} \inflRel \cdots \inflRel u_0 \inflRel w$, we have $\status(u_j) = \untouched$.

	Consider an invocation of $\FindInfima(w)$.
	On line~\ref{line:setIncomplete}, we set $\status(w) \gets \incomplete$\,.
	Then, for all $u \in N$, there is a chain $w \inflRel u_{d-1} \inflRel \cdots \inflRel u_0 = u$ such that $\status(w) = \incomplete$ and $\status(u_j) = \untouched$ for $0 \le j < d$.
	If the invocation $\FindInfima(w)$ terminates without setting $v$ to an element of $S$ in the loop from lines~\ref{line:predLoop}\thru\ref{line:predLoopEnd}, this is because the condition on line~\ref{line:recursiveTestCircuit} was met, in which case the invocation terminates with $\foundcircuit = \true$.
	Otherwise, let $u_0$ be the first vertex in $S$ which is assigned to $v$ in the loop on lines~\ref{line:predLoop}\thru\ref{line:predLoopEnd}.
	By the very fact that $v \notin N$ in the preceding iterations, we do not have $w \preceq v$ for those iterations, and in particular there are not any vertices $u'$ such that $w \preceq u' \preceq v$\,; then the values of $\status(u')$ are unchanged throughout the recursive invocation $\FindInfima(v)$ for these preceding iterations.
	In particular, for all vertices $u' \preceq u_0$, we have $\status(u') = \untouched$, $\status(u') = \complete$, or $\status(u') = \incomplete$.
	As remarked above, any set of vertices $S$ consisting of vertices $u'$ for which $\status(u') = \incomplete$ is noxious.
	Then, by Lemma~\ref{lemma:noxiousChain}, an invocation of $\FindInfima(u_0)$ will return with $\foundcircuit = \true$: so the invocation of $\FindInfima(w)$ will terminate with $\foundcircuit = \true$ as well.
	Thus, under these conditions, the set of circuitous vertices is noxious.
\end{proof}

Lemmas~\ref{lemma:noxiousChain} and~\ref{lemma:circuitNoxious} provide a sufficient condition for an invocation $\FindInfima(w)$ to return with $\foundcircuit = \true$.
The following will provide a useful partial converse:

\begin{lemma}
	\label{lemma:inductComplete}
	Suppose that, for some vertex $w \in V(G)$ every vertex $v \preceq w$ is either untouched or complete with respect to $\Inf$.
	If there are no circuitous vertices $s \preceq w$ and $\foundcircuit = \false$, an invocation $\FindInfima(w)$ will return with $\foundcircuit = \false$, and only change the values of $\status$ and $\Inf$ to make $w$ complete with respect to $\Inf$.
\end{lemma}
\begin{proof}
	If $w$ is complete with respect to $\Inf$, then $\FindInfima$ terminates immediately with no changes to any variables; otherwise, we may assume that $w$ is untouched with respect to $\Inf$.
	We will prove the claim by induction on the \emph{depth} of $w$, which we define as the length of the longest chain $v_0 \inflRel v_1 \inflRel \cdots \inflRel v_d = w$ in $V(G)$: if there are no circuitous vertices $s \preceq w$, $d$ is guaranteed to be bounded above by $n = \card{V(G)}$.

	Consider the base case $d = 0$, 
	The conditions on lines~\ref{line:testCircuit} and~\ref{line:testStatus} are not fulfilled as $\status(w) = \untouched$, so the first operation performed is to set $\status(w) \gets \incomplete$\,.
	No vertex $z$ satisfies the conditions of line~\ref{line:predLoop}, by assumption of the minimality of $w$, and so the instructions within that for loop are never executed; we then set $\status(w) \gets \complete$, and return.
	As well, because $\foundcircuit$ is left unaltered, we still have $\foundcircuit = \false$.
	Then, the only change performed by $\FindInfima(w)$ is to set $\status(w) \gets \complete$\,; but because $v \preceq w \iff v = w$ in this case, and because we had
	\begin{gather}
			\Inf(w,p)
		\;\;=\;\;
			\begin{cases}[r]
				0\,,	&	\text{for $p = \Path(w)$}	\\
				-1\,,	&	\text{otherwise}
			\end{cases}
	\end{gather}
	prior to the invocation, this is sufficient to cause $w$ to be complete with respect to $\Inf$.

	Suppose then that the claim holds for all vertices for with some depth $d \in \N$, and that $w$ has depth $d+1$.
	The conditions on lines~\ref{line:testCircuit} and~\ref{line:testStatus} are not fulfilled, as $\status(w) = \untouched$; we then set $\status(w) \gets \incomplete$.
	In the loop on lines~\ref{line:predLoop}\thru\ref{line:predLoopEnd}, we iterate over all $v \inflRel w$\,: by assumption, every such $v$ is either untouched or complete with respect to $\Inf$, as are all vertices $u \preceq v$ for such $v$.
	It is clear that $\FindInfima(v)$ simply returns without changing $\Inf$ or $\status$ when initially we have $\status(v) = \complete$; then by induction, after each recursive call to $\FindInfima(v)$ on line~\ref{line:recursiveFindInfima}, each vertex $v$ is complete with respect to $\Inf$ and $\foundcircuit = \false$.
	The condition on line~\ref{line:recursiveTestCircuit} is then unfulfilled, so we execute the for loop on line~\ref{line:augmentInfLoop} and proceed to the next vertex $z$.

	Initially, for every $p \in \ens{1,\ldots,k} \setminus \ens{\Path(w)}$, we have $\Inf(w,p) = -1$: then, for such $p$, the initial value of $\Inf(w,p)$ is bounded above by the values of $\Inf(v,p)$ for each $v \inflRel w$ after the recursive calls $\FindInfima(v)$.
	Then by induction, it is easy to show that after the loop on lines~\ref{line:predLoop}\thru\ref{line:predLoopEnd},
	\begin{align}
			\Inf(w,p)
		\;\;=&\;\;
			\max_{v \:\!\inflRel\:\! w} \;\,	\Inf(v,p)	\;,
		\quad
			\text{for $p \ne \Path(w)$}.	
	\end{align}
	For any $u \in V(G)$ with $\Path(u) = p \ne \Path(w)$, we have in particular $u \ne w$, and so $u \preceq w$ if and only if $u \preceq v$ for some $v \inflRel w$; and this holds if and only if $\Dist(u) \le \Inf(v, \Path(u)) \le \Inf(w, \Path(u))$ for that same $v$ after the loop on lines~\ref{line:predLoop}\thru\ref{line:predLoopEnd}, by the induction hypothesis.
	Finally, because there are no circuitous vertices $u \preceq w$, we have in particular $w \not\prec v$ for all $v \inflRel w$\,; then $\Inf(v, \Path(w)) < \Dist(w)$ for all $v \inflRel w$, in which case we have $\Inf(w, \Path(w)) = \Dist(w)$ after line~\ref{line:predLoopEnd}.
	Therefore, we also have $u \preceq w \,\iff\, \Dist(u) \le \Inf(w, \Path(u))$ when $\Path(u) = \Path(w)$\,: thus, the claim holds for $w$ as well.
	By induction, the Lemma holds.
\end{proof}

Lemmas~\ref{lemma:noxiousChain},~\ref{lemma:circuitNoxious}, and~\ref{lemma:inductComplete} will form the basis of our analysis of Algorithm~\ref{alg:testNaturalPreorder} in the next section.

\subsubsection{Analysis of Algorithm~\ref{alg:testNaturalPreorder}}

In the previous section, we described the behavior of the subroutine $\FindInfima$ under conditions of the vertices being either untouched or complete with respect to $\Inf$.
The role of the procedure $\TestNaturalPreorder$ in Algorithm~\ref{alg:testNaturalPreorder} (page~\pageref{alg:testNaturalPreorder}) is to establish a path-labelling $(\Dist,\Path)$ of the orbits of $f$, initialize arrays $\status$ and $\Inf$ so that every vertex is untouched with respect to $\Inf$, and then repeatedly invoke $\FindInfima$ in an attempt to change the status of every vertex to that of being complete with respect to $\Inf$. 
In doing so, we will either determine that $\sI_f$ has circuits (and so $\preceq$ is not a partial order), or assign values of $\Inf(w,p)$ for all $w \in V(G)$ and $p \in \ens{1,\ldots,k}$ so that $(\Dist,\Path,\Inf)$ forms a system of infima for $\preceq$\,.

\begin{algorithm}[p]

	\PROCEDURE \TestNaturalPreorder(G,I,O,f,g):

		\Input{%
			$(G,I,O)$, a geometry with $\card{O} = k$
		\\
			$f: V(G) \partialto V(G)$, a successor function for $(G,I,O)$
		\\
			$g: V(G) \partialto V(G)$, the inverse function of $f$
		}
		\Output{%
			Either a triple $(\Dist,\Path,\Inf)$ representing the natural pre-order for $f$ (in the case that $f$ is a flow function), or $\nil$ (in the case that $f$ is not a flow function).
		}

	\BEGIN
		\LET $\status: V(G) \to \ens{\untouched, \incomplete, \complete}$\;
		\LET $\Dist: V(G) \to \N$\;
		\LET $\Path: V(G) \to \N$\;
		\LET $\Inf: V(G) \x \ens{1,\ldots,k} \to \N$\;

		\medskip
		\LET $\nextpath \gets 1$\;
		\FORALL $w \in O$ \DO																							\nllabel{line:initPathLoop}
			$\InitializePath(w,\nextpath)$\;
			$\INCR \nextpath$\;
		\ENDFOR																												\nllabel{line:initPathLoopEnd}

		\medskip
		\LET $\foundcircuit \gets \false$\;
		\FORALL $w \in O$ \DO																					\nllabel{line:findInfimaLoop}
			$\FindInfima(w)$\;																					\nllabel{line:findInfima}
			\oneline
			\IF $\foundcircuit$ \THEN \RETURN \nil\;
		\ENDFOR																										\nllabel{line:findInfimaLoopEnd}

		\RETURN $(\Dist,\Path, \Inf)$\;
	\END

	\bigskip

	\SUBROUTINE \InitializePath(w,P):

	\BEGIN
		\oneline
		\IF $g(w) = \nil$ \THEN $\Dist(w) \gets 0$\;
		\ELSE {
			$\InitializePath(g(w))$\;
			$\Dist(w) \gets \Dist(g(w)) + 1$\;
		}
		$\Path(w) \gets P$\;

		\medskip
		\FORALL $p \in \ens{1,\ldots,k}$ \DO																		\nllabel{line:infimaInitLoop}
			\short
			\IF $p = P$ \THEN { $\Inf(w,p) \gets \Dist(w)$} \ELSE { $\Inf(w,p) \gets -1$ }
		\ENDFOR																												\nllabel{line:infimaInitLoopEnd}
	
		$\status(w) \gets \untouched$\;

		\medskip

		\RETURN
	\END

	\caption[An algorithm to test whether or not the natural pre-order of a successor function $f$ for a geometry $(G,I,O)$ is a partial order.]{\label{alg:testNaturalPreorder}%
		An algorithm to test whether or not the natural pre-order of a successor function $f$ for a geometry $(G,I,O)$ is a partial order.
		The subroutine $\FindInfima$ is described in Algorithm~\ref{alg:findInfima} on page~\pageref{alg:findInfima}.}
\end{algorithm}

To establish a path-labelling $(\Dist,\Path)$ for the orbits of $f$, we use a subroutine $\InitializePath$.
It is straightforward to verify that $\InitializePath(w,P)$ sets $\Path(v) = P$ for $v = w$ and for all vertices $v$ which precede $w$ on the same path, and correctly assigns their distance along the path from the first vertex of the path.
As well, for all such vertices $v$, it initializes $\status(v)$ and $\Inf(v,p)$ for each $p \in \ens{1,\ldots,k}$ so that $v$ is untouched with respect to $\Inf$.
In doing so, it is easy to see that it does not affect the status of any other vertex, or the labelling of any other paths.
$\TestNaturalPreorder$ calls $\InitializePath(v,p)$ for each vertex $w \in O$,\footnote{%
For an arbitrary path cover $\cC$ for a geometry $(G,I,O)$, the initial point of some paths of $\cC$ may not be a vertex in $I$.
While we are primarily interested in the case where $\card{I} = \card{O}$, in which case there is a one-to-one correspondence between paths and elements of $I$, we may make the algorithm more general by labelling paths from their endpoint.
In doing so, this algorithm may also be used to construct the natural pre-order for $f$ in cases where $\card{I} < \card{O}$.}
setting the path-label $p = \nextpath$ of each vertex in $O$ to a distinct integer $1 \le p \le k$.
This initializes every vertex to be untouched with respect to $\Inf$, and establishes a path-labelling $(\Dist,\Path)$.

Next, we set $\foundcircuit \gets \false$, and we invoke $\FindInfima(w)$ for each vertex $w \in O$, aborting in the case that we discover $\foundcircuit = \true$.
We return the triple $(\Dist, \Path, \Inf)$ if the loop on lines~\ref{line:findInfimaLoop}\thru\ref{line:findInfimaLoopEnd} terminates without ending the procedure call. 

Consider a loop precondition on the values of $\foundcircuit$, $\status$, and $\Inf$ where, prior to some iteration of the loop on lines~\ref{line:findInfimaLoop}\thru\ref{line:findInfimaLoopEnd}, we have
\begin{romanum}
\item
	$\foundcircuit = \false$,

\item
	every vertex is either untouched or complete with respect to $\Inf$, and

\item
	any circuitous vertices in particular are untouched with respect to $\Inf$.
\end{romanum}
By Lemma~\ref{lemma:circuitNoxious}, the set of circuitous vertices is noxious under these conditions; then, by Lemma~\ref{lemma:noxiousChain}, $\FindInfima(w)$ will return with $\foundcircuit = \true$ in the case that there exists some circuitous vertex $v \preceq w$.
If such a circuitous vertex $v$ exists, $\TestNaturalPreorder$ terminates with a return value of $\nil$.
Otherwise, by Lemma~\ref{lemma:inductComplete}, $\TestNaturalPreorder$ terminates with $\foundcircuit = \false$ and $w$ complete with respect to $\Inf$.
In particular, $\status$ and $\Inf$ in such a way as to make $w$ complete with respect to $\Inf$, in which case it is still the case that every vertex $v \in V(G)$ is untouched or complete with respect to $\Inf$.

By induction on the iteration over $w \in O$ on line~\ref{line:findInfimaLoop}, we may then prove that an invocation of $\TestNaturalPreorder(G,I,O,f,g)$ returns $\nil$ if the natural pre-order $\preceq$ is not antisymmetric, and returns with a triple $(\Dist, \Path, \Inf)$ forming a system of infima for $\preceq$ in the case that $\preceq$ is antisymmetric.
If there are any circuitous vertices $v$, eventually there will be an invocation $\FindInfima(w)$ for which $v \preceq w$\,; because the loop precondition above is maintained until such a vertex $w$ is found, it follows that $\TestNaturalPreorder(G,I,O,f,g)$ will terminate with return value $\nil$.
Otherwise, it will eventually terminate with return value $(\Dist,\Path,\Inf)$ such that $(\Dist,\Path)$ is a path-labelling of the orbits of $f$, and where every vertex $w \in O$ is complete with respect to $\Inf$\,.
Because every $v \in V(G)$ is bounded above by some $w \in O$, it follows that $\status(v) = \complete$\,, $\Inf(v,p) \ge -1$ for all $p \in \ens{1,\ldots,k}$\,, and
\begin{gather}
	\label{eqn:infimaCriterionRevisited}
	u \preceq v \;\iff\; \Dist(u) \le \Inf(v, \Path(u))	\qquad \text{for all $u,v \in V(G)$}.
\end{gather}
Thus, the triple $(\Dist,\Path,\Inf)$ forms a system of infima for $\preceq$.

In $\TestNaturalPreorder(G,I,O,f,g)$, the subroutine $\InitializePath(w,P)$ is run once for each vertex along each path, and so is invoked $n = \card{V(G)}$ times; and in each invocation, the run-time is dominated by the loop on lines~\ref{line:infimaInitLoop}\thru\ref{line:infimaInitLoopEnd}, which performs $k$ iterations of a constant-time assignment.
Thus, the loop on lines~\ref{line:initPathLoop}\thru\ref{line:initPathLoopEnd} runs in time $O(kn)$.
However, we will show that this time requirement is swamped by cumulative run-time of the invocations of $\FindInfima$.

For vertices $v \in V(G)$ in general, the subroutine $\FindInfima(v)$ is invoked on line~\ref{line:findInfima} for $w \in O$, and line~\ref{line:recursiveFindInfima} of Algorithm~\ref{alg:findInfima} for $v \in V(G)$ such that $v \inflRel w$ for some $w \in V(G)$.
It is easy to show that every $w \in O$ is maximal in $\preceq$, in which case it is invoked only once for such vertices; for any other vertex $v$, $\FindInfima(v)$ is invoked once for each $w \in V(G)$ such that $v \inflRel w$; because this holds precisely for the set of vertices $w = f(v)$ or $w \sim z = f(v)$, this is bounded above by the degree of $f(v)$.
The first time that $\FindInfima(v)$ is invoked for a particular vertex $v$, $v$ is untouched with respect to $\Inf$; for any subsequent invocation made, $v$ is complete with respect to $\Inf$, and the invocation terminates in time $O(1)$.
Then, the time requirements of all invocations of $\FindInfima(v)$ for a particular vertex $v$ (but not including the time consumed by recursive invocations for vertices $u \inflRel v$) is bounded above by the time spent in the first invocation of $\FindInfima(v)$, plus $\deg(f(v))$ for all subsequent invocations combined. 
The time required by the first invocation is dominated by by the loop at line~\ref{line:augmentInfLoop}, inside the loop from lines~\ref{line:predLoop}\thru~\ref{line:predLoopEnd} (of Algorithm~\ref{alg:findInfima}).
The inner loop performs $k$ iterations of condition tests and assignments requiring $O(1)$ time, which is repeated for each $u \inflRel v$ by the outer loop: as the number of such vertices $u$ is bounded above by $\deg(v)$, the outer loop then requires time $O(k \deg(v))$.

By a temporary abuse of notation, let $\deg(f(v)) = 0$ for vertices $v \in O$\,.
The total run-time of all invocations of $\FindInfima(v)$ is then $O(k \deg(v) + \deg(f(v)))$ for any $v \in V(G)$\,: summing over all vertices $v$, the cumulative run-time of the invocations of $\FindInfima$ is then
\begin{align}
		O\paren{ \sum_{v \in V(G)} \Big[ k \deg(v) + \deg(f(v)) \Big] }
	\;=&\;\;
		O\paren{ \sum_{v \in V(G)} k \deg(v) \;\;+ \sum_{v \in V(G)} \deg(f(v)) }
	\notag\\[1ex]=&\;\;
		O\Big(2 k \abs{E(G)} \,+\, 2 \abs{E(G)}\Big)
	\;=\;
		O(km)\,.
\end{align}
We have therefore shown:
\begin{lemma}
	\label{lemma:testNaturalPreorder}
	Let $(G,I,O)$ be a geometry with $\card{E(G)} = m$ and $\card{O} = k$, $f : V(G) \partialto V(G)$ be a successor function for $(G,I,O)$, and let $g: V(G) \partialto V(G)$ be the inverse of $f$.
	Then the procedure $\TestNaturalPreorder(G,I,O,f,g)$ terminates in time $O(km)$, and returns either
	\begin{itemize}
	\item
		$\nil$, if the natural pre-order $\preceq$ of $f$ is not antisymmetric; or

	\item
		a system of infima $(\Dist,\Path,\Inf)$ for $\preceq$, where in particular $(\Dist, \Path)$ is a path-labelling of the orbits of $f$.
	\end{itemize}
\end{lemma}
Restricting to those geometries satisfying the extremal bound of Theorem~\ref{thm:extremalUpperBound}, this provides a proof of Theorem~\ref{thm:findPreorder} on page~\pageref{thm:findPreorder}.

\subsection{The complete flow-finding algorithm}

As a summary of the results of the previous sections, we now describe an algorithm for discovering whether a geometry $(G,I,O)$ has a flow $(f,\preceq)$ in the case that $\card{I} = \card{O}$, and to construct one if so.

\begin{algorithm}[t]

\PROCEDURE \FindFlow(G,I,O):

\Input{%
		$G$, a graph;
	\\
		$I,O \subset V(G)$, sets of vertices of equal size.
}

\Output{%
	Either $\nil$, if $(G,I,O)$ does not have a flow; or a successor function $f: V(G) \partialto* V(G)$ for $(G,I,O)$, its inverse function $g: V(G) \partialto* V(G)$, and a system of infima $\partialOrd = (\Dist,\Path,\Inf)$ for the natural pre-order $\preceq$ for $f$, in the case where $(G,I,O)$ has a flow.
}

\BEGIN
	
	\LET $n \gets \card{V(G)}$\;
	\LET $m \gets \card{E(G)}$\;
	\LET $k \gets \card{O}$\;

	\oneline
	\IF $m > kn - \binom{k+1}{2}$ \THEN \RETURN \nil\;												\nllabel{line:extremalTest}

	\medskip
	\LET $(f,g) \gets \ObtainMaxFamilyPaths(G,I,O)$\;												\nllabel{line:constructFamilyPaths}
	\FORALL $v \in V(G)$ \DO																				\nllabel{line:testPathCoverLoop}
		\oneline
		\IF $v \notin I \inter O$ \ANDALSO $f(v) = g(v) = \nil$ \THEN \RETURN \nil\;	
	\ENDFOR
	
	\medskip
	\LET $\partialOrd \gets \TestNaturalPreorder(G,I,O,f,g)$\;									\nllabel{line:constructPreorder}
	\short
	\IF $\partialOrd = \nil$ \THEN {\nllabel{line:testPartialOrder}%
		\RETURN \nil
	}
	\ELSE { \RETURN $(f,g,\partialOrd)$ }

\END

	\caption[A complete algorithm for determining if a geometry $(G,I,O)$ has a flow in the case that $\card{I} = \card{O}$.]{\label{alg:completeFlowAlg}%
		A complete algorithm for determining if a geometry $(G,I,O)$ has a flow in the case that $\card{I} = \card{O}$, using the procedures defined in Algorithms~\ref{alg:iterFindPaths} and~\ref{alg:testNaturalPreorder} (pages~\pageref{alg:iterFindPaths} and~\pageref{alg:testNaturalPreorder} respectively).}

\end{algorithm}

On line~\ref{line:extremalTest}, we eliminate any geometry which violates the extremal bound given by Theorem~\ref{thm:extremalUpperBound}, thus restricting to geometries for which $m < kn$.
The invocation of $\ObtainMaxFamilyPaths(G,I,O)$ constructs mutually inverse partial functions on $V(G)$ whose orbits describe a maximum-size family of vertex disjoint paths, by Lemma~\ref{lemma:obtainMaxFamilyPaths}; the loop starting at line~\ref{line:testPathCoverLoop} then determines whether or not this family is a path cover, by seeing if at least one of $f$ and $g$ are defined for all vertices not in $I \inter O$.
If there are vertices $v \notin \dom(f) \union \dom(g) \union (I \inter O)$, then the family of paths obtained is in particular not a causal path cover: by Theorems~\ref{thm:uniqueFamilyPaths} and~\ref{thm:graphth-charn}, $(G,I,O)$ then has no flow.
Otherwise, $f$ is a successor function for $(G,I,O)$, and we may test whether or not its natural pre-order $\preceq$ is a partial order by invoking $\TestNaturalPreorder(G,I,O,f,g)$ on line~\ref{line:constructPreorder}.
By Lemma~\ref{lemma:testNaturalPreorder}, if $\partialOrd = \nil$ on line~\ref{line:testPartialOrder}, then $\preceq$ is not a partial order, and subsequently $\cP_f$ is not a causal path cover; again by Theorems~\ref{thm:uniqueFamilyPaths} and~\ref{thm:graphth-charn}, $(G,I,O)$ then has no flow.
Otherwise, $\partialOrd$ contains a system of infima for $\preceq$, and we return $(f,g,\partialOrd)$.

The run time for $\FindFlow$ is dominated by those of $\ObtainMaxFamilyPaths$ and $\TestNaturalPreorder$, which are both $O(km)$.
Due to the restriction imposed at line~\ref{line:extremalTest}, we then have the following:
\begin{theorem}
	\label{thm:findFlow}
	For a geometry $(G,I,O)$ with $\card{V(G)} = n$ and $\card{I} = \card{O} = k$, the procedure $\FindFlow(G,I,O)$ terminates in time $O(k^2 n)$, returns $\nil$ if $(G,I,O)$ has no flow, and a triple $(f,g,\partialOrd)$ otherwise; where $f: V(G) \partialto V(G)$ is a successor function for $(G,I,O)$, $g: V(G) \partialto V(G)$ is the inverse of $f$, and $\partialOrd$ is a system of infima for the natural pre-order $\preceq$ of $f$.
	In particular, if $(G,I,O)$ has a flow, then $(f, \preceq)$ is a flow for $(G,I,O)$.
\end{theorem}

\section{A semantic map for the DKP representation}
\label{sec:semanticDKP}

The purpose of this Section is to show how to obtain a right-inverse to the DKP representation map $\cR$, from unitary circuits to one-way measurement based algorithms.
We show how this may be done for a one-way measurement patterns with an underlying geometry $(G,I,O)$, and whose operations satisfy certain constraints related to a flow.

\subsection{Influencing walks and measurement adaptation}
\label{sec:dependencies}

For a geometry $(G,I,O)$ with a flow $(f, \preceq)$, the partial order $\preceq$ is meant to describe an order of measurements which allow a unitary embedding to be performed.
However, it represents a \emph{sufficient} condition, not a necessary one: and in particular, as shown in~\cite{DK05u}, it is a sufficient condition under the constraints that a measurement of a qubit $v \in O\comp$ affects the following operations:
\begin{romanum}
\item
	it induces a sign dependency on the measurement of its successor $f(v)$, and a $\pi$-dependency on the measurements of its other neighbors $w \sim f(v)$, and affects no other measurements; and

\item
	it induces an $X$ correction on $f(v)$ if this is an element of $O$, and a $Z$ correction on any neighbor $w \sim f(v)$ which are elements of $O$, and induces no other corrections.
\end{romanum}
These constraints essentially arise from the simplified DKP construction, as described in Section~\ref{sec:flowDefn}.

As we noted in Section~\ref{sec:DKPstandardization}, we may eliminate $\pi$-dependencies in a one-way measurement pattern based computation by performing signal shifting: this may result in more qubits depending on a given measurement result, but relaxes the measurement dependencies to allow transformations to be done with less depth complexity (neglecting the cost of computing boolean expressions for adapting the measurements).
As well, measurements along Pauli axes are not affected by any previous measurements (although their results may be affected, in the counterfactual sense described in Section~\ref{sec:flowDefn}).
We will see that, as a result, the appropriate tool for considering the measurement order for standardized one-way measurement patterns is not the partial order $\preceq$, but rather a special class of influencing walks.\footnote{%
The importance of influencing walks for depth-complexity in one-way measurement based computation was first realized by Broadbent and Kashefi, who first coined the term \emph{influencing path} while working on the results of~\cite{BK2007}.
The recognition of influencing paths as objects of interest helped to inspire simplifications in the work which I've presented in this Chapter form their original forms, and were incorporated in the published version of these results in~\cite{Beaudrap08}.
We will consider here a different approach to the analysis of depth of one-way measurement based procedures in terms of influencing walks than the work in~\cite{BK2007}.}

For a unitary circuit $C$, we may perform the DKP construction by first producing a pattern $\tilde P$ via the \emph{simplified} DKP construction.
As we remarked in Section~\ref{sec:flowDefn}, we have $v \preceq w$ for two distinct qubits involved in $\tilde P$ if the measurement of $v$ induces $X$ or $Z$ corrections onto $w$, which are absorbed into the measurement basis for $w$ if $w \in O\comp$.
These corrections are induced if and only if either $w = f(v)$, in which case the induced correction is $\Xcorr{}{\s[v]}$, or $w \sim f(v)$, in which case the induced correction is $\Zcorr{}{\s[v]}$.
We will refer to these dependencies as \emph{type \iti} and \emph{type \itii} dependencies.
Note that the relationships of qubits with dependencies of type~\iti\ or type~\itii\ are slight generalizations\footnote{%
A segment of type~\itii\ with respect to the path cover $\cP_f$ is a path of the form $v \arc z \arc w$, where the edge $zw$ is not covered by a path of $\cP_f$\,; measurement dependencies of type~\itii, by contrast, include the case $w = f(f(v))$.}
of the relationships occurring in segments of type~\iti\ and type~\itii\ of influencing walks; what is significant about these types is in particular whether the dependency arises from an $X$ or a $Z$ correction induced by the measurement of a qubit.

Consider the further operations required to standardize $\tilde P$, into a standardized pattern:
\begin{enumerate}
\item
	As we noted in Section~\ref{sec:DKPstandardization}, measurements along a Pauli axis can be performed independently of any previous measurement; then, for qubits $v \in O\comp$ and $w = f(v) \in O\comp$, we may eliminate any dependency of type~\iti\ of $w$ on $v$ if the former is to be measured with an angle $\alpha \in \frac{\pi}{2} \Z$.

\item
	For a measurement along the $Y$ axis in particular (\ie\ for measurement angles $\alpha \in \pi \Z + \pion 2$), the way in which type~\iti\ dependencies are eliminated are to replace sign-dependencies with $\pi$-dependencies, which is the effect of a dependency of type~\itii.
	Thus, when $w = f(v)$ for some $v \in O\comp$ and $w \in O\comp$ measured with an odd multiple of $\pion 2$, we may replace the type~\iti\ dependency of $w$ on $v$ with a dependency of type~\itii.

\item
	Whenever a qubit $w \in O\comp$ has a type~\itii\ dependency on a qubit $v$, the effect of the measurement result of $v$ is not to change the basis of measurement for $w$, but rather the significance of the measurement result of $w$, which is formalized in the process of signal shifting.
	Specifically, for any qubit $w$ which has a type~\itii\ dependency on another qubit $v$, we may eliminate the dependency of $w$ on $v$ if we make every qubit $x$ with a type~\iti\ dependency on $w$ also depend on $v$, and similarly for qubits $y$ with a type~\itii\ dependency on $w$.
\end{enumerate}
Recursively applying the last rule, we may describe walks in $G$ similar to influencing walks, of the kind described below:
\begin{definition}
	For a geometry $(G,I,O)$, a successor function $f: O\comp \to I\comp$, and a vector $\vec a = (\alpha_v)_{v \in O\comp}$ of default measurement angles for each qubit, a \emph{segment of an adapting walk} for $f$ is any directed path of one of the following three forms:
	\begin{romanum}
	\item
		$v \arc w$, where $w = f(v)$, and $w$ is not to be measured with an angle $\alpha \in \frac{\pi}{2} \Z$\,;

	\item
		$v \arc z \arc w$, where $v \in O\comp$, $z = f(v)$, and $w \sim z$\,;

	\item
		$v \arc w$, where $w = f(v)$, and $\alpha_w \in \pi\Z + \pion 2$\,.
	\end{romanum}
	An \emph{adapting walk} for $f$ is any directed walk $W$ in $G$ which may be formed concatenation of such segments, subject to the constraint that a type~\iti\ segment may only occur at the end of the walk, and that either $W$ ends with a segment of type~\iti\ or at a vertex in $O$.
\end{definition}
Note that an adapting segment for $f$ of type~\itiii\ is also a segment of type~\iti\ of an influencing walk of $\cP_f$\,.
Thus, all adapting walks for $f$ are also influencing walks of $\cP_f$.

We will show that adapting walks describe the propagation of measurement or correction dependency on previous measurement results for one-way measurement based patterns arising from the DKP construction, with the type of adapting walk determining the nature of the dependency, and that these dependencies may cancel when an even number of them converge on a given qubit:

\begin{theorem}
	\label{thm:dependenciesDKP}
	Let $P = \cR(C)$ be the image of a unitary circuit $C$ in the DKP construction $\cR$\,, where $C$ is given as a stable index tensor expression.
	Let $(G,I,O)$ be the geometry consisting of the graph $G$ of entangling operations on qubits of $P$ (which are indexed by wire-segments of $C$), the set $I$ of non-prepared qubits, and the set $O$ of non-measured qubits, and let $\vec a = (\alpha_v)_{v \in O\comp}$ be the vector of default measurement angles for each qubit.
	If $f: O\comp \to I\comp$ is the function mapping each deprecated index in $C$ to the corresponding advanced index, then $f$ is a flow function for $(G,I,O)$; and for two distinct qubits $v,w \in V(G)$, the following holds:
	\begin{romanum}
	\item
		there is a measurement on $w \in O\comp$ where the sign of the measurement angle depends on $\s[v]$, or a potential $X$ correction on $w \in O$ depending on $\s[v]$, if and only if there are an odd number of adapting walks for $f$ from $v$ to $w$ which end in a segment of type~\iti;

	\item
		there is a potential $Z$ correction on $w \in O$ depending on $\s[v]$ if and only if there are an odd number of adapting walks for $f$ from $v$ to $w$ which end in a segment of type~\itii.
	\end{romanum}
\end{theorem}
\begin{proof}
	Consider the one-way measurement pattern $\tilde{P}$ obtained from $C$ by the \emph{simplified} DKP construction.
	From the discussion in Section~\ref{sec:flowDefn}, the function $f$ is a flow-function by construction.
	We may obtain the one-way measurement pattern  $P$ by standardizing the procedure $\tilde P$\,.

	Consider how the process of standardization propagates the dependencies of qubits $w$ on a particular qubit $v$.
	Let us say that a qubit $w$ has an \emph{$X$-dependency} on a qubit $v$ in a measurement procedure if either the measurement of $w \in O\comp$ has a sign dependency on $\s[v]$, or if $w \in O$ may be subject to an $X$ correction depending on the value of $\s[v]$; and similarly, a qubit $w$ has a \emph{$Z$-dependency} on a qubit $v$ in a measurement procedure if either the measurement of $w$ has a $\pi$-dependency on $\s[v]$, or if $w$ may be subject to a $Z$ correction depending on the value of $\s[v]$.
	If a qubit $w$ has a $Z$-dependency on another qubit $v$, performing signal shifting on the measurement of $w$ will cause any qubits with an $X$-dependency on $w$ to
	\begin{inlinum}
	\item
		accrue an $X$-dependency on $v$, if it had none previously, or

	\item
		lose its $X$-dependency on $v$, if it already had such a dependency, due to cancellation modulo $2$\,;
	\end{inlinum}
	and similarly for $Z$-dependencies.

	We may then describe how standardization transformations cause dependencies to propagate, in terms of ``locally describable'' dependencies.
	Starting from the measurement procedure $\tilde P$, the only $X$-dependencies which exist are of vertices $f(v)$ on their respective qubits $v \in O\comp$\,; then we may reduce the problem to tracking $Z$ dependencies.
	Because the $Z$-dependencies of a qubit $w$ may always be removed unless $w \in O$, we may model this propagation in the form of a ``binary walk'' of a measurement result $\s[v]$, starting from $v$ and ending at vertices which have unremovable $Z$-dependencies.
	We do this as follows:
	\begin{enumerate}
	\item
		For a qubit $w \in O\comp$ and $z \sim f(w)$ distinct from $w$, signal shifting on the measurement at $w$ will propagate the $Z$-dependencies of $w$ to $z$ (possibly cancelling modulo $2$ as described above).

	\item
		For a qubit $w \in O\comp$ and a qubit $z = f(w)$ which is to be measured with an angle $\alpha \in \pi\Z + \pion 2$, the $X$-dependency of $z$ on $w$ is equivalent to a $Z$-dependency; signal shifting on the measurement at $w$ will then propagate the $Z$-dependencies of $w$ to $z$ (possibly cancelling modulo $2$ as described above).
	\end{enumerate}
	Then, we may describe the propagation of the effects of a measurement result $\s[v]$ by an $V(G) \x V(G)$ matrix $T$ over $\Z$, given by
	\begin{align}
		\label{eqn:shiftPropagator}
			T_{wv}
		\;\;=\;\;
			\begin{cases}
				1\,,	&	\text{if $v \in O\comp$, $w \sim f(v)$, and $w \ne v$}	\\[1ex]
				1\,,	&	\begin{minipage}[t]{15em}
								if $v \in O\comp$, $w = f(v)$, and $w$ is to be measured with angle $\alpha \in \pi\Z + \pion 2$
								\\[1ex]
				    	 	\end{minipage}	\\[-2ex]
				0\,,	&	\text{otherwise}
			\end{cases}\;.
	\end{align}
	Note that $T_{wv} = 1$ only if $v$ and $w$ are connected by a segment of an influencing walk for $\cP_f$\,, and in particular, an adapting segment for $f$ of type~\itii\ or~\itiii.
	Then, $\unit_v \cdot (T^\ell\, \unit_w) \ne 0$ only if $v$ and $w$ are connected by an influencing walk for $\cP_f$ consisting of $\ell$ segments.
	Because influencing walks of $\cP_f$ are of bounded length (in particular because there are no vicious circuits), there exists an $\ell$ for which $T^\ell = 0$\,.
	In particular, there are no eigenvectors of $T$ with eigenvalue $1$: then, the matrix $\idop_n - T$ has trivial kernel, and is therefore invertible (and in particular, invertible modulo $2$).

	The matrix $T$ represents the propagation of a $Z$-measurement dependency due to signal shifting or simplification of Pauli measurements, but neglects $X$-dependencies of qubits not to be measured along a Pauli axis, as well as the persistent $Z$-dependencies of elements of $O$, which cannot be signal-shifted --- these are dependencies which cannot be eliminated by pattern transformation.
	If after some number of standardization operations, a qubit $z$ has an unremovable dependency on a qubit $w$ with a (removable) $Z$-dependency on some qubits, the dependencies of $w$ are accumulated into the dependencies of $z$ by a signal shifting on the measurement at $w$.
	We describe this accumulation separately for $X$- and $Z$-dependencies.
	\begin{romanum}
	\item
		Consider another $V(G) \x V(G)$ matrix $F$ over $\Z$, given by
		\begin{align}
		\label{eqn:flowPropagator}
				F_{wv}
			\;\;=\;\;
				\begin{cases}
					1\,,	&	\begin{minipage}[t]{17em}
									if $v \in O\comp$, $w = f(v)$, and $w$ is not to be measured with an angle $\alpha \in \pi\Z + \pion 2$
									\\[1ex]
					    	 	\end{minipage}	\\[-2ex]
					0\,,	&	\text{otherwise}
				\end{cases}\;.
		\end{align}
		Then $F$ represents the effect of a signal shift on the $X$-dependencies of vertices $f(v)$ on their predecessors $v \in O\comp$.
		For qubits of the form $w = f(z)$ for $z \in O\comp$ arbitrary, the contribution of a possible dependency of $w$ on a qubit $v$ arising from the $\ell\th$ signal shift for $\ell \ge 0$ is then given by
		\begin{gather}
				\unit_w \cdot \paren{ F T^\ell \,\unit_v };
		\end{gather}
		and the cumulative dependency over all signal shifting operations is then given by
		\begin{align}
				\unit_w \cdot \paren{\sum_{\ell \in \N} F T^\ell\, \unit_v}
			\;\;=&\;\;
				\unit_w \cdot F \paren{\,\sum_{\ell \in \N} \, T^\ell } \unit_v
			\notag\\=&\;\;
				\unit_w \cdot F \paren{\idop_n - T}\inv\, \unit_v	\,,
		\end{align}
		as $\idop - T$ is non-singular.
		If this evaluates to zero modulo $2$, there is no $X$-dependency of $w$ on $v$ in the pattern $P$.\\[-2ex]
		
	\item
		For $w \in O$, the $Z$-dependency of $w$ on an arbitrary qubit $v$ can be computed as the cumulated contributions from each possible number $\ell$ of signal shifts, which for each $\ell \ge 1$ is given by
		\begin{gather}
				\unit_w \cdot (T^\ell \,\unit_v);
		\end{gather}
		then, similarly to the previous case, the cumulative dependency over all signal shifting operations is then given by
		\begin{align}
				\unit_w \cdot \paren{\sum_{\ell \in \N} T^\ell\, \unit_v}
			\;\;=&\;\;
				\unit_w \cdot \paren{\idop_n - T}\inv\, \unit_v	\,,
		\end{align}
		as $\idop - T$ is non-singular.
		Again, if this evaluates to zero modulo $2$, there is no $Z$-dependency of $w$ on $v$ in the pattern $P$.
	\end{romanum}
	After an ``infinite'' number of such dependency propagations performed on the initial procedure $\tilde{P}$ (equivalent to some finite number of shifts, with subsequent propagations having no effect), we obtain the measurement procedure $P$.

	Note that the coefficients of $F$ corresponds to adapting segments for $f$ of type~\iti, and those of $T$ correspond to adapting segments of type~\itii\ and~\itiii.
	In the procedure $P$, possible dependencies of the signs of measurement angles for a qubit $w$ depending on $\s[v]$, of potential $X$ corrections on a qubit $w$ depending on $\s[v]$, exist then if there are an odd number of walks from $v$ to $w$ consisting of some number of type~\itii\ and~\itiii\ adapting segments followed by a type~\iti\ segment; and possible $Z$ corrections on $w$ depending on $\s[v]$ exist if there are an odd number of walks from $v$ to $w \in O$ consisting entirely of type~\itii\ and~\itiii\ adapting segments.
	This proves the Theorem.
\end{proof}

\begin{algorithm}[t]

\PROCEDURE \TestDependencies(P, f) :

\Input{%
	$P$, a standardized one-way measurement pattern;
	\\
	$f$, a flow-function for the geometry $(G,I,O)$ underlying $P$
}

\Output{%
	A boolean result indicating if $P$ is consistent with the constraints of Theorem~\ref{thm:dependenciesDKP}.
}

\BEGIN
	\algline
	
	\LET $T \gets $ [\,matrix over $\Z_2$ depending on $f$ described in Eqn~\eqref{eqn:shiftPropagator}\,]\;
	\LET $F \gets $ [\,matrix over $\Z_2$ depending on $f$ described in Eqn~\eqref{eqn:flowPropagator}\,]\;
	\LET $D \gets (\idop_n - T)\inv$\;
	\LET $E \gets F D$\;
	\FOREACH measurement or correction operation $\Op{w}{B}$ in $P$ \DO
		\algline
		\LET $\xType \gets ((\Op{}{} = \Meas{}{}) ~\ORELSE~ (\Op{}{} = \Xcorr{}{}))$\;
		\LET $\zType \gets (\Op{}{} = \Zcorr{}{})$\;
		\LET $S \gets \ens{v \in V(G) \,\big|\, \text{$\s[v]$ occurs in $B$}}$\;		\nllabel{line:scanDependencies}

		\medskip
		{\FOREACH $v \in V(G)$ \DO
			\IF $v \in S$ \THEN {
				\short
				\IF $(\xType ~\ANDALSO~ E_{wv} = 0) ~\ORELSE~ (\zType ~\ANDALSO~ D_{wv} = 0)$ \THEN {\RETURN \false}
				\relax
			}
			\ELSE {
 				\short
				\IF $(\xType ~\ANDALSO~ E_{wv} = 1) ~\ORELSE~ (\zType ~\ANDALSO~ D_{wv} = 1)$ \THEN {\RETURN \false}
				\relax
			}
		\ENDFOR}
	\ENDFOR

	\medskip
	\RETURN \true\;
\END

	\caption[An algorithm to test whether the dependencies of a standardized one-way measurement pattern based measurement procedure is consistent with those produced by the DKP construction.]{\label{alg:testDependencies}%
		An algorithm to test whether the dependencies of a standardized one-way measurement pattern based measurement procedure is consistent with those produced by the DKP construction (see Theorem~\ref{thm:dependenciesDKP}).
		}

\end{algorithm}

\label{discn:testDependencies}
Algorithm~\ref{alg:testDependencies} describes an algorithm which, provided a one-way measurement pattern $P$ in standard form (and without shift operators) and a flow $f$ for the geometry underlying $P$, tests if the dependencies of $P$ are consistent with the flow-function those described in Theorem~\ref{thm:dependenciesDKP}.
Let $N$ be the length of the measurement pattern $P$, and let $n = \card{V(G)}$, $m = \card{E(G)}$, and $k = \card{O}$.
There will then be exactly $m$ entangling operations, $n -k$ measurement operations, and $k$ measurement operations in $P$, where the latter two classes of operations may in principle depend on arbitrary subsets of $V(G)$; then, the length of the pattern is $N \in O(m + n^2) = O(n^2)$.
The time required to construct the matrices $T$ and $F$ is $O(n^2)$, and the time to compute $D$ and $E$ (using \eg\ the algorithm of~\cite{PMH08} to perform matrix inversion modulo $2$) is $O(n^3 / \log n)$.
The set $S$ constructed on line~\ref{line:scanDependencies} may be constructed in time $O(n)$\,; then, the run-time of Algorithm~\ref{alg:testDependencies} is dominated by the nested for loops and repeated construction of $S$, both of which requires $O(Nn) \subset O(n^3)$ time in total to execute.

The characterization of dependencies between qubits in a pattern $P \in \img(\cR)$ is the penultimate step towards obtaining a semantic map for $\cR$.
Together with an algorithm to find flows, Algorithm~\ref{alg:testDependencies} provides an efficient algorithm for testing membership in $\img(\cR)$\,: however, the proof of this claim will await the construction of a complete semantic map for $\cR$.

\subsection{Star decomposition of geometries with flows}
\label{sec:starDecomposition}

For a geometry $(G,I,O)$ with a flow-function $f$, the influence relation $\inflRel$ for $f$ suggests a decomposition of $(G,I,O)$ into smaller geometries similar to the star decomposition of circuits described in Section~\ref{sec:flowDefn}.
For each vertex $v \in V(G)$, define the vertex-set
\begin{align}
		V_{v \star}
	\;=&\;
		\ens{w \in V(G) \,\Big|\, w = v \text{~or~} \big[v \inflRel w \text{~and~} w \ne f(f(v)) \big]} \,,
\end{align}
and from these, define the geometry
\begin{subequations}
\label{eqn:starGeometry}
\begin{gather}
			G_{v \star}
		\;\;=\;\;
			G[V_{v \star}]	\,-\, \ens{xw \,\big|\, x,w \ne f(v)}\,,
		\\[1.5ex]
			I_{v \star}
		\;\;=\;\;
				V_{v \star} \setminus \ens{f(v)}	\,,
		\\[1ex]
			O_{v \star}
		\;\;=\;\;
				V_{v \star} \setminus \ens{v}	\,.
\end{gather}
\end{subequations}
The graph $G_{v\star}$ consists of a star graph consisting of a single edge $v w$ for $w = f(v)$, and some additional edges $wx$ for $w = f(v)$ and $v \inflRel x$.
We will call the geometry $(G_{v \star}\,, I_{v \star}\,, O_{v \star})$ a \emph{star geometry}\footnote{%
Star geometries as we have defined them are similar to, but distinct from, the geometries underlying \emph{star patterns} in~\cite{DK06}.
The decomposition result of Lemma~\ref{lemma:starDecomp} was motivated by the discussion of star geometries in that article.}
with \emph{root} $v$, or with \emph{center} $f(v)$.

\begin{lemma}
	\label{lemma:starDecomp}
	Let $(G,I,O)$ be a geometry with flow-function $f$.
	Then there is a decomposition of $(G,I,O)$ into star geometries on each of the vertices $v \in O\comp$, together with the geometry $\big(G\big[\,\tilde{I}\,\big], I, \tilde{I}\,\big)$, where $\tilde{I} = V(G) \setminus \img(f)$.
\end{lemma}
\begin{proof}
	We proceed by induction on the size of $\img(f)$.
	If $\img(f) = \vide$, then $\dom(f) = \vide$ as well, in which case $I \subset V(G) = O$; we then have $(G,I,O) = \big(G\big[\,\tilde{I}\,\big], I, \tilde{I}\,\big)$.
	Otherwise, suppose that the claim holds for geometries $(G,I,O)$ with flow-functions $f$ in which $\card{\img(f)} \le N$ for some $N \in \N$, and that $\card{\img(f)} = N+1$.
	Because $f$ is a flow function, $\sI_f$ is acyclic; then, there is a vertex $v \in V(G)$ such that $v \inflRel w$ only for $w$ maximal in the natural pre-order $\preceq$.
	Let $\cG_{v\star} = (G_{v\star}\,, I_{v\star}\,, O_{v\star})$, and $\cG' = (G', I, O')$, where
	\begin{gather}
		\label{eqn:geometryReduce}
			G'
		\;=\;
			G \setminus f(v)	\,,\quad\text{and}
		\qquad	
			O'
		\;=\;
			O \symdiff \ens{v, f(v)}	\,.
	\end{gather}
	Note that $O_{v\star} \subset O$ by the choice of $v$, and $I_{v\star} = O_{v \star} \symdiff \ens{v, f(v)} \subset O \symdiff \ens{v, f(v)}$\,.
	Because $V(G_{v\star}) = I_{v\star} \union \ens{f(v)}$, we then have $V(G') \inter V(G_{v\star}) \subset I_{v\star} \inter O'$\,: then the composition $(\bar{G}, \bar{I}, \bar{O})  \,=\, \cG_{v\star} \circ \cG'$ is well-defined, and we have
	\begin{subequations}
	\begin{align}
			\bar{G}
		\;\;=&\;\;
			G_{v\star} \union G' 
		\notag\\=&\;\;
			\ens{xw \in G \,\big|\, w = f(v)} \union \Big[ G \setminus f(v) \Big] 
		\;\;=\;\;
			G	\,;
		\\[2ex]
			\bar{I}
		\;\;=&\;\;
			\big( I_{v\star} \setminus O' \big) \union I
		\;\;=\;\;
			\vide \union I
		\;\;=\;\;
			I	\,;
		\\[2ex]
			\bar{O}
		\;\;=&\;\;
			O_{v\star} \union \big( O' \setminus I_{v\star} \big)
		\;\;=\;\;
			O_{v\star} \union \big( O \setminus O_{v\star} \big)
		\;\;=\;\;
			O	\,.
	\end{align}
	\end{subequations}
	Then, we have $(G,I,O) = \cG_{v\star} \circ \cG'$\,.
	The restriction $f'$ of the flow-function $f$ to $\cG'$ does not include $f(v)$, and therefore has cardinality at most $N$\,; then, $\cG'$ has a decomposition into star geometries and the geometry $(G' \setminus \img(f'), I, V(G') \setminus \img(f'))$.
	Finally, note that $V(G) \setminus \img(f') = V(G) \setminus \img(f) = \tilde{I}$, by the definition of $f'$\,: the lemma then holds by induction.
\end{proof}

\begin{figure}[p]
	\begin{center}
		\includegraphics{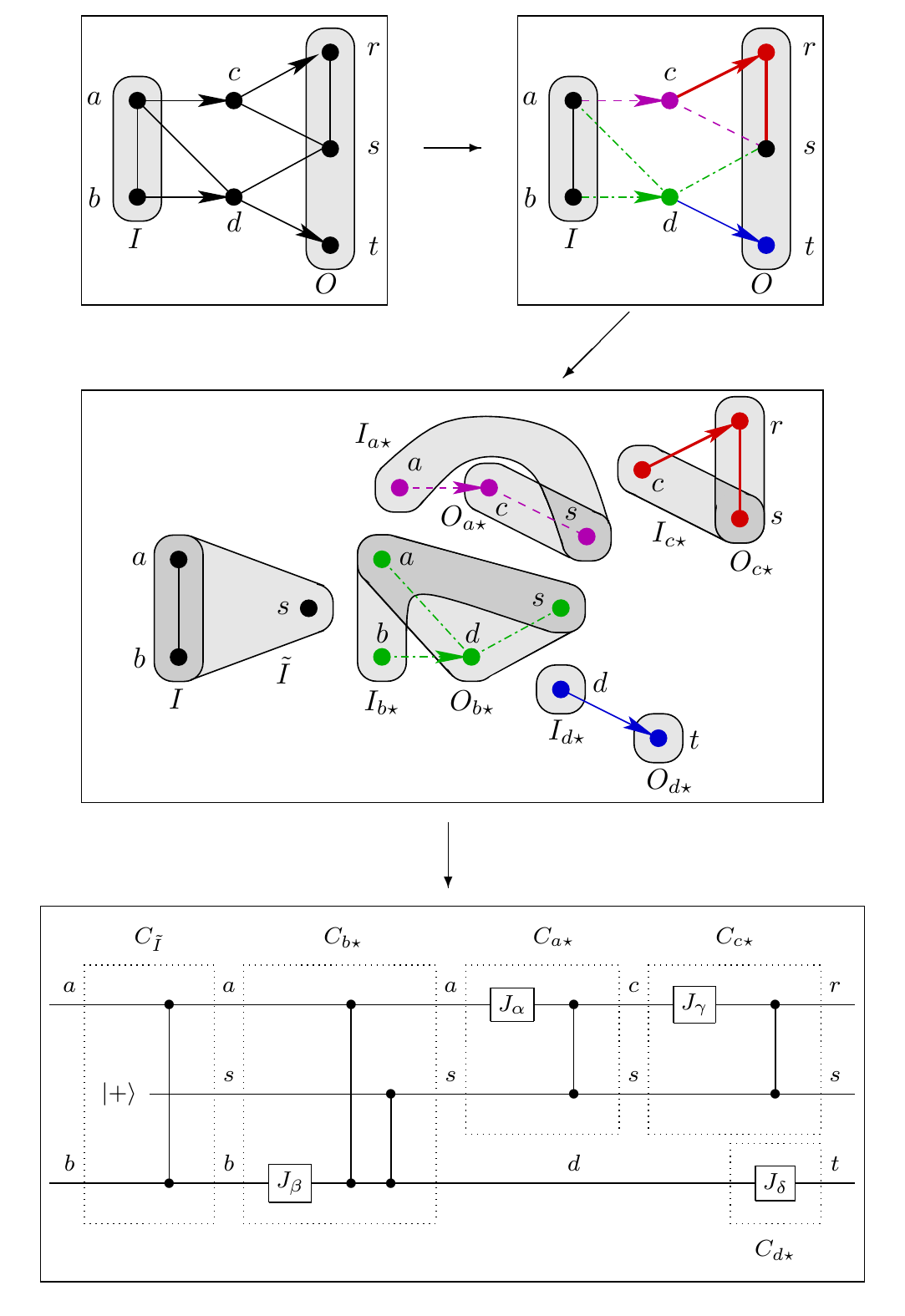}
	\end{center}
	\caption[Illustration of a decomposition of a geometry with a flow-function into star geometries.]{\label{fig:starDecomp}\colouronline
		Illustration of a decomposition of a geometry $(G,I,O)$ with a flow-function $f$ (denoted by arrows in the upper three panels) into star geometries for each $v \in O\comp$, and a reduced geometry $\big(G\big[\,\tilde I\,\big], I, \tilde I\big)$ for $\tilde{I} = V(G) \setminus \img(f)$.
		The middle panel illustrates each star geometry separately, located next to the geometries with which it may be composed to form the original geometry $(G,I,O)$; vertices are repeated for each geometry in which they occur.
		The bottom panel illustrates the unitary circuits corresponding to these star geometries (including a labelling of wire-segments, \ie\ of stable tensor indices), and the complete circuit obtained from composing them.}
\end{figure}
Figure~\ref{fig:starDecomp} illustrates a decomposition of a geometry with a flow into star geometries in the manner described above.

For a geometry $(G,I,O)$ with a given flow-function $f$, we will say that a star geometry $(G_{v\star}\,, I_{v\star}\,, O_{v\star})$ is \emph{maximal} in $(G,I,O)$ if, as in the proof of Lemma~\ref{lemma:starDecomp} above, every $w \in V(G)$ such that $v \inflRel w$ is maximal in $\preceq$\,; then, we may decompose $(G,I,O)$ by extracting maximal star geometries.
The maximal elements of $V(G)$ in the natural pre-order are exactly the elements of $O$ (as $z \preceq f(z)$ for any $z \in O\comp$).
Thus, the root $v$ of a maximal star geometry in $(G,I,O)$ is one such that all neighbors $z \sim f(v)$ are elements of $O$ or equal to $v$ itself.
There always exists such a maximal star for a geometry with a flow: an independent observation of a similar nature has since given rise to an improved algorithm for finding flows, which we describe in Section~\ref{sec:improvedFlowAlgs}.

\label{discn:maximalStarVertex}
Algorithm~\ref{alg:maximalStarVertex} describes an procedure to obtain a vertex $w = f(v) \in V(G)$ which is the center of a maximal star geometry in some geometry $(G[V'], V' \setminus W', O')$, using the fact that $w$ is the center of such a star geometry if and only if all neighbors of $w$ in the graph $G[V']$ are either equal to $v$ or are elements of $O'$.
Having found a vertex at the center of a maximal geometry of $(G,I,O)$, we may obtain the decomposition $(G,I,O) = \cG_{v\star} \circ \cG'$ by constructing these geometries according to~\eqref{eqn:starGeometry} and~\eqref{eqn:geometryReduce}, respectively.
The run-time of Algorithm~\ref{alg:maximalStarVertex} is taken up by constant-time comparisons for up to all of the neighbors of every element of $O'$, and so requires time
\begin{align}
		O\paren{\sum_{w \in O'} \deg(w)}	
	\;\;\subset\;\;
		O\paren{\sum_{w \in V(G)} \deg(w)}
	\;\;=&\;\;\;
		O(2\card{E(G)})
	\notag\\=&\;\;
		O(m)
	\;\;\subset\;\;
		O(k n)	\;,
\end{align}
by the extremal bound of Theorem~\ref{thm:extremalUpperBound}.

\begin{algorithm}[t]
	\PROCEDURE \MaximalStarVertex(G, V', W', O', f):

	\Input{%
		$G$, a graph underlying a geometry $(G,I,O)$;
	\\
		$V', W', O'$, sets describing a sub-geometry $(G[V'], V' \setminus W',O')$;
	\\
		$f: O\comp \to I\comp$, a flow-function function for $(G,I,O)$
	}

	\Output{%
		A vertex $w \in W'$ such that every neighbor $v$ of $w$ such that $v \in V'$ either satisfies $v = g(w)$ or $v \in O'$
	}

	\BEGIN
		\algline
		\FOREACH $w \in W'$ \DO
			\LET $\maximalStar \gets \true$\;
			{\FORALL $v \sim w$ \DO
				\IF $v \in V' \setminus O'$ \ANDALSO $w \ne f(v)$ \THEN {%
					$\maximalStar \gets \false$\;
					\ESCAPE\;
				}\relax
			\ENDFOR}

			\smallskip
 			\oneline
			\IF $\maximalStar$ \THEN \RETURN $w$\;
		\ENDFOR
	\END

	\caption{\label{alg:maximalStarVertex}%
		An algorithm to obtain a vertex $v$ which is the center of a maximal star geometry of a specified geometry.}

\end{algorithm}

\subsection{A semantic map from star decompositions}

Star geometries $(G_{v\star}\,, I_{v\star}\,, O_{v\star})$ correspond to simple unitary circuits in the following manner.
We may construct a circuit $C_{v\star}$ with an input wire for each qubit $a \in I_{v \star}$\,, and which performs the transformation
\begin{gather}
		C_{v\star}
	\;\;=\;\;
		\paren{\prod_{\substack{w \in O_{v\star} \\[0.25ex] w \sim f(v)}} \cZ \pseu[w, f(v)] } J(\alpha) \pseu[:f(v)/v]
\end{gather}
expressed in stable index notation (pages~\pageref{sec:nonstandardRepnsEnd}\thru\pageref{sec:nonstandardRepnsEnd}); we map the labels of the vertices in the geometries to the tensor index names in the expansion, and the angle $\alpha$ is a parameter which we may relate to the measurement angle for the qubit $v$.
Similarly, we may define a circuit $C_{\tilde I}$\ which corresponds to the geometry $(G[\,\tilde{I}\,], I, \tilde{I})$\,:
\begin{gather}
		C_{\tilde I}
	\;\;=\;\;
		\paren{\prod_{\substack{vw \in E(G[\,\tilde I\,])}} \cZ \pseu[v,w] } \paren{\prod_{v \in \tilde I \setminus I} \ket{+} \pseu[:v/] }	\;.
\end{gather}
We will call such circuits the \emph{star circuits} corresponding to these geometries; these are the same circuits described in Section~\ref{sec:flowDefn} in the star decomposition of circuits.
Examples of such circuits for a particular geometry $(G,I,O)$ are illustrated in the bottom panel of Figure~\ref{fig:starDecomp}, in which they are composed together to form a larger circuit.
In that example, we may recover the geometry in the upper left-hand panel by composing
\begin{align}
	\begin{split}
		(G,I,O)
	\;=\;
		(G_{c\star}\,, I_{c\star}\,, O_{c\star})
		\circ
		(G_{d\star}\,, I_{d\star}\,,& O_{d\star})
		\circ
		(G_{a\star}\,, I_{a\star}\,, O_{a\star}) \circ
	\\&
		(G_{b\star}\,, I_{b\star}\,, O_{b\star})
		\circ
		\big(G\big[\,\tilde{I}\,\big], I, \tilde{I}\big)	\,;
	\end{split}
\end{align}
and similarly, the larger circuit $C$ in Figure~\ref{fig:starDecomp} is obtained by composing
\begin{align}
		C
	\;=\;
		C_{c\star} \circ C_{d\star} \circ C_{a\star} \circ C_{b\star} \circ C_{\tilde I}	\;.
\end{align}
Algorithm~\ref{alg:semanticDKP} describes a procedure $\SemanticDKP(G,I,O,P)$ which provides a semantic map $\cS$ for the DKP representation map $\cR$.
\begin{algorithm}[p]

\PROCEDURE \SemanticDKP(G,I,O,P) :

\Input{%
		$P$, a standardized one-way measurement pattern, with an equal number of input and output qubits;
	\\
		$(G,I,O)$, the geometry underlying $P$ (in particular, $\card I = \card O$).
}

\Output{%
	Either \nil, if $(G,I,O)$ has no flow or the dependencies of the operations of $P$ are not consistent with the constraints of Theorem~\ref{thm:dependenciesDKP}; or a unitary circuit $C$ performing the same transformation as $P$, otherwise.}

\BEGIN
	\LET $\flow \gets \FindFlow(G,I,O)$\;
	\short
	\IF $\flow = \nil$ \THEN {%
		\RETURN \nil																									\nllabel{line:flowFail}
	}
	\ELSE {
		\LET $(f,g,\partialOrd) \gets \flow$
	}

	\medskip

	\LET $\theta: V(G) \to \R$\vspace{-1ex}\;
	\oneline
	\FOREACH measurement operation $\Meas{v}{\alpha;B}$ in $P$ \DO
		$\theta(v) \,\gets\, -\alpha$\;
	\ENDFOR
	\oneline
	\IF $\TestDependencies(P,f) = \false$ \THEN \RETURN \nil\;										\nllabel{line:dependenciesFail}

	\medskip

	\LET $O' \gets O$\;																							\nllabel{line:initGeometry}
	\LET $V' \gets V(G)$\;
	\LET $W' \gets V(G) \setminus I$\;																		\nllabel{line:initGeometryEnd}
	\LET $C 	\gets $ [\,the empty circuit \,]\;
	\medskip

	\algline
	\WHILE $W' \ne \vide$ \DO																					\nllabel{line:circuitConstrLoop}
		\LET $w \gets \MaximalStarVertex(G,V',W',O',f)$\vspace{0.1ex}\;
		\LET $v \gets g(w)$\vspace{0.3ex}\;
		\LET $C' \gets  J(\theta(v)) \pseu[:w/v]$\,\;
		\oneline
		\FORALL $z \sim w$ \SUCHTHAT $z \ne v$ \DO
			$C' \gets \cZ \pseu[w,z] \circ C'$\,\;												\nllabel{line:czNeighbors-1}
		\ENDFOR
		\medskip
		$C \gets C \circ C'$\,\;
		$V' \gets V' \setminus \ens{w}$\;																	\nllabel{line:reduceVprime}
		$W' \gets W' \setminus \ens{w}$\;																	\nllabel{line:reduceWprime}
		$O' \gets \big(O' \setminus \ens{w}\big) \union \ens{v}$\; 
			
	\ENDWHILE																										\nllabel{line:circuitConstrLoopEnd}
	\medskip

	\LET $G' \gets G[V']$\;
	\FORALL $v \in V'$ \DO																						\nllabel{line:circuitInputLoop}
		\oneline
		{\FORALL $w \in V'$ \SUCHTHAT $vw \in E(G')$ \DO
			\vspace{0.3ex}
			$C \gets C \circ \cZ \pseu[v,w]$\,\vspace{-0.2ex}\;								\nllabel{line:czNeighbors-2}
		\ENDFOR}
		$V' \gets V' \setminus \ens{v}$\,;
	\ENDFOR																									\nllabel{line:circuitInputLoopEnd}

	\medskip
	\RETURN $C$\;																								\nllabel{line:constructedCircuit}
\END

	\caption[An algorithm for the semantic map corresponding to the DKP representation map.]{\label{alg:semanticDKP}%
		An algorithm for the semantic map $\cS$ corresponding to the DKP representation map $\cR$, in the case of one-way measurement patterns with an equal number of input and output qubits, using the procedures defined in Algorithms~\ref{alg:completeFlowAlg}, \ref{alg:testDependencies}, and~\ref{alg:maximalStarVertex} (pages~\pageref{alg:completeFlowAlg}, \pageref{alg:testDependencies}, and~\pageref{alg:maximalStarVertex} respectively).}
\end{algorithm}%
We will show:
\begin{lemma}
	\label{lemma:semanticDKP}
	Let $P$ be a standardized one-way measurement pattern with underlying geometry $(G,I,O)$, where $\card{I} = \card{O}$.
	Then $\SemanticDKP(G,I,O,P)$ returns a unitary circuit $C$ such that $\cR(C) \cong P$ if $P$ is congruent to some element of $\img(\cR)$, and returns $\nil$ otherwise.
\end{lemma}
\begin{proof}
As noted before, because we can perform the DKP construction by first applying the \emph{simplified} DKP construction, we know that $(G,I,O)$ has a flow if $P \in \img(\cR)$\,.
Similarly, if $P \in \img(\cR)$, then the dependencies of $P$ will be consistent with the constraints imposed by Theorem~\ref{thm:dependenciesDKP} for some flow function.
In the case where $\card I = \card O$, there is at most one flow-function; so if $P \in \img(\cR)$, its dependencies will in particular be consistent with the constraints of Theorem~\ref{thm:dependenciesDKP} for any flow-function discovered for $(G,I,O)$ by $\FindFlow$.
Neither of these are affected by commuting independent operations of $P$ past one another.
Therefore, $\SemanticDKP(G,I,O,P)$ returns $\nil$ at lines~\ref{line:flowFail} or~\ref{line:dependenciesFail} of Algorithm~\ref{alg:semanticDKP}, $P$ is not equivalent to any $P' \in \img(\cR)$.

On lines~\ref{line:initGeometry}\thru\ref{line:initGeometryEnd}, we initialize vertex sets $V'$, $W'$, and $O'$ which will represent various sub-geometries of $(G,I,O)$ of the form
\begin{gather}
	(G[V'], V' \setminus W', O')	\,.
\end{gather}
(More precisely, $W'$ represents $\img(f')$, for successive restrictions $f'$ of the flow-function $f$ to $V'$\,; in the case $\card I = \card O$, this is equivalent to $V' \setminus I$.)
We also initialize an empty circuit $C$ to which we pre-compose other circuits in the loops on lines~\ref{line:circuitConstrLoop}\thru\ref{line:circuitConstrLoopEnd} and lines~\ref{line:circuitInputLoop}\thru\ref{line:circuitInputLoopEnd}, represented in stable index notation; we will sometimes refer to the tensor indices as ``wire segments'' (see the discussion on page~\pageref{discn:stableIndexPathIntegral}) for the sake of clarity.
We will consider the circuits which we compose to $C$ in reverse order, considering sub-circuits proceeding from the inputs to the outputs of the final circuit returned on line~\ref{line:constructedCircuit}.

\begin{itemize}
\item
	In the loop on lines~\ref{line:circuitInputLoop}\thru\ref{line:circuitInputLoopEnd}, we compose to $C$ one controlled-$Z$ operation between wire segments labelled by the vertices contained in $V'$ at that stage of the algorithm, for each edge in the graph $G' = G[V']$.
	This loop is only entered once $W' = \vide$\,; as we only transform $W'$ by removing one vertex $w$ at a time on line~\ref{line:reduceWprime} (which we also remove from $V'$ on line~\ref{line:reduceVprime}), the value of $V'$ just before line~\ref{line:circuitInputLoop} is $V' = V(G) \setminus \img(f)$.
	Thus, the cumulative effect of lines~\ref{line:circuitInputLoop}\thru\ref{line:circuitInputLoopEnd} is to compose $C_{\tilde I}$ to the beginning of $C$, for $\tilde{I} = V(G) \setminus \img(f)$.

\item
	Using the procedure $\MaximalStarVertex$ defined in Algorithm~\ref{alg:maximalStarVertex}, each iteration of the loop on lines~\ref{line:circuitConstrLoop}\thru\ref{line:circuitConstrLoopEnd} produces a vertex $w$ which is the center of a maximal star geometry of $(G[V'], V' \setminus W', O')$.
	For each such $w$, we obtain $v = g(w)$, and construct the star circuit $C'$ corresponding to the geometry $(G_{v\star}\,, I_{v\star}\,, O_{v\star})$, which we append to the input of $C$.
	We then reduce the geometry $(G[V'], V' \setminus W', O')$ by setting $V' \gets V' \setminus \ens{w}$, and $O' \gets O' \symdiff \ens{v, w}$, corresponding to the reduction described in~\eqref{eqn:geometryReduce}; we also reduce $W' \gets W' \setminus \ens{w}$, corresponding to a further restriction of $f$ to the new value of $V'$.

	By induction on $\img(f)$, as in the proof of Lemma~\ref{lemma:starDecomp}, this sequence of star geometries leads to a decomposition of the original geometry $(G,I,O)$ into star geometries, leaving behind the geometry $(G[\tilde I], I, \tilde I)$\,.
	For each such star geometry, we obtain the star circuit and compose it to the input end of $C$\,; then, prior to line~\ref{line:circuitInputLoop}, $C$ consists of a composition of the star circuits corresponding to the geometries in a star decomposition of $(G,I,O)$.
\end{itemize}
Thus, the value of $C$ on line~\ref{line:constructedCircuit} is a composition of the star circuits corresponding to the decomposition of $(G,I,O)$ into star geometries, where in particular for each gate $J(\theta(v))\pseu[:w/v]$\,, we have $\theta(v) = -\alpha_v$, for $\alpha_v$ the angle of the measurement on $v$ in the case where all of the preceding measurements $\Meas{z}{}$ produced the result $\s[z] = 0$.

By construction, the output is a circuit $C$ constructed with the elementary gate set $\Jacz(\A)$ for some $\A \subset \R$.
Such a circuit is in the domain of $\cR$\,; consider then the value of $\bar{P} = \cR(C)$ for such a circuit $C$, and let $(\bar G, \bar I, \bar O)$ be the geometry underlying $\bar P$.
\begin{itemize}
\item
	The DKP construction preserves the existing wire segment labels of $C$ as the labels of vertices in $\bar G$, $\bar I$, and $\bar O$, for the geometry $(\bar G, \bar I, \bar O)$ underlying the measurement pattern $\bar P$.
	As the labels of the wire-segments in $C$ are precisely the vertex-labels of $G$, with the non-advanced indices arising from $I$ and the non-deprecated indices arising from $O$, we then have $V(\bar G) = V(G)$, $\bar I = I$, and $\bar O = O$.

\item
	Because the flow function $f$ for $(G,I,O)$ is used to define the correspondence between deprecated and advanced indices in $C$, and (as we noted in Section~\ref{sec:flowDefn}) this correspondence is a flow-function for $(\bar G, \bar I, \bar O)$ arising from $C$ by the DKP construction, the same function $f$ is then a flow-function for $(\bar G, \bar I, \bar O)$ as well.
	In particular, for each pair of deprecated/advanced indices $v, w$ in $C$, there is an edge $vw \in E(\bar G)$.

\item
	For every operation $\cZ \pseu[v,w]$ in $C$, there will be an edge $vw \in E(\bar G)$ which is not a flow-edge for the function $f$.
	Because each such controlled-$Z$ operation can be attributed to a star circuit for a unique star geometry in a decomposition of $(G,I,O)$ --- as well as for the flow edges arising from the $J(\theta)$ operations --- we then have $E(\bar G) = E(G)$, so that $\bar G = G$.

\item
	Because the angle $\theta$ of the $J(\theta)$ gate between wire segments $v$ and $f(v)$ is the negative of the default measurement angle of $v$ in the pattern $P$ (by the construction of $\theta$) and the negative of the default measurement angle of $v$ in the pattern $\bar P$ (by the construction of $\bar P$), each qubit $v \in O\comp$ has the same default measurement angle in $\bar P$ as in $P$.

\item
	By Theorem~\ref{thm:dependenciesDKP}, the measurement dependencies in $\bar P$ are described by the matrices $F(\idop_n - T)\inv$ and $(\idop_n - T)\inv$, for matrices $T$ and $F$ given by~\eqref{eqn:shiftPropagator} and~\eqref{eqn:flowPropagator} respectively, depending on the flow-function $f$.
	However, $P$ satisfies the same constraints by the test on lines~\ref{line:dependenciesFail}.
	It follows that the measurement operations on $v$ in both $P$ and $\bar P$ are the same for $v \in O\comp$, and the correction operations are similarly identical for $v \in O$, including dependencies on other measurement results.
\end{itemize}
Thus, $P$ and $\bar P$ contain the same operations, albeit possibly in a different order.
Finally, because $P$ is well-formed by assumption, and $\bar P$ is so by construction, the only possible difference in the ordering of corresponding operations between $P$ and $\bar P$ is in the commuting of independent terms by the type composition constraints of Lemma~\ref{lemma:typeCompose}.
Thus, $P$ is congruent to $\bar P$\,; and in particular, $\SemanticDKP(G,I,O,P)$ returns $\nil$ only if $P$ is not congruent to an element of $\img(\cR)$.
\end{proof}

The above Lemma then allows us to show:

\begin{theorem}
	Let $\tilde \cR$ be the restriction of the DKP construction map $\cR$ to to unitary circuits without terminal measurements or trace-out, and with the same number of output wires as input wires; and let $\cS$ be the map taking one-way measurement pattern $P$ to the circuit $\SemanticDKP(G,I,O,P)$, where $(G,I,O)$ is the geometry underlying $P$.
	Then $\cS$ is the semantic map for $\tilde\cR$\;.
\end{theorem}
\begin{proof}
	Taken as a partial function on its input data --- and neglecting the trivial detail that its domain is on tuples $(G,I,O,P)$, rather than just on patterns $P$ --- the domain of $\SemanticDKP$ is the set of well-formed measurement procedures which are congruent to elements of $\img(\tilde\cR)$, which in particular contains $\img(\tilde\cR)$; and each element of $\img(\tilde\cR)$ will itself be mapped by $\tilde\cR \circ \cS$ to a congruent one-way measurement pattern.
	Thus, $\SemanticDKP$ satisfies property~\ref{condition:retraction} on page~\pageref{condition:retraction}.

	For a circuit $C$ for a unitary bijection composed from the gate set $\Jacz(\A)$ for some $\A \subset \R$, let $\tilde P = \tilde\cR(C)$ and $\tilde C = \cS(\tilde P)$\,, and let $(G,I,O)$ be the geometry underlying $\tilde P$; then, consider the correspondence between the elements of $C$, $\tilde P$, and $\tilde C$.
	There is by construction a bijective correspondence between the wire segments of $C$ to the vertices of $V(G)$, and from $V(G)$ to the wire segments of $\tilde C$\,.
	By construction, this bijective correspondence is an isomorphism of stable index expansions:
	\begin{romanum}
	\item
		Successive wire-segments in $C$ correspond to paths $v, f(v), f^2(v)$ in a path-cover $\cP_f$ in $G$, which in turn correspond to successive path segments in $\tilde C$\,.

	\item
		Wire-segments in $C$ which interact via a $\cZ$ gate in $C$ correspond to neighbors $v \sim w$ in $G$ such that $v \ne f(w)$ and $w \ne f(v)$\,, which in turn give rise (via the contribution of some star geometry in the decomposition of $(G,I,O)$, as remarked in the analysis of the preceding section) to wire segments in $\tilde C$ which interact via a $\cZ$ gate.
		As there are no other operations in $\tilde C$ than this, and as every operation in $C$ gives rise to an edge in $G$, the interaction graphs of $C$ and $\tilde C$ are the same.
		Because $C$ and $\tilde C$ are both defined on the same gate-set, which in particular is parsimonious (in which any multi-qubit gate preserves standard-basis states), there is a homomorphism of stable-index expressions from $C$ to $\tilde C$ (where in particular the index labels are identical).

	\item
		The wire segments in $C$ as in $\tilde C$ are separated by gates of the form $J(\theta)$\,; and by construction, in both $C$ and $\tilde C$, the angles $\theta$ of the $J(\theta)$ gates are the negative of the default measurement angle of the qubit $v$ in $\tilde P$, where $v$ is the label of the deprecated index for the $J(\theta)$ gate for $C$ and $\tilde C$ alike.
		As there is only one variety of two-qubit gate, the homomorphism between $C$ and $\tilde C$ is then an isomorphism.
	\end{romanum}
	Thus, $C$ is isomorphic to $\tilde C$ as a quantum circuit, and differ only in the ordering of commuting gates.
	Thus, the map $\cS$ is a semantic map for $\tilde\cR$ in the sense defined on page~\pageref{discn:representationMap}.
\end{proof}

\subsubsection{Remarks and run-time analysis}

The run-time for $\SemanticDKP(G,I,O,P)$ can be accounted for as follows, for $\card{V(G)} = n$, $\card{I} = \card{O} = k$, and $\card{P} = N$:
\begin{itemize}
\item
	As we showed on pages~\pageref{thm:findFlow} and~\pageref{discn:testDependencies}, $\FindFlow(G,I,O)$ requires time $O(k^2 n)$ and $\TestDependencies(P,f)$ requires time $O(Nn) \subset O(n^3)$.

\item
	For each iteration of the loop on lines~\ref{line:circuitConstrLoop}\thru\ref{line:circuitConstrLoopEnd}, we require time $O(kn)$ to run $\MaximalStarVertex(G,V',W',O',f)$, by the discussion on page~\pageref{discn:maximalStarVertex}, and time $O(\deg(v))$ (for a different $v \in O\comp$ in each iteration) for the loop on line~\ref{line:czNeighbors-1}; all other operations require constant time.
	The cumulative run time of the loop is then
	\begin{align}
			O\paren{\sum_{v \in V(G) \setminus \img(f)} \Big[ kn + \deg(v) \Big]}
		\;\;\subset&\;\;
			O\paren{\sum_{v \in V(G)} kn \quad \!\!+ \sum_{v \in V(G)} \deg(v) }
		\notag\\[1ex]=&\;\;
			O(kn^2 + 2 \card{E(G)})
		\notag\\[1ex]=&\;\;
			O(kn^2 + 2m)
		\;\;=\;\;
			O(kn^2)	\,,
	\end{align}
	by the extremal result of Theorem~\ref{thm:extremalUpperBound}.

\item
	For each iteration of the loop on lines~\ref{line:circuitInputLoop}\thru\ref{line:circuitInputLoopEnd}, we perform at most $\deg(v)$ iterations of the loop on line~\ref{line:czNeighbors-2}, plus one constant-time set removal.
	As noted above, just prior to line~\ref{line:circuitInputLoop}, we have $V' = I$\,.
	The cumulative run time of this loop is then
	\begin{align}
			O\paren{\sum_{v \in I} \deg(v)}	
		\;\;\subset\;\;
			O\paren{\sum_{v \in V(G)} \deg(v)}
		\;\;=&\;\;\;
			O(2\card{E(G)})
		\notag\\=&\;\;
			O(m)
		\;\;\subset\;\;
			O(k n)	\;,
	\end{align}
	again by the extremal bound of Theorem~\ref{thm:extremalUpperBound}.
\end{itemize}
The time required by $\SemanticDKP(G,I,O,P)$ is then dominated by the time required by $\TestDependencies(P,f)$, which is $O(n^3)$.

\section{Further developments}
\label{sec:flowsFurtherResults}

The Algorithm~\ref{alg:completeFlowAlg} of Section~\ref{sec:flowAlgs} is (a revision of) the first efficient algorithm to determine whether a broad class of geometries have a flow.
Since the presentation of the original version of that algorithm in~\cite{Beaudrap06}, improved algorithms --- and extensions to the concept of a flow itself --- have been introduced.
As well, flows provide us additional tools to compare the DKP and RBB constructions of one-way measurement patterns.
In this penultimate section of Chapter~\ref{Semantics for unitary one-way patterns}, we present an overview of these results to put the earlier results of this chapter into context.

\subsection{Flows in relation to more general measurement-based constructions}
\label{sec:extendedConstructions}

In Sections~\ref{sec:RBB} and~\ref{sec:otherConstrs}, we presented the simplified RBB construction representing the building blocks of the patterns in~\cite{RBB03}, as well as two other patterns from the more elaborate constructions which are not obviously reproducible by the DKP construction.
Using the graph-theoretic characterization of flows in this chapter, we can show that they cannot be produced by the DKP construction.

\subsubsection{Qubit reversal pattern}

The pattern described in Section~\ref{sec:qubitReversal} for reversing the order of a sequence of ``logical qubits'' in the cluster-state model can be shown not to have a flow, using the characterization of Section~\ref{sec:graphth-charn}, because the underlying geometry does not admit a causal path cover.
We may show this by sketching various constructions of families of vertex-disjoint $I$\thru$O$ paths, in the case of $\card{I} = \card{O} = k$, as follows.

In the following, we label vertices $(a,b)$ by rows and columns; the $j\th$ input vertex is in the row $2(j-1)$, and the zeroth column.
\begin{itemize}
\item
	Suppose the path $P_k$ starting on the row $2(k-1)$ leaves that row, so that it covers the vertex $(2k-2,\ell)$ but not the edge between $(2k-2,\ell)$ and $(2k-2, \ell+1)$.
	Either it returns to the row $2(k-1)$ to cover the vertex $(2k-2,\ell+1)$, or it does not cover $(2k-2,\ell+1)$.
	In the former case, consider the function $\tilde f$ mapping each vertex covered by $P_k$ to its successor in $P_k$; then we have $(2k-2,\ell+1) = f^h(2k-2,\ell)$ for some $h \in \N$.
	Then, the path $(2k-2,\ell) \arc f(2k-2,\ell) \arc \cdots \arc f^{h}(2k-2,\ell) \arc (2k-2,\ell)$ is a vicious circuit, in which case any collection of paths including $P_k$ is not a causal path cover.
	In the latter case where $P_k$ does not cover $(2k-2, \ell+1)$, it then also separates that vertex from any other $I$\thru$O$ path, as a path starting in any other row must intersect $P_k$ to reach $(2k-2,\ell+1)$.
	These two cases are illustrated in Figure~\ref{fig:swapRBBpaths}.
	\begin{figure}[t]
		\begin{center}
			~\hfill
			\includegraphics[width=58mm]{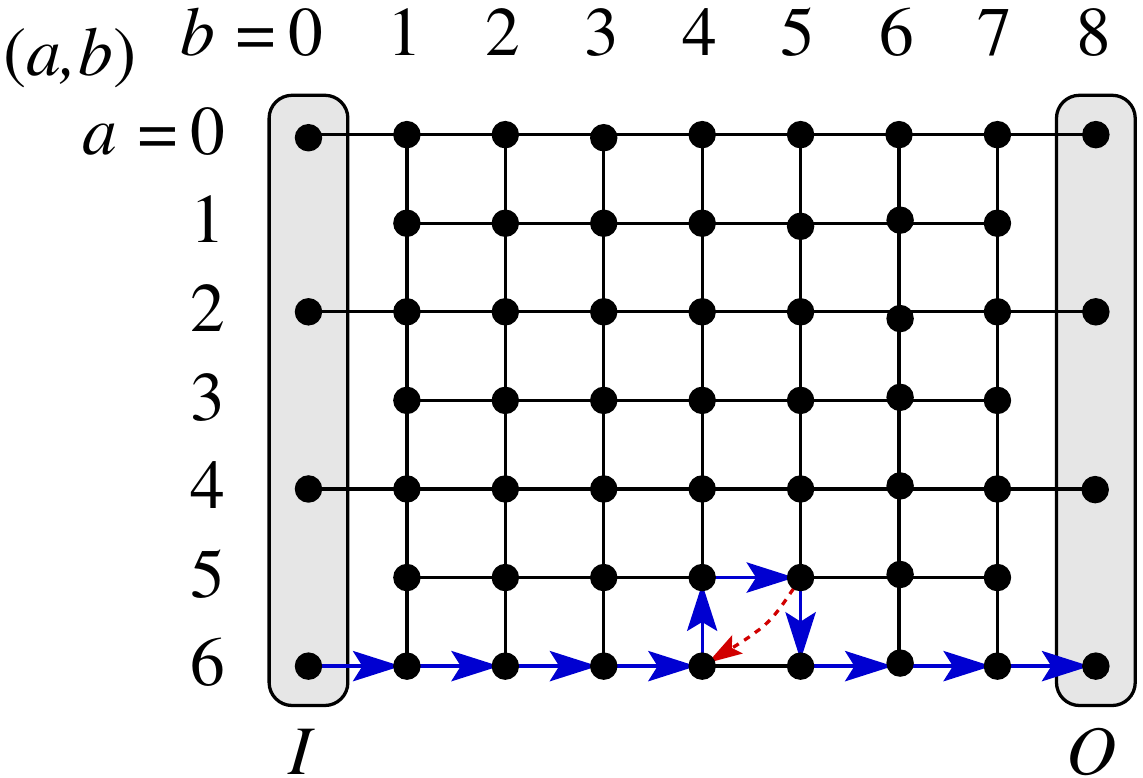}
			\hfill
			\includegraphics[width=58mm]{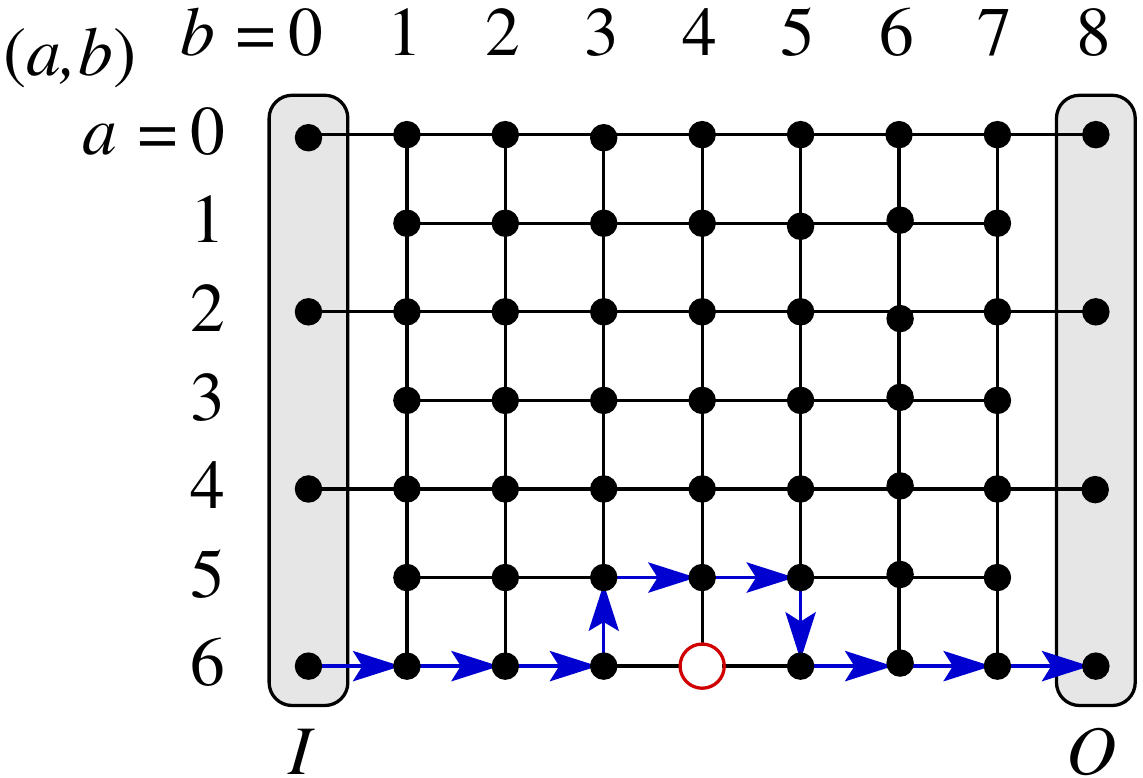}
			\hfill~\hfill~
		\end{center}
		\caption[Examples of an $I$\protect\thru$O$ path in the geometry for the qubit-reversal pattern illustrated in Figure~\protect{\ref{fig:swapRBB}}.]{\label{fig:swapRBBpaths}\colouronline
			Examples of an $I$\thru$O$ path in the geometry for the qubit-reversal pattern illustrated in Figure~\ref{fig:swapRBB}.
			Arrows on the grid lines in these diagrams represent orientations of edges of the geometry corresponding to a single $I$\thru$O$ path $P_k$.
			In the example on the left, a path leaves the row $2(k-1)$ and returns to cover all of the vertices in that row; this induces a vicious circuit (emphasized by the dotted arrow, representing an arc from the influencing digraph induced by an arc of $P_k$ and an edge not covered by $P_k$).
			In the example on the right, a path leaves the row $2(k-1)$ and returns, but fails to cover all of the vertices in that row, leaving a vertex uncovered (shown as a large hollow circle).}
	\end{figure}	

	Thus, if a path starting in the row $2(k-1)\st$ does not remain constrained to that row, any family of paths containing that path is not a causal path cover; and a similar analysis also holds for the path beginning in the first row.
	(This suffices then to show that the geometry does not have a flow \eg\ for $k = 2$.)

\item
	Suppose the path starting in the row $2(k-1) = 2k-2$ remains in that row.
	Then, the only path in a collection of vertex-disjoint $I$\thru$O$ paths that can cover vertices in the row $2k-3$ is the path $P_{k-1}$ starting in the row $2k-4$ (as the odd numbered rows do not have input vertices, and the path $P_{k-1}$ will separate any other path from the row $2k - 3$).
	In order to cover all of the vertices in the row $2k-3$, the path $P_{k-1}$ must cover the first vertex $(2k-3,1)$ in that row; and once in that row, by a similar analysis as applied for the row $2(k-1)$ above, it must remain in that row until it reaches the final vertex $(2k-3,2k-1)$ in that row.
	In order for the output vertex in row $2k-4$ to be covered, the path $P_{k-1}$ must then terminate there, determining that path.

	By induction, we can show that the paths in every lower-numbered row must follow a staircase pattern to higher-numbered rows if those higher-numbered rows are to be completely covered.
	(Such a family of paths is illustrated in Figure~\ref{fig:swapRBBstaircase}).
	\begin{figure}[t]
		\begin{center}
			\includegraphics[width=58mm]{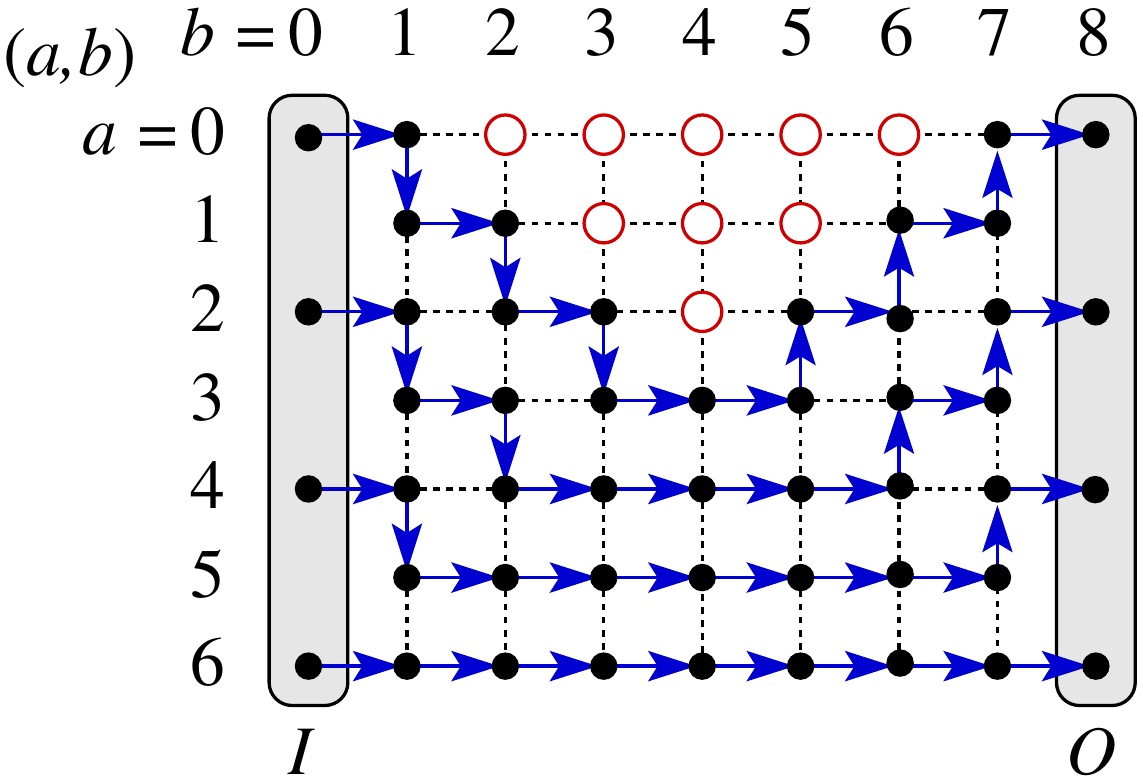}
		\end{center}
		\caption[Illustration of a family of $I$\protect\thru$O$ paths in the geometry for the qubit-reversal pattern illustrated in Figure~\protect{\ref{fig:swapRBB}}.]{\label{fig:swapRBBstaircase}\colouronline
			Illustration of a family of $I$\thru$O$ paths in the the geometry for the qubit-reversal pattern illustrated in Figure~\ref{fig:swapRBB}.
			Arrows on the grid lines in these diagrams represent orientations of edges of the geometry corresponding to a family of vertex-disjoint $I$\thru$O$ paths; the other edges in the grid have been made dashed to emphasize the flow edges.
			This family of paths is constructed to cover all of the vertices in the higher-numbered rows; as a result, vertices in the lower-numbered rows are necessarily left uncovered (represented here by large hollow circles).}
	\end{figure}
	In particular, the path $P_1$ will then leave the first row, so that by the previous case the resulting family of paths will not be a path cover.
\end{itemize}
Thus, the geometry of the qubit reversal pattern of~\cite{RBB03} does not have a causal path cover, by Theorem~\ref{thm:graphth-charn}, does not have a flow.
As a result, it cannot be produced by the DKP construction.

Despite this, we may gain some information about the qubit-reversal pattern using the smenatic map for the DKP construction by an appropriate extension of the input and output systems, as exhibited in~\cite{CLM2005}.
It is easy to see that if we include the left-most element of each row into $I$, and the right-most into $O$, the resulting altered geometry has a flow with a successor function which maps each vertex to the one immediately to the right.
Chains in the natural pre-order $\preceq$ have monotonically increasing column numbers: and therefore $\preceq$ is antisymmetric.
Given that the default measurement angle for each qubit is zero, and assuming that the dependencies of the measurement pattern are consistent with the those of the DKP construction given the flow function described, we may then obtain a measurement pattern which is in the range of the DKP construction.
The geometry for this measurement pattern, and the corresponding unitary circuits in the $\Jacz(\A)$ model and a related unitary circuit involving $\cX$ gates, are illustrated in Figure~\ref{fig:swapRBBredux}.
\begin{figure}[p]
	\begin{center}
			\includegraphics{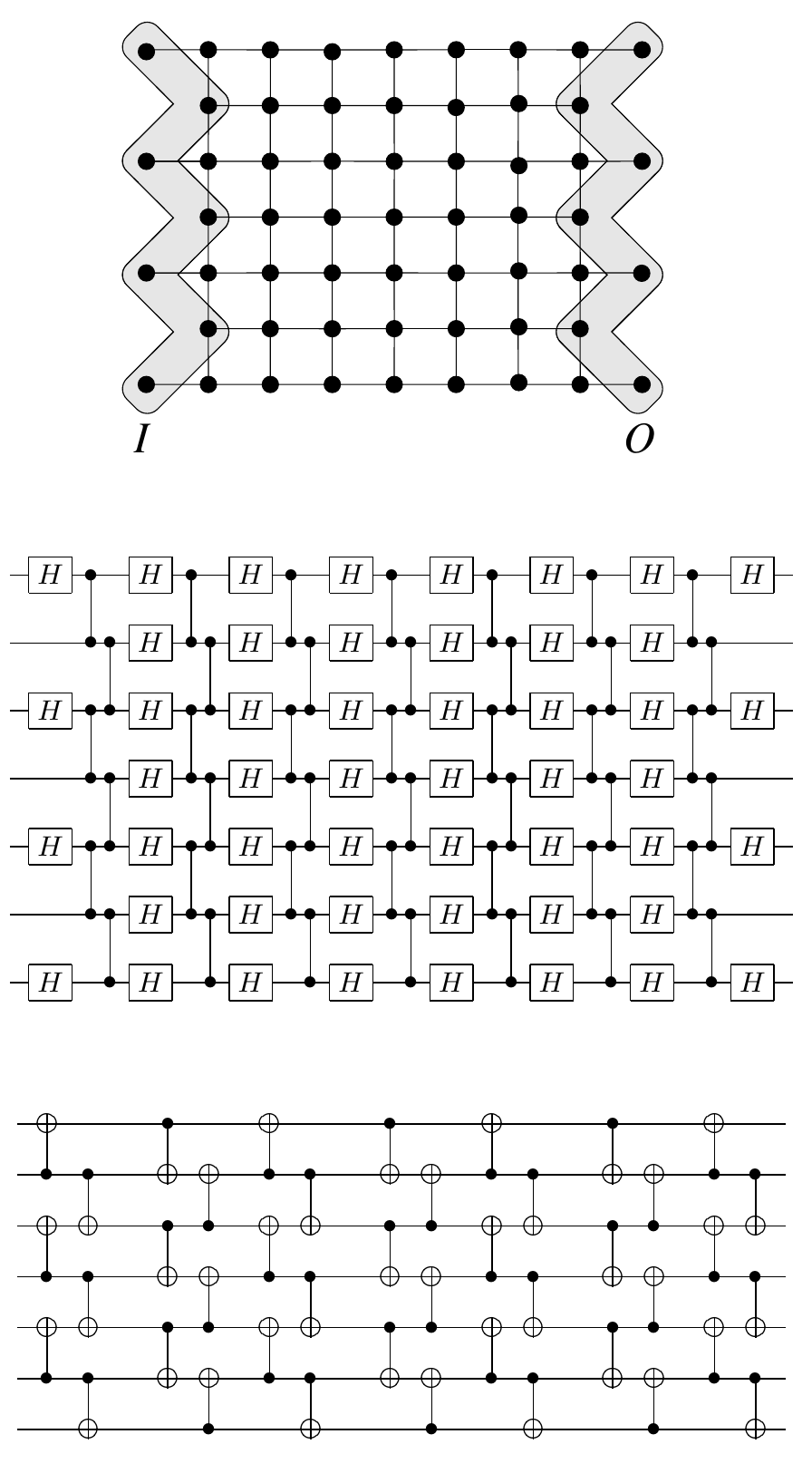}
	\end{center}
	\caption[Illustration of the geometry for the qubit-reversal pattern illustrated in Figure~\protect{\ref{fig:swapRBB}} with augmented input and output subsystems, and corresponding unitary circuits.]{\label{fig:swapRBBredux}%
		Illustration of the geometry for the qubit-reversal pattern illustrated in Figure~\ref{fig:swapRBB} with augmented input and output subsystems, and corresponding unitary circuits.
		The circuit diagram in the middle represents the circuit produced by the procedure $\SemanticDKP$, assuming that each default measurement is zero, and that the dependencies are those consistent with the DKP construction, given a flow function mapping each vertex in the geometry to the one immediately to the right.
		The circuit diagram on the bottom represents an equivalent circuit resulting from gathering the Hadamard operations of the circuit above.}
\end{figure}
The final circuit illustrated there can easily be described as a reversible classical computation: for seven input bits $(v_0, \ldots, v_6)$, the computation performed by that circuit is
\begin{align}
	\lefteqn{(v_0,\; v_1,\; v_2,\; v_3,\; v_4,\; v_5,\; v_6)}\qquad\qquad&\notag\\
	\mapsto&\quad
	(v_6,\; v_4 \oplus v_5 \oplus v_6,\; v_4,\; v_2 \oplus v_3 \oplus v_4,\; v_2,\; v_0 \oplus v_1 \oplus v_2,\; v_0)	\;.
\end{align}
If we apply this circuit instead to qubits, with the odd-indexed qubits initialized to the $\ket{\splus}$ state (\ie\ a uniform superposition of the bit-values described above), we the result is that the odd-indexed qubits remain in the $\ket{\splus}$ state, and the even-indexed qubits are reversed, as required.

We may then understand the qubit-reversal pattern in~\cite{RBB03} as corresponding to a circuit which prepares auxiliary qubits in the $\ket{\splus}$ state, and subsequently removes them by an $\axisX$ measurement.
The fact that we are able to obtain the circuit in this case is possible because the highly structured geometry makes it easier to guess which qubit measurements correspond to terminal measurements in a unitary circuit.
Without exploiting this structure, however, there are no known approaches to \emph{algorithmically} provide semantics for such a measurement pattern in terms of unitary circuits.

\subsubsection{Concise pattern for $Z \ox \cdots \ox Z$ Hamiltonians}

As we noted in Section~\ref{sec:zzMany}, the pattern $\mathfrak {Zz}_{u,v}$ on two qubits $u$ and $v$ of the simplified RBB construction (Section~\ref{sec:RBB}) is a special case of the family of patterns $\mathfrak {Zz\cdots z}^\theta_{v_1, \ldots, v_k}$ on many qubits.
The geometry for that pattern consists of the $k$ qubits $I = O = \ens{v_1, \ldots, v_k}$, and a single auxiliary qubit $a \notin I, O$.
For this geometry, there is exactly one injective function $f: O\comp \to I\comp$\,, which is the identity function on $\ens{a}$\,, which does not satisfy the constraint that $a \sim f(a)$.
(For the sake of uniformity, we may wish to restrict ourselves to patterns performing $\XY$ plane measurements only, in which case we would consider the pattern for this CPTP map given by~\eqref{eqn:zzManyXY}; however, the identity function on $\ens{a_1, a_2}$ suffers from the same problem as that on $\ens{a}$ for the pattern~\eqref{eqn:zzMany}, and the natural pre-order for the involution on $\ens{a_1, a_2}$ is not antisymmetric.)
Thus, this pattern cannot be produced via the DKP construction.

The $\YZ$-plane measurement performed by the pattern $\mathfrak {Zz\cdots z}^\theta$, with respect to the $\ket{\spm \sfy\sfz_{\theta}}$ basis defined in~\eqref{eqn:defYZtheta}, is equivalent to a measurement of the observable
\begin{gather}
	\label{eqn:observableYZ}
		\sO^{\YZ,\theta}
	\;\;=\;\;
		\ket{\splus\sfy\sfz_{\theta}}\bra{\splus\sfy\sfz_{2\theta}}
		\;\;-\;\;
		\ket{\sminus\sfy\sfz_{\theta}}\bra{\sminus\sfy\sfz_{2\theta}}
	\;\;=\;\;
		\cos({\theta}) Z + \sin({\theta}) Y	\;;
\end{gather}
this is a Hermitian operator which anticommutes with the Pauli $X$ operator for any $\theta$.
We can use this to obtain an extended sense of a flow which arises from the simplified RBB construction in the same way that flows arise from the DKP construction.

\subsection{Flows as a certificate of unitarity, and extended flows}
\label{sec:eflow}

In this section, we will consider the significance of flows not only as a structure which may underly one-way patterns, but also as a certificate which makes it possible to verify that a one-way pattern performs a unitary transformation.
In doing so, we will arrive naturally at an extension of the definition of flows which describes structures underlying the simplified RBB construction, and more generally constructions of one-way patterns from $\mathfrak J^\alpha$ and $\mathfrak{Zz\cdots z}^\theta$ patterns.

Recall that for the \emph{simplified} DKP construction, for each qubit $v \in O\comp$, the dependencies which arise are an $X$-dependency on $f(v)$, and a $Z$-dependency on every qubit adjacent to $f(v)$ except for $v$ itself; this may be considered to arise from a correction operation
\begin{gather}
		B_v
	\;\;=\;\;
		X_{f(v)} \paren{ \prod_{\substack{w \sim f(v) \\ w \ne v}} Z_w }
\end{gather}
which is absorbed into the measurements on those respective qubits, to be applied in the case that $\s[v] = 1$.
This unitary operation closely corresponds to the generator $K_{f(v)}$ of the open graphical encoding performed by the preparation map and entanglement graph at the beginning of the measurement pattern,
\begin{gather}
		K_{f(v)}
	\;\;=\;\;
		X_{f(v)} \paren{ \prod_{w \sim f(v)} Z_w }	\;:
\end{gather}
the state of the system just prior to the measurement phase in the simplified DKP construction is stabilized by the group $\gen{K_w}_{w \in I\comp}$ of such operators.
Flows may then be understood as a way of certifying that a measurement pattern performs a unitary transformation by the stabilizer formalism.
We may observe that the analysis of measurements by a single-qubit observable $P_v$ in the stabilizer formalism (as presented in Section~\ref{sec:stabilizerFormalism}) does not at any point require that $P_v$ be a Pauli operator, so long as it either commutes or anticommutes with every generator of the subgroup of the Pauli group which stabilizes the state of the system.
Note that, just as we have described above for $\YZ$-plane measurements, $\XY$-plane measurements can be described by measurements with respect to observables of the form
\begin{gather}
	\label{eqn:observableXY}
		\sO^{\XY,\theta}
	\;\;=\;\;
		\ket{\psub\theta}\bra{\psub\theta}
		\;\;-\;\;
		\ket{\msub\theta}\bra{\msub\theta}
	\;\;=\;\;
		\cos(\theta) X + \sin(\theta) Y	\;,
\end{gather}
and that this is an observable which anticommutes with the Pauli $Z$ operator.
Then, for a flow function $f: O\comp \to I\comp$ for a geometry $(G,I,O)$, an observable $\sO^{\XY,\theta}_v$ anticommutes with $K_{f(v)}$; and $\preceq$ describes an order in which stabilizer generators may be ``removed'' so that the entire computation may be treated via the stabilizer formalism.

We may then show that if $(f,\preceq)$ is a flow for a geometry $(G,I,O)$ that $\preceq$ is a partial specification of a total ordering of the qubits $v \in O\comp$, and $f$ a map associating each qubit $v \in O\comp$ to a generator $K_w$ for $w \in I\comp$, such that the measurement of each such qubit $v$ can be certified to perform an isometry by the stabilizer formalism via successive removals of the generators $K_{f(v)}$ --- provided that the corresponding corrections are performed in response to each measurement, and that the measurements are performed in an ordering extending $\preceq$.
In order to do this, we require that $\sO^{\XY, \theta}_v$ commutes for any $v \in O\comp$ with any remaining generator $K_w$\,; for this to hold for arbitrary $\theta$, this entails that such a generator $K_w$ does not act on $v$ --- which can be guaranteed by requiring that any such $K_w$ in the original stabilizer group be eliminated prior to the measurement of $v$.
This can be done by requiring the measurement a qubit $z$ such that $w = f(z)$ prior to measuring $v$.
Thus, it is sufficient for all $z$ to that $z \preceq f(z)$ and that $v \sim f(z) \,\implies\, z \preceq v$\,, which the flow conditions guarantee. 
We can then interpret flows as a way of connecting the DKP construction to the stabilizer formalism, by identifying the generators $\ens{K_w}_{w \in I\comp}$ in some order as the generators by which each measurement may be treated.

The converse statement --- that any ordering the individual generators $\ens{K_w}_{w \in I\comp}$ as a means of treating measurement-based computation by the stabilizer formalism corresponds to a flow --- is not true, however; and the $\mathfrak{Zz\cdots z}^\theta$ pattern is a counterexample.
We may define an \emph{extended flow}\footnote{%
	This extended sense of flow was described in~\cite{DK05u} for the special case of $\axisY$ measurements, but does not seem to have been seriously examined, having been superceded by the concept of generalized flows (or \emph{glows})~\cite{BKMP07} which we will discuss in Section~\ref{sec:gflows}.} 
(or \emph{eflow}) as follows:
\begin{definition}
	\label{def:eflow}
	For a tuple $(G,I,O,T)$ consisting of a geometry $(G,I,O)$ and a set $T \subset O\comp \inter I\comp$ of qubits to be measured in the $\YZ$-plane (all other qubits assumed to be measured in the $\XY$-plane), an \emph{extended flow} is an ordered pair $(f, \preceq)$ consisting of a function $f: O\comp \to I\comp$, and a partial order $\preceq$ on $V(G)$, such that the conditions
	\begin{subequations}
	\begin{gather}
			\label{eflow:a}		f(v) \sim v	\;, \;\;\text{for $v \in T\comp$}\,;
		\\
			\label{eflow:b}		f(v) = v	\;,\;\;\text{for $v \in T$}\,;
		\\
			\label{eflow:c}		v \preceq f(v)\;;\;\;\text{and}
		\\
			\label{eflow:d}		w \sim f(v)	\;\implies\; v \preceq w
	\end{gather}
	\end{subequations}
	hold for all $v \in O\comp$ and $w \in V(G)$.
\end{definition}
The special case of qubits indicated as being measured in the $\YZ$-plane applies in the case of $\mathfrak {Zz\cdots z}^\theta$ as defined in~\eqref{eqn:zzMany}: the identity function on $\ens{a}$ is an ``eflow function'', with equality being an ``eflow order''.
This corresponds to an identification of the operator $K_a$ as a generator of the open graphical encoding with which $\sO^{\YZ,\theta}$ anticommutes for any $\theta$.
This indicates that the unitarity of the pattern $\mathfrak {Zz\cdots z}^\theta$ may also be certified by the stabilizer formalism.

In the same way as for flows, an eflow for a tuple $(G,I,O,T)$ certifies that any pattern on with the geometry $(G,I,O)$ and performing the appropriate types of measurement will perform a unitary transformation, provided also that it performs the appropriate corrections (or measurement adaptations) in response to the measurements.
Again as for flows, eflows do this by attributing generators $\ens{K_w}_{w \in I\comp}$ to qubit $v \in O\comp$ to be measured.
To show this, it suffices to show that the observables for each measurement commutes with the generators which remain at the point that the measurement is performed, except for a single one with which it anticommutes.
The argument is similar to the one for flows: it is necessary and sufficient for $K_{f(v)}$ to be the only generator which acts on $v$ at the time of the measurement of $v$\,; which holds if and only if $v$ is measured no earlier than any qubit $z$ for which $v \sim f(z)$ or $v = f(z)$\,, which are guaranteed for eflows as for flows.

Any attribution of generators $K_{f(v)}$ for $f(v) \in I\comp$ (as opposed to nontrivial products of such operators) to qubits $v \in O\comp$ must satisfy $f(v) = v$ or $f(v) \sim v$ in order for $K_{f(v)}$ to act on $v$\,; and in order to anticommute with the appropriate measurement observable describing the measurement on the qubit $v$ for \emph{arbitrary} default measurement angles, we require precisely the eflow conditions to hold.
Thus, eflows characterize those geometries, together with specifications of measurement planes, which may underlie measurement patterns which can be certified to perform unitary transformations by the stabilizer formalism.

The simplified RBB construction give rise to tuples $(G,I,O,T)$ with eflows just as the DKP construction gives rise to geometries with flows: the $Z$ corrections in the pattern $\mathfrak{Zz}$ of~\eqref{eqn:zz} used for two-qubit interactions commute with all of the entangler maps of subsequent measurement patterns which are composed with it, and induce $\pi$-dependencies on the qubits they act upon; the same informal meaning of influencing the measurement result applies as that described in Section~\ref{sec:flowDefn}.
After Pauli simplifications and signal shifting, we may arrive at a similar characterization of the dependencies arising in the simplified RBB construction as that described in Theorem~\ref{thm:dependenciesDKP} for the DKP construction: the same matrices $T$ and $F$ describe the propagation of $X$- and $Z$-dependencies in that case.
As well, the analysis of star decompositions may similarly be generalized: we may augment the definition of a star geometry to include those, like that of the pattern $\mathfrak {Zz\cdots z}^\theta$, for which $I = O$ and for which $I\comp$ is a singleton set consisting of a qubit to be measured in the $\YZ$-plane; the decomposition of geometries $(G,I,O)$ into maximal geometries then generalizes straightforwardly.
Then, provided an algorithm to efficiently find eflows, we may similarly arrive at a semantic map for the simplified RBB construction --- or for any augmentation of the DKP construction by including patterns of the form $\mathfrak{Zz\cdots z}^\theta$ --- just as we have above for the DKP construction.

\subsection{A faster and more general algorithm for finding flows}
\label{sec:improvedFlowAlgs}

The algorithm $\FindFlow$ for finding flows presented in Section~\ref{sec:flowAlgs} runs in time $O(k^2 n)$, where $k = \card{I} = \card{O}$ and $n = \card{V(G)}$\,; in particular, it is constrained to the case where the input and output subsystems are of the same size.
In~\cite{MP08}, Mhalla and Perdrix present an algorithm which is both more general than $\FindFlow$, in that it can also determine the presence of a flow for geometries such that $\card{I} < \card{O}$,\footnote{%
	Note that a geometry with $\card{I} > \card{O}$ cannot have a flow, as the set of input vertices must be a subset of the initial vertices of a path cover, and the set of output points must coincide with the set of final vertices of a path cover.}
and which runs in time $O(kn)$.

The speed-up associated with the algorithm of~\cite{MP08} essentially arises out of the observation that the flow function $f$ and partial order $\preceq$ can be found simultaneously, by considering the length of the longest chains of each qubit from the output set, and constructing ``layer sets'' consisting of vertices of similar depth.
The main analytical tool for the algorithm is that of a ``maximally delayed'' flow:
\begin{definition}[{\cite[Definitions~4~\&~5]{MP08}}]
	\label{def:delayLayers}
	For a geometry $(G,I,O)$ and a flow $(f,\le)$, let $V^<_0(G)$ be the set of vertices of $V(G)$ which are maximal in $\le$, and let $V^<_{j+1}(G)$ be the maximal set of vertices of the set
	\begin{gather}
				W^<_j
			\;\;=\;\;
				V(G) \;\setminus\; \paren{\bigcup_{\ell = 0}^j \; V^<_j}
	\end{gather}
	in the order $\preceq$, for any $j > 0$.
	A flow $(f,\preceq)$ is \emph{more delayed} than another causal order $(f',\le)$ if we have $\card{W^\prec_j} \le \card{W^<_j}$ for all $j \in \N$\,, where this inequality is strict for at least one such $j$\,; and $(f,\preceq)$ is \emph{maximally delayed} if no flow for $(G,I,O)$ is more delayed.
\end{definition}
Mhalla an Perdrix show that a maximally delayed flow is also a flow of minimum depth, \ie\ that the causal order of such a flow has the same depth as the natural pre-order for the given flow-function $f$.
For a maximally delayed flow $(f,\preceq)$, let $\ell: V(G) \to \N$ be the function mapping each vertex $v \in V(G)$ to the integer $j$ such that $v \in V^\prec_j$.
Then, the partial order $\le$ characterized by $v \le w \iff [v = w \text{\;or\;} \ell(v) > \ell(w)]$ also yields a minimum-depth flow by construction.
To find a flow of minimum depth, it then suffices to find the sets $V^\prec_j$ of a maximally delayed flow $(f,\preceq)$.

Mhalla and Perdrix show for a maximally delayed flow $(f,\preceq)$ for $(G,I,O)$ that $V^\prec_0 = O$, and $V^\prec_1$ is the set of elements of $W^\prec_0$ such that, for some $w \in O$, $v$ is the only neighbor of $w$ which is contained in $W^\prec_0$.
By induction (by augmenting the set $O$ to include successive ``layers'' $V^\prec_j$), one may show that we then have
\begin{gather}
		V^\prec_{j+1}
	\;\;=\;\;
		\ens{v \in W^\prec_j \;\Big|\; \exists w \in V(G) \setminus W^\prec_j\,:\, \ngbh[G]{w} \inter W^\prec_j = \ens{v}}	\;,
\end{gather}
where again $\ngbh[G]{w}$ is the set of vertices adjacent to $w$ in $G$.
By a depth-first search from $O$, we may find in a finite number of iterations of $j$ those qubits $w \in V(G) \setminus W^\prec_j$ which have such a neighbor $v$, and accumulate them into new layers $V^\prec_{j+1}$.

This decomposition of the geometry $(G,I,O)$ into layers $V^\prec_j$ corresponds strongly to the star decomposition of the geometry $(G,I,O)$ into maximal star geometries; if (as in~\cite{MP08}) we accumulate sets $V^\prec_j$ to produce successively larger output subsystems $O\supp{j} = V(G) \setminus W^\prec_j$, the set $V^\prec_{j+1}$ as described above are precisely those vertices which are the root of a maximal geometry.\footnote{%
	While the algorithm of~\cite{MP08} predates the analysis of star geometries presented in this thesis, the two results were developed independently.}
Thus, in addition to finding flows more efficiently, the algorithm of~\cite{MP08} may also be used to unify the stages of finding a flow for the geometry underlying a pattern $P$, and obtaining the corresponding unitary circuit for $P$, in a semantic map for the DKP construction when $\card{I} = \card{O}$.\footnote{%
	In the case that $\card{I} < \card{O}$, there are additional difficulties which arise due to the fact that there need not be a unique flow-function --- which implies that there is not a unique set of $X$- and $Z$-dependencies which must arise for measurement patterns with such geometries.
	We discuss these issues in Section~\ref{sec:flowsReview}.}

\subsection{An efficient algorithm for finding extended flows}
\label{sec:eflowAlgs}

We may easily extend the analysis and the algorithm of~\cite{MP08} for flows to also efficiently find eflows.
For an eflow $(f,\preceq)$ for a tuple $(G,I,O,T)$\,, define the sets $V^\prec_j$ and $W^\prec_j$ as in Definition~\ref{def:delayLayers}.
Then for each $j \in \N$, $V^\prec_{j+1}$ consists of those vertices $v \in W^\prec_j$ such that, for any $w \in V(G)$ such that $w = f(v)$ or $w \sim f(v)$, we have $w = v$ or $w \in V(G) \setminus W^\prec_j$.
For $v \notin T$\,, this implies that (as in the analysis of~\cite{MP08}) the only neighbor of $f(v)$ which lies in $W^\prec_j$ is $v$ itself; and for $v \in T$, this implies that $v$ has no neighbors in $W^\prec_j$.
Algorithm~1 of~\cite{MP08} can then be easily extended to test for the latter condition, yielding a procedure to construct an eflow.
We will now describe such an algorithm.
\begin{algorithm}[t]
	\SUBROUTINE \AddToLayer(v,w) :
	
	\Data{%
		$\ell$, an array $V(G) \to \N$ for storing layers of vertices of a graph $G$
	\\
		$W$, the set of vertices for which $\ell$ is not yet well-defined
	\\
		$\layer$, the value of the current layer
	\\
		$\emptyLayer$, a flag indicating if the present layer is empty
	\\
		$D$, a directed graph such that vertices with non-zero in-degree are in $W$
	\\
		$T$, a set of vertices representing qubits measured in the $\YZ$-plane
	\\
		$f$, an array $V(G) \partialto V(G)$ for storing ``successors'' of vertices
	\\
		$S'$, a set of vertices to examine as possible ``successors'' in the next layer
	\\
		$v \in V(G)$, a vertex to add to the current layer
	\\
		$w \in V(G)$, a vertex to be designated the ``successor'' of $v$
	}

	\Effect {%
		Sets $f(v) \gets w$, $\ell(v) \gets \layer$, removes all of the arcs in $D$ into $v$, and checks the neighbors of $v$ in $D$ for terminal vertices in $T$ to include into $S'$ (for inclusion into the next layer)
	}

	\BEGIN
		$\emptyLayer \gets \false$\;
		$f(v) \gets w$~;
		$\ell(v) \gets \layer$\;
		$W \gets W \setminus \ens{v}$\;
		\FORALL $z \in \neighbors(D,v)$ \DO
			$\neighbors(D,z) \gets \neighbors(D,z) \setminus \ens{v}$\;
			\oneline
			\IF $z \in T$ \ANDALSO $\neighbors(D,z) = \vide$ \THEN $S' \gets S' \union \ens{z}$\;
		\ENDFOR
	\END
	\caption[A subroutine to attribute a vertex to the current layer, for use in an iterative procedure for finding the layer sets of a maximally-delayed eflow.]{\label{alg:addToLayer}%
		A subroutine to attribute a vertex to the current layer, for use in an iterative procedure for finding the layer sets of a maximally-delayed eflow.
		This subroutine is used in particular to maintain the arcs of a digraph $D$ (a ``sub-diagraph'' of a graph $G$, interpreted as a symmetric digraph) from which we remove any arc ending in a vertex with a well-defined layer.
	}
\end{algorithm}

Algorithm~1 of~\cite{MP08} operates on the basis of maintaining a set of ``correcting vertices'' which, for each new layer, may potentially be the successor of some vertex in the new layer.
We take a similar approach, but must somehow accomodate the vertices of $T$, whose neighbors we must check \emph{before} including them in a layer.
In order to ensure that this modification does not substantially increase the runtime of the flow-finding algorithm of~\cite{MP08}, it will prove helpful to maintain a digraph whose arcs point in either direction along each edge of $G$, and from which we delete any arc ending at a vertex with a defined layer.
As a result, we may easily determine whether a vertex with a defined layer has a single neighbor which does not yet have a defined layer, and also easily determine whether a vertex without a defined layer has any neighbors without a defined layer.
Algorithm~\ref{alg:addToLayer} defines a subroutine $\AddToLayer$ to perform the necessary edge deletions whenever we define the layer of a vertex, as well as several other operations useful to an iterative procedure for building an eflow by layer-sets, by operating on several global variables.

Using the subroutine $\AddToLayer$, we may easily define a procedure similar to Algorithm~1 of~\cite{MP08} to build an eflow by layer sets.
\begin{algorithm}[p]

	\PROCEDURE \FindEflow(G,I,O,T):

	\Input{%
		$G$, a graph with subsets $I, O, T$, where $T \subset V(G) \setminus O$.
	}

	\Output{%
		Either $\nil$, if $(G,I,O,T)$ does not have an eflow; or an ordered pair $(f,\ell)$ consisting of a partial function $f: V(G) \partialto* V(G)$ and a function $\ell: V(G) \to* \N$ such that $(f,\preceq)$ is an eflow, for a partial order $\preceq$ given by $v \preceq w \;\iff\; [\, v = w \text{~or~} \ell(v) > \ell(w) \,]$.
	}

	\BEGIN
		\LET $f: V(G) \partialto V(G)$\;
		\LET $\ell: V(G) \to \N$\;
		\LET $W, S \gets \vide$\;
		\LET $D \gets \text{[\,completely disconnected digraph on $V(G)$\,]}$\;
		\algline
		\FORALL $v \in V(G)$ \DO																						\nllabel{line:eflowInitLoop}
			$\neighbors(D, v) \gets \neighbors(G, v) \setminus O$\;
			\IF $v \in O$ \THEN { 
				$\ell(v) \gets 0$~;
				$f(v) \gets \nil$\;
				\oneline
				\IF $v \notin I$ \THEN $S \gets S \union \ens{v}$\; }
 			\ELSE {
				$W \gets W \union \ens{v}$\;
			}
		\ENDFOR																												\nllabel{line:eflowInitEnd}
		\medskip
		\LET $\layer \gets 0$\;
		\LET $\emptyLayer \gets \false$\;
		\REPEAT																											\nllabel{line:layerLoop}
			$\emptyLayer \gets \true$~;
			$\INCR \layer$\;
			\LET $S' \gets \vide$\;

			\FORALL $w \in S$ \DO {
				\IF $\neighbors(D,w) = \ens{v}$ \SUCHTHAT $v \notin T$ \THEN {							\nllabel{line:checkSuccessor}
					$S' \gets S' \union \ens{v}$\;																\nllabel{line:newSuccessor}
					$\AddToLayer(v,w)$\;
				} \ELSEIF $w \in T$ \THEN {																		\nllabel{line:checkYZmeas}
					$\AddToLayer(w,w)$\;
				} \ELSE {
					$S' \gets \ens{w}$\;																				\nllabel{line:propSuccessor}
				}
			} \ENDFOR
		\smallskip
		$S \gets S'$\;																									\nllabel{line:successorCopy}
	\UNTIL { $\emptyLayer$ }																						\nllabel{line:layerLoopEnd}

	\medskip
		
	\short
	\IF $W = \vide$
		\THEN {\RETURN $(f,\ell)$ }
		\ELSE { \RETURN \nil }
	\END

	\caption[An iterative algorithm to obtain an eflow.]{\label{alg:findEflow}%
		An iterative algorithm to obtain an eflow, using the maximally-delayed layers technique of~\cite{MP08} and the subroutine $\AddToLayer$.}
\end{algorithm}
We begin by creating the digraph $D$, copying the neighborhood relations of $G$, but removing any arcs ending in $O$ (which is necessarily the zeroeth layer).
Any output vertex which is in the range of a ``successor'' function $f: O\comp \to I\comp$ is a potential successor: we define a set of candidate successors from the elements of $O \setminus I$.
Vertices without a defined layer (initially the elements of $O\comp$) are included in the set $W$.
For each layer, we iterate through the set of potential successors, which initially includes no elements of $T$; we maintain the invariant that $v \in T$ becomes a candidate successor only if it has out-degree zero (\ie\ all of its neighbors are in some defined layer).
For any viable successor $w$ --- one satisfying the eflow constraints that $w \in T$ and $w \prec u$ implies that $u \in V^\prec_j$ for some $j < \layer$, or that $w = f(v)$ for $v$ the unique vertex not in $V^|prec_j$ for some $j < \layer$ --- we assign $w$ as the successor of the appropriate vertex $v$, and define the layer of that vertex $v$.
If this vertex $v$ is different from $w$, it then becomes a potential successor for the next layer; and any neighbors of $v$ which are in $T$, and which (subsequent to the removal of any arcs ending in $v$) have out-degree zero, also become viable successors for the next layer.
Otherwise, if $w$ was not a viable successor for this layer, we carry it forward for the next layer.
We assign the list of new candidate successors to $S$ before the end of the iteration; and we repeat until we cannot assign any new vertices to a layer.

If we cannot assign any vertices to new layers, this is either because we have exhausted the vertices (in which case $W = \vide$), or because there are no vertices which are reachable from the existing layers which have viable successors.
In the latter case, there is no eflow; and in the former, we return the eflow ``sucessor'' function and the layer function.

\subsubsection{Analysis of Algorithm~\ref{alg:findEflow}}
\label{sec:findEflowAnalysis}

In the subroutine $\AddToLayer(v,w)$, the run-time is dominated by the loop over the neighbors $z$ of $v$\,, which requires time at most $O(\deg_G(v))$\,.
This subroutine is called at most once for every vertex $w \in V(G)$\,: then, the cumulated run-time of $\AddToLayer(v,w)$ over the execution of $\FindEflow$ is $O\big(\sum\limits_v \deg_G(v) \big) = O(m)$, where $m = \card{E(G)}$.

Let $n = \card{V(G)}$ and $k = \card{O}$.
Initially, there are $k = \card{O}$ elements of $S$, none of which are elements of $T$\,; and as every element $w \in S \setminus T$ corresponds to a unique element of $S' \setminus T$ (either itself, or the vertex $v$ for which $w$ is the successor), there always remain at most $k$ elements of $S \setminus T$ at each iteration of the loop on lines~\ref{line:layerLoop}\thru\ref{line:layerLoopEnd}.
The comparisons on lines~\ref{line:checkSuccessor} and~\ref{line:checkYZmeas} can both be performed in constant time using the set data structures described at the beginning of Section~\ref{sec:flowAlgs}, as can the set-inclusions on lines~\ref{line:newSuccessor} and~\ref{line:propSuccessor}.
The assignment on line~\ref{line:successorCopy} can be performed simply by re-attributing the element list of $S'$ to $S$, as we will immediately re-assign $S'$ to the empty set in any subsequent iteration.
There are at most $n$ iterations of the loop on lines~\ref{line:layerLoop}\thru\ref{line:layerLoopEnd}, as it only repeats as long as at least one vertex has been assigned to a new layer.
Then, setting aside the work performed for vertices $w \in S \inter T$ in the loop on lines~\ref{line:layerLoop}\thru\ref{line:layerLoopEnd} and the work performed by $\AddToLayer$, the work performed by that loop is $O(kn)$.

As elements of $T$ are only included into $S'$ when they are guaranteed to be attributed to the next layer, the total work performed by that loop (again aside from work performed by $\AddToLayer$) for elements of $T$ is then $O(n)$.
The run-time of $\FindFlow(G,I,O,T)$ is then $O(kn + m)$, dominated by the work performed by $\AddToLayer(v,w)$ for each $w \in V(G)$, and by the iteration of the loop on lines~\ref{line:layerLoop}\thru\ref{line:layerLoopEnd} for at most $n$ layers.

This is precisely the run-time described for Algorithm~1, except that Mhalla and Perdrix also use the extremal result of Theorem~\ref{thm:extremalUpperBound} presented in~\cite{BP08} to simplify this to $O(kn)$.
It is an open question whether there is a similar extremal result for eflows.

As we remarked following Definition~\ref{def:eflow}, an efficient algorithm for eflows allows us to obtain a semantic map for the simplified RBB construction using similar techniques as we have presented for the DKP construction throughout the Chapter, as well as for any extension of the DKP or simplified RBB constructions which makes use of the more general $\mathfrak{Zz\cdots z}^\theta$ pattern.

\subsection{Generalized flows}
\label{sec:gflows}

In Definition~\ref{def:eflow}, we have described one sense in which flows may be generalized, by observing that the role of flows is essentially to permit the stabilizer formalism to certify that the measurement pattern performs a unitary operation by attributing some generator $K_w$ to each qubit $v \in O\comp$ to be measured.
A more general approach than this was defined in~\cite{BKMP07} in \emph{generalized flows} (or \emph{gflows}), which extends this both to arbitrary generators of the initial stabilizer group $\gen{K_w}_{w \in I\comp}$ (\ie\ to non-trivial products of the operators $K_w$) and to multiple planes of measurement.

Analogously to \XY-plane measurements and \YZ-plane measurements as we have described above, we may define \emph{\XZ-plane measurements} to be measurements with respect to a basis of the form
\begin{gather}
		\ket{\spm\sfx\sfz_\theta}
	\;\;=\;\;
		\sfrac{1}{\sqrt 2} \ket{\splus \sfy}
	\;\pm\;
		\sfrac{\e^{i\theta}}{\sqrt 2} \ket{\sminus \sfy}	\;.
\end{gather}
(For the duration of this section, we will write $\ket{\spm \sfx\sfy_\theta} = \ket{\pmsub\theta}$ for the sake of uniformity.)
Generalized flows encode a set of conditions imposed on measurements in the \XY, \YZ, and \XZ\ planes so that, as we described in Section~\ref{sec:extendedConstructions}, each observable to be measured anticommutes with exactly one of the remaining generators of the initial stabilizer group, commuting with the rest.
Let $\lambda: O\comp \to* \ens{\XY, \YZ, \XZ}$ be a function which designates, for each qubit to be measured, the plane in which it is to be measured.
The generalized flow conditions are as follows:
\begin{definition}[(paraphrased from~\cite{BKMP07})]
	\label{def:gflow}
	Let $(G,I,O)$ be a geometry and $\lambda$ be a function defined on $O\comp$ indicating measurement planes as above; and let $\cP$ be the power-set function, and $\odd(S)$ be the set of vertices in $G$ which are adjacent to an odd number of elements of $S$.
	Then a \emph{gflow} for $(G,I,O,\lambda)$ is a tuple $(g,\preceq)$ consisting of a function $g: O\comp \to \cP(I\comp)$ and a partial order $\preceq$ which satisfy the following conditions:
	\begin{subequations}
	\begin{align}
			w \in g(v) \;\implies\;&	v \preceq w ,
		\\
			w \in \odd(g(v)) \;\implies\;& v \preceq w ,
		\\
			\lambda(v) = \XY \;\implies\;& v \in \odd(g(v)) \setminus g(v)	, 
		\\
			\lambda(v) = \YZ \;\implies\;& v \in g(v) \setminus \odd(g(v)) ,
		\\
			\lambda(v) = \XZ \;\implies\;& v \in g(v) \inter \odd(g(v)) .
	\end{align}
	\end{subequations}
\end{definition}
The conditions above are sufficient conditions for there to exist a measurement pattern, with underlying geometry $(G,I,O)$ and performing measurements in the planes specified by $\lambda$, which performs a unitary for all specific choices of measurement angles. 
We may show this as follows.

We represent measurements in the \XY, \YZ, and \XZ\ planes by observables $\sO^{\XY,\theta}$, $\sO^{\YZ,\theta}$, and $\sO^{\XZ,\theta}$ respectively, which (like Pauli operators) have eigenvalues $\pm 1$, with corresponding eigenvectors $\ket{\spm\sfx\sfy_\theta}$, $\ket{\spm\sfy\sfz_\theta}$, and $\ket{\spm\sfx\sfz_\theta}$ respectively:
\begin{subequations}
\label{eqn:pauliPlaneObs}
\begin{gather}
		\sO^{\XY,\theta}
	\;\;=\;\;
		\ket{\splus\sfx\sfy_\theta}\bra{\splus\sfx\sfy_\theta}
		\;\;-\;\;
		\ket{\sminus\sfx\sfy_\theta}\bra{\sminus\sfx\sfy_\theta}
 	\;\;=\;\;
		\cos(\theta) X + \sin(\theta) Y	\;;
	\\[1ex]
		\sO^{\YZ,\theta}
	\;\;=\;\;
		\ket{\splus\sfy\sfz_\theta}\bra{\splus\sfy\sfz_\theta}
		\;\;-\;\;
		\ket{\sminus\sfy\sfz_\theta}\bra{\sminus\sfy\sfz_\theta}
	\;\;=\;\;
		\cos(\theta) Z + \sin(\theta) Y	\;,
	\\[1ex]
		\sO^{\XZ,\theta}
	\;\;=\;\;
		\ket{\splus\sfx\sfz_\theta}\bra{\splus\sfx\sfz_\theta}
		\;\;-\;\;
		\ket{\sminus\sfx\sfz_\theta}\bra{\sminus\sfx\sfz_\theta}
	\;\;=\;\;
		\cos(\theta) X + \sin(\theta) Z	\;.
\end{gather}
\end{subequations}
After applying the preparation and entangling operations $\Ent{G} \New{I\comp}{}$ in a one-way measure\-ment pattern, the state-space is stabilized by the group $\sS_{G,I} \,=\, \gen{K_v}_{v \in I\comp}$\,.
As we noted in Section~\ref{sec:extendedConstructions}, to describe how measurement of these observables transform the space stabilized by $\sS_{G,I}$, the analysis of measurements by a single-qubit observable $P_v$ in the stabilizer formalism requires only that each measurement observable either commutes or anticommutes with every generator of the subgroup of the Pauli group which stabilizes the state of the system.
If in particular, for all $1 \le t \le \card{O\comp}$, the state of the system after the $j\th$ measurement is stabilized by a group $\sS\supp{t}$\,, such that the $t+1\st$ measurement operator anticommutes with at least one element of $\sS\supp{t}$, the evolution of the system at each stage will be unitary, and the state post-measurement (selecting for the $+1$ measurement result) will be stabilized by a subgroup $\sS\supp{t+1}$ of the group $\sS\supp{t}$.

Suppose we require that the system at all times be in a joint $+1$-eigenstate of some subgroup of $\sS$ after each measurement.
It then suffices to find a generating set $\ens{S_1, \ldots, S_{\card I\comp}}$ for $\sS_{G,I}$ such that the state of the system after the $t\th$ measurement (and a classically controlled correction, if necessary) is stabilized by $\gen{\,S_j\,}_{j > t}$\,.
In particular, we require $S_t$ to be the generator which anticommutes with the observable of the $t\th$ measurement (\ie\ with 
$\sO_v^{\XY,\theta}$, $\sO_v^{\YZ,\theta}$, or $\sO_v^{\XZ,\theta}$ as appropriate), and for every $S_j$ to commute with that observable for $j > t$, \emph{regardless} of the value of $\theta$.
For a measurement on a qubit $v$, this imposes the following constraints:
\begin{itemize}
\item
	for an $\XY$-plane measurement, we require that $S_k$ acts on $v$ with a $Z$ operation, and $S_j$ acts on $v$ with the identity operation for $j > k$;

\item
	for a $\YZ$-plane measurement, we require that $S_k$ acts on $v$ with an $X$ operation, and $S_j$ acts on $v$ with the identity operation for $j > k$;

\item
	for an $\XZ$-plane measurement, we require that $S_k$ acts on $v$ with an $Y$ operation, and $S_j$ acts on $v$ with the identity operation for $j > k$.
\end{itemize}
In particular, we require that the stabilizer groups $\gen{S_j}_{j > k}$ for each $1 \le k  \le \card{O\comp}$ do not act on the first $k$ qubits which were measured.

From these constraints, we can re-derive the gflow conditions, as follows.
Consider a fixed choice of generating set $\ens{S_j}_{j = 1}^{\card{I\comp}}$ for which the conditions above may be satisfied, for \emph{some} ordering of the generators: each such operator consists of a product of operators $K_v$ for $v \in I\comp$\,.
Define the function $\gamma: \ens{1,\ldots,\card{O\comp}} \to \cP(I\comp)$ such that
\begin{align}
	\label{eqn:presentGenerators}
		S_j
	\;\;=\;\;
		\prod_{v \in \gamma(j)} K_v
	\;\;=&\;\;
		\prod_{v \in \gamma(j)} \paren{ X_v \prod_{w \sim v} Z_w }
	\notag\\\propto&\;\;
		\sqparen{\prod_{v \in \gamma(j)} \!\! X_v \;} \sqparen{\prod_{w \in \odd(\gamma(j))} \!\! Z_w \;} .
\end{align}
Let $\sigma: O\comp \to \N$ describe an arbitrary sequence in which the qubits of $O\comp$ are measured; let $\preceq$ describe the coarsest partial order (relative to the choice of generators $\ens{S_j}_{j = 1}^{\card{I\comp}}$) consistent with every such ordering of $O\comp$\,; and let $g = \gamma \circ \sigma$.
In order for the commutation/anticommutation properties above to hold, we then require the following for the qubit $v$ which is the $k\th$ qubit measured:
\begin{romanum}
\item
	As $S_k$ does not act on any qubits yet to be measured, any qubit $w \in \gamma(k)$ must be measured later than $v$\,, and similarly for any qubit $w \in \odd(\gamma(k))$.
	Therefore, if either $w \in g(v)$ or $w \in \odd(g(v))$, we require $v \preceq w$.

\item
	If $v$ is to be measured in the \XY\ plane, $S_k$ must act on $v$ with a $Z$ operation, in order for the observable $\sO_\theta^{\XY}$ to anticommute with $S_k$; this holds if and only if $v \in \odd(g(v))$ and $v \notin g(v)$.

\item
	If $v$ is to be measured in the \YZ\ plane, $S_k$ must act on $v$ with an $X$ operation, in order for the observable $\sO_\theta^{\YZ}$ to anticommute with $S_k$; this holds if and only if $v \in g(v)$ and $v \notin \odd(g(v))$.

\item
	If $v$ is to be measured in the \XZ\ plane, $S_k$ must act on $v$ with an $Y$ operation, in order for the observable $\sO_\theta^{\XZ}$ to anticommute with $S_k$; this holds if and only if $v \in g(v)$ and $v \in \odd(g(v))$.
\end{romanum}
Then, the conditions for applying the extended stabilizer formalism independently of the measurement angles are precisely those of of Definition~\ref{def:gflow}.

As noted above, if each measurement results in a collapse to their respective $+1$-eigenvalues, the resulting transformation of the state-space is an isometry; and for any $t$ such that the $t\th$ measurement results in the $-1$-eigenvalue of the measurement observable, we may apply $S_t$ as a correction operation to steer it into the post-measurement state corresponding to the $+1$ measurement result.

Therefore, the existence of a gflow for $(G,I,O,\lambda)$ acts as a certificate that a measurement pattern which performs the appropriate corrections (or measurement adaptations) performs a unitary transformation, via the stabilizer formalism.
This fully generalizes the extent to which the stabilizer formalism (relaxed to admit non-Pauli observables) can be used to define certificates for unitarity without requiring precise information about individual measurement observables.\label{discn:gflowsSpecialCases}
In particular, eflows are a special case for gflows, where we require that $g(v)$ is a singleton set for all $v \in O\comp$\,; and flows are a further restriction where we require $v \ne f(v)$ for all $v \in O\comp$.

It is not known whether gflows correspond to any construction of interest for one-way measurement patterns (in the sense that flows correspond to the DKP construction and eflows to the simplified RBB construction, as described in Sections~\ref{sec:flowDefn} and~\ref{sec:extendedConstructions} respectively); however, as an extension of flows, generalized flows seem to extend the semantic map $\SemanticDKP$ to accommodate measurement patterns which may be translated into circuits using closed time-like curves~\cite{Kashefi07priv} (a notion introduced into quantum information theory by Deutsch~\cite{Deutsch91}).\footnote{%
	In general, the availability of closed time-like curves are a powerful computational resource for both classical and quantum computation~\cite{AW08}; however, the extension of $\SemanticDKP$ to patterns with gflows suggested by Kashefi does not seem to produce arbitrary circuits which use closed time-like loops (\cf\ the discussion in the introduction to Chapter~\ref{Measurement Pattern Interpolation} on postselection).}
By virtue of having an eflow, the pattern $\mathfrak{Zz\cdots z}$ of~\eqref{eqn:zzMany} has a gflow.
However, for any measurement pattern whose corresponding tuple $(G,I,O,\lambda)$ has a gflow, every measurement will produce independent and maximally random measurement results; therefore, the qubit-reversal pattern of Section~\ref{sec:qubitReversal} (which, as we noted on page~\pageref{discn:qubitReversalCertainty}, yields some measurement results with certainty conditioned on prior results) does not have a gflow.

In~\cite{MP08}, Mhalla and Perdrix describe an $O(n^4)$ algorithm (where $n = \card{V(G)}$) for determining whether a geometry $(G,I,O)$ has a gflow, conditioned on every qubit being measured in the $\XY$-plane; it is not known whether there exists an efficient algorithm for finding a gflow under more general conditions, although this seems likely.

\subsection{Ignoring measurement dependencies when inferring unitary circuits}

In the preceding discussion of flows, eflows, and gflows as certificates of unitarity in terms of the stabilizer formalism, we have quickly brushed over the issues of the appropriate measurement adaptations and corrections depending on measurement results.
The stabilizer formalism itself furnishes a choice of immediate corrections which suffice to guarantee unitarity, which can then be absorbed into measurements and further refined by Pauli simplifications and signal shifting; however, in situations where there exist more than one ``flow-like'' certificate (which may occur for flows when $\card{I} < \card{O}$, or for gflows), it may be difficult to verify whether a particular measurement pattern meets conditions which are sufficient to show that it performs a unitary transformation.
However, if the geometry underlying a particular measurement pattern (together with additional information such as the planes of measurement for each qubit) admits a flow-like certificate, we may if we wish consider the scheme of dependencies which arise from that certificate in order to ``force'' the pattern to represent of a unitary transformation, changing the measurement/correction dependencies as required.
Because the dependencies serve only to simulate the selection of a particular post-measurement result, the particular dependency scheme does not matter (except possibly for complexity reasons): then one one arising from an arbitrary certificate of this kind will suffice.
We will consider a potential application of this approach in Chapter~\ref{Measurement Pattern Interpolation}.

If we are willing to ignore any existing dependencies in a measurement pattern and simply impose the dependencies arising from some flow-like certificate, we may streamline any algorithm for semantic maps corresponding to constructions of one-way measurement patterns by omitting the step for checking the dependencies, which \eg\ dominates the run-time of Algorithm~\ref{alg:semanticDKP}.

\section{Review and open problems}
\label{sec:flowsReview}

In this chapter, we have reviewed the concept of flows presented in~\cite{DK06}, identifying it as arising out of combinatorial constraints on expressions for unitary circuits.
We have presented a characterization of flows in terms of path covers, and presented uniqueness and extremal results, which contributed to the discovery of the first efficient algorithm for finding flows for geometries of one-way measurement patterns, and also help to bound the run-time of the best existing algorithms for finding flows.
We used these results to describe a function $\cS$ which provides semantics for mapping measurement patterns congruent to those produced by the DKP construction, in terms of unitary circuit models $\Jacz(\A)$ for $\A \subset \R$ defined in Section~\ref{sec:unitaryCircuits}, when those patterns have input and output subsystems of the same size.

The final constraint above in the semantic map $\cS$, that $\card{I} = \card{O}$, does not merely arise out of the corresponding constraint for the procedure $\FindFlow$ of Algorithm~\ref{alg:completeFlowAlg}.
In order to verify that the measurement and correction dependencies for a pattern suffice for the pattern to perform a unitary, the condition $\card{I} = \card{O}$ provides the guarantee (by Theorem~\ref{thm:uniqueFamilyPaths}) that there is at most one flow-function, and therefore at most one set of dependencies which are possible after Pauli simplifications and signal shifting.
For a pattern where $\card{I} < \card{O}$ whose geometry admits more than one flow-function, it becomes necessary to verify that there exists a flow function which gives rise to the dependencies observed in the pattern.
It is possible that the dependency information in the pattern may prove useful for discovering whether such a flow-function exists, but no such algorithm is known.

As we have observed at many points throughout this Chapter, with regards to uniqueness, extremal results, and decompositions in terms of star geometries, the presence of a flow gives rise to many structural constraints, some of which limit the possible complexity of the measurement pattern.
It seems plausible that these may lead to interesting bounds on the size and depth complexity of measurement patterns arising from the DKP construction, or on the time required to compute the dependency matrices described in Theorem~\ref{thm:dependenciesDKP}.

	\chapter[Measurement Pattern Interpolation]{
		Measurement Pattern Interpolation (from Quadratic Form Expansions)}

Can measurement-based quantum computation provide new idioms for designing quantum algorithms?
That is: in addition to being universal for quantum computation by \eg\ representing unitary circuits via the DKP or RBB constructions, can it also provide analytical tools which suggest how currently \emph{unsolved} problems might be solved on a quantum computer?
There are currently no obvious high-level concepts of the sort provided by the circuit model (such as eigenvalue estimation or amplitude amplification) or quantum walks (such as deflection by engineered energy barriers) which may lead to interesting quantum algorithms; however, the ability to describe novel solution techniques is part of what distinguishes a potentially useful model of computation from a model which is ``merely'' a potentially useful scheme for physical implementation.

The largest obstacle to high-level descriptions of algorithms in the measurement based model --- without depending on semantics being provided by another model, \eg\ a unitary circuit model --- seems to be in the details of the classical feed-forward which specifies how measurements are to be adapted based on previous measurement results.
To perform a unitary transformation with a measurement-based algorithm requires in practise a case analysis at nearly every measurement, which is a distraction from any potential ``uniform'' description of the computation.
Therefore, a likely prerequisite for developing intuitions about how to solve problems in measurement-based models is to abstract away this detail of measurement-based computation.

By definition, in a measurement-based procedure which performs a unitary transformation, the final quantum state is independent of the measurement results, even if the operations performed to obtain that state were not.
This suggests an approach where we omit all details of the measurement results or the probabilities of them occurring, except for some particular ``distinguished'' sequence of measurement results which can occur with non-zero probability.
For instance, we may consider the measurements performed by the ``default'' measurement angles (with respect to which all of the measurement adaptations are performed): if the measurement results $\s[v] = 0$ occur for each $v \in O\comp$ with non-zero probability, the adaptation in a procedure which performs a unitary serves only to select for the same transformation as would occur for that branch.
In this case, a measurement-based procedure for a unitary transformation may be regarded as \emph{simulating postselection} of this ``distinguished'' sequence of results, which is feasible due to the special choice of initial state and measurements involved.\footnote{%
	Aaronson~\cite{Aaronson05} shows that allowing postselection even of single-qubit measurement results, together with a set of (approximately) universal unitary operations and preparation of the $\ket{0}$ state, leads to circuit models which can efficiently solve problems in \PP: this is taken as a sign of the unreasonability of models allowing postselection.
	In contrast, the way in which measurements are simulated in the stabilizer formalism (and in the context of the flow/eflow/gflow techniques outlined in Section~\ref{sec:flowsFurtherResults}) allows the evolution of the state-space to be efficiently modelled due to the existence of a highly structured system of correlations of the states being measured, and the relationships between the measurement observables and those correlations.}

If we could actually postselect the distinguished measurement results, then we would not need to perform any adaptation of the measurement procedure, and dispense with classical feed-forward.
This would leave us with a set of data from which one could plausibly form an analytical intuition:
\begin{enumerate}
\item
	a set of qubits $V$\,, with an input subsystem $I \subset V$ supporting an arbitrary input state, with the others prepared initially in the $\ket{\splus}$ state\,;

\item
	a graph $G$ on $V$, describing a set of entangling operations performed between pairs of qubits;

\item
	a \emph{set} (without any ordering) of measurements to perform on the qubits of $V$; and

\item
	the subset of qubits $O \subset V$ on which no measurements are to be performed, in order to leave a residual quantum state as output.
\end{enumerate}
Such data represents a procedure to solve a problem using a \emph{postselection-based} (rather than measurement-based) model of quantum computing. 

Supposing that one can form intuitions about quantum computation in these terms, and even derive tuples $(G,I,O,M)$ consisting of  a geometry $(G,I,O)$ and a set of default measurements $M$ encoding a solution to a computational problem, it would then remain to produce from this a description of a realizable algorithm.
In this case, it is sufficient to specify an ordering of the measurements in the postselection-based ``algorithm'', as well as a specification of how those measurements are to be adapted and any final corrections to be applied in case the distinguished measurement results do not occur.
If this is possible, the postselection-based procedure is meaningfully an algorithm, which may be extended in principle to a realizable procedure to perform the operation with bounded error.

In the case where every measurement is performed in the \XY\ plane, the basis of each measurement is of the form $\ket{\pmsub\alpha} \propto \ket{0} \pm \e^{i\alpha} \ket{1}$ for some angle $\alpha$, as defined in~\eqref{eqn:pmsubAlpha}.
Suppose that we are given a geometry $(G,I,O)$, and a vector $\vec{a} = (\alpha_v)_{v \in O\comp}$ of the default measurement angles for each $v \in O\comp$, which together define a postselection-based transformation $\Upphi: \D(I) \to \D(O)$ in the case where all of the measurements on the qubits $v$ yield $\s[v] = 0$.
Such a postselection-based procedure will also be a special case of some measurement-based procedure, with different computational branches for each sequence of measurement results: we refer to the branch which performs the desired postselection procedure, given by the measurement sequence $\big(\s[v]\big)_{v \in O\comp} \equiv 0$, as the \emph{positive branch}.

Under these constraints, we may formalize the problem of mapping postselection-based procedures to measurement-based algorithms as follows:
\begin{problem}[Measurement Pattern Interpolation (\MPI)]
	\label{problem:MPI}
	Let $(G,I,O,\vec{a})$ be a partial specification of a measurement pattern $\mathfrak P$ in the $\OneWay(\A)$ model for some $\A \subset \R$, where $\mathfrak P$ has underlying geometry $(G,I,O)$ and default measurement angles $\vec{a} = (\alpha_v)_{v \in O\comp} \in \A^{O\comp}$\,.
	Given that $\mathfrak P$ performs a unitary transformation $U: \cH[I] \to \cH[O]$ conditioned on its' positive branch, determine a measurement pattern $\overline{\mathfrak P}$ with underlying geometry $(G,I,O)$ which performs $U$ unconditionally, if such a pattern exists.
\end{problem}
(This problem may readily be extended beyond measurements performed in the \XY-plane, to which $\OneWay(\A)$ is restricted; however, we will be primarily concerned with the formulation presented above.)

In this chapter, I will describe solutions for special cases of the Measurement Pattern Interpolation problem, which revolve around generalized flow conditions on the geometry $(G,I,O)$, or on the special case of ``Pauli measurements'', \ie\ when we have $\ket{\spm_{\alpha_v}} = \ket{\spm \sfx}$ or $\ket{\spm_{\alpha_v}} = \ket{\spm \sfy}$ for each $v \in O\comp$.

In the absence of ``algorithmic insights'' provided by measurement-based computation --- and setting aside theoretical motivations, such as obtaining a more robust understanding of the scope of those measurement-based procedures which perform unitary transformations --- solving \MPI\ is only of practical importance if there exists a source of ``problem instances'' $(G,I,O,\vec a)$ for which we wish to obtain measurement patterns. 
In this chapter, I also present \emph{quadratic form expansions}, which is a format in which unitaries may be expressed concisely.
These are plausibly a ``natural'' source of instances of the Measurement Pattern Interpolation problem, in that these expressions (and similar ones) have arisen multiple times in the course of research in quantum information independently of the formulation of \MPI.

\paragraph{Previous appearances of this work.}

Some of the work in this chapter are developments original to this thesis, including the substitution described in~\eqref{eqn:diagonalExpansions}, and the solved case of the \MPI\ for geometries derived from eflows described in Section~\ref{sec:synthEflows}.
The rest of the work of this chapter is the result of a collaboration with Elham Kashefi, Vincent Danos, and Martin Roetteler, and first appeared in~\cite{BDKR08}.

\section{Quadratic Form Expansions}

\subsection{Notation and definitions}

We first present a notational device which will prove most useful:
\begin{notation}
	For an index set $V$, and for a vector $\vec{x} \in \ens{0,1}^V$, and for a subset $S \subset V$, the vector $\vec{x}_S$ denotes the restriction of $\vec{x}$, considered as a function $\vec x: V \to \ens{0,1}$, to the set $S$.
	In particular, for an operator $\ket{\vec x} \in \cH^{\ox V}$ given by
	\begin{gather}
			\ket{\vec x}
		\;\;=\;\;
			\bigotimes_{v \in V} \;\ket{x_v}\,,
	\end{gather}
	we define the corresponding standard-basis vector $\ket{\vec x_S} \in \cH^{\ox S}$ by
	\begin{gather}
			\ket{\vec x_S}
		\;\;=\;\;
			\bigotimes_{v \in S} \;\ket{x_v}\,.
	\end{gather}
\end{notation}
\begin{definition}
	\label{def:quadFormExpan}
	Let $V$ be a set of $n$ indices, and $I, O \subset V$ be (possibly intersecting) subsets.
	Then a \emph{quadratic form expansion} for an operator $M: \cH^{\ox I} \to \cH^{\ox O}$ is an expression of the form
	\begin{align}
		\label{eqn:quadFormExpansion}
			M
		\;=\;\;
			C\! \sum_{\vec{x} \in \ens{0,1}^V} \!\e^{i Q(\vec{x})} \;\ket{\vec{x}_O}\bra{\vec{x}_I}	\;,
	\end{align}
	where $Q$ is a quadratic form\footnote{%
	A \emph{quadratic form} is a polynomial where every term has degree $2$.}
	in $\R^V$, and $C$ is a complex scalar.
\end{definition}
For an operator expressed in matrix form, a quadratic form expansion provides an expression for each matrix coefficient as a scalar multiple of a sum of complex numbers with modulus $1$ (or \emph{complex units}), where the scalar factor and the arguments of these complex units are given by a uniform formula.
As such, it is essentially an interpretation of an operator as an abstract ``sum over paths'', where each path has the same amplitude, but a different phase: each such ``path'' is given by a particular string $\vec x$, and the coefficient for a transition $\vec{a} \mapsto \vec{b}$ is given by summing over all paths satisfying the constraints $\vec x_I = \vec a$ and $\vec x_O = \vec b$.

\subsection{Related work in the literature}
\label{sec:relatedWork}

Similar expressions to those in~\eqref{eqn:quadFormExpansion} have occurred several times previously in quantum information or quantum mechanics literature (albeit without the terminology or notation presented here).
We will provide a brief overview of some examples.

\subsubsection{Stabilizer codes and Clifford group operations}

Perhaps the earliest presentation of quadratic form expansions is by Schlingemann and Werner~\cite{SW2001}, where they are used to describe a special class of stabilizer codes (but using significantly different notation, and where the expansions are extended to $\vec x \in \Z_d^V$ for $d \ge 2$ rather than only for $\vec x \in \ens{0,1}^V$).
Their analysis corresponds in the boolean case to quadratic form expansions where $Q$ is obtained from the adjacency matrix of a graph, and where $I, O$ form a partition of the index set $V$.
This treatment was extended to describe generalized stabilizer codes and Clifford group operations over $\U(d)$ in Schlingemann~\cite{Schl04}.
The link between the work of Schlingemann and Werner (as well as their notations), and quadratic form expansions as they are defined and treated in this thesis, is perhaps best hilighted by the result of~\cite{GKR2002}, which proved that all stabilizer codes are locally equivalent to some graphical code.

Dehaene and de Moor~\cite{DM03} also present an analysis of Clifford group operations over $\cH$, in which they present an operator formula for a generic Clifford group operation which may be obtained from a description of how it transforms the Pauli group by conjugation.
We will describe this matrix formula, and specify how it may be computed, in Section~\ref{sec:synthClifford}.

\subsubsection{The quantum Fourier transform}

One concrete example of a unitary matrix which is naturally expressed as a quadratic form expansion is the quantum Fourier transform over the cyclic group $\Z_{2^k}$, which we may write
\begin{align}
		\cF_{2^k}
	\;\;=&\;\;
		\sfrac{1}{\sqrt{2^k}} \, \sum_{x,y \in \Z_{2^k}} \e^{2\pi i x y/2^k} \ket{y}\bra{x}
	\notag\\[2ex]\cong&\;\;
		\sfrac{1}{\sqrt{2^k}}
		\sum_{\substack{\vec{x} \in \ens{0,1}^k \\ \vec{y} \in \ens{0,1}^k}}
		\exp\!\paren{\sfrac{i\pi}{2^{k-1}}
			\sqparen{ \sum\limits_{\phantom{h}\!\!j=0}^{k-1} 2^h x_j }
			\sqparen{ \sum\limits_{\phantom{j}\!\!h=0}^{k-1} 2^j y_h }}
			\ket{\vec y} \bra{\vec x},
\end{align}
identifying the integers $x,y \in \Z_{2^k}$ with their usual representations as binary strings; this is a quadratic form expansion with $V$ partitioned into disjoint index sets associated with the bits of $\vec{x}$ and $\vec{y}$, and a quadratic form $Q: \R^{2k} \to \R$ given by
\begin{align}
	Q(\vec x, \vec y) \;\;=\;\; \sum_{h = 0}^{k-1} \; \sum_{j = 0}^{k-h-1} \;\sfrac{2^{(h+j)}}{2^{k-1}} \;\pi \,x_j \,y_h \;.
\end{align}
On page~\pageref{sec:fourierTransformEg}, we will show how this structure can be exploited to reproduce the linear-nearest neighbor implementation of $\cF_{2^k}$ described by Fowler, Devitt, and Hollenberg in~\cite{FDH04} by essentially automatic techniques.

\paragraph{Remark.} The operation $\cF_{2^k}$ is an instance of a generalized Clifford group operation over $\SU(2^k)$\,; however, the description of $\cF_{2^k}$ above differs from the treatment of \eg~\cite{Schl04} in that we expand it here in terms of operations on qubits, rather than leaving it as an atomic operation on $\cH[2^k]$.

\subsubsection{Sums over paths}
\label{sec:sumsOverPaths}

A similar representation of operators as a sum over paths as in~\eqref{eqn:quadFormExpansion} was used by Dawson \etal\ in~\cite{DHHMNO04} to provide a succinct proof of $\BQP \subset \PP$, using cubic polynomials over $\ens{0,1}$ rather than quadratic polynomials; however, comments made in Section~VI of their work anticipate quadratic form expansions with discretized coefficients, and imply a method (which we will describe in Section~\ref{sec:quadFormExpansionsFromCircuits}) of obtaining these expansions from unitary circuits using the gate-set $\Phases(\frac{\pi}{4}\Z)$ defined on page~\pageref{eqn:elemGateSets}.

The expressions considered in~\cite{DHHMNO04} are explicitly described as a sum over computational paths which arise in the evolution of a quantum state through a unitary circuit: the evolution of the system in each path is described by a time-ordered sequence of standard basis states given by boolean strings, where these states differ in at most one bit position for consecutive time-steps.
The ``paths'' in this case are walks in the boolean cube $\ens{0,1}^k$ (for a circuit acting on $k \le n$ qubits) between boolean strings which differ in zero or one bit position (\ie\ including self-loops).
The bit whose value may change at each step is the same across all paths; these walks can then be described by the initial position, followed by one bit for each edge traversed, which may be described by a boolean function $\ens{0,1}^V$ describing the history of the computation in that path.
Summing over all computational paths gives rise to the unitary transition matrix.

The analogy of such matrix formulae to a sum-over-paths, suggested both in~\cite{DHHMNO04} and in this thesis, is prompted in part by the formal similarity between \eg\ quadratic form expansions and \emph{path integrals}.
Introduced by Feynman~\cite{Feynman48}, the path integral formalism is an alternative approach to describing the evolution of quantum states by associating amplitudes to trajectories of the system (\eg\ through space-time or phase space), and describing the probabilities of events upon measurement using the cumulated amplitude over all paths starting from the initial state and ending at a given configuration.
This ``summation over paths'' can be arrived at by taking integrals over all possible configurations at intermediate times (taking the limit as the intervals between these intermediate times tend to zero), which amounts to integrating over a collection of ``virtual events'' for these intermediate times.
This leads to a formulation where the transition amplitude between the initial and final configurations depends on some quantity with the units of energy (usually the \emph{action} of the system) depending on physical parameters associated with these virtual events, which can then be attributed to trajectories in space.
Further details can be found in~\cite{Feynman48, FH65, Schulman81}.
In a quadratic form expansion, the indices of the vector $\vec x$ which are neither in $I$ or $O$ may be considered to correspond to these virtual events, and the quadratic form associated with the ``computational paths'' of the computation seems to correspond to the action (which in most applications of path integration determines the phase of each path).
This suggests that a quadratic form expansion, or similar matrix expression, may be a means of discretizing a path integral in a way which may be amenable to study for quantum algorithms.

There are some technical discrepancies on a formal level, most notably that the phase of a system with multi-body interactions generally is not a quadratic polynomial in the dynamical variables of the system;\label{discn:higherDegree} or we may wish to evaluate path integrals where different paths have amplitudes of different weights (\eg\ with lower amplitudes away from the stationary points of the action, due to cancellation of contributions of similar paths) in an attempt to obtain an approximation which is easier to estimate.
We will show in in Section~\ref{sec:higherDegree} how matrix expressions similar to quadratic form expansions, with \emph{unequally weighted} paths or where the action is of degree higher than $2$, may be transformed into quadratic form expansions.
The simplicity of these transformations suggest that quadratic form expansions are not unduly restrictive from the point of view of representing the discretization of a path integral.
Thus, an examination of when we can synthesize quantum algorithms from quadratic form expansions would plausibly lead to automatic techniques for efficiently estimating path integrals on a quantum computer.

\subsubsection{Remarks on quadratic form expansions as a representation of operators}

Given the possibility of quadratic form expansions arising generically from path-integral descriptions of physical systems, as well as the fact that these and similar matrix formulae have repeatedly arisen in the literature, I conjecture that quadratic form expansions can be expected to arise naturally as mathematical expressions of interest for operators in quantum information theory.
Based on this, I will suppose for the purposes of this thesis that a quadratic form expansion is a reasonable format in which a unitary operator $U$ may be specified, in the context of the problem of obtaining a succinct decomposition of a CPTP map performing $U$ in some standard model of quantum computing (\eg\ a unitary circuit model or measurement based model).

Setting aside the special case of the quantum Fourier transform $\cF_{2^k}$, we will see in Section~\ref{sec:synthClifford} how such an expression of a unitary $U$ may be obtained (without first having a unitary circuit or measurement-based algorithm for $U$) in the case of Clifford group operations.
While this represents a limited class of operations, it nevertheless represents a generic family of operators $U$ for which quadratic form expansions may be readily obtained.

\subsection{Constructing quadratic form expansions}

As noted above, we will be primarily interested in the problem of transforming a quadratic form expansion into a quantum circuit or measurement-based procedure.
However, to illustrate the generality of quadratic form expansions, and how they relate to unitary circuit models and the measurement based model of quantum computing, we will show how to perform the reverse transformation, from unitary circuits and measurement-based procedures to quadratic form expansions.

\subsubsection{Construction as a sum over computational paths in circuits}
\label{sec:quadFormExpansionsFromCircuits}

As we remarked in Section~\ref{sec:sumsOverPaths}, Dawson~\etal\ describe a method for obtaining quadratic form expansions from unitary circuits in~\cite[Section~VI]{DHHMNO04}; the particular set of unitary transformations they use is the set $\ens{H, R_\sfz(\pion 4), \cX}$.
The construction is closely related to the stable index circuit notation presented in Chapter~\ref{Fundamentals of quantum computation} (on page~\pageref{def:stableIndex}: we will sketch this construction using the somewhat more general parameterized gate set $\ens{H, Z^t, \cZ^t}$, where
\begin{align}
		Z^t
	\;=&\;
		\left[\begin{matrix} \;1 & 0 \\[0.5ex] \;0 & \;\e^{i\pi t} \end{matrix}\,\right]
	\;\propto\;
		R_\sfz(\pi t),
	&
		\cZ^t
	\;=&\;
		\left[\begin{matrix} \;1 & \;0 & \;0 & \!0 \\[0.5ex] \;0 & \;1 & \;0 & \!0 \\[0.5ex]
									 \;0 & \;0 & \;1 & \!0 \\[0.5ex] \;0 & \;0 & \;0 & \e^{i\pi t} \end{matrix}\,\right],
\end{align}
and where $t$ is allowed to range over some subset of $(-1,\, 1\;\!]$.

Consider a quantum circuit $C$ implementing $U$ exactly, using the operations $H$, $\cZ^t$, and $Z^t$.
As in a stable index representation of $C$, we assign indices (or \emph{path labels}, to use the terminology of~\cite{DHHMNO04}) to the segments of each wire in the circuit, with each segment bounded by an input terminal, and output terminal or a Hadamard operation.
In particular:
\begin{enumerate}
\item
	We enumerate the ``input'' segments of the circuit from $1$ to $k$, and for each wire $1 \le j \le k$ introduce a path label $x_j$ corresponding to an input bit $x_j \in \ens{0,1}$; and we set $I = \ens{1,\ldots,k}$.

\item
	Reading the circuit from the inputs to the outputs, we introduce new path variables $x_j$ for $j > k$ for each wire segment ``introduced'' by the presence of a Hadamard operation, corresponding to the ``random'' bit which is introduced by applying the Hadamard to the ``previous value'' of the bit.
\end{enumerate}
We may then describe computational paths described by the circuit by a walk in the boolean cube $\ens{0,1}^k$, with one bit possibly being flipped each time a Hadamard gate is performed.

If the total number of path labels introduced is $n$, a single computational path can therefore be described by setting all of the the path variables $x_1 \cdots x_n$ collectively to some particular binary string in $\ens{0,1}^n$.
For a path which is given by a string $\vec x \in \ens{0,1}^n$, the phase of that path (which governs how it interferes with other paths in the unitary evolution described by $C$) is given by a function $\varphi(\vec x)$ which may be iteratively constructed from the gates of $C$ as follows:
\begin{romanum}
\item
	For every Hadamard gate between two wire segments labelled with path variables $x_h$ and $x_j$, we include a term $x_h x_j$\,.

\item
	For every $\cZ^t$ operation between two wires, with a path variable $x_h$ labelling the segment of one wire and $x_j$ labelling the segment of the other wire in which the $\cZ^t$ operation is performed, we add a term $t\, x_h x_j$\,.

\item
	For every $Z^t$ operation on a wire segment labelled with a path variable $x_j$\,, we add a term $t \, x_j^2$.
	(As the path variable $x_j$ ranges over $\ens{0,1}$, the extra power of $2$ has no effect.)
\end{romanum}
We may collect $\cZ^t$ and $Z^t$ terms so that there is at most one $\cZ^t$ gate between segments labelled with $x_h$ and $x_j$ (for $h \ne j$) for some value of $t$, and at most one $Z^t$ gate on a segment labelled with $x_h$, for any $h,j \in \ens{1,\ldots,n}$.
As there is at most one Hadamard operation between two wire segments by construction of the circuit, this implies that the construction of $\varphi$ will include at most one $x_h x_j$ term for each $\ens{h,j} \subset \ens{1,\ldots,n}$.
The phase of a given path, described by a bit-string $\vec{x} \in \ens{0,1}^n$, is then given by $(-1)^{\varphi(\vec x)} = \e^{i\pi\varphi(\vec x)}$.
Each path also has an associated amplitude of $2^{-r/2}$, where $r = n - k$ is the number of Hadamard gates in the circuit.\footnote{%
Although it is quite reasonable to consider $\varphi$ to be simply a polynomial over $\R$, in terms of the descriptions used in Section~VI of~\cite{DHHMNO04}, one may consider $\varphi$ to be a polynomial over the ring $\R / 2\Z$.
If we restrict to $t \in \frac{\pi}{4} \Z$, we may simplify this to the finite ring $\Z_8$ by multiplying all of the coefficients by $4$, and using it to describe powers of $\sqrt{i}$ rather than of $-1$ (but still labelling each path by a vector $\vec x \in \ens{0,1}^n \subset \Z_8^n$).}

Let $O$ be the set of indices $j$ such that some wire is labelled by the path-variable $x_j$ at its' output end.
Then, the initial points of computational paths are described by bit-vectors $\vec{a} \in \ens{0,1}^I$\,, and the
terminal points of paths are described by $\vec{b} \in \ens{0,1}^O$\,. The coefficients
${U_{\vec a}}^{\vec b}$ can then be given as the sum of the contributions of all paths beginning at $\vec{x}_I = \vec{a}$ and ending
at $\vec{x}_O = \vec b$\,:
\begin{align}
		{U_{\vec a}}^{\vec b}
	\;\;=\;\;
		\bra{\vec b} U \ket{\vec a}
	\;\;=&\;\;
		\frac{1}{\sqrt {2^r}}
		\sum_{\substack{\vec{x} \in \ens{0,1}^n \\[0.5ex] \vec{x}_I = \vec{a} \\ \vec{x}_O = \vec{b}}} \e^{i \pi \varphi(\vec x)}	\;.
\end{align}
As $\varphi(\vec{x})$ is a quadratic form over $\vec x$ by construction, this is an expression of the coefficients of $U$ as a quadratic form expansion.

Any stable index tensor expression for $U$ (using any set of elementary gates) can be interpreted as a sum-over-paths formulation of $U$, identifying paths with a setting of all of the indices in the expression.
However, this will not in general yield a \emph{quadratic form expansion}, as the modulus of the amplitude of each path need not be uniform, as in a quadratic form expansion: we then cannot express $U$ a uniformly normalized sum.
As well, gates with generic three-body or higher-degree interactions will give rise to cubic terms in the phase function $\varphi$.
Thus, as a construction for quadratic form expansions, the construction outlined above cannot be extended to arbitrary gate sets (nor even arbitrary ``parsimonious'' gate sets, in the sense of Definition~\ref{def:parsimonious}), but apparently only to gate sets closely related to $\Phases(\A)$.

\subsubsection{Inductive construction from circuit decompositions}

An equivalent, and perhaps more illustrative, construction to the above can be made by supposing that one has a unitary circuit $C$ decomposed into unitaries, each of which has a known quadratic form expansion.
In such a case, we can compose the quadratic form expansions by matrix multiplication.

Suppose we are given a stable index tensor representation of a unitary $U$ in terms of any set of unitary transformations, provided that every unitary transformation $W$ in the decomposition has a known quadratic form expansion,
\begin{gather}
		W
	\;\;=\;\;
		C_W\!\!
		\sum_{\vec x \in \ens{0,1}^{V_{\smaller W}}} \e^{i Q_{\smaller W}(\vec x)} \ket{\vec x_{O_{\smaller W}}}\bra{\vec x_{I_{\smaller W}}}	,
\end{gather}
for some constant $C_W \in \C$, quadratic form $Q_W$, and sets $V_W, I_W, O_W$ depending on $W$. 
Relabelling the elements of the particular sets $V_W$, $I_W$, and $O_W$ in this expansion does not substantially change the operator, except in changing the labels of the space on which it operates.
Consider an instance $W\supp{j}$ of the unitary $W$ in the stable index tensor expression for $U$, with input indices $I\supp{j}$ and output indices $O\supp{j}$\,: we may then make a copy of the above expansion by bijectively mapping $I_W \to I\supp{j}$ and $O_W \mapsto O\supp{j}$, and defining an extended index set $V_W \mapsto V\supp{j}$ of indices such that, for any other set $V\supp{h}$ corresponding to any other unitary in the circuit, we have $v \in V\supp{j} \inter V\supp{h}$ only if $v \in I\supp{j}$ or $v \in O\supp{j}$ (in which case we also have $v \in O\supp{h}$ or $v \in I\supp{h}$ respectively).

Having obtained a quadratic form expansion for \emph{each} unitary in the decomposition of $U$, we may then apply the following Lemma, whose proof is trivial:
\begin{lemma}
	\label{lemma:quadFormExpanComposition}
	Let $W_1$, $W_2$ be operators given by quadratic form expansions of the form
	\begin{align}
			W_j
		\;=&\;\;
			C_j \!\! \sum_{\vec x \in \ens{0,1}^{V_j}} \e^{i Q_j(\vec x)} \; \ket{\vec x_{O_j}} \bra{\vec x_{I_j}}	\;.
	\end{align}
	In the following, $C = C_1 C_2$\,, and sums are over $\ens{0,1}^{V_1 \union V_2}$.
	\begin{romanum}
	\item
		If $V_1 \inter V_2 = I_2 = O_1$\,, then $U_2 U_1 \,=\,
			C\, \sum\limits_{\vec x} \; \e^{\,i Q_1(\vec x) \,+\, i Q_2(\vec x)} \, \ket{\vec x_{O_2}} \bra{\vec x_{I_1}} .
		$\vspace{1ex}
	\item
		If $V_1$ and $V_2$ are disjoint, then $\smash{U_1 \ox U_2 \,=\,
			C\, \sum\limits_{\vec x} \; \e^{\, i Q_1(\vec x) \,+\, i Q_2(\vec x)} \, \ket{\vec x_{O}} \bra{\vec x_{I}} ,
		}$
		where $I = I_1 \union I_2$ and $O = O_1 \union O_2$\,.
	\end{romanum}
\end{lemma}
By construction, the quadratic form expansions derived for each of the unitaries in the decomposition of $U$ satisfy the above properties, and so we may compose them to obtain a quadratic form expansion for the whole.

For the gates $H$, $Z^t$, and $\cZ^t$ in particular, described in the sum-over-paths construction, we have the following expansions:
\begin{subequations}
\begin{align}
		H
	\;=&\;\;
		\sfrac{1}{\sqrt 2}\!
		\sum_{\vec x \in \ens{0,1}^2}
			\e^{i\pi x_1 x_2}
			\ket{x_2}\bra{x_1},
			\quad
	&
		V =& \ens{1,2},& I =& \ens{1},& O =& \ens{2};
	\\[1.5ex]
		Z^t
	\;=&
		\sum_{\vec x \in \ens{0,1}^1}
			\e^{i\pi t x_{\:\!\!1}^{\,2}}
			\ket{x_1}\bra{x_1},
			\quad
	&
		V =& \ens{1},& I =& \ens{1},& O=& \ens{1};
	\\[1.5ex]
		\cZ^t
	\;=&
		\sum_{\vec x \in \ens{0,1}^2}
			\e^{i\pi t x_1 x_2}
			\ket{x_1 x_2}\bra{x_1 x_2},
			\quad
	&
		V =& \ens{1,2},& I =& \ens{1,2},& O=& \ens{1,2}.
\end{align}
\end{subequations}
None of these unitaries involve ``internal'' indices in the sense described above, and so a stable index tensor expression for a unitary $U$ in terms of these unitaries may be easily transformed into a quadratic form expansion for $U$.
(The quadratic form expansions constructed with this set of unitaries will be equivalent to those obtained from the sum-over-paths construction of the previous section.)

\paragraph{Remark.}
We may also decompose unitaries $U$ as a product of \emph{non-unitary} operators with known phase-map decompositions as well.
For instance, we may define the matrix
\begin{align}
	\label{eqn:nonunitary}
		L
	\;\;=&\;\;
		\e^{-i \pi/6} \left[ \begin{matrix}
                        \;1 & \e^{2 \pi i/3} \\ \;1 & -1
                                        \end{matrix} \,\right]
	\;\;=\;\;
 		\e^{-i \pi/6} \!\!\!
 		\sum_{x_1,x_2 \in \ens{0,1}}\!
 			\e^{i\pi(2 x_1 \,+\, x_1 x_2)/3}
 			\ket{x_2}\bra{x_1}	\;;
\end{align}
it is easy to verify that $L$ is self-inverse, in which case the above expansion gives rise to another quadratic form expansion decomposition for $\idop_2$\,:
\begin{gather}
		\idop_2
	\;\;=\;\;
 		\e^{-i \pi/3} \!
 		\sum_{\vec{x} \in \ens{0,1}^3}\!
 			\e^{i\pi \!\!\:\big[(2 x_1^{\,2} \,+\, x_1 x_2) \,+\, (2x_2^{\, 2} + x_2 x_3)\big]\!\!\:/3}
 			\ket{x_3}\bra{x_1}	\;.
\end{gather}

\subsubsection{Construction from one-way measurement patterns}

The qubit-reversal pattern and the many-qubit measurement pattern $\mathfrak{Zz\cdots z}^\theta$, which we have considered in Sections~\ref{sec:qubitReversal} and~\ref{sec:zzMany}, do not have any decomposition into smaller elements due to (currently known) generic constructions from unitary circuits into the one-way model.
This motivates considering quadratic form expansions in terms of measurement-based quantum computing, independently of unitary circuit models.

Suppose we have a measurement-based procedure with underlying geometry $(G,I,O)$ which performs a unitary transformation $U: \cH[I] \to \cH[O]$, and that there is a sequence of measurement results which occurs with non-zero probability for which there are no final corrections to be performed on the qubits in $O$.
Let $(\alpha_v)_{v \in O\comp}$ be the choice of measurement angles for this branch of the computation: we then designate this branch of the computation as the positive branch.
Then we may express $U$ (modulo a scalar factor due to the probability of postselecting the positive branch) in the form
\begin{align}
		U
	\;\propto&\;\;
		\sqparen{\bigotimes_{v \in O\comp} \bra{+_{\ssstyle \alpha_v}}_v}
		\sqparen{\prod_{uv \in E(G)} \!\!\cZ_{u,v}\;}
		\sqparen{\bigotimes_{v \in I\comp} \ket{+}_v},
\end{align}
which represents the operations of preparing the qubits of $I\comp$, performing the entangling operations, and postselecting the states $\ket{\splus_{\alpha_v}}$ for $v \in O\comp$ which occur in the positive branch.
Note that we can factor the postselection operators $\bra{\splus_\alpha}$ into a $Z$-rotation and postselection of the state $\ket{\splus}$,
\begin{gather}
		\bra{+_{\alpha_v}}
	\;\;=\;\;
		\bra{+} R_z(-\alpha_v)	;
\end{gather}
factoring these rotations into the middle, we may collect them together with the $\cZ$ operations corresponding to the edges of the entanglement graph $G$ into a single diagonal unitary operation arising from the exponential of a diagonal Hermitian matrix,
\begin{subequations}
\begin{gather}
		\sqparen{\bigotimes_{v \in O\comp} R_\sfz(-\alpha_v)}
		\sqparen{\prod_{uv \in E(G)} \cZ_{u,v}}
	\;\;=\;\;
		\e^{i J},	\qquad\text{where}
	\\[2ex]
	\label{eqn:phaseMapHamiltonian}
		J
	\;\;\;=
		\sum_{\substack{\ens{u,v} \subset V(G) \\ u \ne v}} \!\!\pi \bigg[ \ket{1}\bra{1}_u \ox \ket{1}\bra{1}_v \bigg]
		\;\;-\;\;
		\sum_{v \in O\comp} \alpha_v \ket{1}\bra{1}_v	\;.
\end{gather}
\end{subequations}
Define a quadratic form $Q: \R^{V(G)} \to \R$ by 
\begin{align}
		Q(\vec x)
	\;\;\;\;=\;
		\sum_{\substack{\ens{u,v} \subset V(G) \\ u \ne v}} \!\!\pi x_u x_v
		\;\;-\;\;
		\sum_{v \in O\comp} \alpha_v x_v^2 \;,
\end{align}
and note that, for any $S \subset V(G)$, we can expand
\begin{align}
		\bigotimes_{v \in S\comp} \ket{\splus}_v
	\;\;\;\propto\;
		\sum_{\vec x' \in \ens{0,1}^{S\comp}} \idop_2^{\,\ox S} \ox \ket{\vec x'}
	\;\;\;\;=&\;
		\sum_{\vec x \in \ens{0,1}^{V(G)}} \Big[ \ket{\vec x_S}\bra{\vec x_S} \Big] \ox \ket{\vec x_{S\comp}}
	\notag\\[0.5ex]=&\;
		\sum_{\vec x \in \ens{0,1}^{V(G)}} \ket{\vec x}\bra{\vec x_S} 	\;.
\end{align}
Then, we may obtain
\begin{align}
		U
	\;\;\propto&\;\;
		\sqparen{\bigotimes_{v \in O\comp} \bra{+_{\ssstyle \alpha_v}}_v}
		\e^{i J}
		\sqparen{\bigotimes_{v \in I\comp} \ket{+}_v}
	\notag\\[1ex]\propto&\;\;
		\sqparen{\sum_{\vec y \in \ens{0,1}^{V(G)}} \!\! \ket{\vec y_O}\bra{\vec y}}
		\e^{i J}\!
		\sqparen{\sum_{\vec x \in \ens{0,1}^{V(G)}} \!\! \ket{\vec x}\bra{\vec x_I}}
	\notag\\[1ex]=&\;\;\;
		\sum_{\vec x, \vec y \in \ens{0,1}^{V(G)}} \!\! \ket{\vec y_O} \bigg[ \e^{i \, \bra{\vec y}\!\!\; J \ket{\vec x}} \bigg] \bra{\vec x_I}
	\;\;\;\;=
		\sum_{\vec x \in \ens{0,1}^{V(G)}} \!\! \e^{iQ(\vec x)} \; \ket{\vec x_O}\bra{\vec x_I},
\end{align}
which is a quadratic form expansion for $U$.

\subsection{Interpretation as a description of a postselection procedure}
\label{sec:postselectionProcedure}

In the context of \MPI\ (page~\pageref{problem:MPI}), the construction above of quadratic form expansions from measurement-based computations hints at an approach for doing the reverse: to turn a quadratic form expansion into a measurement-based procedure, and more precisely into an instance of the \MPI.
Consider a unitary $U$ given by a quadratic form expansion as in~\eqref{eqn:quadFormExpansion}, where the quadratic form $Q$ is given by
\begin{align}
	\label{eqn:quadForm}
		Q(\vec x)
	\;\;=&\;
	\sum_{\ens{u,v} \subset V} \theta_{uv} x_u x_v
\end{align}
for any $\vec x \in \ens{0,1}^V$, for some real parameters $\ens{\theta_{uv}}_{u,v \in V}$\,; and where the sum includes singleton sets corresponding to the case $u = v$.
We may express $Q(\vec x)$ as an inner product $\bra{\vec x} \mspace{-3mu} J \mspace{-3mu} \ket{\vec x}$\,, where $J$ is a $2$-local diagonal operator similar in construction to that of~\eqref{eqn:phaseMapHamiltonian}:
\begin{subequations}
\begin{gather}
		J
	\;\;=\;\;
		J_O \,+\, J_1 \,+\, J_2	\;,\qquad\text{where}
	\\[2ex]
	\begin{aligned}
		J_O
	\;\;=&\;\;
		\sum_{w \in O} \theta_{ww} \ket{1}\bra{1}_w	\;,
	&\qquad
		J_1
	\;\;=&\;\;
		\sum_{v \in O\comp} \theta_{vv} \ket{1}\bra{1}_v	\;,
	\end{aligned}
	\\[1ex]
	\!\!\!\!\!
		J_2
	\;=
		\sum_{\substack{\ens{u,v} \subset V \\ u \ne v}} \!\!\theta_{uv} \bigg[\ket{1}\bra{1}_u \ox \ket{1}\bra{1}_v \bigg].		
\end{gather}
\end{subequations}
Then we may decompose $U$ as
\begin{align}
	\label{eqn:phaseMapDecomp}
		U
	\;\;\propto\;
		\sum_{\vec{x} \in \ens{0,1}^V} \!\!\ket{\vec{x}_O} \bra{\vec x}\;\! \e^{i J} \ket{\vec x} \bra{\vec{x}_I}\,
	\;\;\propto&\;\;
			\sqparen{\bigotimes_{w \in O\comp} \bra{\splus}}
			\e^{i J_O} \e^{i J_1} \e^{i J_2}\!\!\:
			\sqparen{\bigotimes_{v \in I\comp} \ket{\splus}}	\;.
\end{align}
Note that $\e^{iJ_1}$ consists simply of single-qubit $Z$ rotations applied to the elements of $O\comp$,
\begin{align}
		\e^{i J_1}
	\;\;=\;\;
		\prod_{v \in O\comp} \e^{i \theta_{vv} \ket{1}\bra{1}_v}
	\;\propto&\;
			\bigotimes_{v \in O\comp} R_\sfz(\theta_{vv})\;,
\end{align}
and similarly for $\e^{iJ_O}$; and we may express $\e^{iJ_2}$ by a product of (potentially fractional) powers of controlled-$Z$ on distinct pairs of qubits:
\begin{align}
		\e^{i J_2}
	\;\;=\;\;
		\prod_{\substack{\ens{u,v} \subset V \\ u \ne v}} \e^{i \theta_{uv} \ket{1}\bra{1}_u \ox \, \ket{1}\bra{1}_v}
	\;\;=&\;\;
		\prod_{\substack{\ens{u,v} \subset V \\ u \ne v}} {\cZ_{u,v}}\big.^{\!\!\!\!\!\!\theta_{uv}/\pi}	\;.
\end{align}
Let $\alpha_v = -\theta_{vv}$ for each $v \in O\comp$\,: then, commuting $\e^{iJ_O}$ leftwards and (similarly) absorbing $\e^{iJ_1}$ into the projection operations, we obtain
\begin{align}
		U
	\;\;\propto\;
			\e^{i J_O}
			\sqparen{\bigotimes_{w \in O\comp} \bra{+_{\alpha_v} }}
			\sqparen{\prod_{\substack{\ens{u,v} \subset V \\ u \ne v}} {\cZ_{u,v}}\big.^{\!\!\!\!\!\!\theta_{uv}/\pi}}
			\sqparen{\bigotimes_{v \in I\comp} \ket{\splus}}	\;,
\end{align}
where we have applied the relation $\ket{\splus_\alpha} \propto R_\sfz(\alpha) \ket{\splus}$.
This is an expression of $U$ as the preparation of some number of $\ket{+}$ states, followed by a diagonal unitary operator consisting of two-qubit (fractional) controlled-$Z$ operations, followed by post-selection of the states $\ket{\splus_{\alpha_v}}$ for $v \in O\comp$ (which succeeds with non-zero probability, provided the initial scalar factor of the quadratic form expansion was nonzero), and finally applying some single-qubit $Z$ rotations on the remaining qubits.

If $\theta_{uv} \in \ens{0,\pi}$ for all distinct $u,v \in V$, and $\theta_{vv} = 0$ for $v \in O$, the above describes precisely the action of the positive branch of a measurement-based computation in which the default measurement angles on the qubits $v \in O\comp$ are given by $-\theta_{vv}$, and where we condition on all measurements yielding the result $\s[v] = 0$.
Thus, such a quadratic form expansion encodes an instance of \MPI.

Furthermore, we may transform any quadratic form expansion into such a form.
For quadratic form expansions with $\theta_{vv} \ne 0$ for $v \in O$, we may obtain a modified geometry by composing to such output qubits a quadratic form expansion for the identity operation,
\begin{align}
	\label{eqn:idopQuadFormExpan}
		\idop_2
	\;\;=\;\;
		\sum_{x_v, x_{v''}} \delta_{x_v,x_{v''}} \ket{x_{v''}} \bra{x_v}
	\;\;=\;\; 
		\sfrac{1}{2} \sum_{x_v, x_{v'}, x_{v''}} (-1)^{x_v x_{v'} + x_{v'} x_{v''}} \ket{x_{v''}} \bra{x_v}
\end{align}
which corresponds to the pattern $\mathfrak{Id}^2$ of~\eqref{eqn:identityPatterns}; this yields a quadratic form expansion with output indices $O'$ such that $v \notin O'$, but where $v'' \in O'$ and $\theta_{v'' v''} = 0$.
To eliminate cross-term coefficients which are not integer multiples of $\pi$, we may use the equality
\begin{gather}
\label{eqn:diagonalExpansions}
		\e^{i \phi (x_u \oplus x_v)}
	\;\;=\;\;
		\sfrac{1}{2}\! \sum_{\vec a \in \ens{0,1}^2} \e^{i\big[\pi(x_u a_1 + x_v a_1 + a_1 a_2) \,+\, \phi a_2^{\, 2}\big]}
\end{gather}
where $x_u \oplus x_v = x_u + x_v - 2x_u x_v$ is the sum of $x_u$ and $x_v$ modulo $2$; we may eliminate factors of $\e^{i \theta_{uv} x_u x_v}$ such that $\theta_{uv} \notin \pi \Z$ for distinct $u,v$ 
by substituting the equality above.
Using this substitution amounts to decomposing the operator $\cZ^t$ using the construction for $\mathfrak{Zz\cdots z}^\theta$ illustrated in Figure~\ref{fig:zzMany}, as Figure~\ref{fig:diagonalYZ} illustrates.
\begin{figure}[tp]
	\begin{center}
		\includegraphics{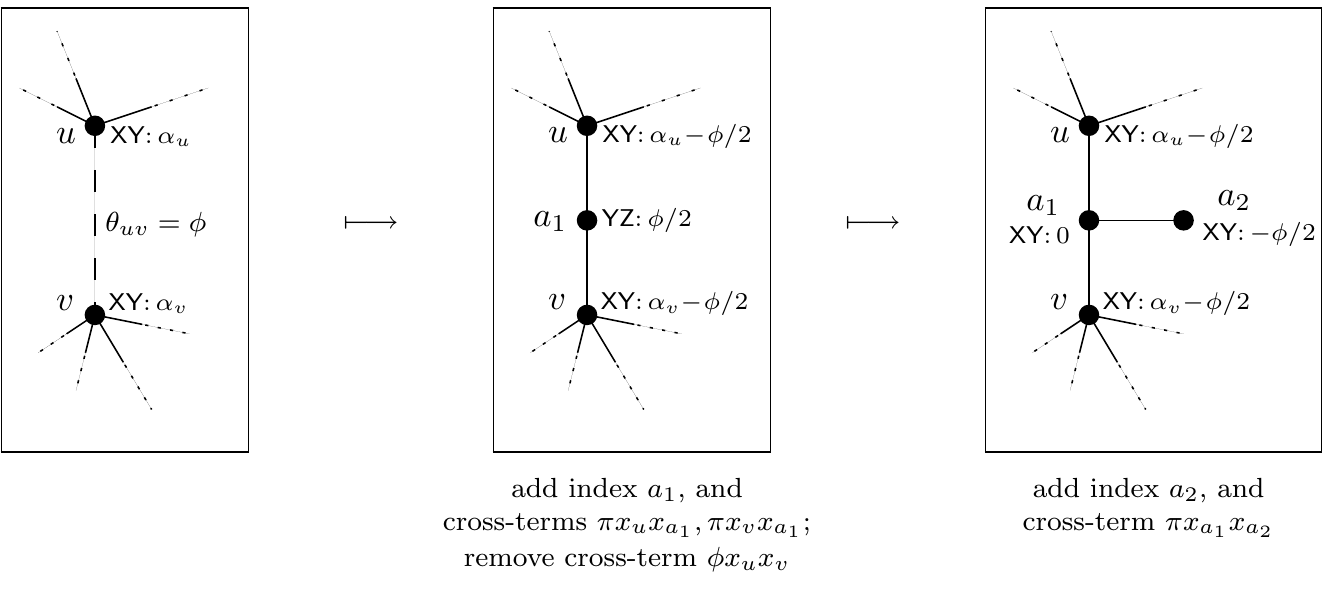}
	\end{center}
	\caption[A transformation of quadratic form expansions to eliminate cross-term angles $\theta_{uv} \notin \ens{0,\pi}$.]{%
		\label{fig:diagonalYZ}%
		A transformation of quadratic form expansions to eliminate cross-term angles $\theta_{uv} \notin \ens{0,\pi}$ which would correspond to fractional entangling operations, illustrated here in terms of geometries and measurement operations of a postselection procedure.
		This transformation essentially consists of applying the decomposition of $\cZ$ into $\mathfrak{Zz}$ and single-qubit rotations described in~\eqref{eqn:clusterEntanglerSubst}, extended to make use of the description of $\mathfrak{Zz\cdots z}^{\phi/2}$ in terms of \XY-plane measurements for more general $\phi$, which is illustrated in Figure~\ref{fig:zzMany}.
		The coefficients of the quadratic form expansion in the left and right panels are given by the corresponding measurement angles in the \XY\ plane; the center panel does not directly correspond to any quadratic form expansion, as it involves a measurement outside of the \XY\ plane.}
\end{figure}

In this way, we may translate any quadratic form expansion into a postselection procedure which uses only the entangling and measurement operations available to the $\OneWay(\A)$ model, where $\A$ is the set of angles which occur in the quadratic form.
In particular, for a quadratic form expansion as above where $\theta_{uv} \in \ens{0,\pi}$ for distinct $u,v$ and $\theta_{vv} = 0$ for $v \in O$, we may obtain a tuple $(G,I,O,\vec a)$ describing an instance of \MPI\ by defining $G$ to be the graph whose vertices are the indices of $V$ and whose edges are those pairs $uv$ such that $\theta_{uv} \ne 0$ and by defining $\vec{a} = (-\theta_{vv})_{v \in O\comp}$\,.
Thus, any such tuple for which the \MPI\ has an efficient solution corresponds to a quadratic form expansion which we may translate to a measurement-based procedure performing a unitary transformation.

\section{Solved cases of Measurement Pattern Interpolation}

At present, there is no general solution to \MPI, which seems to be a difficult problem.
Current approaches towards a solution involve analyzing those measurements which can be treated essentially by the stabilizer formalism, where (as in Sections~\ref{sec:extendedConstructions} and~\ref{sec:gflows}) we may relax the requirement that the measurement observables be Pauli operators.
These lead to the following solved cases:
\begin{enumerate}
\item
	Tuples $(G,I,O,\vec a)$ such that, for the constant function $\lambda: O\comp \to \ens{\XY}$, the tuple $(G,I,O,\lambda)$ has a gflow (Definition~\ref{def:gflow});

\item
	Tuples $(G,I,O,\vec a)$ such that, by applying certain simple graph transformations on $G$ to obtain a smaller graph $\tilde G$ and a vertex set $T \subset V(\tilde G)$, we obtain a tuple $(\tilde G,I,O,T)$ having an eflow (Definition~\ref{def:eflow});

\item
	Tuples $(G,I,O,\vec a)$ such that $\alpha_v \in \frac{\pi}{2}\Z$ for every angle $\vec a = (\alpha_v)_{v \in O\comp}$ (\ie\ where each measurement is performed with respect to either the $\ket{\spm \sfx}$ basis or $\ket{\spm \sfy}$ basis).
\end{enumerate}
In this section, we describe these cases and illustrate how \MPI\ may be solved for each.

As we described in the preceding section, any quadratic form expansion can be transformed into an instance of \MPI\ by first transforming it into another quadratic form expansion whose cross-terms $\theta_{uv}$ are integer multiples of $\pi$.
However, it will in some cases be convenient to synthesize implementations for unitaries directly from a quadratic form expansion having cross-terms which are fractional multiple of $\pi$, corresponding to postselecting in an extended measurement-based model which may use fractional powers of $\cZ$.
This motivates the corresponding extension of the notion of a ``geometry'', which we define here for quadratic form expansions:
\begin{definition}
	\label{def:quadraticFormGeometry}
	For a quadratic form expansion	
	\begin{align}
		\label{eqn:quadFormExpansionAgain}
		C\!\! \sum_{\vec{x} \in \ens{0,1}^V} \e^{i Q(\vec{x})} \;\ket{\vec{x}_O}\bra{\vec{x}_I}
		&&
		\text{where}\quad
		Q(\vec x)
	\;=&\;
		\sum_{\substack{\ens{u,v} \subset V}} \theta_{uv} x_u x_v \;\,,
	\end{align}
	the \emph{geometry induced by the quadratic form} is a triple $(G,I,O)$\,, where $G$ is a \emph{weighted} graph
	with vertex-set $V$, edge-set $\ens{uv \mid u \ne v ~\text{and}~ \theta_{uv} \ne 0}$\,, and edge-weights
	$W_G(uv) = \frac{\theta_{uv}\!}{\pi}$.
	Without loss of generality, we may require $W_G(uv) \in (-1,\, 1\,]$ for all $uv$; we may then refer to edges $uv$ being of \emph{unit weight} if $W_G(uv) = 1$, and \emph{fractional weight} if $\theta_{uv} \notin \ens{0,1}$.
\end{definition}
Geometries with only edges of unit weight correspond to geometries as described in Definition~\ref{def:geometry} in Section~\ref{sec:geometriesOneWay}; we will refer to such geometries as \emph{unit weight geometries}.
Geometries having one or more fractional edge-weights we refer to as \emph{fractional weight geometries}.
We will note in a given context whether or not we are including geometries with fractional weight edges in the context of any given instance of \MPI; however, unless otherwise noted, a reference to a ``geometry'' should be understood as referring to the unit weight geometries which arise in the one-way measurement model.

\subsection{Geometries with gflows}
\label{sec:interpolateGflows}

The approach taken for \MPI\ based on gflows is to ignore all angle information, and consider those tuples $(G,I,O,\vec a)$ which represent postselection-schemes which may be interpolated regardless of the value of $\vec a$.
By doing so, the objective is to consider postselection schemes which \emph{generically} perform unitary transformations, in the following sense:
\begin{problem}[Generic Measurement Pattern Interpolation (\GMPI)]
	\label{problem:GMPI}
	Given an input geometry $(G,I,O)$, determine if there exists a family of measurement patterns $\mathfrak P(\vec{a})$ in the model $\OneWay(\R)$ having the underlying geometry $(G,I,O)$ and default measurement angles $\vec{a} \in \R^{O\comp}$, such that every pattern $\mathfrak P(\vec a)$ performs a unitary embedding.
\end{problem}
For a tuple $(G,I,O,\vec a)$, if the geometry $(G,I,O)$ is a ``\emph{yes}'' instance of \GMPI, we can then solve \MPI\ by constructing the particular pattern $\mathfrak P(\vec a)$.

The genericity of the measurement angles corresponds to the constraint that each measurement may be performed with respect to an arbitrary observable which anticommutes with $Z$.
The approach taken by analysis with gflows, as we described in Section~\ref{sec:gflows}, is to determine whether this suffices to ensure unitarity of the post-selection procedure via the stabilizer formalism.
In the case that the stabilizer formalism does guarantee the unitarity of the transformation, we also obtain a correction strategy which provides an interpolation of the measurement procedure to all possible measurement sequences.
Recall the definition of a gflow:
\begin{definition}
	Let $(G,I,O)$ be a geometry and $\lambda$ be a function defined on $O\comp$ indicating measurement planes as above; and let $\cP$ be the power-set function, and $\odd(S)$ be the set of vertices in $G$ which are adjacent to an odd number of elements of $S$.
	Then a \emph{gflow} for $(G,I,O,\lambda)$ is a tuple $(g,\preceq)$ consisting of a function $g: O\comp \to \cP(I\comp)$ and a partial order $\preceq$ which satisfy the following conditions:
	\begin{subequations}
	\begin{align}
			w \in g(v) \;\implies\;&	v \preceq w ,
		\\
			w \in \odd(g(v)) \;\implies\;& v \preceq w ,
		\\
			\lambda(v) = \XY \;\implies\;& v \in \odd(g(v)) \setminus g(v)	, 
		\\
			\lambda(v) = \YZ \;\implies\;& v \in g(v) \setminus \odd(g(v)) ,
		\\
			\lambda(v) = \XZ \;\implies\;& v \in g(v) \inter \odd(g(v)) .
	\end{align}
	\end{subequations}
\end{definition}
As we saw in Section~\ref{sec:gflows}, the function $g: O\comp \to \cP(I\comp)$ essentially encodes a list of generators $\ens{S_j}_{j = 1}^{n-k}$ for the state-space prepared by the map $\Ent{G} \circ \New{I\comp}{}$\,.
Provided that the qubits $v \in O\comp$ are measured in an order extending $\preceq$, the set $g(v)$ describes the set of qubits acted on with the $X$ operations of some stabilizer generator; qubits $w \in \odd(g(v))$ are acted on with $Z$ operations, and qubits in $g(v) \inter \odd(g(v))$ are acted on by both (or equivalently, by a $Y \propto X Z$ operation).
The generator $S_j$ described then anticommutes with the measurement observable $\sO_v^{\lambda(v),\alpha_v}$ defined in~\eqref{eqn:pauliPlaneObs}, for $\alpha_v$ arbitrary.
The post-measurement state corresponding to the result $\s[v] = 0$ can then be selected by performing $S_j$ as a post-measurement correction, which we may re-express as an adaptation the measurement observables for the qubits on which it acts.

\subsubsection{Geometries of unit weight with \XY-plane gflows}

In the case where $\lambda$ is a constant function $\lambda: O\comp \to \ens{\XY}$ and where $\card{V(G)} = n$, Mhalla and Perdrix~\cite{MP08} present an efficient algorithm to find a gflow $(g,\preceq)$, where $\preceq$ is defined in terms of a layer function $\ell: V(G) \to \N$ such that $v \preceq w \iff [\,v = w \text{~or~} \ell(v) > \ell(w)\,]$.
This leads immediately to a solution to \MPI\ in the case of unit weight geometries $(G,I,O)$ when the geometry has a gflow, as follows.
We may obtain a linear order $\le$ extending $\preceq$ by constructing level sets $\sV_j$ for each $j \in \N$\,, inserting each vertex $v \in V(G)$ into the set $\sV_{\ell(v)}$\,.\label{discn:linearExtension}
Using the representation of sets described in Section~\ref{sec:flowAlgs}, the sets $\sV_j$ are themselves linearly ordered by construction, leading to a linear ordering of $V(G)$ by merging these linear orders in such a way that any $v \in \sV_{j+1}$ precedes any $w \in \sV_j$ for each $j$.
We may then simulate the postselection of the residual state corresponding to the result $\s[v] = 0$ for each qubit $v \in O\comp$, by composing patterns of the form
\begin{gather}
	\sqparen{\prod_{u \in g(v)} \!\! \Xcorr{u}{\s[v]} \;} \sqparen{\prod_{\substack{w \in \odd(g(v)) \\ w \ne v}} \!\! \Zcorr{w}{\s[v]} \;} \!\Meas{v}{\alpha_v}
\end{gather}
in the ordering given by $\le$.
(The operations used here are those of the $\OneWay(\R)$ model defined by Definition~\ref{def:oneWayModel}.
The restriction of $w \ne v$ in the right-most product is due to the fact that $v \in \odd(g(v))$ has been discarded by the measurement.)
Then, the following CPTP map performs a unitary embedding for any choice of angles $\vec a = (\alpha_v)_{v \in O\comp}$\,:
\begin{align}
	\label{eqn:gflowPattern}
	\paren{
		\ordprod[\ge]_{v \in O\comp}
			\sqparen{\prod_{u \in g(v)} \!\! \Xcorr{u}{\s[v]} \;} \sqparen{\prod_{\substack{w \in \odd(g(v)) \\ w \ne v}} \!\! \Zcorr{w}{\s[v]} \;} \!\Meas{v}{\alpha_v} }
	\paren{
		\prod_{uv \in E(G)} \Ent{u,v} }
	\paren{
		\prod_{v \in I\comp} \New{v}{\sfx} },
\end{align}
where the left-most product is ordered left-to-right according to the order $\ge$.
By the analysis of Section~\ref{sec:postselectionProcedure}, if $(G,I,O, \vec a)$ is an instance of the \MPI\ derived from a quadratic form expansion for an operator $U$, and $(G,I,O)$ is a unit weight geometry, the unitary operation which is performed is $U$.

\paragraph{Run-time analysis.}

Let $n = \card{V(G)}$, $m = \card{E(G)}$, and $k = \card{O} \ge \card{I}$.
Determining whether a geometry has a gflow for the special case where all measurements are performed in the \XY-plane, and constructing the gflow $(g,\preceq)$ if so, can be done in time $O(n^4)$ by Algorithm~2 of~\cite{MP08}.
Having a gflow $(g, \preceq)$, the measurement procedure in~\eqref{eqn:gflowPattern} is well-defined and requires $O(n^3)$ time to construct, dominated by the product over $v \in O\comp$ of the correction operations for $u \in g(v)$ and $w \in \odd(g(v))$, where the latter require time $O(m) \subset O(n^2)$ to compute.
Bringing this measurement procedure into a standard form by the standardization procedure of Section~\ref{sec:DKPstandardization} also requires time $O(n^3)$\,; the resulting measurement pattern contains $n-k$ measurement operations and up to $k$ correction operations, each of which involves $O(n)$ measurement dependencies.
We may therefore conclude:

\begin{theorem}
	\label{thm:synthGflowsExact}
	For a unitary embedding $U$ given as a quadratic form expansion, with unit weight geometry $(G,I,O)$ such that $\card{V(G)} = n$, there is an $O(n^4)$ algorithm which either determines that $(G,I,O)$ has no gflow, or constructs pattern in the $\OneWay(\R)$ model whose default measurement angles are given by the square terms of the quadratic form and consisting of $O(n^2)$ operations and which performs the transformation $U$.
\end{theorem}

As we note in Theorem~\ref{thm:synthGflowsExact}, the one-way measurement pattern constructed by this technique is in the model $\OneWay(\A)$, where $\A \subset \R$ contains the coefficients for the square terms of the quadratic form.
We may strengthen this to obtain an algorithm producing one-way patterns in a fixed-precision model of measurement-based quantum computing using any set $\A$ sufficient for generating single-qubit unitaries, such as $\A = \frac{\pi}{4}\Z$.
A measurement in the \XY-plane with respect to the basis $\ket{\pmsub\alpha}$ can be transformed into a \axisZ-axis measurement by a $J(-\alpha)$ rotation~\eqref{eqn:basicGates}: more generally, we have
\begin{gather}
	\Meas{v}{\alpha;\beta}	\;\;=\;\;	\Meas{v}{\sfz} \; \sfJ_v^{(- 1)^{1 + \beta} \cdot \alpha}	\;.
\end{gather}
By the universality of the $\Jacz(\frac{\pi}{4}\Z)$ gate set defined on page~\pageref{eqn:elemGateSets}, and by the Solovay-Kitaev Theorem, we may approximate the unitary operator $J(\alpha) \in \U(2)$ to any precision $\epsilon > 0$ by some product of $N$ operations of the form $J(\theta_j)$ for $\theta_j \in \frac{\pi}{4}\Z$\,, where $N \in \polylog(1/\epsilon)$:
\begin{gather}
	\label{eqn:approxJ}
	\norm{J(\alpha) \Big.\;-\;\Big. J(\theta_{N-1}) \cdots J(\theta_1) J(\theta_0)}_\infty <\; \sfrac{\epsilon}{2}	\;.
\end{gather}
This approximation can be found in time $\polylog(1/\epsilon)$ time by the Solovay-Kitaev theorem.
In order to represent the conditional change of sign, note that
\begin{subequations}
\begin{gather}
		J(-\alpha)
	\;\;=\;\;
		H R_z(-\alpha)
	\;\;=\;\;
		H X R_z(\alpha) X
	\;\;=\;\;
		Z J(\alpha) X \;.
\end{gather}
\end{subequations}
Then, we may approximate the effect of a measurement $\Meas{v}{\alpha}$ for arbitrary $\alpha \in \R$ by a (non-standardized) measurement pattern of the form
\begin{align}
		\Meas{v}{\alpha;\beta}
	\;\;\cong\;&\;\;
		\Meas{v'}{\sfz}
		\sfJ^{(-1)^{1+\beta}\cdot\alpha}_{v/v'}
	\notag\\\approx_\epsilon&\;\;
		\Meas{v'}{\sfz}
		\;\Zcorr{v'}{1 + \beta}
		\;\mathfrak J^{\theta_{N-1}}_{v'/a_{N-1}}
		\;\cdots\,
		\;\mathfrak J^{\theta_2}_{a_2/a_1}
		\;\mathfrak J^{\theta_1}_{a_1/v}
		\;\Xcorr{v}{1 + \beta}
	\notag\\=\;&\;\;
		\Meas{v'}{\sfz}
		\;\mathfrak J^{\theta_{N-1}}_{v'/a_{N-1}}
		\;\cdots\,
		\;\mathfrak J^{\theta_2}_{a_2/a_1}
		\;\mathfrak J^{\theta_1}_{a_1/v}
		\;\Xcorr{v}{1 + \beta}	\;,
\end{align}
using the $\mathfrak{J}^\theta_{u/v}$ patterns defined in~\eqref{eqn:elemOpsDKP}; we introduce the auxiliary qubits $a_1, \ldots, a_{N-1}$ in order to approximate $\mathfrak J^\alpha_{v/v'}$ to precision $\epsilon$ using the approximation of~\eqref{eqn:approxJ}, and substituting the measurement result $\s[v']$ throughout the rest of the measurement pattern wherever $\s[v]$ occurs.

To obtain an approximation of the CPTP map $\rho \mapsto U \rho U\herm$ with precision $\epsilon$, it suffices to approximate each of the $n - k$ measurement operations with precision $\frac{\epsilon}{n - k}$ by the development of~\eqref{eqn:errorAdditiveCPTP}.
We then obtain the following corollary:
\begin{corollary}
	\label{cor:synthGflowsApprox}
	For a unitary embedding $U$ given as a quadratic form expansion, with unit weight geometry $(G,I,O)$ such that $\card{V(G)} = n$, for each $\epsilon > 0$ and $\delta > 0$ there exists an algorithm running in time $O(n^4 + n \log(n/\epsilon)^{3 + \delta})$ which either determines that $(G,I,O)$ has no gflow, or constructs a pattern in the $\OneWay(\frac{\pi}{4}\Z)$ model consisting of $O(n^2 + n \log(n/\epsilon)^{3 + \delta})$ operations and which approximates the CPTP map $\rho \mapsto U \rho U\herm$ to precision $\epsilon$.
\end{corollary}

\subsubsection{Geometries of fractional weight with flows}

Geometries with fractional edges correspond, as we remarked earlier, to extended one-way measurement procedures with include entangling operations of the form $\Ent[t]{u,v}(\rho) \;=\; \cZ^t\, \rho \;\cZ^{-t}$.
The state space of the system is then not necessarily a stabilizer code, in which case the approach given above for gflows then do not directly apply.

As we noted in Section~\ref{sec:postselectionProcedure}, we may attempt to treat such quadratic form expansions by applying the substitutions of~\eqref{eqn:diagonalExpansions} in order to obtain an equivalent expansion which only involved cross-term angles in $\ens{0,\pi}$.
However, this approach still fails to produce a problem instance approachable by the technique of gflows, as it produces a quadratic form with an extra auxiliary index $a_1$ which has no square term.
It is easy to see that the absence of an $a_1^{\, 2}$ term with a generic coefficient in this expansion is crucial, by considering the corresponding formula where the quadratic form includes a $\theta a_1^{\, 2}$ term:\label{discn:squareTermConstrained}
\begin{align}
		\sfrac{1}{2}\! \sum_{\vec a \in \ens{0,1}^2} &\e^{i\big[\pi(x_u a_1 + x_v a_1 + a_1 a_2) \,+\, \theta a_1^{\, 2} \,+\, \phi a_2^{\, 2}\big]}
	\notag\\[1ex]=&\;\;
		\sfrac{1}{2} \sqparen{
				\e^{i [ 0 ]}
			\;+\;
				\e^{i [\pi (x_u + x_v) \,+\, \theta]}
			\;+\;
				\e^{i [\phi]}
			\;+\;
				\e^{i [\pi (x_u + x_v + 1) \,+\, \theta \,+\, \phi]}
		}
	\notag\\[2ex]=&\;\;
		\sfrac{\e^{i\phi/2}}{2} \sqparen{
				\paren{
					\e^{-i \phi/2} \,+\, \e^{i\phi/2}
				}
			\;+\;
				\e^{i\theta} (-1)^{x_u \oplus x_v} \paren{
					\e^{-i \phi/2} \,-\, \e^{i \phi/2}
				}
		}
	\notag\\[2ex]=&\;\;
		\e^{i\phi/2} \sqparen{
				\cos(\tfrac{\phi}{2})
			\;-\;
				i \, \e^{i\theta} (-1)^{x_u \oplus x_v} \sin(\tfrac{\phi}{2})
		}	\;.
\end{align}
This expression is not a complex unit for all values of $x_u, x_v \in \ens{0,1}$ and angles $\phi \in (-\pi, \pi]$; in particular, for $\phi = \pion 2$\,, we may see that
\begin{gather}
	\label{eqn:nonuniformity}
		\sfrac{1}{2}\! \sum_{\vec a \in \ens{0,1}^2} \e^{i\big[\pi(x_u a_1 + x_v a_1 + a_1 a_2) \,+\, \frac{\pi}{2} a_1^{\, 2} \,+\, \frac{\pi}{2} a_2^{\, 2}\big]}
	\;\;=\;\;
		\begin{cases}
			1 + i	\,,	&	\text{if $x_u \oplus x_v = 0$}	\\[1ex]
			0\,,					&	\text{if $x_u \oplus x_v = 1$}	
		\end{cases}	\;.
\end{gather}
This is a very different expression from the expression $\e^{i \phi (x_u \oplus x_v)}$ in the case $\theta = 0$\,: thus, even if the result is a unitary in the latter case, it is by no means clear that quadratic form expansions with the same geometry should be ``generically'' unitary, regardless of the coefficients of their square terms.
(As an extreme example, where $u,v \in I \inter O$, the resulting operation would be singular for $\theta = \phi = \pion 2$, even though the operator may be unitary for $\theta = 0$.)
However, this genericity is necessary for the geometry to be a ``yes'' instance of the \GMPI.

We can attempt to resolve this by noting that the substitutions in~\eqref{eqn:diagonalExpansions} are a result of an extended measurement-based procedure which allows arbitrary measurements in the \YZ\ plane: measurements on the two auxiliary qubits $a_1$ and $a_2$ of~\eqref{eqn:diagonalExpansions}, where $a_1$ is measured with an \axisX-measurement and $a_2$ with an arbitrary \XY-plane measurement, are equivalent to an arbitrary \YZ-plane measurement on a single auxiliary qubit, as illustrated in the middle panel of Figure~\ref{fig:diagonalYZ}.\label{discn:simulateYZmeas}
In particular, as we noted in Section~\ref{sec:extendedConstructions}, this construction has an eflow, which is a special case of a gflow where $g(v)$ are singleton sets for each $v \in O\comp$.
We may then treat such quadratic form expansions by performing a transformation of the underlying \emph{geometry} $(G,I,O)$, to yield a unit-weight geometry $(G',I,O)$ with additional information indicating sets of vertices $T$ to be measured in the \YZ\ plane, and try to find an eflow for $(G',I,O,T)$.
We will use a similar technique in Section~\ref{sec:synthEflows} to synthesize one-way measurement patterns for certain cases involving auxiliary vertices of degree $1$.

For a unitary $U$ which is specified by a quadratic form expansion, whose geometry has fractional weight edges but which can be transformed into an equivalent unit-weight geometry $(G',I,O)$ and set $T$ such that $(G',I,O,T)$ has an eflow $(f,\preceq)$, we may apply the same techniques as outlined above for gflows to efficiently arrive at an implementation (or approximation) of $U$.
In this case, the vertices of $T$ can be identified with the edges of fractional weight in $G$; conversely, any qubit $v \in V(G) \setminus O$ lies outside of $T$, and so the edges $v \,f(v)$ in $G$ are of unit weight.
By the comments at the end of Section~\ref{sec:extendedConstructions}, the star geometry rooted at vertices $v \in T$ correspond to operations $\cZ^t$ in a unitary circuit via the simplified RBB construction, with the edges of the form $v \, f(v)$ representing $J(\theta)$ gates as in the analysis of flows in terms of the DKP construction.

There is, however, a simpler approach to synthesizing an implementation for $U$ by observing that the above also essentially describes an extension of the semantic map $\SemanticDKP$ constructed in Section~\ref{sec:semanticDKP}, where we ignore the absence of information about the adaptation of measurements, and allow non-flow edges to have fractional weight to represent fractional powers of $\cZ$ in the resulting unitary circuit.
First, if necessary, we transform the quadratic form expansion into a form where each $v \in O$ has $\theta_{vv} = 0$, using the technique described on page~\pageref{eqn:idopQuadFormExpan}: it is easy to verify that the flow function can be extended to the new vertex set arising from this transformation, so we may without loss of generality suppose that the input geometry has $\theta_{vv} = 0$ for $v \in O$.\label{discn:synthFlows}
Then, we may straightforwardly interpret the quadratic form expansion as a unitary circuit, by noting that it exactly represents the matrix formula produced by making explicit the summations in a stable index tensor product:
\begin{romanum}
\item
	the edges $v w$ for $w = f(v)$ correspond to a pair of deprecated/advanced tensor indices in an operation $J(\theta_{vv})\pseu[:w/v]$ operation;

\item
	any non-flow edge $vw$ corresponds to a pair of stable tensor indices in a $\cZ\pseu[v,w]$ operation.
\end{romanum}
In the above, we still obtain a unitary circuit by replacing non-flow edges with edges of fractional weight; these correspond to $\cZ^t\pseu[v,w]$ operations for fractional powers $t$.
This motivates an extension of the definition of flows to fractional-weight geometries:
\begin{definition}
	\label{def:flowFractional}
	Suppose $(G,I,O)$ is a fractional-weight geometry.
	A flow for $(G,I,O)$ is then an ordered pair $(f,\preceq)$ satisfying properties \eqref{flow:a}\thru\eqref{flow:c}, such that for all $ab \in E(G)$ with $W_G(ab) < 1$, we also have $f(a) \ne b$ and $f(b) \ne a$ (\ie, all flow-edges in $G$ have unit weight).
\end{definition}
If a fractional-weight geometry $(G,I,O)$ has a flow, and if the square terms for $v \in O$ are all zero, we may synthesize a circuit from the quadratic form expansion for $U$ using the description above.

We may verify the correctness of this circuit construction by inducting on the number of edges of fractional weight.
For the base case, if $(G,I,O)$ has a known flow $(f,\preceq)$ and no fractional-weight edges, we may synthesize a circuit for $U$ using the algorithm $\SemanticDKP$ (using the flow $(f,\preceq)$ rather than searching for one using the procedure $\FindFlow$).
The resulting circuit has a gate $J(\theta_{vv})\pseu[:x_{f(v)}/x_v]$ for each $v \in O\comp$ and a gate $\cZ\pseu[x_v, x_w]$ for each non-flow edge $vw \in E(G)$\,, which can be expressed as
\begin{gather}
	\label{eqn:unitWeightFlowCircuit}
		C
	\;\;=\;\;
		\Bigg(\prod_{v \in O\comp} J(\theta_{vv})\pseu[:x_{f(v)}/x_v]\Bigg)
		\Bigg(\prod_{\substack{vw \in E(G) \\ v \ne f(w) \\ w \ne f(v)}} \!\!\cZ\,\pseu[x_v , x_w ]\Bigg)
\end{gather}
in stable index notation.
We may then induct for geometries with fractional edge-weights if we can show we can decompose the geometry into ones with fewer fractional edge-weights.

For an arbitrary fractional edge $ab \in E(G)$ and each each $z \in O$, we may define $m(ab,z)$ to be the maximal vertex $v \in V(G)$ in the ordering $\preceq$ subject to $z$ being in the orbit of $v$ under $f$ (that is, $z = f^\ell(v)$ for some $\ell \ge 0$), such that at least one of $v \preceq a$ or $v \preceq b$ holds.
For a set $S \subset V(G)$, let $G[S]$ represent the subgraph of $G$ induced by $S$ (\ie\ by deleting all vertices in $G$ not in $S$). Then, define the following subgraphs of $G$, and corresponding geometries:
\begin{itemize}
\item
	Let $V_2$ be the set of vertices $m(ab,z)$ for each $z \in O\comp$: it is easy to show that $a,b \in V_2$. Let $G_2 = G[V_2]$,
	and let $\cG_2 = (G_2, V_2, V_2)$.

\item
	Let $V_1$ be the set of vertices $u \in V(G)$ such that $u \preceq v$ for some $v \in V_2$; let $G_1 = G[V_1] \setminus \ens{uv \,\big|\, u,v \in V_2}$\,; and let $\cG_1 = (G_1, I, V_2)$.

\item
	Let $V_3$ be the set of vertices $u \in V(G)$ such that $u \succeq v$ for some $v \in V_2$; let $G_3 = G[V_3] \setminus \ens{uv \,\big|\, u,v \in V_2}$; and let $\cG_3 = (G_3, V_2, O)$.
\end{itemize}
This decomposes the geometry $(G,I,O)$ into three (possibly still fractional-weight) geometries with flows, as illustrated in Figure~\ref{fig:decompose}.
\begin{figure}[t]
	\begin{center}
		\includegraphics{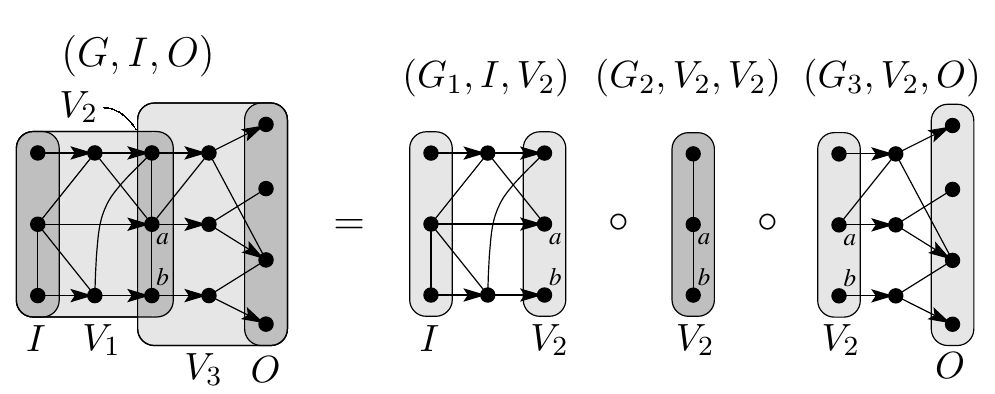}
		\caption[Illustration of the decomposion of a quadratic form expansion about an edge, expressed in terms of geometries.]{\label{fig:decompose}%
		Illustration of the decomposion of a quadratic form expansion about an edge $ab$, expressed in terms of geometries. 
		(Again, the composition of geometries $(G_3, V_2, O) \circ (G_2, V_2, V_2) \circ (G_1, I, V_2)$ is illustrated in the opposite order in the diagram above.)
		$V_2$ is a set of maximal vertices under the constraint of being bounded from above, by the vertices $a$ and $b$, in a partial order $\preceq$ associated with a flow for the fractional-edge geometry $(G,I,O)$. 
		Arrows represent the action of the corresponding flow function $f$.
		}
	\end{center}
\end{figure}

Let $Q_1$ be a quadratic form on $\ens{0,1}^{V_1}$ consisting of the terms $x_u x_v$ of $Q$ for $u \in V_1$ or $v \in V_1$,
but not both; $Q_2$ be a quadratic form on $\ens{0,1}^{V_2}$ consisting of the terms $x_u x_v$ of $Q$ for \emph{distinct}
$u, v \in V_2$; and similarly let $Q_3$ be defined on $\ens{0,1}^{V_3}$, and consist of the remaining terms of $Q$.
Then $Q_1$, $Q_2$, and $Q_3$ define quadratic form expansions for some operations $U_1$, $U_2$, and $U_3$ (respectively)
with geometries $\cG_1$, $\cG_2$, and $\cG_3$ (respectively).
\begin{itemize}
\item
	$U_2$ in particular will be a product of operations $\cZ^{W_G(uv)}$ for distinct $u,v \in V_2$\,, as it is a quadratic form expansion whose input and output indices coincide. Then $U_2$ can be represented as a circuit with a wire for each $u \in V_2$, with fractional controlled-$Z$ gates $\cZ^{W_G(uv)}$ for each edge $uv \in E(G)$.
\item
	Both $\cG_1$ and $\cG_3$ have flows, and fewer fractional edges than $(G,I,O)$.
	By induction, $U_1$ and $U_3$ are also unitary embeddings, and have stable index expressions including terms $J(\theta_{vv}) \pseu[:f(x_v)/x_v]$ for each $v \in V(G_j) \setminus O_j$ and $\cZ^{W_G(uv)} \pseu[x_u \,, x_v]$ for each non-flow edge $uv \in E(G_j)$, for $j \in \ens{1,3}$.
\end{itemize}
Because $Q_1(\vec{x}_{\text{\raisebox{-0.3ex}{$V_1$}}}) + Q_2(\vec{x}_{\text{\raisebox{-0.3ex}{$V_2$}}}) + Q_3(\vec{x}_{\text{\raisebox{-0.3ex}{$V_3$}}}) = Q(\vec{x})$ for all $\vec{x} \in \ens{0,1}^{V}$ by construction, the composite operation $U_3 U_2 U_1$ can differ from $U$ by at most a scalar factor by Lemma~\ref{lemma:quadFormExpanComposition}.
We then obtain a composite circuit
\begin{gather}
		C
	\;\;=\;\;
		\Bigg(\prod_{v \in O\comp} J(\theta_{vv})\pseu[:x_{f(v)}/x_v]\Bigg)
		\Bigg(\prod_{\substack{vw \in E(G) \\ v \ne f(w) \\ w \ne f(v)}} \!\!\cZ^{W_G(vw)}\pseu[x_v , x_w ]\Bigg)
\end{gather}
performing the operation $U$, generalizing the construction given for unit-weight geometries described in~\eqref{eqn:unitWeightFlowCircuit}.
A direct way to verify that this circuit is correct is to expand the tensor product in terms of the coefficients of each operator $J(\theta)$ and $\cZ^t$\,:
\begin{subequations}
\begin{align}
	\label{eqn:coeffsJalpha}
		J(\theta_{vv})\pseu[:x_w/x_v]
	\;\;=&\;\;
		\sfrac{1}{\sqrt 2} \, \e^{i (-\theta_{vv}/2 \,+\, \theta_{vv} x_v \,+\, \pi x_v x_w)}
	\notag\\[0.5ex]\propto&\;\;
		\sfrac{1}{\sqrt 2} \, \e^{i (\theta_{vv} x_v^2 \,+\, \pi x_v x_w)}	\;,
	\\[1.5ex]
		\cZ\pseu[x_v, x_w]
	\;\;=&\;\;
		(-1)^{x_v x_w}	\;;
\end{align}
\end{subequations}
by summing over indices which are both advanced and deprecated, we may then verify that the coefficients of the unitary operation performed by $C$ are the same as those of $U$ (up to a product of the scalar factors $\e^{-i\theta_{vv}/2}$ in the proportionality of~\eqref{eqn:coeffsJalpha} above).

\paragraph{Run-time analysis.}

Let $n = \card{V(G)}$, $m = \card{E(G)}$, and $k = \card{O}$.
Determining whether a geometry has a flow, and constructing the flow $(f,\preceq)$ if so, can be done in time $O(kn)$ by Algorithm~1 of~\cite{MP08}.
For each edge $uv$, we may check whether one of $W_G(uv) = 1$ or $\big[ u \ne f(v)$ and $v \ne f(u) \big]$ holds: if all edges satisfy this constraint, the unitary circuit described above is well-defined, and may be constructed in time $O(m)$ by iterating through the edges of $V(G)$.
By the extremal result of Theorem~\ref{thm:extremalUpperBound}, any geometry with a flow has $m \le kn - \binom{k+1}{2}$: by eliminating geometries with more edges than this, we may reduce the total running time of this algorithm to $O(kn)$.

In the case $\card{I} = \card{O}$, a flow function $f$ is unique if it exists by Theorems~\ref{thm:graphth-charn} and~\ref{thm:uniqueFamilyPaths} (pages~\pageref{thm:graphth-charn} and~\pageref{thm:uniqueFamilyPaths}, respectively); so in this case, if $(G,I,O)$ has a flow when we neglect the edge-weights of $G$, but there is an edge $v \, f(v)$ of fractional weight, there is no flow for $(G,I,O)$ considered as a fractional-weight geometry.
Finally, note that there is an operation in the circuit for each edge in the geometry: if the original quadratic form expansion had non-zero coefficients $\theta_{vv}$ for $v \in O$, the \emph{expanded} geometry $(G',I,O')$ obtained by applying the technique on page~\pageref{eqn:idopQuadFormExpan} will have at most $m + 2k \le kn - \binom{k+1}{2} + 2k \le kn + 1$ edges (where $m = \card{E(G)}$ is the number of edges of the \emph{original} geometry), by the upper bound of Theorem~\ref{thm:extremalUpperBound}.
Therefore, we have the following result:
\begin{theorem}
	\label{thm:synthFlowsExact}
	For a transformation $U \in \U(2^k)$ given as a quadratic form expansion with geometry $(G,I,O)$ such that $\card{V(G)} = n$ and $\card{E(G)} = m$, there is an $O(kn)$ algorithm which either determines that $(G,I,O)$ has no flow, or constructs a stable index expression representing a unitary circuit consisting of at most $m + 2k \le kn + 1$ operations which performs $U$.
\end{theorem}

In order to obtain a circuit using a finite elementary gate set, we may again approximate each gate $J(\theta_{vv})$ and each gate $\cZ^t$ to precision $\frac{\epsilon}{m}$ by a product of $N$ gates in the $\Jacz(\frac{\pi}{4})$ model, where $N \in \polylog(m/\epsilon)$ for arbitrary $\epsilon > 0$, by the Solovay-Kitaev Theorem; and by the analysis of~\eqref{eqn:errorAdditiveCPTP}, this suffices to approximate $U$ to precision $\epsilon$.
We then obtain:\label{discn:synthFlowsEnd}
\begin{corollary}
	\label{cor:synthFlowsApprox}
	For a transformation $U \in \U(2^k)$ given as a quadratic form expansion with geometry $(G,I,O)$ such that $\card{V(G)} = n$ and $\card{E(G)} = m$, for each $\epsilon > 0$ and $\delta > 0$ there exists an algorithm running in time $O(kn \log(kn/\epsilon)^{3 + \delta})$ which either determines that $(G,I,O)$ has no flow, or constructs a stable index expression representing a unitary circuit in the $\Jacz(\frac{\pi}{4}\Z)$ model, which consists of $O(kn \log(kn/\epsilon)^{3 + \delta})$ operations and approximates $U$ to precision $\epsilon$.
\end{corollary}

\subsubsection{The quantum Fourier transform as an example}
\label{sec:fourierTransformEg}

For the most part, quadratic form expansions have occurred in the literature as analytical tools rather than as the preferred means of representing unitary transformations of interest.
While there is a possible source in path integrals, as we remarked on page~\pageref{sec:sumsOverPaths}, it is not yet known whether expressions resulting from discretizing path integrals would yield tuples $(G,I,O,\lambda)$ with generalized flows.
At the same time, the study of flows and gflows has thus far been essentially restricted to their potential as tools to analyze procedures in the one-way model, in an attempt to characterize procedures which perform unitaries under certain constraints (as in the \GMPI) or provide automatically verifiable semantics for measurement-based computations (as in Chapter~\ref{Semantics for unitary one-way patterns}).
Therefore, there are few known ``natural'' examples of a unitary whose definition lends itself to the techniques of this section.

However, as we remarked in Section~\ref{sec:relatedWork}, there is one very prominent example of a unitary transformation which is naturally defined as a quadratic form expansion: the quantum Fourier transform $\cF_{2^k}$ over $\Z_{2^k}$\,.
Recall that $\cF_{2^k}$ has an expansion with the quadratic form
\begin{align}
	\label{eqn:qftGeom}
	Q(\vec x, \vec y) \;\;=\;\; \sum_{h = 0}^{k-1} \; \sum_{j = 0}^{k-h-1} \;\sfrac{2^{(h+j)}}{2^{k-1}} \;\pi \,x_j \,y_h \;,
\end{align}
where the vectors $\vec x$ and $\vec y$ correspond to the qubits of the input system $I$ and the output system $O$, respectively.
The (fractional-edge) geometry $\cG$ arising from this quadratic form expansion is illustrated\footnote{%
The problem of finding suitable graphical presentations of graphs is an open research problem in computational geometry; the presentation of the geometry in Figure~\ref{fig:qftSynth} was chosen by hand to best exhibit the symmetries of the geometry.}
in Figure~\ref{fig:qftSynth} for the case $n = 5$.

We may show that the fractional-weight geometry $\cG$ has a flow without having to perform a flow-finding algorithm, as follows.
Because the only edges which have unit weight in $\cG$ are those of the form $x_j y_{\text{\raisebox{-0.5ex}{$k \minus j \minus 1$}}}$ for each $0 \le j < k$, these must be the flow edges if $\cG$ has a flow.
To show that the function $f(x_j) = y_{\text{\raisebox{-0.5ex}{$k \minus j \minus 1$}}}$ is a flow-function, consider the influence relation $\inflRel$ induced by $f$ (Definition~\ref{def:inflRel}, page~\pageref{def:inflRel}).
By construction, the graph of the geometry $\cG$ is a bipartite graph, with partitions $\ens{x_j}_{j = 0}^{k-1}$ and $\ens{y_h}_{h = 0}^{k-1}$\,; then, for each vertex $x_j$, aside from $f(x_j)$, the only vertices $w$ such that $x_j \inflRel w$ are the other vertices $x_\ell$ adjacent to $y_{\text{\raisebox{-0.5ex}{$k \minus j \minus 1$}}}$.
The set of vertices adjacent to a given vertex $y_h$ are the vertices $x_\ell$ for $0 \le \ell \le k - h - 1$\,, by~\eqref{eqn:qftGeom}.
Thus, the vertices $x_\ell$ such that $x_j \inflRel x_\ell$ satisfy $\ell \ne j$ and $0 \le \ell < k - (k-j-1) - 1$, which holds if and only if $0 \le \ell < j$.
Then, every walk in the influence digraph $\sI_f$ (Definition~\ref{def:inflDigraph}, page~\pageref{def:inflDigraph}) either ends at a vertex $y_h$ for some $h$, or visits a sequence of vertices $x_{s_1} x_{s_2} \cdots$ for some monotonically decreasing sequence $\ens{s_j}_{j \ge 1}$ of indices.
In particular, $\sI_f$ contains no circuits; thus, by Lemma~\ref{lemma:causalOrderFlow} (page~\pageref{lemma:causalOrderFlow}), the geometry $\cG$ has a fractional-edge flow.

\begin{figure}[tb]
	\begin{center}
		\includegraphics{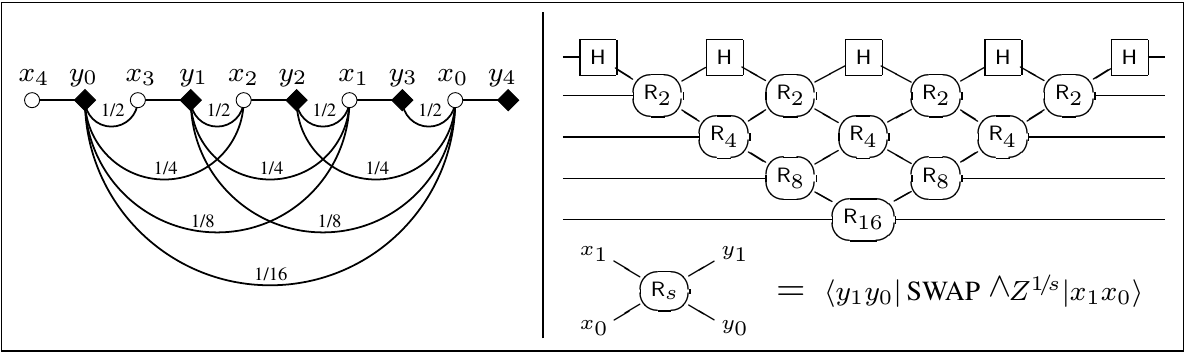}
	\end{center}
	\caption[The geometry for the quadratic form expansion of the QFT for $\Z_{32}$, and the corresponding circuit.]{\label{fig:qftSynth}
		The geometry for the quadratic form expansion of the QFT for $\Z_{32}$, and the corresponding circuit due to~\cite{FDH04}.
		In the geometry (on the left), input vertices are labelled by circles, output vertices by lozenges, and fractional edges are labelled with their edge-weights.}
\end{figure}
Illustrated in Figure~\ref{fig:qftSynth} alongside the presentation of $\cG$ is a representation of the circuit that may be constructed from $\cG$ together with the flow-function $f$.
Note that the static index tensor representation of unitary circuits in the $\Jacz(\R)$ model does not impose any \emph{a priori} restrictions on the layout or interactions of the qubits (in particular, it does not explicitly represent {\small SWAP} gates) --- partly because the gate-set $\Jacz(\R)$ itself places no restrictions on interactions between qubits (and doesn't contain {\small SWAP} gates).
However, by
\begin{alphanum}
\item
	representing the input and output wires in the conventional way, with vertical position increasing in order of the significance of each bit;

\item
	identifying the edges of unit weight in the geometry with $H = J(0)$ operations\footnote{%
	The presence of Hadamards in the circuit is due to the lack of any square terms in the quadratic form, corresponding to measurement angles of $0$ in the post-selection description of the unitary operation; this in turn leads to $J(0)$ gates in the constructed circuit.}
	in the unitary circuit, and arranging the other elements of the circuit diagram about those gates; and

\item
	allowing the circuit wires to cross each other, and run in directions other than horizontally (in particular, producing a crossing for each fractional edge illustrated),
\end{alphanum}
we are led to a presentation of this circuit as above, where we have collected the wire-crossings and fractional $\cZ$ operations into a family of compound gates as illustrated.

The circuit illustrated in Figure~\ref{fig:qftSynth} was first described by Fowler, Devitt, and Hollenberg in~\cite{FDH04}, and is presented in a similar graphical format in~\cite{RBB03}: therefore, the above presentation is neither new nor an independent discovery.
This example is meant only to illustrate simple heuristics which, together with the techniques of the preceding sections, can at least reproduce what is known for the quantum Fourier transform, and in a model which places spatial restrictions on multi-qubit operations (unlike the $\Phases(\A)$ or $\Jacz(\A)$ models).

\subsection{Geometries with simple modifications admitting eflows}
\label{sec:synthEflows}

As we noted on pages~\pageref{discn:squareTermConstrained}\thru\pageref{discn:simulateYZmeas}, there are instances of quadratic form expansions which represent unitary operations, and where this unitarity may depend crucially on the value of some of the coefficients $\theta_{vv}$ for $v \in O\comp$\,.
However, such quadratic form expansions can nevertheless be related to generalized flows, if we can identify it with a postselection procedure in which constrained postselections in the \XY\ plane can be regarded as allowing the postselection of states in some other plane (such as the \YZ\ plane) to be simulated.

In a one-way measurement pattern, a qubit $v \in O\comp \inter I\comp$ with degree $1$ may represent (considered as a simulation of some element of a unitary circuit) either the introduction of an auxiliary qubit prepared in the $\ket{\splus}$ state which undergoes further transformations, or a terminal measurement on some qubit which has undergone some transformations prior to measurement.
In the latter case, the preceding transformations may be interpreted as a change of reference frame prior to measurement, in order to simulate measurement in some other plane by an \XY-plane measurement.
In particular, to simulate a \YZ-plane measurement with an \XY-plane measurement, it suffices (as we noted on page~\pageref{eqn:simulateYZwXY}) to perform a Hadamard transformation prior to the \XY-plane measurement.

It is easy to verify that the only single-qubit rotations of the form $J(\theta)$ which do this are for $\theta \in \pi \Z$\,: measuring an observable $\sO^{\XY,\alpha} = \cos(\alpha) X + \sin(\alpha) Y$ as defined in~\eqref{eqn:pauliPlaneObs} after performing a $J(\theta) = H R_\sfz(\theta)$ rotation is equivalent to measuring the observable
\begin{gather}
	\label{eqn:changeFrameXY}
	J(\theta)\herm \; \sO^{\XY,\alpha} \, J(\theta)
	\;\;=\;\;
	\cos(\alpha) \, Z \;\;-\;\; \sin(\alpha) \, R_\sfz(\theta)\herm Y R_\sfz(\theta)	\;,
\end{gather}
which is of the form $\sO^{\YZ,\alpha'}$ only for $\theta$ a multiple of $\pi$.
Considering a geometry of unit weight (possibly obtained by performing the substitution described in~\eqref{eqn:diagonalExpansions}), any vertex $w \in V(G)$ of degree $1$ which is adjacent to a vertex $v \in V(G)$ with $\theta_{vv} \in \ens{0,\pi}$ may then be considered to be expressing a postselection procedure which selects for some state $\ket{\splus \sfy\sfz_\alpha}_v$ as defined in~\eqref{eqn:defYZtheta}.
Furthermore, from the fact that
\begin{subequations}
\label{eqn:telescopingJ}
\begin{align}
		J(\alpha) J(0) J(\beta) J(\gamma) \;\;=&\;\; H R_\sfz(\alpha) R_\sfz(\beta) J(\gamma) \;\;=\;\; J(\alpha + \beta) J(\gamma)	\;,
	\\
		J(\alpha) J(\pi) J(\beta) J(\gamma) \;\;=&\;\; H R_\sfz(\alpha) X R_\sfz(\beta) H R_\sfz(\gamma) \;\;=\;\; J(\alpha - \beta) J(\gamma + \pi)	\;,
\end{align}
\end{subequations}
we may extend this to arbitrary paths of odd length: if $G$ contains a path of the form $v_0 v_1 v_2 \cdots v_{N\minus1} v_N$ for $N$ odd such that $v_N$ has degree $1$, $v_j$ has degree $2$ for $1 \le j < N$, and $v_j \in \pi\Z$ for $j$ even, the corresponding postselection procedure may be regarded as simulating the postselection of some state in the \YZ\ plane state on the qubit $v_0$.
More generally, we may consider \emph{arbitrary} products of operators $J(\theta)$ which map measurement observables of interest (\eg, observables of the form $\sO^{\YZ,\alpha}$ or $\sO^{\XZ,\alpha}$) to \XY-plane observables.

By detecting when chains of vertices in a geometry represent postselections in either the \YZ\ or \XZ\ plane, we can obtain smaller geometries and a measurement-plane function $\lambda$ which may represent ``\emph{yes}'' instances of \GMPI\ (where we extend the domain of that problem in the natural way to include measurement operations in each of the \XY, \YZ, and \XZ\ planes).
Thus, for any algorithm for finding gflows for non-constant maps $\lambda$, this yields another approach to synthesizing measurement-based patterns from quadratic form expansions, by contracting paths which may be regarded as representing the simulation of a postselection of observables in the \YZ, or \XZ\ planes.

Currently, the only efficient algorithm for obtaining gflows for $(G,I,O,\lambda)$ for $\lambda$ not a constant function is Algorithm~\ref{alg:findEflow}, which determines whether or not a tuple $(G,I,O,T)$ has an eflow.
As we noted on page~\pageref{discn:gflowsSpecialCases}, this is equivalent to determining whether the tuple $(G,I,O,\lambda)$ has a gflow, for $\lambda$ given by
\begin{gather}
		\lambda(v)
	\;\;=\;\;
		\begin{cases}
			\YZ	\,,	&	\text{if $v \in T$}	\\
			\XY	\,,	&	\text{otherwise}
		\end{cases}	\;,
\end{gather}
and where we require that $g(v)$ be a singleton set for each $v \in O\comp$.
We may then attempt to map a quadratic form expansion provided as input to a unit-weight geometry $(G,I,O)$ and a set of vertices $T \subset V(G)$ such that $(G,I,O,T)$ has an eflow.

\subsubsection{Treating fractional-weight geometries}

As we have noted before, by applying the substitution of~\eqref{eqn:diagonalExpansions}, we may transform any geometry with fractional edges, corresponding to coefficients $\theta_{uv} \notin \pi \Z$ into a geometry with unit edge-weights.
As noted in Figure~\ref{fig:diagonalYZ}, this will yield a certain number of pairs of qubits which represent postselections in the \YZ\ plane: this should be immediately apparent from the fact that the substitution amounts to a decomposition of $\cZ^t$ into a pattern of the form $\mathfrak{Zz\cdots z}^{\pi t/2}$\,, which itself has an eflow.

In the case where the original geometry has a flow (in the sense defined for fractional-edge geometries in Definition~\ref{def:flowFractional}), rather than attempting to obtain a unit-weight geometry with an eflow, it may prove more convenient (\eg\ if the objective is to obtain a circuit representation of a unitary) to follow the analysis of pages~\pageref{discn:synthFlows}\thru\pageref{discn:synthFlowsEnd}.

\subsubsection{Transformations of quadratic form expansions to find eflows}

In order to consider whether a unit-weight quadratic form expansion can be mapped to a one-way measurement pattern using eflows, we may first simplify the expansion by taking advantage of identities of the same form as~\eqref{eqn:telescopingJ} above.
Consider once more a quadratic form expansion
\begin{align}
	\label{eqn:quadFormExpansion-3}
	C\!\! \sum_{\vec{x} \in \ens{0,1}^V} \e^{i Q(\vec{x})} \;\ket{\vec{x}_O}\bra{\vec{x}_I}
	&&
	\text{where}\quad
	Q(\vec x)
\;=&\;
	\sum_{\substack{\ens{u,v} \subset V}} \theta_{uv} x_u x_v \;\,,
\end{align}
and suppose there are indices $a,b,c,v \in V \setminus (I \union O)$ such that $Q(\vec x)$ contains the terms
\begin{gather}
	\begin{split}
		&\pi x_a x_b + \pi x_b x_c + \pi x_c x_v	\\
		&\quad + \theta_{aa} x_a^2 + \theta_{bb} x_b^2 + \theta_{cc} x_c^2 + \theta_{vv} x_v^2	\;,
	\end{split}
\end{gather}
and no other terms in $x_b$, $x_c$, or $x_v$ (though it may contain many other terms in $x_a$).
In the geometry induced by this quadratic form expansion, $v$ has degree $1$, and $b$ and $c$ have degree $2$.
If $\theta_{cc} \in \pi\Z$, we may then apply one of the following equalities:
\begin{subequations}
\label{eqn:telescopingQuadFormExpan}
\begin{align}
		\sum_{b,c,v} &
			\exp\paren{i\sqparen{\pi x_a x_b + \pi x_b x_c + \pi x_c x_v + \theta_{aa} x_a^2 + \theta_{bb} x_b^2 + 0 \!\cdot\! x_c^2 + \theta_{vv} x_v^2}}
	\notag\\[0.5ex]=&\;\;\;
		\sum_{b,v} \paren{
			\e^{i\sqparen{\pi x_a x_b + \theta_{aa} x_a^2 + \theta_{bb} x_b^2 + \theta_{vv} x_v^2}}
			\;\;+\;\;
			(-1)^{x_b \oplus x_v}
			\e^{i\sqparen{\pi x_a x_b + \theta_{aa} x_a^2 + \theta_{bb} x_b^2 + \theta_{vv} x_v^2}}}
	\notag\\[0.5ex]=&\;\;\;
		\sum_{b,v} \; 2 \, \delta_{b,v} \,
			\e^{i\sqparen{\pi x_a x_b + \theta_{aa} x_a^2 + \theta_{bb} x_b^2 + \theta_{vv} x_v^2}}
	\notag\\[0.5ex]=&\;\;\;
		2 \sum_{v} \e^{i\sqparen{\pi x_a x_v + \theta_{aa} x_a^2 + (\theta_{bb} + \theta_{vv}) x_v^2}}	\;;
	\\[-4.3em]\notag
	\intertext{}
		\sum_{b,c,v} &
			\exp\paren{i\sqparen{\pi x_a x_b + \pi x_b x_c + \pi x_c x_v + \theta_{aa} x_a^2 + \theta_{bb} x_b^2 + \pi \!\cdot\! x_c^2 + \theta_{vv} x_v^2}}
	\notag\\[0.5ex]=&\;\;\;
		\sum_{b,v} \paren{
			\e^{i\sqparen{\pi x_a x_b + \theta_{aa} x_a^2 + \theta_{bb} x_b^2 + \theta_{vv} x_v^2}}
			\;\;-\;\;
			(-1)^{x_b \oplus x_v}
			\e^{i\sqparen{\pi x_a x_b + \theta_{aa} x_a^2 + \theta_{bb} x_b^2 + \theta_{vv} x_v^2}}}
	\notag\\[0.5ex]=&\;\;\;
		\sum_{b,v} \; 2 \, \delta_{b,(1 - v)} \,
			\e^{i\sqparen{\pi x_a x_b + \theta_{aa} x_a^2 + \theta_{bb} x_b^2 + \theta_{vv} x_v^2}}
	\notag\\[0.5ex]=&\;\;
		2 \sum_{v} \e^{i\sqparen{\pi x_a (1 - x_v) + \theta_{aa} x_a^2 + \theta_{bb}(1 - 2x_v + x_v^2) + \theta_{vv} x_v^2}}
	\notag\\[0.5ex]=&\;\;
		2 \sum_{v} \e^{i\sqparen{\pi x_a x_v + (\pi + \theta_{aa}) x_a^2 + (-\theta_{bb} + \theta_{vv}) x_v^2}}	\;.
\end{align}
\end{subequations}
This yields a quadratic form expansion with two fewer indices, and where $v$ is still an index of degree $1$.
We may repeat this substitution as long as $v$ remains adjacent to a vertex $c'$ of degree $2$, which is itself adjacent to a vertex of degree $2$ and such that $\theta_{c'c'} \in \pi\Z$.
Any such chain will of vertices in the quadratic form will then be reduced to a vertex $v$ which is adjacent to some qubit $a$ which either has $\theta_{aa} \notin \pi\Z$, or which is involved in more than two cross-terms of the original quadratic form.

Performing the substitutions above can only help to obtain a tuple $(G',I,O,T)$ which can be analyzed using eflows.
By removing additional indices, we reduce the complexity of the quadratic form expansion, and also (in the corresponding instance of \MPI) reduce the number of qubits whose particular measurement angles may affect whether the transformation performed by the positive branch is an isometry.
For a given vertex $v$ of degree $1$ in the geometry induced by the quadratic form expansion, provided both the quadratic form and also a representation of the geometry in terms of adjacency lists, determining whether one of the substitutions of~\eqref{eqn:telescopingQuadFormExpan} can be applied can be determined in constant time.
Simplifying the quadratic form expansion provided as input as much as possible, and updating the coefficients of the square terms, is therefore a relatively inexpensive procedure.
We may assume for the remainder of this section that the geometry $(G,I,O)$ induced by the quadratic form provided as input has unit weight, and that for any vertex $v \in V(G) \setminus (I \union O)$ with degree $1$ that is adjacent to a vertex $a \in O\comp$, either the degree of $a$ is at least $3$ or $\theta_{aa} \notin \pi\Z$.

Provided a quadratic form expansion whose geometry $(G,I,O)$ satisfies the above constraints, define $\vec a = (\alpha_v)_{v \in O\comp}$ given by $\alpha_v = -\theta_{vv}$.
Consider the set $S$ of vertices $v$ of degree $1$: for such vertices $v$ and for the unique neighbor $a \sim v$, as we noted in the analysis of~\eqref{eqn:changeFrameXY}, the postselection of the states $\ket{\psub{\alpha_a}}$ and $\ket{\psub{\alpha_v}}$ is equivalent to post-selecting some state $\ket{\splus\sfy\sfz_{\alpha'}}$ if and only if $\alpha_a \in \pi\Z$\,.
Let $S' \subset S$ be the subset of those degree $1$ vertices whose unique neighbors $a$ have $\theta_{aa} \in \pi\Z$\,: then, using the relations
\begin{subequations}
\begin{align}
		\bra{\psub{\alpha_v}}_v \bra{\splus}_a \cZ_{av} \ket{\splus}_v
	\;\;\propto&\;\;\;
		\bra{0}_v \bra{\splus}_a  \ket{\splus}_v
	\;+\;
		\e^{-i\alpha_v} \bra{1}_v \bra{\sminus}_a \ket{\splus}_v
	\notag\\=&\;\;\;
		\sfrac{1}{\sqrt 2} \bra{\splus \sfx}_a \;\;+\;\;\; \sfrac{\e^{-i\alpha_v}\!}{\sqrt 2} \bra{\sminus \sfx}
	\notag\\=&\;\;\;
		\bra{\splus\sfy\sfz_{(\minus \alpha_v)}}_a	\;,
	\\[2ex]
		\bra{\psub{\alpha_v}}_v \bra{\sminus}_a \cZ_{av} \ket{\splus}_v
	\;\;\propto&\;\;\;
		\bra{0}_v \bra{\sminus}_a  \ket{\splus}_v
	\;+\;
		\e^{-i\alpha_v} \bra{1}_v \bra{\splus}_a \ket{\splus}_v
	\notag\\\propto&\;\;\;
		\sfrac{1}{\sqrt 2} \bra{\splus \sfx}_a \;\;+\;\;\; \sfrac{\e^{i\alpha_v}\!}{\sqrt 2} \bra{\sminus \sfx}
	\notag\\=&\;\;\;
		\bra{\splus\sfy\sfz_{\alpha_v}}_a	\;,
\end{align}
\end{subequations}
we may infer that the instance of the \MPI\ represented by the tuple $(G,I,O,\vec a)$ is equivalent to an instance $(G', I, O, T, \vec a')$, defined as follows:
\begin{romanum}
\item
	$G' = G \setminus S'$\,;

\item
	$T$ consists of the neighbors $a$ of the elements of $v \in S'$ in the graph $G$, which are to be measured in the \YZ\ plane; and

\item
	$\vec a' = (\alpha'_v)_{v \in V(G') \setminus O}$\,, where $\alpha'_a = -\e^{i \alpha_a} \alpha_v$ for vertices $a \in T$ which are adjacent in $G$ to some vertex $v \in S'$, and $\alpha'_w = \alpha_w$ otherwise.
\end{romanum}
(Again, this is an instance of an \emph{extension} of \MPI, in which measurements in the \YZ\ plane are permitted.)
We may then test whether the triple $(G',I,O,T)$ has an eflow using Algorithm~\ref{alg:findEflow}.

As in the case of gflows described on page~\pageref{eqn:gflowPattern}, an eflow leads immediately to a solution to (the original version of) \MPI\ for the unit weight geometries $(G,I,O)$ which we have assumed we are given as input.
Algorithm~\ref{alg:findEflow} represents the partial order $\preceq$ of an eflow $(f,\preceq)$ on $(G', I, O, T)$ in terms of a layer function $\ell: V(G') \to \N$ such that $v \preceq w \iff [\,v = w \text{~or~} \ell(v) > \ell(w)\,]$.
We may obtain a linear order $\le$ extending $\preceq$ by constructing level sets $\sV_j$ for each $j \in \N$\,, inserting each vertex $v \in V(G')$ into the set $\sV_{\ell(v)}$\,.
By the remarks on page~\pageref{discn:linearExtension}, this allows us to obtain a linear ordering of $V(G)$ extending $\preceq$.
For qubits $v \in V(G') \setminus T$, we may select for the result $\s[v] = 0$ using the pattern
\begin{subequations}
\begin{align}
	\label{eqn:eflowPatternXYmeas}
		\mathfrak {Ps}_v
	\;\;=\;\;
		\Xcorr{f(v)}{\s[v]} \sqparen{\prod_{\substack{w \sim f(v) \\ w \ne v}} \!\! \Zcorr{w}{\s[v]} \;} \!\Meas{v}{\alpha_v} 	\;;
\end{align}
for qubits $a \in T$, using the approach described in the analysis of~\eqref{eqn:zzManyXY}, we may select for the result $\s[a] = 0$ of a \YZ-plane measurement in the $\ket{\spm\sfy\sfz_{\alpha'_a}}$ basis in the extended instance of \GMPI, by performing the pattern
\begin{gather}
	\label{eqn:eflowPatternYZmeas}
		\mathfrak {Ps}_a
	\;\;=\;\;
		\sqparen{\prod_{\substack{u \sim a \\ u \ne v}} \Zcorr{u}{\s[v]}}
		\Meas{v}{\alpha_v;\s[a]}
		\Meas{a}{\,\alpha_a}	\;,
\end{gather}
\end{subequations}
where $v \in V(G) \setminus V(G')$ is the corresponding vertex of degree $1$ adjacent to $a$.
Then, the following CPTP map performs a unitary embedding for any choice of angles $\vec a = (\alpha_v)_{v \in O\comp}$ such that $\alpha_a \in \pi\Z$ for $a \in T$\,:
\begin{align}
	\label{eqn:eflowPattern}
	\paren{
		\ordprod[\mspace{-30mu}\ge\mspace{15mu}]_{v \in V(G') \setminus O}
			\mathfrak{Ps}_v }
	\paren{
		\prod_{uv \in E(G)} \Ent{u,v} }
	\paren{
		\prod_{v \in I\comp} \New{v}{\sfx} },
\end{align}
where the left-most product is ordered left-to-right according to the order $\ge$.
By the analysis of Section~\ref{sec:postselectionProcedure}, if $(G,I,O, \vec a)$ is an instance of the \MPI\ derived from a quadratic form expansion for an operator $U$, the unitary operation which is performed is $U$.

\paragraph{Run-time analysis.}

Let $n = \card{V(G)}$, $m = \card{E(G)}$, and $k = \card{O} \ge \card{I}$.
Determining whether a geometry has an eflow can be done by Algorithm~\ref{alg:findEflow} in time $O(kn + m)$ by the analysis on page~\pageref{sec:findEflowAnalysis}.
Having an eflow $(f, \preceq)$, the measurement procedure in~\eqref{eqn:eflowPattern} is well-defined and requires $O(m)$ time to construct, dominated by the product over $v \in V(G') \setminus O$ of the patterns $\mathfrak {Ps}_v$, each of which have complexity at most $O(\deg_G(v))$. 
Bringing this measurement procedure into a standard form by the standardization procedure of Section~\ref{sec:DKPstandardization} (also requires time $O(m)$\,, by a similar analysis as applies for patterns constructed by the DKP construction.
The resulting measurement pattern contains $n-k$ measurement operations and up to $k$ correction operations, each of which involves $O(n)$ measurement dependencies.
Finally, each substitution in~\eqref{eqn:diagonalExpansions} and~\eqref{eqn:telescopingQuadFormExpan} requires $O(1)$ time to perform provided a graph-based representation of the geometry underlying the original quadratic form expansion; the former expansion can be applied at most $m$ times (once for each edge), and the latter at most $m + n$ times (one for each vertex in the geometry, after the addition of vertices due to fractional edges) after spending $O(n)$ time finding all vertices of degree $1$.
We may therefore conclude:

\begin{theorem}
	\label{thm:synthEflowsExact}
	For a unitary embedding $U$ given as a quadratic form expansion such that $\card{V(G)} = n$, $\card{E(G)} = m$, and $\card{O} = k$, there is an $O(kn + m)$ algorithm which determines whether or not the quadratic form expansion cannot be transformed into a set of default measurement angles $\vec a'$ and a tuple $(G',I,O,T)$ which has an eflow.
	If so, the algorithm also constructs a pattern in the $\OneWay(\R)$ model consisting of $O(kn + m)$ operations which performs the transformation $U$.
\end{theorem}

As for the analysis in terms of gflows on page~\pageref{thm:synthGflowsExact}, we may strengthen this result to obtain an algorithm producing one-way patterns in a model with a finite gate set such as $\A = \frac{\pi}{4}\Z$\,: given that $m \ge n - k$ for a geometry with an eflow (from which we may infer that there are $O(m)$ measurement operations in the exact measurement pattern we must be approximated), using the same techniques as in the case for gflows, we then obtain the following corollary:

\begin{corollary}
	\label{cor:synthEflowsApprox}
	For a unitary embedding $U$ given as a quadratic form expansion such that $\card{V(G)} = n$, $\card{E(G)} = m$, and $\card{O} = k$, for each $\epsilon > 0$ and $\delta > 0$ there exists an algorithm running in time $O(kn + m \log(m / \epsilon)^{3+\delta})$ algorithm which determines whether or not the quadratic form expansion cannot be transformed into a set of default measurement angles $\vec a'$ and a tuple $(G',I,O,T)$ which has an eflow.
	If so, the algorithm also constructs a pattern in the $\OneWay(\frac{\pi}{4}\Z)$ model consisting of $O(kn + m \log(m/\epsilon)^{3 + \delta})$ operations which approximates the CPTP map $\rho \mapsto U \rho U\herm$ to precision $\epsilon$.
\end{corollary}

\subsection{Synthesizing measurement patterns for the Clifford group}
\label{sec:synthClifford}

If a quadratic form expansion has a geometry whose edges all have unit weight, and its' other coefficients are multiples of $\pion 2$, then it corresponds to the positive branch of a one-way measurement pattern which measures only $X$ or $Y$ observables.
A measurement pattern of this sort, if it performs a unitary operation, performs a Clifford group operation, as we may show (using the stabilizer formalism) that it transforms the Pauli group to itself.

An algorithm of Aaronson and Gottesman~\cite{AG04} can produce a Clifford circuit of size $O(n^2 / \log(n))$ in classical deterministic time $O(n^3 / \log(n))$ for a Clifford group operation $U$ acting on $n$ qubits, from a description of how $U$ transforms Pauli operators by conjugation.\label{discn:AG04}
By converting the circuit into a measurement-based algorithm, and performing the graph transformations of~\cite{HEB04} to remove auxiliary qubits, we may obtain a pattern of at most $3n$ qubits\footnote{%
In~\cite{BB06}, Clifford operations on $n$ qubits are described as having minimal patterns for are described as requiring at most $2n$ qubits; however, this only holds up to local Clifford operations on the output qubits.
These local Clifford operations may be used to adapt subsequent measurements, if the output state is to be operated on by more measurement based procedures, or simply measured to obtain a classical result; however, in the context of a decomposition of unitary transformations in \eg\ the $\OneWay(\frac{\pi}{2}\Z)$ model, these local Clifford operations should be explicitly described.}
in time $O(n^4 / \log(n))$.
Building on the results of~\cite{DM03}, we show how to classically compute such a minimal pattern in time $O(n^3 / \log(n))$ by solving \MPI\ for a quadratic form expansion for $U$.

\subsubsection{Obtaining a quadratic form expansion}

We now outline the relevant results of~\cite{DM03}.
Define the following notation for bit-flip and phase-flip operators on a qubit $t$ out of a collection $\ens{1,\ldots,n}$:
\begin{align}
		P_t
	\;\;=&\;\;
		X_t	\,,
	&
		P_{n+t}
	\;\;=&\;\;
		Z_t	\,.
\end{align}
Let $\diag(M) \in \Z_2^{\, m}$ represent the vector of the diagonal elements of any square boolean matrix $M$; and let $\vec{d}(M) = \diag\left(M\trans\! \left[\begin{smallmatrix} 0 & \idop_n\! \\ 0 & 0 \end{smallmatrix} \right] M\right) \in \Z_2^{\, 2n}$
for a $2n \x 2n$ matrix $M$ over $\Z_2$. Then, we may represent an $n$ qubit Clifford operation $U$ by a $2n \x 2n$ boolean matrix $C$ and a vector $\vec{h} \in \ens{0,1}^{2n}$, whose coefficients are jointly given by
\begin{align}
	\label{eqn:LeuvenTableau}
		U P_t U\herm
	\;\;=&\quad
		i\big.^{d_t(C)} \big(-1\big)^{h_t}
		\bigotimes_{j = 1}^n \sqparen{ Z_j^{\, C_{(n+j)t}} X_j^{\, C_{jt}} }
\end{align}
for each $1 \le t \le 2n$.~ (Note that the factor of $i^{d_t(C)}$ is only necessary to ensure that the image of $P_t$ is Hermitian, and does not serve as a constraint on the value of $C$.)
We will call an ordered pair $(C,\vec h)$ a \emph{Leuven tableau}\footnote{%
This terminology is motivated by the similarity between the block matrix $\big[\;\! C\trans \, \vec{h} \;\!\big]$ and \emph{destabilizer tableaux}, as defined in~\cite{AG04}.}
for a Clifford group element $U$ if it satisfies~\eqref{eqn:LeuvenTableau}.

Provided a Leuven tableau $(C,\vec h)$ for a Clifford group operation $U$, \cite{DM03}~provides a matrix formula for $U$ which we may obtain for $U$.
We outline the derivation of this formula as follows.
Decompose $C$ as a block matrix $C = \left[ \begin{smallmatrix} \,E\, & \,F\, \\ \,G\, & \,H\, \end{smallmatrix} \right]$ with $n \x n$ blocks, and then find invertible matrices $\tilde{R}_1, \tilde{R}_2$ over $\Z_2$ such that $\tilde{R}_1\inv G \tilde{R}_2 = \left[ \begin{smallmatrix} \,0\; & \!0\; \\ \,0\; & \idop_r \end{smallmatrix}\right]$ for some $r < n$~ (using \eg\ the decomposition algorithm of~\cite{PMH08} to obtain $\tilde{R}_1$ and $\tilde{R}_2$ in terms of $O(n^2/\log(n))$ elementary row operations).
Then, define the matrices
\begin{align}
		\left[ \begin{matrix}
			\,\tilde{E}_{11}\, & \,\tilde{E}_{12}\, \\[1ex]
			\,\tilde{E}_{21}\, & \,\tilde{E}_{22}\,
		\end{matrix} \right]
	\,=&\;\;
		\tilde{R}_1\trans E \tilde{R}_2	\,,
	&
		R_1
	\,=&\,\,
		\tilde{R}_1	\;,
	&
		R_2
	\;=&\,\,
		\left[\begin{matrix}
			\,\tilde{E}_{11}\inv\, & 0 \\[1ex]
			\;0\; & \;\idop_r\;
		\end{matrix}\right]\trans\!\!
		\tilde{R}_2\trans	\;,
\end{align}
where $\tilde{E}_{11}$ is taken to be a block of size $(n-r) \x (n-r)$. We may then obtain the block matrices
\begin{align}
		\left[\begin{smallmatrix}
			\idop_{n \minus r} & 		E_{12}		& F_{11} & F_{12} \\[1ex]
					E_{21}		& 		E_{22} 		& F_{21} & F_{22} \\[1ex]
					0				&			0			& H_{11} & H_{12}	\\[1ex]
					0				&	\idop_r	& H_{21} & H_{22}
				 \end{smallmatrix}\right]
	\;\;=&\;\;
		\left[\begin{smallmatrix} R_1\trans & 0 \\ 0 & R_1\inv \end{smallmatrix}\right]
		C
		\left[\begin{smallmatrix} R_2\trans & 0 \\ 0 & R_2\inv \end{smallmatrix}\right]	\,,
\end{align}
and use these to construct the $n \x n$ boolean matrices
\begin{align}
		M_{br}
	\;&=\;\;
		\left[\begin{matrix} \;F_{11} + E_{12} H_{21}\; & \;E_{12}\; \\[1ex] \;E_{12}\trans\; & \;E_{22}\; \end{matrix}\right]\,,
	&
		M_{bc}
	\;=&\;\;
		\left[\begin{matrix} \;0\; & \;H_{21}\trans\; \\[1ex] \;H_{21}\; & \;H_{22}\; \end{matrix}\right]\,.
\end{align}
Next, define 
\begin{align}
	\begin{split}
			\vec{d}_{br}
		&=\;
			\diag(M_{br}) \,,
		\\
			L_{br}
		\;&=\;
			\Lower\!\big(M_{br} + \vec{d}_{br} \vec{d}_{br}\trans\big) \,,
	\end{split}
	&
	\begin{split}
			\vec{d}_{bc}
		&=\;
			\diag(M_{bc}) \,,
		\\
			L_{bc}
		\;&=\;
			\Lower\!\big(M_{bc} + \vec{d}_{bc} \vec{d}_{bc}\trans\big) \,,
	\end{split}
\end{align}
where $\Lower(M)$ is the strictly lower-triangular part of a square matrix $M$ (with all other coefficients set to $0$).
Finally, define $\Pi_r = \left[\begin{smallmatrix} 0 & 0 \\ 0 & \idop_r\! \end{smallmatrix}\right]$ and $\Pi_r\dual
= \idop_n - \Pi_r$ for the sake of brevity, and let\footnote{%
These formulae for $\vec{t}$ and $\vec{h}_{bc}$ may be obtained by repeated application of Theorem~2 of~\cite{DM03} to the data specified by Theorem~6.}
\begin{subequations}
\begin{align}
		\vec{t}
	\;\;=&\;\;
		\left[\begin{matrix} \,\idop_n \,&\, 0 \;\end{matrix}\right] \vec{h}
		\;+\;	\diag\Big( \sqparen{ R_2\inv \Pi_r  } \!L_{br} \sqparen{ R_2\inv \Pi_r  }\trans \Big)		\;,
	\\[0.5ex]
	\begin{split}
		\vec{h}_{bc}
	\;\;=&\;\;
		 \left[\begin{matrix}\; 0 \,&\, R_2\intrans \,\end{matrix}\right] \vec{h}
		\;+\;	R_2\intrans \diag\Big(
					R_2\trans \Big[ L_{bc} \,+\, \Pi_r M_{bc} 
	\\[-0.5ex]&\mspace{100mu}
					\,+\, \paren{\Pi_r\dual + \Pi_r M_{bc}} L_{br} \paren{\Pi_r\dual +  M_{bc} \Pi_r} \Big] R_2
				\Big)	\;.
	\end{split}
\end{align}
\end{subequations}
Then Theorem~6 of~\cite{DM03} states that the unitary operation $U$ for the Clifford operation characterized by $(C, \vec{h})$ is given by the matrix formula
\begin{align}
	\label{eqn:cliffordFormula}
	\begin{split}
		U
	\,=\:\!
		\sfrac{1}{\sqrt{2^r}}\!\!\!
		\sum_{\substack{\vec{x}_b \in \ens{0,1}^{n \minus r} \\ \vec{x}_c, \vec{x}_r \in \ens{0,1}^r}}\!
		\bigg[\,
			(&-1)^{\paren{
									\vec{x}_{br}\trans L_{br} \vec{x}_{br} \,+\, 
									\vec{x}_r \trans \vec{x}_c \,+\,
									\vec{x}_{bc}\trans L_{bc} \vec{x}_{bc}	\,+\,
									\vec{h}_{bc}\trans \vec{x}_{bc} 
									}} \;\x
	\\[-4ex]&
			(-i)^{\paren{\vec{d}_{br}\trans \vec{x}_{br} \,+\, \vec{d}_{bc}\trans\vec{x}_{bc}}}	\;
			\ket{R_1\big. \vec{x}_{br}}\bra{R_2\inv \vec{x}_{bc} + \vec{t}}
		\,\bigg]	\,,
	\end{split}
\end{align}
where $\vec{x}_{br} = \left[\begin{smallmatrix} \vec{x}_b \\ \vec{x}_r \end{smallmatrix}\right]$ and
$\vec{x}_{bc} = \left[\begin{smallmatrix} \vec{x}_b \\ \vec{x}_c \end{smallmatrix}\right]$ are $n$ bit boolean
vectors.

The formula in~\eqref{eqn:cliffordFormula} shows strong similarities to a quadratic form expansion.
In particular, consider disjoint sets of indices $V_b$, $V_r$, and $V_c$, with $\card{V_b} = n-r$ and $\card{V_r} = \card{V_c} = r$.
Let $V = V_b \union V_c \union V_r$, $I = V_b \union V_c$, and $O = V_b \union V_r$, and define the following notation for $\vec{x} \in \ens{0,1}^V$\,:
\begin{align}
		\vec{x}_I
	\;=&\;
		\left[\begin{matrix} \vec{x}_b \\ \vec{x}_c \end{matrix}\right]
	\;=\;
		\left[\begin{matrix} \vec{x}_{\text{\raisebox{-0.3ex}{$V_b$}}} \\ \vec{x}_{\text{\raisebox{-0.3ex}{$V_c$}}} \end{matrix}\right] \in \ens{0,1}^{I}	\;,
	&
		\vec{x}_O
	\;=&\;
		\left[\begin{matrix} \vec{x}_b \\ \vec{x}_r \end{matrix}\right]
	\;=\;
		\left[\begin{matrix} \vec{x}_{\text{\raisebox{-0.3ex}{$V_b$}}} \\ \vec{x}_{\text{\raisebox{-0.3ex}{$V_r$}}} \end{matrix}\right] \in \ens{0,1}^{O}	\;,
\end{align}
\begin{align}
	\begin{split}
		Q(\vec x)
	\;=&\;\;
		\pi\Big( \vec{x}_O\trans L_{br} \vec{x}_O \,+\,
							\vec{x}_O\trans \Pi_r \vec{x}_I \,+\, \vec{x}_I\trans L_{bc} \vec{x}_I \,+\,
							\vec{x}_I\trans \vec{h}_{bc} \vec{h}_{bc}\trans \vec{x}_I	\Big)
	\\&\mspace{170mu}-\;
		\sfrac{\pi}{2} \Big(	\vec{x}_O\trans \vec{d}_{br}\vec{d}_{br}\trans \vec{x}_O \,+\,
									\vec{x}_I\trans \vec{d}_{bc}\vec{d}_{bc}\trans \vec{x}_I \Big)	\;.
	\end{split}
\end{align}
Then,~\eqref{eqn:cliffordFormula} is equivalent to
\begin{align}
	\label{eqn:cliffordAlmostQuadFormExpan}
		U
	\;\;=\;\;
		\sfrac{1}{\sqrt{2^r}}
		\sum_{\vec{x} \in \ens{0,1}^V}
			\e^{i Q(\vec x)} \ket{R_1\big. \vec{x}_O}\bra{R_2\inv \vec{x}_I + \vec{t}}	\;,
\end{align}
\begin{figure}[tb]
	\begin{center}
		\includegraphics{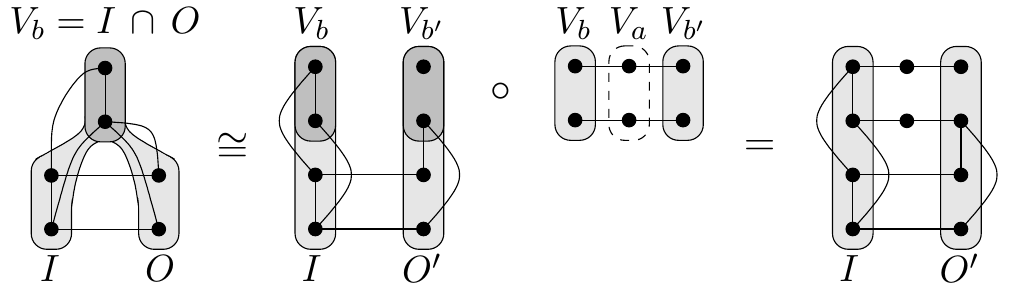}
		\caption[Illustration of an equivalence of two quadratic form expansions, via their geometries.]{\label{fig:idxSplit}%
		Illustration of an equivalence of two quadratic form expansions, via their geometries.
		On the left is a geometry whose inputs and output intersect; on the right is a geometry from an equivalent quadratic form expansion, constructed so that the input and output indices are disjoint.} 
	\end{center}
\end{figure}%
which is essentially a quadratic form expansion sandwiched between two networks of controlled-not and $X$ gates.
To obtain a simple quadratic form expansion, we would like to perform a change of variables on $\vec{x}_I$ and $\vec{x}_O$; but this cannot be done as $I$ and $O$ intersect at $V_b$, and the changes of variables do not necessarily respect the partitioning of $I$ and $O$ with respect to this intersection.
However, we may add auxiliary variables in order to produce an expansion with disjoint input and output indices.
Recall the expansion of the identity of~\eqref{eqn:idopQuadFormExpan},
\begin{align}
		\idop_2
	\;\;=\;\;
		\sum_{\vec{x} \in \ens{0,1}^2} \delta_{x_1,x_2} \ket{x_2} \bra{x_1}
	\;\;=\;\; 
		\sfrac{1}{2} \sum_{\vec{x} \in \ens{0,1}^3} (-1)^{x_1 x_3 + x_2 x_3} \ket{x_2} \bra{x_1}	\;,
\end{align}
where $\delta_{x,y}$ is the Kronecker delta.
Let $V_a$ and $V_{b'}$ be disjoint copies of $V_b$, and set $V' = V \union V_a \union V_{b'}$ and $O' = V_{b'} \union V_r$.
Writing $\vec{x}_a$ and $\vec{x}_{b'}$ for the restriction of $\vec{x} \in \ens{0,1}^{V'}$ to $V_a$ and $V_{b'}$, we then define
\begin{align}
		\vec{x}_I
	\;=&\;
		\left[\begin{matrix} \vec{x}_b \\ \vec{x}_c \end{matrix}\right] \in \ens{0,1}^{\, I}	\;,
	&
		\vec{x}_{O'}
	\;=&\;
		\left[\begin{matrix} \vec{x}_{b'} \\ \vec{x}_r \end{matrix}\right] \in \ens{0,1}^{\, O'}	\;,
\end{align}
\vspace{-1ex}
\begin{align}
	\mspace{-10mu}
		Q'(\vec{x}_I\,,\, \vec{x}_a\,,\, \vec{x}_{O'})
	\;=&\;\;
		\pi\Big(	\vec{x}_{O'}\trans L_{br} \vec{x}_{O'} \,+\,
							\vec{x}_{O'}\trans \Pi_r \vec{x}_I \,+\, \vec{x}_I\trans L_{bc} \vec{x}_I \,+\,
							\vec{h}_{bc}\trans \vec{x}_I	\Big)
	\notag\\&\mspace{10mu}
		\;+\;	\pi \vec{x}_I\trans \!\left[\begin{smallmatrix} \idop_{n \minus r} \\[0.5ex] 0 \end{smallmatrix}\right]\! \vec{x}_a
		\;+\;	\pi \vec{x}_{O'}\trans \!\left[\begin{smallmatrix} \idop_{n \minus r} \\[0.5ex] 0 \end{smallmatrix}\right]\! \vec{x}_a
		\;-\;	\sfrac{\pi}{2} \Big(	\vec{d}_{br}\trans \vec{x}_{O'} \,+\,	\vec{d}_{bc}\trans \vec{x}_I \Big)	\;.
\end{align}
Note that the difference between the expressions for $Q'$ and $Q$ is essentially that all instances of $\vec{x}_O$ have been replaced with $\vec{x}_{O'}$ (which is independent from $\vec{x}_I$), and the presence of the terms involving $\vec{x}_a$.
We therefore have
\begin{align}
	\label{eqn:quadFormTransform}
		\sum_{\vec{x} \in \ens{0,1}^V}
			\e^{i Q(\vec x)} &\ket{R_1\big. \vec{x}_O}\bra{R_2\inv \vec{x}_I + \vec{t}}
	\notag\\=&\;\;\;
		\sum_{\vec{x}_I,\, \vec{x}_{O'}}
			\delta_{\vec x_b, \vec x_{b'}} \, \e^{i Q'(\vec x_I,\, \vec{0},\, \vec x_{O'})} \ket{R_1\big. \vec{x}_{O'}}\bra{R_2\inv \vec{x}_I + \vec{t}}
	\notag\\=&\;\;\;	
		\sfrac{1}{2^{n-r}} \sum_{\vec{x} \in \ens{0,1}^{V'}}
			\e^{i Q'(\vec{x}_I,\, \vec{x}_a,\, \vec{x}_{O'})} \ket{R_1\big. \vec{x}_{O'}}\bra{R_2\inv \vec{x}_I + \vec{t}} \,.
\end{align}
This transformation is illustrated in Figure~\ref{fig:idxSplit} as a transformation of the geometries underlying $Q$ and $Q'$\,.
Substituting the final expression of~\eqref{eqn:quadFormTransform} into~\eqref{eqn:cliffordAlmostQuadFormExpan} and performing the appropriate change of variables, we have
\begin{align}
	\label{eqn:cliffordQuadFormExpan}
		U
	\;\;=\;\;
		\sfrac{\sqrt{2^r}}{2^n}
		\sum_{\vec{x} \in \ens{0,1}^{V'}}
			\e^{i Q'(R_2(\vec{x}_I + \vec{t}),\, \vec{x}_a,\, R_1^{\minus 1} \vec{x}_{O'})} \ket{\vec{x}_{O'}}\bra{\vec{x}_I}	\;:
\end{align}
expanding the expression in the exponent, it is easy to see that this is a quadratic form expansion for $U$.

\subsubsection{Interpolating the measurement pattern}

Note that the quadratic form of the expansion in~\eqref{eqn:cliffordQuadFormExpan} has only angles $\theta_{uv}$ which are multiples of $\pion 2$, with $\theta_{uv} \in \ens{0, \pi}$ for $u \ne v$.
This then represents the positive branch of a one-way measurement pattern using only \axisX- and \axisY-axis measurements, and having only $n-r$ auxiliary vertices.
From the analysis of the stabilizer formalism in Section~\ref{sec:stabilizerFormalism}, the fact that the result of the post-selection procedure is in particular a \emph{unitary} transformation implies that each individual post-selection induces a unitary transformation; then, we can then obtain a correction procedure to obtain a complete measurement pattern for $U$ via the stabilizer formalism.

First, we may produce a complete measurement procedure in which we perform correction operations after each measurement.
After applying the preparation and entangling operations $\Ent{G} \New{I\comp}{}$ before the measurements, the state-space is stabilized by the group $\sS_{G,I} \,=\, \gen{K_v}_{v \in I\comp}$\,.
Because post-selecting the measurement results $\s[v]$ for $v \in O\comp$ performs an isometry of the encoded space, there must be a list of generators $(S_v)_{v \in O\comp}$ for $\sS_{G,I}$ such that, for some ordering $v_1, v_2, \ldots$ of the qubits of $V(G)$,
\begin{romanum}
\item
	$S_{v_j}$ anticommutes with the measurement observable for the measurement on $v_j$\,, where we define
	\begin{gather}
			\sO_{v_j}
		\;\;=\;\;\
			\begin{cases}
				X	\,,	&	\text{if $\theta_{v_j v_j} \in \pi\Z$}	\\
				Y	\,,	&	\text{if $\theta_{v_j v_j} \in \pi \Z + \pion 2$}
			\end{cases}\;;
	\end{gather}

\item
	$S_{v_j}$ commutes with the measurement observable $\sO_{v_h}$ for any $h > j$\,.

\end{romanum}
We may produce such a list of generators for any given enumeration $(v_j)_{j = 1}^n$ of the elements of $V(G)$ in polynomial time, essentially by row-reduction.

We arrange the generators $K_v$ for $v \in O\comp$ in rows in arbitrary order, with the columns denoting the qubits on which each operator acts.
We then decompose each operator $K_v$ into a product of operators $\sO_w$ for $w \in O\comp$ and $X_w$ for $w \in O$ (which we call the \emph{commuting} part of $K_v$) and a product of $Z_w$ operators for $w \in V(G)$ (which we call the \emph{anticommuting} part of $K_v$).
It is obvious that the anticommuting part of each operator $K_v$ will anticommute with any $\sO_w$ for a qubit $w \in O\comp$ on which it acts, and similarly the commuting part of $K_w$ commutes with every observable $\sO_w$.
We may use these components to form a stabilizer tableau, as illustrated in Figure~\ref{fig:stabReduce}, with the ``anticommuting'' $Z$ part of the operator in the leading columns, and the ``commuting'' part in the trailing columns.
Row transformations of this matrix then produce different generators of $\sS_{G,I}$, where the commutation relations with the observables $\sO_w$ for $w \in O\comp$ can be easily read from the $Z$ component part of the table.
By putting this table into reduced row-echelon form, we may then obtain a collection of generators $\ens{S_v}_{v \in O\comp}$ for $\sS_{G,I}$ which not only satisfy the criteria above, but such that $S_v$ anticommutes with $\sO_w$ only for $v = w$\,.
As each correction operation only acts on the qubit $v \in O\comp$ whose measurement it depends upon (and which we are considering to be measured destructively), these generators $S_v$ also represent the final correction operations for the measurement pattern: at the end of the measurement sequence, it suffices to perform the correction $S_v^{\s[v]}$ for each $v \in O\comp$ to obtain a unitary transformation.
\begin{figure}[p]
	\begin{center}
		\includegraphics[width=120mm]{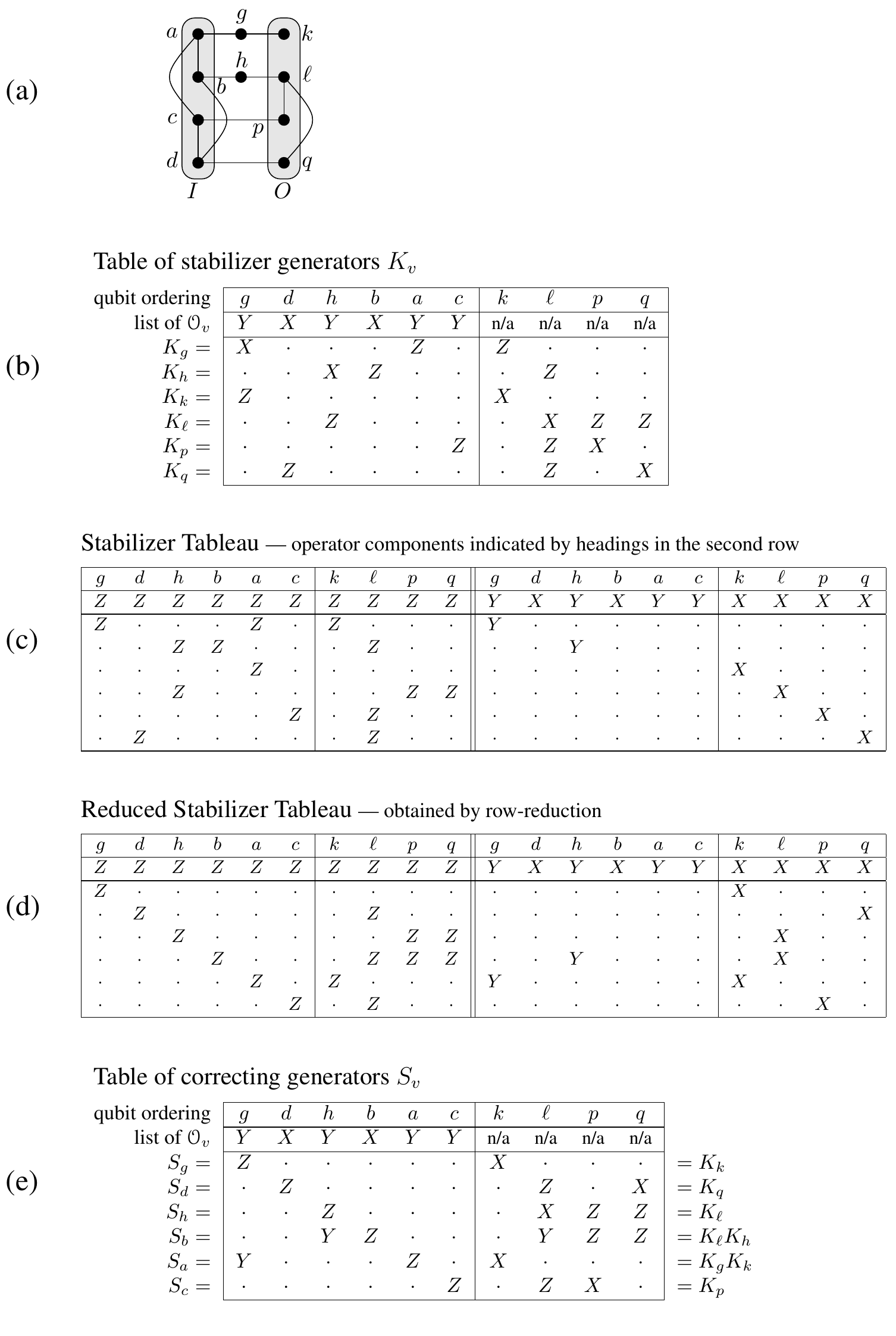}
		\vspace{-1ex}
	\end{center}
	\caption[Illustration of an algorithm for obtaining the correction operations for a postselection-based procedure for a unitary $U$ using measurement observables $X$ and $Y$.]{\label{fig:stabReduce}%
		Illustration of an algorithm for obtaining the correction operations for a postselection-based procedure for a unitary $U$ using measurement observables $X$ and $Y$.
		\begin{inlinum}
		\item The input geometry.
		\item	The stabilizer generators $K_v$ arising from the geometry, as well as an arbitrary measurement order of the qubits of $O\comp$ and a specification of the measurement observables $\sO_v$\,; ``dots'' represent identity operations $\idop_2$.
		\item The same generators presented in a ``stabilizer tableau'', in which each generator is decomposed into a $Z$ component (which anticommutes with every measurement observable) and a non-$Z$ component (chosen to commute with the observable $\sO_v$ for each qubit $v \in O\comp$).
		Note that this tableau is essentially an $(n - \card{I}) \x 2n$ boolean matrix.
		\item The reduced row-echelon form arising from the previous table.
		For each $j \ge 1$, the $j\th$ row represents an operator which anticommutes with the $j\th$ measurement observable, and commutes with every other measurement observable.
		\item A more explicit representation of the preceding table.
		\end{inlinum}
		}
\end{figure}

\paragraph{Run-time analysis.}

For a Clifford operation $U \in \U(2^n)$ given as a Leuven tableau $(C, \vec h)$, the quadratic form of~\eqref{eqn:cliffordAlmostQuadFormExpan} can be found in time $O(n^3 / \log( n))$, which is dominated by the time required to compute $R_1$ and $R_2$, using the techniques of~\cite{PMH08}.\footnote{%
The algorithm of~\cite{PMH08} obtains an LU decomposition for matrix over $\Z_2$\,: while the result is described as applying to invertible matrices, it in fact can be applied to arbitrary boolean matrices.}
From this, the quadratic form expansion of~\eqref{eqn:cliffordQuadFormExpan} can be obtained by a constant number of matrix multiplications, and by inverting $R_1$ and $R_2$ (which may be computed alongside $R_1$ and $R_2$ themselves).
The boolean row-reductions required to compute the correcting generators $S_v$ for $v \in O\comp$ also requires time $O(n^3 / \log(n))$ by using a modified version of the algorithm of~\cite{PMH08} to do simple row-reduction.\footnote{%
It is sufficient to modify the algorithm of~\cite{PMH08} to eliminate ones above the diagonal as well as below, and to forego the column-reduction phase (consisting of the transposition and all operations afterwards).}
The resulting measurement pattern contains at most $3n$ qubits, and therefore requires $O(n^2)$ entangling operations, $O(n)$ measurement operations, and $O(n)$ correction operations each of complexity at most $O(n)$, yielding a total size of $O(n^2)$.
Therefore, we have:

\begin{theorem}
	\label{thm:synthClifford}
	For an $n$-qubit Clifford group operation $U$ given in the form of a Leuven tableau, there is an $O(n^3 / \log(n))$ algorithm which produces a pattern of size $O(n^2)$ in the $\OneWay(\frac{\pi}{2})$ model which performs $U$.
\end{theorem}

\subsubsection{Minimality of patterns synthesized from quadratic form expansions}

Hein, Eisert, and Briegel present techniques in~\cite{HEB04} for the elimination of Pauli observable measurements in graph states by
\begin{inlinum}
\item
	transformations of the underlying geometry, and

\item
	local unitary transformations of the qubits, which amount to changes in the observable to be performed for some of the measurements. 
\end{inlinum}
Paraphrasing~\cite{HEB04} and extending it to open graphical encodings, these transformations are as follows:
\begin{lemma}[{\cite[Proposition~1]{HEB04}}]
	\label{lemma:HEB04}
	Let $G$ be a graph: throughout the following, let $\sim$ denote adjacency in $G$.
	Define a unitary embedding
	\begin{gather}
		\cE_{G,I} = \paren{\prod_{u \sim v} \cZ_{uv} } \paren{ \bigotimes_{v \in I\comp} \ket{\splus}_v}
	\end{gather}
	performing an open graphical encoding.
	For a vertex $u \in V(G)$, let $G \ast u$ represent the \emph{local complementation} of $G$ about $u$, which is the graph on $V(G)$ where
	\begin{gather}
			vw \in E(G \ast u)
		\;\;\iff\;\;
			\begin{cases}
				vw \in E(G)		\;,	&	\text{if $v \not\sim u$ or $w \not\sim u$}	\\
				vw \notin E(G)	\;,	&	\text{if $v \sim u$ and $w \sim u$}
			\end{cases}	.
	\end{gather}
	If an $X_a$, $Y_a$, or $Z_a$ observable measurement is performed on some qubit $a \in I\comp$ for a joint state of the form $\cE_{G,I} \ket{\psi}$ for any $\ket{\psi} \in \cH[I]$, then the resulting pure state, conditioned on the outcome $\pm 1$, are as follows:
	\begin{subequations}
	\begin{romanum}
	\item
		 for a $Z_a$ observable measurement, the post-measurement state is
		\begin{gather}
			\ket{\spm \sfz}_a \ox U_{a,\pm\sfz} \, \cE_{G'_\sfz,I} \ket{\psi}
		\end{gather}
		where $G'_\sfz = G \setminus a$, $U_{a,\splus\sfz} = \idop_{V(G)}$, and $U_{a,\sminus\sfz} = \prod\limits_{v \sim a} Z_v$\;;

	\item
		for a $Y_a$ observable measurement, the post-measurement state is
		\begin{gather}
			\ket{\spm \sfy}_a \ox U_{a,\pm\sfy} \, \cE_{G'_\sfy,I} \ket{\psi}
		\end{gather}
		where $G'_\sfy = (G \ast a) \setminus a$, and $U_{a,\spm\sfy} = \prod\limits_{v \sim a} R_\sfz(\pm\pion 2)_v$\;;

	\item
		for an $X_a$ observable measurement, for an arbitrarily chosen $b \in I\comp$ adjacent to $a$, the post-measurement state is
		\begin{gather}
			\ket{\spm \sfx}_a \ox U_{a,\pm\sfx} \, \cE_{G'_\sfx,I} \ket{\psi}
		\end{gather}
		where $G'_\sfx = (G \ast a \ast b \ast a) \setminus a$, and where we define
		\begin{align}
				U_{a,\splus\sfx}
			\;=&\;\;
				R_\sfy(\mpion 2)_b \prod_{\substack{v \sim a \\ v \not\sim b_0 \\ v \ne b_0}} Z_v	\;;
			&
				U_{a,\sminus\sfx}
			\;=&\;\;
				R_\sfy(\pion 2)_b \prod_{\substack{v \sim b_0 \\ v \not\sim a \\ v \ne a}} Z_v	\;.
		\end{align}
	\end{romanum}
	\end{subequations}
\end{lemma}
It can be easily verified that~\cite[Proposition~1]{HEB04} can be extended as above, as the analysis of that article depends only on the presence of the stabilizer generators $K_v$ for $v \in I\comp$.\footnote{%
The input subsystem $I$ does not contribute any generators $K_v$ for $v \in I$, precisely because the input subsystem is not in general stabilized by any non-trivial Pauli operators.
Despite this, local complementations still affect adjacency relations which are entirely restricted to the input subsystem.
This may be verified by noting that an open graphical encoding can be characterized by $2^{\card I}$ stabilizer groups in parallel, with each group describing a scenario where each qubit $v \in I$ is stabilized by $X_v$ or $Z_v$ prior to encoding.
Collectively, these groups characterize any unitary transformations performed on the state, by linearity.
For each such group, any qubit $v \in I$ which is initially stabilized by $X_v$ does contribute a generator $K_v$ to the stabilizer group in that case; the generators of such groups are transformed in the same way as in graph states.}
The connection with local complementations of graphs (which is not made explicit in the presentation of Proposition~1) can be verified by application of Proposition~8 in the same article.

The results of~\cite{HEB04} would seem to be pertinent for the measurement patterns produced from quadratic form expansions, as described by the constructions above.
This suggests that we may be able to apply them to obtain measurement patterns involving fewer qubits than the ones produced by our construction.
Despite this, we may show that the measurement patterns produced by the techniques of this section cannot be reduced using Lemma~\ref{lemma:HEB04}.

Let $A$ denote the set of auxiliary qubits, corresponding to the indices of the vector $\vec{x}_a$ adjoined in~\eqref{eqn:quadFormTransform}.
In the resulting measurement pattern, these are all to be measured with the observable $X$, as there is no square term in the quadratic form $Q'$ for any of these indices.
As well, they are adjacent only to the input/output vertices (corresponding to the indices of $\vec{x}_I$ and $\vec{x}_O$).
These properties still hold in the final quadratic form expansions after the final change of variables, which do not act on the indices of $A$.
To eliminate a vertex $v \in A$ using Lemma~\ref{lemma:HEB04} in the open graphical encoding of the resulting one-way pattern, we must identify an output qubit $b' \in O$ adjacent to $v$, and transform the graph $G$ of the underlying geometry by to the graph $G'_{\sfx} \,=\, (G \ast v \ast b' \ast v) \setminus a$.
This would result in a geometry where $b'$ has the former neighbors of $v$ in $G$ (excepting itself): in particular, it will only be adjacent to one other qubit $b \in I$.
As well, a $R_\sfy(\pm\pion 2)$ rotation must be applied to $b$ after the entangling procedure.
This rotation is essentially unavoidable, in the following sense:
\begin{itemize}
\item
	Because $b' \in O$, it will not be measured, and so the $R_\sfy(\pm\pion 2)$ operation cannot be absorbed into a measurement.

\item
	Because $b'$ is not adjacent to any elements of $V(G'_{\sfx}) \setminus I$, the only stabilizer generator of the output space of the encoding $\cE_{G'_\sfx, I}$ is $K'_{b'} = X_{b'} Z_b$, which does not commute with $R_\sfy(\pm\pion 2)$.
	Then, the $R_\sfy(\pm\pion 2)$ is not equivalent (in the open graphical encoding) to any operator which does not act on $b'$.
\end{itemize}
Then, to represent the result of the transformation described by Lemma~\ref{lemma:HEB04} on a qubit $v \in I\comp$ which is to be measured, we must perform a local Clifford operation outside of the Pauli group to one of the qubits of the resulting encoding.
In the $\OneWay(\frac{\pi}{2}\Z)$ model, this local unitary will be represented by a pattern involving at least one additional qubit; thus, the number of qubits in the total pattern are not decreased.

Given a Clifford operation $U$ with a one-way measurement pattern synthesized from a quadratic form expansion, the techniques of~\cite{HEB04} may be applied if $U$ is to be composed with some other one-way pattern, or if the output subsystem is to be measured to produce a final classical result; however, as a representation of a unitary transformation, the above analysis shows that the number of qubits in the measurement pattern cannot be reduced using those techniques.

\subsubsection{Comparison with the algorithm of Aaronson and Gottesman}

As we noted on page~\pageref{discn:AG04}, an approximately optimal Clifford circuit for a Clifford group operation (consisting of $O(n^2 / \log(n))$ gates) can be found in time $O(n^3 / \log(n))$.
This may be done from a Leuven tableau by transforming it into a destabilizer tableau (another representation of transformations of Pauli operators by $U$ which differs trivially from a Leuven tableau), and then applying the algorithm of~\cite{AG04}.
To obtain a measurement pattern from such a circuit by composing the patterns for each gate, opportunistically removing vertices using the result of Lemma~\ref{lemma:HEB04} (with each removal taking $O(n^2)$ time), leads instead to an $O(n^4 / \log(n))$ time algorithm.
Thus, by Theorem~\ref{thm:synthClifford}, we may synthesize a minimal one-way measurement pattern more efficiently via quadratic form expansions than from the circuits produced by the algorithm of~\cite{AG04}.

The upper bound in Theorem~\ref{thm:synthClifford} on the complexity of the measurement pattern in $O(n^2)$, in contrast with the complexity of $O(n^2/\log(n))$ achieved by the circuits of~\cite{AG04}.
It is an open question whether the complexity of the patterns produced by the algorithm outlined above can be bounded more strongly than $O(n^2)$.
It is not obvious, however, that the number of edges in the entanglement graph or (nearly equivalently) the complexity of the final correction operations should saturate theupper bound of $O(n^2)$, or that the one-way measurement model itself imposes enough structural constraints to cause an asymptotically significant inflation in the optimal representation of a unitary transformation.
I therefore conjecture that the complexity of the measurement patterns synthesized by the algorithm above can also be improved to $O(n^2 / \log(n))$, as in the optimal bound for Clifford circuits.

\section{Further Remarks}

In this Chapter, we have focused largely on how we can transform or construct quadratic form expansions in such a way as to obtain an instance of \MPI, or a related problem such as \GMPI, which we are able to efficiently solve.
Before closing, we will remark on some of the ways in which the work in this Chapter may be extended.

\subsection{Reductions from more complex operator formulae}
\label{sec:higherDegree}

As we remarked on page~\pageref{discn:higherDegree}, path integrals are a potential ``natural'' source of quadratic form expansions, for which we may wish to obtain one-way measurement patterns.
In the case of path integrals, the most natural analogue of the quadratic form is often the integral of the action along a trajectory through space; however, these trajectories usually manifest as a polynomial of degree higher than $2$ for non-trivial systems.
Direct attempts to discretize a path integral may then lead to formulae similar to quadratic form expansions, except where the quadratic form is substituted by a higher degree polynomial.
As well, natural approaches to efficiently estimating the path integral of a physical system may involve attributing different weights to the discretized paths, corresponding to different amounts of constructive/destructive interference of the amplitudes of similar paths which differ from the trajectories of minimum action.
We outline here how variations on quadratic form expansions, involving higher degree polynomials and amplitudes which are non-uniform across the path variables $\vec{x} \in \ens{0,1}^n$, may be reduced to quadratic form expansions.

\subsubsection{Higher degree polynomials governing phases}

In order to treat matrix formulae with higher degree polynomials governing the phase contributions of the terms in the sum, consider a generalization of the substitution described by~\eqref{eqn:diagonalExpansions} to more than two bits.
Considering the construction of the pattern $\mathfrak{Zz\cdots z}^\theta$ described in~\eqref{eqn:zzManyXY}, it is possible to show that
\begin{gather}
	\label{eqn:diagonalExpansionBig}
		\e^{i\phi(x_1 \oplus \cdots \oplus x_d)}
	\;\;=\;\;
		\sfrac{1}{2}\! \sum_{\vec a \in \ens{0,1}^2} \e^{i\big[\pi(x_1 a_1 + \cdots + x_d a_1 + a_1 a_2) \,+\, \phi a_2^{\, 2}\big]}
\end{gather}
for any $d \ge 1$.
We may easily prove by induction that
\begin{gather}
		x_1 \oplus \cdots \oplus x_d
	\;\;=\;\;
		\sum_{\substack{S \subset \ens{1,\ldots,d} \\ S \ne \vide}} \; \prod_{j \in S} (-2)^{\card S - 1} x_j	\;;
\end{gather}
equivalently, we have
\begin{gather}
		\prod_{j = 1}^d x_j
	\;\;=\;\;
		\paren{\sfrac{-1}{2}}^{d-1}
		\sqparen{
			-\paren{x_1 \oplus \cdots \oplus x_d}
			\;\;\;+\!\!
			\sum_{\substack{S \prsubset \ens{1,\ldots,d} \\ S \ne \vide}} \prod_{j \in S} (-2)^{\card S - 1} x_j	\;
		}\;,
\end{gather}
which we can apply recursively to products of the bits $x_1$ through $x_d$ to obtain the formula
\begin{gather}
		\prod_{j = 1}^d x_j
	\;\;=\;\;
		-\paren{\sfrac{-1}{2}}^{d-1}
		\sum_{\substack{S' \subset \ens{1,\ldots,d} \\ S' \ne \vide}} \paren{\,\bigoplus_{j \in S} \; x_j}	\;.
\end{gather}
Thus a monomial of degree $d$ may be expanded into a sum of $2^d - 1$ parities; we may then use the substitution of~\eqref{eqn:diagonalExpansionBig} to express any factor of the form $\e^{i \theta x_1 \cdots x_d}$ in terms of a sum of $2^{d+1} - 2$ auxiliary indices, with two such indices for each sum modulo $2$ to be computed.
If the degree of the terms of the action of a physical system is bounded by some constant, this then leads to a constant factor overhead in the size of a quadratic form expansion, in order to represent a similar expression whose ``paths'' have phases given by higher degree polynomials.

As a result of the analysis above, the degree of the action in the dynamical variables of a physical system need not be an overriding concern in the formulation of a quadratic form expansion, provided that the degree is bounded above by a small constant.
A comparison between this and techniques for ``decoupling'' higher degree terms in path integrals in the physics literature (such as the Hubbard--Stratonivich transformation: see~\cite{Auerbach94,Gubernatis85} for expositions) may provide insights into a high-level physical interpretation of this substitution.

\subsubsection{Unequal amplitude contributions across paths}

Quadratic form expansions are a normalized sum of contributions of operators $\ket{\vec x_O}\bra{\vec x_I}$ for $\vec x \in \ens{0,1}^n$\,, each of which has one singular value of $1$ and the rest equal to zero.
That is to say, as a sum over paths, each path is given equal weight; only the phases differ.
However, it is natural to consider operator sums where terms may have different weights for different boolean strings $\vec x$\,.
One reason may be that, in order to discretize a path integral, contributions of many continuous paths are attributed to each discrete path string $\vec x$\,, and these will in general result in contributions with different amplitudes.
Therefore, matrix expressions in which each coefficient is a sum of complex numbers with different moduli (instead of complex units), normalized over all matrix coefficients, may be of interest.

In case the amplitudes of the paths depend uniformly on the value of a single bit $x_v$ in the path string $\vec{x}$ (\ie\ if all paths with $x_v = 0$ have the same weight, and similarly for $x_v = 1$), we may represent this bias in the amplitudes simply by means of a choice of measurement basis for $v$.
Considering the relationship between quadratic form expansions and instances of \MPI\ shown in Section~\ref{sec:postselectionProcedure}, we may attribute the uniform modulus of the paths in that case to the fact that every measurement is performed with a state which is a uniform-weight superposition of the standard basis: the projection of the states $\ket{0}$ and $\ket{1}$ onto $\ket{\psub\alpha}$ for any $\alpha$ has the same modulus.
In order to achieve a systematic bias towards either $x_v = 0$ or $x_v = 1$, it suffices to consider a measurement of $v$ where the preferred measurement result is similarly biased towards the corresponding standard basis state.

More generally, we may express the bias in the amplitudes of the paths in terms of a boolean polynomial on $\vec{x}$\,.
To achieve the necessary biases, we may consider substitutions of the sort shown in~\eqref{eqn:nonuniformity}.
The non-uniformity of the sum which presented problems for an analysis in terms of \GMPI\ in Section~\ref{sec:interpolateGflows}, here allows us to express amplitudes for paths in terms of the parity of two path variables.
By using parities of boolean expressions, we may produce biases depending on the value variable $\vec x$ consisting of a monomial of arbitrary degree, using the techniques of the preceding page.
By summing over such monomial variations, we may obtain arbitrary dependencies of path amplitudes on the path variables.
Again, if the dependency of the amplitudes on the path variables is of bounded degree, this yields a constant factor increase in the complexity of the resulting matrix expression, and ultimately yields a sum with uniform weights across all terms.

\subsection{Techniques for simulating postselection via error correction}

The results of this chapter have revolved about rather special conditions on the structure of a quadratic form expansion, under which the simulation of postselection is essentially possible generically due to restrictions on the graphical encoding performed by the preparation and entangling phases (in the case of ``\emph{yes}'' instances of different versions of the \GMPI) or of the default measurement angles (in the case of Clifford group unitaries).
In order to achieve a deeper understanding of how postselection-based procedures may be extended to measurement-based patterns, it would be useful to arrive at characterization of sequences of measurement observables on the one hand, and graphical encodings on the other, such that postselection of the $+1$ result of each measurement in sequence performs an isometry.

In~\cite{MCSB98}, necessary and conditions are laid out for conditions under which a CPTP map (such as a measurement operation) performs a reversible operation on a given subspace of the possible states of a physical system.
Considering the special case of single-qubit measurements on open graphical encodings, and supplemented by an explicit account of the evolution of the state-space under such measurements, may plausibly lead to a fuller understanding of \MPI.

\section{Review and open Problems}

In this chapter, we have considered the possibility of developing new techniques for developing quantum algorithms via the one-way measurement model by presenting techniques for transforming quadratic form expansions, and for synthesizing realizable procedures for unitary transformations by considering solvable instances of \MPI.
We have also presented quadratic form expansions themselves as interesting expressions in their own right, predicting that they are likely to recur as an important form in which unitary operations may be expressed, and suggesting in particular that they may be obtained by discretizing path integral formulations of propagators for physical systems.

The major open problem of this chapter which may be clearly addressed is whether a meaningful link between path integrals and quadratic form expansions is possible, and whether the \MPI\ could be solved for typical instances which would be generated in this way.
Affirmative answers for both would represent the most likely way in which the results of this chapter (or extensions of them for more general instances of the \MPI) could see concrete application.

From a purely theoretical point of view, a subject which has not been addressed in this chapter is the question of when a quadratic form expansion represents a unitary transformation.
Also of potential interest is what transformation is performed on the input subsystem by the postselection procedure described by a quadratic form expansion, when that operator is not a unitary transformation.
Answers to these questions may also shed light on the one-way measurement based model, as a potentially useful model of computation.

Finally, note that the construction of the one-way measurement patterns produced for Clifford group operations in Section~\ref{sec:synthClifford} does not hint at any obvious circuit structure.
In particular, while the depth of the measurement pattern is constant, an asymptotically optimal circuit for a Clifford group operation has size $O(n^2/\log(n))$, and therefore depth similar to $n/\log(n)$.
Is it possible to arrive at a semantic map for Clifford patterns in some sense similar to that which was done for the DKP and simplified RBB constructions in Chapter~\ref{Semantics for unitary one-way patterns}?
	\chapter[Further research directions]{
		Further research directions ()}

This thesis has been concerned with the one-way measurement model as a means of representing unitary transformations.
This has largely revolved around \emph{flows} on one hand, as a combinatorial structure arising out of a convenient representation of unitary circuits with certain ``simple'' gate sets, and generalizations of flows, which arise from links between flows and the techniques of the stabilizer formalism.
From this analysis, we have obtained efficient algorithms for recognizing certain one-way measurement patterns which perform unitary transformations, and translating them into unitary circuit models.
These techniques then proved useful for describing how we might be able to automatically obtain realizable procedures for unitary transformations which are specified by a certain type of operator formula which has recurred in the literature, and which are similar in form to a path integral formulation of propagators for a physical system.

The most natural (if perhaps ambitious) research direction from a mathematical standpoint is to characterize those one-way measurement patterns which perform unitary transformations.
This problem seems to demand an extension of the techniques of the stabilizer formalism.
The most likely necessary ingredient is one which is contained in the analysis of certain measurement patterns in~\cite{RBB03}: given an open graphical encoding stabilized by some subset of the Pauli group, we may obtain Hamiltonians which have a non-trivial kernel, in which the state of the system lies.
This allows us to consider the continuous group of unitary operators which also stabilize the state, which may be derived from continuous-time evolution of these Hamiltonians.
Judiciously chosen operators of this form can be used to transform measurement observables into Pauli operators whose measurement can be treated by the stabilizer formalism.
Connecting these rotation operations to the conditions of~\cite{MCSB98} for reversibility of quantum operations may yield an analysis of sequences of measurements, for which a characterization may be obtained.
Solving this problem will require a characterization of how stabilizer codes transform after successive information-preserving single-qubit measurements; this amounts to finding a more general class of codes than stabilizer codes, which is closed under such operations.

After the problem of characterizing measurement patterns performing unitary transformations, the next natural problem is to recognize instances of \MPI\ which may be ``interpolated'' to satisfy such a characterization (where we extend \MPI\ to admit arbitrary ``default measurement observables'' in place of the vector of default measurement angles).
A successful analysis of this problem would represent essentially complete low-level knowledge of the one-way measurement model.

While the above problems may have interesting practical dividends, the immediate question from the point of view of application is where one may expect to obtain instances of \MPI, in order to have something to apply these techniques to.
I have tried to motivate quadratic form expansions as a recurring form of operator expression in the quantum computing literature, but this does not describe a programme for actually obtaining such expressions.
The natural question is then whether or not it is possible to transform path integrals for physical systems of interest into quadratic form expansions, for which we may solve \MPI\ and thereby obtain efficient algorithms for simulating those systems.
At the same time, an exploration of a possible connection between quadratic form expansions and path integrals may itself yield interesting theoretical insights into the one-way measurement model, or into quantum computation in general.

\renewcommand{\bibname}{References}
\addcontentsline{toc}{chapter}{\textbf{\bibname}}

\bibliographystyle{plain}

\def\articleIndex{$\langle\textrm{\itshape article index}\rangle$}
\newcommand\BibliographyPreamble{%
	Many of the articles below are freely avaliable on the Cornell E-print Archive: such articles are indicated by the notation [\myurlstyle{arXiv:\articleIndex}] in their bibliographical entries.
	For such articles, an abstract and links to Postscript/PDF versions of the article can be found at the URL  [\myurlstyle{www.arXiv.org/abs/\articleIndex}] corresponding to the article's index.
	
	\ifElectronicVersion\ifFinalVersion
	After each bibliographical citation, the list of numbers indicate the page numbers on which the article is cited.
	\fi\fi
}

\bibliography{\BibliographyFile}

\nocite{*}

}

\newcommand		\DedicationFile			{part-0c-dedication}

\newcommand		\BibliographyFile			{thesis}

\newcommand\ThesisTitle{%
	Theory\ of\ measurement-based\ quantum\ computing%
}

\newcommand\ThesisAuthor{%
	Jonathan\ Robert\ Niel\ de\ Beaudrap%
}

\newcommand\ThesisProgramme{%
	Doctor\ of\ Philosophy
}

\newcommand\ThesisSubject{%
	Combinatorics~\&~Optimization
}

\newcommand\ThesisYear{%
	2008
}

\newif\ifElectronicVersion
\ElectronicVersiontrue

\newif\ifFinalVersion
\FinalVersionfalse

\documentclass[
					letterpaper, 11pt,
					titlepage,
 					oneside, 
 					final
					]{report}
\pdfoutput=1

\usepackage{color}
\usepackage{graphicx}
\usepackage{geometry}
\usepackage{framed}

\definecolor{RefColor}{rgb}{0,0,0}
\definecolor{CiteColor}{rgb}{0,0.4,0}
\definecolor{UrlColor}{rgb}{0,0,0.5}

\ifElectronicVersion
	\usepackage[letterpaper=true,pdftex,bookmarks,pagebackref,
			colorlinks=true,	
			linkcolor=RefColor,
			citecolor=CiteColor,
			urlcolor=UrlColor,
			plainpages=false, 
     			pdfpagelabels=true, 
      			pdftitle=\ThesisTitle,
      			pdfauthor=\ThesisAuthor
      			]%
			{hyperref}
\fi

\let\origdoublepage\cleardoublepage
\newcommand{\clearemptydoublepage}{%
  \clearpage{\pagestyle{empty}\origdoublepage}}
\let\cleardoublepage\clearemptydoublepage

\usepackage[margin=1.5em,font=small,labelfont=sc,labelsep=period]{caption}
\usepackage[mathscr]{euscript}
\usepackage{upgreek}
\usepackage{amsthm}
\usepackage{jrnbsymb}
\usepackage{textcomp}

\makeatletter
\def\texttilde{\raise.2ex\hbox{\m@th$\scriptstyle\sim$}}
\def\@nbsp{~}
\newcommand\myurlstyle[1]{{\small \ttfamily #1}}
\newcommand\arXiv[1][]{[\href{http://www.arXiv.org/abs/#1}{\myurlstyle{arXiv:#1}}]}
\ifElectronicVersion
  	\let\URL=\url
  	\renewcommand\url[1]{[\myurlstyle{\URL{#1}}]}
\else
	\newcommand\url[1]{\bgroup
		\expandafter\let\@nbsp\texttilde
			[\myurlstyle{#1}]%
	\egroup}
\fi

\makeatother

\makeatletter
\renewcommand\@chapapp{{\sc \chaptername}}

\let\@old@chapter\@chapter
\def\@chapter@w@subtitle[#1][#2(#3)#4]{%
	\def\@tempa{#3}%
	\ifx\@tempa\empty\def\@tempa{\@old@chapter[#1]{#2}}%
	\else\def\@tempa{\@old@chapter[#1]{#2 \\\Large\it #3}}\fi
	\@tempa}
\def\@chapter[#1]#2{%
	\@chapter@w@subtitle[#1][#2()]%
	\label{#1}%
	\bigcapsword}

\long\def\bigcapsword#1{%
	\ifcat\noexpand#1\noexpand\@endline \def\@tempa{\bigcapsword}\else
	\ifcat\noexpand#1\noexpand\@space \def\@tempa{\bigcapsword}\else
	\def\@tempa{\@bigcapsword#1}\fi\fi
	\@tempa}

\def\@bigcapsword#1#2 {{\uppercase{\LARGE #1}}\hskip-0.1ex\textsc{#2} }

\renewcommand\section{\@startsection {section}{1}{-7mm}%
                                   {-3.5ex \@plus -1ex \@minus -.2ex}%
                                   {2.3ex \@plus.2ex}%
                                   {\normalfont\Large\bfseries}}
\renewcommand\subsection{\@startsection{subsection}{2}{-7mm}%
                                     {-3.25ex\@plus -1ex \@minus -.2ex}%
                                     {1.5ex \@plus .2ex}%
                                     {\normalfont\large\bfseries}}
\renewcommand\subsubsection{\@startsection{subsubsection}{3}{-7mm}%
                                     {-3.25ex\@plus -1ex \@minus -.2ex}%
                                     {1.5ex \@plus .2ex}%
                                     {\normalfont\normalsize\bfseries}}

\makeatother

\def\todo[#1]{\noindent\texttt{[--- #1 ---]}}

\renewcommand\vec\mathbf
\newcommand\unit[1][e]{\vec{\hat{#1}}}

\def\Kdelta[#1][#2]{\delta_{#1,#2}}

\newcommand\sfrac[2]{\frac{\text{\raisebox{-0.5ex}{$\textstyle \smaller{#1}$}}}{\text{\raisebox{0.3ex}{$\textstyle \smaller{#2}$}}}}

\newcommand\ssstyle{\scriptscriptstyle}

\makeatletter
\newcommand\ordprod[1][<]{%
	\@ifnextchar_{\@ordprod[#1]}{\@ordprod[#1]_{}}}
\def\@ordprod[#1]_#2{%
	\@ifnextchar^{\@@ordprod[#1]_{#2}}{\@ordprod[#1]_{#2}^{}}}
\def\@@ordprod[#1]_#2^#3{%
	\Bigg.\smash{\prod\limits_{#2}^{#3}}\big.^{#1}\:\:}
\makeatother

\newlength\spacewidth
\newcommand\colouronline{\hspace{-\spacewidth}{\footnotesize [\emph{colour online}]}\hspace{\spacewidth}}

\usepackage[algochapter, linesnumbered, boxed, longend]{algorithm2e}

\renewcommand\thealgocf{\thechapter-\arabic{algocf}}

\newcommand\captionstyle{\small}

\setalcapskip{0ex}
\SetAlTitleFnt{captiontitlestyle}
\SetAlCapFnt{\captionstyle}

\Setnlsty{}{\textit\bgroup}{\egroup \:\:\:\:}

\makeatletter

\renewcommand{\algocf@caption@boxed}{\vskip 1.5em\box\algocf@capbox}%
\renewcommand{\algocf@captiontext}[2]{#1\algocf@typo. \AlCapFnt{}#2} 

\makeatother

\makeatletter
\newif\if@alg@oneline@
\@alg@oneline@false
\newcommand\oneline{\@alg@oneline@true}
\newif\if@alg@short@
\@alg@short@false
\newcommand\short{\@alg@short@true}

\newcommand\algline{\SetLine}

\SetCommentSty{textit}
\SetKwComment{Comment}{--- }{}

\SetKwInOut{Input}{Input}
\SetKwInOut{Output}{Output}

\SetKwFor{ForAll}{for\,all}{do}{endfor}
\SetKwFor{ForEach}{for\,each}{do}{endfor}
\SetKwFor{While}{while}{do}{endwhile}

\SetKwIF{@For}{alg@e}{alg@f}{for}{do}{}{}{}
\SetKwIF{@ForAll}{alg@g}{alg@g}{for\,all}{do}{}{}{}
\SetKwIF{@ForEach}{alg@h}{alg@i}{for\,each}{do}{}{}{}
\SetKwIF{@While}{alg@j}{alg@k}{while}{do}{}{}{}
\SetKwIF{Short@If}{alg@a}{alg@b}{if}{}{else if}{else}{}
\SetKwIF{Short@Then}{alg@c}{alg@d}{then}{}{else}{}

\long\def\INDENTIF#1\THEN#2#3{%
	\ifx #3\ELSE
		\def\@tempa##1{%
			\bgroup
				\SetNoline\dontprintsemicolon
				\uShort@If{#1}{\lShort@Then{}{#2; \\ \lElse {##1;}}}%
			\egroup\@alg@short@false}\else
		\def\@tempa{%
				\bgroup
					\SetNoline\dontprintsemicolon
					\uShort@If{#1}{\lShort@Then{}{#2;}}%
			\egroup\@alg@short@false #3}\fi
	\@tempa}

\long\def\INDENTELSEIF#1\THEN#2#3{%
	\ifx #3\ELSE
		\def\@tempa##1{%
			\bgroup
				\SetNoline\dontprintsemicolon
				\uElseShort@If{#1}{\lShort@Then{}{#2; \\ \lElse {##1;}}}%
			\egroup\@alg@short@false}\else
		\def\@tempa{%
				\bgroup
					\SetNoline\dontprintsemicolon
					\uElseShort@If{#1}{\lShort@Then{}{#2;}}%
			\egroup\@alg@short@false #3}\fi
	\@tempa}

\long\def\IF#1\THEN#2#3{%
	\if@alg@oneline@
		\def\@tempa{\lIf{#1}{#2}#3}%
		\ifx #3\ELSE\else\@alg@oneline@false\fi
	\else
		\if@alg@short@\def\@tempa{\INDENTIF{#1}\THEN{#2}#3}\else		
			\ifx #3\ELSE
				\def\@tempa{\uIf{#1}{\SetNoline #2}#3}\else
				\ifx #3\ELSEIF
					\def\@tempa{\uIf{#1}{\SetNoline #2}#3}\else
					\def\@tempa{\If{#1}{\SetNoline #2}#3}%
				\fi
			\fi
		\fi
	\fi
	\@tempa}
\long\def\ELSEIF#1\THEN#2#3{%
	\if@alg@oneline@
		\def\@tempa{\lElseIf{#1}{#2}}%
		\ifx #3\ELSE\else\@alg@oneline@false\fi
	\else
		\if@alg@short@\def\@tempa{\INDENTELSEIF{#1}\THEN{#2}#3}\else		
			\ifx #3\ELSE
				\def\@tempa{\uElseIf{#1}{\SetNoline #2}#3}\else
				\ifx #3\ELSEIF
					\def\@tempa{\uElseIf{#1}{\SetNoline #2}#3}\else
					\def\@tempa{\ElseIf{#1}{\SetNoline #2}#3}%
				\fi
			\fi
		\fi
	\fi
	\@tempa}

\long\def\ELSE#1{%
	\if@alg@oneline@
		\def\@tempa{\lElse{#1}}\else
		\def\@tempa{\Else{\SetNoline #1}}\fi
	\@alg@oneline@false\@tempa}

\long\def\FOR#1\DO#2\ENDFOR{%
	\if@alg@oneline@
		\def\@tempa{\lFor{#1}{#2}}\else
		\if@alg@short@
			\def\@tempa{\SetNoline\u@For{#1}{#2}}\else
			\def\@tempa{\For{#1}{\SetNoline #2}}\fi\fi
	\@alg@oneline@false\@tempa}

\long\def\FORALL#1\DO#2\ENDFOR{%
	\if@alg@oneline@
		\def\@tempa{\lForAll{#1}{#2}}\else
		\if@alg@short@
			\def\@tempa{\SetNoline\u@ForAll{#1}{#2}}\else
			\def\@tempa{\ForAll{#1}{\SetNoline #2}}\fi\fi
	\@alg@oneline@false\@alg@short@false\@tempa}

\long\def\FOREACH#1\DO#2\ENDFOR{%
	\if@alg@oneline@
		\def\@tempa{\lForEach{#1}{#2}}\else
		\if@alg@short@
			\def\@tempa{\SetNoline\u@ForEach{#1}{#2}}\else
			\def\@tempa{\ForEach{#1}{\SetNoline #2}}\fi\fi
	\@alg@oneline@false\@alg@short@false\@tempa}

\long\def\WHILE#1\DO#2\ENDWHILE{%
	\if@alg@oneline@
		\def\@tempa{\lWhile{#1}{#2}}\else
		\if@alg@short@
			\def\@tempa{\SetNoline\u@While{#1}{#2}}\else
			\def\@tempa{\While{#1}{\SetNoline #2}}\fi\fi
	\@alg@oneline@false\@alg@short@false\@tempa}

\long\def\REPEAT#1\UNTIL#2{%
	\if@alg@oneline@
		\def\@tempa{\lRepeat{#2}{#1}}\else
		\def\@tempa{\Repeat{#2}{\SetNoline #1}}\fi
	\@alg@oneline@false\@tempa}

\def\BEGIN{\@ifstar{\@BEGIN[\SetNoline]}{\@BEGIN[\SetLine]}}

\long\def\@BEGIN[#1]#2\END{\par\bgroup #1\Begin{\SetNoline #2 {}}\egroup}

\SetKw{Procedure}{procedure}
\SetKw{Subroutine}{subroutine}

\long\def\PROCEDURE#1:#2\BEGIN{%
	\ResetInOut{Output}%
	#2%
	\medskip\Procedure $#1$:
	\BEGIN}

\long\def\SUBROUTINE#1:#2\BEGIN{%
	\ResetInOut{Effect}%
	#2
	\medskip\Subroutine $#1$:
	\BEGIN}

\makeatother

\let\P\relax
\let\H\relax

\makeatletter
\newcommand\smaller{%
	\ifmmode\def\@tempa{\small@scalemath}\else\def\@tempa{\small@scaletext}\fi
	\@tempa}

\newcommand\larger[1][6/5]{%
	\ifmmode\def\@tempa{\@scalemath[#1]}%
	\else\def\@tempa{\@scaletext[#1]}\fi
	\@tempa}

\newdimen\scaled@size

\long\def\small@scalemath#1{\mathchoice{%
	\mbox{\small@scaletext{$\displaystyle #1$}}}{%
	\mbox{\small@scaletext{$\textstyle #1$}}}{%
	\mbox{\small@scaletext{$\scriptstyle #1$}}}{%
	\mbox{\small@scaletext{$\scriptscriptstyle #1$}}}}

\long\def\small@scaletext#1{%
	\scaled@size=\f@size pt\relax
	\ifdim\scaled@size<6pt\scaled@size=5pt\relax\else
		\ifdim\scaled@size<7pt\scaled@size=6pt\relax\else
			\ifdim\scaled@size<8pt\scaled@size=7pt\relax\else
				\ifdim\scaled@size<9pt\scaled@size=8pt\relax\else
					\ifdim\scaled@size<10pt\scaled@size=9pt\relax\else
						\multiply\scaled@size by 5\relax
						\divide\scaled@size by 6\relax
					\fi
				\fi
			\fi
		\fi
	\fi
	\edef\@tempa{\curr@fontshape}%
	\edef\@tempb{\strip@pt\scaled@size}%
	\bgroup
		\edef\font@name{%
			\csname\@tempa/\@tempb\endcsname}%
		\edef\f@size{\@tempb}%
	   \pickup@font
	   \font@name
  		\@@enc@update
		#1%
	\egroup}
\makeatother

\newcommand\ssfStyle[1]{\textmd{\textup{\textsf{#1}}}}
\newcommand\srmStyle[1]{\textup{\textrm{\smaller{#1}}}}
\newcommand\parrmStyle[1]{\textrm{\textmd{\textup{(#1\kern0.1ex)}}}}
\newcommand\paritStyle[1]{\textrm{\textmd{\textup{(\textit{#1}\kern0.1ex)}}}}
\newcommand\parbfitStyle[1]{\textrm{\textbf{\textup{(\textit{#1}\kern0.1ex)}}}}

\newcommand\complexityStyle[1]{\ssfStyle{#1}}
\newcommand\gateStyle[1]{\textsc{#1}}
\newcommand\problemStyle[1]{\srmStyle{\bf #1}}
\newcommand\pauliStyle[1]{\ssfStyle{#1}}

\NewNameCommandStyle{ComplexityClass}{\complexityStyle{#1}}
\NewNameCommandStyle{GateSet}{\gateStyle{#1}}
\NewNameCommandStyle{Problems}{\problemStyle{#1}}
\NewNameCommandStyle{Planes}{\pauliStyle{#1}}
\NewNameCommandStyle{rm}{\parrmStyle{#1}}
\NewNameCommandStyle{it}{\paritStyle{#1}}
\NewNameCommandStyle{bfit}{\parbfitStyle{#1}}
\NewNameCommandStyle{s}{\mathscr{#1}}

\let\L\relax

\namecommands{\cA, \cC, \cE, \cF, \cG, \cK, \cM, \cP, \cR, \cS, \cU, \cW}
\namecommands{\sI, \sT, \sG, \sD, \sS, \sO, \sP}
\namecommands{\sfC, \sfT, \sfF, \sfH, \sfJ, \sfP, \sfProb, \sfN, \sfQ, \sfR, \sfS, \sfz, \sfx, \sfy, \sfn, \sfm, \sfs}

\namecommands[bb]{\A}
\namecommands[sf]{\E, \L, \D, \U, \SU, \Herm}

\newcommand\ParityL{\complexityStyle{\raisebox{0.2ex}{$\oplus$}L}}

\namecommands[ComplexityClass]{\BQP, \BPP, \PP, \P, \NP, \co}
\namecommands[Problems]{\MPI, \GMPI}
\namecommands[GateSet]{\M, \Gates, \Phases, \Clifford, \Cluster, \Graph, \OneWay, \Jacz, \CoherCtrl, \Cond}
\namecommands[Planes]{\XY, \YZ, \XZ}

\NameMathOperators{\poly, \polylog, \Inp, \Outp, \odd, \Lower}

\namecommands{\iti, \itii, \itiii, \bfiti, \bfitii, \bfitiii, \bfitiv, \rma, \rmb, \rmc, \rmd, \rme, \rmf}

\let\oldtr\tr
\renewcommand\tr[1][]{\oldtr_{\text{\raisebox{-0.3ex}{$#1$}}}}

\newcommand\sV[1][\prec]{\mathscr{V}^{#1}}

\newcommand\deffunction[1]{\SetKwFunction{#1}{#1}}

\deffunction{MaximalStarVertex}
\deffunction{TestDependencies}
\deffunction{SemanticDKP}

\deffunction{FindFlow}
\deffunction{FindEflow}
\deffunction{GetLayers}
\deffunction{AddToLayer}

\deffunction{TestNaturalPreorder}
\deffunction{FindInfima}
\deffunction{InitializePath}

\deffunction{VisitVertex}
\deffunction{ObtainMaxFamilyPaths}
\deffunction{AlternPathSearch}

\SetKw{LET}{let}
\SetKw{SUCHTHAT}{such\,that}
\SetKw{NEXT}{skip to next}
\SetKw{RETURN}{return}

\SetKw{ANDALSO}{and}
\SetKw{ORELSE}{or}
\SetKw{NOT}{not}

\newcommand\ESCAPE{\textmd{escape this loop}}

\ResetInOut{Input}
\SetKwInOut{Data}{Data}
\SetKwInOut{Effect}{Effect}

\newcommand\INCR[1]{#1 \textup{\texttt{++}}\;\!}

\namecommands[tt]{
	\true, \false,
	\iter,
	\foundpath, 
	\fixed, \nil, \marked,
	\status, \untouched, \incomplete, \complete,
	\foundcircuit,
	\nextpath,
	\partialOrd,
	\xType, \zType, 	
	\Inf, \Path, \Dist,
	\flow, \maximalStar,
	\emptyLayer, \layer, \neighbors
 }

\namecommand[c]{\H}	
\newcommand\cH[1][2]{{\H_{#1}}}
\def\Prob{\sfProb}

\newcommand\axisX{\ssfStyle{X}}
\newcommand\axisY{\ssfStyle{Y}}
\newcommand\axisZ{\ssfStyle{Z}}

\newcommand\symdiff{\text{\raisebox{0.2ex}{\,$\scriptstyle \triangle$\,}}}

\newcommand\splus{\smaller+}
\newcommand\sminus{\smaller-}
\newcommand\spm{\smaller\pm}
\newcommand\smp{\smaller\mp}

\newcommand\psub[1]{\splus_{\ssstyle #1}}
\newcommand\msub[1]{\sminus_{\ssstyle #1}}
\newcommand\pmsub[1]{\spm_{\ssstyle #1}}
\newcommand\mpsub[1]{\smp_{\ssstyle #1}}

\newcommand\eg{\textit{e.\nolinebreak\hspace{-0.2ex}g.}\nolinebreak}
\newcommand\ie{\textit{i.\nolinebreak\hspace{-0.2ex}e.}\nolinebreak}
\newcommand\cf{\textit{c.\nolinebreak\hspace{-0.1ex}f.}\nolinebreak}
\newcommand\etc{\textit{e\nolinebreak\hspace{-0.1ex}t\nolinebreak\hspace{-0.1ex}c.}\nolinebreak}
\newcommand\etal{\textit{et~al.\nolinebreak}}

\newcommand\fn{\emph{\small f$^{\text{\raisebox{0.09ex}{\kern0.32ex\underline{\kern-0.2ex{n}}}}}$}}
\newcommand\fns{\emph{\small f$^{\text{\raisebox{0.09ex}{\kern0.32ex\underline{\kern-0.2ex{n\:\!\!s}}}}}$}}

\makeatletter
	\newcommand\partialto{%
		\@ifstar{\rightharpoonup}{\DOTSB\relbar\joinrel\rightharpoonup}}
\makeatother

\newcommand\arc{\protect\to*}
\newcommand\diwalk{{\,\arc\cdots\arc\,}}
\newcommand\inflRel{\vartriangleleft}
\newcommand\thru{\,--\:\:\!}

\newcommand\minus{\text{--}}

\newcommand\comp{^{\textsf{c}}}				

\newcommand\New[2]{\:\!\mathsf{N}_{#1}^{#2}\:\!}
\newcommand\Ent[2][]{\:\!\mathsf{E}^{\;\!#1}_{#2}\:\!}

\newcommand\Xcorr[2]{\:\!\mathsf{X}_{#1}^{#2}\:\!}
\newcommand\Zcorr[2]{\:\!\mathsf{Z}_{#1}^{#2}\:\!}
\newcommand\Shift[2]{\:\!\mathsf{S}_{\s[#1]}^{#2}\:\!}
\newcommand\Op[2]{\:\!\mathsf{Op}_{#1}^{#2}\:\!}

\makeatletter
\newcommand\Meas[3][]{\@Meas[#1]{#2}{#3}}

\def\@Meas[#1]#2#3{%
	\:\!%
	\def\@tempa{}
	\@ifempty{#1}{%
		\@tempa{\mathsf{M}_{#2}^{#3}}%
	}{%
		\@tempa{\mathsf{M}_{#2}^{(#1, #3)}}%
}\:\!}

\newcommand\s[1][]{%
	\mathsf{s}\@ifempty{#1}{}{\smaller{[#1]}}}

\let\ringaccent\r
\def\r{\@ifstar{\ringaccent}{\meas@result}}
\newcommand\meas@result[1][]{%
	\mathsf{r}\@ifempty{#1}{}{{\mathchoice
		{\text{\footnotesize $[#1]$}}%
		{\text{\small $[#1]$}}%
		{\scriptscriptstyle [#1]}%
		{\text{\tiny $[#1]$}}%
}}}
\makeatother

\newcommand\ctrl{\text{\raisebox{0.3ex}{\small $\wedge$}}\mspace{-1.5mu}}	
\newcommand\cZ{\ctrl Z}																		
\newcommand\cX{\ctrl X}																		

\makeatletter
\newcommand\@pion[2][]{
	\mathchoice%
		{{\text{\small\:\!\raisebox{0.3ex}{\raisebox{0.5ex}{$#1\pi$}$\!\!\!\;/\!_{#2}$}}}}
		{{\text{\small\:\!\raisebox{0.3ex}{\raisebox{0.5ex}{$#1\pi$}$\!\!\!\;/\!_{#2}$}}}}
		{{\scriptscriptstyle \text{\raisebox{0.2ex}{\raisebox{0.5ex}{$#1\pi$}$\!\!\!\:/\!_{#2}$}}}}
		{\big.\text{\raisebox{0.2ex}{\raisebox{0.5ex}{$#1\pi$}$\!\!\!\:/\!_{#2}$}}}}

\newcommand\pion{\@pion[]}
\newcommand\mpion{\@pion[\text{--}\:\!]}
\makeatother

\newcommand\paren[1]		{\left( #1 \right)}		     						 	
\newcommand\sqparen[1]	{\left[ #1 \right]}		   					  		
\newcommand\ens[1]		{\left\{#1\right\}}										
\newcommand\ket[1]		{\left\lvert#1\right\rangle\mspace{-1.5mu}}		
\newcommand\bra[1]		{\mspace{-1.5mu}\left\langle#1\right\rvert}		
\newcommand\bracket[2]	{\left\langle#1\mspace{-1mu}\left|\mspace{1mu}#2\right\rangle\right.}	
\newcommand\card[1]		{\left\lvert #1 \right\rvert}							
\newcommand\norm[1]		{\left\| #1 \right\|}
\newcommand\anglebr[1]	{\left\langle #1 \right\rangle}
\newcommand\gen\anglebr

\newcommand\ngbh[2][]{N_{\ssstyle #1}(#2)}

\let\abs\card
\newcommand\supp[1]		{^{(#1)}}

\newcommand\intrans{^{\,-\!\top}}		

\makeatletter
\def\pseu[#1]{%
	\mspace{-5mu}%
	\mathchoice
		{\@pseu\@iden\scriptstyle[#1:]}%
		{\@pseu{\big.\smash}\scriptstyle[#1:]}%
		{\@pseu{\big.\smash}\scriptscriptstyle[#1:]}%
		{\@pseu{\big.\smash}{\scriptscriptstyle\smaller}[#1:]}%
	\mspace{2mu}}
\def\@pseu#1#2[#3:#4]{%
	\def\@tempa{#4}
	\ifx\@tempa\@empty
		\left[\mspace{2mu}#1{\text{\raisebox{0.2ex}{$#2{#3}$}}}\mspace{2mu}\right]%
	\else
		\@@pseu{#1}{#2}[#3:#4]%
	\fi
}
\def\@@pseu#1#2[#3:#4/#5:]{%
	\def\@tempa{#3}%
	\ifx\@tempa\@empty\else
		\def\@tempa{\raisebox{0.2ex}{$#2{#3}$}\,}%
	\fi
	\left[#1{\@tempa\tfrac{#2{#4}}{#2{#5}}}\right]%
}
\makeatother

\newcommand\exprbr[1]	{\langle\mspace{-5mu}\langle \, #1 \, \rangle\mspace{-5mu}\rangle}

\makeatletter
	\def\matrix@check{\@gobble}
\makeatother

\renewenvironment{cases}[1][c]{%
 	\left\{ \begin{array}{#1@{\quad}l}%
}{%
	\end{array} \right\}%
}

\makeatletter

\let\c@definition\relax

\makeatother

\makeatletter

\newcommand\deftheoremstyle[4]{%
	\@oparg{\@defthmstyle{#1}{#2}{#3}{#4}}[]}

\def\@defthmstyle#1#2#3#4[#5]#6#7{%
	\@xp\def\csname th@#1@numwithin\endcsname##1##2{#7}%
	\@xp\def\csname th@#1@numdef\endcsname[##1]##2{%
		\ifx\relax##1\relax
			\toks@{#6{##2}}%
		\else
			\toks@\@xp{\csname the##1\endcsname}%
			\xdef\@tempb{\csname th@#1@numwithin\endcsname{\the\toks@}{\@nx#6{##2}}}%
			\@xp\toks@\@xp{\@tempb}%
		\fi
		\def\@tempa{\@xp\def\csname the##2\endcsname}%
		\@xp\@tempa\@xp{\the\toks@}}%
	\@@defthmstyle{#1}{#2}{#3}{#4}[#5]}

\newtoks\thm@prebody
\newtoks\thm@postbody

\def\@@defthmstyle#1#2#3#4[#5]#6#7#8#9{%
	\toks@{#3}%
	\def\@tempa{#6}\ifx\space\@tempa
		\toks@\@xp{\the\toks@ \thm@headsep\fontdimen\tw@\font\relax}%
	\else
		\def\@tempb{\newline}%
		\ifx\@tempb\@tempa
			\toks@\@xp{%
				\the\toks@ \thm@headsep\z@skip
				\def\thmheadnl{~\newline}
			}%
		\else
			\toks@\@xp{\the\toks@ \thm@headsep#6\relax}%
		\fi
	\fi
	\begingroup
		\thm@space@setup
		\@defaultunits\@tempskipa#2\thm@preskip\relax\@nnil
		\@defaultunits\@tempskipb#9\thm@postskip\relax\@nnil
		\xdef\@gtempa{\thm@preskip\the\@tempskipa
			\thm@postskip\the\@tempskipb\relax}%
	\endgroup
	\@temptokena\@xp{\@gtempa
		\thm@headfont{#4}\thm@headpunct{}\thm@prebody{#7}\thm@postbody{#8}%
	}%
	\@ifempty{#5}{%
		\let\thmhead\thmhead@plain
		}{%
		\@namedef{thmhead@#1}##1##2##3{#5}%
		\@temptokena\@xp{\the\@temptokena
			\@xp\let\@xp\thmhead\csname thmhead@#1\endcsname}%
	}%
	\@xp\xdef\csname th@#1\endcsname{%
		\the\toks@ \the\@temptokena
		\advance\linewidth -\leftmargin%
 		\advance\@totalleftmargin \leftmargin%
		\parshape \@ne \@totalleftmargin \linewidth
	}%
}

\def\@xthm#1#2[#3]{%
  \ifx\relax#3\relax
    \newcounter{#1}%
  \else
    \newcounter{#1}[#3]%
  \fi
  \csname th@\the\thm@style @numdef\endcsname[#3]{#1}%
  \toks@{#2}%
  \@xp\xdef\csname#1\endcsname{%
    \@nx\@thm{%
      \let\@nx\thm@swap
        \if S\thm@swap\@nx\@firstoftwo\else\@nx\@gobble\fi
      \@xp\@nx\csname th@\the\thm@style\endcsname}%
      {#1}{\the\toks@}}%
}

\def\@endtheorem{\the\thm@postbody\@endpefalse\endtrivlist}

\def\@begintheorem#1#2[#3]{%
	\def\reserved@a{#1}\def\reserved@b{\adhoclabel}%
	\ifx\reserved@a\reserved@b
      \@ifempty{#2}{\let\thmnumber\@gobble}{\let\thmnumber\@iden}%
      \@ifempty{#3}{\@jrnbthm@err{Name argument missing for named-theorem environment}}%
      \let\thmnote\@gobble
      \deferred@thm@head{\the\thm@headfont\thmhead{#3}{{#2}}{}}%
   \else
      \@ifempty{#1}{\let\thmname\@gobble}{\let\thmname\@iden}%
      \@ifempty{#2}{\let\thmnumber\@gobble}{\let\thmnumber\@iden}%
      \@ifempty{#3}{\let\thmnote\@gobble}{\let\thmnote\@iden}%
      \deferred@thm@head{\the\thm@headfont\thmhead{#1}{{#2}}{{#3}}}%
   \fi
	\the\thm@headpunct
	\hskip\thm@headsep
	\thmheadnl 
	\the\thm@prebody 
	\ignorespaces}

\def\deferred@thm@head#1{%
	\if@inlabel \indent \par \fi 
	\if@nobreak
		\adjust@parskip@nobreak
	\else
		\addpenalty\@beginparpenalty
		\addvspace\@topsep
		\addvspace{-\parskip}%
	\fi
	\global\@inlabeltrue
	\everypar\dth@everypar
 	\@item[\normalfont#1]
 	\global\@newlistfalse%
	\ignorespaces
}

\newlength\theoremmargin
\newcommand\newthesistheoremstyle[4]{%
	\deftheoremstyle{#1}%
		{#2}{%
 			\setlength\topsep{0mm}%
			\setlength\partopsep{0mm}%
			\setlength\listparindent{\parindent}%
			\setlength\leftmargin{\theoremmargin}%
 			\setlength\labelwidth{\leftmargin}%
			\addtolength\labelwidth{-\labelsep}%
		}%
		{\bfseries}
		[##1\thmnumber{~##2}\thmnote{\textmd{~##3}}\:\!.]%
		{#3}%
		{##1-##2}%
      { }%
      {#4}%
      {}%
      {#2}}

\newthesistheoremstyle{definition}{1em}{\arabic}{\rmfamily\mdseries\upshape}
\newthesistheoremstyle{theorem}{1em}{\arabic}{\rmfamily\mdseries\itshape}
\newthesistheoremstyle{example}{1em}{\arabic}{\rmfamily\mdseries\upshape}

\makeatletter
\deftheoremstyle{proof}
		{1em}{%
 			\setlength\topsep{0mm}%
			\setlength\partopsep{0mm}%
			\setlength\listparindent{\parindent}%
			\setlength\leftmargin{\theoremmargin}%
 			\setlength\labelwidth{\leftmargin}%
			\addtolength\labelwidth{-\labelsep}%
		}
		{\itshape}%
		[#1\thmnote{~#3}~---]
		{}
		{}
		{\newline}
		{\rmfamily\mdseries\upshape}
		{\hfill\qed}
		{1em}

\makeatother

\deftheoremstyle{problem}%
	{1em}{%
 			\setlength\topsep{0mm}%
			\setlength\partopsep{0mm}%
			\setlength\listparindent{\parindent}%
			\setlength\leftmargin{\theoremmargin}%
 			\setlength\labelwidth{\leftmargin}%
			\addtolength\labelwidth{-\labelsep}%
	}%
	{\bfseries}%
	[\thmnote{#3~}#1:]%
	{}
	{}
	{1.5ex}%
	{\rmfamily\mdseries\upshape}%
	{}%
	{1em}

\theoremstyle{definition}
\newtheorem{definition}{Definition}[chapter]
\newtheorem*{notation}{Notation}
\newtheorem*{notation*}{\adhoclabel}

\theoremstyle{theorem}
\newtheorem{theorem}{Theorem}[chapter]
\newtheorem*{theorem*}{\adhoclabel}
\newtheorem{lemma}[theorem]{Lemma}
\newtheorem*{lemma*}{\adhoclabel}

\newtheorem{corollary}{Corollary}[theorem]

\theoremstyle{example}
\newtheorem{example}{Example}%
\newtheorem{observation}{Observation}%
\newtheorem*{observation*}{Observation}%

\makeatletter
\@addtoreset{observation}{theorem}
\makeatother

\def\qed{$\blacksquare$}

\theoremstyle{proof}
\newtheorem*{proof}{Proof}

\theoremstyle{problem}
\newtheorem{problem}{Problem}

\makeatletter
\newcommand\textbalance[2][]{%
	\@ifnextchar[{\@textbalance[#1]{#2}}{\@textbalance[#1]{#2}[]}}
\def\@textbalance[#1]#2[#3]{\phantom{#3}#1#2#3\phantom{#1}}
\makeatother

\makeatletter
\newcounter{inlinum}
\renewcommand\theinlinum{\textup{(\alph{inlinum})}}
\newenvironment{inlinum}{%
	\bgroup
		\def\item{\refstepcounter{inlinum}\theinlinum~\ignorespaces}%
		\catcode`\^^M=10%
		\setcounter{inlinum}{0}%
		\ignorespaces
}{%
	\egroup\aftergroup\@after@inlinum
}
\def\@after@inlinum{%
	\catcode`\^^M=10%
	\gdef\@tempa{\catcode`\^^M=5}%
	\expandafter\@tempa\ignorespaces}%
\makeatother

\newcounter{romanum}
\renewcommand\theromanum{\paritStyle{\roman{romanum}}}
\newenvironment{romanum}
          {\setcounter{romanum}{0}
           \begin{list}{\theromanum}{%
					\usecounter{romanum}%
					\setlength\labelsep{.5em}%
					\setlength\listparindent{\parindent}
					\settowidth\labelwidth{\paritStyle{viii}}%
					\setlength\leftmargin{\labelwidth}%
					\addtolength\leftmargin{\labelsep}%
					\setlength\itemindent{0ex}
					\setlength\itemsep{4pt}%
				}}{\end{list}}

\newcounter{alphanum}
\newenvironment{alphanum}{%
	\vspace{-1.4ex}%
	\begin{list}{\textbf{\upshape (\alph{alphanum})}}{%
	\usecounter{alphanum}%
	\setlength\labelwidth{2em}%
	\setlength\labelsep{1.5ex}%
	\setlength\parsep{0.5ex plus0.2ex minus0.1ex}%
	\setlength\itemsep{0.5ex}}
	\setlength\leftmargin{3em}
	\setlength\itemindent{0ex}}%
	{\end{list}\vspace{-1.4ex}}

\begin{document}

%



\pagenumbering{roman}

\begin{titlepage}
        \begin{center}
        \vspace*{1.0cm}

        \Huge
        \textbf{\ThesisTitle}

        \vspace*{1.0cm}

        \normalsize
        by \\

        \vspace*{1.0cm}

        \Large
        \ThesisAuthor \\

        \vspace*{3.0cm}

        \normalsize
        A thesis \\
        presented to the University of Waterloo \\ 
        in fulfillment of the \\
        thesis requirement for the degree of \\
        \ThesisProgramme \\
        in \\
        \ThesisSubject \\

        \vspace*{2.0cm}

        Waterloo, Ontario, Canada, \ThesisYear \\

        \vspace*{1.0cm}

        \copyright\ \ThesisAuthor\ \ThesisYear\\
        \end{center}
\end{titlepage}

\pagestyle{plain}
\setcounter{page}{2}

\ifFinalVersion\ifElectronicVersion
  \noindent
  I hereby declare that I am the sole author of this thesis.  This is a true copy of the thesis, including any required final revisions, as accepted by my examiners.

  \bigskip
  
  \noindent
  I understand that my thesis may be made electronically available to the public.
\else
 \noindent
 I hereby declare that I am the sole author of this thesis.

 \smallskip

 \noindent
 I authorize the University of Waterloo to lend this thesis to other
 institutions or individuals for the purpose of scholarly research.

 \bigskip

 \noindent
 I further authorize the University of Waterloo to reproduce this
 thesis by photocopying or by other means, in total or in part,
 at the request of other institutions or individuals for the purpose
 of scholarly research.
\fi
\fi
\newpage


\begin{center}\textbf{Abstract}\end{center}
In the study of quantum computation, data is represented in terms of linear operators which form a generalized model of probability, and computations are most commonly described as products of unitary transformations, which are the transformations which preserve the quality of the data in a precise sense.
This naturally leads to \emph{unitary circuit models}, which are models of computation in which unitary operators are expressed as a product of ``elementary'' unitary transformations.
However, unitary transformations can also be effected as a composition of operations which are not all unitary themselves: the \emph{one-way measurement model} is one such model of quantum computation.

In this thesis, we examine the relationship between representations of unitary operators and decompositions of those operators in the one-way measurement model.
In particular, we consider different circumstances under which a procedure in the one-way measurement model can be described as simulating a unitary circuit, by considering the combinatorial structures which are common to unitary circuits and two simple constructions of one-way based procedures.
These structures lead to a characterization of the one-way measurement patterns which arise from these constructions, which can then be related to efficiently testable properties of graphs.
We also consider how these characterizations provide automatic techniques for obtaining complete measurement-based decompositions, from unitary transformations which are specified by operator expressions bearing a formal resemblance to path integrals.
These techniques are presented as a possible means to devise new algorithms in the one-way measurement model, independently of algorithms in the unitary circuit model.
\newpage


\ifFinalVersion
\begin{center}\textbf{Acknowledgements}\end{center}

	Many people have helped to contribute to the writing of this thesis.
	The following deserve special recognition:

	\begin{description}
	\item[Elham Kashefi,]
		for introducing me to the one-way measurement model and to the questions about it which she was interested in; for the many stimulating discussions in Waterloo, Oxford, Grenoble, and Edinburgh; and for being a patient hostess.

	\item[Robert Raussendorf,]
		for the handful of discussions in Waterloo that we had about the model which he helped to define.
		His feedback helped me to better understand the differences in perspectives about how to present the problems which Elham and I were interested in, at a time when I was unsure of how to communicate them to the wider community.

	\item[Michele Mosca and Ashwin Nayak,]
		for being very liberal supervisors, and highly supportive of international voyages to conferences and collaborators alike.

	\item[Anne Broadbent and Joe Fitsimons,]
		for insights which helped to enrich my perspective of flows and measurement-based quantum computation.

	\item[Martin Pei]
		for being, quite simply, an excellent collaborator.

	\item[John Watrous and Daniel Gottesman,]
		for being the instructors wose provided me with (more than anyone else) the \emph{conceptual} framework necessary to study and describe the subject of this thesis.

	\item[Marg Feeney,]
		for being very patient in reminding me of deadlines and requirements, and helpful in answering my questions about administrivia; she is the most excellent administrative staff member that I've ever encountered.
	\end{description}

	Finally, I'd like to thank Donny Cheung, Dave Evans, Alma Juarez, Cory Kasper, Ben Korvermaker, Mike LaCroix, Eva M$\upiota \upchi \upalpha \uplambda$, Judy Nowak, Mike Patterson, Carlos P\'erez-Delgado, Joel Reardon, Bill Rosgen, Lana Sheridan, Craig Sloss, Devin Smith, Jason Skomorowski, Stephanus du Toit, Rob Warren, Greg Zaverucha, and Joanna Zimbiecka for helping to make my time as a student in Waterloo as pleasant as it was.

\newpage
\fi

\tableofcontents
\newpage

\addcontentsline{toc}{chapter}{\textbf{List of Figures}}
\listoffigures

\addcontentsline{toc}{chapter}{\textbf{List of Algorithms}}
\listofalgorithms
\newpage

\newpage

	\definecolor{RefColor}{rgb}{0.5,0,0}

\bgroup
\addcontentsline{toc}{chapter}{\textbf{Preamble}}

\makeatletter
\let\@chapter\@old@chapter
\makeatother

\renewcommand\thesection{\arabic{section}}

\chapter*{Preamble \\\Large\it and overview of the thesis}


\section{Introduction}
\bigcapsword

Quantum computation developed as an attempt to respond to two questions:
\begin{enumerate}
\item
	How can we build a computing device to efficiently simulate physical systems?

\item
	What class of problems can be efficiently solved by physical systems, when we regard them as performing computations?
\end{enumerate}
These questions have helped to spur research, continuing in the tradition of Landauer~\cite{Landauer61,Landauer91}, to study information and computations in physical terms.

Richard Feynman~\cite{Feynman82} put forward the thesis that a model of computation founded on quantum mechanics was necessary and fruitful, in order to more efficiently solve problems such as the simulation of quantum mechanical systems.
Because of the interesting interplay between wave mechanics and discrete event detection at work in quantum mechanics, it seems to defy efficient simulation by ``classical'' models of computation, such as Turing machines~\cite{Turing36}.
Feynman proposed a model for a computer which exploited quantum mechanical effects as an antidote to this situation.
In 1985, David Deutsch proposed a programmable model of universal computation~\cite{Deutsch85}, and posited that the difficulty of efficiently simulating quantum mechanics with classical models of computation was a sign of a fundamental computational advantage of models founded on quantum mechanical principles.
This prompted the study of quantum/classical computational separations~\cite{BV93,Simon97}, which eventually led to Peter Shor's discovery of efficient algorithms for the discrete logarithm and integer factoring problems in such a model~\cite{Shor94}.
As these problems are widely regarded as being ``difficult'' to solve by classical models of computation, the discovery of these algorithms sparked significant interest in the question of how a quantum computer might be physically realized.

Due to the link with Schr\"odinger evolution in quantum mechanics, quantum computation is most often (and usually conveniently) described in terms of \emph{unitary transformations}, which are transformations of those configurations of the system which preserve the property of yielding point-mass probability distributions for a suitably chosen measurement operation.
The ``standard'' model of quantum computation is therefore that of \emph{unitary circuits} (described in Section~\ref{sec:unitaryCircuits}), which expresses unitary transformations by products of other unitary transformations, reducing the evolution of quantum states to some set of elementary operations which are considered to be likely to be physically realizable.

In 2001, Robert Rauss\-endorf and Hans Briegel advanced a model~\cite{RB01} for a quantum computer by performing controlled measurement operations performed in a fixed state of a regular spin network.
This gave rise to an alternative model of quantum computation, called the \emph{one-way measurement model} (described in Chapter~\ref{One-way measurement-based computation}).
This model contrasted sharply with the picture of quantum computation via unitary circuits in that measurement operations were used to effectively simulate unitary transformation on a fixed initial state, whereas measurements are often portrayed as being prototypical examples of non-unitary operations.
While it is not exceedingly difficult (having the results of~\cite{RB01} in hand) to describe simple principles by which universal quantum computation may be driven by measurements on a spin network, this contrast of perspectives on quantum computation between transformation by unitaries, and transformation by measurements, makes it difficult to interpret measurement based procedures as ``first class'' descriptions of unitary transformations of quantum states.

In this thesis, we explore the relationship between the one-way measurement model and unitary transformations, with the aim of exploring, on the one hand, techniques to recognize one-way measurement procedures which represent unitary transformations and translate them into the circuit model, and on the other hand to identify when expressions for unitary transformations may be easily translated to a one-way measurement procedure.
A secondary, underlying theme is the subject of combinatorial structures which unify different representations of unitary transformations: throughout the thesis, it is by identifying and recognizing such combinatorial structures that the task of translating between different representations of unitary transformations (such as unitary circuits and one-way measurement-based procedures) becomes tractable.

\section{Structure of the thesis}

This thesis contains five chapters, discounting the present one.

The first, introductory chapter, introduces the mathematical tools for describing quantum computation, including three individual models of importance:
\begin{enumerate}
\item
	\emph{Unitary circuit models} --- more properly speaking, this is a family of computational models, in which elementary unitary operations are composed to produce unitary operations of greater complexity.
	In practise, however, almost all models of unitary circuits considered are essentially equivalent to a ``standard'' choice of elementary gates, which we introduce here.
	We also define a non-standard, but closely related, unitary circuit model which will come to dominate the analysis of unitary circuits throughout the thesis; and a non-standard representation of unitary circuits which also plays an important role in the analysis of the later chapters.

\item
	\emph{Classically controlled unitary circuits} --- an elaboration of unitary circuit models by explicitly representing a classical control, we present these models essentially in order to concretely describe the way in which they are usually translated back into ``fully quantum'' unitary circuits.

\item
	\emph{Clifford circuits} --- a model of computation with restricted power, the limited operations of this model nevertheless provides useful analytical tools for describing transformations of a special class of states by measurements.
\end{enumerate}

The second chapter introduces measurement-based computation, with emphasis on the \emph{one-way measurement-based model} in particular.
This model is the main subject of the thesis.
In particular, we describe how universal quantum computation is effected in both the original \emph{cluster-state} model, and the less restrictive model based on \emph{open graphical encodings}, presenting two standard constructions for measurement-based computation to do so.

The third chapter is concerned with algorithms to recognize a certain class of one-way measurement-based procedures that perform unitary transformations, and to translate these procedures into unitary circuits.
This is done by the graph-theoretical characterization and examination of a combinatorial structure arising out of the non-standard circuit representation introduced in Chapter~\ref{Fundamentals of quantum computation}.
Using this combinatorial structure, we may reduce the problem of recognizing and translating such patterns to previously solved problems in graph theory.
We then consider how the structure can be generalized to extend the scope of these recognition and translation algorithms.
We also consider how combinatorial structures underlying one-way measurement-based computations may allow the certification that they perform unitary transformations, even in the absence of efficient algorithms to translate them to unitary circuits.

The fourth chapter takes advantage of the certifying structures described in the preceding chapter, to suggest techniques to automatically synthesize decompositions of unitary transformations (as unitary circuits, or as measurement-based procedures) from concise matrix expressions for such transformations.
These techniques are presented as a stepping stone towards devising ``native'' idioms for solving problems in the one-way measurement model, which would contribute to the usefulness of the one-way measurement model as a useful model of computation, in addition to being a potentially useful proposal for implementation.
At the same time, we consider the matrix expressions themselves, in particular noting their recurring role in quantum computation, as well as the possibility that they may naturally arise from representations of propagators of for physical systems as path integrals.

In the final chapter, we consider natural research directions, which essentially consist of the ways that the analysis of Chapter~\ref{Measurement Pattern Interpolation} may be most naturally extended.

\section{Contributions presented in this thesis}

The following is a list of the contributions made by myself, as presented in this thesis, to the understanding of one-way measurement-based computation, the context in which these results arose, and the subsequent work in the literature which followed.
The description of where the results have been previously published, in each case, may be found in the pre-amble of the corresponding chapters.

\subsection*{Graph-theoretic descriptions of flows, and flow-finding algorithms}

Sections~\ref{sec:graphth-charn} and~\ref{sec:flowAlgs} present work by myself (in the case of~\ref{sec:extreme}, with Martin Pei) on ``flows'', which may be interpreted as a certificate for the unitarity of a one-way measurement-based procedure~\cite{DK06}, arising from a particular construction of measurement-based procedures (which is described in Section~\ref{sec:DKPstandardization} as the \emph{simplified DKP construction}).

The characterizations of Section~\ref{sec:graphth-charn} led to the first efficient algorithm for determining whether a geometry has a flow, albeit conditioned on the input and output subsystems of the corresponding measurement-based procedure being of the same size (corresponding to the case of a unitary bijection).
These algorithms were made possible by an identification of a uniqueness result in~\ref{sec:unique}.
The result of~\ref{sec:extreme} then helps to further bound the running time.

Subsequently, a faster and more general algorithm for finding flows unconditionally has been presented in~\cite{MP08}.
This algorithm is likely to be optimal; bounds on its running time are also improved by the result of Section~\ref{sec:extremal}.
An examination of the algorithm of~\cite{MP08} imply a second graph-theoretic characterization of flows, which I briefly describe in Section~\ref{sec:improvedFlowAlgs} in relation to other work in Chapter~\ref{One-way measurement-based computation}.

\subsection*{Translation of one-way measurement procedures to unitary circuits}

One application of flows as a certificate of unitarity is to obtain a unitary circuit which is equivalent~\cite{DK06}, both in complexity (in a sense described in Section~\ref{sec:semanticsReduce}) and in the transformation which it performs.
However, an explicit description of such an algorithm has not yet appeared in the literature, and involves a non-trivial amount of combinatorial analysis.

Section~\ref{sec:semanticDKP} presents an explicit, efficient algorithm to translate one-way measurement procedures with flow certificates to unitary circuits in this way.
In particular, Section~\ref{sec:dependencies} characterize the dependency relations of measurement patterns with flows which have been fully standardized: this completes the work begin in~\cite{BK2007} on the dependency relations in measurement patterns, which is pertinent to the topic of depth complexity.
Section~\ref{sec:starDecomposition} then presents a new decomposition result for geometries with flows, which is pertinent to the faster flow-finding algorithm of~\cite{MP08}.
These results allow for the verification of the correctness of the measurement dependencies in the measurement procedure, and then construction an appropriate circuit.

\subsection*{Extension of flows}

In Sections~\ref{sec:eflow}, I present an extension of the ``flow'' certificate, which I call ``extended flows''.
These extended flows may be taken as a certificate of unitarity for a different construction for one-way procedures (which is described in Section~\ref{sec:RBB} as the \emph{simplified RBB construction}).
Exploiting the description given in Section~\ref{sec:improvedFlowAlgs} of the algorithm of~\cite{MP08}, I present an efficient algorithm to detect extended flows, with essentially no change to the running time of~\cite{MP08}.
I also present a sketch of how the efficient mapping to unitary circuits, which was presented for flows, may be similarly extended.

\subsection*{Measurement Pattern Interpolation}

In Chapter~\ref{Measurement Pattern Interpolation}, as part of joint work with Elham Kashefi, Vincent Danos, and Martin Roetteler, I present the definition of a quadratic form expansion.
Similar objects have occurred previously in the literature; ours was the first to explicitly connect them with measurement-based computation, although a link through graph-states~\cite{Schl04} was previously known.
The contribution of Chapter~\ref{Measurement Pattern Interpolation} as a whole is to outline how algorithms may be naturally compiled for the one-way measurement model, and to propose that techniques along these lines may be applicable to the problem of approximating path integrals for physical propagators.
There does not appear to be prior work on compiling for the one-way measurement model, except for the standard constructions via the circuit model.

The algorithms presented in Section~\ref{sec:interpolateGflows} are essentially applications of a principle forwarded by Drs. Kashefi and Danos: that for any certificate that a measurement pattern performs a unitary transformation which is based on the stabilizer formalism, there exists an efficient algorithm to produce such a measurement pattern from a description of how it acts conditioned on one particular sequence of measurement results.
The contribution of Section~\ref{sec:interpolateGflows} consists of the analysis of applying this principle to each ``instance'', including making the necessary connections with the implementation of diagonal operations presented in~\cite{BB06}, and with the concept of ``fractional weight'' flows.

Section~\ref{sec:synthClifford} presents a refinement of the formulae for Clifford group operations presented in~\cite{DM03}, which allows them to be presented as quadratic form expansions.
Having found this connection, I present an algorithm to obtain a measurement pattern, essentially via an application of the stabilizer formalism~\cite{GotPhD}.
This construction of one-way measurement procedures for the Clifford group elements is shown to be minimal with respect to the reduction techniques of~\cite{HEB04}, and is proven to run in time comparable to the algorithm of~\cite{AG04} for producing unitary circuits for Clifford group (and strictly faster than an algorithm using~\cite{AG04} as a subroutine to produce minimal one-way measurement procedures for the Clifford group).

\section{Conventions}

\subsection*{Notation and Pseudocode}

Throughout the thesis, Dirac (``bra-ket'') notation is used, along with conventional notations for Hilbert spaces: see \eg~\cite{NC00} for an introduction to how it is used conventionally.
However, owing to the regular discussion of operations in terms of the measurement-based model, transformations of quantum states are often represented as completely positive trace preserving linear superoperators (or \emph{CPTP maps}, described on page~\pageref{dfn:CPTP}) rather than as unitary matrices acting on column vectors.
Other notations pertaining to linear operations are a minor variation on that of~\cite{Watrous07}.

Special notations for operators, states, and other variables are usually introduced where they are used.
However, it may be useful to note the following conventions, which may help to avoid possible ambiguities of interpretation:
\begin{itemize}
\item
	The constant $i$ in italics always represents a square root of $-1$ in the complex numbers $\C$\,.
	In particular, it is never used as a variable to  iterate over the terms of a sum, or the vertices of a path.
	(For the latter two purposes, variables such as $h, j, \ell, \ldots$ are most often used.) 

\item
	The constant $\e$ in an upright typeface always represents Euler's constant, the base of natural logarithms.
	(This should be contrasted with an italic $e$, with may represent a generic element of a finite ground set $V$.)

\item
	The identity operator (or superoperator) for a physical system is always represented as $\idop$\,.
	(This should be contrasted with the variable $I$, which always represents a subset of some larger ground-set $V$.)

\item
	Superoperators are always presented as capital sans-serif letters such as \,$\Ent{}{}$, or fraktur letters such as $\mathfrak P$.
	Operations in fraktur font in particular always represent procedures in a measurement-based model of computation.
	(This should be contrasted with capital italic letters, which usually represent unitary operations.)

\item
	Vectors over $\Z_2$ (the integers modulo $2$) are represented in a serif bold-face font, usually as a variable $\vec x \in \Z_2^V$ for some set $V$.
	(This should be contrasted with the sans-serif letters $\sfx$, $\sfy$, $\sfz$, which are generic labels common to the notation for certain state vectors, measurements, and unitary transformations.)

\item
	In the presentation of an algorithm, $a = b$ always denotes an equality comparison.
	(This should be contrasted with $a \gets b$, which denotes an assignment.)

\item
	In the presentation of an algorithm, the notation $\INCR{a}$ represents an instruction to increment the integer variable $a$.

\item
	In the presentation of an algorithm, a \emph{procedure} takes as input \emph{only} those variables which are explicitly specified in its invocation: \ie\ it has no ``side effects''.
	(This should be contrasted with a \emph{subroutine}, which may affect the values of data which are presented as being ``global variables'' available to the subroutine.)

\subsection*{Colours and Hyperlinking}

In the figures, I have tried wherever appropriate to emphasize distinct elements using grayscale shading rather than colour.
Where I found it important to provide additional emphasis with colour, I have tried also to supplement this with non-coloured visual cues (\eg\ different styles of dashed lines).

This thesis was written using PDF\LaTeX, and the electronic version is hyperlinked.
Links are coloured, I hope unobtrusively, according to the nature of the reference:
\begin{description}
\item[\textcolor{RefColor}{red}] represents a link to another part of the main text, \eg\ Theorem~\ref{thm:churchHilbertSpace};
\item[\textcolor{CiteColor}{green}] represents a link to the bibliography, \eg\ \cite{HEB04};
\item[\textcolor{UrlColor}{blue}] represents a link to a URL on the internet, \eg\ \arXiv[0806.1972].
\end{description}
The entries of the Table of Contents, List of Figures, and List of Algorithms are also hyperlinked, although they are not coloured.

\end{itemize}

\egroup

	\pagenumbering{arabic}
	\pagestyle{headings}

	\chapter[Fundamentals of quantum computation]{
		Fundamentals (of quantum computation)}


Computation is the act of transforming a piece of data by physical processes, into a form which may be interpreted as representing \eg\ the solution to a mathematical problem.
Data may be stored as any reliably measurable (and manipulable) property of a physical system, in which case computations are most appropriately described in terms of operations on that system.
Observations along these lines made by Landauer~\cite{Landauer61} spurred research into computation as a physical process, culminating in a series of papers by Benioff~\cite{Benioff80,Benioff81,Benioff82} proposing an implementation of Turing machines~\cite{Turing36} along quantum mechanical principles.

In~\cite{Feynman82}, Feynman considered the complementary question of whether or not quantum mechanical phenomena could be efficiently simulated by the existing models of computation, and concluded that a new type of computer was required in order to intrinsically model quantum mechanical phenomena.
A model of a computation operation using quantum mechanical states as data was presented by Deutsch~\cite{Deutsch85}, and the theory of \emph{quantum computation} was subsequently developed by Bernstein and Vazirani~\cite{BV93,BV97} and Yao~\cite{Yao93}; however, the mathematical tools for this description of information were largely developed by von Neumann~\cite{Neumann32}.

Quantum computation is most often described in terms of \emph{unitary circuits}, in which quantum data is represented as unit $\ell_2$-norm \emph{state vectors} over $\C^d$ for some $d \ge 2$, which generalize the notion of a point-mass probability distribution; transformations of the state are given by \emph{unitary matrices}, which are the linear transformations which preserve the set of these vectors.
In this model, measurements are postponed to the end of the computation, or (in the case of a computation which is to be used as a subroutine of a larger quantum computation) not performed at all, and so the description of measurements and post-measurement states is often simplified.
However, it will prove essential for the discussion of the \emph{one-way measurement-based model} to establish conventions and definitions for the more general description of quantum computation in terms of \emph{density operators} which generalize both state vectors and arbitrary probability distributions, and the transformations which preserve this set of objects, which are \emph{completely positive and trace-preserving superoperators} (or CPTP maps).

In this chapter, I will introduce the principal mathematical objects of quantum computation and standard results concerning universal quantum computation, abstracted away from particular applications or implementations.
I also introduce the conventions used in this thesis to describe \emph{unitary circuits}, the principal language for describing quantum computation.

\paragraph{New results.}

Unsurprisingly, the vast majority of the work in this chapter is prior art; however, the section ``Non-standard representations of unitary circuits'' (pages~\pageref{sec:nonstandardRepns}\thru\pageref{sec:nonstandardRepnsEnd}) presents new results of an elementary nature, which fit most naturally in this chapter.


\section{Quantum data, and quantum transformations}
\label{sec:quantumData}

\subsection{Qualitative remarks on the nature of quantum data}

We begin with some remarks on quantum states, and the differences between them and classical probability, as a pre-amble to the technical definitions that follow.

The utility of probabilistic theories of evolution and computation is that they allow the transformation of data when one does not have perfect information about a system.
A probability distribution (usually) represents some amount of information about a physical system, but (usually) not complete information; and transformations of such a distribution represents the information that can be obtained about that system under certain conditions.
In particular, a probability distribution is a \emph{model} of the state of a physical system, or of the distribution of inputs to a problem, and stochastic transformations represent a model of the evolution of a system, without committing to any answer as to whether a more precise model for the same physical system is possible.
However, there is an implicit assumption that there is no penalty for coarseness of resolution of events compared to the possible states of the system in a traditional probabilistic model, except in the coarseness of the corresponding predictions.

Let us say that one set of events $R$ is a \emph{refinement} of another set $S$ if any system which is determined with respect to $R$ (\ie, such that there is an event $r \in R$ which occurs with certainty), then the system is also determined with respect to $S$.
Conceptually, there if a function from each $r \in R$ to some $s \in S$, which determines the events in $S$ which occur conditioned on the events in $R$.
The more refined a set of events is, the more precisely it allows description of a physical system; and a \emph{maximally} refined set of events $R$ for a system is one where any property which can be tested simultaneously with the events of $R$ can in principle be \emph{defined} in terms of events $r \in R$.
\label{discn:pureState}
Then, let us say that a system is in a \emph{pure state} if, for some maximally refined set of events $R$ for the system, the system can be described by the event $r \in R$.
Classical probability theory is then a means of dealing with probabilities of events and stochastic processes on the premise that there may be in principle a unique set of distinguishable pure states, and that analysis of other distributions may be performed with respect to this unique set of states of complete knowledge.

By contrast, quantum mechanics presents us with physical systems in which there seems to be a mismatch between observable events (such as the outcomes of measurements in experiments) and the possible physical states of a system.
It is possible to prepare physical systems such as the polarization of light signals, or the orientation of the spin of ions, in different ways which correspond to point-mass distributions of certain events, but where these events cannot be simultaneously tested.
This results in states which correspond to maximal information regarding certain physical properties, but seem only to be distinguishable from one another statistically, through multiple preparations.

What is more, the way in which such physical systems evolve demonstrate that the inability to resolve between different maximally refined sets of events in such systems is crucial: they may exhibit different evolutions depending on whether they interact with other systems, even when those interactions may only be used to test for some event.
An example is with light filtered with a polarizing lens: vertically polarized light cannot pass through a horizontally polarized lens.
However, there is some probability of the light passing through the horizontal filter if another filter (aligned diagonally to the horizontal axis) is placed between the source of polarized light and the horizontal filter.
The intermediate filter causes an increase in the probability of light being detected after the original blocking filter, suggesting that the light has been transformed \emph{en route}.
However, unless the physical system is sensitive to the context in which the intermediate filter is being used, there is no reason why this filter should have a different effect than it would when used to either let diagonally polarized light pass through, or filter out light with the opposite diagonal polarization, in an attempt to measure in a different experiment which of the two diagonal polarizations it may have.

The example of diagonally versus horizontally or vertically polarized light illustrates that physical systems can be prepared in different pure states which can only be distinguished statistically, and that these states are sensitive to testing for properties of the system, and are generally changed by measurement of a property to a state consistent with having a definite value of that property.
This suggested to Einstein, Podolsky, and Rosen~\cite{EPR35} that quantum mechanics is incomplete, as they posited that there must be a single underlying maximally refined set of events, which contained complete information corresponding to what they referred to as \emph{elements of reality} (\ie\ point-mass functions over any testable set of events).
This corresponds to the some of the current interpretation of quantum mechanics (\eg\ the Bohmian interpretation~\cite{Bohm52}) involving \emph{hidden variables} which we are prevented from testing precisely.
However, the more common view has been that there do not exist properties which we are prevented for fundamental reasons from testing, and that there does not exist a \emph{unique} maximally refined set of events, although there do exist multiple \emph{incompatible} such maximally refined sets, in the sense that (by necessity) they cannot be tested for simultaneously, and that testing for one will disturb point-mass functions over the others.

The apparent absence of a maximally refined set of events for certain physical systems prompted the development of early quantum theory, in which tests for events (and the set of events which is being tested for) must be specified explicitly.
This led to the specialized concept of \emph{measurement} in quantum mechanics (and consequently in quantum computation as well) which will play a central role in this thesis.

\subsection{Density operators}

Quantum states may be described by \emph{density operators}, and property testing on these states may be represented as \emph{projections} on these operators~\cite{Neumann55}.
We will illustrate this formalism of quantum states and measurements as an extension of classical probability theory, as follows.

\subsubsection{Description of classical probability distributions}

We describe probability in terms of an event space $\E(S)$ consisting of the set of functions $f: S \to \R$\,, for some finite set of events $S$.
The set of probability distributions $\Prob(S)$ in this space are the non-negative functions $p \in \E(S)$ of unit $\ell_1$-norm,
\begin{gather}
		\norm{p}_1
	\;\;=\;\;
		\sum_{s \in S} p(s)
	\;\;=\;\;
		1	\;.
\end{gather}
Point-mass functions are given by characteristic functions $\bigchi_{\ens{s}}$ for $s \in S$, where for $E \subset S$, $\bigchi_E$ is the characteristic function of $E$\,:
\begin{gather}
		\bigchi_E(s)
	\;\;=\;\;
		\begin{cases}
			1\,,	&	\text{if $s \in E$}	\\
			0\,,	&	\text{otherwise}
		\end{cases}\,;
\end{gather}
we may then decompose $p \in \Prob(S)$ as a \emph{convex combination}\footnote{%
A \emph{convex combination} of a set $X$ is a linear combination of the form $\smash{\sum\limits_{x \in X} \lambda(x) x}$, for some non-negative function $\lambda: X \to \R$ with unit $\ell_1$-norm.}
of operators $\bigchi_{\ens{s}}$ for $s \in S$,
\begin{gather}
	p \;=\; \sum\limits_{s \in S} p(s) \bigchi_{\ens{s}}	\;\;.
\end{gather}
Events correspond to subsets $E \subset S$, and the probability of $E$ corresponds to the sum of $p(s)$ for $s \in E$\,; equivalently, we can express it in terms of the adjoint functional $\bigchi_E\herm$\,, which maps characteristic functions $\bigchi_A$ to their ``overlap'' $\card{E \inter A}$\,:
\begin{gather}
	\label{eqn:simpleProbEvent}
		\Pr_{p}(E)
	\;=\;
		\bigchi_E\herm \; p
	\;=\;
		\anglebr{\bigchi_E, p}	\;,
\end{gather}
where $\anglebr{p,q}$ is the usual inner product over Euclidean space, and $A\herm$ is the Hermitian adjoint for an operator $A$.
Joint probability mass functions $g \in \Prob(S \x T)$ are maps from pairs of events $(s,t) \in S \x T$ to their probabilities $g(s,t)$\,, and can be described as convex combinations of the point-mass functions $\bigchi_{(s,t)}$\,; for independently distributed variables $p \in \Prob(S)$ and $q \in \Prob(T)$\,, we may describe such a joint distribution as a tensor product of the distributions $p$ and $q$\,, by identifying distributions in $\Prob(S \x T)$ with the unit $\ell_1$-norm functions in the event space $\E(S) \ox \E(T)$.
The probability distributions $\Prob(S \x T)$ then correspond to the non-negative, unit trace distributions of $\E(S) \ox \E(T) \,\cong\, \E(S \x T)$.

Transformations $U$ of probability distributions are given by the linear transformations which preserve $\ell_1$-norm on $\E(S)$, \ie\ by \emph{stochastic transformations}.
By virtue of their linearity, they may be characterized by conditional probabilities, \ie\ how $U$ transforms point-mass functions $\bigchi_{\ens{s}}$\,.
Independent transformations $U$ on an event space $\E(R)$ and $V$ on an event space $\E(T)$ then given by the tensor product $U \ox V$; in particular, the marginal distribution over $S$ of a probability distribution over $S \x T$ can be recovered by setting the map $V$ to be the unique ``stochastic'' transformation $V: \E(T) \to \R$, using the equivalence $\E(S) \ox \R \,\cong\, \E(S \x \ens{1}) \,\cong\, \E(S)$\,.

An isomorphic choice of mathematical structure would be to substitute real-valued diagonal matrices, and the Euclidean inner product with the Hilbert--Schmidt inner product,
\begin{gather}
		\anglebr{A,B}_{\Tr{}}
	\;\;=\;\;
		\Tr(A\herm B)	\;.
\end{gather}
For a maximally refined set of measurable events $R$ on a system, let $\ens{\ket{s}\big.}_{s \in S}$ represent the standard basis for the Hilbert space $\cH[S]$\,.
Then, we may use the following constructions for $\E(S)$ and $\Prob(S)$:
\begin{subequations}
\label{eqn:embedHilbertSchmidt}
\begin{align}
		\Pi_{\ens{s}}
	\;\;=&\;\;
		\ket{s}\bra{s}
		\;\;\text{is a rank-$1$ projector $\big($\cf\ point-mass functions $\bigchi_{\ens{s}}\big)$;}
	\\[1ex]
		\E(S)
	\;\;=&\;\;
		\Span_{\R}\ens{\ket{s}\bra{s} \,\big|\, s \in S}
		\;\;\big(\text{\cf\ span of the functions $\bigchi_{\ens{s}}$}\big);
	\\[1ex]
	\begin{split}
			\Prob(S)
		\;\;=&\;\;
			\ens{p \in \E(S) \,\big|\, \text{$p$ non-negative, with unit trace}}
		\\[-0.1ex]&
			\;\;\;\text{(\cf\ non-negative functions with unit $\ell_1$-norm);}
	\end{split}
	\\[1ex]
	\begin{split}
			\smash{\Pr_{p}(E)}
		\;\;=&\;\;
			\Tr(\Pi_E \; p)
 		\;\;\text{for $p \in \Prob(S)$ and $E \subset S$,\; where~} \smash{\Pi_E \;=\; \sum_{s \in E} \ket{s}\bra{s}}
 		\\[-0.3ex]&\;\;(\text{\cf\ Equation~\eqref{eqn:simpleProbEvent} above}).
	\end{split}
\end{align}
\end{subequations}
In particular, distinct point-mass distributions are orthogonal with respect to the Hilbert--Schmidt inner product, and probabilities of events $E \subset R$ are given by the \emph{operator functionals} $\Pi_E\conj$ characterized by mapping other projectors $\Pi_S$ to the overlap $\Pi_E\conj \Pi_A \,=\, \anglebr{\Pi_E, \Pi_A}_{\Tr{}} \,=\, \card{E \inter A}$.

\subsubsection{Generalization to arbitrary quantum distributions}

As we noted above, a single maximally refined set of events does not seem to suffice to describe the pure states accessible to quantum systems.
For a quantum system which admits an event space $\E(S)$ as constructed above, we will generalize the set of pure states to all rank-$1$ projectors $\ket{\psi}\bra{\psi}$\,: a maximally refined set of events is then a maximal collection
\begin{gather}
	\label{eqn:rankOneOrthProjectors}
	B = \ens{\ket{\psi_j}\bra{\psi_j}\Big.}_{j = 1}^{\card{S}}
\end{gather}
of rank-$1$ projections on $\cH[S]$\,, subject to these projectors being orthogonal with respect to the Hilbert--Schmidt inner product.
We may then consider events of which this set is a refinement, which correspond to sets of projectors which all commute with the projections in $S$; and we may consider states of the system which are convex combinations of elements of $S$.

Considering all such sets of orthogonal rank-$1$ projectors prompts us to generalize the event space of the system from $\E(S)$ to the set of all matrices which can be decomposed as a real-valued linear combination of orthogonal rank-$1$ projectors: that is, the Hermitian operators (normal operators with real eigenvalues) $\Herm(S)$ over $\cH[S]$.
Distributions describing states of the system are then as follows:

\begin{definition}
	The set of \emph{density operators} over $S$ is the set of positive semidefinite Hermitian operators over $\cH[S]$ with unit trace,
	\begin{align}
			\D(S)
		\;\;=&\;\;
			\ens{\rho \in \Herm(S) \,\Big|\, \rho \ge 0 \text{~and~} \Tr(\rho) = 1}\,.
	\end{align}
\end{definition}
As before, joint distributions are formed by tensor products of the event space: we have
\begin{gather}
	\Herm(S \x T) \;=\; \Herm(S) \ox \Herm(T) \,,
\end{gather}
which can be easily proven by considering a basis for $\Herm(S \x T)$ over $\R$ consisting of the tensor product of two such bases for $\Herm(S)$ and $\Herm(T)$.
For joint distributions $\tau \in \D(S \x T)$, we then take the non-negative, unit-trace elements of $\Herm(S \x T)$.

\subsubsection{Transformations of density operators}

The natural extension of a transformation between probability distributions is a \emph{linear superoperator} --- that is, a linear operation that itself acts on linear operations $\rho \in \Herm(R)$ --- which in particular maps density operators to density operators.
It follows that valid transformations of quantum data consist of \emph{trace-preserving, positive} superoperators $\Upphi: \D(\cH[S]) \to \D(\cH[S'])$\,, \ie\ that $\Upphi$ maps operators of $\L(\cH[S])$ to operators with the same trace, and positive semidefinite operators to other positive semidefinite operators.
In order for such a superoperator $\Upphi$ to be valid when applied to part of a joint distribution of two systems, we also require that $\Upphi \ox\, \id_{\D(T)}$ be a valid transformation (and therefore a positive operator) on $\D(S \ox T)$ for any $T$\,: we say that $\Upphi$ is \emph{completely positive}.
\begin{definition}
	\label{dfn:CPTP}
	A \emph{transformation} of a set of density operators $\D(S)$ is a completely positive and trace preserving mapping (or \emph{CPTP map}) $\Upphi: \D(S) \to \D(S')$ for some state-space $\D(S')$.
\end{definition}
It is possible to show (see \eg~\cite{Watrous07}) that for $\Upphi: \L(\cH[S]) \to \L(\cH[S'])$ completely positive and trace preserving, there exist linear operators $\ens{A_j}_{j = 1}^{n}$, for some $n \in \N$ and $A_j: \cH[S] \to \cH[S']$ non-zero, such that
\begin{gather}
	\label{eqn:KrausCondition}
		\Upphi(\rho)
	\;\;=\;\;
		\sum_{j = 1}^n A_j \,\rho\, A_j\herm	\;,
	\quad\text{~where~}\quad
		\sum_{j = 1}^n
		A_j\herm A_j
	\;\;=\;\;
		\idop_{\cH[S]}	\;.
\end{gather}
Such operators $A_j$ are called \emph{Kraus} operators, and are non-unique. 
For another set $T$, if $\Uppsi: \L(\cH[T]) \to \L(\cH[T'])$ is another CPTP map given by
\begin{align}
		\Uppsi(\sigma)
	\;\;=&\;\;
		\sum_{k = 1}^m B_k \,\sigma\, B_k\herm	\;,
\end{align}
we may form the transformation performed by independent transformations $\Upphi: \D(S) \to \D(S')$ and $\Uppsi : \D(T) \to \D(T')$ on a \emph{joint} distribution over $\D(S \x T)$ by taking tensor products of the Kraus operators: we write
\begin{align}
		(\Upphi \ox \Uppsi)(\tau)
	\;\;=&\;\;
		\sum_{j = 1}^n \sum_{k = 1}^m \,
			(A_j \ox B_k) \, \tau \, (A_j \ox B_k)\herm
\end{align}
for $\tau \in \L(\cH[S \x T]) \to \L(\cH[S \x T])$.

We may use this to determine the marginal distributions of joint density operators $\tau \in \D(S \x T)$\,.
Consider the map $\tr[T] : \Herm(T) \to \R$ which maps every Hermitian operator over $\cH[T]$ to its trace, given \eg~by
\begin{gather}
	\label{eqn:traceOut}
		\tr[T](\sigma)
	\;\;=\;\;
		\sum_{t \in T} \bra{t} \sigma \ket{t}	\;:
\end{gather}
note that $\tr[T]$ is the unique CPTP map from $\Herm(T)$ to $\R$.
Then, we may obtain the marginal distribution $\tau_S$ over $\D(S)$ of a density operator $\tau \in \D(S \x T)$ by the \emph{partial trace} over $T$, defined by
\begin{gather}
		\Big(\id_{\D(S)} \,\ox\, \tr[T]\Big)(\tau)
	\;\;=\;\;
		\sum_{t \in T} \Big( \idop_{\cH[S]} \ox \bra{t} \Big) \,\tau\, \Big( \idop_{\cH[S]} \ox \ket{t} \Big)	\;.
\end{gather}
We refer to the process of applying such a map as \emph{tracing out} the system represented by the state space $\D(T)$; and we commonly write such a map by $\tr[T]$, leaving the identity on any other subsystems implicit in the description of the map.

\subsubsection{Quantum and classical registers}

Based on the preceding, we will make the following definitions:
\begin{definition}
	\label{def:register}
	A \emph{quantum register} $\sfQ$ over a finite set $Q$ is an abstract physical system with state-space $\D(Q) \le \Herm(Q)$, consisting of unit trace operators $\rho \ge 0$ supported on a Hilbert space $\cH[Q]$.
	Similarly, a \emph{classical register} $\sfC$ over a finite set $C$ is an abstract physical system with state-space $\Prob(C) \le \D(C)$; and where furthermore, we require that for any joint state $\rho \in \D(C \x R_1 \x \cdots \x R_n)$ of $\sfC$ together with any collection of registers $\ens{\sfR_1, \ldots, \sfR_n}$, $\rho$ is a convex combination of operators $\vec{p}_j \ox \sigma_j$, for $\vec{p}_j \in \Prob(C)$ and $\sigma_j \in \D(R_1 \x \cdots \x R_n)$.
	For either a classical register with state-space $\P(R)$ or quantum register with state-space $\D(R)$, the \emph{dimension} of the register is $\card{R}$.
\end{definition}
The motivation for defining classical registers in addition to quantum registers is to abstractly represent physical systems which we cannot (either for fundamental reasons or practical limitations) put into general quantum states: examples include read-out devices in laboratory settings, and hand-written notes on paper, whose states do not exhibit obvious quantum behavior but which still may not be determinable in advance.

\begin{notation}
	Let $X$ and $Y$ be convex sets in complex Hilbert spaces.
	In analogy to the linear extension $V \x W \incl V \ox W$ for vector spaces, we will let $X \ox Y$ represent the \emph{convex} extension of the set $X \x Y$, which consists of convex combinations of tensor products $\rho \ox \sigma$ for $\rho \in X$ and $\sigma \in Y$.
\end{notation}
In particular, for $\rho$ a joint state of a classical register $\sfC$ together with any collection of registers $\ens{\sfR_1, \ldots, \sfR_n}$ as in Definition~\ref{def:register} above, we have $\rho \in \Prob(C) \ox \D(R_1 \x \cdots \x R_n)$.

Given the definition of a classical register above, the following Lemma allows us to interpret aggregates of classical registers as a single, compound classical register:
\begin{lemma}
	Let $\rho$ be a joint state of a collection of quantum registers $\ens{\sfQ_1,\ldots,\sfQ_n}$ and classical registers $\ens{\sfC_1,\ldots,\sfC_m}$.
	Then we have
	\begin{gather}
		\rho \in \D(Q_1 \x \cdots \x Q_n) \ox \Prob(C_1 \x \cdots \x C_m),
	\end{gather}
	where $Q_j$ (respectively $C_j$) are the finite sets over which the registers $\sfQ_j$ (respectively $\sfC_j$) are defined.
\end{lemma}
\begin{proof}
	We prove this by induction on $m$.
	If $m = 0$, there is nothing to prove.
	Otherwise, suppose the statement holds for all $m < M$ for some integer $M \ge 1$.
	As $\sfC_m$ is a classical register, $\rho$ is a convex combination of the form
	\begin{align}
			\rho
		\;\;=\;\;
			\sum_{c_m \in C_m} \lambda_m(c_m) \;\sigma_{c_m} \ox \ket{c_m}\bra{c_m}	\,,
	\end{align}
	where $\lambda_m \in \Prob(C_m)$ is the marginal distribution of $\rho$ on $\sfC_m$\,. 
	For $\lambda_m(t) > 0$, the density operators $\sigma_t$ are the states of the other registers conditioned on $\sfC_m$ being in the state $\ket{t}\bra{t}$: we may verify this for events $E \in \Herm(Q_1 \x \cdots \x Q_n \x C_1 \x \cdots \x C_{m-1})$ by
	\begin{align}
			\Pr_{\rho}(E|\sfC_m = t)
		\;\;=&\;\;
			\frac{\Pr_{\rho}(E,t)}{\Pr_{\rho}(t)}
		\;\;=\;\;
			\frac{\Tr\Big(\Big[E \ox \ket{t}\bra{t}\Big] \rho\Big)}{\Tr\Big(\ket{t}\bra{t} \,\tr[R](\rho)\Big)}
		\notag\\[2ex]=&\;\;
			\frac{%
				\Tr\paren{\sum\limits_{c_m} \, \lambda_m(c_m) \; E \, \sigma_{c_m} \ox \ket{t}\!\bracket{t}{c_m}\!\bra{c_m} \;}
			}{%
				\Tr\paren{\sum\limits_{\vec r} \sum\limits_{c_m} \, \lambda_m(c_m) \; \bra{\vec r} \sigma_{c_m} \ket{\vec r} \,\ox\, \ket{t}\!\bracket{t}{c_m}\!\bra{c_m} }
			}
		\notag\\[2ex]=&\;\;
			\frac{\lambda_m(t) \Tr(E \, \sigma_t)}{\lambda_m(t)}
		\;\;=\;\;
			\Tr(E \, \sigma_t)	\;;
	\end{align}
	we may let $\sigma_t$ be arbitrary for $\lambda_m(t) = 0$.
	Let $\tilde C = C_1 \x \cdots \x C_{m-1}$\,: by hypothesis, we can decompose each $\sigma_{c_m}$ as a combination
	\begin{gather}
			\sigma_{c_m}
		\;\;=\;\;
			\sum_{\tilde{\vec c} \in \tilde C} \tilde\lambda_{c_m}(\tilde{\vec c}) \; \tau_{(\tilde{\vec c},c_m)} \ox \ket{\tilde{\vec c}}\bra{\tilde{\vec c}}
	\end{gather}
	again for some marginal distribution $\tilde \lambda_{c_m} \in \Prob(\tilde{C})$ depending on $c_m \in C_m$, and for conditional states $\tau_{(\tilde{\vec c},c_m)} \in \D(Q_1 \x \cdots Q_n)$.
	As before, we may identify tuples and standard basis vectors as follows:
	\begin{subequations}
	\begin{gather}
			\vec{c} \;=\; (\tilde{\vec c}, c_m) \,\in\, C_1 \x \cdots \x C_{m-1} \x C_m \;=\; C \,,
		\\[1ex]
			\ket{\vec c} \;=\; \ket{\;\! \big.(\tilde{\vec c}, c_m) \,} = \ket{\tilde{\vec c}} \ox \ket{c_m} \;\in\; \cH[C]	\;;
	\end{gather}
	\end{subequations}
	then, if we let $\lambda(\vec c) = \lambda_m(c_m) \tilde{\lambda}_{c_m}(c_1, \ldots, c_{m-1})$, we have
	\begin{align}
			\rho
		\;\;=&
			\sum_{\substack{\tilde{\vec c} \in \tilde C \\[0.2ex] c_m \in C_m}}  \lambda_m(c_m) \tilde\lambda_{c_m}(\tilde{\vec c}) \;\;\tau_{(\tilde{\vec c},c_m)} \ox \ket{\tilde{\vec c}}\bra{\tilde{\vec c}} \ox \ket{c_m}\bra{c_m}
		\notag\\[1ex]=&\;\;
			\sum_{\vec c \in C} \;\lambda(\vec c) \;\tau_{\vec c} \ox \ket{\vec c}\bra{\vec c}	\,,
	\end{align}
	where for each $\vec c \in C_1 \x \cdots \x C_m$, we have $\tau_{\vec c} \in \D(Q_1 \x \cdots \x Q_n)$\,.
	Then $\rho$ is a convex combination of the form described by the Lemma.
\end{proof}
This allows us to ``factor'' a composite system, consisting of classical and quantum registers, into classical and quantum \emph{subsystems} which we may treat as a single (composite) quantum register, together with a single (composite) classical register; and the states of these subsystems as consisting of ``classically probabilistic'' and ``irreducibly quantum'' data.

\subsection{Measurement of quantum data}

As we noted before, quantum systems are sensitive to testing for events, which requires that we describe how they evolve under event testing.
We will define three senses of measurement, starting with the following:
\begin{definition}
	\label{def:projectiveMeas}
	For a finite set $Q$, a set of operators $\ens{\Pi_r}_{r \in R} \subset \Herm(Q)$ is a \emph{complete set of orthogonal projectors} if
	\begin{gather}
			\sum_{r \in R} \;\! \Pi_j
		\;\;=\;\;
			\idop_{Q}
		\qquad\text{~and~}\qquad
			\Pi_j \Pi_k
		\;\;=\;\;
			\Kdelta[j][k]\, \Pi_j\,,
	\end{gather}
	where
	\begin{gather}
			\Kdelta[j][k]
		\;\;=\;\;
			\begin{cases}
				1	\,,	&	\text{if $j = k$}	\\
				0	\,,	&	\text{otherwise}	
			\end{cases}
	\end{gather}
	is the Kronecker delta.
	A \emph{projective measurement} with respect to $\ens{\Pi_r}_{r \in R}$ on a quantum register $\sfQ$ over $Q$ is a mapping $\sfP: \D(Q) \to \D(Q)$ given by
	\begin{gather}
			\sfP(\rho)
		\;\;=\;\;
			\sum_{r \in R} \Pi_r\;\! \rho \,\Pi_r	\;.
	\end{gather}
	We call $\sfP$ a \emph{complete} projective measurement if each $\Pi_r$ has rank $1$.
\end{definition}
Note that for any state $\rho$ with which the projectors $\Pi_j$ all commute, we have $\sfP(\rho) = \rho$\,: that is, if there is a set of orthogonal point-mass distributions (pure states) on the system which can be used to decompose both $\rho$ and the projectors $\Pi_s$, the measurement does not affect the state of the system.
However, for any other $\rho \in \D(R)$, a projective measurement will disturb the state.

We usually suppose that measurement corresponds to a process of actually measuring a physical parameter, the result of which is recorded in the state of another system used for read-out (represented by a \emph{classical} register).
This corresponds to a sense of measurement defined by von Neumann~\cite{Neumann55}:
\begin{definition}
	\label{def:vonNeumannMeas}
	A \emph{von Neumann measurement} with respect to a complete set of orthonormal projectors $\ens{\Pi_r}_{r \in R} \subset \Herm(Q)$, on a quantum register $\sfQ$ over a finite set $Q$, is a mapping $\sfN: \D(Q) \,\to\; \D(Q) \ox \Prob(R)$ given by
	\begin{align}
			\sfN(\rho)
		\;\;=&\;\;
			\sum_{r \in R}
				\Big(\Pi_r \ox \ket{r}\Big) \rho \Big(\Pi_r \ox \bra{r}\Big)
		\;\;=\;\;
			\sum_{r \in R}
				\, \Pi_r \:\!\rho\, \Pi_r \,\ox\, \ket{r}\bra{r}	\;.
	\end{align}
	We call $\sfN$ a \emph{complete} von Neumann measurement if each $\Pi_r$ has rank $1$.
\end{definition}
The output space consists of joint distributions across the original quantum system $\sfQ$ and a classical \emph{result register} $\sfR$, whose marginal distribution may be interpreted as simply a probability distribution.

A special case of a von Neumann measurement which it will be often useful to discuss is a measurement \emph{with respect to an observable}.
In quantum mechanics, Hermitian operators $A$ may correspond to physical properties of a system (such as energy or angular momentum), where the property described by $A$ has the expected value $\lambda$ for any system in a state $\rho$ such that $\Tr(A \rho) = \lambda$.
We may describing measurement of an observable $A$ with a von Neumann measurement, as follows:
\begin{definition}
	\label{def:measObservableVN}
	Let $A \in \Herm(Q)$ be an operator with spectrum given by
	\begin{gather}
		\Lambda = \ens{\lambda \in \R \,\big|\, \det(A - \lambda \idop_Q) = 0}\,.
	\end{gather}
	A von Neumann measurement \emph{of the observable $A$} on a quantum register $\sfQ$ over $Q$ is a mapping $\sfN: \D(Q) \,\to\; \D(Q) \ox \Prob(\Lambda)$ given by
	\begin{align}
			\sfN(\rho)
		\;\;=&\;\;
			\sum_{\lambda \in \Lambda}
				\, \Pi_\lambda \:\!\rho\, \Pi_\lambda \,\ox\, \ket{\lambda}\bra{\lambda}	\;,
	\end{align}
	where each operator $\Pi_\lambda$ is the projector onto the $\lambda$-eigenspace of $A$.
\end{definition}
As a mathematical concept, this can be trivially extended to measurement of any normal operator on a quantum register, but this will not be necessary for this thesis.

We will regard the state of a classical subsystem (such as a result register) to be known in practise --- that is, without an infinite regress of measurements --- even if it is not determined in advance; then, we may consider the state of the quantum registers conditioned on a particular state of the classical subsystem.
For a post-measurement state $\bar\rho$ resulting from a von Neumann measurement, we may consider post-measurement distributions $\bar\rho_R \in \D(R)$ conditioned on a particular value of the result register after measurement $t$, which for an arbitrary projector $E$ over $\cH[R]$ describing (by a conventional abuse of notation) an event $E$, is given by
\begin{align}
		\Tr(E \, \bar\rho_{\text{\raisebox{-0.2ex}{$R$}}}) 
	\;\;=\;\;
		\Pr_{\bar\rho}(E|t)
	\;\;=&\;\;
		\frac{\Pr_{\bar\rho}(E,t)}{\Pr_{\bar\rho}(t)}
	\;\;=\;\;
		\frac{\Tr\Big(\big[E \ox \ket{t}\bra{t}\big] \bar\rho\Big)}{\Tr\Big(\ket{t}\bra{t} \, \tr[R](\bar\rho)\Big)}
	\notag\\[2ex]=&\;\;
		\frac{%
			\Tr\Big(E\, \Pi_t \rho\, \Pi_t \,\ox\, \ket{t}\bra{t}\Big) 
		}{%
			\Tr\paren{\sum\limits_{r \in R} \bra{r}\Pi_t \rho \Pi_t\ket{r} \cdot \ket{t}\bra{t}}
		}
	\notag\\[2ex]=&\;\;
		\Tr\paren{ E \sqparen{\frac{\Pi_t \rho\, \Pi_t}{\Tr\paren{\Pi_t \rho\, \Pi_t}}}}	\;,
\end{align}
which entails by linearity that
\begin{gather}
		\bar\rho_{\text{\raisebox{-0.2ex}{$R$}}}
	\;\;=\;\;
		\frac{\Pi_t \rho\, \Pi_t}{\Tr\paren{\Pi_t \rho\, \Pi_t}}
\end{gather}
conditioned on the result register being in the state $t$.

A final variety of measurement which we consider is one in which the system being measured is in a sense destroyed by the measurement, as in the case of photons: this occurs \eg\ in the detection of photons.
In this case, a measurement is performed and a result obtained, but no conditional state of the original system meaningfully exists.
We may describe such a measurement by:
\begin{definition}
	\label{def:destructiveMeas}
	Let $\ens{\big.\ket{\psi_j}\big.}_{j = 1}^n \subset \cH[Q]$ be a collection of (not necessarily normalized) vectors such that
	\begin{gather}
			\sum_{j = 1}^n \ket{\psi_j}\bra{\psi_j} \;=\; \idop_Q	\,,
	\end{gather}
	and consider a mapping $r: \ens{1, \ldots, n} \to R$.
	Then a \emph{destructive measurement}\footnote{%
	Destructive measurements as defined here correspond to \emph{positive operator-valued measurements}, or POVMs (see \eg~\cite{NC00}): we may define
	\begin{gather*}
			E_t
		\;\;=
			\sum\limits_{j\;\!:\;\!r(j) = t}	\ket{\psi_j}\bra{\psi_j}	\,,
	\end{gather*}
	from which we obtain
	\begin{gather*}
			\Tr(E_t \, \rho)
		\;=\;
			\Tr\paren{\sum\limits_{j = 1}^n \ket{\big.t} \bracket{\big.t}{\big.r(j)}\!\bra{\big.\psi_j} \rho \ket{\big.\psi_j}\bra{\big.r(j)}}
		\;=\;
			\Tr\paren{\ket{t}\bra{t} \Big.\Meas{}{}(\rho)}	\;; 
	\end{gather*}
	the primary difference between POVMs and destructive measurements as presented here is in the explicit description of the classical register retaining the measurement result.}
	on a state space $\D(Q)$ with respect to the states $\ens{\big.\ket{\psi_j}\big.}_{j = 1}^n$ and mapping $r$ is a transformation $\Meas{}{} \,:\, \D(Q) \to \E(R)$ of the form
	\begin{align}
			\Meas{}{}(\rho)
		\;=\;
			\sum_{j = 1}^n \,\ket{r(j)\big.}\bra{\psi_j\big.} \,\rho\, \ket{\big.\psi_j}\bra{\big.r(j)}	\,.
	\end{align}
	We say that $\Meas{}{}$ is an \emph{orthonormal} (destructive) measurement if the states $\ket{\psi_j}$ are orthonormal, and that it is also \emph{complete} if $r$ is injective.
\end{definition}%
We will generally be interested in the special case of complete orthonormal (destructive) measurements, in which case the vectors $\ket{\psi_j}$ form an orthonormal basis.
In particular, we may define a sense of the destructive measurement of an observable:
\begin{definition}
	\label{def:measObservableDestructive}
	Let $A \in \Herm(Q)$ be an operator with spectrum given by
	\begin{gather}
		\Lambda = \ens{\lambda \in \R \,\big|\, \det(A - \lambda \idop_Q) = 0}\,,
	\end{gather}
	and for each $\lambda \in \Lambda$, let $\ens{\big.\ket{\psi_{\lambda, j}}\big.}_{j = 1}^{m_\lambda}$ be an orthogonal basis of the $\lambda$-eigenspace of $A$.
	Then a destructive measurement \emph{of the observable $A$} on a quantum register $\sfQ$ over $Q$ is a mapping $\Meas{}{}: \D(Q) \,\to\; \Prob(\Lambda)$ given by
	\begin{align}
			\Meas{}{}(\rho)
		\;\;=&\;\;
			\sum_{\lambda \in \Lambda} \sum_{j = 1}^{m_\lambda}
				\, \ket{\lambda}\bra{\psi_{\lambda,j}} \:\!\rho\, \ket{\psi_{\lambda,j}} \bra{\lambda}
		\;\;=\;\;
			\big(\tr[Q] \,\circ\, \sfN\,\big)(\rho)	\;,
	\end{align}
	where $\sfN: \D(Q) \to\, \D(Q) \ox \Prob(\Lambda)$ is a von Neumann measurement of the observable $A$, and $\tr[Q]$ is a trace-out operation on $Q$.
\end{definition}

Orthonormal destructive measurements and projective measurements can both can be recovered from von Neumann measurements: projective measurements represent the marginal distribution of the system which is measured, and projective destructive measurements represent the marginal distribution of the result register.

\subsection{Quantum state vectors and qubits}

Although it will often be important to describe quantum computation in terms of the more general mathematical apparatus described so far, we now provide some specialized definitions and notation which will help to streamline discussion in many cases.

\subsubsection{State vectors for pure states}

Recall the definition of \emph{pure state} on page~\pageref{discn:pureState} as a point-mass distribution over a maximally refined set of events.
As we described on page~\pageref{eqn:rankOneOrthProjectors}, a point-mass function corresponds to a choice of one projector $\ket{\psi_j}\bra{\psi_j}$, which is characterized by vector $\ket{\psi_j}$ out of an orthonormal basis.
In those cases when we restrict our attention to pure states, we may simplify our analysis by considering the following in place of density operators:
\begin{definition}
	For a quantum register $\sfR$ system in a pure state $\rho \in \D(R)$, the \emph{(pure) state vector} of $\sfR$ is a unit $\ell_2$-norm vector $\ket{\psi} \in \cH[R]$ such that $\rho = \ket{\psi}\bra{\psi}$.
\end{definition}
Note that state vectors are defined only up to scalar factors, as we have $\big[\;\! \omega \ket{\psi} \:\!\big] \big[\;\! \omega\conj \!\bra{\psi} \:\!\big] = \ket{\psi}\bra{\psi}$ for any vector $\ket{\psi}$ and scalar factor $\omega \in \C$ with $\abs{\omega} = 1$.
As well, we may more generally represent some pure state $\rho$ by any non-zero vector $\ket{\psi}$ by first ``normalizing'' $\ket{\psi}$ to obtain a unit (pure state) vector for $\rho$, as follows:
\begin{gather}
		\rho
	\;\;=\;\;
		\sqparen{\sfrac{\ket{\psi}}{\,\norm{\,\ket{\psi}\,}_2}}
		\sqparen{\sfrac{\bra{\psi}}{\,\norm{\,\ket{\psi}\,}_2}}
	\;\;=\;\;
		\sfrac{\ket{\psi}\bra{\psi}}{\bracket{\psi}{\psi}}	\,.
\end{gather}
Thus, we will primarily be interested in proportionality rather than equality when treating pure-state vectors.

\subsubsection{Qubits and qubit state vectors}

We will be primarily interested in a specific type of quantum register, which defines the units of quantum data that we will manipulate:
\begin{definition}
	A \emph{qubit} is a quantum register over the set $\ens{0,1}$\,: we may write $\D(2)$ for the state-space of a qubit, and $\cH = \Span_\C\big\{ \ket{0}, \ket{1} \big\}$ for the Hilbert space on which the operators of $\D(2)$ are supported.
	(We will occasionally write $\D(2^n) = \D(2 \x \cdots \x 2)$ for the state space of $n$ generic qubits.)
	By analogy, we consider a \emph{classical bit} (considered as a physical system) to be a classical register over $\ens{0,1}$.
\end{definition}

The emphasis that will be placed on bits and qubits in this thesis motivate some simplifying notations and conventions:
\begin{notation}
	We will tend to represent qubits by lower-case roman letters.
	We will denote the state space of a qubit $v$ by $\D(v)$\,, the Hilbert space on which it is supported by $\cH[v]$\,, and the identity operator on $\cH[v]$ by $\idop_v$\,, to distinguish these from spaces/operators corresponding to different qubits.
\end{notation}
\begin{notation}
	In a mild abuse of notation, the joint state spaces of multiple qubits $u,v,\ldots$ will be denoted $\D(u, v, \ldots)$, the Hilbert space which supports it by $\cH^{\ox \ens{u,v,\ldots}}$, and the identity operator of that Hilbert space variously as $\idop_2^{\ox \ens{u,v,\ldots}}$ and $\idop_{u,v,\ldots}$\,.
	(We will also tend to represent the identity superoperator $\id_{\D(u,v,\ldots)}$ acting on $\D(u,v,\ldots)$ by the latter two symbols.)
\end{notation}
We will also extend the above conventions to state-spaces and operations on bits, although these will be less common in practise.

\begin{notation}
	We will define the following alternative and short-hand notations for various qubit state vectors:
	\begin{align}
		\begin{split}
				\ket{\splus \sfz} \;=&\; \ket{0}\Big.,
			\\[1ex]
				\ket{\sminus \sfz} \;=&\; \ket{1}\Big.,
		\end{split}
			&\quad\;
		\begin{split}
				\ket{\splus \sfx} \;=&\; \tfrac{1}{\sqrt 2}\Big[ \ket{0} + \ket{1} \Big],
			\\[1ex]
				\ket{\sminus \sfx} \;=&\; \tfrac{1}{\sqrt 2}\Big[ \ket{0} - \ket{1} \Big],
		\end{split}
			&\quad\;
		\begin{split}
				\ket{\splus \sfy} \;=&\; \tfrac{1}{\sqrt 2}\Big[ \ket{0} + i \ket{1} \Big],
			\\[1ex]
				\ket{\sminus \sfy} \;=&\; \tfrac{1}{\sqrt 2}\Big[ \ket{0} - i \ket{1} \Big].	
		\end{split}
	\end{align}
	We often further abbreviate $\ket{\splus} = \ket{\splus \sfx}$ and $\ket{\sminus} = \ket{\sminus \sfx}$.
	We may also write $\ket{\pm}$ or $\ket{\spm \sfx}$ to denote one or both of the state vectors $\ket{\splus \sfx}$ or $\ket{\sminus \sfx}$, and similarly for $\ket{\spm \sfy}$ and $\ket{\spm \sfz}$.
\end{notation}

\section{Unitary evolution and elementary gate sets}
\label{sec:unitaryEvolution}

We have already partially described transformations that can be performed on quantum data: by CPTP maps in general, and by measurement operations (projective, von Neumann, and destructive) as a special case.
We now describe the theory by which general CPTP maps may be decomposed into (or approximated by) the product of discrete elementary operations, which are supposed to be physically implementable in principle.
At the same time, we present the theoretical importance of \emph{unitary transformation} of quantum data, which will motivate the questions explored in this thesis.

\subsection{Unitary evolution of quantum states}

Evolution of pure states (described by state-vectors $\ket{\psi_t} \in \cH[Q]$) in time are governed by the Schr\"odinger equation~\cite{Schrodinger26,Neumann55},
\begin{gather}
		i \hbar \sfrac{\d}{\d t}\ket{\psi_t}
	\;\;=\;\;
		H_t \ket{\psi_t}	\,,
\end{gather}
where $H_t$ is the \emph{Hamiltonian} of the system, a linear operator on $\cH[Q]$ which may depend on time, whose eigenvalues are possible energy levels of pure states of the system.
Energy is represented here by real numbers: abstracting away the physical content of this equation, $H_t$ is therefore a Hermitian operator.
Then, we have
\begin{align}
		\sfrac{\d}{\d t} \bracket{\psi_t}{\psi_t}
	\;\;=&\;\;
		\sqparen{\sfrac{\d}{\d t} \bra{\psi_t}} \ket{\psi_t}
		\;+\;
		\bra{\psi_t} \sqparen{\sfrac{\d}{\d t} \ket{\psi_t}}
	\notag\\[1ex]=&\;\;
		\sqparen{ \sfrac{i}{\hbar} \bra{\psi_t} H_t\herm } \ket{\psi_t}
		\;+\;
		\bra{\psi_t} \sqparen{ \sfrac{-i}{\hbar} H_t \ket{\psi_t}}
	\;\;=\;\;
		0\;;
\end{align}
that is, Schr\"odinger evolution preserves the $\ell_2$-norm of vectors (\ie\ it is well-defined as an evolution of pure state vectors).
The evolution of an input state $\ket{\psi_0}$ from $t = 0$ to some $t = \tau$ can then be described by some operator $U_\tau$ on the pure states of the system which preserves the $\ell_2$-norm.
For a Hilbert space $\cH[Q]$ on a finite set $Q$, we may consider the action of $U_\tau$ on the standard basis of $\cH[Q]$\,: 
we have $\bra{q} U_\tau\herm U_\tau \ket{q} = 1$ by the preservation of norm of the states $\ket{q}$ for $q \in Q$; and for any combination $\ket{\phi} = \frac{1}{\sqrt 2} \ket{p} + \frac{1}{\sqrt 2} \e^{i \theta} \ket{q}$ for distinct $p,q \in Q$ and arbitrary angles $\theta \in \R$, we have
\begin{align}
		1
	\;\;=&\;\;
		\bra{\phi} U_\tau\herm U_\tau \ket{\phi}
	\notag\\=&\;\;
		\tfrac{1}{2}\sqparen{\,\bra{p}U_\tau\herm U_\tau \ket{p} \,+\, \e^{i \theta} \bra{p} U_\tau\herm U_\tau \ket{q}
				\,+\, \e^{-i \theta} \bra{q}U_\tau\herm U_\tau \ket{p} \,+\, \bra{q} U_\tau\herm U_\tau \ket{q}\,}
	\notag\\=&\;\;
		1 \;+\; \Re\Big(\e^{i \theta} \bra{p}U_\tau\herm U_\tau \ket{q}\Big)	\;.
\end{align}
Then $\e^{i \theta} \bra{p}U_\tau\herm U_\tau \ket{q}$ has no real component for any angle $\theta$, which implies $\bra{p}U_\tau\herm U_\tau\ket{q} = 0$ for distinct $p,q \in Q$\,.
We therefore have $U_\tau\herm U_\tau = \idop_Q$\,: that is, $U_\tau$ is a \emph{unitary operation}.
Applied to the evolution of density operators, the time-evolution $\Upphi_\tau$ of density operators corresponding to pure states from time $t = 0$ to $t = \tau$ is then
\begin{subequations}
\begin{align}
		\Upphi_\tau\big(\ket{\psi_0}\bra{\psi_0}\big)
	\;\;=\;\;
		\ket{\psi_\tau} \bra{\psi_\tau}
	\;\;=&\;\;
		U_\tau \ket{\psi_0}\bra{\psi_0} U_\tau\herm\,,
\end{align}
which by linearity implies that
\begin{gather}
	\label{eqn:unitaryCPTP}
		\Upphi_\tau(\rho)
	\;\;=\;\;
		U_\tau \,\rho\: U_\tau\herm
\end{gather}
for all $\rho \in \D(Q)$, which is a valid CPTP map.
We say that a CPTP map is unitary if it can be expressed as in~\eqref{eqn:unitaryCPTP}; and we denote the set of unitary operators on $\cH[S]$ by $\U(S)$.
\end{subequations}

Precisely how \emph{measurement processes} (as described by CPTP maps of the kind described in Definitions~\ref{def:projectiveMeas} through~\ref{def:measObservableDestructive}, on pages~\pageref{def:projectiveMeas}\thru\pageref{def:measObservableDestructive}) might arise from unitary evolution, represent an explicit departure from unitary evolution, or represent something entirely subjective, is a subject of ongoing research and debate (see \eg~\cite{BCFFS00,GRW86,Popper82,Wallace02,Zurek03}).
Such considerations are beyond the scope of this thesis: in practise, we adopt the approach of von Neumann~\cite{Neumann55}, and treat unitary evolution and measurements as complementary ways in which states may be transformed; and we supplement this also with maps of the form $\rho \mapsto \rho \ox \sigma$ for density operators $\rho$ and $\sigma$, corresponding to augmenting the system by introducing an auxiliary system in the state $\sigma$.

\paragraph{A remark on continuous time evolution.}

As we have noted above, Schr\"odinger evolution of physical systems can be described as being continuous in time.
Despite this, we will generally describe computation as being performed by \emph{discrete} transformations, \eg\ by applying unitary transition operators, rather than Schr\"odinger evolution.
The reasons for this are practical in nature: because we can only perform operations with finite precision in any particular setting, and because arbitrarily high precision can only be achieved with a commensurate decrease in the speed of computation, we limit ourselves to models of quantum computation where (like measurement and the introduction of auxiliary systems) the evolution of a quantum system is performed in discrete steps.
As well, the same unitary operation can be realized in different ways by Schr\"odinger evolution: for instance, considering time-independent Hamiltonians alone, a unitary $U_\tau$ can be achieved in time $\tau$ by Schr\"odinger evolution by any Hamiltonian $H$ satisfying $U_\tau = \exp(- i \hbar\, H \tau)$, where
\begin{gather}
		\exp(A)
	\;\;=\;\;
		\sum_{n \in \N} \sfrac{A^n}{n!}
\end{gather}
for any Hermitian matrix $A$; the set of such Hamiltonians is an equivalence class of operators with common eigenvectors, and whose eigenvalues differ by integer multiples of $2\pi$.
By considering discrete-time unitary evolution, we abstract away details such as the particular Hamiltonian, which in an implementation may be determined later.

\subsection{Elementary sets of gates}

To perform computations, we suppose that we are provided a set of elementary operations (or \emph{elementary gates}) as follows:
\begin{definition}
	\label{def:elemGatesTypes}
	A set $\Gates$ of elementary gates on a set of classical registers $B$ and a set of quantum registers $V$ is a set of operators $\Op{\sigma:\alpha/\delta}{}$\,, each of which consists of some CPTP map
	\begin{gather}
		\Op{}{} \;:\; \D\big(\ens{0,1}^{n+m}\big) \;\to\; \D\big(\ens{0,1}^{n+\ell}\big)	\;,
	\end{gather}
	which operates on finite sequences of registers given by $\sigma$, $\alpha$, and $\delta$ (where the input and output distributions of $\Op{}{}$ are also consistent with the additional constraints imposed by the state space for bits as classical registers).
	In particular, $\alpha = (a_j)_{j = 1}^\ell$ is a sequence of \emph{allocated} registers which are produced as output but which were not present as input; $\delta = (d_j)_{j = 1}^m$ is a sequence of \emph{discarded} registers which are taken as input but not produced as output; and $\sigma = (s_j)_{j = 1}^n$ is a sequence \emph{stable} registers taken as input and produced as output, yielding an operation
	\begin{gather}
		\Op{\sigma:\alpha/\delta}{} \;:\; \D\big(\sigma;\delta\big) \;\to\; \D\big(\sigma;\alpha\big)	\;.
	\end{gather}
	We say that the \emph{type} of the operation is $[\sigma:\alpha/\delta]$.
\end{definition}
The sequence of registers for an elementary operation, once fixed, is significant.
This may be illustrated with a map $\Meas{a,b:r/\vide}{}$ performing a von Neumann measurement on one of two qubits $a$ and $b$, producing a single result bit $r$, defined by
\begin{gather}
		\Meas{a,b:r/\vide}{}(\rho)
	\;\;=\;\;
		\sum_{\beta \in \ens{0,1}}
			\Big(\idop_a \ox \ket{\beta}\bra{\beta}_b\Big) \,\rho\, \Big(\idop_a \ox \ket{\beta}\bra{\beta}_b\Big)  \ox \ket{\beta}\bra{\beta}_r \,;
\end{gather}
then, for \eg\ two qubits $u$ and $v$ in independent pure states, $\Meas{u,v:r/\vide}{}$ performs a measurement on $v$ (producing the result in $r$) while leaving the state of $u$ undisturbed, while $\Meas{v,u:r/\vide}{}$ does the opposite.

In order for operations on bits and qubits --- or more generally, any register --- to be unambiguous, there must be exactly one ``copy'' of a given register which is to be operated upon; and any register to be operated on by some operation cannot have been explicitly discarded by the preceding operations without being re-allocated.
We formalize these constraints through the concept of the types defined for each gate, as follows:
\begin{figure}[t]
	\begin{center}
		\includegraphics[width=50mm,height=30mm]{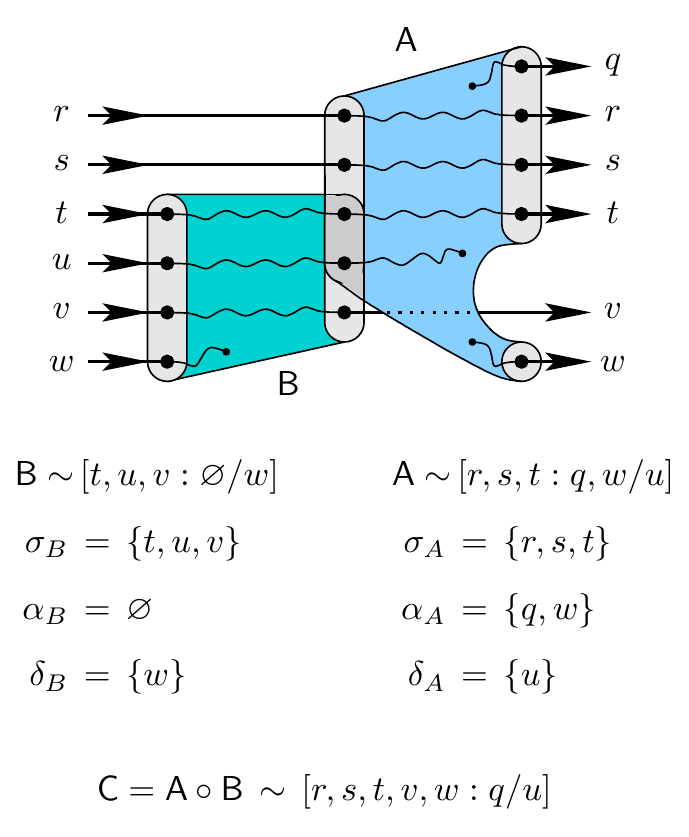}
	\end{center}
	\caption[Illustration of the type of the composition of two operations]{\label{fig:typeCompose}\colouronline%
		Illustration of the type of the composition of two operations $\mathsf A$ and $\mathsf B$ acting on named registers (represented as ``wires'' progressing in time from left to right).
		Registers not explicitly required by the input or output of $\mathsf B$ are assumed to be available to be either allocated, discarded, or transformed by $\mathsf A$ as necessary, and vice-versa.
		Otherwise, the types of $\mathsf A$ and $\mathsf B$ must agree on what bits or qubits are available to be operated on after the application of $\mathsf B$ and before the application of $\mathsf A$.
	}
\end{figure}
\begin{lemma}
	\label{lemma:typeCompose}
	For two operations $\mathsf{A}_{\sigma_A:\alpha_A/\delta_A}$ and $\mathsf{B}_{\sigma_B:\alpha_B/\delta_B}$ with types $[\sigma_A:\alpha_A/\delta_A]$ and $[\sigma_B:\alpha_B/\delta_B]$ respectively, the composition $\mathsf{A}_{\sigma_A:\alpha_A/\delta_A} \,\circ\, \mathsf{B}_{\sigma_B:\alpha_B/\delta_B}$ is well-formed if and only if
	\begin{romanum}
	\item
		there is no intersection between the registers in $\alpha_A$, and the registers in $\alpha_B$ or $\sigma_B$ which are explicitly specified in the output of $\mathsf{B}_{\sigma_B:\alpha_B/\delta_B}$\,;

	\item
		there is no intersection between the registers of $\delta_B$, and the registers in $\delta_A$ or $\sigma_A$ which are explicitly specified in the input of $\mathsf{A}_{\sigma_A:\alpha_A/\delta_A}$\,.
	\end{romanum}
	If the above conditions hold, the composition $\mathsf{C}_{\sigma_C:\alpha_C/\delta_C} = \mathsf{A}_{\sigma_A:\alpha_A/\delta_A} \,\circ\, \mathsf{B}_{\sigma_B:\alpha_B/\delta_B}$ has type $[\sigma_C:\alpha_C/\delta_C]$, where the sequences $\sigma_C$, $\alpha_C$, and $\delta_C$ are arbitrary orderings of the registers in the following ``set'' constructions:
	\begin{subequations}
	\label{eqn:typeCompose}
	\begin{align}
 			\sigma_C
		\;\;=&\;\;
			(\sigma_A \inter \sigma_B) \union (\alpha_A \inter \delta_B)	\,,
		\\[1ex]
			\alpha_C
		\;\;=&\;\;
			(\alpha_A \setminus \delta_B) \union (\alpha_B \setminus \delta_A)	\,,
		\\[1ex]
			\delta_C
		\;\;=&\;\;
			(\delta_B \setminus \alpha_A) \union (\delta_A \setminus \alpha_B)	\,.
	\end{align}
	\end{subequations}
\end{lemma}
The type composition relations of~\eqref{eqn:typeCompose} are illustrated in Figure~\ref{fig:typeCompose}.\footnote{%
The emphasis given here on the composability of CPTP maps is meant largely to prefigure similar discussions for particular, ``non-standard'' representations of quantum computation.}

A particular set $\Gates$ of elementary operations defines a specific \emph{circuit model}, consisting of the well-formed compositions of those gates:
\begin{definition}
	Given a set $\Gates$ of elementary gates on a set $B$ of bits and $V$ of qubits,
	a \emph{circuit in the model defined by $\Gates$} (or \emph{in the $\Gates$ model}) is the composition $C$ of a finite sequence of operations of $\Gates$.
	The \emph{size} of such a circuit is the number of operations it uses, and the \emph{depth} of the circuit is the minimum integer $D$ such that the circuit may be decomposed in the form,
	\begin{gather}
			C \;\;=\;\; \ordprod[>]_{t = 1}^D \big(\Op{t,1}{} \ox \Op{t,2}{} \ox \cdots \ox \Op{t,m_t}{}\big)
	\end{gather}
	\ie\ as an ordered product of $D$ parallel (tensor) products of gates on independent sets of qubits.
\end{definition}

From this point onwards, we will consider only models of quantum computation acting on bits and qubits (\ie\ where the set of elementary gates only acts on aggregates of registers over $\ens{0,1}$).
We also assume that the number of available bits and qubits is infinite, and that the only limitation on the number of qubits which a circuit may operate upon is imposed by the set of elementary gates.

\begin{notation}
	Operators $U$ written in italics represent linear operators (\eg\ unitary transformations), and superscripts denote exponents of these operators.
	Operators $\Upupsilon$ written in upright sans-serif represent CPTP maps.
	Superscripts of CPTP maps denote either real-valued parameters or parameters provided by an interactive classical control, rather than exponents.
	For two CPTP maps $\mathsf A$ and $\mathsf B$, we will often write $\mathsf A \mathsf B$ for $\mathsf A \circ \mathsf B$ when this composition is well-defined (although we will continue to write in terms of compositions for much of this section).
\end{notation}

\subsection{Universality for quantum computation/unitary evolution}
\label{sec:universal}

In order to have a robust model of computation, it is helpful not only to have a set of discrete operations with which to define a sense of computational complexity, but also to have a well-defined notion of the range of operations which that set generates.
This motivates a discussion of senses of \emph{universality} of quantum operations.

\subsubsection{Universality for generating/approximating distributions}

For quantum computation on aggregates of qubits, the natural candidate is the the ability to generate all CPTP maps $\Upphi: \D(2^n) \to \D(2^n)$ for any $n \ge 1$.
Alternatively, if we wish to consider models of computation with only a finite set of elementary operations, we cannot generate the uncountably\footnote{%
``Uncountable'' is used here in the set-theoretic sense, meaning that the cardinality of the set of such maps is greater than that of the natural numbers.}
many CPTP maps from finite compositions from a finite set of operations; however, we may define a sense of \emph{approximate} universality by defining a sense in which a CPTP map may be approximated with some precision $\epsilon > 0$.
We will present a formal definition of approximating CPTP maps below.
However, defining universality in terms of generating or approximating arbitrary CPTP maps misses important models of computation (such as some models of measurement-based quantum computation, which we present in Section~\ref{sec:oneWayModel}) which do not take quantum states as input, but which achieve what is essentially the main practical motivation of quantum computation: a means of efficiently preparing a larger class of probability distributions via measurement than is apparently possible with, say, randomized Turing machines.

The complementary alternative --- considering only probability distributions that may be prepared by a set of gates --- presents no distinction between quantum and classical computation.
While the set of functions computable by classical and by quantum computers from explicitly specified primitives are the same, it is commonly thought that the functions which can be \emph{efficiently} computed by each are not the same.
We might wish to investigate descriptions of universality for quantum computation which strictly subsumes what classical computers can achieve, if we are interested in possible indications of the differences in what can be efficiently done by classical computation and by quantum computation.
One approach could be to explicitly consider computational resources in any discussion of universality: to describe \eg\ pertinent bounds (or lack of bounds) on time, space, multiparty communication, \etc\ when describing senses of universality.
A discussion of different senses of universality is presented in~\cite{NDMB2007}, which raises the issue of addressing universality and computational complexity at the same time.
However, for the sake of simplicity, it is more convenient to completely separate concerns of efficiency and universality.\footnote{%
The rarity of such senses of universality in the literature is likely due, no doubt, to the success of the Strong Church-Turing thesis, which posits that any ``physically realizable'' model of computation is polynomial-time equivalent to the Turing machine model of computation.
Even in light of the apparent failure of the Strong Church-Turing thesis with the advent of quantum computation, there seems to be remarkably few natural senses of universal computation arising out of ``physically realistic'' models of computation which are not provably poly-time equivalent, except for those which are equivalent to constraining another ``physically realistic'' model in some way.}

A reasonable compromise position is to define a sense of universality which strictly extends the sense of generating probability distributions, but which does not presume that arbitrary quantum inputs are possible inputs to the operations of the model --- that is, simply the generation or approximation of arbitrary density operators.
The natural distance on probability distributions is that induced by the $\ell_1$-norm, which --- using the correspondence of~\eqref{eqn:embedHilbertSchmidt} --- is the sum of the absolute values of the entries of a diagonal matrix.
By~\cite[Lemma~10.2]{KSV2002}, the \emph{operator $\ell_1$-norm} (also called the trace norm) on linear operators $M$ between finite-dimensional Hilbert spaces,
\begin{subequations}
\begin{align}
		\norm{M}_1
	\;\;=&\;\;
		\sum_{j} \,\sigma_j	\;,
		&&\text{\small where $\sigma_1 \ge \sigma_2 \ge \ldots \ge 0$ are the singular values of $M$},
	\\=&\;\;
		\Tr\paren{\sqrt{M\herm M\,}}	\;,
		&&\text{\small for $\sqrt{M\herm M\,} = S$ such that $S \ge 0$ and $S^2 = M\herm M$};
\end{align}
and the corresponding metric on density operators,
\begin{gather}
		d(\rho,\sigma)
	\;\;=\;\;
		\norm{\rho - \sigma}_1
\end{gather}
\end{subequations}
are the natural generalization to arbitrary density operators.
Then, we define a sense of universality for quantum computation as follows:
\begin{definition}
	\label{def:universalQC}
	For two density operators $\rho, \tilde\rho$, we say that \emph{$\tilde\rho$ approximates $\rho$ with precision~$\epsilon$} if $0 \le \norm{\tilde\rho - \rho}_1 < \epsilon$.
	A collection of elementary gates $\Gates$ is \emph{universal for quantum computation} if any density operator $\rho \in \D(2^n)$ on $n$ qubits can be generated as a circuit in the $\Gates$ model, for every $n \ge 1$; alternatively, $\Gates$ is \emph{approximately universal for quantum computation} if any such operator $\rho$ can be approximated to arbitrary precision $\epsilon > 0$ by some circuit in the $\Gates$ model.
\end{definition}
This definition of universality for quantum computation corresponds to the definition of ``\textbf{CQ}-universality'' presented in~\cite{NDMB2007}.

\subsubsection{Universality for generating/approximating CPTP maps}

Although generation or approximation of density operators is what we have chosen to define universality, it is certainly sufficient to be able to generate or approximate all CPTP maps on $n$ qubits (corresponding to ``\textbf{QQ}-universality'' as described in~\cite{NDMB2007}).
The ultimate significance of a CPTP map from the point of view of quantum computation is in the probability distributions arising from measurement.
A natural definition of an approximation $\tilde\Upphi$ of a CPTP map $\Upphi$ is then when the result of a measurement (\eg\ a destructive measurement as in Definition~\ref{def:destructiveMeas}) on either $\tilde\Upphi(\rho)$ or $\Upphi(\rho)$ yields similar probability distributions for any input state $\rho$; more generally, we may consider measurements performed on $\big[ \tilde\Upphi \ox \idop \big](\rho)$ and $\big[ \Upphi \ox \idop \big](\rho)$, where $\rho$ is a state of some larger system.
Then, considering upper bounds on the difference of the effects of two maps $\tilde\Upphi : \D(Q) \to \D(Q')$ and $\Upphi : \D(Q) \to \D(Q')$ on arbitrary density operators, we are led to the following metric on superoperators:
\begin{align}
		\Delta\big(\tilde\Upphi, \Upphi\big)
	\;\;=&\;\;
		\sup_{\card{A} \ge 1}\;\;\;
		\sup_{\rho \in \D(Q \x A)}
		\quad
		d\paren{\Big(\tilde\Upphi \ox \idop_A\Big)(\rho)\,,\; \Big(\Upphi \ox \idop_A\Big)(\rho)}
	\notag\\[1.5ex]=&\;\;
		\sup_{\card{A} \ge 1}\;\;
		\sup_{\substack{M \in \Herm(Q \x A) \\ M \ge 0,\, M \ne 0}}
		\frac{\norm{\Big(\tilde\Upphi \ox \idop_A\Big)(M) - \Big(\Upphi \ox \idop_A\Big)(M)}_1}{\norm{M}_1}	\;.
\end{align}
We may simplify this definition slightly by removing the emphasis on positive operators, and defining the \emph{superoperator norm} (introduced as the \emph{diamond norm} in~\cite{AKN98}), in the fashion of~\cite{KSV2002}:\footnote{%
This definition actually corresponds to the description of $\norm{\Uppsi}_\lozenge$ on page~109 and to the statement of Theorem~11.1 on page~110, in~\cite{KSV2002}.
The definition presented here also differs in that the descriptions in~\cite{KSV2002} do not involve a supremum over $\card{A}$.}
\begin{gather}
		\norm{\Uppsi}_\lozenge
 	\;\;=\;\;
 		\sup_{\card{A} \ge 1}\;\;
 		\sup_{\substack{M \in \L(\cH[Q] \ox \cH[A]) \\ M \ne 0}}
 		\frac{\norm{\Big(\Uppsi \ox \idop_A\Big)(M)}_1}{\norm{M}_1}	\;.		
\end{gather}
It is easy to see that the value of the inner supremum is non-decreasing with the size of $\card A$, as we may isometrically embed $\cH[Q] \ox \cH[A]$ into $\cH[Q] \ox \cH[A']$ for any $\card{A'} \ge \card{A}$.
For the outer supremum, it suffices to take $\card{A} = \card{Q}$, as shown in~\cite[Theorem~11.1]{KSV2002}; then, we may equivalently define
\begin{gather}
		\norm{\Uppsi}_\lozenge
 	\;\;=\;\;
 		\sup_{\substack{M \in \L(\cH[Q] \ox \cH[Q]) \\ M \ne 0}}
 		\frac{\norm{\Big(\Uppsi \ox \idop_Q\Big)(M)}_1}{\norm{M}_1}	\;.		
\end{gather}
Therefore, we present the following extension of Definition~\ref{def:universalQC} for the purposes of describing the generation or approximation of CPTP maps:
\begin{definition}
	\label{def:universalCPTP}
	For two CPTP maps $\Upphi, \tilde\Upphi$ with the same domain and range, we say that \emph{$\tilde\Upphi$ performs $\Upphi$} if $\tilde\Upphi = \Upphi$\,, and that \emph{$\tilde\Upphi$ approximates $\Upphi$ with precision~$\epsilon$} if \mbox{$0 \le \norm{\big.\smash{\tilde\Upphi - \Upphi}\big.}_\lozenge < \epsilon$}.
	A collection of elementary gates $\Gates$ is \emph{universal for CPTP transformation} if any CPTP map $\Upphi: \D(2^n) \to \D(2^m)$ from $n$ qubits to $m$ qubits can be performed by a circuit in the $\Gates$ model, for every $n,m \ge 1$; alternatively, $\Gates$ is \emph{approximately universal for CPTP transformation} if any such map $\Upphi$ can be approximated to arbitrary precision $\epsilon > 0$ by some circuit in the $\Gates$ model.
\end{definition}
Because a density operator $\rho \in \D(Q)$ is equivalent to a CPTP map $\D(1) \to \D(Q)$, (approximate) universality for CPTP transformation entails (approximate) universality for quantum computation.

\paragraph{Remark on approximation of composite CPTP maps.}
We can approximate any composition of CPTP maps $\Uppsi_N \circ \cdots \circ \Uppsi_2 \circ \Uppsi_1$ by using approximations to each of the maps $\Uppsi$, with an imprecision which is at worst the sum of the imprecisions of each of the components: that is, imprecision in approximations compose \emph{additively}, as shown in~\cite{AKN98}.
We may show this by first noting that for superoperators $\Uppsi: \L(\cH[Q]) \to \L(\cH[Q'])$ and $\Upphi: \L(\cH[Q']) \to \L(\cH[Q''])$, we have
\begin{align}
	\label{eqn:superoperatorNormSubMultiplic}
		\smaller{\norm{\Upphi \circ \Uppsi}_\lozenge}
		\;\;\smaller=&\quad
		\smaller{
	 		\sup_{\substack{M \in \L(\cH[Q] \ox \cH[Q]) \\ M \ne 0}}
 			\frac{\norm{\Big((\Upphi \circ \Uppsi) \ox \idop_Q\Big)(M)}_1}{\norm{M}_1}
		}
	\notag\\[1.8ex]\smaller=&\quad
		\smaller{
	 		\sup_{\substack{M \in \L(\cH[Q] \ox \cH[Q]) \\ M \ne 0}}
			\;\;
 			\sqparen{
				\frac{\Bigg.\norm{\bigg(\Upphi \ox \idop_Q\bigg)\Big[\Big(\Uppsi \ox \idop_Q\Big)(M)\Big]}_1}{\norm{\Big(\Uppsi \ox \idop_Q\Big)(M)}_1}
			}	
 			\sqparen{
				\frac{\Bigg.\norm{\Big(\Uppsi \ox \idop_Q\Big)(M)}_1}{\norm{M}_1}
			}
		}
	\notag\\[1.8ex]\smaller\le&\;\;
		\smaller{
	 		\sqparen{
				\sup_{\substack{M' \in \L(\cH[Q'] \ox \cH[Q']) \\ M \ne 0}}
 				\frac{\bigg.\norm{\Big(\Upphi \ox \idop_{Q'}\Big)(M')}_1}{\norm{M'}_1}
			}	
 			\sqparen{
				\sup_{\substack{M \in \L(\cH[Q] \ox \cH[Q]) \\ M \ne 0}}
				\frac{\bigg.\norm{\Big(\Uppsi \ox \idop_Q\Big)(M)}_1}{\norm{M}_1}
			}
		}
	\notag\\[1.8ex]\smaller=&\quad
		\smaller{
			\norm{\Upphi}_\lozenge \norm{\Uppsi}_\lozenge	\;,
		}
\end{align}
where we implicitly take the supremum over all auxiliary spaces for the second-last inequality.
It is possible to show that any CPTP map has unit superoperator norm:\footnote{%
Using the definition on page of $\norm{\Uppsi}_\lozenge$ in~\cite[page~110]{KSV2002}, and Theorem~11.1 which follows shortly after, this follows from Theorem~\ref{thm:churchHilbertSpace} below.
The fact that CPTP maps have \emph{at most} unit superoperator norm may alternatively be obtained from Theorem~\ref{thm:churchHilbertSpace} and the sub-multiplicative property shown above, together with the fact that $\norm{\Upphi \ox \Uppsi}_\lozenge = \norm{\Upphi}_\lozenge \norm{\Uppsi}_\lozenge$\,, and from the fact that the trace operation and isometries each have unit superoperator norm.}
then, if $\tilde\Uppsi: \L(Q) \to \L(Q')$ is a CPTP map approximating $\Uppsi$ with precision $\epsilon$ and $\tilde\Upphi: L(Q') \to \L(Q'')$ is a CPTP map approximating $\Upphi$ with precision $\epsilon'$, we have
\begin{align}
	\label{eqn:errorAdditiveCPTP}
		\norm{\tilde\Upphi \circ \tilde\Uppsi - \Upphi \circ \Uppsi}_\lozenge
	\;\;\le&\;\;
		\norm{\tilde\Upphi \circ \big(\tilde\Uppsi - \Uppsi\big)}_\lozenge \;+\;
		\norm{\big(\tilde\Upphi - \Upphi\big) \circ \Uppsi}_\lozenge
	\notag\\[1ex]\le&\;\;
		\norm{\tilde\Upphi}_\lozenge \norm{\tilde\Uppsi - \Uppsi}_\lozenge \;+\;
		\norm{\tilde\Upphi - \Upphi}_\lozenge \norm{\Uppsi\Big.}_\lozenge
	\;\;<\;\;
		\epsilon + \epsilon'\;.
\end{align}
Thus, to approximate a composite CPTP map within a small margin of error, it suffices to approximate some set of gates which is universal for CPTP transformation, with sufficiently good precision to bound the cumulative error.

\subsubsection{Universality for generating/approximating unitary transformations}

One of the primary tools of quantum computation is to reduce this sense of universality for quantum computation to the ability to perform/approximate unitary transformations in particular.
This is possible due to the following corollary (see \eg~\cite[Lecture~5]{Watrous07} and~\cite{KSW2008} for explicit treatments) to a theorem by Stinespring~\cite{Stinespring1955}:
\begin{theorem}
	\label{thm:churchHilbertSpace}
	For Hilbert spaces $\cH[]$ and $\cK$ of finite dimension, a superoperator $\Upphi: L(\cH[]) \to \L(\cK)$ is a CPTP map if and only if, for some auxiliary Hilbert space $\cE$ and some isometry $T: \cH[] \to \cK \ox \cE$, we have
	\begin{gather}
	\label{eqn:churchHilbertSpace}
			\Upphi(\rho)
	 	\;\;=\;\;
			\tr[\cE](T \rho \:T\herm)\;.
	\end{gather}
\end{theorem}
If we define $\sfT(\rho) = T \rho \,T\herm$, this is a decomposition of $\Upphi$ into an isometry and a trace-out operation.
When $\cH[]$, $\cE$, and $\cK$ as above are restricted to tensor products of $\cH$, the isometry $T: \cH[] \to \cK \ox \cE$ can be decomposed into the preparation of an arbitrary state $\ket{\phi} \in \cA \cong \cH^{\ox m}$ (for some number $m$ of auxiliary qubits such that $\cH[] \ox \cA \cong \cK \ox \cE$), and a unitary transformation $U: \cH[] \ox \cA \to \cK \ox \cE$\,:
\begin{gather}
		T
	\;\;=\;\;
		U \big[\idop_{\cH[]} \ox \ket{\phi}_{\cA}\big]	\;.
\end{gather}
Then, the decomposition of a CPTP map $\Upphi$ on qubits as in~\eqref{eqn:churchHilbertSpace} can be further elaborated as being of the form
\begin{subequations}
\begin{gather}
			\Upphi(\rho)
	 	\;\;=\;\;
			\tr[\cE]\Big(U \big[\rho \ox \ket{\phi}\bra{\phi} \big] U\herm\Big)
		\;\;=\;\;
			\Big(\tr[\cE] \circ \Upupsilon \circ \mathsf A\,\Big)(\rho)	\;,
	\\[1ex]
		\text{where}\quad	\Upupsilon(\rho) \;=\; U \rho U\herm,	\qquad \mathsf A(\rho) = \rho \ox \ket{\phi}\bra{\phi}	.
\end{gather}
\end{subequations}
The characterization of CPTP maps by such decompositions, is an additional motivation (apart from the physical motivation of Schr\"odinger evolution) to analyze quantum computation in terms of unitary transformations, even if we are interested in CPTP maps in general.

In order to obtain a similar sense of universality as for CPTP maps, it will be useful to discuss a norm on operators analogous to the superoperator norm described above.
We define the \emph{operator sup-norm} of an operator $M: \cH[Q] \to \cH[Q']$ by
\begin{gather}
		\norm{M}_\infty
	\;\;=\;\;
		\sup_{\substack{\vec v \in \cH[Q] \\ \vec v \ne \vec 0}}
		\frac{\norm{M \vec v}_2}{\norm{\vec v}_2}	\;;
\end{gather}
it is easy to verify that $\norm{M}_\infty$ is the largest singular value of $M$.
By a similar derivation as in~\eqref{eqn:superoperatorNormSubMultiplic}, it is easy to prove that
\begin{subequations}
\begin{align}
		\norm{L M}_\infty
	\;\;\le&\;\;
		\norm{L}_\infty \norm{M}_\infty	\;;
\intertext{it is also easy to verify that the following properties hold:}
		\norm{M}_\infty
	\;\;=&\;\;
		1
		\;\;\;\text{for $U$ a unitary embedding},
	\\[0.3ex]
		\norm{M}_\infty
	\;\;=&\;\;
		\norm{M}_1
		\;\;\text{for $M$ with rank $1$},		
	\\[0.5ex]
		\norm{\big.\smash{M\herm}\big.}_\infty
	\;\;=&\;\;
		\norm{\big.M}_\infty	\;,
	\\[0.5ex]
		\norm{L \ox M}_\infty
	\;\;=&\;\;
		\norm{L}_\infty \norm{M}_\infty	\;.
\end{align}
\end{subequations}
Then for operators $U, \tilde U \in \U(N)$ and $\ket{\psi} \in \cH[N] \ox \cH[M]$ for positive integers $N$ and $M$, we have
\begin{align}
	\lefteqn{
		\smaller{
			\norm{
				(\tilde U \ox \idop_M) \ket{\psi}\bra{\psi} (\tilde U\herm \ox \idop_M)
				\,-\,
				(U \ox \idop_M) \ket{\psi}\bra{\psi} (U\herm \ox \idop_M)
			}_1
		}
	}\qquad\qquad
	\notag\\\smaller\le&\quad
 		\smaller{
			\norm{	(\tilde U \ox \idop_M) \ket{\psi}\bra{\psi} \Big(\big[\tilde U\herm - U\herm\big] \ox \idop_M\Big)	}_1
		\;\;+\;\;\;
			\norm{	\Big(\big[\tilde U - U\big] \ox \idop_M \Big) \ket{\psi}\bra{\psi} (U\herm \ox \idop_M)	}_1
 		}
	\notag\\\smaller=&\quad
 		\smaller{
			\norm{	(\tilde U \ox \idop_M) \ket{\psi}\bra{\psi} \Big(\big[\tilde U\herm - U\herm\big] \ox \idop_M\Big)	}_\infty
		\;\;+\;\;\;
			\norm{	\Big(\big[\tilde U - U\big] \ox \idop_M \Big) \ket{\psi}\bra{\psi} (U\herm \ox \idop_M)	}_\infty
 		}
	\notag\\\smaller\le&\quad
		\smaller{
			\norm{\Big.\, \tilde U \,\Big.}_\infty \; \norm{\Big.\ket{\psi}\bra{\psi}\Big.}_\infty \; \norm{\Big.\tilde U\herm - U\herm \Big.}_\infty
		\;\;+\;\;\;
		\norm{\Big.	\tilde U - U \Big.}_\infty \; \norm{\Big.\ket{\psi}\bra{\psi}\Big.}_\infty \; \norm{\Big. U\herm \Big.}_\infty
		}
	\notag\\\smaller=&\quad
		\smaller{	2 \norm{\tilde U - U}_\infty	\;.	}
\end{align}
By convexity, the same inequality holds if we replace $\ket{\psi}\bra{\psi}$ with any operator $\rho \in \D(N \x M)$.
As a result, for $\Upphi(\rho) = U \rho U\herm$ and $\tilde\Upphi(\rho) = \tilde U \rho \tilde U\herm$ we then have
\begin{gather}
	\label{eqn:unitaryApprox}
		\norm{\tilde\Upphi - \Upphi}_\lozenge
	\;\;\le\;\;
		2 \norm{\tilde U - U}_\infty	\;.
\end{gather}
Finally, note that replacing a unitary $U$ with some operator $U' = \e^{i \theta} U$ (for any real $\theta$) in the definition of the map $\Upphi$ as above yields the same CPTP map $\Upphi$.
Because our interest in unitary operators arises from the CPTP maps which they induce as above, Theorem~\ref{thm:churchHilbertSpace} together with~\eqref{eqn:unitaryApprox} motivates the following definition:%
\begin{definition}
	\label{def:universalUnitary-1}
	For two unitary operations $U, \tilde U \in \U(N)$ for some $N \ge 1$, we say that \emph{$\tilde U$ performs $U$} if $\tilde U$ is proportional to $U$, and that \emph{$\tilde U$ approximates $U$ with precision~$\epsilon$} if there is an angle $\theta \in \R$ such that $\norm{\big.\smash{\tilde U - \e^{i\theta} U}}_\infty \le \frac{\epsilon}{2}$\,. 
	A set $\sS$ of unitary operations is \emph{universal for unitary transformation} if any unitary $U$ on $n$ qubits can be performed by a composition of operators from $\sS$, for every $n \ge 1$; alternatively, $\sS$ is \emph{approximately universal for unitary transformation} if any such map $U$ can be approximated to arbitrary precision $\epsilon > 0$ by some product of operators from $\sS$.
\end{definition}
This definition differs from the standard treatments of approximate universality (see \eg~\cite{KSV2002,NC00}) by the multiplication of $\epsilon$ by $\frac{1}{2}$\,, which has been inserted for consistency with Definition~\ref{def:universalCPTP}, and by the insertion of the scalar factor $\e^{i\theta}$.
The former variation is inconsequential because of the Solovay-Kitaev Theorem, which we discuss below; the latter variation is made to facilitate description of unitary circuits, starting in Section~\ref{sec:unitaryCircuits}.

It is useful to note that a similar derivation as in~\eqref{eqn:errorAdditiveCPTP} allows us to show that errors in approximating unitaries are also additive: if $\norm{\big.\smash{\tilde U - U}}_\infty < \epsilon$ and $\norm{\big.\smash{\tilde V - V}}_\infty < \epsilon'$ for $U,V,\tilde U, \tilde V \in \U(N)$ for some $N \ge 1$, we then have
\begin{align}
	\label{eqn:errorAdditiveUnitary}
		\norm{\tilde U \tilde V - U V}_\infty
	\;\;\le&\;\;
		\norm{\tilde U \big(\tilde V - V\big)}_\infty \;+\;
		\norm{(\tilde U - U) V}_\infty
	\;\;<\;\;
		\epsilon + \epsilon'\;.
\end{align}

Because we will be interested in describing unitary CPTP maps with non-unitary CPTP maps, we may more generally define the following:
\begin{definition}
	\label{def:universalUnitary-2}
	A set $\Gates$ of elementary operations is \emph{universal for unitary transformation} if any unitary CPTP map $\Upphi$ from $n$ qubits to $n$ qubits, for any $n \ge 1$, can be performed by a composition of operations from $\Gates$, and is \emph{approximately universal for unitary transformation} if any such map $\Upphi$ can be approximated to arbitrary precision $\epsilon > 0$ by some composition of operations from $\Gates$.
\end{definition}
Note that for approximate universality for unitary transformation of a set $\Gates$, the approximating compositions may not themselves be unitary; they may only approach arbitrarily close to being unitary.
By Theorem~\ref{thm:churchHilbertSpace}, a set of elementary gates which is (approximately) universal for unitary transformation, allows for the preparation of states on arbitrarily many qubits, and which allows the tracing-out of qubits, is also (approximately) universal for CPTP transformation.

These results and concepts allow us to reduce the notion of universal quantum computation to (approximate) universality for unitary transformation: but there remains a question of how sensitive quantum computational complexity is to the particular choice of elementary operations.
This question was addressed by Kitaev~\cite[Lemma~4.7]{Kitaev97}, who notes that the result was independently discovered by Robert Solovay: detailed treatment of the result can be found in~\cite{DN2006,KSV2002}.
The following is a paraphrasing of the result as described by~\cite{DN2006}:
\begin{theorem*}[Solovay-Kitaev Theorem]
	Let $U \in \U(N)$ be a unitary operator for some $N \ge 1$, and $\sS$ a set of unitary operations which is closed under inversion and generates a dense subgroup of $\U(N)$.
	Then there is a $\polylog(1/\epsilon)$-time deterministic algorithm which produces a product $\tilde U \,=\, \tilde U_L \cdots \tilde U_2 \, \tilde U_1$\,, where $L \in \polylog(1/\epsilon)$ and $\tilde U_j \in \sS$ for each $1 \le j \le L$, such that $\tilde U$ approximates $U$ with precision $\epsilon$.
\end{theorem*}
There are several variations on the algorithm, with different complexities for the run-time and the size of the approximating product; the complexity for both described in~\cite{KSV2002} is $O(\log(1/\epsilon)^{3 + \delta})$ for arbitrary $\delta > 0$.
This result holds for any fixed $N$\,; however, the running-time of the algorithm for producing an approximating unitary $\tilde U \in \U(N)$ (\eg~as described by~\cite[Theorem~8.5]{KSV2002}) is exponential in $N^2$.
Nevertheless, from a theoretical point of view, the particular set of unitaries by which we describe unitary circuits is essentially a matter of convenience, provided that we compare only sets of elementary operations which act on a small, and fixed, number of dimensions.

\section{Unitary circuit models}
\label{sec:unitaryCircuits}

While quantum computation is likely in practise to involve some aspect of classical control, it is often presented in terms of decompositions of CPTP maps
\begin{gather}
	\label{eqn:unitaryCircuitDecomp}
		\Upphi \;\;=\;\; \Meas{}{} \circ \Upupsilon \circ \mathsf A	\;,
\end{gather}
where $\mathsf A$ prepares some number of qubits in a fixed joint state, $\Upupsilon$ is a unitary operation on the set of qubits (and in particular does not involve any explicit interaction with any classical data registers), and $\Meas{}{}$ is a composition of complete orthonormal (destructive) measurements on some of the qubits and trace-outs.
By the discussion following Theorem~\ref{thm:churchHilbertSpace} on page~\pageref{thm:churchHilbertSpace}, every CPTP map can be decomposed in such a form.

\vspace{1ex}\noindent
In this section, we consider models of computation of this type:
\begin{definition}
	\label{def:unitaryCircuitModel}
	A model of quantum computation is a \emph{unitary circuit model} if its elementary gates consist only of preparation of some number of qubits in a fixed initial state, unitary transformations of some number of qubits, complete (orthogonal \& destructive) measurements on some number of qubits, and trace-out operations.
\end{definition}
Because unitary quantum circuits do not have any operations which can act on classical bits, we may assume without loss of generality that all measurements are performed at the end; similarly, we may assume that any auxiliary qubits are introduced at the beginning, leading to a decomposition as in~\eqref{eqn:unitaryCircuitDecomp}.

\subsection{Particular unitary operations of interest}
\label{sec:particularUnitaries}

Throughout the rest of this thesis, we will be interested in referring to the particular unitary operations defined as follows (in matrix form):
\begin{definition}
	\label{def:Pauli}
	The \emph{Pauli operators} are the operators $\ens{\idop_2, X, Y, Z}$, where the latter three operations are defined by
	\begin{align}
			X
		\;\;=&\;\;
			\left[\begin{matrix} \,0 & \,1 \\ \,1 & \,0 \end{matrix}\,\right],
		&
			Y
		\;\;=&\;\;
			\left[\begin{matrix} \,0 & \!-i \\ \,i & \,0 \end{matrix}\,\right],
		&
			Z
		\;\;=&\;\;			
			\left[\begin{matrix} \,1 & \,0 \\ \,0 & \!-1 \end{matrix}\,\right];
	\end{align}
	and we define the $X$-rotation, $Y$-rotation, and $Z$-rotation operators for angles $\alpha \in \R$ by
	\begin{subequations}
	\begin{align}
			R_\sfx(\alpha)
		\;\;=&\;\;
			\exp(-iX\alpha/2)
		\;\;=\;\;
			\left[\begin{matrix}	\;\cos(\tfrac{\alpha}{2}) & -i\sin(\tfrac{\alpha}{2}) \\[1ex]
													 -i\sin(\tfrac{\alpha}{2}) & \;\cos(\tfrac{\alpha}{2}) \end{matrix}\,\right]	,
		\intertext{}
		\notag\\[-2em]
			R_\sfy(\alpha)
		\;\;=&\;\;
			\exp(-iY\alpha/2)
		\;\;=\;\;
			\left[\begin{matrix}	\;\cos(\tfrac{\alpha}{2}) & -\sin(\tfrac{\alpha}{2}) \\[1ex]
													 \;\sin(\tfrac{\alpha}{2}) & \;\cos(\tfrac{\alpha}{2}) \end{matrix}\,\right]	,			
		\intertext{}
			R_\sfz(\alpha)
		\;\;=&\;\;
			\exp(-iZ\alpha/2)
		\;\;=\;\;
			\left[\begin{matrix} \;\e^{\minus i \alpha/2} & \,0 \\[1ex] \,0 & \e^{i \alpha/2} \end{matrix}\,\right].
	\end{align}
	\end{subequations}
\end{definition}
\begin{definition}
	We define the \emph{controlled-$Z$ transform} $\cZ$, the \emph{controlled-$X$ transform} $\cX$, the \emph{Hadamard transform} $H$, and the \emph{$J(\alpha)$ transform} as follows:
	\begin{subequations}
	\label{eqn:basicGates}
	\begin{align}
			\cZ
		\;\;=&\;\;
			\left[\begin{matrix} \,1 & \,0 & \,0 & \,0 \\ \,0 & \,1 & \,0 & \,0 \\ \,0 & \,0 & \,1 & \,0 \\ \,0 & \,0 & \,0 & \!-1 \end{matrix}\right],
		&
			\cX
		\;\;=&\;\;
			\left[\begin{matrix} \,1 & \,0 & \,0 & \,0 \\ \,0 & \,1 & \,0 & \,0 \\ \,0 & \,0 & \,0 & \,1 \\ \,0 & \,0 & \,1 & \,0 \end{matrix}\,\right],
		\\[2ex]
			H
		\;\;=&\;\;
			\sfrac{1}{\sqrt 2} \left[\begin{matrix} \,1 & \,1 \\[0.5ex] \,1 & \!-1 \end{matrix}\right]	,
		&
			J(\alpha)
		\;\;=&\;\;
			\sfrac{1}{\sqrt 2}\left[\begin{matrix} \,\e^{\minus i\alpha/2} & \,\e^{i \alpha/2} \\[0.5ex]
																 \,\e^{\minus i\alpha/2} & \!-\e^{i \alpha/2} \end{matrix}\right] .
	\end{align}
	\end{subequations}
\end{definition}

\subsection{Two related unitary circuit models}
\label{sec:standardCircuitModels}

We may use the unitaries defined in the previous section to describe two useful families of elementary gates, which are universal or approximately universal for unitary transformation for appropriate choices of angle parameters.
We describe how this may be done by summarizing the development presented in~\cite[Chapter~4]{NC00}.
We start with single-qubit unitaries.

First, note that an arbitrary single-qubit unitary $U \in \SU(2)$ can be decomposed as a product of $Y$ and $Z$ transformations,
\begin{gather}
		U
	\;\;=\;\;
		R_\sfz(\alpha) R_\sfy(\beta) R_\sfz(\gamma),
\end{gather}
for some angles $\alpha, \beta, \gamma \in (-\pi, \pi]$.
Note that we may decompose 
\begin{gather}
		R_\sfy(\beta)
	\;\;=\;\;
		R_\sfz(\pion 2) H R_\sfz(\beta) H R_\sfz(\mpion 2)\,,
\end{gather}
which implies that
\begin{gather}
	U \;\;=\;\; R_\sfz(\alpha + \tfrac{\pi}{2}) \,H\, R_\sfz(\beta) \,H\, R_\sfz(\gamma - \tfrac{\pi}{2}),
\end{gather}
so that $\SU(2)$ can be generated with $Z$-rotations and Hadamard transforms alone.
For approximation of single-qubit unitaries, following the development in \cite{BMPRV99}, the operations
\begin{align}
		J(\pion 4)
	\;=&\;\;
		H R_\sfz(\pion 4),
	&
		J(\mpion 4)
	\;=&\;\;
		H R_\sfz(\mpion 4)
	\;=\;
		H J(\pion 4)\herm H,
\end{align}
have eigenvalues $\e^{\pm i \theta}$ for $\theta = \arccos(\frac{2 + \sqrt 2}{4})$, which is an irrational multiple of $\pi$\,.
Such an angle then generates a dense subgroup of $\R / 2\pi\Z$, the group of real angles modulo complete cycles of $2\pi$.\footnote{%
We may effectively prove this by the Pigeon-Hole principle. For any $\epsilon > 0$, let $N$ be the smallest positive integer such that $\frac{\pi}{N} < \epsilon$.
We may divide the interval $[0,2\pi)$ into $N$ partitions $I_m = [\frac{2m\pi}{N}, \frac{2(m+1)\pi}{N})$, which we use to represent a connected subset $I_m + 2\pi\Z$ of the compact group $\R/2\pi\Z$.
Because $\theta$ is an irrational multiple of $\pi$, $(\theta + 2\pi\Z)$ is an element of $\R/2\pi\Z$ with infinite order.
As a result, there is at least one interval $I_{\widetilde m} + 2\pi\Z$ which contains at least two distinct elements $t, t'$ of the group $\gen{\theta + 2\pi\Z} \le \R/2\pi\Z$\,.
Then, there are integers $K$ and $L$ such that $(K\theta - L\theta) = (t - t') \in I_0 + 2\pi\Z$ --- that is, for some $M \in \Z$, there exists an angle $\theta' = K\theta - L\theta + 2M\pi \in [0, \frac{2\pi}{N})$.
Note that any element of $\R$ differs from a multiple of $\theta'$ by at most $\frac{\pi}{N} < \epsilon$.
Thus, for any $\epsilon$ and for any $x \in \R/2\pi\Z$, there exists a multiple of $\theta + 2\pi\Z$ which is within distance $\epsilon$ of $x$.}
Let $\ket{\spm \hat{\sfn}}$ represent the unit $\e^{\pm i \theta}$-eigenvectors of $J(\pion 4)$; then, if we define operators
\begin{gather}
		R_{\hat n}(\alpha)
	 \;\;=\;\;
		\e^{i \alpha} \ket{\splus \hat{\sfn}}\bra{\splus \hat{\sfn}}
		\;\;+\;\;
		\e^{-i \alpha} \ket{\sminus \hat{\sfn}}\bra{\sminus \hat{\sfn}}	\,,
\end{gather}
we may take powers of $J(\pion 4) = R_{\hat \sfn}(\theta)$ to obtain arbitrarily good approximations to $R_{\hat n}(\alpha)$ for any angle $\alpha \in (-\pi, \pi]$.
Similarly, if $\ket{\spm \hat{\sfm}}$ represent the unit $\e^{\pm i \theta}$-eigenvectors of $J(\mpion 4)$, and we define the operators
\begin{gather}
		R_{\hat \sfm}(\alpha)
	 \;\;=\;\;
		\e^{i \alpha} \ket{\splus \hat{\sfm}}\bra{\splus \hat{\sfm}}
		\;\;+\;\;
		\e^{-i \alpha} \ket{\sminus \hat{\sfm}}\bra{\sminus \hat{\sfm}}	\,,
\end{gather}
we may take powers of $J(\mpion 4) = R_{\hat \sfm}(\theta)$ to obtain arbitrarily good approximations to $R_{\hat \sfm}(\alpha)$ for any angle $\alpha \in (-\pi, \pi]$.
By~\cite[Section~4.2]{NC00} (but see~\cite{NC00errata} for an erratum), because $J(\pion 4)$ and $J(\mpion 4)$ do not commute, we may then approximate any unitary $U \in \SU(2)$ to arbitrary precision using the group generated by $\ens{\big.J(\pion 4),\, J(\mpion 4)\big.}$.

The set of operations $\ens{\cX, U}_{U \in \U(2)}$ is universal for unitary transformation~\cite[Section~4.5.2]{NC00}.
By the preceding remarks about single-qubit unitaries, we may perform arbitrary $n$-qubit unitaries using controlled-not transformations, Hadamard transformations, and arbitrary $Z$-rotations; and using the remarks on approximation of unitary operations in Section~\ref{sec:universal}, we approximate arbitrary unitaries (up to scalar factors) using controlled-not transformations and $J(\spm\pion 4)$ transformations.
Finally, using the following equivalencies,
\begin{align}
		\cX
	\;=&\;\;
		(\idop_2 \ox H) \,\cZ\, (\idop_2 \ox H)	\,,
	&
		J(0)
	\;=&\;\;
		H	,
	&
		J(\alpha)
	\;=&\;\;
		H R_\sfz(\alpha)	\,,
\end{align}
we obtain the following:
\begin{lemma}
	\label{lemma:universalUnitaries}
	For a set of angles $\A \subset \R$, the two parameterized sets of unitary transformations $\ens{H, R_\sfz(\alpha),\cZ}_{\alpha \in \A}$ and $\ens{J(\alpha), \cZ}_{\alpha \in \A}$ are both universal for unitary transformation when $\A = \R$, and approximately universal for unitary transformation when $\A = \frac{\pi}{4}\Z$.
\end{lemma}
Because $R_\sfz(\mpion 4) = R_\sfz(\frac{7\pi}{4})$ and because we may approximate $H$ to arbitrary precision using $J(\spm \pion 4)$, we have actually shown that the sets $\ens{H, R_\sfz(\pion 4),\cZ}$\,, $\ens{J(0), J(\pion 4),\cZ}$, and $\ens{J(\mpion 4), J(\pion 4),\cZ}$ are approximately universal for unitary transformation; however, the somewhat weaker statement presented in Lemma~\ref{lemma:universalUnitaries} will be more useful in further discussion, as each set of unitaries described can be described by the same family of angles, and can be used to exactly generate the other.

We define the following unitary CPTP maps corresponding to the unitaries described in Lemma~\ref{lemma:universalUnitaries}, acting on density operators $\rho \in \D(2)$ and $\varrho \in \D(2 \x 2)$:
\begin{subequations}
\label{eqn:cptpElemUnitaries}
\begin{align}
		\sfH(\rho)
	\;\;=&\;\;
		H \rho H	\,,
	&
		\sfR^\alpha(\rho)
	\;\;=&\;\;
		R_\sfz(\alpha) \,\rho\, R_\sfz(\alpha)\herm	\,,
	\\
		\sfJ^\alpha(\rho)
	\;\;=&\;\;
		J(\alpha) \,\rho\, J(\alpha)\herm\,,
	&
		\Ent{}(\varrho)
	\;\;=&\;\;
		\cZ\,\varrho\,\cZ	\,.
\end{align}
\end{subequations}
The map $\Ent{}{}$ we will tend to call an \emph{entangling} or \emph{entangler} operation, referring to the fact that it maps states of two independent qubits to \emph{entangled} states (\ie\ joint states which are not independent), provided that neither of the qubits are initially in a state which is a convex combination of the standard basis.\footnote{%
The map $\rho \mapsto \Ent{}{}(\rho)$ is not the only map with this property, of course, but it is the map which will arise most often in this thesis.
In particular, $\Ent{}{}$ is the only entangling operation which is allowed in the one-way measurement-based model of quantum computing described in Section~\ref{sec:oneWayModel}.}
We may define three more non-unitary CPTP maps: preparation maps $\New{}{\sfx}$ and $\New{}{\sfz}$ for the pure states $\ket{\splus} = \ket{\splus \sfx}$ and $\ket{0} = \ket{\splus \sfz}$ respectively,
\begin{align}
	\begin{split}
		\New{v}{\sfx} : \ens{1}& \to \D(v)	\,,
	\\[1ex]
		\New{v}{\sfx}(1) \;\;=&\;\; \ket{\splus \sfx} \bra{\splus \sfx}_v	\,,
	\end{split}
	&
	\begin{split}
		\New{v}{\sfz} : \ens{1}& \to \D(v)	\,,
	\\[1ex]
		\New{v}{\sfz}(1) \;\;=&\;\; \ket{\splus \sfz} \bra{\splus \sfz}_v	\,;
	\end{split}
\end{align}
and complete measurements $\Meas{}{\sfz}$ in the standard basis (discarding the qubit $v$ on which it operates and allocating a bit $\s[v]$ to store the result),
\begin{align}
	\begin{split}
	\begin{gathered}
			\Meas{v}{\sfz}
		\,:\,
			\D(v) \to \Prob(\s[v])
		\\[1ex]
			\Meas{v}{\sfz}(\rho)
		\;\;=\;\;
			\sum_{r \in \ens{0,1}} \bra{r}_v \rho \,\ket{r}_v \,\ox\, \ket{r}\bra{r}_{\s[v]}	\;.
	\end{gathered}
	\end{split}
\end{align}
Then, we define the following sets of gates:
\begin{definition}
	\label{def:elemGateSets}
	For a set $\A \subset \R$, an infinite set $V$ of qubits, and an infinite set $B$ of bits, we define the following parameterized sets of elementary gates,
	\begin{subequations}
	\label{eqn:elemGateSets}
	\begin{align}
			\mspace{-50mu}\Phases(\A)
		\;\;=&\;\;
			\ens{\sfH_v\,,\; \sfR_v^\alpha\,,\; \Ent{u,v}\,,\; \New{v}{\sfz}\,,\; \New{v}{\sfx}\,,\; \Meas{v}{\sfz}\,,\;  \tr[v]
			\,\Big|\, \alpha \in \A \text{~and~} v,w \in V}\,,
		\\[1ex]
			\Jacz(\A)
		\;\;=&\;\;
			\ens{\sfJ_v^\alpha\,,\; \Ent{u,v}\,,\; \New{v}{\sfz}\,,\; \New{v}{\sfx}\,,\; \Meas{v}{\sfz}\,,\; \tr[v]
				\,\Big|\, \alpha \in \A \text{~and~} v,w \in V}\,.
	\end{align}
	\end{subequations}
	where the map $v \mapsto \s[v]$ is an injective map from $V$ to $B$.
\end{definition}	
The circuit models arising from these two sets of elementary operations are both unitary circuit models; by Lemma~\ref{lemma:universalUnitaries}, both are (approximately) universal for quantum computation for an appropriate choice of angles $\A \subset \R$.
As well, circuits over each gate set may be easily transformed to the other using the equivalences $\sfJ^\alpha = \sfH \circ \sfR^\alpha$ and $\sfR^\alpha = \sfH \circ \sfJ^\alpha$.

\subsection{Representations of unitary circuits}
\label{sec:repnsUnitaryCircuits}

One of the topics of this thesis is a description of transformations between unitary circuits, and the \emph{one-way measurement-based model} of quantum computation, which we describe in Chapter~\ref{One-way measurement-based computation}.
As well, we will often be interested in presenting unitary circuits diagrammatically.
We now present different ways in which unitary circuits may be represented.

\subsubsection{Standard representations of unitary circuits}

The two conventional ways of representing unitary circuits are as products of unitary operators (as we have been doing so far), and by \emph{circuit diagrams}.

When writing unitary circuits as products of operators, we conventionally fix representation of unitary operators with respect to the standard basis.
For instance, for an $n$-qubit unitary, we would write something of the form
\begin{gather}
	\label{eqn:genericTensor}
		U
	\;\;=\;\;
		\Big[\;{U_{\vec x}}^{\vec y}\;\Big]_{\vec x, \vec y \in \ens{0,1}^n}
\end{gather}
where superscripts denote row indices and subscripts denote column indices.
We identify the bit-positions in the strings $\vec x, \vec y \in \ens{0,1}^n$ with particular qubits:\footnote{%
This identification is an injection from the set $\ens{1,\ldots,n}$ to some subset $L$ of $n$ distinct qubits among the set of all qubits in the model of computation being considered; this induces a particular isometry between $\U\big(\cH^{\!\!\ox n}\big)$ and the unitary group $\U\big(\cH^{\!\!\ox V}\big)$, where $V$ is the set of \emph{all} qubits admitted by the model of computation.}
for a sequence $(v_j)_{j = 1}^n$ of such qubits, we may then write a particular unitary operation acting on these qubits as $U_{v_1,\ldots,v_n}$.
This describes a transformation of pure state vectors, which is how quantum computation is conventionally described for unitary circuit models.
Products of such operators are then defined as performing the appropriate operation on the qubits listed as operands, and performing the identity otherwise: for example, we may write
\begin{gather}
		W_{b,c} V_a U_{a,b,c} T_b \ket{\psi}_{b,c}
	\;\;=\;\;
		(\idop_2 \ox W) (V \ox \idop_2 \ox \idop_2) U (\idop_2 \ox \ket{\psi})
\end{gather}
for generic unitary operators $W \in \U(8)$, $V,T \in \U(2)$, $U \in \U(8)$, and $\ket{\psi} \in \cH^{\ox 2}$, labelling the tensor factors as $a$, $b$, and $c$ in order.
The operator specified by this notation clearly depends on the order of the composition, but only up to permutations of commuting operators.

\begin{figure}[t]
	\begin{center}
		\includegraphics{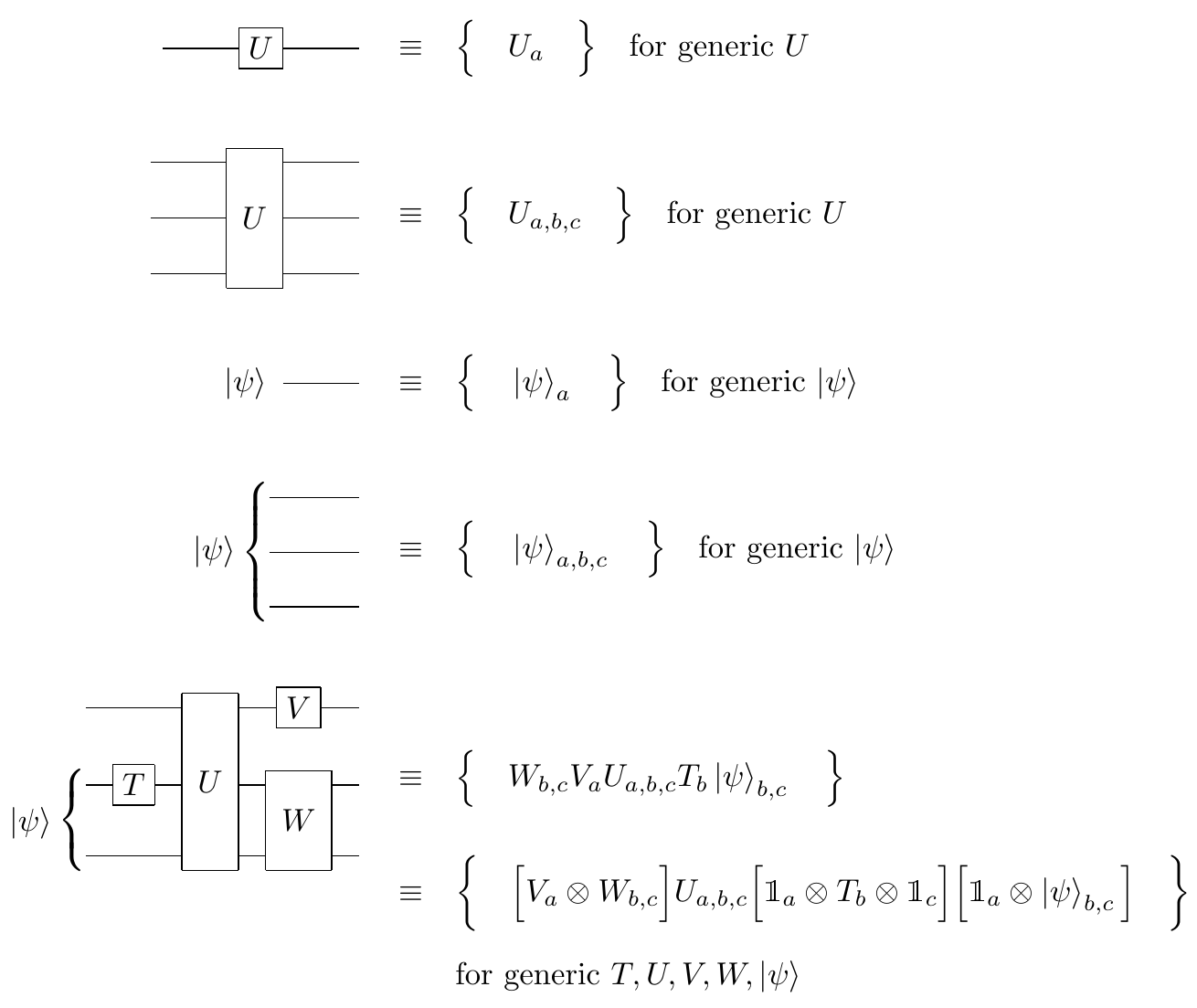}
	\end{center}
	\caption[Examples of simple quantum circuit diagrams.]{\label{fig:qcircuit-1}%
		Examples of simple quantum circuit diagrams.
		In the operator expressions for these examples, we arbitrarily label the qubits of each circuit from top to bottom in alphabetical order ($a, b, c, \ldots$).
		Diagrams compose from left to right, in contrast to the right-to-left composition in written notation; this represents an arrow of time in the diagrams from left to right.}
\end{figure}
\begin{figure}[tp]
	\begin{center}
		\includegraphics{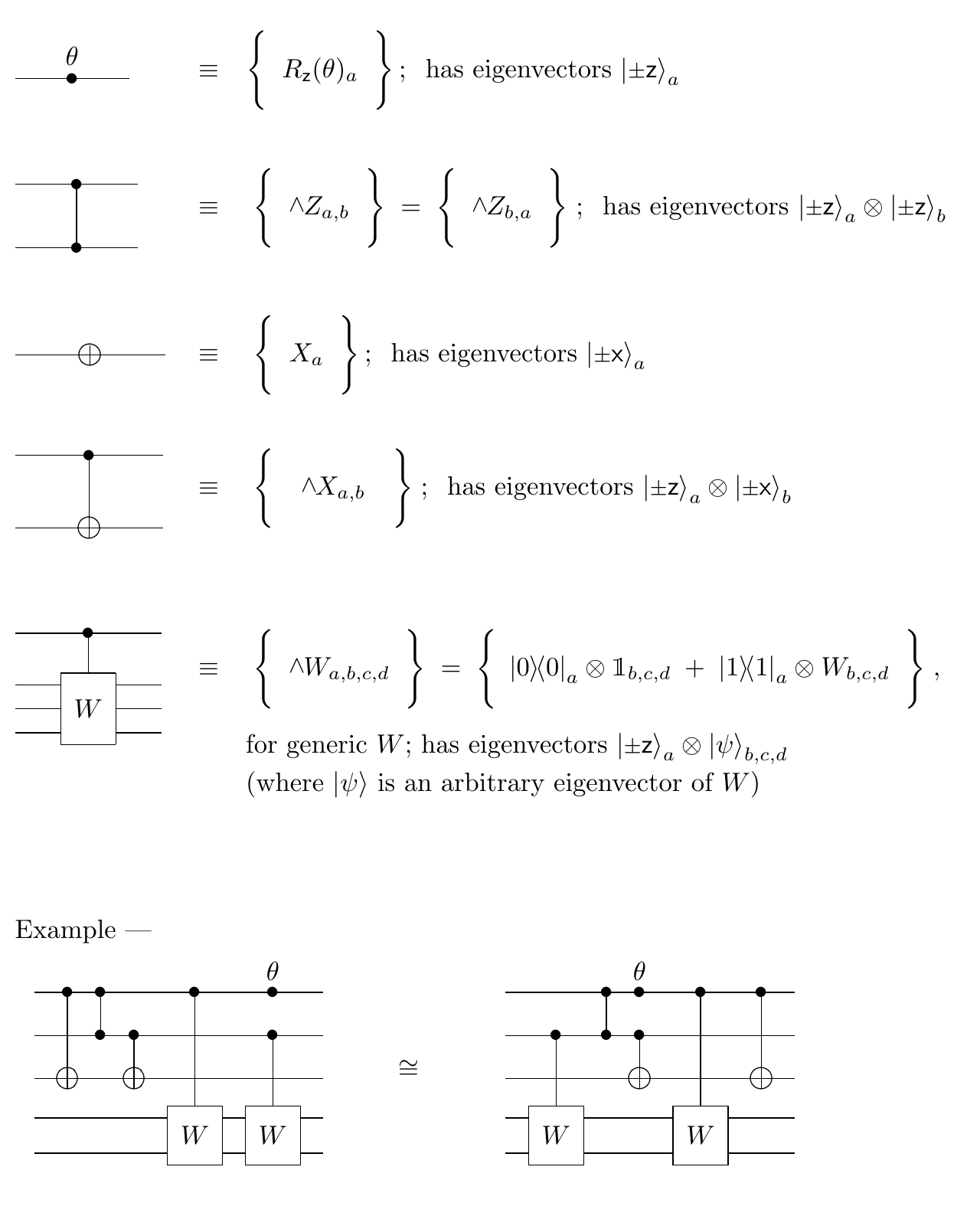}
	\end{center}
	\caption[Examples of special notations for quantum circuit diagrams.]{\label{fig:qcircuit-2}%
		Examples of special notations for quantum circuit diagrams.
		In the operator expressions for these examples, we again arbitrarily label the qubits of each circuit from top to bottom in alphabetical order ($a, b, c, \ldots$).
		The bottom-most circuits illustrate two equivalent circuits under commutation relations, where these relations are made visually apparent by common symbols on collections of wires.}
\end{figure}
Circuit diagrams are an alternative method of representing quantum circuits, as in Figures~\ref{fig:qcircuit-1} and~\ref{fig:qcircuit-2}.
Distinct qubits are represented by distinct \emph{wires} (often drawn as progressing from left to right, as already illustrated in Figure~\ref{fig:typeCompose}), schematically representing the trajectory of a qubit.
Operations on qubits are then drawn with boxes or other symbols representing multi-qubit interactions.
Figure~\ref{fig:qcircuit-1} illustrates simple examples of quantum circuit diagrams for some generic unitary operators and state vectors in terms of products of operators on specified qubits.
This notation allows operators which can be taken in tensor product to be represented more clearly as being simultaneously applicable, but is still strongly dependent on the sequence of operations specified --- and has difficulty expressing operations on qubits which are not adjacent in the linear arrangement of qubits in the diagram.

Circuit diagrams readily admit an improvement over the operator notation, by making commutation relations more apparent through the following notational devices.
Given an operator $U$ which operates on some collection of $n$ qubits, if the eigenvectors of $U$ can be expressed by vectors $\ket{\phi_1} \ox \cdots \ox \ket{\phi_m}$ where each $\ket{\phi_j}$ ranges over (possibly different) orthonormal bases for $\cH^{\!\!\ox n_j}$, where $n_1 + \cdots + n_m = n$ is a partition of $n$, we may represent $U$ using special symbols on each of the partition classes of qubits corresponding to the tensor decomposition of the eigenvectors.
Then, for different operators represented in this way, the operators commute whenever they act on common sets qubits and are represented by common symbols.
Standard notational devices are illustrated in Figure~\ref{fig:qcircuit-2}, along with an example which exploits these devices to show the equivalence of two circuits.
Such devices can also partially overcome the problem of specifying operations which act on non-adjacent qubits.

\subsubsection{Non-standard representations of unitary circuits}
\label{sec:nonstandardRepns}

One of the main topics of this thesis is translations between different representations of quantum computations.
Therefore, it will be useful to have a notation for unitary circuits in which irrelevant details such as the order of commuting gates are essentially absent, while being at the same time easy to produce by an algorithm.
This motivates the definition of an alternative representation for unitary circuits. 

One alternative notation (albeit one which is rare in the quantum computing literature) is Einstein summation notation~\cite[Section~B5]{Einstein1916}, where we provide different indices for the domain and range of each operator (\eg\ as in~\eqref{eqn:genericTensor} above, for the rows and columns), and identify the output indices of operators with the input indices of any subsequent operator acting on the same qubit.
We implicitly sum over any indices which are repeated (representing matrix multiplication of the matrix representation of the operators).
Thus, for example, for a composition of operators
\begin{subequations}
\begin{align}
	\label{eqn:arbitraryOperatorProduct}
		C
	\;\;=\;\;
		W_{b,c} V_a U_{a,b,c} T_b \ket{\psi}_{b,c}
\end{align}
for generic unitary operators $W \in \U(8)$, $V,T \in \U(2)$, $U \in \U(8)$, and $\ket{\psi} \in \cH^{\ox 2}$, we may equivalently write
\begin{align}
	{C_{a_0}}^{\!\!\!\!a_2,b_3,c_2}
	\;\;=\;\;
	\Big[{W_{b_2,c_1}}^{\!\!\!\!b_3,c_2}\Big]
	\Big[{V_{a_1}}^{\!\!\!\!a_2}\Big]
	\Big[{U_{a_0,b_1,c_0}}^{\!\!\!\!a_1,b_2,c_1}\Big]
	\Big[{T_{b_0}}^{\!\!\!\!b_1}\Big]
	\Big[\ket{\psi}^{b_0,c_0}\Big] .
\end{align}
Indices which are repeated are \emph{bound} variables over which we sum, and indices which are not repeated correspond to \emph{free} variables corresponding to rows (for subscripted indices) or columns (for superscripted indices) of a matrix representation of an operator.
(The brackets here are provided for additional emphasis.)
Because the sense of composition of tensors here is provided by the indices, we may freely rearrange the factors of such an expression without changing the meaning of the expression, which implies that the following composition of operators
\begin{align}
		\Big(\bra{a_2} \ox& \bra{b_3} \ox \bra{c_2}\Big) C \ket{a_0}
	\notag\\=&\;\;
		\sum_{a_1} \sqparen{\sum_{b_0,b_1,b_2} \paren{\sum_{c_0,c_1} 
			{W_{b_2,c_1}}^{\!\!\!\!b_3,c_2} \;\;
			{V_{a_1}}^{\!\!\!\!a_2} \;\;
			{U_{a_0,b_1,c_0}}^{\!\!\!\!a_1,b_2,c_1} \;\;
			{T_{b_0}}^{\!\!\!\!b_1} \;\;
			\ket{\psi}^{b_0,c_0}
		}} 
\end{align}
\end{subequations}
(where each index $a_h, b_j, c_k$ ranges over $\ens{0,1}$) can be expressed as a sum over a commuting product of scalars.
Thus, this notation has added flexibility over the conventional operator notation, in that the order of the factors is not significant.
However, because the meaning is preserved by indicating the order of the multiplication in the tensor indices, circuits which are equivalent up to rearrangements of commuting operations still give rise to inequivalent expressions: for instance, although it is easy to verify that the operations $\cZ_{a,b}$, $\cZ_{a,c}$, and $\cZ_{b,c}$ commute, the two products 
\begin{subequations}
\label{eqn:tripleCZproduct}
\begin{gather}
		\cZ_{b,c} \; \cZ_{a,c} \; \cZ_{a,b}
	\;\;\equiv\;\;
		{\cZ_{b_1,c_1}}^{\!\!\!\!b_2,c_2} \;\; {\cZ_{a_1,c_0}}^{\!\!\!\!a_2,c_1} \;\; {\cZ_{a_0,b_0}}^{\!\!\!\!a_1,b_1}	\,,
	\\[1ex]
		\cZ_{a,c} \; \cZ_{b,c} \; \cZ_{a,b}
	\;\;\equiv\;\;
		{\cZ_{a_1,c_1}}^{\!\!\!\!a_2,c_2} \;\; {\cZ_{b_1,c_0}}^{\!\!\!\!b_2,c_1} \;\; {\cZ_{a_0,b_0}}^{\!\!\!\!a_1,b_1}	\,,	
\end{gather}
\end{subequations}
still give rise to expressions which are not equivalent in Einstein notation by a relabeling of the indices and re-ordering of the terms.

A compromise between Einstein summation notation, and the approach of using special symbols in circuit diagrams, would be an Einstein-like tensor notation in which an index can be repeated multiple times when it corresponds to elements of an orthonormal basis which are preserved by several tensors --- that is, where the tensor indices may be interpreted as indicating elements of an orthonormal basis $\ens{\ket{\phi_j}}$ of some subset of the qubits, such that for several successive unitary operators $U$ in succession in the circuit, there exist a collection of unitaries $V_j$ for which $U \big( \ket{\phi_j} \ox \ket{\psi} \big) = \ket{\phi_j} \ox V_j \ket{\psi}$ holds for all states $\ket{\psi}$ of the remaining qubits.

\vspace{1ex}
\noindent To this end, we may consider a variant of Einstein summation notation, as follows:
\vspace{-0.8ex}
\begin{notation*}[Stable index tensor notation]
	\label{def:stableIndex}
	For a unitary operator $U$ acting on qubits $v_1, \ldots, v_n$, let $\sigma$ and $\delta$ be sequences of $n_1$ and $n_2$ qubits respectively (where $n = n_1 + n_2$) such that the eigenvectors of $U$ consist entirely of vectors $\ket{\vec{s}}_\sigma \ox \ket{\psi}_\delta$, for standard basis vectors $\ket{\vec s}$ on the qubits $\sigma$ and vectors $\ket{\psi}$ which are \emph{not known} to be decomposable as tensor products over the standard basis for qubits in $\delta$.
	We then write the matrix coefficients of $U$ in the standard basis by
	\begin{subequations}
	\label{eqn:stableIndex}
	\begin{gather}
			{U_{\vec x}}^{\vec y}
		\;\;=\;\;
				U\pseu[\vec s:\vec a/\vec d]	\;,
	\end{gather}
	where $\vec{x} = (\vec s ; \vec d)$ and $\vec{y} = (\vec s; \vec a)$ are decompositions of $\vec x$ and $\vec y$ into the appropriate substings.
	That is, we define $U\pseu[\vec s: \vec a / \vec d]$ by the operator equality
	\begin{align}
			U
		\;\;=&
			\sum_{\substack{\vec s \in \ens{0,1}^{n_1} \\ \vec{a},\vec{d} \in \ens{0,1}^{n_2}}}		\!\!
				U\pseu[\vec s:\vec a/\vec d]
				\;\; \Big( \ket{\vec s}\bra{\vec s}_\sigma \ox \ket{\vec a}\bra{\vec d}_\delta \Big)	\;.
	\end{align}
	\end{subequations}
	We call the indices of $\vec s$ \emph{stable}, the indices of $\vec d$ \emph{deprecated}, and the indices of $\vec a$ \emph{advanced}.
	The sequence of the advanced indices correspond to the same qubits being operated on as the sequence of deprecated qubits, in order.
	A product of such expressions is well-formed if each index is advanced at most once and deprecated at most once; and we sum implicitly over indices which are both advanced and deprecated in a product expression.
\end{notation*}

\begin{notation*}[A further shorthand notation]
	For a generic unitary operator $U \in \U(2^n)$ which is not known to have eigenvectors of the form $\ket{s}_{v_1} \ox \ket{\psi}_{v_2, \ldots, v_n}$ for arbitrary standard basis states $\ket{s} = \ket{\spm \sfz}$, for any qubit $v_1$ that $U$ acts upon, we may represent $U$ by the symbol $U\pseu[:\vec a/\vec d]$ (\ie\ without stable indices).
	Conversely, if the eigenvectors of such an operator $U$ are all standard basis vectors, we may represent $U$ by the symbol $U\pseu[\vec s]$ (\ie\ without advanced or deprecated indices).
	Finally, for a unitary embedding, there may be advanced indices which do not correspond to any deprecated indices; we may pad out the sequence of deprecated indices by a space, dot, or similar placeholder in this case.
\end{notation*}

\begin{example}
	For a product of generic operators $C = W_{b,c} V_a U_{a,b,c} T_b \ket{\psi}_{b,c}$ as described in~\eqref{eqn:arbitraryOperatorProduct}, we may write
	\begin{gather}
			C\pseu[:a_2,b_3,c_2/a_0,\;\cdot\;,\;\cdot\;]
		\;\;=\;\;
			W\pseu[:b_3,c_2/b_2,c_1]V\pseu[:a_2/a_1]U\pseu[:a_1,b_2,c_1/a_0,b_1,c_0]T\pseu[:b_1/b_0]\ket{\psi}\pseu[:b_0, c_0/] .
	\end{gather}
	Because the operators are not specified as preserving standard basis vectors over any of their operands, there are no stable indices; then, the meaning of this expression essentially reduces to that of Einstein notation, where we just sum over repeated indices (\ie\ indices which are advanced in some factor and deprecated in some factor).
\end{example}

\begin{example}
	\label{eg:commutingControls}
	For some generic unitary $W \in \U(4)$, we may define the operator 
	\begin{gather}
		\ctrl W \;=\; \ket{0}\bra{0} \ox \idop_2^{\,\ox 2} \;+\; \ket{1}\bra{1} \ox W
	\end{gather}
	analogously as in the example of Figure~\ref{fig:qcircuit-2}; then, the stable index representations of the circuits $\cX_{a,d} \,  \ctrl W_{a,b,c}$ and $\ctrl W_{a,b,c} \, \cX_{a,d}$ on four qubits may be given by
	\begin{align}
		\cX\pseu[a:d_1/d_0] \ctrl W\pseu[a:b_1,c_1/b_0,c_0]
		&&
		\text{and}
		&&
		\ctrl W\pseu[a:b_1,c_1/b_0,c_0] \cX\pseu[a:d_1/d_0] \;,
	\end{align}
	respectively.
	Note that these are equivalent up to permutations of the factors.
\end{example}
The informal semantics of stable index tensor notation is to provide a description of circuits in terms of computational paths.%
\label{discn:stableIndexPathIntegral}
In particular, stable index notation for circuits was partially inspired by the \emph{path-labelling} of circuits in~\cite{DHHMNO04}, which is explicitly concerned with descriptions of circuits in terms of computational paths: the tensor indices here correspond to labels of ``wire segments'' in that article, which are separated by \eg\ Hadamard gates, or other operations which do not preserve the standard basis.
For a given operator, the stable indices then represent qubits for which each vector of the standard basis is preserved, with coefficients being assigned (conditioned on a particular value of the stable indices) to transitions from certain values of the deprecated indices to various values of the advanced indices.
For products of operators which preserve standard basis vectors on a common set of qubits, it is not necessary (as is done in Einstein notation) to introduce distinct indices for the domain and the range of the operator, given that the coefficients for all of the cross-terms will be zero.
Stable index tensor notation then consists simply of elaborating Einstein notation by so-called ``stable'' indices.

By construction, there is a close analogy between \emph{stable, deprecated, and advanced indices} in the stable index notation, and \emph{stable, discarded, and allocated qubits} as described in Definition~\ref{def:elemGatesTypes}.\label{discn:typeCorrespondance}
This correspondence could be made exact if we imposed the constraint for any CPTP map $\Upphi$, we discard any qubit for which $\Upphi$ does not preserve the marginal distributions $\ket{0}\bra{0}$ and $\ket{1}\bra{1}$, substituting it if desired by a differently distributed qubit, and never re-allocating any qubit which have already been discarded.
This correspondence will play a useful role in our discussion of the one-way measurement-based model in Chapter~\ref{One-way measurement-based computation}.

\paragraph{Constructing stable index representations.}
We may automatically generate a stable index representation $S$ for a unitary circuit from a representation as a composition of operators $C$ on specified qubits, as follows.
Let $L$ be the set of qubit labels in $C$.
We may then transliterate the terms in $C$, defining a set of tensor indices incrementally as we do so, as follows:
\begin{enumerate}
\item
	For the first operator in $C$ to act on a given qubit $v$, we define a tensor index $v_0$\,, and designate $v_0$ as \emph{the current tensor index} for $v$.

\item
	For any operator $U_{a,b,\ldots}$ in $C$, we translate $U_{a,b,\ldots}$ into a stable index term $U\pseu[\vec s_U:\vec a_U/\vec d_U]$ with tensor indices as follows.
	For any qubit $v$ upon which $U$ acts, let $v_j$ be the current tensor index for $v$ (for some $j \in \N$).
	\begin{itemize}
	\item
		If $U_{a,b,\ldots}$ is a unitary embedding which allocates $v$, then $v_0$ is an advanced index in $\vec{a}_U$, without a corresponding deprecated index.

	\item
		If $U_{a,b,\ldots}$ is a unitary embedding which takes a state on $v$ as an input, and $U_{a,b,\ldots}$ does not preserve standard basis states on $v$, then we define a \emph{new} current tensor index $v_{j+1}$ for $v$; then $v_{j+1}$ is an advanced index in $\vec{a}_U$, corresponding to the deprecated index $v_j$ in $\vec{d}_U$.

	\item
		If $U_{a,b,\ldots}$ is a unitary embedding which takes a state on $v$ as an input, and $U_{a,b,\ldots}$ preserves standard basis states on $v$, then $v_j$ remains the current tensor index for $v$, and is a stable index in $\vec{s}_U$.
	\end{itemize}
	In particular, every new index which is advanced by a gate $U\pseu[\vec s:\vec a/\vec d]$ must be distinct from any other index being advanced, as well as from all of the indices which have occurred to that point.
	Performing this transliteration of operators, sequentially for all terms in $C$, produces the corresponding stable index tensor expression $S$.
\end{enumerate}
For a given circuit decomposition $C$ of a unitary operation, we will refer to the above construction (or one which differs from it only by a relabelling of indices) as the stable index tensor expression $S$ corresponding to $C$.
This construction leads to a natural (partial) mapping $f$ on the set $V(S)$ of indices of a stable index tensor expression $S$; for each index $v_j$, we define $f(v_j) = v_{j+1}$ if the latter is well-defined.
More generally, for each index $v$ which is deprecated in some term of $S$, we define $f(v)$ to be the corresponding index which is advanced.
If $I$ is the set of ``input indices'' (ones which are not advanced by any term in $S$) and $O$ the set of ``output indices'' (which are not deprecated by any term in $S$), $f$ is then an injective map from $V(S) \setminus O$ to $V(S) \setminus I$ by construction.

Example~\ref{eg:commutingControls} on page~\pageref{eg:commutingControls} illustrates the intuitive purpose of stable index notation: by allowing stable indices to be repeated multiple times while remaining ``free'' variables, we also discard some redundant combinatorial information about the order of the product of operators.
If we allow the terms of the product to be permuted arbitrarily, information about order of commuting operators is then lost completely.
Circuits which are congruent up to the re-ordering of commuting gates may then give rise to ``equivalent'' stable index tensor expressions, in the following sense:
\begin{definition}
	A \emph{homomorphism} of stable index tensor expressions is a bijection $\lambda$ of the \emph{index labels} of one stable index expression $S_1$ to those of another such expressions $S_2$, such that for any vector $(\vec s; \vec a; \vec d)$ of index labels for which there is an operator $U\pseu[\vec s:\vec a/\vec d]$ in $S_1$, there is an operator $W\pseu[\lambda(\vec s):\lambda(\vec a)/\lambda(\vec d)]$ in $S_2$, where we write $\lambda(\vec s) = \lambda(s_j)_{j = 1}^n = \big(\lambda(s_1), \ldots, \lambda(s_n)\big)$ , and similarly for $\lambda(\vec a)$ and $\lambda(\vec d)$.
	An \emph{isomorphism} of stable index tensor expressions is such a homomorphism where there is also a bijective mapping $\tau$ of \emph{terms} of $S_1$ to those of $S_2$, where we require
	\begin{gather}
			\tau\paren{\bigg.U\pseu[\vec s:\vec a/\vec d]}
		\;\;=\;\;
			U\pseu[\lambda(\vec s):\lambda(\vec a)/\lambda(\vec d)]
	\end{gather}
	for each term $U\pseu[\vec s:\vec a/\vec d]$ in $S_1$.
\end{definition}
\paragraph{Remark.}
Homomorphisms of stable index representations are \emph{combinatorial} homomorphisms, not \emph{algebraic} ones.
In particular, for two isomorphic stable index representations $S = U_n\pseu[\ast] \cdots U_2\pseu[\ast] U_1\pseu[\ast]$ and $S' = U'_n\pseu[\ast] \cdots U'_2\pseu[\ast] U'_1\pseu[\ast]$, whose isomorphism is witnessed by an appropriate bijection $\tau$ of the terms, it need not be the case that $\tau\big( U_j\pseu[\ast] \big) = U'_j\pseu[\ast]$.

Isomorphic stable index expressions represent not only equivalent unitary operations (which follows from the implicit summation convention), but equivalent circuit decompositions as well:
\begin{lemma}
	Let $C_1$ and $C_2$ be two sequences of unitary operators on a common set of qubits, composed from a common set of unitaries.
	If the stable index tensor representations of $C_1$ and $C_2$ are isomorphic, then $C_1$ and $C_2$ are congruent up to re-ordering of commuting operators.	
\end{lemma}
\begin{proof}
	Consider the coarsest equivalence relation $\cong$ on circuits which consist of permutations of the gates of $C_1$, such that $C \cong C'$ if $C$ differs from $C'$ by a transposition of commuting gates.
	By definition, the equivalence classes of $\cong$ are sets of circuits which are congruent up to re-ordering of commuting operators.
	Consider stable index tensor expressions $S$ and $S'$ arising from two unitary circuits $C$ and $C'$, where $S$ acts on the same index set as $S'$, and $S$ differs from $S'$ only by a transposition of two terms $U_1\pseu[\vec s_1: \vec a_1/\vec d_1]$ and $U_2\pseu[\vec s_2:\vec a_2/\vec d_2]$, where without loss of generality the former precedes the latter in $C$, and vice-versa in $C'$.
	Then, the corresponding gates $U_1$ and $U_2$ in $C$ occur with $U_1$ preceding $U_2$, and occur in $C'$ in the opposite order.
	Consider the index sets for these two operations in the stable index tensor expression:
	\begin{itemize}
	\item
		If the sequences of indices $(\vec s_1; \vec a_1; \vec d_1)$ and $(\vec s_2; \vec a_2; \vec d_2)$ do not overlap, the operations $U_1$	and $U_2$ act on disjoint sets of qubits, and so they commute.
	
	\item
		Suppose the sequences of indices $(\vec s_1; \vec a_1; \vec d_1)$ and $(\vec s_2; \vec a_2; \vec d_2)$ overlap, and let $v$ be an index occurring in both sequences.
		In the construction of $S$, because $v$ occurs in $U_2\pseu[\vec s_2:\vec a_2/\vec d_2]$, it cannot be a deprecated index in $U_1\pseu[\vec s_1:\vec a_1/\vec d_1]$\,; and because it occurs in $U_1\pseu[\vec s_1:\vec a_1/\vec d_1]$, it cannot be an advanced index in $U_2\pseu[\vec s_2:\vec a_2/\vec d_2]$.
		By the construction of $S'$, we similarly have that $v$ cannot be a deprecated index in $U_2\pseu[\vec s_2:\vec a_2/\vec d_2]$ or an advanced index in $U_1\pseu[\vec s_1:\vec a_1/\vec d_1]$.
		Then, $v$ is a stable index of both terms.
	\end{itemize}
	Therefore, the only qubits $v$ on which $U_1$ and $U_2$ both act are those whose standard basis states are preserved by both $U_1$ and $U_2$.
	Then, $U_1$ and $U_2$ commute, in which case $C \cong C'$.

	Let $S_1$ and $S_2$ be the stable index tensor representations of $C_1$ and $C_2$, where the isomorphism is given by an index relabelling map $\lambda$ and mapping of the terms $\tau$.
	Let $S'_2$ be the representation obtained by applying $\lambda$ to all of the indices in $S_2$\,: then $S_1$ and $S'_2$ consist of a product of the same terms in different orders.
	Without loss of generality, we will then suppose that $S_2$ and $S_1$ differ only by a reordering of terms.
	Because the terms of $S_1$ differ from those of $S_2$ by a permutation, by induction on the decomposition of this permutation into transpositions, we therefore have $C_1 \cong C_2$.
\end{proof}
Rearranging the terms of a stable index tensor expression for a circuit not only preserves the meaning (by summation over indices which are both advanced and deprecated) as a unitary operator, but also preserves structural information about the circuit itself, while discarding information about the ordering of commuting operators.
As a consequence, stable index notation facilitates identification of congruent circuits; and we may freely re-arrange the terms of such an expression while preserving a congruence class of unitary circuits.

However, not all congruent circuits have isomorphic tensor index expansions: congruent circuits may still yield non-isomorphic stable index expressions when the common operands of two commuting gates do not have their standard basis states preserved.
\begin{example}
	The two circuits $\cX_{c,b} \; \cX_{a,b} \; H_a$ and $\cX_{a,b} \; \cX_{c,b} \; H_a$ on three qubits $a,b,c$ have the following respective stable index expressions:
	\begin{align}
			\cX\pseu[c:b_3/b_2] \cX\pseu[a_1:b_2/b_1] H\pseu[:a_1/a_0]
		&&
			\text{and}
		&&
			\cX\pseu[a_1:b_3/b_2] \cX\pseu[c:b_2/b_1] H\pseu[:a_1/a_0]	\;.
	\end{align}
	These two expressions are non-isomorphic: in the first expression, the leftmost factor contains the indices $c$ (which is neither advanced nor deprecated in the entire expression) and $b_3$ (which is not deprecated in the entire expression); the second expression does not contain any factor with these properties.
	The inequivalence of the two expressions above is due to the fact that the two operations $\cX_{a,b}$ and $\cX_{b,c}$ preserve the basis $\ket{\spm \sfx}$ on $b$, rather than the basis $\ket{\spm \sfz}$.
\end{example}

One solution to this would be to diagonalize every element of $U$ of a circuit, decomposing them into diagonal operations flanked by ``virtual'' unitaries performing a some change of basis from the standard basis to the eigenbasis of $U$, essentially representing changes in reference frame rather than ``real'' unitary transformations performed by the circuit.\footnote{%
It is plausible that the ``virtual'' gates may represent in some sense \emph{actual} work that would be required to change the Hamiltonian of a physical system, in order to perform time-dependent unitary evolution in some physical implementations.
However, such issues lie outside the scope of this thesis.}
Interpreting the ``virtual'' unitaries strictly as a change of reference frame rather than as part of the evolution of the system, the diagonal representations of each gate $U$ would then represent preservation of the eigenbasis of $U$ in place of the standard basis.
For sequences of operators which commute, a judicious choice of change-of-basis operators would then allow the diagonalized gates to share stable indices in common.
However, we can avoid such decompositions into ``real'' and ``virtual'' unitary operations if we consider elementary gate sets for which the commutation relations\footnote{%
In this context, by ``commutation relations'' we mean the relation of \emph{whether} or not two unitaries commute, for all pairs of unitaries in the set, without any further comment on \eg\ the values of commutators of these operators.} 
are sufficiently restricted:
\begin{definition}
	\label{def:parsimonious}
	A set $S$ of unitary operations is \emph{parsimonious} if the eigenvectors of every multi-qubit gate in $S$ are standard basis states, and if distinct gates in $S$ which act on the same qubits commute if and only if their eigenvectors are standard basis states.\footnotemark
\end{definition}\footnotetext{%
The title of~\cite{DKP06} predates this definition of ``parsimonious'' by four years, and does not refer to the concept we have defined here; however, as we show immediately below, the set of unitary operations described there is parsimonious in this sense.}
\label{discn:elemParsimonious}
For example, for any set $\A \subset \R$, the two universal sets of unitaries $\ens{H, R_\sfz(\alpha),\cZ}_{\alpha \in \A}$ and $\ens{J(\alpha), \cZ}_{\alpha \in \A}$ described in Section~\ref{sec:particularUnitaries} are parsimonious.
This is evident for the former set of unitaries; for the latter, it is sufficient to note that
\begin{align}
		J(\alpha)\herm J(\beta)\herm J(\alpha)J(\beta)
	\;\;=&\;\;
		R_\sfz(-\alpha) \,H R_\sfz(-\beta) \,H \;\! H \;\! R_\sfz(\alpha) \,H R_\sfz(\beta)
	\notag\\=&\;\;
		R_\sfz(-\alpha) R_\sfx(\alpha-\beta) R_\sfz(\beta)	\;,
\end{align}
which is equal to $\idop_2$ if and only if $\alpha = \beta$.

The restrictions imposed on parsimonious sets of unitary operations make it simple to determine whether two operations on some collection of qubits commute.
This allows us to prove the following:
\begin{lemma}
	Let $C_1$ and $C_2$ be two unitary circuits composed from a common parsimonious set of unitaries.
	If $C_1$ is congruent to $C_2$ up to re-ordering of commuting operators, then the stable index representations of $C_1$ and $C_2$ will be isomorphic.
\end{lemma}
\begin{proof}
	Let $K_0 \cong K_1 \cong \cdots \cong K_\ell$ be a sequence of circuits which differ only by a transposition of two commuting gates, where $C_1 = K_0$ and $C_2 = K_\ell$.
	We may show by induction on $\ell$ that the stable index tensor expressions for $C_1$ and $C_2$ are isomorphic by induction on $\ell$ by proving the case $\ell = 1$.

	Let $S_0$ and $S_1$ be the stable index representations of $K_0$ and $K_1$.
	Up to a relabeling, we may assume that the indices of $S_0$ and $S_1$ are all of the form $v_0, v_1, \ldots$ for different qubits $v$ acted on by $K_0$ and $K_1$: we set $v_0$ to be the first index corresponding to a given qubit in either $S_0$ or $S_1$, and $v_{j+1}$ to be an advanced index corresponding to each deprecated index $v_j$ for $j \in \N$.
	Let $U_1$ and $U_2$ be the pair of gates whose order differs between the two circuits; and consider the truncation $S'_0$ of $S_0$ just prior to the terms corresponding to $U_1$ and $U_2$, and similarly for $S'_1$.
	Because $K_0$ and $K_1$ differ only by a transposition of commuting gates $U_1$ and $U_2$, $S'_0 = S'_1$.
	Let $S''_0$ and $S''_1$ be the truncations of $S_0$ and $S_1$ just after the terms corresponding to $U_1$ and $U_2$\,:
	\begin{itemize}
	\item
		If $U_1$ and $U_2$ act on disjoint sets of qubits, the terms corresponding to $U_1$ and $U_2$ in $S''_0$ may differ from those in $S''_1$ only by the indices which are advanced in each; and as the indexing scheme we have fixed is the same for each, and the indices which will be advanced for $U_1$ and $U_2$ are independent of each other in both expressions, $S''_0$ differs from $S''_1$ only by a transposition of those two terms.

	\item
		If $U_1$ and $U_2$ both have the standard basis as their eigenbases, their corresponding terms in $S_0$ and $S_1$ only have stable indices, and therefore do not advance any new indices.
		Then the term in $S_0$ and in $S_1$ corresponding to $U_1$ are the same, and similarly for $U_2$\,; the difference between $S''_0$ and $S''_1$ is a transposition of those two terms, and so they are isomorphic.

	\item
		If $U_1$ and $U_2$ perform the same single-qubit operation, there will be some common tensor index $u_j$ which is current for that qubit prior to the application of $U_1$ and $U_2$ in both $S'_0$ and $S'_1$. 
		Then, the terms corresponding to $U_1$ and $U_2$ in $S_0$ will be $U_2\pseu[:u_{j+2}/u_{j+1}]U_1\pseu[:u_{j+1}/u_j]$, and the corresponding terms in $S_1$ will be $U_1\pseu[:u_{j+2}/u_{j+1}]U_2\pseu[:u_{j+1}/u_j]$ for some indices $v''$ and $w''$.
		Given that $U_1$ and $U_2$ perform the same operation, we then have $S''_0 = S''_1$.
	\end{itemize}
	Because the operations of $K_0$ and $K_1$ are identical after $U_1$ and $U_2$, the differences between $S_0$ and $S_1$ after the terms corresponding to $U_1$ and $U_2$ can only arise from the indices advanced by each term; because we have fixed the indexing scheme, the only difference between $S_0$ and $S_1$ is then the (possibly trivial) transposition of those two terms.
	Thus, $S_0$ and $S_1$ are isomorphic.
\end{proof}
Thus, for unitary circuits constructed from a parsimonious set of gates, isomorphism of stable index tensor expressions is equivalent to congruence up to rearrangements of commuting gates.
We will take advantage of this feature of this tensor notation in later chapters of the thesis.

\noindent We conclude our discussion of stable tensor index expressions with two remarks.
\begin{enumerate}
\item
	\label{item:nonstandardRepsOrderRecovery}%
	\textbf{Recovering the order of unitary operations.}~~
	As we have noted, isomorphism classes of stable tensor index expressions (for a parsimonious set of operations) correspond to congruency classes of unitary circuits under permutations of commuting gates.
	In any such circuit, the order of \emph{non-commuting} gates can be determined from the indexing: for a tensor index $v_j$ which is deprecated in a given stable index expression and for $v_{j+1} = f(v_j)$, any operation acting on $v_{j+1}$ must occur after any (other) operation in the circuit acting on $v_j$\,, with the only operator acting on both being the operator which deprecates $v_j$ and correspondingly advances $v_{j+1}$\,.
	The complement of the relation ``$U$ must occur after $V$'' is then a pre-order\footnote{%
		\label{fn:preorder}%
		A pre-order on a class $S$ is a binary relation which is reflexive ($x \preceq x$ for all $x \in S$) and transitive ($x \preceq y$ and $y \preceq z$ implies $x \preceq z$ for all $x,y,z \in S$).
 		An instructive example from quantum information processing is the binary relation ``$\to*$'' on bipartite pure states $\ket{\Psi} \in \cH[A] \ox \cH[B]$, where $\ket{\Phi} \to* \ket{\Psi}$ if and only if $\ket{\Psi}$ can be obtained by LOCC on $\cH[A]$ and $\cH[B]$ from $\ket{\Phi}$. 
	 	Every state can be obtained via LOCC from itself, and if $\ket{\Phi} \to* \ket{\Psi}$ and $\ket{\Psi} \to* \ket{\Upsilon}$, then $\ket{\Phi} \to* \ket{\Upsilon}$ by composing protocols.
	 	There are also pairs of states with $\ket{\Phi} \to* \ket{\Psi}$ and $\ket{\Psi} \not\to* \ket{\Phi}$, \eg\ for $\ket{\Phi}$ an entangled state and $\ket{\Psi}$ a product state; and there are distinct states with $\ket{\Phi} \to* \ket{\Psi} \to* \ket{\Phi}$, \ie\ if $\ket{\Phi}$ and $\ket{\Psi}$ only differ by local unitaries.
	 	(A result of Nielsen~\cite{Nielsen99} shows that that there also exist pairs of states such that $\ket{\Phi} \not\to* \ket{\Psi}$ and $\ket{\Psi} \not\to* \ket{\Phi}$, so this pre-order exhibits the complete spectrum of possible relations between pairs of elements.)}
	which characterizes the \emph{possible} orderings of the operations in a stable-index expression, up to commuting operations.

\item
	\textbf{Induced combinatorial structures on the indices.}
	The operations of a stable-index tensor expression also induce structures on the tensor indices.
	From any stable-index tensor expression $S$, we may construct a graph $G$ whose vertices are the indices of $S$, and where $v w \in E(G)$ for two indices $v$ and $w$ if and only if there exists an operator in $S$ involving both $v$ and $w$\,: we may call such a graph an \emph{interaction graph} for a stable index expression.
	If $I$ is the set of non-advanced indices of $S$, and $O$ is the set of non-deprecated indices, the function $f$ which maps each deprecated index to an advanced index then describes vertex-disjoint $I$\thru$O$ directed paths in $G$.

	The interaction graph $G$ encodes much of the information of the original expression $S$: in particular, from $G$ and the function $f$, we may recover the order in which tensor indices are deprecated (if they are deprecated at all) in any ordering of the terms of of $S$ consistent with a conventional circuit notation.
	Expressed as a partial order $\preceq$, we clearly have $v \preceq f(v)$, as $f(v)$ is advanced in the same term which deprecates $v$; and for any $w$ which is adjacent to $f(v)$, we have $v \preceq w$, as $f(v)$ and $w$ are acted on by an operator $U$ in common and therefore must be current at the same time at the stage where $U$ is performed.
	These same graph and ordering structures arise in the one-way model, and form the core of the topic of Chapter~\ref{Semantics for unitary one-way patterns}.
\end{enumerate}
\label{sec:nonstandardRepnsEnd}

\section{Classically controlled unitary circuits}

We may extend unitary circuits to obtain a description of quantum computation which involves \emph{classical control}: that is, dependency of operations on classical bits, which in particular may be the results of measurements.
In this section, we briefly consider such models of quantum computation, and their relationship to unitary circuits without classical control.
This will also allow us to define operations which will be necessary to describe the \emph{one-way measurement model} in Chapter~\ref{One-way measurement-based computation}.

We may define a ``classically controlled'' CPTP operation using the following construction:
\begin{definition}
	For a sequence of bit-registers $\vec c = (c_j)_{j = 1}^n$ and a map $f: \ens{0,1}^n \to* S$ for some set $S$, define the completely positive (but \emph{not} trace-preserving) operators $\Cond^{f,s} : \Prob(\vec c) \to \Prob(\vec c)$ for each $s \in S$, given by
	\begin{gather}
			\Cond^{f:s}(\rho)
		\;\;=\;\;
			\Pi_s \,\rho\, \Pi_s \;,
		\qquad\text{where}\quad
			\Pi_s
		\;\;=
			\sum_{\vec x \in f\inv(s)} \ket{\vec x}\bra{\vec x}	.
	\end{gather}
	For sequence of generic registers $\sigma$, $\alpha$, and $\delta$, a \emph{classically controlled} CPTP map is a sum of superoperators, of the form
	\begin{subequations}
	\begin{gather}
			\Upphi_{\sigma:\alpha/\delta}^{f(\vec c)} : \Prob(\vec c) \ox \D(\sigma; \delta) \to \Prob(\vec c) \ox \D(\sigma; \alpha)
		\\[1ex]
			\Upphi^{f(\vec c)}_{\sigma:\alpha/\delta}(\rho)
		\;\;\;=
			\sum_{\vec x \in \ens{0,1}^n} \Cond^{f:\vec x}_{\vec c} \ox \Uppsi_{\sigma:\alpha/\delta}^{(\vec x)}	\;,
	\end{gather}
	\end{subequations}
	for some family of \emph{conditionally applied CPTP maps} $\ens{\Uppsi\supp{s}}_{s \in S}$ which all have the same domain and range.
\end{definition}
A common example of a classically controlled operation is a \emph{classically-controlled not} operation, where we perform an $X$ operation on a qubit depending on the value of a classical bit:
\begin{align}
		\Upphi_{c,q}(\rho)
	\;\;=&\;\;
		\Big[ \ket{0}\bra{0}_c \ox \idop_q \Big] \rho \Big[ \ket{0}\bra{0}_c \ox \idop_q \Big]
		\;+\;
		\Big[ \ket{1}\bra{1}_c \ox X_q \Big] \rho \Big[ \ket{1}\bra{1}_c \ox X_q \Big]
	\notag\\=&\;\;
		\bigg[\Big( \Cond^{\id:0}_c \ox \idop_q \Big) + \Big( \Cond^{\id:1}_c \ox \Xcorr{q}{}\Big) \bigg](\rho)	,
\end{align}
where $\Xcorr{}{}: \D(2) \to \D(2)$ is given by $\Xcorr{}{}(\rho) = X \rho X$.
It is easy to see that classically controlled CPTP maps $\Upphi$ are indeed CPTP maps, by forming the Kraus operators of $\Upphi$ from the tensor products of the $\Cond^{f:s}$ operations and the conditionally applied maps $\Uppsi\supp{s}$ which define $\Upphi$.

We may distinguish two classes of conditionally controlled CPTP maps, which will play an important role in this thesis:
\begin{definition}
	A \emph{classically controlled unitary} operation is a conditionally controlled CPTP map $\Upphi$ whose conditionally applied maps $\ens{\Uppsi\supp{s}}_{s \in S}$ are all unitary operations.
	Similarly, a \emph{classically controlled measurement} operation is a conditionally controlled CPTP map $\Upphi$ whose conditionally applied maps $\ens{\Uppsi\supp{s}}_{s \in S}$ are all measurement operations.
\end{definition}
Our interest in unitary circuit models (as opposed to models with classical control) is largely due to the convenience of omitting explicit reference to classical control in the analysis of quantum algorithms.
However, as instances of computational problems are usually described as being represented with classical information, a robust experimental set-up which is capable of solving different instances of a problem will very probably control which particular unitary evolution occurs by classical control of the unitary evolution.

\begin{definition}
	A model of quantum computation is a \emph{classically controlled unitary circuit model} if its elementary gates consist only of operations which
	\begin{alphanum}
	\item
		prepare some number of qubits in a fixed initial state;

	\item
		perform a (possibly classically-controlled) unitary operation on some number of qubits; or

	\item
		perform a (possibly classically-controlled) complete (orthogonal \& destructive) measurement on some number of qubits.
	\end{alphanum}
\end{definition}

\subsection{Classically controlled extensions, and deferred measurement}

For a given unitary circuit model defined by some gate-set \Gates, it is common to extend to a classically controlled unitary circuit model by including classically controlled versions $\Upupsilon^{c}$ of the unitary gates $\Upupsilon \in \Gates$, which perform either $\Upupsilon$ or the identity superoperator depending on the value of a single classical bit $c$.
We may also consider classically controlled versions $\mathsf K^c$ of measurement gates, where $\mathsf K^{\ssstyle 0}, \mathsf K^{\ssstyle 1} \in \Gates$ are different measurement gates which are performed on a given set of qubits depending on the value of a classical bit $c$.\label{discn:classicalControl}
Such an extended gate set, including controlled versions of unitaries and measurements in $\Gates$, we will call $\textsc{cc}(\Gates)$.\footnote{\label{fn:classicalControlFn}%
We may easily extend this definition to $\textsc{cc}(\Gates)$ when $\Gates$ does not define a unitary circuit model, by considering any pair of conditionally applied CPTP maps $\Uppsi^0, \Uppsi^1 \in \Gates$\,; however, when $\Uppsi^0$ and $\Uppsi^1$ are unitaries, it is conventional to require that one of them be the identity superoperator.}
In such a circuit model, classically controlled CPTP maps may either perform the identity or perform a non-trivial operation on a set of qubits, depending on the results of measurements performed in the middle of the circuit.
This is the primary qualitative difference between unitary circuit models and classically controlled versions of those models.

It is also common to transform classically-controlled circuits into unitary circuits without classical control, following a folklore result reported in~\cite{BV97} and attributed to \cite{Bernstein97}.
For a gate-set $\Gates$, we define a gate-set $\textsc{qc}(\Gates)$ as follows:
\begin{itemize}
\item
	If $\Gates$ doesn't include the measurement operator $\Meas{}{\sfz}$, we include it in $\textsc{qc}(\Gates)$\,;

\item
	For any unitary $\Upupsilon \in \Gates$ performing $\rho \mapsto U \rho U$ for $\rho \in \D(2^n)$, we include a (\emph{coherently}) \emph{controlled} version $\ctrl\Upupsilon \in \textsc{qc}(\Gates)$, which performs the mapping
	\begin{gather}
		\label{eqn:controlledU}
			\ctrl\Upupsilon(\rho)
		\;\;=\;\;
			\Big( \ket{0}\bra{0} \ox \idop_2^{\ox n} \;+\; \ket{1}\bra{1} \ox U \Big)
			\rho
			\Big( \ket{0}\bra{0} \ox \idop_2^{\ox n} \;+\; \ket{1}\bra{1} \ox U\herm \Big);
	\end{gather}

\item
	For any complete measurement $\mathsf K$ given by
	\begin{gather}
			\mathsf K(\rho)
		\;\;=\;\;
			\sum_{\vec x \in \ens{0,1}^n} \ket{\vec x}\bra{\psi_{\vec x}} \rho \ket{\psi_{\vec x}}\bra{\vec x}
	\end{gather}
	for some orthonormal basis $\ens{\big.\ket{\psi_{\vec x}}\big.}_{\vec x \in \ens{0,1}^n}$ of $\cH^{\ox n}$, we include the unitary change of basis operation $\mathsf B^{\mathsf K} \in \textsc{qc}(\Gates)$, defined by
	\begin{gather}
			\mathsf B^{\mathsf K{}{}}(\rho)
		\;\;=\;\;
			B \rho B\herm\,,
		\qquad\text{where}\quad
			B
		\;\;=
			\sum_{\vec x \in \ens{0,1}^n} \ket{\vec x}\bra{\psi_{\vec x}}	\,.
	\end{gather}
	We also include a coherently controlled version $\ctrl \mathsf B^{\mathsf K} \in \textsc{qc}(\Gates)$, defined in terms of $B^{\mathsf K}$ by~\eqref{eqn:controlledU}.

\item
	Every state-preparation, unitary operation, or trace-out operation in $\Gates$ is included in $\textsc{qc}(\Gates)$, but not any measurement operation except for $\Meas{}{\sfz}$.
\end{itemize}
Having defined the gate-set $\textsc{qc}(\Gates)$, we may translate any circuit composed from the gate set $\textsc{cc}(\Gates)$ into a ``coherent'' version in the gate-set $\textsc{qc}(\Gates)$ by replacing classical input registers with quantum input registers (thereby extending the domain of the input), and commuting all of the measurement operations to the end of the circuit, as follows:
\begin{enumerate}
\item
	Decompose each measurement operation $\mathsf K$ on $n$ qubits (other than $\Meas{}{\sfz}$) into a change of basis, from the measurement basis of $\mathsf K$ to the standard basis, followed by a measurement in the standard basis:
	\begin{align}
			\mathsf K_{\alpha/\delta}
		\;\;=\;\;
			\sqparen{\bigotimes_{v \in \delta} \Meas{v}{\sfz}} \circ \mathsf B^{\mathsf K}_{\delta}	\;,
	\end{align}
	where we attribute classical bits $\s[v] \in \alpha$ to each $v \in \delta$, in sequence.
	We similarly replace controlled measurement operators with \emph{classically controlled} changes of basis followed by measurement in the standard basis:
	\begin{align}
			\mathsf K^{c}_{\alpha/\delta}
		\;\;=\;\;
			\sqparen{\bigotimes_{v \in \delta} \Meas{v}{\sfz}} \circ \Big(\mathsf B^{\mathsf K^1}\Big)^{\!c}_{\!\delta}	\,
																				\circ \Big(\mathsf B^{\mathsf K^0}\Big)^{\!1-c}_{\!\delta}	\;.
	\end{align}

\item
	We substitute any classically-controlled operation $\Upupsilon^c$ depending on a classical input bit $c$, including instances of such operations arising from measurements in the previous step, by a (coherently) controlled unitary $\ctrl\Upupsilon$ acting on the input qubit which has replaced the input bit;

\item
	For each qubit $v$ and any unitary $\Upupsilon \in \Gates$, we perform the following substitution repeatedly until there are no further classically controlled unitaries remaining:
	\begin{gather}
			\Upupsilon^{\s[v]}_{\sigma} \; \Meas{v}{\sfz} 
		\;\;\mapsto\;\;
			\Meas{v}{\sfz} \; \ctrl\Upupsilon_{v, \sigma}
	\end{gather}
\end{enumerate}
Because each classically controlled operation in a $\textsc{cc}(\Gates)$ circuit acts either on an input bit or a bit allocated by a measurement, the above procedure will eventually remove every classically controlled unitary (either from the original circuit or from the decomposition of classically controlled measurements), and replace it with a coherently controlled unitary, thereby producing a unitary circuit.
Replacing the classical registers at the input with equivalent quantum registers also leaves the effect of the circuit as a superoperator unchanged on the original input space: we refer to this as a \emph{coherent extension} of the original circuit.
In summary:
\begin{lemma*}[Principle of Deferred Measurement \cite{Bernstein97,BV97}]
	\label{principleDeferredMeas}
	For any gate set $\Gates$ giving rise to a unitary circuit model, every classically controlled circuit $C$ in a gate set $\textsc{cc}(\Gates)$ can be transformed to a unitary circuit $Q$ in the gate set $\textsc{qc}(\Gates)$, where $Q$ is a coherent extension of $C$, and where the difference in the complexity of $Q$ and $C$ is bounded above by the number of measurements in $C$.
\end{lemma*}
The principle of deferred measurement is what allows us in practise to consider quantum computation in terms of unitary circuits.
Further, if $\Gates$ is universal for quantum computation, we may further decompose the operations of $\textsc{qc}(\Gates)$ into operations of $\Gates$; the Solovay-Kitaev theorem then places an upper bound on the factor of increase of the resulting $\Gates$ circuit over the original $\textsc{cc}(\Gates)$ circuit.
Therefore, we do not incur very large computational costs by restricting ourselves to unitary circuit models, rather than always explicitly considering classically controlled unitary circuits.


\section{Clifford circuits and the stabilizer formalism}
\label{sec:cliffordCircuits}

A final model of quantum computation which we will consider in this Chapter is one known \emph{not} to be universal for quantum computation; and furthermore, one which is known to be efficiently simulatable on classical computers.
This model is a controlled-unitary circuit model whose unitary operations are restricted to what is called the \emph{Clifford group}, and whose preparation/measurement operations are similarly restricted.
In this section, we present the results of~\cite{GotPhD} about this model of computation, as it provides an illustrative pre-amble to measurement-based computation.

\subsection{The Pauli and Clifford groups}

\begin{definition}
	The \emph{Pauli group} on $n$ qubits is the group of operators $\cP^{\ox n}$, consisting of $n$-fold tensor products of the group $\cP = \gen{\idop_2\,, X, Y, Z} = \gen{i\idop_2\,, X, Z}$ generated by the Pauli operators (see Definition~\ref{def:Pauli}).
	The \emph{Clifford group} on $n$ qubits is the normalizer of the Pauli group in $\U(2^n)$.
\end{definition}

The following Lemma is proven for the set $\ens{H,R_\sfz(\pion 2), \cX}$ by~\cite{Gottesman98}, and therefore also holds as a corollary for the sets $\smash{\ens{H, R_\sfz(\alpha), \cZ}_{\alpha \in \frac{\pi}{2}\Z}}\,$, and $\smash{\ens{J(\alpha), \cZ}_{\alpha \in \frac{\pi}{2} \Z}}\,$ by the remarks preceding Lemma~\ref{lemma:universalUnitaries}:

\begin{lemma}
	For any $n \ge 1$, the Clifford group on $n$ qubits can be generated by the following sets of unitaries together with tensor products with scalar multiples of the identity: $\ens{H,R_\sfz(\pion 2), \cX}$, $\smash{\ens{H, R_\sfz(\alpha), \cZ}_{\alpha \in \frac{\pi}{2}\Z}}\,$, and $\smash{\ens{J(\alpha), \cZ}_{\alpha \in \frac{\pi}{2} \Z}}\,$.
\end{lemma}

A Clifford group operation $U$ can be characterized (up to an unimportant scalar factor) by the action induced on the Pauli group by conjugation by $U$.
Because $U Y_v U\herm = U (iX_v Z_v) U\herm = i (UX_vU\herm)(UZ_vU\herm)$ for a $Y$ operation on any single qubit $v$, and $U \idop U\herm = \idop$, we may further reduce this to how $U$ transforms tensor products of the operators $X$ and $Z$ on single qubits by conjugation.
This then leads to a natural representation of Clifford group operations as CPTP maps, albeit acting as a transformation of the zero-trace, non-positive tensor products of $X$ and $Z$ operators with tensor powers of $\idop_2$\,.
For instance, it is easy to verify by calculation using the CPTP maps defined by~\eqref{eqn:cptpElemUnitaries} that
\begin{subequations}
\label{eqn:cliffordMapping}
\begin{align}
		\sfH(X)	\;=&\;	Z	\,,								&	\sfH(Y)	\;=&\;	-Y	\,,					&	\sfH(Z)	\;=&\;	X	\,;
	\\[2ex]
		\sfR^{\pi/2}(X)	\;=&\;	Y \,,		&	\sfR^{\pi/2}(Y) \;=&\; -X	\,,			&	\sfR^{\pi/2}(Z)	\;=&\;	Z\,;
	\\[2ex]
		\begin{split}
			\Ent{}(X \ox \idop_2)\;=&\;	X \ox Z\,,
		\\
			\Ent{}(\idop_2 \ox X) \;=&\;	Z \ox X\,,
		\end{split}
	&
		\begin{split}
			\Ent{}(Y \ox \idop_2)\;=&\;	Y \ox Z\,,
		\\
			\Ent{}(\idop_2 \ox Y) \;=&\;	Z \ox Y\,,
		\end{split}
	&
		\begin{split}
			\Ent{}{}(Z \ox \idop_2)	\;=&\;	Z \ox \idop_2\,,
		\\
			\Ent{}{}(\idop_2 \ox Z) \;=&\;	\idop_2 \ox Z\,.
		\end{split}
\end{align}
\end{subequations}
(Note also that the middle column of these equations can be easily computed from the left and right columns by $Y = iXZ = -iZX$.)
The way in which compositions of these CPTP maps transform $\cP^{\ox n}$ can then be easily computed.

The Clifford group acts \emph{transitively} on elements of the Pauli group with eigenvalues $\pm 1$: that is, for any two Hermitian Pauli group elements $P_1, P_2 \in \cP^{\ox n} \setminus \ens{\pm \idop_2^{\ox n}}$\,, there is a Clifford group operation $U$ such that $P_2 = U P_1 U\herm$.
This can easily be seen from the fact that for any Pauli operator $P$, we may apply $\sfR^{\pi/2}$ operations to obtain an operator $P'$ for which at least one qubit $a$ is acted on by $X$, and the others with $Z$\,; and then perform $\Ent{}$ operations on pairs of qubits $(a,v)$ for which $v$ is acted on by $P'$ by a $Z$ operation, which will yield the operator $X_a$ acting only on $a$.
By defining the mapping
\begin{subequations}
\begin{gather}
		\mathsf{HSWAP}_{a,b} \;=\; \Ent{a,b} \sfH_a \sfH_b \Ent{a,b} \sfH_a \sfH_b \Ent{a,b}	\;,
\intertext{which for product operators $\rho \ox \sigma \in \L(\cH[2] \ox \cH[2])$ performs the mapping}
		\mathsf{HSWAP}(\rho \ox \sigma) \;=\; \sfH(\sigma) \ox \sfH(\rho)	\;,
\end{gather}
\end{subequations}
we can see that performing an $\mathsf{HSWAP}_{a,n}$ on $a$ and the $n\th$ qubit, we obtain operation $Z_n$ acting only on the $n\th$ qubit, up to a scalar factor of $\pm 1$ (not directly attributable to an operation on any given qubit).
Finally, we may transform $-Z_n$ to $Z_n$ by conjugation by $X_n \,=\, H_n R_\sfz(\pion 2)_n^2 H_n$\,; and so, any Pauli operator $P \in \cP^{\ox n}$ may be mapped by a Clifford group operation to $Z_n = \idop_2^{\ox n-1} \ox Z$.
We can do this for any Pauli operator; furthermore, by performing such a Clifford transformation in reverse, we may also obtain an arbitrary Pauli operator from a $Z$ on the $n\th$ qubit.
This proves our claim of transitivity.

\subsection{Stabilizer groups and stabilizer codes}

It is easy to verify that any two Pauli operators $P \in \cP = \gen{\idop_2, X, Y, Z}$ either commute or anticommute with each other;  and that therefore the same holds for two elements of $\cP^{\ox n}$.
We may then consider subgroups of $\cP^{\ox n}$ generated by commuting sets of operators.
Any such group may be characterized by a set of independent generators $\ens{S_1,\ldots,S_{n-k}}$ for some $k \le n$\,.
As $\ens{X,Y,Z}$ are all self-inverse, such a group will be of size $2^{n-k}$, and in particular isomorphic to a subgroup of $\Z_2^n$.

We may also show that for any $0 \le k \le n$, the Clifford group acts transitively on abelian subgroups of the Pauli group of size $2^{n-k}$ whose operators all have $+1$ eigenspaces (which we call \emph{Pauli stabilizer groups}, often simply \emph{stabilizer groups}).
We may show this by induction on $n - k$, as follows.
The case of $n = k$ is trivial, as there is only one such group, $\ens{\idop_2^{\ox n}}$.
For $n > k$, let $\ens{S_1,\ldots,S_{n-k}}$ be a set of generators for the group: we may reduce the problem by considering a Clifford group operation $U$ such that $U S_{n-k} U\herm = Z_n$, using the transitivity of the action of the Clifford group on the non-trivial Hermitian elements of $\cP^{\ox n}$
Because two Pauli operators $P_1$ and $P_2$ commute if and only if $U P_1 U\herm$ and $U P_2 U\herm$ commute, it is easy to show that the operators $U S_j U\herm$ must act on the $n\th$ qubit with either the identity or a $Z$ operation.
For those generators $S_j$ such that the latter holds, define $S'_j = S_j S_{n-k}$\,; then, $U S'_j U\herm$ acts on the $n\th$ qubit with the identity.
If we define $S'_j = S_j$ for the other stabilizers, $\{ S'_j \}_{j = 1}^{n-k}$ generates the same abelian group, and by construction is transformed via conjugation by $U$ to a set of operators $\ens{T_1 \ox \idop_2\,,\; \ldots\,,\; T_{n-k-1} \ox \idop_2\,,\; Z_n}$.
The first $n-k-1$ of these generate an abelian subgroup on the first $n-1$ qubits, which by induction we may transform by a Clifford group operation to the group generated by $\ens{Z_{k+1}, \ldots, Z_{n-1}}$.
Again, this transformation may be reversed to yield an arbitrary stabilizer code of size $2^{n-k}$; then any such abelian group of size $2^{n-k}$ to be mapped to any other.

A set of Pauli operators which commutes, by that very fact, shares a common set of eigenvectors.
The group $\gen{Z_{k+1}, Z_{k+2}, \ldots, Z_n}$ in particular admits a $2^k$-dimensional subspace of $\cH^{\ox n}$ which are $+1$-eigenvectors of the entire group (specifically, the set of states of the form $\ket{\psi} \ox \ket{0}^{\ox n-k}$ for arbitrary $\ket{\psi} \in \cH^{\ox k}$); by transitivity of the Clifford group on stabilizer groups of similar size, every stabilizer group of size $2^{n-k}$ which does not include $-\idop_2^{\ox n}$ has a $2^k$-dimensional space of joint $+1$-eigenvectors.
We say that a vector $\ket{\psi}$ is \emph{stabilized} by an operator $M$ if $\ket{\psi} = M\ket{\psi}$\,, \ie\ if it is a $+1$-eigenvector of $M$\,, which motivates the following terminology:

\begin{definition}
	A \emph{stabilizer code} $\cE$ on $n$ qubits is the space of joint $+1$-eigenvectors of a stabilizer group on $n$ qubits.
	In particular, a \emph{stabilizer state} on $n$ qubits is a state $\rho = \ket{\psi}\bra{\psi}$, where $\ket{\psi}$ is a joint $+1$-eigenvector of a commuting group of $2^n$ Pauli operators on $n$ qubits.
	The \emph{stabilizer group of the code} is the group for which the code is the joint $+1$-eigenspace.
\end{definition}

\begin{example}
 	Let $S \in \ens{X,Y,Z}$, and let $\ket{\splus \sfs}$ represent the corresponding state vector in $\ens{\ket{\splus \sfx}, \ket{\splus \sfy}, \ket{\splus \sfz}}$.
 	It is easy to verify that $\ket{\splus \sfs}$ is stabilized by $S$\,; then, for any sequence of non-trivial Pauli operators $\big(S_j\big)_{j = 1}^n$, where each operator $S_j$ also acts on qubit $j$ and stabilizes the state vector $\ket{\splus \sfs_j}$, the product state
 	\begin{gather}
 			\ket{\psi}
 		\;\;=\;\;
 			\bigotimes_{j = 1}^n \;\ket{\splus \sfs_j}
 	\end{gather}
 	is stabilized by the abelian group $\gen{S_1,\,\ldots,\,S_n}$.
 	As a special case, the state $\ket{0}^{\ox n} = \ket{0} \ox \cdots \ox \ket{0}$ is stabilized by the group $\gen{Z_1, \ldots, Z_n}$.
 \end{example}

\subsection{The stabilizer formalism}
\label{sec:stabilizerFormalism}

As stabilizer codes can be characterized by a short list of generators for its stabilizer group, the state space of a system which is initially prepared in a pure state in some stabilizer code and then transformed by Clifford group operations can be efficiently computed.
We may extend this result further: consider the circuits which can be composed of the unitary CPTP maps $\sfH$, $\sfR^{\pi/2}$, and $\Ent{}$; the state preparation maps
\begin{subequations}
\begin{align}
		\New{v}{\sfx}(\rho)
	\;\;=&\;\;
		\rho  \;\ox\; \ket{\splus \sfx}\bra{\splus \sfx}_{v}	\;,
	\\[2ex]
		\New{v}{\sfy}(\rho)
	\;\;=&\;\;
		\rho  \;\ox\; \ket{\splus \sfy}\bra{\splus \sfy}_{v}	\;,
	\\[2ex]
		\New{v}{\sfz}(\rho)
	\;\;=&\;\;
		\rho  \;\ox\; \ket{\splus \sfz}\bra{\splus \sfz}_{v}	\;;
\end{align}
\end{subequations}
the von Neumann measurement operations
\begin{subequations}
\begin{align}
		\sfP_{v:r}^\sfx(\rho)
	\;\;=&\;\;\;
	\text{\small$
		\ket{\splus \sfx}\bra{\splus \sfx}_v \rho \ket{\splus \sfx}\bra{\splus \sfx}_v \;\ox\; \ket{0}\bra{0}_r
		\;\;+\;\;
		\ket{\sminus \sfx}\bra{\sminus \sfx}_v \rho \ket{\sminus \sfx}\bra{\sminus \sfx}_v	\;\ox\; \ket{1}\bra{1}_r \;,
	$}\\[2ex]
		\sfP_{v:r}^\sfy(\rho)
	\;\;=&\;\;\;
	\text{\small$
		\ket{\splus \sfy}\bra{\splus \sfy}_v \rho \ket{\splus \sfy}\bra{\splus \sfy}_v \;\ox\; \ket{0}\bra{0}_r 
		\;\;+\;\;
		\ket{\sminus \sfy}\bra{\sminus \sfy}_v \rho \ket{\sminus \sfy}\bra{\sminus \sfy}_v	\;\ox\; \ket{1}\bra{1}_r  \;,
	$}\\[2ex]
		\sfP_{v:r}^\sfz(\rho)
	\;\;=&\;\;\;
	\text{\small$
		\ket{\splus \sfz}\bra{\splus \sfz}_v \rho \ket{\splus \sfz}\bra{\splus \sfz}_v \;\ox\; \ket{0}\bra{0}_r 
		\;\;+\;\;
		\ket{\sminus \sfz}\bra{\sminus \sfz}_v \rho \ket{\sminus \sfz}\bra{\sminus \sfz}_v	\;\ox\; \ket{1}\bra{1}_r \;;
	$}
\end{align}
\end{subequations}
and the corresponding destructive measurement operations,
\begin{subequations}
\label{eqn:pauliMeas}
\begin{align}
		\Meas{v}\sfx(\rho)
	\;\;=&\;\;\;
	\text{\small$
		\bra{\splus \sfx}_v \rho \ket{\splus \sfx}_v \;\ox\; \ket{0}\bra{0}_{\s[v]}
		\;\;+\;\;
		\bra{\sminus \sfx}_v \rho \ket{\sminus \sfx}_v	\;\ox\; \ket{1}\bra{1}_{\s[v]} \;,
	$}\\[2ex]
		\Meas{v}\sfy(\rho)
	\;\;=&\;\;\;
	\text{\small$
		\bra{\splus \sfy}_v \rho \ket{\splus \sfy}_v \;\ox\; \ket{0}\bra{0}_{\s[v]}
		\;\;+\;\;
		\bra{\sminus \sfy}_v \rho \ket{\sminus \sfy}_v	\;\ox\; \ket{1}\bra{1}_{\s[v]} \;,
	$}\\[2ex]
		\Meas{v}\sfz(\rho)
	\;\;=&\;\;\;
	\text{\small$
		\bra{\splus \sfz}_v \rho \ket{\splus \sfz}_v \;\ox\; \ket{0}\bra{0}_{\s[v]}
		\;\;+\;\;
		\bra{\sminus \sfz}_v \rho \ket{\sminus \sfz}_v	\;\ox\; \ket{1}\bra{1}_{\s[v]} \;.
	$}
\end{align}
\end{subequations}

\refstepcounter{footnote}\footnotetext{\addtocounter{footnote}{-1}%
To be precise, we are using here the slightly extended sense of the function $\Gates \mapsto \textsc{cc}(\Gates)$, described in the footnote (\ref{fn:classicalControlFn}) on page~\pageref{fn:classicalControlFn}, for when $\Gates$ does not define a unitary circuit model.}%
\begin{definition}
	For an infinite set $V$ of qubits and $B$ of bits, the \emph{Clifford circuit model} is the circuit model generated by $\Clifford = \textsc{cc}(\cG)$,\footnotemark\ where
	\begin{gather}
		\cG \;=\;
		\text{\small$
			\ens{ \sfH_v\,,\; \sfR_v^{\pi/2}\,,\; \Ent{u,v}\,,\; \New{v}{\sigma}\,,\; \sfP^\sigma_{v:r}\,,\; \Meas{v}{\sigma}
			\,\Big|\; \sigma \in \ens{\sfx, \sfy, \sfz}, ~r \in B, \text{~and~} v,w \in V}\,.
		$}
	\end{gather}
\end{definition}
Note that circuits in the $\Phases(\frac{\pi}{2}\Z)$ model which do not use trace-out operations are also circuits in the \Clifford\ model.

Assuming an arbitrary input state, the state-space of the output to a Clifford circuit conditioned on particular values of each bit in the circuit (and in particular, conditioned on particulars outcomes of each von Neumann measurement performed) may be efficiently computed by transforming generators of the stabilizer group of the system.
The techniques for doing this, described below, were presented by~\cite{GotPhD} and are referred to collectively as \emph{the stabilizer formalism}.
\label{discn:stabilizerFormalism}

The state preparation operators $\New{v}{\sigma}$ simply adjoin another qubit $v$ on which the existing generators act with the identity, and also add another generator $S_v$ stabilizing the state $\ket{\splus \sigma}_v$.
As noted before, the effect of the unitary operators $\sfH$, $\sfR^{\pi/2}$, and $\Ent{}{}$ may be efficiently computed on each generator.
This leaves the von Neumann measurements $\sfP_{v:r}^\sigma$ and destructive measurement $\Meas{v}\sigma$\,, which we describe following the treatment of~\cite{GotPhD}.
Consider a Pauli operator $P_v$ on a single qubit $v$, for $P \in \ens{X, Y, Z}$.
For the corresponding label $p \in \ens{\sfx, \sfy, \sfz}$, a $\sfP_{v:r}^{p}$ measurement is equivalent to a measurement of the observable $P_v$ (recall Definitions~\ref{def:measObservableVN} and~\ref{def:measObservableDestructive}), and relabelling the measurement results $(-1)^r \mapsto* r$ to obtain measurement results $r \in \ens{0,1}$.
Abusing our earlier terminology, we will momentarily speak of the measurements in the Clifford circuit model as being measurements of the corresponding Pauli operator as an observable in this fashion.

Consider a state $\rho = \ket{\psi}\bra{\psi}$ for $\ket{\psi} \in \cE$, where we let $\cE$ be the space stabilized by a stabilizer group $\sS = \gen{S_1, \ldots, S_{n-k}}$; and define the projectors
\begin{align}
		\Pi_0
	\;\;=\;\;
		\sfrac{1}{2}\paren{\idop_2^{\ox n} + P_v}
	&&
		\text{and}
	&&
		\Pi_1
	\;\;=\;\;
		\sfrac{1}{2}\paren{\idop_2^{\ox n} - P_v}\,.
\end{align}
\begin{itemize}
\item
	If $P_v \in \sS$, then $\ket{\psi}$ is a $+1$-eigenvector of $P_v$, in which case a von Neumann $P_v$ measurement will yield the result $\s[v] = 0$ and leave the state of the quantum part of the system unchanged.

\item
	If $P_v \notin \sS$ but $P_v$ commutes with all of the generators $S_j$, then we may decompose the code $\cE$ stabilized by $\sS$ into two orthogonal subspaces $\cE_0$ and $\cE_1$, stabilized by $\sS_0 = \sS \oplus \gen{P_v}$ and $\sS_1 = \sS \oplus \gen{-P_v}$ respectively.
	(That these subspaces are orthogonal follows from the orthogonality of the $\pm 1$-eigenspaces of $P_v$.)
	Let $\ket{\psi_0} = \Pi_0 \ket{\psi}$ and $\ket{\psi_1} = \Pi_1 \ket{\psi}$\,: then a von Neumann $P_v$ measurement yields the state $\ket{\psi_0}\bra{\psi_0} \ox \ket{0}\bra{0}_r \;+\; \ket{\psi_1}\bra{\psi_1} \ox \ket{1}\bra{1}_r$\,, where $r$ is the bit containing the measurement result.
	For each $r \in \ens{0,1}$, we have
	\begin{gather}
		S_j \ket{\psi_r} \;=\; S_j \Pi_r \ket{\psi} \;=\; \Pi_r S_j \ket{\psi} = \ket{\psi_r}
	\end{gather}
	for every $1 \le j \le n-k$\,; then, the state after measurement conditioned on the result bit having the value $r$ is is in the space $\cE_r$ stabilized by $\sS_r = \sS \oplus \gen{\big.\text{\small$(-1)^r P_v$}}$.
	Because the initial state $\ket{\psi}$ may be an \emph{arbitrary} state in $\cE_0$ or $\cE_1$ (which would then be unaffected by measurement), this fully characterizes the post-measurement state space.
	Because the number of generators of the stabilizer group increases in this case, $\cE_0$ and $\cE_1$ each have half the dimension of $\cE$, and so this operation is non-unitary.

\item
	If $P_v$ anticommutes with a generator $S_j$, without loss of generality we may suppose that $P_v$ anticommutes with $S_1$ in particular.
	Then, define a new set of generators $\{ S'_j \}_{j = 1}^{n-k}$ for $\sS$, such that
	\begin{gather}
			S'_j
		\;\;=\;\;
			\begin{cases}
				S_j		\;,	&	\text{if $j = 1$ or $S_j$ commutes with $P_v$}	\\[0.5ex]
				S_1 S_j	\;,	&	\text{otherwise}
			\end{cases}	\;;
	\end{gather}
	then $S'_1$ anticommutes with $P_v$\,, and $S'_j$ commutes with $P_v$ for $j \ge 2$.
	Let $\sS_0 = \ens{P_v, S'_2, \ldots, S'_{n-k}}$, and $\sS_1 = \ens{-P_v, S'_2, \ldots, S'_{n-k}}$, and let $\cE_0$ and $\cE_1$ be the codes stabilized by these groups.
	Once more, let $\ket{\psi_0} = \Pi_0 \ket{\psi}$ and $\ket{\psi_1} = \Pi_1 \ket{\psi}$\,: then a von Neumann $P_v$ measurement yields the state $\ket{\psi_0}\bra{\psi_0} \ox \ket{0}\bra{0}_r \;+\; \ket{\psi_1}\bra{\psi_1} \ox \ket{1}\bra{1}_r$\,, where $r$ is the bit containing the measurement result.
	For each $r \in \ens{0,1}$, we again have
	\begin{gather}
		S'_j \ket{\psi_r} \;=\; S'_j \Pi_r \ket{\psi} \;=\; \Pi_r S'_j \ket{\psi} = \ket{\psi_r}	\,;
	\end{gather}
	for every $2 \le j \le n-k$\,; however, we have
	\begin{gather}
			S'_1 \ket{\psi_0}
		\;=\;
			\sfrac{1}{2} S'_1 \paren{\idop_2^{\ox n} + P_v} \ket{\psi}
		\;=\;
			\sfrac{1}{2} \paren{\idop_2^{\ox n} - P_v} S'_1 \ket{\psi}
		\;=\;
			\ket{\psi_1}	\;,
	\end{gather}
	and similarly $S'_1 \ket{\psi_1} = \ket{\psi_0}$.
	Then, the state after measurement conditioned on the result bit having the value $r$ is in the space $\cE_r$ stabilized by the group $\sS_r$\,, and the two possible post-measurement results may be mapped to one another by the operation $S'_1$.
	(In particular, if we wish to select for the post-measurement state that would arise for the $0$ state, it suffices to apply the operation $S_1$ if instead the measurement result $1$ arises.)

	Note that the probabilities of the two outcomes in this case are both $\frac{1}{2}$\,: we have
	\begin{align}
			\bracket{\psi_0}{\psi_0}
		\;\;=\;\;
		 	\sfrac{1}{2} \bra{\psi} (\idop_2^{\ox n} + P_v) \ket{\psi}
		\;\;=&\;\;
		 	\sfrac{1}{2} \bra{\psi} S'_1 (\idop_2^{\ox n} - P_v) S'_1 \ket{\psi}
		\notag\\=&\;\;
		 	\bracket{\psi_1}{\psi_1}	\;,
	\end{align}
	which implies that applying a $P_v$ measurement on $\ket{\psi}$ yields $+1$-eigenstates and $-1$-eigenstates of $P_v$ each with equal probability.
	Because $S'_1$ maps the spaces $\cE_0$ and $\cE_1$ into one another, we may consider arbitrary vectors $\ket{\psi_0} \in \cE_0$ and $\ket{\psi_1} \in \cE_1$ such that $\ket{\psi_0} = S'_1 \ket{\psi_1}$\,.
	If $\bracket{\psi_0}{\psi_0} = \bracket{\psi_1}{\psi_1} = \frac{1}{2}$, then $\ket{\psi} = \ket{\psi_0} + \ket{\psi_1}$ is a state vector stabilized by $S'_1$, and also by $S'_j$ for $j \ge 2$.
	Then, for any state vector $\ket{\psi'}$ in either $\cE_0$ or $\cE_1$\,, we may construct a state $\ket{\psi}$ stabilized by $\sS$ for which $\ket{\psi'}$ is a possible post-measurement state.
	This implies that the codes $\cE_0$ and $\cE_1$ characterize the possible post-measurement state-spaces.

	Finally, for any two states $\ket{\psi}, \ket{\phi}$ stabilized by $\sS$, we have
	\begin{gather}
			\bra{\phi} P_v \ket{\psi}
		\;\;=\;\;
			\bra{\phi} S'_1 P_v \ket{\psi}
		\;\;=\;\;
			- \bra{\phi} P_v S'_1 \ket{\psi}
		\;\;=\;\;
			- \bra{\phi} P_v \ket{\psi}	\,,
	\end{gather}
	so that $\bra{\phi} P_v \ket{\psi} = 0$.
	Define the state-vectors
	\begin{gather}
			\ket{\bar\psi_r}
		\;\;=\;\;
			\frac{\Pi_r \ket{\psi}}{\sqrt{\bra{\psi}\Pi_r{\ket{\psi}}\,}\,}
		\;\;=\;\;
			\sqrt{2} \; \Pi_r \ket{\psi}	\;,
		\\[1ex]
			\ket{\bar\phi_r}
		\;\;=\;\;
			\frac{\Pi_r \ket{\phi}}{\sqrt{\bra{\phi}\Pi_r{\ket{\phi}}\,}\,}
		\;\;=\;\;
			\sqrt{2} \;\Pi_r \ket{\phi}
	\end{gather}
	for the states arising after a $P_v$ measurement on $\ket{\psi}$ and $\ket{\phi}$ respectively, conditioned on obtaining the result $r \in \ens{0,1}$.
	Then, the inner product of $\ket{\bar\phi_r}$ with $\ket{\bar\psi_r}$ can be given by
	\begin{align}
			\bracket{\bar\phi_r}{\bar\psi_r}
		\;\;=\;\;
			2 \bra{\phi} \Pi_r \ket{\psi}
		\;\;=&\;\;
			\bra{\phi} (\idop_2^{\ox n} + (-1)^r P_v) \ket{\psi}
		\notag\\=&\;\;
			\bracket{\phi}{\psi}	\;;
	\end{align}
	thus the transformation performed by the $P_v$ measurement (conditioned on either result) preserves inner products, and is therefore a unitary transformation.

\item
	Finally, for any destructive $P_v$ measurement, note that the state of $v$ after a similar von Neumann measurement is characterized by being a $\pm 1$-eigenvector of $P_v$ determined by the measurement result; and in particular is independently distributed from the other qubits.
	Then no information about the rest of the qubits is lost by discarding the qubit.
	If the measurement result is $r \in \ens{0,1}$, let $R_v = (-1)^r P_v$ be the operator stabilizing the state of $v$ in the post-measurement stabilizer group $\sS_r = \{S'_j\}_{j = 1}^{n-k'}$ --- where we may easily verify that $k' = k-1$ in the case where $P_v$ commuted with (but was not generated by) the pre-measurement stabilizer group, and $k' = k$ otherwise.
	Because all the generators $S'_j$ commute, any other operator $S'_j \ne R_v$ must acts on $v$ with either the identity or a $P_v$ operation.
	We may then form a new set of generators
	\begin{gather}
			S''_j
		\;\;=\;\;
			\begin{cases}
				S'_j		\;,	&	\text{if $S'_j = R_v$ or if $S'_j$ acts on $v$ with $\idop_2$}	\\[0.5ex]
				S'_j R_v	\;,	&	\text{otherwise}
			\end{cases}
	\end{gather}
	for $\sS_r$\,, where $R_v \in \{S''_j\}_{j = 1}^{n- k'}$ is the only operator which acts non-trivially on $v$.
	We may then trace $v$ out by removing the generator $R_v$, and tracing out $v$ from the domain of the other generators $S''_j$\,.
\end{itemize}
Thus, the state-space of a system produced by a Clifford circuit, assuming a arbitrary pure input state, can be efficiently computed for any possible value of the measurement results, and even whether the evolution will be unconditionally unitary.
\label{discn:endStabilizerFormalism}

Applied to the special case when there are \emph{no} quantum inputs (\ie\ where all qubits are allocated by the circuit itself, and prepared in one of the states $\ket{\splus \sfx}$, $\ket{\splus \sfy}$, or $\ket{\splus \sfz}$), this yields the following result, reported in~\cite{Gottesman98} and attributed to Emmanuel Knill:\footnote{%
The result stated in~\cite{Gottesman98} is actually an equivalent result, in which the only preparation and measurement operations are $\New{}{\sfz}$ and $\sfP_{v:r}^{\sfz}$ (and the unitary gate $\cX$ is used in place of $\cZ$).}
\begin{theorem*}[Gottesman-Knill Theorem]%
	\label{thm:GottesmanKnill}
	A circuit in the Clifford circuit model which has no quantum inputs can be simulated efficiently on a classical computer.
\end{theorem*}
Note that the \Clifford\ model is not even approximately universal for quantum computation: because there are only finitely many elements of $\cP^{\ox n}$, there are also only finitely many stabilizer groups of size $2^n$ on $n$ qubits, and thus finitely many stabilizer states.
Thus, there exist CPTP maps $\Upphi: \ens{1} \to \D(2^n)$ which cannot be approximated to arbitrary precision with a Clifford circuit.

As we remarked above, circuits in the $\Phases(\frac{\pi}{2}\Z)$ model without trace-out operations are also Clifford circuits; and therefore circuits in the $\Jacz(\frac{\pi}{2}\Z)$ models without trace-out operations can be easily simulated by Clifford circuits; then these models are also not approximately universal for quantum computation.\footnote{\label{fn:parityL}%
A further result of Aaronson and Gottesman~\cite{AG04} show that simulating Clifford circuits is a complete problem for $\ParityL$~\cite{BHMD91}, the class of problems which can be solved by a nondeterministic Turing machine (NTM) which uses only logarithmic workspace, and where in particular the ``\emph{yes}'' instances of the problem have an odd number of accepting paths for that NTM.
Solving linear equations over $\Z_2$ is another problem which is complete for $\ParityL$~\cite{Damm90}, and which is conjectured to be insufficient even to perform universal polynomial-time \emph{classical} computation.}


\section{Conclusion}
\label{sec:introConclusion}

In this Chapter, we have described quantum states as being represented by density operators, and transformations of them as being by CPTP maps, which may be performed or approximated by composing discrete operations.
We have also noted the important role that unitary evolution has in quantum mechanics, and described the role of unitary transformations in defining the most commonly used models of quantum computation.

Other significantly different models of quantum computation exist which are not obviously special cases of either unitary circuit or classically-controlled unitary circuit models, but which can be shown to be polynomial-time equivalent:
\begin{description}
\item[Adiabatic quantum computation]\cite{FGGS00}
	rests on the adiabatic theorem first described by Born and Fock~\cite{BF28} (see~\cite{AR2004} for a modern treatment), and which is a model of computation relying on the evolution of states under a slowly and continuously varying Hamiltonian.
	This model lends itself naturally to optimization problems: an analysis of the strengths and limitations of this model, and its relation to unitary circuits, can be found in~\cite{DMV2001}.

\item[Topological quantum computation]\cite{Kitaev2003}
	is a model which exploits properties of particle statistics to approximate unitaries to arbitrary precision by braiding the trajectories of ``virtual particles'' which can be simulated via the quantum Hall effect~\cite{ASW1984}.
	This is a natural model for describing algorithms for such problems as approximating knot invariants~\cite{AJL2006,LK2006}; and by exploiting the topological properties of the particle trajectories, it is possible to naturally achieve fault-tolerant computation.
	A review of toplogical quantum computation can be found in~\cite{NSSFS2008}.

\item[Continuous-time quantum walks]\cite{FG98}
	represent a generalization of diffusion processes in graphs by random walks by replacing the diffusion equation with Schr\"odinger evolution.
	(Discrete-time quantum walks can also be formulated naturally by generalizing random walks~\cite{Watrous98}, but these may be easily subsumed by classically controlled unitary circuits.)
	This model led to the discovery of a quantum speed-up for evaluating \textsc{nand}-trees~\cite{FGG07}; the universality of this model is shown in~\cite{Childs2008}.
\end{description}

In this thesis, we are primarily interested in one further model of quantum computation, which may be subsumed into classically controlled unitary circuit models, but in which measurements depending on many classical controls play a significant role, resulting in a model with a significantly different character than unitary circuit models.
This model is the \emph{one-way measurement-based model}, which we introduce in the next chapter.
	\chapter[One-way measurement-based computation]{
		One-way measurement-based computation}


In the absence of efficient algorithms for simulating quantum mechanics or for solving difficult number-theoretic problems such as factoring or discrete logarithms, quantum computers represent a potential breakthrough in scientific computing.
However, the technical challenges involved in controlling the evolution of a quantum mechanical system for the purposes of computation have prompted the exploration of different ``computational primitives'' for quantum computation --- meaning essentially different sets of operations which
\begin{inlinum}
\item
	form a model which is universal for quantum computation, and
\item
	may be easier to implement physically.
\end{inlinum}

In order to achieve universal quantum computation via unitary circuits, multi-qubit entangling gates are required: otherwise, only states which are tensor-products of single-qubit states can be produced.
As scalable quantum control of multi-qubit systems is an open research problem, reducing the number of rounds involving multi-qubit interactions is one plausible route to achieving practical scalable quantum computation.
However, to perform or approximate any unitary evolution which cannot be efficiently simulated on a classical computer, many rounds of multi-qubit operations such as $\cX$ or $\cZ$ (as defined in \eqref{eqn:basicGates}) are required in a unitary circuit model.\footnote{%
To be precise, although it is widely believed that the measurement results arising out of quantum mechanics (and arbitrary unitary circuits in particular) cannot be efficiently approximated by a classical model of computation such as Turing machines, any unitary circuit on $n$ qubits in which each qubit is operated on fewer than $O(\log(N))$ times by a two-qubit gate can be simulated in $n \cdot\poly(N)$-time by a classical computer~\cite{Jozsa03}.}

If we are interested in describing unitary transformations of states (or \emph{approximately} unitary transformations, in the sense of approximating a unitary CPTP map with precision $\epsilon$), while allowing classically controlled measurements to play a significant role in the computation, a single round of multi-qubit operations --- or alternatively, access to appropriate multi-qubit resource states --- will suffice to achieve universal quantum computation.
The models which make this possible are the \emph{measurement based models} of quantum computation.

\begin{figure}[t]
	\begin{center}
		\includegraphics{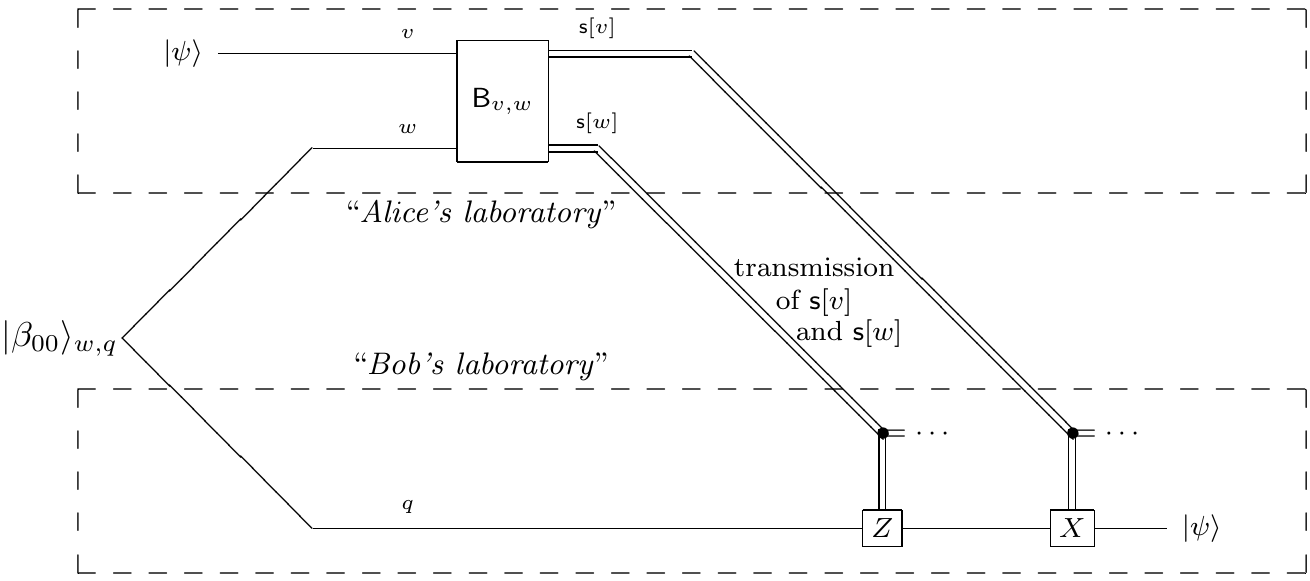}	
	\end{center}
	\caption[An illustration of the ``teleportation'' protocol.]{\label{fig:teleportation}%
		An illustration of the ``teleportation'' protocol defined in~\eqref{eqn:teleportation}.
		We use similar conventions as in Figures~\ref{fig:qcircuit-1} and~\ref{fig:qcircuit-2} to illustrate circuits with measurements and classical control; double-lines represent classical bits produced by measurement.
		The system is divided into two separate ``laboratories'' controlled by separate parties (``Alice'' and ``Bob''), between which we suppose that coherently controlled operations cannot occur.
		Diagonal lines represent the sharing of information resources: the preparation of the joint initial state $\ket{\beta_{00}}$ of $w$ and $q$, and the transmission of the bits $\s[v]$ and $\s[w]$.}
\end{figure}

In~\cite{GC99}, Gottesman and Chuang showed that the preparation of a special set of multi-qubit states, together with teleportation~\cite{BBCJPW93}, was universal for quantum computation.
Teleportation is essentially a means of implementing the identity operation by means of measurement and classical control: if we define the \emph{Bell basis} by the states
\begin{small}
\begin{align}
	\begin{split}
			\ket{\beta_{00}}
		\;=&\;
			\sfrac{1}{\sqrt 2} \ket{0} \ox \ket{0} \;+\; \sfrac{1}{\sqrt 2} \ket{1} \ox \ket{1}	\;,
		\\
			\ket{\beta_{01}}
		\;=&\;
			\sfrac{1}{\sqrt 2} \ket{0} \ox \ket{0} \;-\; \sfrac{1}{\sqrt 2} \ket{1} \ox \ket{1}	\;,
	\end{split}
	&
	\begin{split}
			\ket{\beta_{10}}
		\;=&\;
			\sfrac{1}{\sqrt 2} \ket{0} \ox \ket{1} \;+\; \sfrac{1}{\sqrt 2} \ket{1} \ox \ket{0}	\;,
		\\
			\ket{\beta_{11}}
		\;=&\;
			\sfrac{1}{\sqrt 2} \ket{0} \ox \ket{1} \;-\; \sfrac{1}{\sqrt 2} \ket{1} \ox \ket{0}	\;,
	\end{split}
\end{align}
\end{small}
and a \emph{Bell basis measurement} operation by the map
\begin{gather}
		\mathsf B_{v,w}(\rho_{v,w})
	\;\;=\;
		\sum_{x,z \in \ens{0,1}}	\!
			\bra{\beta_{xz}} \rho \ket{\beta_{xz}}_{\!v,w} \; \Big[ \ket{x}\bra{x}_{\s[v]} \ox \ket{z}\bra{z}_{\s[w]} \Big] \;,
\end{gather}
then teleportation (as described in~\cite{GC99}) is the map $\mathsf T_{q/v} : \D(v) \to \D(q)$ given by
\begin{gather}
	\label{eqn:teleportation}
		\mathsf T_{q/v}(\rho_v)
	\;\;=\;\;
		\Big( \Xcorr{q}{\s[v]} \circ \Zcorr{q}{\s[w]} \circ \,\mathsf B_{v,w} \circ \Big[ \idop_v \ox \:\! \ket{\beta_{00}}\bra{\beta_{00}}_{w,q} \Big]\Big)(\rho_v),
\end{gather}
where $\Zcorr{q}{s}$ and $\Xcorr{q}{s}$ are classically-controlled $Z$ and $X$ operations depending on the bit $s$.
The result of~\cite{BBCJPW93} is that $\mathsf T_{q/v}$ as defined in~\eqref{eqn:teleportation} performs the map $\idop_{q/v}$\,: that is, the state of $v$ is ``teleported'' unchanged into $q$.\footnote{%
The description of the output neglects the classical measurement results $\s[v]$ and $\s[w]$, which are also produced as output and which will be uncorrelated and uniformly randomized bits; it is conventional to implicitly discard classical measurement results when they are no longer required.}
This remains true even though the qubit $q$ may never have interacted with the qubit $v$, and may be separated from $v$ by a large distance: if the qubits $w$ and $q$ are controlled by different parties (conventionally named ``Alice'' and ``Bob'') after their joint preparation in the state $\ket{\beta_{00}}$, it suffices for Alice to communicate the results $\s[v]$ and $\s[w]$ to Bob, and for Bob to perform the appropriate operations depending on those bits.
This is illustrated in Figure~\ref{fig:teleportation}.

Teleportation is thus an important primitive for transmitting quantum information, using the correlations of the Bell state $\ket{\beta_{00}}\bra{\beta_{00}}$ and classical communication; \cite{GC99} showed that by generalizing the initial states used as a resource, teleportation (and in particular Bell measurements) could also be used as a \emph{computational} primitive.
In the three years following Gottesman and Chuang's result, three different proposals for quantum computation based on measurement operations arose.
Nielsen~\cite{Nielsen2003} described a model of quantum computation based on four-qubit measurement operations; this together with the result of~\cite{GC99} which was subsequently refined by Leung~\cite{Leung2001} into what is called the \emph{teleportation-based model of quantum computation} in~\cite{CLM2005}.
The third proposal, the \emph{one-way measurement-based model} presented by Robert Raussendorf and Hans Briegel in~\cite{RB01}, forms the starting point for the main topic of this thesis.

In this chapter, we will describe different variants of the one-way measurement model in terms of the \emph{stabilizer formalism} presented in Section~\ref{sec:stabilizerFormalism}, and provide an overview of applications and potential schemes for implementing one-way measurement based quantum computing.

\paragraph{New results.}

Strictly speaking, none of the results of this Chapter are new.
However, we present explicit definitions for three distinct (but closely related) constructions of measurement-based computation, to which we will regularly refer in later chapters.
These constructions are the \emph{simplified} and \emph{complete DKP constructions} (after Danos, Kashefi, and Panangaden~\cite{DKP07}), defined in Section~\ref{sec:DKPstandardization}, and the \emph{simplified RBB construction} (after Raussendorf, Browne, and Briegel~\cite{RBB03}), defined in Section~\ref{sec:RBB}.

\section{Overview and proposals for implementation}

As the one-way measurement model was originally proposed as a scheme for implementations of quantum computation, we will provide a quick overview of the model in order to hilight the aspects which make it seem reasonable as a candidate for scalable quantum computation.
Before describing the model in full detail, we will give an overview of different proposals for fulfilling the requirements of the one-way measurement model.

\subsection{A brief description of the one-way model} 

The one-way measurement model of quantum computing proposed in~\cite{RB01} is a model of computation which involves no explicit multi-qubit operations: instead, it allows the preparation of arbitrary states from a uniform family of states,\footnote{%
A uniform family of states is one which can be efficiently specified by a classical, deterministic Turing machine in time polynomial in the size of the desired state.}
called \emph{cluster states}.

\begin{definition}
	\label{def:clusterState}
	The \emph{$n \x d$ grid} is a graph $G_{n,d}$ whose vertex set consists of \emph{grid sites} $v \in \ens{1,\ldots,n} \x \ens{1,\ldots,d}$, and whose edges are given by adjacent grid sites, \ie\ pairs $vw$ of vertices $v, w \in V(G)$ for which $\norm{v - w}_1 = 1$\,, where $\norm{x}_1$ is the $\ell_1$-norm.
	The \emph{$n \x d$ cluster state} is then a stabilizer state whose stabilizer group is given by $\cC_{n,d} = \gen{K_v}_{v \in V(G)}$, where
	\begin{gather}
			K_v
		\;\;=\;\;
			X_v \prod_{w \in \ngbh{v}} Z_w	\;,
	\end{gather}
	and where $\ngbh{v}$ denotes the vertices adjacent to $v$ in the graph $G_{n,d}$.
\end{definition}

The $n \x d$ cluster state can be produced by evolution by an Ising interaction operating for an appropriate length of time, where each qubit in the grid is initially prepared in the $\ket{+}$ state~\cite{RB01}: several schemes for how this may be done are described in the original article.

Having a cluster state, computation is then performed using single qubit measurements.
Qubits are destructively measured either using $Z$ observables, or observables of the form $M(\theta) = \cos(\theta) X + \sin(\theta) Y$ which anticommute with $Z$; the qubits measured with a $Z$ observable are effectively removed from the computation, while those measured with an observable $M(\theta)$ are used to control what computation is performed.
It is possible to show that each qubit in the cluster is maximally entangled with its' nearest neighbors; then, each measurement result is maximally random.
The different measurement outcomes corresponds to different transformations of the data in the computation: in order to simulate \eg\ a unitary circuit, it becomes necessary to retain the results of the measurements long enough to counteract the randomness, \eg\ by changing the measurement observables of later measurements.
Thus, the one-way measurement model is a \emph{classically controlled} model of computation.
Furthermore, the number of measurement results that an operation may depend upon may be arbitrarily large, in general scaling with the length of the computation.

We will describe the one-way measurement model more fully in Section~\ref{sec:oneWayComputation}: for the time being, we will remark that the necessary elements of the one-way model as a scheme for for implementations of quantum computation are the possibility of creating a family of large entangled states; the ability to perform classically controlled measurements; and the ability to compute and transmit measurement results quickly enough to act as the classical control of further measurements.
We can broaden the criteria for implementation somewhat by observing the following:
\begin{itemize}
\item
	The large entangled states required do not necessarily have to be cluster states.
	We will show how the model of~\cite{RB01} may be generalized to allow quantum computation with general \emph{graph states}, which we may define in analogy to cluster states by replacing the $n \x d$ grid with an arbitrary graph.

\item
	Rather than adapting the measurements, we may allow classically controlled unitary operations just prior to each measurement, which performs a change of reference frame which maps some ``logical'' measurement observable to a particular ``default'' measurement observable.
	Generalizing still further, we may provide for the preparation of arbitrary density operators if we leave some qubits unmeasured to support a final state; in this case, classically controlled unitaries are necessary to enable the preparation of pure states.
\end{itemize}

These are the primary features which make the one-way measurement model an attractive prospect for implementation: it provides a strict divide between a stage involving the creation of entanglement, and a stage involving only single-qubit operations.
The requirement that these operations depend on the results of previous measurements may present a challenge, but this challenge is of a significantly different character than the control of a many-body Hamiltonian for the purpose of performing multi-qubit transformations.
As well, once a cluster state has been prepared, it can in principle be safely stored until ready for use, provided a sufficiently reliable quantum memory.

\subsection{Proposals for implementation} 

One of the promises of the one-way measurement model is that the state involved is quite simple (in particular, it is a stabilizer state) and can in principle be easily prepared.
The original proposal~\cite{RB01} itself suggested that the state may be possible to produce by controlled Ising interactions on qubits prepared in the $\ket{\splus}$ state.
We now consider the proposals for schemes for producing cluster states and graph states, and for performing one-way measurement-based computation generally.

\subsubsection{Optical proposals}

The first proposal for efficient quantum computing by linear optics (albeit not implementing the one-way measurement model) was presented in~\cite{KLM2001}, which introduced an entangling gate using number counting photo-detectors (thus introducing an effective non-linearity) which failed with finite probability using techniques similar to~\cite{GC99}.
However, in order to perform scalable quantum computation, a significant amount of overhead in error correction is required even in the case of ideal apparatus as there is a significant probability of photon loss due to destructive measurement.

The approach of~\cite{KLM2001} was improved upon in~\cite{Nielsen2004}, which creates a cluster state as an entangled resource by successive ``fusion'' operators, used to append qubits onto an existing graph state to make a larger graph state.
The overhead due to the possibility of failure of the fusion operators is then realized as repeated attempts to create a cluster state: in the case of ideal apparatus, no error correction is required, resulting in a reduction of overhead.
A further improvement is made by~\cite{BR2005}, who employ fusion operators realized via a ``simplified'' application of the Hong-Ou-Mandel effect~\cite{HongOuMandel87}\footnote{%
The simplification entailed in the results of~\cite{BR2005} is that the Hong-Ou-Mandel effect does not occur as a part of a Mach-Zehnder interferometer, which is sensitive to perturbations in path length on the order of one wavelength.} to reduce the sensitivity of the interferometer on path perturbations.

Another approach~\cite{LBK2005} known as ``repeat until success'' uses a simple error correcting code in conjunction with linear optics and measurement, to implement entangling gates in such a way that the quantum information of the system is not destroyed when the logical gate fails: this allows cluster states to be built with a much better expected time, as in the ideal case no losses arise from a failed operation.

In all of the above models, as with all optical set-ups, the primary challenges to be overcome are inefficiencies associated with photon creation and detection: see \eg~\cite{LBBKK2006} for a discussion of these issues; more information on linear optical proposals can be found in~\cite{Kok2007}.

\subsubsection{Matter qubits and ``hybrid'' matter-optics approaches}

In~\cite{BK2005}, a scheme for constructing cluster states by interacting ``stationary'' qubits (specifically, defects in diamonds) is proposed which probabilistically entangles pairs of excited qubits by performing interferometry on the photons emitted in decay processes.
Combining this with the ``repeat until success'' techniques from linear optics led to the proposal of~\cite{LBBKK2006}, which promises to provide the best features of both proposals with regards to resilience against noise: \ie\ stable quantum memory provided by the spatially separated ``stationary'' qubits, whose interactions are mediated by deterministic photon gates provided by the optical elements.

An alternative approach to the previous, explicitly hybrid model between matter and optics is to employ a \emph{broker/client model} as described in~\cite{BBFM2006}, in which fragments of graph states are generated using fast-reacting but unstable subsystems (such as electron spin), and then transferred to more stable sub-systems (such as nuclear spins) to reduce the probability of loss..

Optical lattices form another setting for proposals of one-way measurement-based computation, wherein cold atoms are localized in a periodic electromagnetic field.
One such proposal~\cite{CAJ2005} proposes to represent qubits by the local modes of a linear array of non-interacting fermions, which become entangled simply as a consequence of Fermi statistics as the state of the fermions undergo mirror inversion.
Another proposal~\cite{KPA2006} proposes to mediate interactions by using global control to transport a ``virtual'' mediating qubit, represented by a pair of reserved excited states for each lattice site arising from the tuning of the lattice.

Almost all of the models mentioned in this and the preceding section address only the creation of the cluster state (which is admittedly likely to be the more difficult task in performing one-way measurement based computation, due to the need to control multi-qubit operations).
A departure from this is a proposal for producing cluster states with charge and flux qubits with quantum dots is described in~\cite{TLFHN2006}, which is robust to the nonuniformity of the dot formation process.
Measurement adaptations are explicitly performed by controlled rotations, which are performed by moderating the same continuous time Hamiltonian which gives rise to the cluster state itself.

A result of~\cite{Nielsen2006} is that neither the cluster state itself, (nor any other universal resource for the one-way model, can be the unique ground state of any ``physically realistic'' Hamiltonian; despite this, \cite{BR2006} describes a nearest-neighbor two-local Hamiltonian with a unique ground state on the grid, which may be interpreted as an \emph{encoded} cluster state. 
This indicates that a cluster state (or equivalent) is likely to be physically attainable by cooling an appropriate system to its ground state, and performing measurements to obtain the cluster state itself (equivalently, performing single-qubit measurements in such a way as to \emph{simulate} the a one-way measurement-based computation using a cluster).

Finally, a different approach to the creation of a cluster state is to encode the qubits in different energy modes of a single physical system.
One proposal of this type~\cite{MFP2008} is to implement cluster states in the modes of an optical cavity, using highly peaked (or \emph{squeezed}) distributions over continuous variables to represent computational basis states.
The high degree of coherent control over optical cavities is touted as an advantage for this scheme; however, the degree of ``squeezing'' required to perform any given computation is not yet known.


\subsection{Approaches to fault-tolerance in the one-way model}

A pre-requisite of scalable quantum computation is the ability to protect the state of a system against noise.
An alternative to achieving the ability to directly control the interactions of the system versus the environment is to apply generic techniques for protecting quantum information against uncontrolled operations: this approach is what is generally referred to as \emph{fault tolerance} of quantum computation \cite{KLABVZ2002,LW2003}.

Most of the attention to fault-tolerant approaches to one-way measurement-based computation have focused on the linear-optical proposals for implementation, where the noise operations (aside from those induced by imperfect realizations of logical gates) can be described in terms of loss errors and decoherence errors.
A scheme for loss-tolerance is proposed in~\cite{VBR2006}, which (provided perfect logical gates) can cope with photon losses of up to 50\%, by using stabilizer techniques to allow would-be measurement results of lost photons to be inferred.
However, \cite{RRM2007} describe that this yields an exponential blow-up of error rate for other reasons.
A modification of the scheme of~\cite{VBR2006} is also presented in~\cite{RRM2007}, for which they suggest a trade-off exists between the error rates of different types of error.
An optical protocol protecting against both polarization and loss errors is presented in~\cite{DHN2006}, requiring Bell pairs and employing the fusion operator techniques of~\cite{BR2005}.

An analysis applicable to linear optical set-ups, but also to optical lattices, is presented in~\cite{KRE2007}.
Using renormalization theory, they present a means of treating losses incurred due to non-deterministic entangling gates which eliminates the need to ``route'' states (\ie\ conditionally operate in the case of other, failed, operations).
This analysis yields scaling of physical requirements similar to hypothetical schemes where the entangling gates are deterministic, which may substantially improve the prospects of those physical implementations which rely on lossy, probabilistic operations.

Analyses of one-way based computation in decoherence free subspaces (\ie\ subspaces of the state-space of a system which are naturally protected from noise by symmetries of the system) are presented in~\cite{TPK2007} for phase-damping errors in an optical lattice setting; the authors propose that the construction should be extensible to more general errors.
An experimental demonstration for a corresponding proposal in linear optics is presented in~\cite{PTSPKZ2007}.

Finally, a generic scheme for performing topological quantum computation natively in a three-dimensional cluster state (\ie\ in which two-dimensional grid graphs are replaced by three-dimensional grids) is presented in~\cite{RHG2007}.
The computation is performed in this proposal by simulating the evolutions of ``defects'' in a two-dimensional lattice over discrete time-steps; the trajectories of the defects trace out paths in the three-dimensional lattice, which may be braided to effect quantum computation (as briefly described for topological quantum computation in general, in Section~\ref{sec:introConclusion}).

\section{Computing in the one-way measurement model}
\label{sec:oneWayComputation}

The one-way measurement model can be used to achieve universal quantum computation by performing \emph{adaptive single-qubit measurements} --- classically controlled measurement operations, which may depend on the results of prior measurements --- and classically controlled Pauli operations on the output qubits, which compensate for the randomness of the prior measurement results.
(In the original presentation~\cite{RB01}, the correction or \emph{byproduct} operation is never explicitly performed, and is instead used to adapt the final measurement used to produce the read-out of the computation: in order to describe this model of computation as being universal in the sense of Definition~\ref{def:universalQC}, we include the conditional corrections in the description.\footnote{%
The one-way model, as presented in~\cite{RB01}, is a model of ``\textbf{CC}-universal'' quantum computation in the terminology of~\cite{NDMB2007}: however, the way in which we extend it by including classically controlled unitaries to produce a ``\textbf{CQ}-universal'' model of computation is already implicit in the analysis of~\cite{RB01}.})

Every operation after the preparation of the cluster state acts only on individual qubits --- albeit possibly with classical control --- and (except for the classically controlled unitaries by which we extend the model) discards the qubits it acts on.
The cluster state is therefore consumed as a computational resource via measurement; hence the name of the model, which is irreversible, in contrast with the reversibility of closed-system Schr\"odinger dynamics.

We will spend most of this thesis considering aspects of measurement-based models of quantum computation which have been abstracted from the model of~\cite{RB01}, with emphasis on one such model in particular (Definition~\ref{def:oneWayModel}) which retains what can be regarded as the essence of the original proposal in~\cite{RB01}.
In doing so, we will implicitly describe schemes of translating unitary circuits into various models of quantum computation based on adaptive single-qubit measurements.

\paragraph{A remark on terminology.}

Despite the existence of other models of measurement-based computation, such as teleportation-based computation~\cite{Leung2001} or further relaxations of the one-way model as in~\cite{GE2007}, we will generally refer only to the models presented in this Chapter when we use the term \emph{measurement-based quantum computation}.
Because of the emphasis on \emph{unitary} circuits associated with most research in quantum computation, we will use the terms \emph{pattern} or \emph{procedure} in place of the word ``circuit'' when describing well-defined compositions of operations in measurement-based models of computation, following conventions present in the literature (\eg~\cite{DK06,DKP07,BKMP07}), in order to reinforce the prominence of non-unitary operations in the model.

\subsection{The cluster state model}
\label{sec:clusterStateModel}

We consider a slightly extended version of the model of~\cite{RB01} (which we call the \emph{cluster state model}), as follows.
For any $n,d \ge 1$, we include the map $\New{G_{n,d}}{}$ which prepares the $n \x d$ cluster state on fresh qubits.
The measurement operations used in~\cite{RB01} are $\Meas{}{\sfz}$ measurements as in~\eqref{eqn:pauliMeas} and, \emph{\XY-plane measurements} $\Meas{}{\alpha}$, defined for angles $\alpha \in (-\pi, \pi]$ by
\begin{gather}
	\label{eqn:xyPlaneMeas}
		\Meas{v}{\alpha}(\rho)
	\;\;=\;\
		\bra{\psub\alpha}_v \,\rho\, \ket{\psub\alpha}_v		\;\ox\; \ket{0}\bra{0}_{\s[v]}
		\;\;+\;\;
		\bra{\msub\alpha}_v \,\rho\, \ket{\msub\alpha}_v	\;\ox\; \ket{1}\bra{1}_{\s[v]} \;,
\end{gather}
where the states $\ket{\pmsub\alpha}$ are the $\pm 1$-eigenstates of the operator $\cos(\alpha) X + \sin(\alpha) Y$\,,
\begin{gather}
	\label{eqn:pmsubAlpha}
		\ket{\pmsub\alpha}
	\;\;=\;\;
		\sfrac{1}{\sqrt 2} \ket{0} \;\pm\; \sfrac{\e^{i \alpha}}{\sqrt 2} \ket{1}	\;.
\end{gather}
We refer to $\alpha$ as the \emph{measurement angle} of the measurement operator.\footnote{%
Representing the states $\ket{\pmsub\alpha}$ in the Bloch sphere, $\alpha$ is the angle between of the state $\ket{\psub\alpha}$ and the $\ket{\splus} = \ket{\splus \sfx}$ state, \ie\ the \axisX-axis.} 
The \XY-plane measurements are usually classically controlled, in that the sign of the measurement angle may depend on the parity of some sequence of bits --- in particular, some sequence of prior measurement results.\footnote{%
Note that this is a slight departure from the description of classically controlled measurements on page~\pageref{discn:classicalControl}, where the measurement performed depends explicitly on the value of a \emph{bit}, rather than abstractly on the parity of a set of many bits.
While it is likely that the measurement basis would depend explicitly on some bit which stores the value of such a boolean expression in practical implementations, we will not consider explicitly how this parity is evaluated.
However, in a serious consideration of \eg\ the depth-complexity of quantum operations in the one-way model, how the parities are evaluated is an important detail.}
Therefore, we introduce the abbreviation
\begin{gather}
		\Meas{v}{\alpha; \beta}
	\;\;=\;\;
		\Meas{v}{(\minus 1)^\beta \!\!\;\cdot \alpha}
\end{gather}
for a classically controlled measurement performing either $\Meas{}{\alpha}$ or $\Meas{}{\minus \alpha}$, depending on a boolean expression $\beta = \s[a] + \s[b] + \cdots$ representing the sum (modulo $2$) of some set of measurement results.
\label{discn:defaultMeasAngle}
We refer to $\alpha$ as the \emph{default angle} of the measurement (or the \emph{default measurement angle}).
It will be convenient to extend this further to accommodate changes in angle by a possible addition by $\pi$, which is not a part of the model of~\cite{RB01}, but which will facilitate analysis later: therefore, we will write
\begin{gather}
		\Meas{v}{\alpha; \beta, \gamma}
	\;\;=\;\;
		\Meas{v}{(\minus 1)^\beta \!\!\;\cdot \alpha \,+\, \gamma \pi}
\end{gather}
for boolean expressions $\beta$ and $\gamma$\,; we call $\beta$ a \emph{sign dependency} $\gamma$ a \emph{$\pi$-dependency}.
Similarly, as previously mentioned, we also extend the model of~\cite{RB01} by allowing operations $\Xcorr{v}{\beta}$ and $\Zcorr{v}{\beta}$ on arbitrary qubits $v$, which perform the identity if $\beta = 0$ and an $X$ or $Z$ operation (respectively) on $v$ if $\beta = 1$.
Thus, the extended version of the model of~\cite{RB01} can be described as follows:
\begin{definition}
	For any set of angles $\A \subset \R$ the \emph{cluster state model} is the model of computation with the elementary gate set
	\begin{align}
		\mspace{-50mu}
			\Cluster(\A)
		\;\;=\;\;
			\bigg\{& \New{G_{n,d}}{}\,,\; \Meas{v}{\sfz}\,,\; \Meas{v}{\alpha;\beta,\gamma}\,,\; \Xcorr{v}{\beta}\,,\; \Zcorr{v}{\beta}\,,\; \tr[v]
 				\,\bigg|\, n,d \ge 1, ~\alpha \in \A\,,
		\notag\\[-1.5ex]&\quad\qquad
			~\beta,\gamma \;= {\textstyle\sum\limits_{\s[v] \in S} \s[v]} ~~\text{for any $S \subset B$}, ~\text{and}~ v \in \N \x \N \bigg\}\,,
	\end{align}
	where $B$ is an infinite set of bits and $\s: \N \x \N \to B$ is injective.
	We will say that a $\mathfrak P$ procedure in the $\Cluster(\A)$ model is in \emph{standard form} if it can be decomposed as $\mathfrak P \,=\, \mathsf C \circ \mathsf M \circ \smash{\New{G_{n,d}}{}}$\,, where $\mathsf M$ is a composition of \emph{adaptive measurements} $\Meas{v}{\alpha;\beta}$ without $\pi$-dependencies, and where $\mathsf C$ is a composition of \emph{correction operations} $\Xcorr{v}{\beta}$ and $\Zcorr{v}{\beta}$.
	We then call $\New{G_{n,d}}{}$ the \emph{preparation stage} of the procedure, $\mathsf M$ the \emph{measurement stage} of the procedure, and $\mathsf C$ the \emph{correction stage} of the procedure.
\end{definition}
Because cluster states are stabilizer states, procedures in the $\Cluster(\frac{\pi}{2}\Z)$ model can be efficiently simulated by classical circuits by the Gottesman-Knill Theorem~(page~\pageref{thm:GottesmanKnill}).
However, procedures in $\Cluster(\frac{\pi}{4}\Z)$ fall outside of the domain of the stabilizer formalism; no corresponding result is known in that case.

In order to perform universal quantum computation, the one-way model must be able in particular to prepare (or approximate) arbitrary pure states for the qubits on which it act.
The measurement process, however, produces random outcomes.
In order to describe how the cluster state model can produce pure states by adaptive measurements and single-qubit unitary corrections, we give an account of this model in terms of unitary circuits.
This can be facilitated by considering further extensions beyond the cluster state model.

\subsection{The graph state model}

Consider the varieties of states which may be obtained after we perform all $\axisZ$ measurements in a cluster state measurement pattern, and before we apply any other operation.
Because $\axisZ$ measurements are not subject to adaptations based on prior measurements, we may commute the $\axisZ$ measurements to the beginning of the measurement phase of the procedure (whether or not it is in standard form) by applying the relations
\begin{align}
	\label{eqn:ZmeasCommute}
		\Meas{v}{\sfz} \, \Zcorr{v}{\beta} \;=&\; \Meas{v}{\sfz}	\,,
	&
		\Meas{v}{\sfz} \, \Xcorr{v}{\beta} \;=&\; \Shift{v}{\beta} \, \Meas{v}{\sfz}	\,;
\end{align}
for the latter relation, we define the \emph{shift} operator\footnote{%
The shift operator is usually not defined in accounts of measurement-based computation (with~\cite{DKP07} being an exception), probably because it is a very simple ``classical'' operation.
However, in any implementation of quantum computers based on single-qubit measurements, shift operators are likely to play an important role in describing how the parity expressions $\beta$ are computed in the midst of a computation.
Thus, while much of the analysis later in the thesis follows the convention of eliminating shift operations from measurement-based procedures, I explicitly allow shift operators as elementary gates, and extend the sense of ``standard forms'' found in the literature to allow shift operators to occur between measurements.}
$\Shift{v}{\beta}$,
\begin{subequations}
\label{eqn:shiftOptor}
\begin{gather}
		\Shift{v}{\beta}: \Prob(\s[v]) \to \Prob(\s[v])
	\\[1.5ex]
		\mathsf S^{\beta}(\rho)
	\;\;=
		\sum_{x \in \ens{0,1}} \ket{x \oplus \beta}\bra{x}\, \rho \,\ket{x}\bra{x \oplus \beta}
\end{gather}
\end{subequations}
acting on the bit $\s[v]$, where $a \oplus b$ is the sum of bits $a$ and $b$ modulo $2$.

Because the cluster state is a stabilizer state, we may determine the state of the system after the $\axisZ$ measurements.
For a measurement on a qubit $v$, $K_v$ is the only generator of the stabilizer group $\cC_{n,d}$ (as described in Definition~\ref{def:clusterState}) which does not commute with the observable $Z_v$.
By the stabilizer formalism (Section~\ref{sec:stabilizerFormalism}), the state of the other qubits after a $Z_v$ measurement where we obtain $\s[v] = r$ is then given by
\begin{gather}
		\tilde\cC\supp{r}
	\;\;=\;\;
		\gen{\tilde K\supp{r}_u}_{u \in V(G) \setminus v}	\;,
\end{gather}
where we have
\begin{gather}
		\tilde K\supp{r}_u
	\;\;=\;\;
		\begin{cases}[r@{}l]
				K_u				\;=&\; X_u \prod\limits_{w \sim u} Z_w
														\,,	&	\text{if $u \not\sim v$}	\\[3ex]
				(-1)^r Z_v K_u	\;=&\; (-1)^r \,X_u \!\prod\limits_{\substack{w \sim  u \\ w \ne v}} \!Z_w
														\,,	&	\text{if $u \sim v$}
		\end{cases}\,,
 \end{gather}
and where $\sim$ is the adjacency relation in the grid.
In the case where $\s[v] = 0$, the state of the system is stabilized by a group generated by $\tilde K_u\supp{0}$, which like the generators $K_u$ are tensor products of an $X$ operator and some number of $Z$ operators.
Otherwise, we may convert the state of the system to that stabilized by $\tilde\cC\supp{0}$ by performing the operations $\Zcorr{u}{}$ on every qubit $u$ adjacent to $v$ in $G_{n,d}$\,: because $Z_u$ anticommutes with $X_u$ and commutes with $Z_w$ for any qubit $w$, we have
\begin{gather}
		\Zcorr{u}{}\Big(\tilde K\supp{1}_u\Big)
	\;\;=\;\;
		Z_u \sqparen{ - X_u \prod_{w} Z_w } Z_u
	\;\;=\;\;
		X_u \prod_{w} Z_w 
	\;\;=\;\;
		\tilde K\supp{0}_u	\;.
\end{gather}
Then, we can select for the post-measurement state corresponding to $\s[v] = 0$ by performing the operation
\begin{gather}
	\mathsf B_v \;\;=\;\; \prod_{u \sim v} \Zcorr{u}{\s[v]}	\;.
\end{gather}
We call such an operation a \emph{byproduct} operation.
To perform pure-state computation, there is no loss in generality in selecting for the post-measurement state corresponding to $\s[v] = 0$\,, rather than $\s[v] = 1$\,, as they are equivalent up to conditional unitaries which may be performed in the cluster-state model in any case.

As we noted, the post-measurement state produced in this way is stabilized by operators $\tilde K\supp{0}_u$ for $u \in G_{n,d} \setminus v$, which are each tensor products of $X$ and $Z$ operators, but which are distinct from the stabilizers of $\cC_{n,d}$.
Furthermore, by construction, each $\tilde K\supp{0}_u$ acts on $u$ with an $X$ operation, and acts on a qubit $w \in V(G_{n,d})$ with a $Z$ operation if and only if $w \ne v$ and $w \sim u$: that is, if $w$ is adjacent to $u$ in the graph $G_{n,d} \setminus v$.
Thus, a $\axisZ$ measurement on $v$ (followed by the byproduct $\mathsf B_v$ if required) corresponds to removing a vertex represented by $v$ from the $n \x d$ grid graph, producing a state similar to the cluster state, but defined on a different graph.
This motivates the following definition:
\begin{definition}
	For an arbitrary graph $G$, the \emph{graph state} corresponding to $G$ is a stabilizer state $\rho_{\ssstyle G} = \ket{G}\bra{G}$, where $\ket{G}$ is a unit-vector stabilized by the group $\sS_G = \gen{K_{G,u}}_{u \in V(G)}$ for generators given by 
	\begin{gather}
		\label{eqn:graphStabilizers}
			K_{G,v}
		\;\;=\;\;
			X_v \prod_{w \in \ngbh[G]{v}} Z_w	\;,
	\end{gather}
	and where $\ngbh[G]{v}$ denotes the set of vertices adjacent to $v$ in the graph $G$.
\end{definition}
Applying the same analysis as for performing a $\axisZ$ measurement on a qubit $v$ of the $n \x d$ cluster state, we may show that for any graph $G$ and vertex $v \in V(G)$, we have
\begin{gather}
		\sqparen{\prod_{w \in \ngbh[G]{v}} \Zcorr{w}{\s[v]}} \Meas{v}{\sfz} (\rho_{\ssstyle G})
	\;\;=\;\;
		\rho_{\ssstyle [G \setminus v]}	\;.
\end{gather}
Any $\axisZ$ corrections on qubits $w$ which are also to be measured with a $\axisZ$ measurement can be absorbed into the $\Meas{w}{\sfz}$ operation as in~\eqref{eqn:ZmeasCommute}.
Let $S$ be the set of vertices measured with a $\Meas{}{\sfz}$ operation: $\axisZ$ corrections then accumulate on the neighbors of vertices in $S$, yielding the transformation
\begin{gather}
		\sqparen{\prod_{w \in \ngbh{S}} \Zcorr{w}{\,\big(\!\sum_{v \in S_w} \s[v]\big)}}
		\sqparen{\prod_{v \in S} \Meas{v}{\sfz}}	,
\end{gather}
where $\ngbh{S}$ is the set of qubits in $G_{n,d}$ adjacent to (but not contained in) $S$ in the $n \x d$ grid, and $S_w = \ngbh[G_{n,d}]{w} \inter S$.
This operation serves to transform $\rho_{\ssstyle G_{n,d}}$ to a graph state corresponding to some induced subgraph, $\rho_{\ssstyle [G_{n,d} \setminus S]}$\,.

We may abstract this process of producing graph states from cluster states via vertex removal, to describe a more general measurement-based model.
For an \emph{arbitrary} graph $G$, let $\New{G}{}$ be the map preparing the graph state $\ket{G}\bra{G}$.
Then we may consider a \emph{graph state model}  of computation, generalizing the cluster state model, by allowing preparation maps $\New{G}{}$ for arbitrary $G$ (or, for instance, for $G$ from some broader family of graphs) rather than just for grids $G_{n,d}$\,.

\subsection{Open graph states, geometries, and the general ``one-way'' model}
\label{sec:geometriesOneWay}

As with other models of quantum computing, in order to describe how to achieve universal quantum computation in measurement-based models of quantum computing, it is convenient to reduce to universality for unitary transformations.
This motivates another extension beyond graph-state models --- which can only prepare independent graph states and perform classically controlled single-qubit operations on them --- to a model which permits multi-qubit operations on previously allocated qubits, just as we generalized universality for quantum computation to universality for CPTP transformation.

A natural approach (and one also suggested in~\cite{RB01}) is to extend beyond graph states to maps which embed quantum states $\rho \in \D(I)$ into a stabilizer code similar to a graph state.
Here, $I$ is a set representing an \emph{input system} of qubits, which we include in the vertex-set $I \subset V(G)$ of a graph $G$\,: we then consider the stabilizer code described by a group $\sS_{G,I}$ generated by the operators $K_{G,v}$ as in~\eqref{eqn:graphStabilizers}, but ranging only over qubits $v \in V(G) \setminus I$.
In order to describe an explicit map for this embedding, consider how the operators $K_{G,v}$ may be generated by Clifford group operations from simpler operations.
From~\eqref{eqn:cliffordMapping}, we may observe that
\begin{gather}
		K_{G,u}
	\;\;=\;\;
		\sqparen{\prod_{v \in \ngbh[G]{u}} \!\!\Ent{u,v}} (X_u)	\,,
\end{gather}
where this product may be taken in any order; then, the code which is stabilized by $\sS_{G,I} = \gen{K_{G,u}}_{u \in V(G) \setminus I}$ can be obtained from the code stabilized by $\sS_{T,I} = \gen{X_u}_{u \in V(G) \setminus I}$ by performing the operation
\begin{gather}
		\Ent{G}
	\;\;=
		\prod_{uv \in E(G)}
			\!\!\Ent{u,v}	\;.
\end{gather}
The operators $X_u$ also generate the stabilizer group of a graphical code, for the ``trivial'' graph $T$ on the vertices $V(G)$ but with no edges.
This code consists of states $\ket{\splus}_u$ on each $u \in V(G) \setminus I$, with no constraints imposed on the qubits of $I$.
Thus, the map $\cE_{G,I} : \D(I) \to \D(V(G))$ encoding the qubits of $I$ into the code stabilized by $\sS_{G,I}$ can be given by
\begin{gather}%
	\label{eqn:openGraphEncoding}
		\cE_{G,I}
	\;\;=\;\;
		\sqparen{\prod_{uv \in E(G)} \!\! \Ent{u,v}} \sqparen{\prod_{u \in V(G) \setminus I} \!\!\New{u}{\sfx}}
	\;\;=\;\;
		\Ent{G} \, \New{I\comp}{}	\;,
\end{gather}
where we abbreviate $I\comp = V(G) \setminus I$.
We may refer to $\cE_{G,I}$ as an \emph{open graph state} encoding.\footnote{%
This terminology is similar to, but different from, that used in \eg~\cite{BKMP07,MP08}, 
who use the term ``open graph state'' to refer to what we call a \emph{geometry} in Definition~\ref{def:geometry} (in agreement with~\cite{DK05u}).} 
However, in contrast to the $\Cluster(\A)$ model, we do not make the embedding $\cE_{G,I}$ a primitive operation of the corresponding model of computation, and instead define the following model:
\begin{definition}
	\label{def:oneWayModel}
	For any set of angles $\A \subset \R$, infinite set $V$ of qubits, and infinite set $B$ of bits, the \emph{one-way model} is the model of computation with the elementary gate set
	\begin{align}
			\OneWay(\A)
		\;\;=\;\;
			\bigg\{& \New{v}{\sfx}\,,\; \Ent{u,v}\,,\; \Meas{v}{\sfz}\,,\; \Meas{v}{\alpha;\beta,\gamma}\,,\; \Xcorr{v}{\beta}\,,\; \Zcorr{v}{\beta}\,,\; \Shift{v}{\beta}\,,\; \tr[v]
 				\,\bigg|\, u,v \in V, ~\alpha \in \A, 
		\notag\\[-1ex]&\qquad\qquad
			  ~\text{and}~ ~\beta,\gamma \;= {\textstyle\sum\limits_{\s[v] \in S} \s[v]} ~~\text{for any $S \subset B$} \bigg\}\,.
	\end{align}
	We will say that a procedure $\mathfrak P$ in the $\OneWay(\A)$ model is in \emph{standard form} if it can be decomposed as $\mathfrak P \,=\, \mathsf C \,\circ\, \mathsf M \,\circ\, \Ent{G} \circ \New{S}{}$ for $\mathsf C$ consisting of corrections, $\mathsf M$ consisting of shift operations and measurement operations with no $\pi$-dependencies, $\Ent{G}$ consisting of a product of \emph{entangler} operations $\Ent{u,v}$ for distinct pairs $\ens{u,v} \subset V(G)$ for some graph $G$, and $\New{S}{}$ consisting of preparation operations $\New{v}{\sfx}$ for $v \in S \subset V(G)$.
	We refer to these as the \emph{correction}, \emph{measurement}, \emph{entangling}, and \emph{preparation} stages respectively.
\end{definition}
The model of measurement-based quantum computing which later chapters will be primarily concerned with is the model $\OneWay(\A)$ for sets of angles $\A \subset \R$.
We will refer to models of this form generically as \emph{the one-way model}, and refer to ``one-way procedures'', ``one-way patterns'', \etc\ when describing algorithms in the one-way model.

The purpose of extending to open graph encodings is to make possible the \emph{composition} of one-way patterns.
Therefore, in addition to the extension to admit an input subsystem, we also explicitly identify \emph{output} subsystems of measurement-based procedures, which simply consists of those qubits which are not measured (and the classical results of the measurements).
We may then achieve universality for unitary transformations by performing an appropriate set of elementary unitary transformations --- effectively simulating a unitary circuit in doing so.

\begin{figure}[t]
	\begin{center}
		\includegraphics{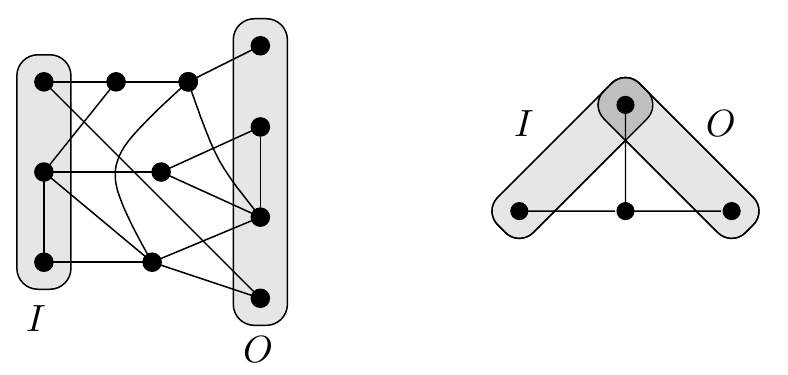}
	\end{center}
	\caption[Illustration of two geometries for one-way procedures.]{\label{fig:genericGeom}%
		Illustration of two geometries for one-way procedures.
		Labels for the vertices have been omitted.
		Note in particular that the sets $I$ and $O$ may overlap.}
\end{figure}

A specification of an open graph state encoding, together with the set of qubits to be left unmeasured, effectively captures the \emph{structural} information (that aspect which is independent of measurement observables and measurement results) of a one-way measurement procedure, which we may describe as follows:

\begin{definition}
	\label{def:geometry}
	For a measurement based procedure $\mathfrak P$ in the one-way model, the \emph{geometry underlying $\mathfrak P$} is a triple $(G,I,O)$ where
	\begin{romanum}
	\item
		$G$ is the \emph{entanglement graph}, whose vertices are labelled by the qubits on which $\mathfrak P$ acts, and whose edges are given by the operations of the entangling phase of $\mathfrak P$\,;

	\item
		$I \subset V(G)$ is the \emph{input subsystem} of $\mathfrak P$, consisting of the qubits which are not allocated by the preparation phase of $\mathfrak P$\,;

	\item
		$O \subset V(G)$ is the \emph{output subsystem} of $\mathfrak P$, consisting of the qubits which are not discarded by the measurement phase of $\mathfrak P$.
	\end{romanum}
\end{definition}
We will often illustrate geometries in the style shown in Figure~\ref{fig:genericGeom}.

\begin{figure}[t]
	\begin{center}
		\includegraphics{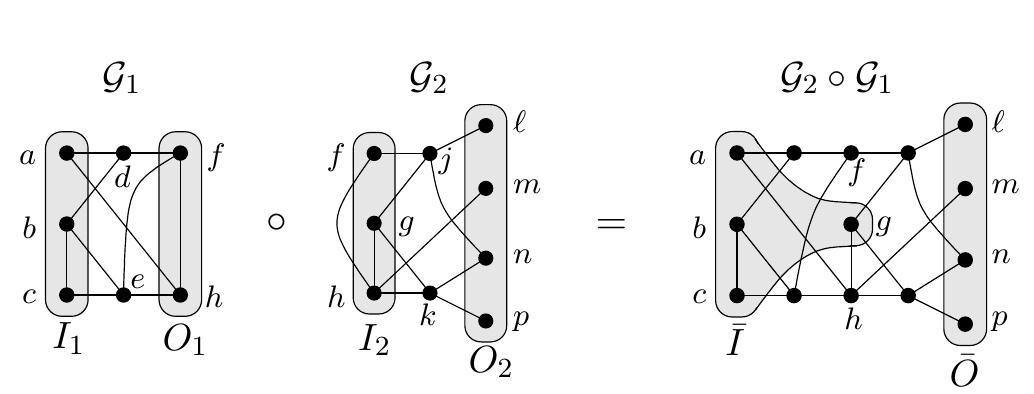}
		\vspace{-2ex}
	\end{center}
	\caption[Illustration of the composition of two generic geometries for one-way procedures.]{\label{fig:geomCompose}%
		Illustration of the composition of two generic geometries for one-way procedures.
		Notice that composition in the diagrams is left-to-right, while the composition in written notation is right-to-left.}
		\vspace{2ex}
\end{figure}

When composing one-way computations, we will often be interested in converting the composition into standard form, in which case no qubit can be allocated after measurement operations have occurred.
Therefore, when composing one-way procedures $\mathfrak P = \mathfrak P_2 \circ \mathfrak P_1$, we will be interested in restricting the qubits which are allocated in $\mathfrak P_2$ to be distinct from any qubits involved in $\mathfrak P_1$.
We can express this in terms of a sense of composing the \emph{geometries} of these patterns, using the constraints described in Lemma~\ref{lemma:typeCompose} (page~\pageref{lemma:typeCompose}):
\begin{definition}
	For two geometries $\cG_1 = (G_1, I_1, O_1)$ and $\cG_2 = (G_2, I_2, O_2)$, we say that $\cG_1$ and $\cG_2$ are \emph{composable} if $\big[V(G_1) \inter V(G_2)\big] \subset \big[I_2 \inter O_1\big]$\,.
	The \emph{composition} $\bar\cG = \cG_2 \circ \cG_1$ is then defined by $\bar\cG = (\bar G, \bar I, \bar O)$, where
	\begin{subequations}
	\begin{align}
			V(\bar G)
		\,\;=&\;\;\;
			V(G_1) \union V(G_2)	\,,
		\\
			E(\bar G)
		\,\;=&\;\;\;
			E(G_1) \symdiff E(G_2)	\,,
		\\
			\bar I
		\;\;=&\;\;\;
			I_1 \,\,\union (I_2 \setminus O_1) \,,
		\\
			\bar O
		\,\;=&\;\;\;
			O_2 \union (O_1 \setminus I_2) \;,
	\end{align}
	\end{subequations}
	where $S \symdiff T = [S \setminus T] \union [T \setminus S]$ is the symmetric difference of $S$ and $T$.
	Two one-way procedures $\mathfrak P_1$ and $\mathfrak P_2$ are said to be \emph{properly composable} if their underlying geometries  $\cG_1$ and $\cG_2$ are composable; the underlying geometry of the procedure $\mathfrak P_2 \circ \mathfrak P_1$ is then $\cG_2 \circ \cG_1$.
\end{definition}
Figure~\ref{fig:geomCompose} illustrates the composition of two geometries.
From this point onwards, we will suppose that two composable patterns $\mathfrak P_1$ and $\mathfrak P_2$ are properly composable: we lose no generality in doing this, as we may ``relabel'' the qubits acted on by the patterns $\mathfrak P_1$ and $\mathfrak P_2$ to achieve this if necessary.

\subsection{Universality of the one-way and graph-state models}

\begin{figure}[t]
	\begin{center}
		\includegraphics{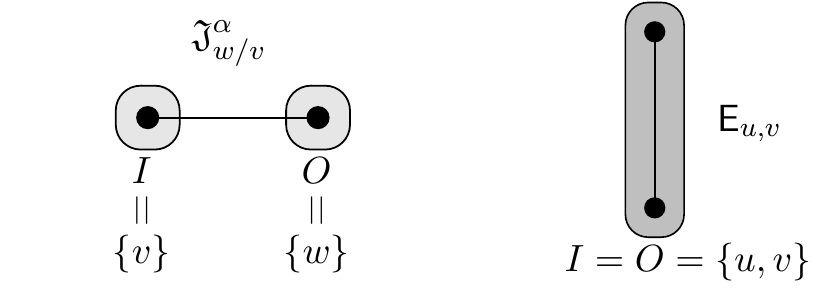}
		\vspace{-2ex}
	\end{center}
	\caption[Illustration of the geometries of two primitive one-way patterns.]{\label{fig:primitiveGeom}%
		Illustration of the geometries of two primitive one-way patterns.
		(Note that the input and output subsystems of the pattern on the right coincide.)}
		\vspace{2ex}
\end{figure}

We may now illustrate how to perform universal quantum computation in the one-way model, using the construction illustrated in~\cite{DKP07}, which constructs measurement patterns which ``simulate'' unitary circuits.
Consider the one-way patterns
\begin{align}
	\label{eqn:elemOpsDKP}
		\mathfrak J_{w/v}^\alpha
	\;\;=&\;\;
		\Xcorr{w}{\s[v]} \Meas{v}{-\alpha} \Ent{v,w} \New{w}{\sfx}	\;,
	&
		\Ent{u,v}	\;,
\end{align}
whose underlying geometries are illustrated in Figure~\ref{fig:primitiveGeom}.
The single entangler operation $\Ent{u,v}$ in particular is a simple one-way procedure in its own right, and can be interpreted as representing the same operation between two wires in a unitary circuit model such as $\Phases(\A)$.
To describe the $\mathfrak J_{w/v}^\alpha$ operation, note that the first two operations put the system into a stabilizer code which is stabilized by the two operator group $\sS \,=\, \ens{\idop_v \ox \idop_w\,,\; Z_v \ox X_w}$.
We can decompose $\Meas{v}{-\alpha}$ into a $\axisZ$-rotation and an $\axisX$ measurement:
\begin{align}
		\Meas{v}{-\alpha}(\rho)
	\;\;=&\;\;
		\bra{\psub{(\minus\alpha)}}_v\, \rho \,\ket{\psub{(\minus\alpha)}}_v	\;\ox\; \ket{0}\bra{0}_{\s[v]}
		\;\;+\;\;
		\bra{\msub{(\minus\alpha)}}_v\, \rho \,\ket{\msub{(\minus\alpha)}}_v	\;\ox\; \ket{1}\bra{1}_{\s[v]}
	\notag\\[1ex]=&\;\;
		\bra{\splus}_v \Big[ R_\sfz(\alpha)_v \;\rho\; R_\sfz(-\alpha)_v \Big] \ket{\splus}_v	\ox \ket{0}\bra{0}_{\s[v]} \;\;+\;\;
	\notag\\&\qquad\qquad
		\bra{\sminus}_v \Big[ R_\sfz(\alpha)_v \;\rho\; R_\sfz(-\alpha)_v \Big] \ket{\sminus}_v	\ox \ket{1}\bra{1}_{\s[v]}
	\notag\\[1ex]=&\;\;
		\Meas{v}{\sfx} \, \sfR_v^{\alpha} (\rho)	\;.
\end{align}
Because the operator $R_\sfz(\alpha)$ is diagonal, $\sfR^\alpha_v$ commutes with the operation $\Ent{v,w}$ as well as $\New{w}{\sfx}$\,.
Then, we have
\begin{align}
	\label{eqn:factorRotationJ}
		\mathfrak J_{w/v}^\alpha
	\;\;=\;\;
		\Xcorr{w}{\s[v]} \Meas{v}{-\alpha} \Ent{v,w} \New{w}{\sfx}
	\;\;=&\;\;
		\Xcorr{w}{\s[v]} \Meas{v}{\sfx} \sfR_v^{\alpha} \Ent{v,w} \New{w}{\sfx}
	\notag\\=&\;\;
		\Xcorr{w}{\s[v]} \Meas{v}{0} \Ent{v,w} \New{w}{\sfx} \sfR_v^{\alpha} 
	\;\;=\;\;
		\mathfrak J_{w/v}^{\,0} \, \sfR_v^{\alpha} \;.
\end{align}
Then, it suffices to consider the effect of $\mathfrak J_{w/v}^{\,0}$, which by linearity we may characterize by the effect on any four linearly independent $2 \x 2$ operators.
We will consider the effect of $\mathfrak J_{w/v}^{\,0}$ on the Pauli operators via the stabilizer formalism, as follows.
As we noted above, the maps $\Ent{v,w} \New{w}{\sfx}$ encodes a single-qubit state into the code stabilized by $\ens{\idop_v \ox \idop_w\,,\; Z_v \ox X_w}$\,.
An $X_v$ observable measurement then performs an isometry, and performing a $\Xcorr{w}{\s[v]}$ correction selects the post-measurement state result corresponding to $\s[v] = 0$.
Thus, we may reduce how $\mathfrak J_{w/v}^{\,0}$ transforms the Pauli operators on $v$ to how it transforms $X_v$ and $Z_v$ in particular.
We may then compute
\begin{subequations}
\label{eqn:stabilizerJ}
\begin{align}
%
%
			\gen{\, X_v \,}
		\;\;\mapstoname{\New{w}{\sfx}}\;\;
			\gen{\begin{array}{c@{\,\ox\,}c@{}l}
					X_v		&	\idop_w	&	\;,	\\
					\idop_v	&	X_w	
			\end{array}}
		\;\;\mapstoname{\Ent{v,w}}\;\;
		&
			\gen{\begin{array}{c@{\,\ox\,}c@{}l}
					X_v	&	Z_w	&	\;,	\\
					Z_v	&	X_w	
			\end{array}}
		\;\;\mapstoname{\Xcorr{w}{\s[v]} \Meas{v}{\sfx}}\;\;
			\gen{\, Z_w	\,}
	\;,
	\\[3ex]
%
%
%
			\gen{\, Z_v \,}
		\;\;\mapstoname{\New{w}{\sfx}}\;\;
			\gen{\begin{array}{c@{\,\ox\,}c@{}l}
					Z_v		&	\idop_w	&	\;,	\\
					\idop_v	&	X_w	
			\end{array}}
		\;\;\mapstoname{\Ent{v,w}}\;\;
		&
			\gen{\begin{array}{c@{\,\ox\,}c@{}l}
					Z_v	&	\idop_w	&	\;,	\\
					Z_v	&	X_w	
			\end{array}}
	\notag\\=\;\;&
			\gen{\begin{array}{c@{\,\ox\,}c@{}l}
					Z_v		&	\idop_w	&	\;,	\\
					\idop_v	&	X_w	
			\end{array}}
		\;\;\mapstoname{\Xcorr{w}{\s[v]} \Meas{v}{\sfx}}\;\;
			\gen{\, X_w	\,}
	\;.
\end{align}
\end{subequations}
In both cases, we have $\mathfrak J_{w/v}^{0}(\sigma_v) \,=\, \sfH_w (\sigma_w)$\,, which then holds for all $\sigma \in \L(\cH[v])$.
Therefore, we have
\begin{align}
	\label{eqn:semanticsJ}
		\mathfrak J_{w/v}^{\alpha}(\rho_v)
	\;\;=\;\;
		\sqparen{\sfH \circ \sfR^{\alpha} (\rho)}_w
	\;\;=\;\;
		\sqparen{\sfJ^{\alpha}(\rho)}_w	\;.
\end{align}
Thus, the operations performed by the patterns $\mathfrak J_{w/v}^{\alpha}$ and $\Ent{v,w}$ correspond to the unitary operations of the $\Jacz(\A)$ unitary circuit model defined in Definition~\ref{def:elemGateSets}.
Using stable index notation, we may then define an operator homomorphism $\Phi$ mapping unitary circuits in the $\Jacz(\A)$ model to (non-standard form) measurement procedures in the $\OneWay(\A)$ model as follows.
For a stable index tensor expression $C$, we may require without loss of generality that the terms of $C$ are in an order corresponding to the order in which the operations are performed in the circuit (\eg~as in an ordinary operator expression for $C$).
Identifying advanced/deprecated tensor indices in $C$ with allocated/discarded qubits in the one way pattern (using the correspondence described on page~\pageref{discn:typeCorrespondance}), we define
\begin{align}
	\label{eqn:circuitPatternHomphm}
		\Phi\Big(J(\alpha)\pseu[:w/v]\Big)
 	\;\;=&\;\;
		\mathfrak J_{w/v}^{\alpha}	\;;
	&
		\Phi\Big(\cZ\pseu[v,w]\Big)
	\;\;=&\;\;
		\Ent{v,w}\;:
\end{align}
the unitary transformation described by a unitary circuit $C$ in stable index notation is then also performed by the one-way procedure $\mathfrak P = \Phi(C)$\,.\footnote{%
	We have deliberately neglected the allocation of the result register $\s[v]$ in this account of how unitaries are performed, in part because we are primarily interested in how the quantum registers are transformed rather than the classical information produced as a result.
	We conventionally ignore classical bits in discussions of the ``transformations performed'' by measurement patterns: this can be formalized by adopting a convention of implicitly tracing out all result registers at the end of a quantum computation.}

\subsection{Standardization of one-way patterns, and the DKP constructions}
\label{sec:DKPstandardization}

The homomorphism $\Phi$ from circuits to one-way patterns is the main instrument in the simplest constructions of standardized measurement patterns, presented by Danos, Kashefi, and Panangaden in~\cite{DKP07}.
In this section, we describe these constructions, which we will refer to as the DKP constructions for one-way patterns.

For any one-way procedure --- and for the one-way procedures arising from the homomorphism $\Phi$ of~\eqref{eqn:circuitPatternHomphm} in particular --- we may obtain a standardized one-way procedure by accumulating correction operations from the beginning of the pattern and moving them to the end.
To do so, we must commute them past preparation operations (which we may easily do) and entangler operations (which induces further corrections), and absorbing them into the measurement operations to describe how measurements are adapted.
We describe how this is done in this section, essentially via the stabilizer formalism, representing the correction operations as a product of Pauli operators, together with boolean expressions representing their classical dependencies.

Starting from the beginning of the pattern, we scan towards the end of the pattern, maintaining a set of correction operations that we encounter.
We attempt to commute each correction operator past the other operators in the pattern, as follows:
\begin{enumerate}
\item
	As the preparation maps $\New{v}{\sfx}$ are always the first operations on any qubit $v \in I\comp$ (in particular because, as we have restricted ourselves to properly composable patterns, no qubit is discarded and subsequently re-allocated), any operations which precede a preparation map also commutes with it.
	Thus, we may trivially commute any correction operations past $\New{v}{\sfx}$ operations.

\item
	Similarly, we may commute correction operations past any entangling operator $\Ent{u,v}$.
	However, operations of the form $\Ent{u,w}$ do not commute with correction operations $\Xcorr{v}{\beta}$\,: in particular, we have
	\begin{align}
			\Ent{u,v} \Xcorr{v}{\beta}
		\;\;=\;\;
			\Zcorr{u}{\beta} \Xcorr{v}{\beta} \Ent{u,v}	\;.
	\end{align}
	Thus, in commuting an $\Xcorr{v}{\beta}$ operation past an entangler acting on $w$ and another qubit $u$, we induce an additional correction $\Zcorr{u}{\beta}$, increasing the complexity of the correction operation.

\item
	Finally, for a measurement on any qubit $v$ that we encounter, we absorb the corrections accumulated on $v$ into the bases of measurement, according to the equations
	\begin{subequations}
	\label{eqn:measAbsorbCorr}
	\begin{gather}
	\begin{alignat}{4}
			\Meas{w}{\alpha;\beta,\gamma} \Xcorr{w}{\beta'}
		\;\;=&\;\;\;
			\Meas{w}{\big[(\minus 1)^{\beta + \beta'} \cdot\, \alpha \,+\, \gamma\pi\big]}
		&\;\;=&\;\;\;
			\Meas{w}{\alpha;(\beta + \beta'),\gamma}	\;,
		\\
			\Meas{w}{\alpha;\beta,\gamma} \Zcorr{w}{\gamma'}
		\;\;=&\;\;\;
			\Meas{w}{\big[(\minus 1)^\beta \cdot\, \alpha \,+\, (\gamma + \gamma')\pi\big]}
		&\;\;=&\;\;\;
			\Meas{w}{\alpha;\beta,(\gamma + \gamma')} \;.
	\end{alignat}
	\end{gather}
	\end{subequations}
	This decreases the complexity of the corrections which we keep track of, transferring the information of the corrections on $v$ to the measurement on $v$.
\end{enumerate}
The result is a measurement procedure with a phase of classically controlled correction operations on the outputs, preceded by a mixed sequence of preparation maps, entanglers, and corrections with both sign- and $\pi$-dependencies.
Again, as the preparations are always the first operation performed on a qubit, and the measurements always the last, the operations in this mixed sequence all commute; we may then separate them into preparation, entangling, and measurement phases.

The complexity of the operations described above may be performed in time polynomial in the number of qubits $n$ operated on in the pattern.
For an arbitrary one-way pattern, at each point in time, we must track at most one $\Xcorr{v}{\beta_v}$ operation and one $\Zcorr{v}{\gamma_v}$ operation per qubit, with both $\beta$ and $\gamma$ being boolean expressions of complexity at most $O(n)$.
Each entangler operation acting on a qubit $v$ then induces a change in the expression $\Zcorr{v}{\gamma_v}$ of complexity $O(n)$\,, and there are $O(\deg_G(v))$ entanglers which may act on $v$, where $G$ is the graph of the geometry underlying the pattern.
Then, the complexity of updating the corrections on each qubit throughout the pattern is at most $O(n \deg_G(v))$\,; accumulated across all qubits, the work required is then $O\big(n \,\smash{\sum\limits_v \deg_G(v)}\big) = O(n m)$, where $m = \card{E(G)}$.

For one-way patterns arising out of the homomorphism $\Phi$, we may achieve a much better bound, arising from the fact that the corrections are not fully general: in particular, the correction on each qubit $w$ is initially just an operation $\Xcorr{w}{\s[v]}$ for some qubit $v$.
Following the same analysis as above, the work required to standardize such a pattern is then at most $O\big(\smash{\sum\limits_v \deg_G(v)}\big) = O(m)$.

In the case of the one-way patterns arising from $\Phi$, the resulting (non-standard form) procedure is one of the constructions for one-way patterns described in~\cite{DKP07}, which we will refer to in later chapters as follows:
\begin{definition}
	\label{def:simplifiedDKP}
	The \emph{simplified DKP construction for one-way patterns} is the procedure for producing $\OneWay(\A)$ patterns from unitary circuits $C$ in the $\Jacz(\A)$ model by
	\begin{enumerate}
	\item
		constructing the measurement pattern $\Phi(C)$, where $\Phi$ is the mapping defined in~\eqref{eqn:circuitPatternHomphm};\footnotemark

	\item
		commuting the preparations and entangler maps of the resulting concatenated procedure to the beginning of the procedure;

	\item
		absorbing the resulting classically controlled corrections $\Xcorr{v}{\beta}$ and $\Zcorr{v}{\gamma}$ to the end of the measurement procedure, absorbing them into the measurements on the qubits in $V(G) \setminus O$ when required by the relations of~\eqref{eqn:measAbsorbCorr}.
	\end{enumerate}
\end{definition}%
\footnotetext{\label{fn:DKPformalSemantics}%
The construction of~\cite{DKP07} does not make use of stable index tensor notation for circuits, but rather through an equivalent consideration of formal semantics of symbols in a measurement calculus corresponding to the CPTP maps in the gate set $\OneWay(\A)$ which we defined above.}
A standardized procedure can be easily obtained from the simplified DKP construction by eliminating $\pi$-dependencies from measurements by observing that $\ket{\pmsub{(\alpha + \pi)}} = \ket{\mpsub\alpha}$.
Then, $\pi$-dependencies may be abstracted away from measurements via shift operators, using the equation
\begin{gather}
	\label{eqn:piElim}
		\Meas{v}{\alpha;\beta,\gamma}
	\;\;=\;\;
		\Meas{v}{(\minus 1)^\beta \cdot\, \alpha \,+\, \gamma \pi}
	\;\;=\;\;
		\Shift{v}{\gamma} \Meas{v}{(\minus 1)^\beta \cdot\, \alpha}
	\;\;=\;\;
		\Shift{v}{\gamma} \Meas{v}{\alpha;\beta}	\;.
\end{gather}
That is, rather than potentially ``rotating the apparatus by 180 degrees'' to exchange the states $\ket{\pmsub\alpha}$ in a given measurement, we conditionally change the measurement result $\s[v]$ by adding (modulo 2) the expression $\gamma$.

There are two further techniques for simplifying a measurement procedure arising from the simplified DKP construction, which are also described in~\cite{DKP07}:

\paragraph{Pauli simplifications.}

It is easy to verify that $\ket{\pmsub0} \;=\; \ket{\spm \sfx}$, and $\ket{\pmsub{\pi\!/\:\!\!2}} = \ket{\spm \sfy}$\,; we may use this to further reduce measurement dependencies for measurements $\Meas{}{\alpha;\beta}$ for $\alpha$ a multiple of $\pion 2$, as follows.
For $\alpha$ any multiple of $\pi$, the measurement operators $\Meas{v}{\alpha}$ and $\Meas{v}{-\alpha}$ on any qubit $v$ are equivalent, as they differ by an integer multiple of $2\pi$\,: changes of the sign of the angle of measurement have no impact, and as a result such a measurement does not have any classical dependencies.
For $\alpha$ an odd multiple of $\pion2$, we have $-\alpha \equiv \alpha + \pi \pmod{2\pi}$, in which case we may also extract the classical dependency into a shift operator
\begin{align}
		\Meas{v}{(\minus 1)^\beta \cdot \,\pion 2}
	\;\;=\;\;
		\Meas{v}{\pion 2 \,+\, \beta \pi}
	\;\;=\;\;
		\Shift{v}{\beta} \Meas{v}{\pion 2}	\;.
\end{align}
These measurements (together with any \axisZ\ measurements $\Meas{v}{\sfz}$) are described as \emph{Pauli measurements}, and the procedure of eliminating dependencies of these measurements on prior measurement results as \emph{Pauli simplifications}.
By performing Pauli simplifications, the Pauli measurements in a one-way pattern may be performed independently of any other measurement result, and in particular they may be commuted to the beginning of the measurement phase of any one-way pattern.

\paragraph{Signal shifting.}

After bringing a one-way pattern from the simplified DKP construction into standard form and performing Pauli simplifications, we may eliminate any shift operators by commuting them to the end of a measurement procedure and discarding them.
For any one-way model operation $\Op{w}{\exprbr{\s[v]}}$ on a qubit $w$ which depends on a measurement result $\s[v]$, we have the relation
\begin{align}
			\Op{u}{\exprbr{\s[w]}} \Shift{w}{\beta}
	\;\;=&\;\;
			\Shift{w}{B} \Op{u}{\exprbr{ \s[w] + \beta }}
\end{align}
where $\exprbr{ \s[v] + \beta }$ is the result of substituting $\s[v]$ in the boolean expression with the value of $\s[v] + \beta$\,.
We may use such relations to commute all shift operators to the end of a one-way procedure.
Because shift operators are purely classical operations, shift operations at the end of a measurement procedure have no effect on the quantum output, and so they may be eliminated.
We call this procedure of eliminating shift operations \emph{signal shifting}.
(If this is performed simultaneously with the standardization procedure described above, we may accumulate shift operators along with the correction operations, and ultimately commute the shift operators past the corrections: an identical analysis for the run-time of this procedure holds as for standardization without signal shifting.)

The measurement pattern that arises from performing these transformations is then a standard form one-way pattern, where the measurement phase begins with a collection of commuting Pauli measurements; and where the operational dependencies of every measurement are represented explicitly in the classical controls of each operation, as a result of eliminating the shift operations.
We will refer to this construction as follows:
\begin{definition}
	\label{def:completeDKP}
	The (\emph{complete}) \emph{DKP construction for one-way patterns} is the procedure for producing $\OneWay(\A)$ patterns from unitary circuits in the $\Jacz(\A)$ model by
	\begin{enumerate}
	\item
		obtaining a measurement pattern from $C$, via the simplified DKP construction of Definition~\ref{def:simplifiedDKP}\,;

	\item
		eliminating $\pi$-dependencies, via the relation~\eqref{eqn:piElim}\,; 

	\item
		performing Pauli simplifications and signal shifting; and

	\item
		commuting Pauli measurements to the beginning of the measurement phase.
	\end{enumerate}
\end{definition}
The analysis of these constructions prove the following result, by the discussion following Definition~\ref{def:elemGateSets} on page~\pageref{def:elemGateSets}:
\begin{theorem}
	The $\OneWay(\A)$ model is
	\begin{inlinum}
	\item
		universal for quantum computation and for unitary transformations when $\A = \R$\,; and

	\item
		approximately universal for quantum computation and for unitary transformations when $\A = \frac{\pi}{4}\Z$\,.
	\end{inlinum}
	Furthermore, we may without loss of generality require that the measurement pattern is standardized, contains no shift operations, and that Pauli measurements are performed before any non-Pauli measurement operation.
\end{theorem}

\subsection{The simplified RBB construction}
\label{sec:RBB}

We may apply the techniques of the preceding sections to show the universality of the cluster-state model, by adapting the construction to the constrained case where the graph is an $n \x d$ grid.
We will do this by showing how the techniques of the preceding sections may be used to describe (a modest extension of) the elementary constructions of Raussendorf, Browne, and Briegel~\cite{RBB03} for quantum computation in the cluster state model.\footnote{%
	The constructions of~\cite[page~5]{RBB03} assume arbitrary precision of measurement angles, allowing the exact implementation of arbitrary single-qubit using a fixed ``horizontal width'' when embedded in the grid.
	For an account of approximate universality, a trivial but useful extension is to describe computations in terms of ``chains'' of $J(\alpha)$ operations on distinct qubits of arbitrary length, and then describe how to compensate for differences in horizontal positions in the grid.
}
We will refer to the resulting construction as ``the simplified RBB construction''.

The main difference between cluster-state based computation and what we have described in Definition~\ref{def:oneWayModel} as general one-way patterns are the topological constraints on the interactions between qubits imposed by the grid graph.
These parallel a commonly imposed constraint in unitary circuit models: a \emph{linear nearest neighbor} unitary circuit model is one where we a impose a linear ordering on the qubits by assigning them to points on a line (possibly corresponding \eg~to physical locations of the qubits as physical systems), and operations are restricted to operate only on adjacent qubits (or more generally, on blocks consecutive qubits, for operations acting on two or more qubits).
Linear nearest neighbor models can simulate more general models by performing swap operations on adjacent qubits, effecting permutations of the qubits in the linear ordering, in order to allow arbitrary collections of qubits to interact.
It is easy to show that the operation
\begin{gather}
	\label{eqn:simpleSwap}
		\mathsf{SWAP}_{u,v}
	\;\;=\;\;
		\sfH_u \sfH_v \Ent{u,v} \sfH_u \sfH_v \Ent{u,v} \sfH_u \sfH_v \Ent{u,v}
\end{gather}
interchanges independent pure states of two neighboring qubits $u$ and $v$: thus, for universality, it suffices to map linear nearest-neighbor unitary circuits into the cluster-state model.

\begin{figure}[t]
	\begin{center}
		\includegraphics{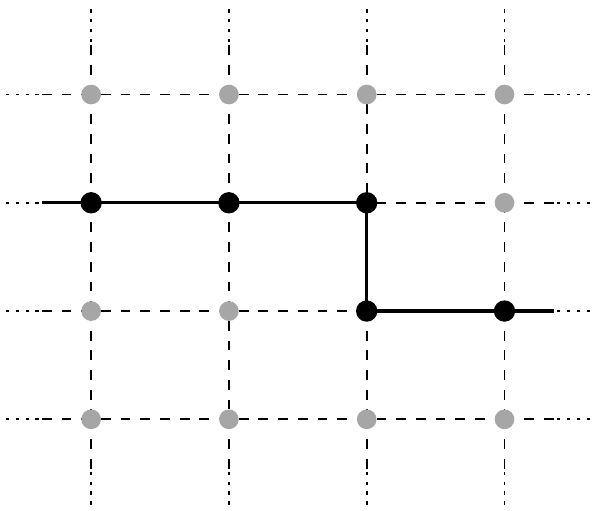}
	\end{center}
	\caption[Illustration of an embedding of a chain in the grid.]{\label{fig:isolatedChain}%
		Illustration of an embedding of a chain in the grid.
		Grey dots and broken lines represent qubits and entanglement relations which are effectively removed from the cluster state by $\axisZ$ measurements, to produce a graph state whose underlying graph contains a sequence of vertices with degree $2$.
	}
\end{figure}

To represent single-qubit gates, we may compose patterns of the form $\mathfrak J_{w/v}^\alpha$ as before.
However, in order to represent a chain of patterns
\begin{gather}
	\mathfrak J_{v_N/v_{N \minus 1}}^{\alpha_{N \minus 1}} \cdots\;\, \mathfrak J_{v_2/v_1}^{\alpha_1} \, \mathfrak J_{v_1/v_0}^{\alpha_0}	\;,
\end{gather}
which we may refer to as a \emph{chain pattern}, we require that the qubits $v_j$ for $0 < j < N$ all have degree $2$ in the corresponding entanglement graph (as any neighbors aside from $v_{j-1}$ and $v_{j+1}$ represents an operation not in the pattern described above).
To embed the corresponding graph state in the grid without additional neighborhood relationships, we must remove every neighbor of these qubits $v_j$ except for those specified by the pattern, which we do using $\axisZ$ measurements.
An illustration of such an embedding is illustrated in Figure~\ref{fig:isolatedChain}.
Typically, such chain patterns would be embedded as a horizontal path through the grid, but as Figure~\ref{fig:isolatedChain} also shows, we may also employ more general paths in the grid.

Independent single-qubit unitaries acting on independent qubits may then be embedded in the grid as a collection of vertex-disjoint paths, where in particular the distance between two vertices of distinct paths is at least $2$ (in order to maintain a buffer of at least one grid site between two chain patterns).
The paths so traced out in the grid then correspond to \eg~the wires of a unitary circuit diagram.
In the construction, we may then represent the logical qubits of the quantum circuit by parallel horizontal paths in the grid whose vertical separation is precisely $2$.
Note that the paths which in general may be of different lengths, as each horizontal step corresponds to a single-qubit $J(\alpha)$ transformation for some given $\alpha$.
However, we may extend any path in the grid by a path of length $2$ or more in a way which represents the identity operation, using chain patterns of the form
\begin{align}
	\label{eqn:identityPatterns}
		\mathfrak {Id}^2_{v_2/v_0}
	\;\;=&\;\;
		\mathfrak J_{v_2/v_1}^{\,0} \, \mathfrak J_{v_1/v_0}^{\, 0}		\;,
	&
		\mathfrak {Id}^3_{v_3/v_0}
	\;\;=&\;\;
		\mathfrak J_{v_3/v_2}^{\pion 2} \, \mathfrak J_{v_2/v_1}^{\pion 2} \, \mathfrak J_{v_1/v_0}^{\pion 2}	\;;
\end{align}
by~\eqref{eqn:circuitPatternHomphm}, described in stable index notation, these operations are equivalent to unitary circuits $H \pseu[:v_2/v_1] H \pseu[:v_1/v_0]$ and $J(\pion 2) \pseu[:v_3/v_2] J(\pion 2) \pseu[:v_2/v_1] J(\pion 2) \pseu[:v_1/v_0]\Big.$ respectively.
The former obviously implements the single-qubit identity operation $\idop \pseu[:v_2/v_0]$\,, as $H$ is self-inverse; for the latter, recalling that $J(\pion 2) = H R_\sfz(\pion 2)\Big.$, we may characterize its effect by conjugation on linear operators via the stabilizer formalism:
\begin{subequations}
\begin{align}
		X
	&\;\mapstoname{\; H R_\sfz(\pion 2) \;}\;
		-Y
	\;\mapstoname{\; H R_\sfz(\pion 2) \;}\;
		Z
	\;\mapstoname{\; H R_\sfz(\pion 2) \;}\;
		X	\,,
	\\[1ex]
		Z
	&\;\mapstoname{\; H R_\sfz(\pion 2) \;}\;
		X
	\;\mapstoname{\; H R_\sfz(\pion 2) \;}\;
		-Y
	\;\mapstoname{\; H R_\sfz(\pion 2) \;}\;
		Z	\,;
\end{align}
\end{subequations}
then $J(\pion 2) J(\pion 2) J(\pion 2)$ performs $\idop_2$ as well.
Thus, both $\mathfrak {Id}^2_{v_2/v_0}$ and $\mathfrak {Id}^3_{v_3/v_0}$ map states $\rho_{v_0}$ to $\rho_{v_j}$ for some $j \ge 2$\,; the first using a path of length $2$ in the grid, and the second using a path of length $3$.
Using combinations of these, we can then implement a path of any length $\ell \ge 2$ in the grid, representing a sequence of operations which performs the identity $\idop_2$ on a given qubit.
Thus, whenever required, we may suppose that the paths in the grid corresponding to any two qubits are of the same length, regardless of the number of (non-trivial) unitary transformations performed on them.

\begin{figure}[t]
	\begin{center}
		\includegraphics[width=146mm]{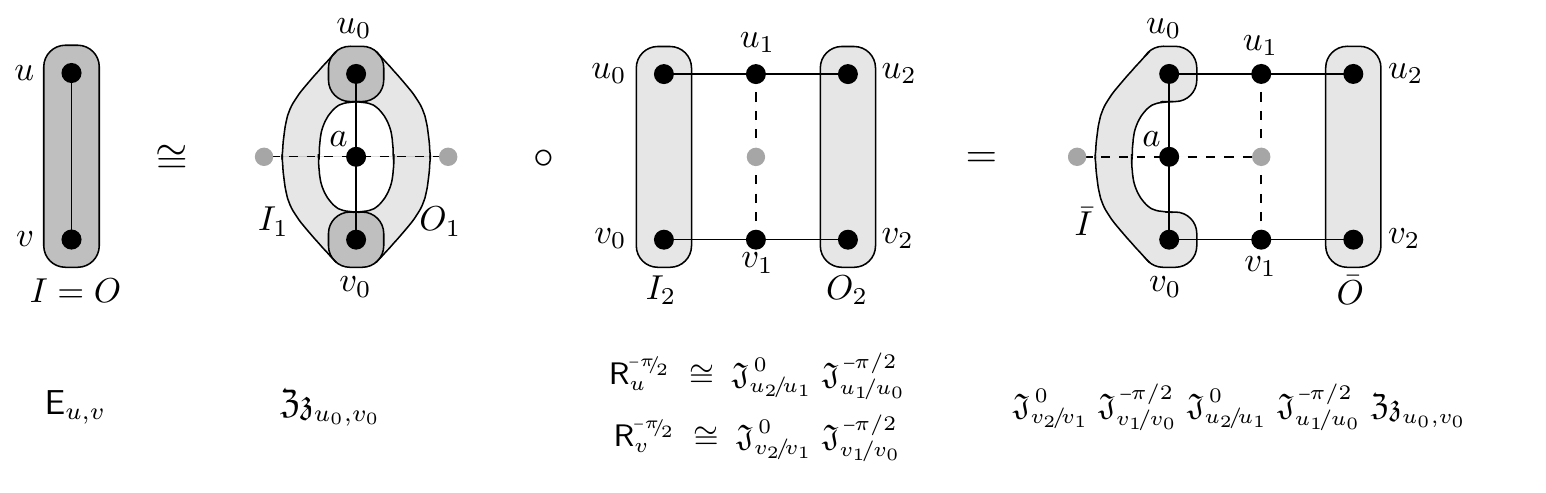}
	\end{center}
	\caption[Illustration of an embedding of \protect{$\protect\ctrl Z_{u,v}$} in the grid in terms of geometries, using the decomposition of~\eqref{eqn:clusterEntanglerSubst}.]{\label{fig:zzConstr}%
		Illustration of an embedding of $\protect\cZ_{u,v}$ in the grid in terms of geometries, using the decomposition of~\eqref{eqn:clusterEntanglerSubst}.
		Congruency relations represent equivalence up to changes of the input/output system.
		Grey dots and broken lines indicate qubits and entanglement relations of the grid state which must be removed via $\axisZ$ measurements in order not to interfere with the corresponding measurement pattern (displayed below each geometry).}
\end{figure}

In order to achieve universality for quantum computation, it then suffices to implement a logical $\cZ$ operation (or $\cX$ operation) in the cluster-state, between two qubits which are at the same horizontal position, and are vertically separated in the grid by a distance of $2$ (equivalently, by one grid site).
One approach to doing this may be described using a technique described in~\cite[Section~3.8]{BB06}.
Consider the pattern
\begin{gather}
	\label{eqn:simpleZzPattern}
		\mathfrak {Zz}_{u,v}
	\;\;=\;\;
		\Zcorr{v}{\s[a]}
		\Zcorr{u}{\s[a]}
		\Meas{a}{\pion 2}
		\Ent{a,v}
		\Ent{a,u}
		\New{a}{\sfx}
\end{gather}
with input and output subsystems $I = O = \ens{u,v}$\,.
The preparation and entangling phases encode the input state into the code stabilized by $\ens{ \idop_u \ox \idop_a \ox \idop_v \,, Z_u \ox X_a \ox Z_v }$\,: the operation $\Meas{a}{\pion 2}$ then represents a $\axisY$ measurement which performs an isometry on this code transformation, with the subsequent corrections effectively selecting for the result $\s[a] = 0$.
Using the stabilizer formalism, we may then determine
\begin{subequations}
\begin{align}
		\gen{ X_u \ox \idop_v }
	\;\mapstoname{\; \Ent{a,v} \Ent{a,u} \New{a}{\sfx} \;}&\;
		\gen{\begin{array}{c@{\,\ox\,}c@{\,\ox\,}c@{}l}
				X_u	&	Z_a	&	\idop_w	&	\;,	\\
				Z_v	&	X_w	&	Z_w		&
		\end{array}}
	\notag\\[-0.2ex]=&\;
		\gen{\begin{array}{c@{\,\ox\,}c@{\,\ox\,}c@{}l}
				Y_u	&	Y_a	&	Z_w		&	\;,	\\
				Z_v	&	X_w	&	Z_w		&
		\end{array}}
	\;\mapstoname{\; \Zcorr{v}{\s[a]}\Zcorr{u}{\s[a]} \Meas{a}{\pi/2} \;}\;
		\gen{ Y_u \ox Z_w }	\;,		
 	\\[2ex]
		\gen{ \idop_u \ox X_v }
	\;\mapstoname{\; \Ent{a,v} \Ent{a,u} \New{a}{\sfx} \;}&\;
		\gen{\begin{array}{c@{\,\ox\,}c@{\,\ox\,}c@{}l}
				\idop_u	&	Z_a	&	X_w		&	\;,	\\
				Z_v		&	X_w	&	Z_w		&
		\end{array}}
	\notag\\[-0.2ex]=&\;
		\gen{\begin{array}{c@{\,\ox\,}c@{\,\ox\,}c@{}l}
				Z_u	&	Y_a	&	Y_w		&	\;,	\\
				Z_v	&	X_w	&	Z_w		&
		\end{array}}
	\;\mapstoname{\; \Zcorr{v}{\s[a]}\Zcorr{u}{\s[a]} \Meas{a}{\pi/2} \;}\;
		\gen{ Z_u \ox Y_w }	\;,		
%
	\intertext{}
		\gen{ Z_u \ox \idop_v }
	\;\mapstoname{\; \Ent{a,v} \Ent{a,u} \New{a}{\sfx} \;}&\;
		\gen{\begin{array}{c@{\,\ox\,}c@{\,\ox\,}c@{}l}
				Z_u	&	\idop_a	&	\idop_w	&	\;,	\\
				Z_v	&	X_w		&	Z_w		&
		\end{array}}
	\;\mapstoname{\; \Zcorr{v}{\s[a]}\Zcorr{u}{\s[a]} \Meas{a}{\pi/2} \;}\;
		\gen{ Z_u \ox \idop_w }	\;,		
	\\[2ex]
		\gen{ \idop_u \ox Z_v }
	\;\mapstoname{\; \Ent{a,v} \Ent{a,u} \New{a}{\sfx} \;}&\;
		\gen{\begin{array}{c@{\,\ox\,}c@{\,\ox\,}c@{}l}
				\idop_u	&	\idop_a	&	Z_w	&	\;,	\\
				Z_v		&	X_w		&	Z_w	&
		\end{array}}
	\;\mapstoname{\; \Zcorr{v}{\s[a]}\Zcorr{u}{\s[a]} \Meas{a}{\pi/2} \;}\;
		\gen{ \idop_u \ox Z_w }	\;;		
\end{align}
\end{subequations}
in particular, it is a symmetric operation on $u$ and $v$ which leaves $Z_u$ and $Z_v$ invariant, so it applies a two qubit diagonal operation.
It is then easy to verify that the CPTP map which $\mathfrak {Zz}$ performs is 
\begin{gather}
	\label{eqn:zz}
		\mathfrak {Zz}_{u,v}(\rho)
	\;\;=\;\;
		\sfR^{\pion 2}_u \sfR^{\pion 2}_v \Ent{u,v} (\rho)
	\;\;=\;\;
		\e^{-i\pi Z_u \ox Z_v / 4} \,\rho\,\, \e^{i\pi Z_u \ox Z_v / 4}	\;;
\end{gather}
equivalently, we may substitute any entangler operation $\Ent{u,v}$ in the DKP construction with the composite CPTP map
\begin{gather}
	\label{eqn:clusterEntanglerSubst}
		\Ent{u,v}
	\;\;=\;\;
		\sfR^{\mpion 2}_v \; \sfR^{\mpion 2}_u \; \mathfrak {Zz}_{u,v}	\;.
\end{gather}
To obtain an equivalent one-way pattern, we may decompose $\sfR^{\mpion 2} = \sfJ(0)\sfJ(\mpion 2)$, and then substitute the rotations in~\eqref{eqn:clusterEntanglerSubst} with chain patterns $\mathfrak J_{u''/u'}^{\,0} \, \mathfrak J_{u'/u}^{\mpion 2}$ (and similarly for $v$); this construction is illustrated in Figure~\ref{fig:zzConstr}.
Alternatively, if the logical entangler operation $\Ent{u,v}$ precedes $\mathfrak J^{\alpha_u}_{u'/u}$ and $\mathfrak J^{\alpha_v}_{v'/v}$  operations in a measurement pattern, we may employ the relations
\begin{align}
		\mathfrak J^{\alpha_v}_{v'/v} \; \mathfrak J^{\alpha_u}_{u'/u} \; \Ent{u,v}
	\;\;=&\;\;
		\mathfrak J^{\,0}_{v'/v} \; \sfR_v^{\alpha_v} \; \sfR_v^{-\pion 2} \; \mathfrak J^{\,0}_{u'/u} \; \sfR_u^{\alpha_u} \sfR_u^{-\pion 2} \; \mathfrak{Zz}_{u,v}
	\notag\\=&\;\;
		\mathfrak J^{\alpha_v - \pi/2}_{v'/v} \; \mathfrak J^{\alpha_u - \pi/2}_{u'/u} \; \mathfrak{Zz}_{u,v}	\;,
\end{align}
which follow from~\eqref{eqn:factorRotationJ}.

\paragraph{A remark on terminology.}

We have taken some license in referring to the above construction as ``the simplified RBB construction'', in that some of the constructions above (\eg~the construction of the pattern $\mathfrak {Zz}$) do not appear explicitly in~\cite{RBB03}.
However, all of these constructions are clearly implicit in that work.
The construction above also completely omits the cleverer constructions of~\cite{RBB03}, including compact patterns for reversing an entire consecutive block of qubits, non-nearest neighbor $\cX$ operations, the quantum Fourier transform over $\Z_{2^k}$ for arbitrary $k$, and addition circuits for $\Z_{2^k}$.

\subsection{Universality of the cluster state model}

We may use the constructions above, and the pattern transformations described for the DKP constructions in Section~\ref{sec:DKPstandardization}, to prove the (approximate) universality of the cluster-state model pattern in standard form, as follows.

By composing the elementary patterns of the simplified RBB construction above, we may construct a one-way pattern in grid graphs $G_{n,d}$ for sufficiently large $n,d \in \N$ which implement arbitrary unitary circuits in a nearest-neighbor version of the $\Jacz(\A)$ model.
Restricting to one-way patterns which initialize each input qubit in the $\ket{+}$ state and performing the appropriate trace-out operations, we may then prepare arbitrary density operators by Theorem~\ref{thm:churchHilbertSpace}.
This yields (approximate) universality for the model $\Cluster(\A)$, for any $\A \subset \R$ for which $\Jacz(\A)$ is (approximately) universal.

Using the standardization procedures of Section~\ref{sec:DKPstandardization}, we may obtain a one-way pattern in which the first stage is to apply an open graph state encoding on the input subsystem: the preparation and entanglement phases of the resulting pattern can then be replaced with the preparation of a sufficiently large cluster state, and removing the qubits which do not play a role in the pattern via $\axisZ$ measurements.
Thus, we may prepare arbitrary density operators in the cluster-state model; and by the approximate universality of $\Jacz(\frac{\pi}{4}\Z)$, the cluster-state model is approximately universal for quantum computing for measurement angles which are multiples of $\pion 4$.
Furthermore, by performing signal shifting and Pauli simplifications then yield a measurement pattern involving no $\pi$-dependencies or shift operations, thus yielding a pattern in the cluster-state model in standard form.

\vspace{1ex}
\noindent The above then reproduces the following result of~\cite{RB01, RBB03}:
\vspace{-0.5ex}
\begin{theorem}
	The $\Cluster(\A)$ model is
	\begin{inlinum}
	\item
		universal for quantum computation when $\A = \R$\,; and

	\item
		approximately universal for quantum computation when $\A = \frac{\pi}{4}\Z$\,.
	\end{inlinum}
	Furthermore, we may without loss of generality require that the measurement pattern is standardized, and that Pauli measurements are performed before any non-Pauli measurement operation.
\end{theorem}

\section{Other constructions in the one-way model}
\label{sec:otherConstrs}

Before closing this chapter, we remark upon two other constructions in the one-way measurement model which we will refer to in later chapters, which lie beyond the DKP and the simplified RBB construction schemes described above.

\subsection{Qubit reversal pattern in the grid.}
\label{sec:qubitReversal}

\begin{figure}
		\begin{center}
			\includegraphics[width=58mm]{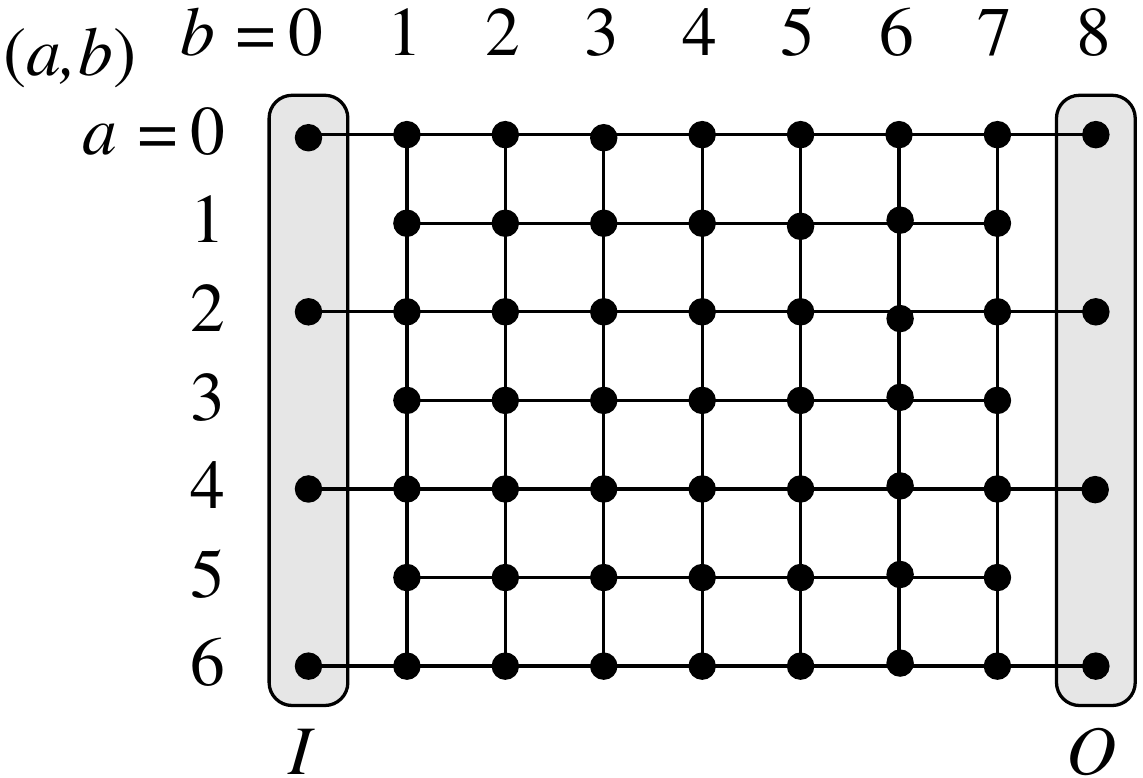}
		\end{center}
		\caption[Geometry for the measurement pattern of~\protect{\cite[Fig.~10]{RBB03}} for reversing a sequence of qubits in the cluster-state model.]{\label{fig:swapRBB}%
		Geometry for the measurement pattern of~\protect{\cite[Fig.~10]{RBB03}} for reversing a sequence of qubits in the cluster-state model.
		Every qubit (except for the elements of $O$) is measured with an $\axisX$ measurement, selecting for the state $\ket{\splus}$: the corresponding correction operations may be determined by the stabilizer formalism.
		Numbering the rows and columns of grid-sites from $0$ at the upper-left, the resulting transformation maps states $\rho_{(0,0),(0,2),(0,4),(0,6)}$ to $\rho_{(8,6),(8,4),(8,2),(8,0)}$\,, \ie\ exchanging the logical qubits of rows $0$ and $6$ as well as rows $2$ and $4$. 
	}
	\end{figure}

	Figure~\ref{fig:swapRBB} illustrates a geometry for a one-way pattern for reversing a sequence of consecutive logical qubits in a linear nearest neighbor model, using only $\axisX$ measurements, illustrated in~\cite[Fig.~10]{RBB03}.
	(This pattern is a special case of a more general transformation, which performs a unitary $\e^{i \varphi Z \ox \cdots \ox Z}$ transformation in addition to the qubit reversal by measuring a particular qubit in a different basis.)
	Numbering grid sites $(a,b)$ according to the rows and columns of the grid (starting from $0$ at the upper-left), it is possible to show that the resulting transformation is unitary, and in particular, it reverses order of the the ``logical'' qubits of rows $\ens{0,2,4,6}$.

	This may be shown using \eg~the stabilizer formalism: to do so, however, it is useful to note the presence of correlations in the measurement results of the qubits in the interior block of columns 1\thru7.
	Consider the twelve-qubit set $S$ given by
	\begin{gather}
			S
		\;\;=\;\;
			\ens{ (a,b) ~\text{adjacent to}~ (d,d) \,\big|\, d \in \ens{2,3,4,5,6}\, }	\;:
	\end{gather}
	it is easy to verify that the product of $K_{(a,b)} = X_{(a,b)} Z_{(a\sminus1,b)} Z_{(a\splus1,b)} Z_{(a,b\sminus1)} Z_{(a,b\splus1)}$ for $(a,b) \in S$ is a tensor product of $X_{(a,b)}$ operators, also for $(a,b)$ ranging over $S$.
	If we measure the qubits in this set column-by-column with $X$ observable measurements, this implies that the unique qubit in $S$ in the seventh column will yield the $\pm 1$ result with certainty if the product of the preceding eleven measurements is $\pm 1$.
	\label{discn:qubitReversalCertainty}%
	We may similarly show that every qubit in the seventh column (except for $(0,7)$ and $(6,7)$) will yield measurement results correlated in a similar way with a subset of qubits in the preceding six columns; the measurements on those qubits then perform the identity, while the other measurements will anticommute with some generators of the stabilizer group $\gen{K_v}_{v \notin O}$.

	Note that these correlations arise from cancelling out $Z$ operators in the interior block; by a similar technique of cancelling such $Z$ operators, it is possible to show that $X$ and $Z$ observables on each of the input qubits get mapped by measurements (and selecting for the $\ket{\splus}$ measurement result) to the same observable on the ``reversed'' output qubit.
	By scaling this measurement pattern in the obvious way, we may obtain a qubit-reversal pattern for an arbitrary number of qubits.

	This measurement pattern is noteworthy in that it falls outside of the domain of automated approaches to inferring the semantics of a measurement-based pattern (one of which we describe in Chapter~\ref{Semantics for unitary one-way patterns}), precisely because of the correlations which produce measurement results with certainty.
	In the special case of swapping two consecutive qubits, it is also more compact than the construction of $\mathsf{SWAP}_{u,v}$ described in~\eqref{eqn:simpleSwap} using the construction for $\Ent{u,v}$ in the simplified RBB construction above.

\subsection{Concise patterns for $Z \ox \cdots \ox Z$ Hamiltonians.}
\label{sec:zzMany}

\begin{figure}[t]
		\begin{center}
			\includegraphics{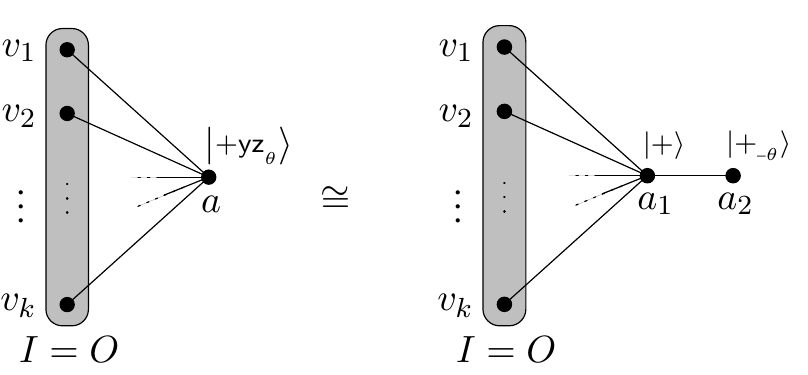}
		\end{center}
		\caption[Illustration of the geometry for the pattern $\protect\mathfrak {Zz\cdots z}^{\theta}$ applied to $k$ qubits.]{\label{fig:zzMany}%
			Illustration of the geometry for the pattern $\mathfrak {Zz\cdots z}^{\theta}$ applied to $k$ qubits, $v_1, \ldots, v_k$\,.
			In the left-most figure, the auxiliary qubit is to be measured in the basis \protect{$\ket{\spm \sfy\sfz_{\theta}}$}, with corrections to be performed to effectively select the positive result.
			The right-most figure illustrates an equivalent pattern using only $\XY$-plane measurements, in which the two auxiliary qubits are to be measured, performing corrections if necessary to select for the measurement results given.
			}	
	\end{figure}

	The construction of the $\mathfrak{Zz}$ pattern in~\eqref{eqn:simpleZzPattern} is a special case of a more general construction in~\cite{BB06} in the more general one-way model for applying unitaries arising from $Z^{\ox k}$ Hamiltonians for $k$ arbitrarily large.
	We may define the states $\ket{\spm \sfy\sfz_{\theta}}$ as follows:
	\begin{gather}
		\label{eqn:defYZtheta}
			\ket{\spm \sfy \sfz_\theta}
		\;\;=\;\;
			\sfrac{1}{\sqrt 2} \ket{\splus \sfx} \;\pm\; \sfrac{\e^{-i\theta}}{\sqrt 2} \ket{\sminus \sfx}	\;;			
	\end{gather}
	it is possible to verify that these are $\pm 1$ eigenvectors of the observable $\cos(\theta) Z + \sin(\theta) Y$.
	Define the measurement operator
	\begin{gather}
			\mspace{-20mu}
			\Meas[\YZ]{a}{\theta}(\rho)
		\;\;=\;\;
			\bra{\splus\sfy\sfz_\theta}_a \rho\, \ket{\splus\sfy\sfz_\theta}_{\!a} \ox \ket{0}\bra{0}_{\s[a]}
			\;+\;\;
			\bra{\sminus\sfy\sfz_\theta}_a \rho\, \ket{\sminus\sfy\sfz_\theta}_{\!a} \ox \ket{1}\bra{1}_{\s[a]}
	\end{gather}
	which performs a $\YZ$-plane measurement similar to the $\XY$-plane measurements defined by~\eqref{eqn:xyPlaneMeas}.
	Then, the measurement pattern
	\begin{subequations}
	\begin{gather}
			\label{eqn:zzMany}
			\smash{\underbrace{\mathfrak{Zz\cdots z}}_{\text{$k$ times}}}_{v_1,\cdots,v_k}^{\,\theta}
		\;\;=\;\;
			\sqparen{\prod_{j = 1}^k \Zcorr{u_j}{\s[a]}}
			\Meas[\YZ]{a}{\theta}
			\sqparen{\prod_{j = 1}^k \Ent{a, u_j}}
			\New{a}{\sfx}
	\end{gather}
	performs a unitary rotation
	\begin{gather}
			\smash{\underbrace{\mathfrak{Zz\cdots z}}_{\text{$k$ times}}}_{v_1,\cdots,v_k}^{\,\theta}(\rho)
		\;\;=\;\;
			\e^{\minus i \theta Z_{v_1} \ox \cdots \ox Z_{v_k}/2} \,\rho \,\;\e^{i \theta Z_{v_1} \ox \cdots \ox Z_{v_k}/2}	\;.
		\\[-1ex]\notag
	\end{gather}
	\end{subequations}%
	\label{discn:reduceZz}
	The operation $\mathfrak{Zz}_{u,v}$ is a special case for $k = 2$ and $\theta = \pion 2$; the geometry for the more general measurement pattern is illustrated in Figure~\ref{fig:zzMany}.

	It is easy to verify the behavior of $\mathfrak{Zz\cdots z}$ on standard basis state vectors, as follows.
	Using the fact that $\cZ \big( \ket{\splus} \ox \ket{0} \big) \,=\, \ket{\splus} \ox \ket{0}$ and $\cZ \big( \ket{\splus} \ox \ket{1} \big) \,=\, \ket{\sminus} \ox \ket{1}$\,, the state vector produced by the preparation and entanglement procedure of $\mathfrak{Zz\cdots z}^{\theta}_{v_1,\ldots,v_k}$ for an input state $\ket{\vec x}_{v_1,\ldots,v_k}$ (for $\vec x \in \ens{0,1}^k$ arbitrary) is
	\begin{align}
			\ket{\psi_{\vec x}}
		\;\;=\;\;
			\sqparen{\prod_{j = 1}^k \cZ_{a, u_j}}
			\mspace{-40mu}&\mspace{40mu}
			\ket{\splus}_a \ox \ket{\vec x}_{v_1,\ldots,v_k}
		\notag\\[1ex]=&\;\;
			\begin{cases}
				\ket{\splus}_a \ox \ket{\vec x}_{v_1,\ldots,v_k}	\,,	&	\text{if $x_1 \oplus \cdots \oplus x_k \equiv 0 \pmod{2}$}	\\[1ex]
				\ket{\sminus}_a \ox \ket{\vec x}_{v_1,\ldots,v_k}	\,,	&	\text{if $x_1 \oplus \cdots \oplus x_k \equiv 1 \pmod{2}$}
			\end{cases}
		\notag\\[2ex]=&\;\;
			\ket{\spm}_a \ox \ket{\vec x}_{v_1,\ldots,v_k}	\quad\text{for $\ket{\vec x}$ a $\pm 1$-eigenvector of $Z^{\ox k}$}.
	\end{align}
	Note that $\bracket{\splus \sfy\sfz_{2\theta}}{\splus} = \frac{1}{\sqrt 2}$ and $\bracket{\splus \sfy\sfz_{\theta}}{\sminus} = \frac{\e^{i\theta}}{\sqrt 2}$\,; then, if we apply the projection $\ket{\splus \sfy\sfz_{\theta}}\bra{\splus \sfy\sfz_{\theta}}$ to $\ket{\psi}$ as described above and renormalize, we obtain
	\begin{align}
			\ket{\splus \sfy\sfz_{2\theta}}\bra{\splus \sfy\sfz_{\theta}}_a \ket{\psi}_{a,v_1,\ldots,v_k}
		\;=&\;\;
			\begin{cases}
				\frac{1}{\sqrt 2} \ket{\vec x}_{v_1,\ldots,v_k}	\,,					&	\text{if $\ket{\vec x}$ a $+1$-eigenvector of $Z^{\ox k}$}	\\[1.5ex]
				\frac{\e^{i\theta}}{\sqrt 2} \ket{\vec x}_{v_1,\ldots,v_k}	\,,	&	\text{if $\ket{\vec x}$ a $-1$-eigenvector of $Z^{\ox k}$}
			\end{cases}
		\notag\\[2ex]\propto&\;\;
			\bigg\{ \e^{\mp i\theta/2} \ket{\vec x}_{v_1,\ldots,v_k}	\,,
				\quad\text{for $\ket{\vec x}$ a $\pm 1$-eigenvector of $Z^{\ox k}$} \bigg\}
		\notag\\[2ex]=&\;\;\;
			\Big[ \e^{-i \theta Z^{\ox k}/2} \, \ket{\vec x} \Big]_{v_1,\ldots,v_k}	\;;
	\end{align}
	a similar analysis for the projection $\ket{\splus \sfy\sfz_{\theta}}\bra{\splus \sfy\sfz_{\theta}}$ yields the result
	\begin{gather}
			\ket{\sminus \sfy\sfz_{\theta}}\bra{\sminus \sfy\sfz_{\theta}}_a \ket{\psi}_{a,v_1,\ldots,v_k}
		\;\;\propto\;\;
			\Big[ \e^{-i \theta Z^{\ox k}/2} \, Z^{\ox k} \ket{\vec x} \Big]_{v_1,\ldots,v_k}	\;,
	\end{gather}
	so that the pattern $\mathfrak{Zz\cdots z}_{v_1,\cdots,v_k}^{\,\theta}$ performs the rotation described, by linearity.

	In order to perform such a measurement-based procedure in the one-way model as described above, we must represent the measurement $\Meas[\YZ]{a}{\theta}$ in terms of $\XY$-plane measurements.
	We may simulate a $\YZ$-plane measurement using $\XY$-plane measurements by noting that
	\begin{gather}
			\sfH\big(\cos(\theta)Z + \sin(\theta)Y\big)
		\;\;=\;\;
			\cos(-\theta) X + \sin(-\theta) Y	\;;
	\end{gather}
	that is, a Hadamard transformation effects a change of reference frame, transforming $\YZ$-plane measurement observables to some corresponding $\XY$-plane measurement observables.
	Then, we have the following congruency,
	\begin{gather}
		\label{eqn:simulateYZwXY}
			\Meas[\YZ]{a}{\theta}
 		\;\;\cong\;\;
			\Meas[\YZ]{\s[a_2]/a_1}{\theta}
		\;\;=\;\;
			\Meas{a_2}{-\theta}
			\Xcorr{a_2}{\s[a_1]}
			\Meas{a_1}{\,0}			
			\Ent{a_1,a_2}
			\New{a_2}{\sfx}	\;,
	\end{gather}
	where congruency is up to changes in the input and output subsystems; the measurement on $a_1$ performs a Hadamard transform to effect the necessary change of reference frame to perform the $\YZ$-plane measurement via $\XY$-plane measurements.
	Absorbing the correction on $a_2$ into the measurement, we may then transform $\mathfrak {Zz\cdots z}^{\theta}$ into a pattern in the one-way model as described in this section, by the congruence
	\begin{gather}
		\label{eqn:zzManyXY}
			\smash{\underbrace{\mathfrak{Zz\cdots z}}_{\text{$k$ times}}}_{v_1,\cdots,v_k}^{\,\theta}
		\;\;\cong\;\;
			\sqparen{\prod_{j = 1}^k \Zcorr{u_j}{\s[a_2]}}
			\Meas{a_2}{-\theta;\s[a_1]}
			\Meas{a_1}{\,0}
			\sqparen{\prod_{j = 1}^k \Ent{a_1, u_j}}
			\Ent{a_1,a_2} \New{a_2}{\sfx} \New{a_1}{\sfx}	\;.
	\end{gather}
	The geometry for this measurement pattern is also illustrated in Figure~\ref{fig:zzMany}.
	Nevertheless, the pattern $\mathfrak{Zz\cdots z}^{\theta}$ described in~\eqref{eqn:zzMany} illustrates the potential usefulness of extending beyond $\XY$-plane measurements in the one-way model; and we will also refer to this construction in later Chapters.

\section{Conclusion}

In this section, we have described the one-way measurement model in detail, by way of defining constructions for measurement procedures which simulate unitary circuit models; and we have also given an overview of approaches to robustly implementing quantum computers via the one-way model. 

Simulation of unitary circuits is the standard approach to obtaining one-way measurement procedures to perform quantum computation: however, it is possible to construct measurement patterns which do not clearly correspond to the simulation of a unitary circuit.
An example of one technique for doing so are the simplification techniques of~\cite{HEB}, which allows the transformation of any one-way measurement procedure involving measurements of Pauli observables into an equivalent procedure without Pauli observable measurements, on fewer qubits.
The way in which this is done takes advantage of a correspondence between \emph{local Clifford operations} (\ie\ products of single-qubit unitaries from the Clifford group) on graph state, and \emph{local complementation} of the graphs which describe the graph states: we describe this correspondence in Lemma~\ref{lemma:HEB04}.
This gives rise to transformations of graph states, and subsequently of measurement-based procedures, where the resulting procedures no longer correspond to a simulation of a unitary circuit in any clear way (see for example \cite[Figure~16]{HEB04}).

This raises the question of the conditions under which we may identify when a measurement pattern may be meaningfully considered to simulate some unitary circuit: this problem is the one which we consider in Chapter~\ref{Semantics for unitary one-way patterns}.
	
	\chapter[Semantics for unitary one-way patterns]{
 		Semantics (for unitary one-way patterns)
}


One-way measurement based quantum computation describes unitary transformations of quantum states as a composition of CPTP maps, many of which are not themselves unitary.
The fact that the resulting transformation is unitary is due to an appropriate combination of measurement observables and entanglement operations which, as in the analysis of the pattern $\mathfrak J^\alpha$ in~\eqref{eqn:stabilizerJ}, may be treated with the stabilizer formalism (Section~\ref{sec:stabilizerFormalism}).
The post-measurement residual state in either case will be an isometric image of the pre-measurement state; and the possible post-measurement states corresponding to the different measurement results may be mapped to one another by single-qubit unitary transformations.
As a result, we may perform a correction after each measurement to simulate postselection of the $+1$ measurement result, yielding a mixed one-way measurement pattern; equivalently, we may absorb the byproduct operations into future measurements, interpreting them as changes of reference frame, in order to obtain a one-way measurement based computation in standard form.

As we are interested in unitary transformations, we are presented with natural decision problems when considering quantum computation in the one-way measurement model. 
Measurements do not transform quantum states unitarily --- is it possible to tell whether a one-way measurement based computation performs a unitary transformation between its input and output subsystems?
Is it possible to describe precisely \emph{how} such computations transform quantum states, by translation to \eg\ a quantum circuit of comparable complexity, using a reasonable set of elementary gates?
If not, is it at least possible to efficiently verify whether it performs some given transformation?

The problem of determining when a measurement problem performs an isometry seems similar to to the more general question of when a classically controlled circuit in general performs an isometry.
By~\cite{MCSB98}, this may be rephrased as asking whether each measurement performed has the same bias towards one measurement result or another, across all possible input states.
This problem is \NP-hard if we require exponential precision, as we can encode the evaluation of arbitrary boolean functions $f: \ens{0,1}^n \to \ens{0,1}$ to produce the state
\begin{gather}
		\ket{\textsc{sat}_f}
	\;\;=\;\;
		\sfrac{1}{\sqrt {2^n}} \sum_{\vec x \in \ens{0,1}^n} (-1)^{f(x)}	\ket{x}
\end{gather}
and perform measurements in the $\ket{\spm}$ basis.
If any such measurement has non-zero probability of yielding the $\ket{\sminus}$ state, this would indicate that $f$ is satisfiable but not a tautology.
However, it is easy to verify that the natural approach to producing a state such as $\ket{\textsc{sat}_f}$ requires the use of non-Clifford group operations, whereas the open graphical encoding procedures described in~\eqref{eqn:openGraphEncoding} consist of Clifford group operations, and prepare states in a known stabilizer code.
As well, we require that the post-measurement residual states be related to each other by local unitary operations in order to be able to adaptively perform the (single-qubit) measurements of a one-way procedure.
It may be hoped that the additional structure imposed by these constraints yields a tractable problem.

These questions arise because simple unitary circuit models such as $\Phases(\A)$ defined in Section~\ref{sec:unitaryCircuits} are the \emph{de facto} standard models of quantum computation: they seem to provide the most natural set of idioms for uniformly expressing algorithms independently of architecture.
(We will commonly refer ``the'' unitary circuit model, by which we will mean the $\Phases(\A)$ or $\Jacz(\A)$ models for some $\A \subset \R$.)
While alternative models are promoted as abstractions of proposals for physically implementing a quantum computer (\eg\ measurement-based models~\cite{GC99,RB01} and the adiabatic model~\cite{FGGS00}), and ``purely physical'' idioms seemingly divorced from uniform circuit families have led to new quantum algorithms (as with the \textsc{nand} tree algorithm of~\cite{FGG07}, inspired by transmission and reflection rates associated with energy barriers), the unitary circuit model provides what seems to be the simplest mathematical model for describing quantum computation.
Because of the present uncertainty as to \emph{which} implementation proposal will prove successful, the unitary circuit model represents a natural candidate for an \emph{intermediate language}\footnote{%
\emph{Intermediate language} refers here to a language used to facilitate the design of a suite of compilers, which translate programs from one or many source language (or from many source languages) to one of several different target architectures.}
for representing algorithms independently of the target architecture, even if alternative models suggest new algorithms.

Therefore, I argue that how effectively one can translate decompositions of unitaries to and from the unitary circuit model is currently an important topic for any proposed alternative model of quantum computation.
Unless some one proposal for implementation comes to dominate over the others, or an unlikely breakthrough (\eg\ the discovery of a proof that $\BQP = \BPP$) is made, translation to a ``simple'' model of quantum computation such as the unitary circuit model is the best that can be reasonably hoped for as a \emph{uniform} way of obtaining low-level semantics of a quantum computation.
In other words: the most reasonable means of understanding the behavior of a quantum algorithm is currently by reduction to the unitary circuit model.
This motivates the problem of effective translations from those one-way pattern which perform unitary embeddings, into the unitary circuit model.

The topic of this chapter is one such reduction, from those one-way measurement patterns which arise out of the DKP constructions to unitary circuits.
We consider this topic by examining a \emph{flow} property which emerges from that construction, which gives rise to efficiently detectable combinatorial structures and combinatorial decompositions of the corresponding measurement pattern.

\vspace{-1ex}
\paragraph{Previous appearances of this work.}

Earlier versions of many of the results of this chapter appear in~\cite{Beaudrap06,Beaudrap08,BP08}.
The work presented in Section~\ref{sec:graphth-charn} (except for~\ref{sec:extreme}) and Section~\ref{sec:flowAlgs} first appeared in a preliminary form in~\cite{Beaudrap06}; the results of the former was published as~\cite{Beaudrap08} in essentially the form presented here.
The results of Section~\ref{sec:extreme} are joint work with Martin Pei, and were published in~\cite{BP08}.
The rest of the work in this Chapter, except where indicated, are developments original to this thesis.

\vspace{-1ex}
\paragraph{Graph theoretic notations.}

We will assume basic familiarity with graph theory: an introduction to the subject and basic definitions can be found in~\cite{Diestel}.
Whenever a graph $G$ is clear from context, $\sim$ will denote the adjacency relation of $G$.
In graphs, we will denote an edge between $v$ and $w$ by $vw$, and in directed graphs, we will let $v \arc w$ represent an arc between $v$ and $w$\,; and in graphs or directed graphs, we will represent (directed) paths by concatenated sequences of edges/arcs, $v_1 v_2 \cdots v_n$ or $v_1 \arc v_2 \arc \cdots \arc v_n$.
We adopt the convention that graphs do not have self-loops on vertices.

\section{Defining semantics by reduction to unitary circuits}
\label{sec:semanticsReduce}

Before proceeding, we will put forward a precise definition of how the semantics of a measurement-based computation may be reduced to another model, such as the unitary circuit model.

We are generally not interested in \emph{arbitrary} translations to and from the unitary circuit models: in this instance, the particular circuit model becomes significant.
For instance, in the case of the one-way measurement model, we may easily obtain a circuit in the $\Phases(\A)$ or $\Jacz(\A)$ models by describing the measurement pattern as a circuit in a classically controlled unitary circuit model, and applying the principle of deferred measurement~(page~\pageref{principleDeferredMeas}) to obtain a unitary circuit in some unitary circuit model $\CoherCtrl$ which includes (coherently) controlled unitary transformations performing conditional change of basis operations to simulate adaptive measurements.
Such a circuit can then be decomposed in \eg\ the $\Jacz(\A)$ model by decomposing the individual transformations of the model $\CoherCtrl$\:\!.
However, such a direct translation scheme may require many more qubits of workspace, and many more multi-qubit operations, than is actually necessary.
For instance, if we translate a circuit of the form $J(\alpha_n) \cdots J(\alpha_2) J(\alpha_1)$ on a single qubit into the one-way model using the DKP construction, and then obtain a unitary circuit from that pattern via the principle of deferred measurement as above, we will obtain a circuit involving $n + 1$ qubits and $O(n)$ two-qubit operations, whereas the original circuit required only operations on a single qubit.

In contrast with the above, we may hope to obtain a translation scheme to unitary circuit models which in some sense preserves the complexity of the computation.
In order to make this sense of complexity-conservation well defined despite the differences in how resources are used in different models of computation, we may make reference (as we have done above for the DKP construction) to another function which translates from unitary circuits to the computational model of interest.
In particular, suppose we have a map $\cR$ which maps unitary circuits in some gate model $\Gates$ to some target model $\M$ of quantum computation.
We would like to consider a mapping $\cS$ such that
\begin{romanum}
\item
	\label{condition:superset}
	$\cS$ is a map from a superset of $\img(\cR)$ to unitary circuits using the gate-set \Gates\ (where in particular, $\dom(\cS)$ includes any procedures in $\M$ which may be in some sense ``congruent'' to the image of a circuit in $\cR$);


\item
	\label{condition:noninflation}
	$\cS \circ \cR$ maps each circuit in $\dom(\cR)$ to an equivalent circuit of lesser or equal gate complexity;

\item
	\label{condition:retraction}
	$\cR \circ \cS$ maps procedure in $\img(\cR)$ to a ``congruent'' procedure in $\cR$.
\end{romanum}
We will not formally define here when two computational procedures are ``congruent'' in an arbitrary model of computation: however, for both the unitary circuit model and for the one-way measurement based model, we will adopt the convention of Section~\ref{sec:unitaryCircuits} that computations are congruent in either model if they only differ by transpositions of commuting operations (and by a possible relabelling of the qubits).

It may seem unproductive to consider a map $\cS$ from a model $\M$ as above to unitary circuits, which is explicitly defined with reference to a \emph{existing} translation procedure $\cR$ from unitary circuits to $\M$, as the most obvious source of a computation in $\img(\cR)$ is to take an existing unitary circuit $C$ and to produce $\cR(C)$.
However, given the current prominence of the unitary circuit model, I would argue that a mature understanding of a model $\M$ of quantum computation currently requires the ability to perform meaningful translations \emph{in both directions} between $\M$ and the unitary circuit model.
Should new insights arise for performing algorithms in the model $\M$, these insights can then also be translated back to the unitary circuit model.

In practise we will be largely interested in mappings between congruency classes of unitary circuits and of other models of computation.
In the case where $\cR$ is an injective map from a unitary circuit model to some model $\M$, we may simply wish for $\cS$ to be an inverse for the action of $\cR$ on congruency classes of unitary circuits.
However, for more general transformations $\cR$ which may map incongruent circuits (but which perform equivalent CPTP maps) to congruent procedures in some other model of computation, the conditions described above may be of interest in the absence of a left inverse for $\cR$.

In this thesis, we are primarily interested in the one-way measurement model, so we consider the case of $\M = \OneWay(\A)$ for some $\A \subset \R$.
\label{discn:representationMap}
A function $\cR$ as above provides an image of unitary circuit ``idioms'' --- structural formulas for circuits, which are used to perform certain tasks --- in the one-way model: a one-way measurement based computation which is in the image of such a function $\cR$ may be said to be a representation of some unitary circuit in the one-way model.
We may therefore call $\cR$ a \emph{representation} map, and the one-way measurement based routines in $\img(\cR)$ \emph{representations} of circuits.

The map $\cS$ is then a means of recognizing one-way measurement patterns which simulate circuits in this sense, revealing semantics for these one-way measurement patterns in terms of unitary circuits, in a way which
\begin{inlinum}
\item
	preserves the structure of some representative of each class of circuits which are translated by $\cR$ into a given one-way measurement based procedure, and

\item
	such that the one-way measurement based algorithm can be recovered from the circuit through the ``representation'' map.
\end{inlinum}
We may call a map $\cS$ of this sort a \emph{semantic} map, and say that a circuit in $\img(\cS)$ provides the \emph{semantics} of a one-way pattern in terms of some unitary circuit model.\label{discn:semanticMap}

It is not difficult to imagine more properties which we would like the map $\cS$ to have beyond the ones described above, for instance:
\begin{itemize}
\item
	It is obviously desirable for $\cS$, and the membership predicate for $\dom(\cS)$, to be efficiently computable.

\item
	It would be convenient for a semantic map $\cS$ to also be able to recognize \emph{parts} of one-way measurement based computations which resemble ``idioms'' from the circuit model, and be able to translate them into components of the corresponding circuit.
	This would allow us to describe the semantics of a one-way measurement pattern by reduction to smaller pieces.

\item
	We might also like $\cS$ to be well defined on the set of \emph{all} one-way measurement based computations performing unitary transformations, which may be a proper superset of the range of $\cR$\,: this may suggest which extensions to the unitary circuit model can be easily translated into a one-way measurement pattern, and which may therefore be useful tools to adopt for designing quantum algorithms.
\end{itemize}
We are of course primarily interested in efficient algorithms, and will restrict our attention without further comment to semantic maps which can be performed in polynomial time in the size of the specification for a one-way measurement based computation.
The second of the extensions above, while beyond the scope of this thesis to formalize, is easy to achieve in an intuitive sense in some instances which will be described in this chapter.
The third of these extensions is more difficult, in part because it is not yet known how to determine whether a one-way measurement based algorithm realizes a unitary transformation or not: for any such ``extended semantics'' map $\cS$, a polynomial-time algorithm for the membership predicate of $\dom(\cS)$ is not yet known.

We have defined semantic maps $\cS$ above relative to a map $\cR$ for representing unitary circuits in the one-way measurement based model, and it is conceivable that for some interesting maps $\cR$, it may be difficult or impossible to find a map $\cS$ which satisfies all of the properties~\ref{condition:superset}\thru\ref{condition:retraction} on pages~\pageref{condition:superset}\thru\pageref{condition:retraction}.
In this case, we may be satisfied by an $\cS$ which \emph{approximately} achieves these properties for a given $\cR$.
For instance, we may require that for a unitary circuit $c$, the number of single-qubit and two-qubit gates in the circuit $(\cS \circ \cR)(C)$ are increased beyond those of $C$ only by small scalar factors (\eg\ less than $1 + \epsilon$ for a suitably small $\epsilon$), or that even if $\cR$ is not a left-inverse of $\cS$, that there is an efficiently computable reduction map $f$ on $\dom(\cS)$ such that $f \circ \cR \circ \cS$ is the identity on $\dom(\cR)$.
However, the less closely we adhere to the conditions that $\cR \circ \cS$ is equivalent to the identity and that $\cS \circ \cR$ does not inflate the complexity of unitary circuits, the less useful the mapping $\cS$ is as a means of providing semantics for a one-way measurement based computation in terms of the unitary circuit model.

The main result of this chapter is to describe a mapping $\cS$ which acts as a semantic map with respect to the DKP construction described in Section~\ref{sec:oneWayComputation}, where we further restrict to those patterns produced from circuits without terminal measurements or trace-out operations.
This allows us to identify a class of one-way patterns which perform unitary transformations, and provide semantics for them in terms of unitary circuits.
The semantic map $\cS$ in this case is defined in terms of a \emph{flow} property of the geometry underlying the one-way pattern, which will be the main subject of this chapter.
Towards the end of the Chapter, we sketch similar results for a semantic map for the simplified RBB construction by using a slight extension of flows, building on the analysis for flows and on improved flow-finding algorithms to do so.

\subsection*{A remark on terminal measurement and trace-out operations}

As we have described it in Section~\ref{sec:oneWayComputation}, the model $\OneWay(\A)$ includes trace-out operations.
When considering how to transform a pattern $\mathfrak P$ in the $\OneWay(\A)$ model to a unitary circuit model, given that the pattern is congruent to a pattern in the image of the DKP construction, we may augment the output subsystem $O$.
This results in a measurement pattern $\mathfrak P'$ which also performs a unitary embedding, and is also congruent to a pattern from the DKP construction; the unitary circuit $C$ corresponding to the original pattern $\mathfrak P$ may be obtained by adding trace-out operations to the circuit $C'$ corresponding to $\mathfrak P'$.
However, patterns produced by the DKP construction from circuits with terminal measurements pose a more difficult problem, as it seems difficult to distinguish between qubits which are measured in order to drive the transformation of data and qubits which are measured in order to yield a classical probabilistic output of interest (speaking of the measurements in their roles of representing operations of the unitary circuit).

Given that we are primarily interested in how unitary transformations may be described by measurement-based computation, we set aside the issue of how to provide semantics for arbitrary patterns which may be produced by the DKP construction, and focus on those which arise from circuits which do not perform measurements or trace-out operations.
Fortunately, by the comments made above, semantics for patterns in $\img(\cR)$ which perform trace-outs may still be efficiently recovered; but for patterns arising from circuits with measurements, it seems that more general tools will be required.
We will pass over this subject for the remainder of the thesis.

\section{Flows: structure underlying the DKP construction}
\label{sec:flowDefn}

As we defined it in Definition~\ref{def:simplifiedDKP}, the simplified DKP construction produces a measurement pattern from a circuit in the $\Jacz(\A)$ model (for some $\A \subset \R$) by transforming each gate $J(\alpha)$ and $\cZ$ into a simple measurement pattern via the mapping $\Phi$ defined in~\eqref{eqn:circuitPatternHomphm}, composing those patterns in the appropriate manner.
The resulting measurement dependencies exhibits a simple structure which may be described in terms of local properties of the entanglement graph, as follows.

For a unitary circuit $C$ in the $\Jacz(\A)$ model, consider the last gate $J(\alpha)$ performed.
(If there is more than one such gate which may be performed in parallel, we may choose an arbitrary one.)
We may decompose $C$ into circuits $C_v C'$, where $C_v$ consists of the final $J(\alpha)$ gate on $v$, and any controlled-$Z$ operations on $v$ which follow it, and $C'$ consists of the rest of the operations of $C$.
We can recursively apply this procedure to decompose $C$ into layers of $\cZ_{u,v}$ gates separated by $J(\alpha)_v$ gates for various qubits $u$ and $v$, where the $\cZ$ gates of each layer following a $J(\alpha)_v$ gate all act on the qubit $v$.
We will refer to this as a \emph{star decomposition} of a $\Jacz(\A)$ circuit.
Consider the patterns produced by the DKP construction, for each sub-circuit $C_v$ consistsing of a $J(\alpha)_v$ gate and the subsequent layer of $\cZ$ gates: writing $C_v$ in stable-index tensor notation (pages~\pageref{sec:nonstandardRepns}\thru\pageref{sec:nonstandardRepnsEnd}), we have
\begin{align}
		\Phi(C_v)
	\;\;=&\;\;
		\Phi\paren{\paren{\prod_u \cZ\pseu[u,v']} J(\alpha)\pseu[:v'/v]}
	\notag\\[1ex]=&\;\;
		\paren{\prod_u \Ent{u,v'}{}} \mathfrak J^\alpha_{v'/v}
	\notag\\[1ex]=&\;\;
		\paren{\prod_u \Ent{u,v'}{}} \Xcorr{v'}{\s[v]} \Meas{v}{-\alpha} \Ent{v,v'} \New{v'}{\sfx}	\;,
	\notag\\[1ex]\cong&\;\;
		\paren{\prod_u \Zcorr{u}{\s[v]}} \Xcorr{v'}{\s[v]} \paren{\prod_u \Ent{u,v'}{}} \Meas{v}{-\alpha} \Ent{v,v'} \New{v'}{\sfx}	\;.
\end{align}
where in the last step we commute the correction operation $\Xcorr{v'}{\s[v]}$ to the right.
The $J(\alpha)$ gate in the original circuit $C$ gives rise to a qubit $v'$ which is a ``successor'' of the qubit $v$\,, in the sense that the pattern $\mathfrak J^\alpha_{v'/v}$ transfers the state of $v$ to $v'$ (up to the given unitary transformation); and every other neighbor of $v'$ in the entanglement graph will be subject to a correction operation which depends on the result $\s[v]$ of the measurement on $v$.

If we compose measurement patterns of this form together, this structure of correction is (almost) preserved: because the $\Zcorr{u}{\s[v]}$ operations commute with any entangling operations acting on the qubits $u$, the only additional correction which can be induced is a correction $\Zcorr{v''}{\s[v]}$ on a possible successor of $v'$ in cases where the same qubit is later acted on by another $J(\alpha')$ gate in $C$.
Absorbing these correction into measurements where appropriate, these give rise to sign- and $\pi$-dependencies on $v$ of the measurements on qubits adjacent to $v'$ in the entanglement graph (other than $v$ itself).

In the case of qubits measured with $\axisX$ measurements (\ie\ whose measurement angle is zero), the ``effect'' of $X$ corrections are trivial; in every other case, however, there is either a non-trivial effect on the basis of measurement, or (speaking somewhat counterfactually) on the result obtained upon measurement in the case of a $\pi$-dependency.
In particular, while $\pi$-dependencies could be removed by performing signal-shifting, \ie\ by introducing an operator which conditionally toggles the measurement result, we could also speak informally of the ``measurement result that would occur'' being affected by the change in the measurement by the addition of $\pi$ to the measurement angle.\footnote{%
	This intuition of ``counterfactual'' change of measurement results could perhaps be given an ontological foundation in terms of hidden variables, thereby removing its counterfactual nature; a formal discussion of this is beyond the scope of this thesis.}
Thus, we interpret the angle adaptations given by the simplified DKP construction as potential ``influences'' on the measurement outcomes, neglecting the imperviousness of $\axisX$ measurements to sign dependencies for the sake of uniformity.

Signal shifting and Pauli measurement simplifications can be applied to reduce the logical depth of a measurement-based computation, and so produce a pattern of the \emph{complete} DKP construction.
However, the dependencies described by the \emph{simplified} DKP construction are local in nature, and can be described simply by the geometry $(G,I,O)$ of the measurement pattern (Definition~\ref{def:geometry}), together with a function $f: O\comp \to I\comp$ mapping each qubit $v$ in the pattern to its ``successor'' (where $O\comp = V(G) \setminus O$, and $I\comp = V(G) \setminus I$), arising from the $\mathfrak J^{\alpha}$ patterns.
In particular, the measurement of each qubit $v \in O\comp$ influences the measurement of its successor $f(v)$ by a sign dependency, and the measurements of the other neighbors of $f(v)$ by a $\pi$-dependency.
Furthermore, these dependencies entail that both $f(v)$ and every qubit adjacent to $f(v)$ --- except for $v$ itself --- would be measured strictly after $v$ in the measurement order arising from the simplified DKP construction.

\begin{figure}[t]
	\begin{center}
			\includegraphics{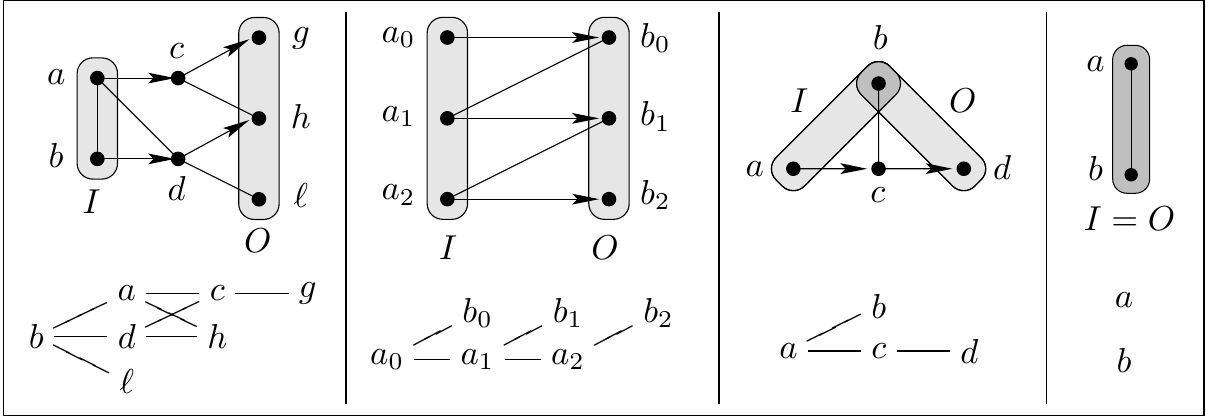}
			\vspace{-2ex}
	\end{center}
	\caption[Examples of geometries with flows.]{\label{fig:examplesFlow}
		Examples of geometries with flows.
		Arrows indicate the action of a function $f: O\comp \to I\comp$ along otherwise undirected edges.
		Corresponding partial orders $\preceq$ for each example are given by Hasse diagrams below the graphs (with minimal elements on the left and maximal elements on the right).
		In the right-most example, the two vertices $a$ and $b$ are incomparable, \ie\ there is no order relation between them.}
		\vspace{1em}
\end{figure}

\begin{figure}[t]
	\begin{center}
			\includegraphics{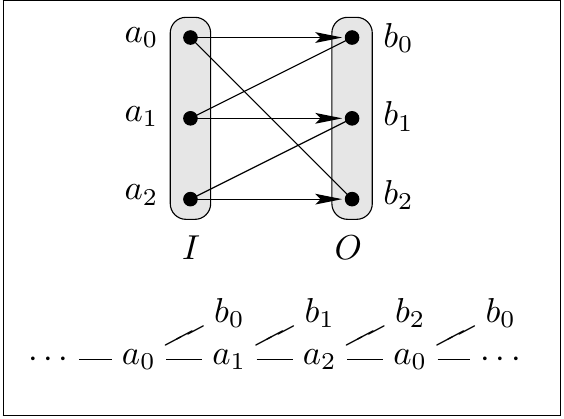}
			\vspace{-2ex}
	\end{center}
	\caption[Example of a geometry with no flow.]{\label{fig:exampleNoFlow}
		Example of a geometry with no flow (\cf\ the second geometry from the left in Figure~\ref{fig:examplesFlow}).
		Arrows indicate the action of an injective function $f: O\comp \to I\comp$ along otherwise undirected edges.
		Also given is a reflexive and transitive binary relation $\preceq$ which satisfies conditions~\eqref{flow:b} and \eqref{flow:c} for this function $f$, but which is not antisymmetric.}
\end{figure}

``Flows'' were defined by Danos and Kashefi~\cite{DK06} in an approximate\footnote{%
The presentations of the one-way model in~\cite{RB01,RBB03} already contained several constructions which do not have ``flows'', which were formulated later.
However, at least in the case of the the simplified RBB construction presented in the previous chapter, there is a simple correspondence to the DKP construction which allows these to be related to geometries with flows.
It was thought that a modest extension of the definition of flows would also capture those one-way measurement based algorithms~\cite{DK05priv}.}
early attempt to characterize those one-way measurement patterns which perform unitary transformations, based on this structure of dependencies arising from this construction.
Without reference to a pattern known to be obtained from the DKP construction, a flow is defined as follows:
\begin{definition}
	\label{def:flow}
	For a geometry $(G,I,O)$, a \emph{flow} is an ordered pair $(f, \preceq)$ consisting of a function $f: O\comp \to I\comp$, and a partial order\footnote{%
		A partial order $\preceq$ is a pre-order (see footnote~(\ref{fn:preorder}) on page~\pageref{fn:preorder}) which is anti-symmetric: that is, for which $x \preceq y$ and $y \preceq x$ imply $x = y$.
		Examples include the divisibility relation $x \,|\, y \,\iff\,  (y \div x \in \N)$ for complex numbers $x$ and $y$\,, and the subset relation $X \subset Y \,\iff\; \forall z: \paren{z \in X \implies* z \in Y}$ on sets $X$ and $Y$.}
	$\preceq$ on $V(G)$, such that the conditions
	\begin{subequations}
	\begin{gather}
			\label{flow:a}			v \sim f(v)	\;,
		\\
			\label{flow:b}		\textbalance{v \preceq f(v)\;,}[\;\;\text{and}]
		\\
			\label{flow:c}		w \sim f(v)	\;\implies\; v \preceq w
	\end{gather}
	\end{subequations}
	hold for all $v \in O\comp$ and $w \in V(G)$.
\end{definition}
Examples of geometries with and without flows are illustrated in Figures~\ref{fig:examplesFlow} and~\ref{fig:exampleNoFlow}.

The function $f$ and the partial order $\preceq$ capture the essential structure of the dependencies in the simplified DKP construction: $f$ represents the mapping of qubits to their successors, implementing logical ``wires'' via single-qubit state transfer, and the partial order $\preceq$ represents \emph{one possible} order (not necessarily one of minimum depth) in which the qubits may be measured to perform a unitary computation.

By~\cite{DK06}, any geometry which has a flow underlies some one-way measurement pattern which can be obtained by the DKP construction, and which therefore performs a unitary transformation: moreover, the proof in~\cite{DK06} is by reduction to unitary circuits.
We may then use flows to obtain a semantic map $\cS$ (as described on page~\pageref{discn:semanticMap}), corresponding to the choice of the \emph{complete} DKP construction as a choice of representation map $\cR$ from unitary circuits to the one-way measurement model, if
\begin{inlinum}
\item
	we can find a flow for a geometry $(G,I,O)$; and
\item
	we can verify whether the \emph{dependencies and corrections} of a measurement pattern are consistent with one arising from the DKP construction.
\end{inlinum}
The main subject of this Chapter is essentially how to perform these tasks efficiently.

\subsection{Direct interpretation of flows in terms of unitary circuit structure}
\label{sec:flowsNonstandardRepn}

On pages~\pageref{sec:nonstandardRepns}\thru\pageref{sec:nonstandardRepnsEnd}, we presented the stable-index representation of unitary circuits, and in particular described the combinatorial structure of the ``interaction graph'', whose vertices consist of tensor indices (corresponding to ``wire segments'') of the unitary circuit.
The edges of this graph consist of pairs of indices which are involved in a single unitary gate; and in particular, there is a function $f$ mapping each deprecated index to a corresponding advanced index, imposing an arrow of time on the indices.

The map $\Upphi$ of~\eqref{eqn:circuitPatternHomphm} from the $\Jacz(\A)$ model to the $\OneWay(\A)$ model, which forms the first part of the simplified DKP construction of Section~\ref{sec:DKPstandaridzation}, identifies the interaction graph of a circuit in the $\Jacz(\A)$ model with the geometry $(G,I,O)$ of the corresponding one-way pattern.
In particular, advanced indices correspond to prepared (or non-input) qubits, and deprecated indices to measured (or non-output) qubits; and the function $f$ on the indices of the stable index expression may then be identified with a function $f: O\comp \to I\comp$ on the geometry $(G,I,O)$.

A flow consists of a function $f$ and a partial order $\preceq$, where $f$ may be considered to correspond to the mapping of deprecated indices to advanced indices in a stable index representation of a unitary circuit, and the partial order $\preceq$ to correspond to the order described on page~\pageref{item:nonstandardRepsOrderRecovery} in the discussion on ``Recovering the order of unitary operations''.
Flows may then also be interpreted as describing the structure of a unitary circuit, expressed as a stable index tensor product, from the geometry $(G,I,O)$ of a measurement pattern.

\section{Graph constructions and characterizations for flows}
\label{sec:graphth-charn}

The main result of this chapter is to characterize flows in graph-theoretic terms, which allows us to reduce the problem of deciding whether a geometry has a flow to problems in graph theory with efficient solutions, and to examine (and solve) an extremal problem which further bounds the running time of those algorithms.
This will ultimately enable us to obtain an efficiently computable semantic map $\cS$ (as described on page~\pageref{discn:semanticMap}) for the DKP construction described in Section~\ref{sec:DKPstandarization}, and provide a strong upper bound on its running time.

In this section, we consider useful constructions of directed graphs, culminating in the aforementioned graph-theoretic characterizations.
These constructions presented in this section will also allow us to more easily describe the generality of the geometries which have flows.
We also make use of these constructions to obtain uniqueness results, and bounds on the number of edges of graphs with flows, which ultimately lead to the first efficient algorithm for determining when a geometry $(G,I,O)$ has a flow.

\subsection{Generalizing to path covers/successor functions}

We will begin by exploring how to relax the notion of a flow to admit a larger class of objects, starting by noting the most important properties of a flow.
\begin{lemma}
	\label{lemma:injectiveFlow}
	If $(f, \preceq)$ is a flow for a geometry $(G,I,O)$, then $f$ is injective.
\end{lemma}
\begin{proof}
	Suppose $x,y \in O\comp$ are such that $f(x) = f(y)$.
	Then $y \sim f(y) = f(x)$, so that $x \preceq y$, and $x \sim f(x) = f(y)$, so that $y \preceq x$\,.
	It follows that $x = y$.
\end{proof}
The limitation to injective functions $f$ motivates a description of flows in terms of vertex-disjoint collections of paths (or more precisely, an adaptation of the concept of a ``path cover''~\cite{Diestel}):
\begin{definition}
	\label{def:pathCover}
	Let $(G,I,O)$ be a geometry. A collection $\cC$ of (possibly trivial\footnote{%
	A (directed) path or walk is \emph{trivial} if it is of length zero, \ie\ it starts and ends at a single vertex without traversing any edges (respectively, arcs).})
	directed paths in $G$ is a \emph{path cover} of $(G,I,O)$ if the following conditions hold:
	\begin{romanum}
	\item
		each $v \in V(G)$ is contained in exactly one path (i.e. the paths cover $G$ and are vertex-disjoint);
	\item
		each path in $\cC$ is either disjoint from $I$, or intersects $I$ only at its initial point;
	\item
		each path in $\cC$ intersects $O$ only at its final point.
	\end{romanum}
\end{definition}
In the case $\card{I} = \card{O}$, a path cover of $(G,I,O)$ is a collection of vertex-disjoint paths from $I$ to $O$ which covers all the vertices of $G$.
For a flow $(f,\preceq)$, there is a natural connection between the function $f$ and path covers for the geometry $(G,I,O)$:
\begin{lemma}
	\label{lemma:orbitPathCover}
	Let $(f, \preceq)$ be a flow on a geometry $(G,I,O)$.
	Then there is a path cover $\cP_f$ of $(G,I,O)$ such that, for all vertices $v,w \in V(G)$, $v \arc w$ is an arc in some path of $\cP_f$ if and only if $w = f(v)$.
\end{lemma}
\begin{proof}
	Let $(f,\preceq)$ be a flow on $(G,I,O)$.
	Define a digraph $P$ on the vertices of $G$, and with arcs $v \arc f(v)$ for $v \in O\comp$.
	Because $f$ is both a function and injective, every vertex in $P$ has maximal out-degree and maximal in-degree $1$.
	Thus, $P$ is a collection of vertex-disjoint dipaths and closed walks.\footnote{%
	A closed walk is one which begins and ends at the same vertex.}
	Furthermore, for every arc $(v \arc w) \in A(P)$, we have $v \preceq w$; by induction, $v \preceq z$ whenever there is a dipath from $v$ to $z$ in $P$.
	Then if $v$ and $z$ are such that there are dipaths from $v$ to $z$ and from $z$ to $v$, then $v \preceq z$ and $z \preceq v$, in which case $x = z$ and the dipaths are trivial.
	Thus, $P$ is acyclic, so $P$ consists entirely of vertex-disjoint dipaths.

	Let $\cP_f$ be the collection of maximal dipaths in $P$.
	We show that $\cP_f$ satisfies each of the criteria of Definition~\ref{def:pathCover}:
	\begin{romanum}
	\item
		Any vertex $v$ which is neither in $\dom(f)$ nor $\img(f)$ will be isolated in $P$: then, the trivial path on $v$ is an element of $\cP_f$.
		All other vertices are in either $\dom(f)$ or $\img(f)$, and so are contained in a non-trivial path of $\cP_f$.
		As these paths are vertex-disjoint, each vertex is contained in exactly one path.

	\item
		Each vertex in $I$ has in-degree $0$, and so may only occur at the beginning of any path in $\cP_f$.

	\item
		The vertices in $P$ which have out-degree $0$ are precisely the output vertices $O$: therefore one occurs at the end of every path, and they may only occur at the end of paths in $\cP_f$.
	\end{romanum}
	Then $\cP_f$ is a path cover, whose paths contain only arcs $v \arc f(v)$, as required.
\end{proof}
It will be occasionally be convenient to alternate between descriptions of problems in terms of path covers for a given geometry, and functions $f$ such that $\cP_f$ is a path cover.
In particular, we may define:
\begin{definition}
	\label{def:successorFn}
	A \emph{successor function} for a geometry $(G,I,O)$ is an injective function $f: O\comp \to I\comp$ such that the maximal orbits of $f$ form a path cover for $(G,I,O)$.
	A successor function $f$ for $(G,I,O)$ is a \emph{flow function} if there exists a partial order $\preceq$ on $V(G)$ such that $(f, \preceq)$ is a flow.
\end{definition}
The terminology of ``successor function'' is meant to be suggestive of the mapping $v_j \mapsto* v_{j+1}$ between successive indices in a stable index tensor expression for a single qubit, which in turn correspond (by the comments made on page~\pageref{discn:stableIndexPathIntegral}) to successive segments of a wire (separated by single-qubit gates which do not preserve standard basis states) in a quantum circuit diagram.
The generalization from flow functions to successor functions will prove very helpful in the analysis below.

\subsection{Generalizing from partial orders to pre-orders}

Figure~\ref{fig:exampleNoFlow} (on page~\pageref{fig:exampleNoFlow}) illustrates a geometry with successor functions, but which has no flow functions.
There are only two possible successor functions $f$ and $f'$ for that geometry, given by
\begin{align}
		f(a_j) \;=&\; b_j	\;\;,
	&
		f'(a_j) \;=&\; b_{(j \minus 1) \bmod{3}}
\end{align}
(\ie\ taking the appropriate representative modulo $3$ for the index in the case of $f'$).
The successor function $f$ is the function actually illustrated in Figure~\ref{fig:exampleNoFlow}, and $f': O\comp \to I\comp$ corresponds to instead directing the diagonal edges of the graph illustrated there.
To show that neither of these are flow functions, we may show that any reflexive and transitive binary relation satisfying the conditions~\eqref{flow:b} and~\eqref{flow:c} will fail to be anti-symmetric; in which case there can be no partial order $\le$ such that either $(f, \le)$ or $(f', \le)$ is a flow for that geometry.
Specifically, to satisfy the flow conditions for $f$, such a binary relation would satisfy the constraints $a_0 \le a_1 \le a_2 \le a_0$\,, while for $f'$ it would have to satisfy the constraints $a_0 \ge a_1 \ge a_2 \ge a_0$\,.
In either case, the vertices $a_j \in I$ are distinct but mutually related vertices: so any such relation fails to be a partial order.
Because no partial order can satisfy the conditions~\eqref{flow:b} and~\eqref{flow:c} for either $f$ or $f'$, that geometry has no flow.

Analysis of the form above will be generally useful, and motivates a generalization from partial orders to pre-orders
	similar to our generalization from flow functions to successor functions.
\begin{definition}
	\label{def:inflRel}
	Let $(G,I,O)$ be a geometry and $f: O\comp \to I\comp$ be a successor function for $(G,I,O)$.
	The \emph{influence relation} $\inflRel$ for $f$ is the binary relation such that $v \inflRel w$ for $v \in O\comp$ and $w \in V(G)$ if and only if $v \ne w$ and either $w = f(v)$ or $w \sim f(v)$.
	An \emph{influencing pre-order} $\preceq$ for $f$ is a pre-order on $V(G)$ which extends $\inflRel$\,, \ie\ a reflexive and transitive relation such that
	\begin{subequations}
	\begin{gather}
 			\label{natural:a}
			v \preceq f(v)
		\\
 			\label{natural:b}
			w \sim f(v) \;\;\implies\;\; v \preceq w
	\end{gather}
	\end{subequations}
	holds	for all $v \in O\comp$ and $w \in V(G)$.
	A \emph{causal order} for $f$ is a influencing pre-order for $f$ which is also a partial order, \ie\ a binary relation such that $(f,\preceq)$ is a flow.
\end{definition}
Note that the influence relation $v \inflRel w$ describes exactly the conditions~\eqref{flow:b} and~\eqref{flow:c} on a causal order: the definition of an influencing pre-order then relaxes nothing more of the flow conditions than the condition that $\preceq$ be anti-symmetric.
The proof given at the beginning of this section that the geometry of Figure~\ref{fig:exampleNoFlow} has no flows is essentially that every influencing pre-order of $f$ or $f'$ must somehow fail to be a full-fledged partial order.

\subsection{Deciding if a successor function is a flow function}
\label{sec:influencing}

Every successor function has an influencing pre-order, and for each $f$, we identify one in particular:
\begin{definition}
	\label{def:naturalPreorder}
	Let $(G,I,O)$ be a geometry and $f: O\comp \to I\comp$ be a successor function for $(G,I,O)$.
	The \emph{natural pre-order} $\preceq$ for $f$ is the transitive and reflexive closure\footnote{%
	That is, $\preceq$ is the coarsest reflexive and transitive relation which extends $\inflRel$, and so may be described as the logical conjunction of all influencing pre-orders.
	As there exists at least one influencing pre-order, namely the relation $R$ such that $v R w$ for all $v, w \in V(G)$, this closure is guaranteed to exist.}
	of the relation $\inflRel$\,. 
\end{definition}
The natural pre-order $\preceq$ for a successor function $f$ is by definition an influencing pre-order, and in particular the coarsest influencing pre-order for $f$.
As a result, it can be used to determine whether or not $f$ is a flow function:
\begin{lemma}
	\label{lemma:causalOrderFlow}
	A successor function $f$ is a flow function if and only if its natural pre-order $\preceq$ is a partial order.
\end{lemma}
\begin{proof}
	If $\preceq$ is a partial order, then $(f, \preceq)$ is a flow, and so $f$ is a flow function.
	For the converse, suppose that there is \emph{some} partial order $\le$ such that $(f, \le)$ is a flow: then $\le$ is an influencing pre-order for $f$.
	Being a partial order, for each distinct pair of vertices $v,w \in V(G)$, at least one of $v \nleqslant w$ or $w \nleqslant v$ holds; then, for such pairs of vertices, at least one of $v \not\preceq w$ or $w \not\preceq v$ holds as well, as $\preceq$ is coarser than $\le$\,.
	Thus $\preceq$ is a partial order.
\end{proof}
We can use this to efficiently decide when a successor function $f$ is a flow function by reduction to the \emph{transitive closure} problem on directed graphs: specifically, by constructing $\preceq$ from the influence relation $\inflRel$ for $f$.
While the algorithms will be simpler to describe in terms of the binary relations $\inflRel$ and $\preceq$ themselves, we will describe the graph-theoretic presentation of this problem: this is the presentation used in the relevant literature, and the graph constructions we describe here will also prove convenient for analysis in the following sections.

\begin{definition}
	\label{def:inflDigraph}
	For a successor function $f$ on a geometry $(G,I,O)$, the \emph{influence digraph} is a directed graph with vertex-set $V(G)$, and with an arc $v \arc w$ between $v, w \in V(G)$ if and only if $v \inflRel w$.
\end{definition}
\begin{figure}[t]
	\begin{center}
		\includegraphics{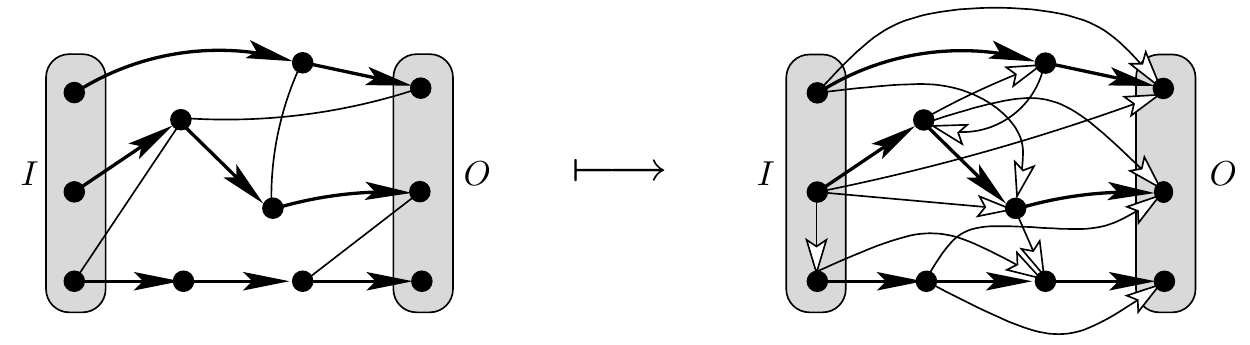}
	\end{center}
	\caption[Illustration of the influence digraph construction.]{\label{fig:inflDigraph}%
		Illustration of the influence digraph construction.
		On the left: a geometry $(G,I,O)$ with a successor function $f$, represented by the corresponding path cover $\cP_f$\,.
 		On the right: the corresponding influence digraph $\sI_f$\,.
 		Solid arrows represent arcs of the form $v \arc f(v)$\,, and hollow arrows represent arcs $v \arc w$ for $w \sim f(v)$ in the graph $G$\,.}
\end{figure}
An example of this construction is illustrated in Figure~\ref{fig:inflDigraph}.
We may then characterize the natural pre-order $\preceq$ of $f$ in terms of the \emph{transitive closure} of $\sI_f$\,:
\begin{definition}
	The \emph{transitive closure of a digraph $D$} is a directed graph $T(D)$ with $V(T(D)) = V(D)$, and such that $(v \arc w) \in A(T(D))$ if and only if there is a non-trivial directed path from $v$ to $w$ in the digraph $D$.
\end{definition}
If we let $\sT_f$ be the transitive closure of $\sI_f$, the natural pre-order of a successor function $f$ is then the relation $\preceq$ such that $v \preceq w$ if and only if either $v = w$ or $(v \arc w) \in A(\sT_f)$ for two vertices $v,w \in V(\sI_f)$.
This allows us to reduce the construction of $\preceq$ to the following problem applied to $\sI_f$:

\begin{problem}[Transitive Closure]
 	For a digraph $D$, construct its transitive closure $T(D)$.
\end{problem}
The transitive closure problem is solvable in polynomial time by the Floyd--Warshall algorithm~\cite{Floyd62,CLRS}.
We will present a faster algorithm using techniques presented in~\cite[Chapters 3\thru4]{Nuutila95},
\label{disc:stronglyConnected}
based on Tarjan's algorithm~\cite{Tarjan72, CLRS} for determining the \emph{strongly connected components} of a digraph (equivalence classes of vertices in a digraph which are mutually reachable by directed paths).
We will describe this algorithm in detail in Section~\ref{sec:flowAlgs}, where we will prove the following:
\begin{theorem}
	\label{thm:findPreorder}
	Let $f$ be a successor function of a geometry $(G,I,O)$, and let $n = \card{V(G)}$ and $k = \card{O}$.
	Then there is an algorithm constructing the natural pre-order $\preceq$ for $f$, or which determines that $\preceq$ is not a partial order, in time $O(k^2 n)$.
\end{theorem}

\subsection{Influencing walks and causal path covers}

While a detailed analysis of the algorithms described above will wait until Section~\ref{sec:flowAlgs}, it will be useful to consider the graph-theoretic structures which are involved in those algorithms.
These structures also present a means of proving uniqueness results which led to the first algorithms for determining whether a geometry has a flow, and extremal results which allow refinements in their running times.

In the previous section, we observed that not only does the geometry of Figure~\ref{fig:exampleNoFlow} have a cyclic underlying graph, but also a complete cycle of relationships between some of its vertices in any influencing pre-order $\le$ for a given successor function for that geometry.
These cycles of relationships will manifest themselves as directed cycles in the corresponding influence digraphs $\sI_f$, but can be traced also to circuit\footnote{%
	In this context, we are using the graph-theoretic term \emph{circuit}, which usually refers to closed walks of length $3$ or more.
}
in the original graph $G$, of the following sort:
\begin{definition}
	\label{def:viciousCircuit}
	Let $G$ be a graph, and $\cP$ a family of vertex-disjoint directed paths in $G$.
	A walk $W = u_0 u_1 \cdots u_\ell$ is an \emph{influencing walk} for $\cP$ if it is a concatenation of zero or more paths (\emph{segments} of the influencing walk) of the following two types:
	\begin{romanum}
	\item
		$v \arc w$, where this is an arc in some path of $\cP$;
	\item
		$v \arc z \arc w$, where $v \arc z$ is an arc in some path of $\cP$ and $w z \in E(G)$ is an edge not covered by $\cP$.
	\end{romanum}
	A \emph{vicious circuit} for $\cP$ is a closed influencing walk for $\cP$ with at least one segment.
\end{definition}
It is easy to show (by induction) that we may also characterize an influencing walk $W$, for a family of vertex-disjoint paths $\cP$, as one whose arcs 
\begin{inlinum}
\item
	never traverse an edge of $G$ in a direction opposite that of an arc of $\cP$, and

\item
	do not traverse two edges in a row which are not covered by $\cP$.
\end{inlinum}
Influencing walks were first identified as objects of interest by Broadbent and Kashefi~\cite{BK2007}, who examine their role in the depth complexity of unitaries in the one-way measurement model.

We will occasionally refer to the two types of segments as \emph{type \iti} and \emph{type \itii} respectively.
The motivation for the two types of segment is in the two ways in which distinct vertices $v, w \in V(G)$ may be related by the influence relation $\inflRel$ for a successor function:
\begin{lemma}
	\label{lemma:influencingNatlPreorder}
	Let $\cP_f$ be a path cover for $(G,I,O)$ with successor function $f$, and let $\inflRel$ and $\preceq$ be the influence relation and the natural pre-order of $f$ respectively.
	Then $v \preceq w$ if and only if there is an influencing walk $W = v \diwalk w$ for $\cP_f$ from $v$ to $w$, and in particular, $v \inflRel w$ if and only if there is an influencing walk $W$ for $\cP_f$ from $v$ to $w$ which consists of one segment.
\end{lemma}
\begin{proof}
	Suppose $W$ is an influencing walk for $\cP_f$ from $v$ to $w$.
	A segment of type \iti\ is a dipath $v \arc w$ between vertices $w = f(v)$, and a segment of type \itii\ is a dipath $v \arc z \arc w$ for vertices $v$, $z$, and $w$ such that $w \sim z$ and $z = f(v)$, and where $w \ne v$ (because the edge $v z$ is covered by the path cover $\cP_f$).
	In both cases, we have $v \inflRel w$\,.
	Similarly, if $v \inflRel w$, then there is either an influencing walk $W = v \arc w$ or an influencing walk $W' = v \arc z \arc w$ for $z = f(v)$.
	The Lemma then holds for $\inflRel$ and influencing walks of one segment.

	An influencing walk in general (of \emph{zero} or more segments) then corresponds to the reflexive and transitive closure of the influence relation, \ie\ to the natural pre-order $\preceq$ of $f$.
	In particular, an influencing walk $v \diwalk w$ of \emph{zero} segments is the trivial walk on $v = w$, in which case $v \preceq w$.
	Influencing walks $W$ of $n \ge 1$ segments from $v$ to $w$ we may decompose into an influencing walk $W'$ of $n-1$ segments from $v$ to some vertex $u$, and a single segment from $u$ to $w$.
	By induction, we have $v \preceq u \inflRel w$, so $v \preceq w$.
	For the converse, suppose $v \preceq w$ for some vertices $v,w \in V(G)$.
	By definition, there is then a sequence of vertices $\paren{u_j}_{j = 0}^\ell$ for some $\ell \in \N$, such that $v = u_0 \inflRel u_1 \inflRel \cdots \inflRel u_\ell = w$.
	Then, for each $0 \le j < \ell$\,, we have $u_j \ne u_{j+1}$, and either $u_{j+1} = f(u_j)$, or $u_{j+1} \sim f(u_j)$, and we may define an dipath $W_j = u_j \arc u_{j+1}$ or $W_j = u_j \arc f(u_j) \arc u_{j+1}$, respectively.
	In the latter case, if $u_{j+1} = f(f(u_j))$, the path $W_j$ is a concatenation of two influencing walk segments of type~\iti; otherwise, $W_j$ is itself a single influencing walk segment of either type~\iti\ or type~\itii.
	In any case, the concatenation $W = W_0 \cdots W_{\ell-1}$ of these paths is an influencing walk for $\cP_f$ from $v$ to $w$.
\end{proof}
Because of the correspondence between segments of an influencing walk and the influence relation $\inflRel$, a single segment influencing walk essentially corresponds to a single arc of the influence digraph $\sI_f$, and an influencing walk in general to a directed walk in $\sI_f$.

The close correspondence between influencing walks for $\cP_f$ and the natural pre-order for $f$ motivates the following definition:
\begin{definition}
	A \emph{causal path cover} $\cC$ for a geometry $(G,I,O)$ is a path cover which has no vicious circuits.
\end{definition}
Using this further definition, we may then easily characterize flows in terms of path covers:
\begin{theorem}
	\label{thm:graphth-charn}
	Let $f$ be a successor function of a geometry $(G,I,O)$ and let $\cP_f$ be the path cover induced by $f$.
	Then $f$ is a flow function if and only if $\cP_f$ is a causal path cover.
\end{theorem}
\begin{proof}
	Let $\preceq$ be the natural pre-order of $f$.
	By Lemma~\ref{lemma:influencingNatlPreorder}, we have $v \preceq w$ if and only if there is an influencing walk for $\cP_f$ from $v$ to $w$\,: then we have $v \preceq w$ and $w \preceq v$ for distinct $v, w \in V(G)$ if and only if there are influencing walks $W' = v \diwalk w$ and $W'' = w \diwalk v$ for $\cP_f$\,.
	If there are such walks, then $C = W' W''$ is a vicious circuit for $\cP_f$.
	Conversely, a vicious circuit $C = \sigma_1 \cdots \sigma_\ell$ for $\cP_f$ (concatenated from some non-zero number of segments $\sigma_j$) can be decomposed into an influencing walk $W' = \sigma_1 \cdots \sigma_{\ell - 1}$ between two distinct vertices $v$ and $w$, and another influencing walk $W'' = \sigma_\ell$ from $w$ to $v$.
	Then if there is such a circuit $C$, we have $v \preceq w$ and $w \preceq v$ for some two distinct vertices $v$ and $w$, so that $\preceq$ is not a partial order.
	Thus, $\preceq$ is a partial order if and only if $\cP_f$ lacks vicious circuits.
	By Lemma~\ref{lemma:causalOrderFlow}, $f$ is a flow function if and only if its natural pre-order $\preceq$ is a partial order; the theorem then holds.
\end{proof}
This completes our graph-theoretic characterization of flows.
The main utility of these constructions and characterizations is to reduce the properties of the natural pre-order $\preceq$ of a function $f$ to influencing walks for $\cP_f$\,, or to walks in the influence digraph $\sI_f$\,.
This change in emphasis will serve to clarify the graph-theoretic results which follow.



\subsection{Bounding the number of edges of a geometry with a flow}
\label{sec:extreme}

For a geometry $(G,I,O)$ with a flow-function $f$, let us call an edge $vw \in E(G)$ a \emph{flow edge} if either $w = f(v)$ or $v = f(w)$, \ie\ if $vw$ is an edge covered by the path-cover $\cP_f$\,; every other edge of $G$ is a \emph{non-flow} edge.
We have seen how influencing walks for path covers $\cP_f$ correspond to the natural pre-order $\preceq$ for $f$\,: and in this terminology, from the remarks made just after Definition~\ref{def:viciousCircuit}, an influencing walk $W$ is essentially a walk which follows the paths of $\cP_f$, but which occasionally jumps across paths by traversing a non-flow edge (under the constraint that it does not traverse two such edges in a row).
Thus, the more non-flow edges there are, the more freedom there is in how an influencing walk may be constructed.

It is reasonable to suppose that the more non-flow edges there are, the more likely there are to be influencing walks $W' = v \diwalk w$ and and $W'' = w \diwalk v$ for some distinct vertices $v, w \in V(G)$\,, producing a vicious circuit for $\cP_f$.
We might then ask if there is an upper bound on the number of non-flow edges that a geometry with a flow may have.
Note that, whether or not a geometry $(G,I,O)$ has a flow, a path cover $\cC$ for $(G,I,O)$ must have exactly $k = \card{O}$ paths, as all the paths of $\cC$ must terminate at an output vertex, and all output vertices terminate such a path.
These paths will collectively cover exactly $n - k$ edges of $G$, where $n = \card{V(G)}$, as there is a one-to-one correspondence between elements of $O\comp$ and the arcs of $\cC$ which leave those vertices; all other edges of $G$ are non-flow edges relative to this path cover $\cC$.
Then, we may describe this question without reference to any particular path cover in terms of the following:

\begin{definition}
	For $n \ge k \ge 1$, define $\Gamma(n,k)$ to be the maximum number of edges in a geometry $(G,I,O)$ which has a flow, subject to $\card{V(G)} = n$ and $\card{O} = k$.
\end{definition}

For the question of whether \emph{there exists} a geometry on $n$ vertices and with $m$ edges which admits a flow, we can consider any graph $G$ with $n$ vertices, under the constraint that it admits some family of vertex-disjoint (directed) paths $\cP = \ens{P_1,\ldots,P_k}$ covering the entire graph.
Without loss of generality, we may then let $O$ be the final points of those paths, and $I$ be (an arbitrary subset of) the initial points of those paths.
Because we may consider the outputs and inputs to be given by a family of paths in this way, we will often refer to the successor function of a family of paths $\cP$ which covers the vertices of a graph, whether or not it is explicitly a path cover for a geometry.

\begin{figure}[tb]
	\begin{center}
		\qquad\includegraphics[width=120mm]{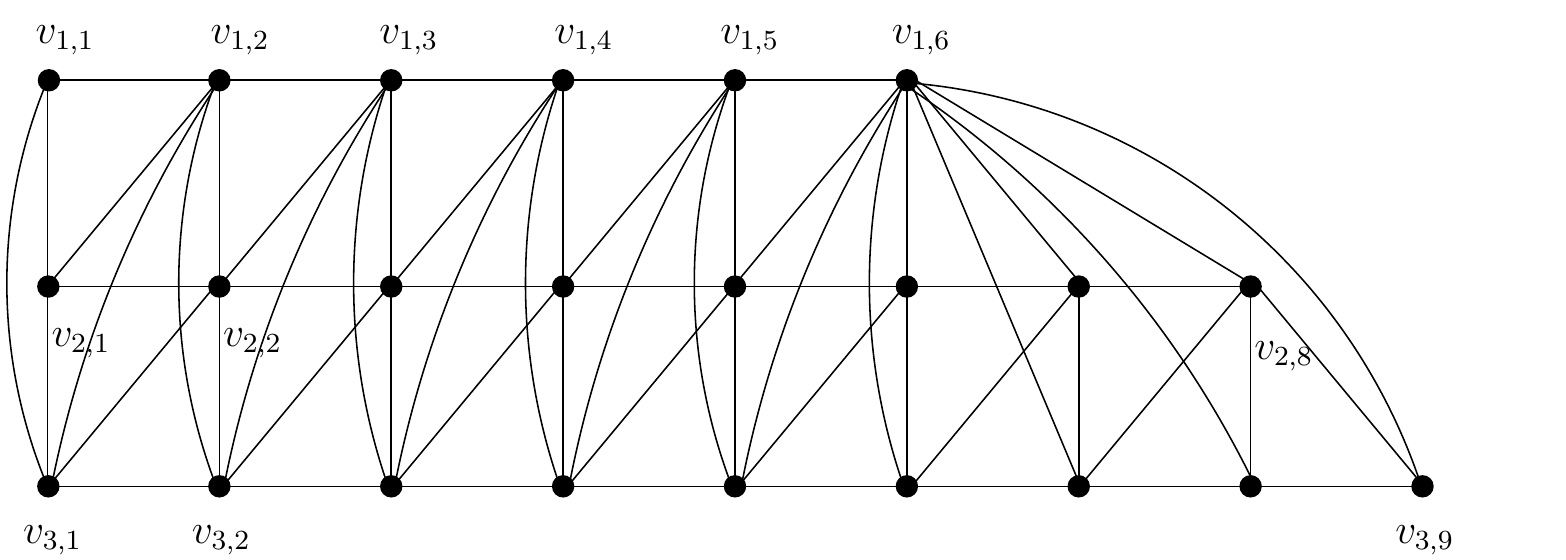}
	\end{center}
	\caption{\label{fig:extremalEg}%
		The graph $\sG(n_1,n_2,n_3)$ for $n_1=6$, $n_2=8$, $n_3=9$.}
\end{figure}

As we noted on page~\pageref{disc:stronglyConnected}, we may reduce this question to the question of whether the strongly-connected components of the influence digraph $\sI_f$ (Definition~\ref{def:inflDigraph}, page~\pageref{def:inflDigraph}) all consist of exactly one vertex.
This holds if and only if $\sI_f$ lacks directed circuits, \ie\ if $\sI_f$ is \emph{acyclic}.
We may then consider whether the successor function $f$ for $\cP$ has a causal order, and consider the problem in terms of adding non-flow edges to a graph $G$, which initially contains only the paths of $\cP$, under the constraint of keeping $\sI_f$ acyclic.
This leads to a simple upper bound on $\Gamma(n,k)$\,:

\begin{theorem}
	\label{thm:extremalUpperBound}
	$\Gamma(n,k) \le kn - \binom{k+1}{2}$ for all integers $n \ge k \ge 1$.
\end{theorem}
\begin{proof}
	We may base this proof on two simple observations about a path cover $\cP_f = \ens{P_1, \ldots, P_k}$ such that $\sI_f$ is acyclic.
	For each $1 \le j \le k$, let $n_j$ denote the number of vertices in the path $P_j$.
	\begin{observation}
		\label{obs:extr1}
		Consider a path $P_j = v_1 \arc v_2 \diwalk v_{n_j}$\,.
		If $v_a v_b \in E(G)$ for $b > a+1$, then $\sI_f$ will contain the directed cycle $v_a \arc v_{a+1} \diwalk v_{b-1} \arc v_a$\,.
		Then, if $\sI_f$ is acyclic, we have $v_a v_b \in E(G)$ for $a < b$ if and only if $b = a+1$, that is if $v_a v_b$ is a flow edge.
	\end{observation}
	\begin{observation}
		\label{obs:extr2}
		Consider any two distinct paths $P_h = v_1 \arc v_2 \diwalk v_{n_h}$ and $P_j = w_1 \arc w_2 \diwalk w_{n_j}$\,.
		If there are edges $v_a w_b\,,\, v_c w_d \in E(G)$ for some $a < c$ and $b > d$, then $\sI_f$ will contain the directed cycle $v_a \diwalk v_{c-1} \arc w_d \diwalk w_{b-1} \arc v_a$.
		Then, if $\sI_f$ is acyclic, we have non-flow edges $v_a w_b\,,\, v_c w_d \in E(G)$ only if the relation
		\begin{gather}
				a < c \;\implies\; b \le d
		\end{gather}
		holds. (The condition $b > d \;\implies\; a \ge c$ is equivalent, up to relabelling).
	\end{observation}
	The first observation implies that the only non-flow edges that $(G,I,O)$ may contain if it has a flow are edges between distinct paths $P_h$ and $P_j$\,; and the second imposes a useful constraint on the endpoints of pairs of non-flow edges.
	We may then assume that these constraints hold for all non-flow edges without loss of generality.

	We define a function $\lambda$ from the non-flow edges to $\N$ as follows.
	For a non-flow edge $v_a w_b$ between the $a\th$ vertex of a path $P_h = v_1 \diwalk v_{n_h}$ and the $b\th$ vertex of a path $P_j = w_1 \diwalk w_{n_j}$, we let
	\begin{gather}
		\lambda(v_a w_b) = a+b	\;.
	\end{gather}
	Suppose $\sI_f$ is acyclic.
	For two distinct edges $v_a w_b$ and $v_c w_d$ are non-flow edges between the two paths $P_h$ and $P_j$.
	Observation~\ref{obs:extr2} then implies that $\lambda(v_a w_b)$ and $\lambda(v_c w_d)$ differ:
	\begin{itemize}
	\item
		If $a < c$, we have $b \le d$, in which case $\lambda(v_a w_b) = a + b < c + d = \lambda(v_c w_d)$\,;

	\item
		If $c < a$, we have $d \le b$, in which case $\lambda(v_c w_d) = c + d < a + b = \lambda(v_a w_b)$\,.
	\end{itemize}
	Then, restricted to the non-flow edges between $P_h$ and $P_j$\,, $\lambda$ is injective.
	Because $2 \le \lambda(e) \le n_h + n_j$ for all non-flow edges $e$ between these paths, there are at most $n_h + n_j - 1$ non-flow edges between those two paths of $\cP$.

	Applying this to all pairs of paths $P_h$ and $P_j$, the number of non-flow edges in $G$ is then bounded above by
	\begin{align}
			\sum_{1\le h < j \le k} (n_h+n_j-1)
		\;\;=&\;\;
			\sfrac{1}{2}\left[ \sum_{h = 1}^k \sum_{j = 1}^k (n_h + n_j - 1) \;\;-\;\; \sum_{h = 1}^k (2n_h - 1) \right]
		\notag\intertext{}
			=&\;\;
			\sfrac{1}{2}\left[ \sum_{h = 1}^k (k n_h + n - k) \;\;-\;\; \sum_{h = 1}^k (2n_h - 1) \right]
 		\notag\\[1ex]=&\;\;
			\sfrac{1}{2}\left[2 k n - k^2 - 2 n + k \right]
		\notag\\[1ex]=&\;\;
 			kn - n - \tfrac{1}{2}(k^2 - k)	\;.
	\end{align}
	As the number of edges in the paths $P_j$ themselves is collectively $n - k$, the total number of edges $G$ may have if $\sI_f$ is acyclic is at most $kn - k - \tfrac{1}{2}(k^2 - k) \;=\; kn - \binom{k+1}{2}$\,.
\end{proof}
This result in itself will prove sufficient to strongly bound the running time of the algorithms presented in this chapter.
For completeness, however, we will consider a construction which saturates this upper bound.
Consider the following construction for any $n \ge k \ge 1$:

\begin{definition}
	\label{def:extremalGraph}
	Let $n_1,n_2,\ldots,n_k$ be an integer partition of $n$ such that $n_1 \le n_2 \le \cdots \le n_k$\,.
	For each $1 \le j \le k$\,, let $P_j \,=\, v_{j,1}\, v_{j,2}\,\cdots\,v_{j,n_j}$\,.
	Define $\sG(n_1,\ldots,n_k)$ to be the graph containing these paths, as well as the following non-flow edges for each $1 \le h < j \le k$\,:
	\begin{romanum}
	\item	
		If $n_h > 1$, then for each $1 \le a < n_h$\,, we include the edge $v_{h,a} v_{j,a}$;
	
	\item
		If $n_j > 1$, then for each $1 \le a < n_h$\,, we include the edge $v_{h,a+1} v_{j,a}$;

	\item
		For each $n_h \le a \le n_j$\,, we include the edge $v_{h,n_h} v_{j,a}$.
	\end{romanum}
	We will describe \iti\thru\itiii\ as the \emph{types} of non-flow edges in $\sG(n_1,\ldots,n_k)$\,.
	We also define $\sD(n_1, \ldots, n_k)$ to be the influence digraph $\sI_f$ for the successor function $f$ of the family of paths $\ens{P_1, \ldots, P_k}$.
\end{definition}
Figure~\ref{fig:extremalEg} illustrates an example of this graph construction for $\sG(6,8,9)$. 
(A diagram of the corresponding digraph $\sD(6,8,9)$ has a rather large number of arcs, and is not shown.)

For the sake of brevity, let $G = \sG(n_1, \ldots, n_k)$.
Define $\cP = \ens{P_1, \ldots, P_k}$, and let $f$ be the successor function for this family of paths.
Then each non-flow edge in $G$ induces up to two arcs in the associated digraph $\sI_f$\,.
For each $1 \le h < j \le k$\,, the arcs of $\sI_f$ induced by non-flow edges between the paths $P_h$ and $P_j$ are of six different types, labelled \rma\thru\rmf, which we group together by the type of the non-flow edge which induces them:
\begin{subequations}
\label{eqn:extremalDigraphArcTypes}
\begin{align}
	\iti \;\mapsto
	\;&\;
	\begin{brace}
		\;\rma\;\;	v_{h,a-1} \arc v_{j,a}	&\;\;\;\:\text{for $1 < a \le n_h$ (if $n_h > 1$), and}	\\
		\;\rmb\;\;	v_{j,a-1} \arc v_{h,a}	&\;\;\;\:\text{for $1 < a \le n_h$ (if $n_j > 1$);}
	\end{brace}
	\qquad\\[2ex]
	\itii	\;\mapsto
	\;&\;
	\begin{brace}
		\;\rmc\;\;	v_{h,a} \arc v_{j,a}		&\text{for $1 \le a < n_h - 1$ (if $n_h > 1$), and}	\\
		\;\rmd\;\;	v_{j,a-1} \arc v_{h,a+1}	&\text{for $1 < a \le n_h - 1$ (if $n_h > 1$);}
	\end{brace}
	\qquad\\[2ex]
	\itiii	\;\mapsto
	\;&\;
	\begin{brace}
		\;\rme\;\;	v_{h,n_h-1} \arc v_{j,a}		&\;\:\,\text{for $n_h \le a \le n_j$, and}	\\
		\;\rmf\;\;	v_{j,a-1} \arc v_{h,n_h}		&\;\:\,\text{for $\max \{ n_h \,, 2 \} \le a \le n_j$ (if $n_j > 1$).}
	\end{brace}
	\qquad
\end{align}
\end{subequations}
We may refer to \rma\thru\rmf\ as \emph{rules} for inclusion of arcs in $\sD(n_1, \ldots, n_k)$.
In addition to these arcs, $\sD(n_1, \ldots, n_k)$ also contains arcs $v_{j,a} \arc v_{j,a+1}$ arising from orienting the paths $P_j$ themselves, and arcs $v_{j,a} \arc v_{j,a+2}$ from traversing two path edges in a row (corresponding to the adjacency relation $v_{j, a+2} \sim f(v_{j,a})$, which yields $v_{j,a} \inflRel v_{j,a+2}$).

\begin{lemma}
	\label{lemma:acyclicConstr}
	For any $n \ge k \ge 1$ and any integer partition $n_1 \le \cdots \le n_k$ of $n$, the digraph $\sD(n_1, \ldots, n_k)$ is acyclic.
\end{lemma}
\begin{proof}
	Let $D = \sD(n_1, \ldots, n_k)$ for the sake of brevity.
	All of the arcs in $D$ produced by the rules \rma\thru\rme\ are either of the form $v_{h,a} \arc v_{j,b}$ with $a < b$ and no
	constraints on $h$ and $j$\,, or $v_{h,a} \arc v_{j,a}$ with $h < j$\,.
	In either case, if an arc $v_{h,a} \arc v_{j,b}$ is of one of the types \rma\thru\rme, we have $(a,h) < (b,j)$ in the lexicographic ordering on ordered pairs of integers.
	The same also holds for the arcs $v_{h,a} \arc v_{h,a+1}$ and $v_{h,a} \arc v_{h,a+2}$ yielded by the paths $P_h$.
	Then, if there are arcs in $D$ of the form $v_{j,b} \arc v_{h,a}$ where $(b,j) > (a,h)$\,, they must arise from the rule~\rmf, in which case $a = n_h$\,.

	Note that none of the rules \rma\thru\rmf\ produce arcs which leave the final vertex $v_{h,n_h}$ of any path $P_h$\,, so there are no non-trivial walks in $D$ which leave such a vertex.
	Then, it is easy then to show by induction that if there is a directed walk in $D$ between distinct vertices $v_{h,a}$ and $v_{j,b}$\,, either $(a,h) < (b,j)$ in the lexicographic order, or $b = n_j$\,.

	Let $v_{h,a}$ and $v_{j,b}$ be two vertices, with a directed walk $W$ from $v_{h,a}$ to $v_{j,b}$\,.
	Because of the existence of $W$, we know that $a \ne n_h$\,; then, there is a directed walk from $v_{j,b}$ to $v_{h,a}$ only if $(b,j) < (a,h)$\,.
	We would then have $b = n_j$\,, in which case there are no directed walks from $v_{j,b}$ to \emph{any} other vertices in $D$\,.
	So, for any two distinct vertices $v_{h,a}$ and $v_{j,b}$\,, there cannot both be a directed walk $v_{h,a} \diwalk v_{j,b}$ and also a directed walk $v_{j,b} \diwalk v_{h,a}$\,.
	Thus, $D$ is acyclic.
\end{proof}
As well as giving rise to an acyclic digraph $D(G,P_1,\ldots,P_k)$\,, we also have:
\begin{lemma}
	\label{lemma:constrMaximal}
	$\big\lvert E\big(\sG(n_1,\ldots,n_k)\big) \big\rvert \,=\, kn - \binom{k+1}{2}$\,,\, for any $n \ge k \ge 1$ and integer
	partition $n_1 \le \cdots \le n_k$ of $n$.
\end{lemma}
\begin{proof}
	Between any pair of paths $P_h$ and $P_j$ in $\sG(n_1,\ldots,n_k)$\,, there are $n_h - 1$ edges of type
	\iti, $n_h - 1$ edges of type \itii, and $n_j - n_h + 1$ edges of type \itiii.
	There are then $n_h + n_j - 1$ non-flow edges between $P_h$ and $P_j$\,.
	This saturates the upper bound for non-flow edges between pairs of paths in Theorem~\ref{thm:extremalUpperBound}: summed over all pairs of paths, and including the edges in the paths $P_h$, the total number of edges in $\sG(n_1,\ldots,n_k)$ is then $kn - \binom{k+1}{2}$.
\end{proof}
Lemmas~\ref{lemma:acyclicConstr} and~\ref{lemma:constrMaximal}, together with Theorem~\ref{thm:extremalUpperBound}, prove:
\begin{theorem}
	\label{thm:extremalBound}
	$\Gamma(n,k) = kn - \binom{k+1}{2}$ for all integers $n \ge k \ge 1$.
	That is, for every $n \ge k \ge 1$ and any geometry with $\card{V(G)} = n$ and $\card{O} = k$, the geometry $(G,I,O)$ has a flow only if $m \le nk - \binom{k+1}{2}$, where $m = \card{E(G)}$; and there exist geometries $(G,I,O)$, \eg\ with $G = \sG(n_1, \ldots, n_k)$, for any integer partition $n_1 \le \cdots \le n_k$ of $n$, which saturate this bound.
\end{theorem}
In later sections, we will only need the result of Theorem~\ref{thm:extremalUpperBound}: the matching lower bound only serves to prove that no improvement in the upper bound is possible.

\subsection{Uniqueness in the case \protect{$\card I = \card O$}}
\label{sec:unique}

Influencing walks for path-covers are closely related to \emph{walks which alternate with respect to $\cP$}, as defined by Diestel~\cite{Diestel} in his presentation of a clever proof of Menger's Theorem (a forerunner of the Min-Flow/Max-Cut theorem) by~\cite{BGH2001}.
\begin{definition}[\protect{\cite[page~64]{Diestel}}]
	Let $\cP = \ens{P_1, \ldots, P_k}$ be a collection of vertex-disjoint $I$\thru$O$ paths $P_j = v_{j,1} \cdots v_{j,n_j}$.
	A walk $W = u_0 u_1 \cdots u_\ell$ \emph{alternates with respect to $\cP$} if it starts at a vertex in $I$ not covered by $\cP$\,, and if the following three conditions hold for all $0 \le j \le \ell$\,:
	\begin{romanum}
	\item
		if $u_j u_{j+1}$ is covered by $\cP$, then $u_j = v_{h,a+1}$ and $u_{j+1} = v_{h,a}$ for some $1 \le h \le k$ and $1 \le a < n_h$\,;
	\item
		if $u_j = u_h$ for some $h \ne j$, then $u_j$ is covered by $\cP$\,;
	\item
		if $u_j$ is covered by $\cP$ for some $0 < j < \ell$, then at least one of $u_{j-1} u_j$ or $u_j u_{j+1}$ is covered by $\cP$.
	\end{romanum}
\end{definition}
Such walks share with influencing walks the property that, for any vertex covered by $\cP$, at least one of the two edges incident to the vertex are covered by $\cP$ as well.
By relaxing the requirement that $W$ must start at a vertex not covered by $\cP$, we may also consider ``alternating walks'' with respect to path covers: it is easy to see that closed ``alternating walks'' of this sort correspond to vicious circuits whose arcs have been reversed.

The role that alternating walks play in the proof of Menger's Theorem in~\cite{BGH2001}, which we present as Lemma~\ref{thm:BGH2001} on page~\pageref{thm:BGH2001}, provides some intuition of the importance of the link with alternating walks.
For completeness, the statement of the Theorem is as follows (see \eg~\cite{Diestel}):
\begin{theorem*}[Menger's Theorem]
	Let $G$ be a graph and $I, O \subset V(G)$.
	Then the minimum size of a set $S$ such that $V \setminus S$ is disconnected, with $I$ and $O$ in different components, is equal to the maxmum number of vertex-disjoint $I$\thru$O$ paths in $G$.
\end{theorem*}
Alternating walks arise as a sort of ``difference'' of distinct collections of vertex-disjoint $I$\thru$O$ paths.\footnote{%
The role of ``alternating walks'' and ``augmenting walks'' for vertex-disjoint $I$\thru$O$ paths, phrased in terms of differences of sets of paths, is similar to the analogous notions for \eg\ matchings in graphs, and have a basis (so to speak) in matroid theory.
The analogy here is used here only for illustrative purposes: a thorough discussion of augmenting walks in the context of matroid theory is beyond the scope of this thesis.}
Specifically, an $I$\thru$O$ walk $W$ which alternates with respect to a collection of vertex-disjoint paths $\cP$ can be used to obtain a new collection $\cP'$ with strictly more paths by using $W$ as an \emph{augmenting walk}, as in a network flow:\footnote{%
For a discussion of augmenting walks in the context of network flows, see \eg~\cite{CLRS}.}
interpreting $W$ as a ``difference'' of the two families $\cP'$ and $\cP$, we can obtain $\cP'$ by ``adding'' $W$ to $\cP$.
These addition and subtraction analogies can be made formal by considering the skew-symmetric directed adjacency matrices $A$ and $A'$ of the families of paths $\cP$ and $\cP'$ as directed subgraphs of $G$, given by
\begin{gather}
	\label{eqn:adjDigraphForSubtraction}
		A_{uv}	\;=\;	\begin{cases}
								\;\;1,		&	\text{if $u \arc v$ is an arc of $\cP$}	\\
									-1,		&	\text{if $v \arc u$ is an arc of $\cP$}	\\
								\;\;0,		&	\text{otherwise}
		      	     	\end{cases}
\end{gather}
(and similarly for $A'$ and $\cP'$), and considering the digraphs yielded by taking sums and differences of these adjacency matrices when those sums/differences are $\ens{-1,0,1}$-matrices.

Maximal collections of vertex-disjoint paths do not have such ``augmenting'' walks, and path covers for geometries $(G,I,O)$ are by necessity a maximal collection of such paths.
A difference of two path covers would then not contain any augmenting walks.
However, it is not difficult to see that the symmetric difference will contain walks which are very much like alternating walks, except that they must begin and end at the same point, which is necessarily covered by the path cover.
This intuition suggests that path covers which lack vicious circuits should be unique.
In fact, we may prove something somewhat stronger:

\begin{theorem}
	\label{thm:uniqueFamilyPaths}
	Let $(G,I,O)$ be a geometry such that $\card{I} = \card{O}$.
	If $(G,I,O)$ has a causal path cover $\cC$, then $\cC$ is also the only maximum-size collection of vertex-disjoint dipaths from $I$ to $O$.
\end{theorem}
\begin{proof}
	Suppose that $\cC$ is a path cover for $(G,I,O)$ with successor function $f: O\comp	\to I\comp$, and suppose there is a maximum-size collection $\cF$ of vertex-disjoint	$I$ -- $O$ dipaths which differs from $\cC$.
	Let $S \subset V(G)$ be the set of vertices not covered by $\cF$: because $\card{\cF} = \card{\cC} = \card{I} = \card{O}$, we have $S \inter I = \vide$ and $S \inter O = \vide$, in which case $\cF$ is a path cover for the geometry $(G \setminus S, I, O)$.
	Let $\tilde f$ be the successor function of $\cF$ on that geometry: then $\tilde f$ is injective.
	Because $\card{I} = \card{O}$, $\tilde f$ will also be a bijection from $V(G) \setminus (O \union S)$ to $V(G) \setminus (I \union S)$.
	Then, let $\tilde g$ be the inverse function of $\tilde f$.
	
	Because $\cC$ and $\cF$ differ, there must exist a vertex $v \in O\comp$ such that $v \arc f(v)$ is not an arc of any path of $\cF$. Note also that for $z \in \dom(f)$, $f(z) \notin \dom(f)$ holds only if $f(z) \in O \setminus I \subset \dom(\tilde g)$; that is, $f(z) \in \dom(f) \union \dom(\tilde g)$.
	Then, define a vertex sequence $\paren{u_j}_{j \in \N}$ in $G$ by setting $u_0 = v$, $u_1 = f(v)$, and 
	\begin{align}
			\label{eqn:inflWalkRecurrance}
			u_{j+1}
		\;=&\;
			\begin{cases}
				\; f(u_j)	\;,			&\!\!	\text{$u_j \in S$ or $u_j \ne f(u_{j-1})$}	\\[.5ex]
				\; \tilde{g}(u_j)	\;,	&\!\!	\text{$u_j \notin S$ and $u_j = f(u_{j-1})$}
			\end{cases}
	\end{align}
	for all $j \ge 1$.
	We then have $u_j \sim u_{j+1}$ for all $j \in \N$, so this defines a directed walk $u_0 \arc u_1 \arc \cdots$ in $G$.
	Figure~\ref{fig:exampleInflConstr} illustrates this construction.

\begin{figure}[t]
	\begin{center}
		\includegraphics{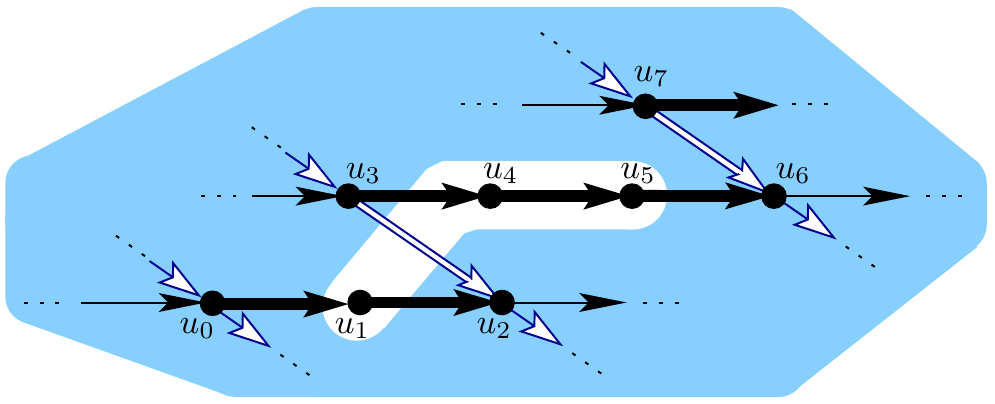}
	\end{center}
	\caption[Illustration of an unbounded influencing walk.]{\label{fig:exampleInflConstr}\colouronline
 		Illustration of an unbounded influencing walk, for a path cover $\cC$ (solid arrows), induced by the presence of another maximum-size collection $\cF$ of paths from $I$ to $O$ (hollow arrows).
		The shaded area represents the set of vertices covered by $\cF$, which in this illustration does not include all of the vertices.
		The arcs of $\cC$ and of $\cF$ which are thicker are those involved in the influencing walk.}
\end{figure}

	\begin{observation}
		\label{obs:consecutiveDistinct}
		Any three consecutive vertices $u_j, u_{j+1}, u_{j+2}$ are distinct.
		That $u_{j+1}$ differs from $u_j$ and $u_{j+2}$ follows from $u_j \sim u_{j+1}$ and $u_{j+1} \sim u_{j+2}$, as we do not permit $G$ to have self-loops.
		We have either $u_2 = f(u_1)$ or $u_2 = \tilde{g}(u_1)$, with the latter being equivalent to $\tilde{f}(u_2) = u_1$.
		In the former case, $u_0$ and $u_2$ lie along on a common path, so $u_0 \ne u_2$; and because $\tilde{f}(u_0) \ne f(u_0) = u_1$ by the choice of $u_0$, we have $u_2 \ne u_1$ in the latter case.
		We may induct similarly, if the proposition holds up to some particular $j \ge 0$:\
		\begin{itemize}
		\item
			Suppose $u_{j+3} = f(u_{j+2})$.
			If we also have $u_{j+2} = f(u_{j+1})$, then $u_{j+3}$ and $u_{j+1}$ lie along a common path of $\cC$.
			Otherwise, $u_{j+1} u_{j+2}$ is an edge not covered by $\cC$ while $u_{j+2} u_{j+3}$ is.
			In either case, $u_{j+3} \ne u_{j+1}$.

		\item
			Suppose $u_{j+3} = \tilde{g}(u_{j+2})$.
			Then $u_{j+2} \notin S$ and $u_{j+2} = f(u_{j+1})$; the latter of which implies that either $u_{j+1} \in S$ or $u_{j+1} = \tilde{g}(u_j)$.
			In the former case, $u_{j+1} \notin \img(\tilde{g})$, so that $u_{j+1} \ne u_{j+1}$\,; and if $u_j = \tilde{f}(u_{j+1})$, we have
			\begin{gather}
				u_{j+3} = \tilde{g}(u_{j+2}) \ne \tilde{g}(u_j) = u_{j+1}	\;.
			\end{gather}
		\end{itemize}
		In any case, $u_{j+1}$, $u_{j+2}$, and $u_{j+3}$ all differ, so the claim holds by induction.
		\end{observation}

	\begin{observation}
		For any $\ell \ge 0$, the walk $W_\ell = u_0 \diwalk u_\ell$ is an influencing walk.
		By definition, $u_0 \arc u_1$ is an arc in some path of $\cC$, so this is an influencing walk, as is the trivial walk on $u_0$.
		Otherwise, suppose that for some particular $\ell \ge 0$ we have both $W_\ell = u_0 \diwalk u_\ell$ and $W_{\ell+1} = u_0 \diwalk u_{\ell+1}$ are influencing walks.
		\begin{itemize}
		\item
			If $u_\ell \arc u_{\ell+1}$ is not an arc of $\cC$, then we have $u_{\ell+1} \ne f(u_\ell)$, in which case we have $u_{\ell + 2} = f(u_{\ell+1})$.
			Similarly, if $u_{\ell+1} \in S$, we also have $u_{\ell 2 1} = f(u_\ell)$.
			In either case, $u_{\ell+1} \arc u_{\ell+2}$ is a segment of type~\iti.

		\item
			If $u_\ell \arc u_{\ell+1}$ is an arc of $\cC$ and $u_{\ell+1} \notin S$, then we have $u_{\ell + 2} = \tilde{g}(u_{\ell+1})$.
			Because $u_{\ell + 2} \ne u_\ell$ by Observation~\ref{obs:consecutiveDistinct}, the arc $u_{\ell+1} u_{\ell+2}$ is not an arc covered by $\cC$, in which case $u_\ell \arc u_{\ell+1} \arc u_{\ell + 2}$ is a segment of type~\itii.
		\end{itemize}
		Then, $W_{\ell + 2} = u_0 \diwalk u_{\ell + 2}$ is either a concatenation of $W_{\ell + 1}$ with a segment of type~\iti, or a concatenation of $W_\ell$ with a segment of type~\itii, and is in either case also an influencing walk.
		The claim then holds by induction.	
	\end{observation}
	Because $G$ is a finite graph, the Pigeon Hole Principle implies that there must be integers	$\ell, \ell' \in \N$ with $\ell < \ell'$, $u_\ell = u_{\ell'}$, and $u_{\ell+1} = u_{\ell'+1}$.
	Because $W_{\ell'+1}$ is an influencing walk, either $u_\ell$ is at the beginning of a segment, or $u_{\ell + 1}$ is (or both).
	if $u_\ell$ is at the beginning of a segment, the closed walk $u_\ell \diwalk u_{\ell'}$ is a concatenation of segments from $W$, and so is a closed influencing walk for $\cC$.
	Otherwise, $u_{\ell+1} \diwalk u_{\ell'+1}$ is a concatenation of segments from $W$, and is in this case a closed influencing walk for $\cC$.
	In either case, there exists a vicious circuit for $\cC$, in which case $\cC$ is not a causal path cover.

	Thus, if $\cC$ is a causal path cover, there can be no such vertex sequence $\paren{u_j}_{j \in \N}$ as defined above, and so there can be no maximum family of vertex-disjoint $I$ -- $O$ paths $\cF$ which differs from $\cC$.
	$\cC$ is then the unique such family of paths.
\end{proof}

In the case where $\card{I} = \card{O}$, \ie\ for geometries of one-way patterns corresponding to unitary circuits without measurements or trace-out, we may then reduce the problem of finding a flow to that of finding an arbitrary collection $\cC$ of vertex-disjoint $I$\thru$O$ paths of maximum size.
If this collection of paths is not a path cover, we know that any path cover for $(G,I,O)$ cannot be causal, and so the geometry has no flow; otherwise, we may test the successor function $f$ of $\cC$ to see if it is a flow function, which by Lemma~\ref{lemma:causalOrderFlow} and Theorem~\ref{thm:findPreorder} (page~\pageref{lemma:causalOrderFlow} and following) can be determined in polynomial time.
This will allow us to prove the following in Section~\ref{sec:flowAlgs}:
\begin{theorem}
	Let $(G,I,O)$ be a geometry with $\card{I} = \card{O} = k$ and $\card{V(G)} = n$.
	Then there is an algorithm which determines that $(G,I,O)$ does not have a flow, or constructs a flow $(f, \preceq)$ for $(G,I,O)$ such that $\preceq$ is the coarsest causal order for $f$, in time $O(k^2 n)$.
\end{theorem}

\section{Flow-finding algorithms}
\label{sec:flowAlgs}

Throughout this section, we will suppose that we are given as input some particular geometry $(G,I,O)$, and define $n = \card{V(G)}$, $m = \card{E(G)}$, and $k = \card{O}$.
In some parts of our analysis, we will further suppose that $\card{I} = \card{O}$.

\paragraph{Partial function data types.}

A mutable variable for a partial function $f$ between two sets $A$ and $B$ (which we will denote $f: A  \partialto B$) will be represented as arrays $f: A \to B \union \ens{\nil}$, where $\nil$ is a reserved value not contained in any set of vertices or indices.
The domain of $f$ is then the set of $x \in A$ for which $f(x) \ne \nil$.
We will also adopt the convention that a partial function $f$ is injective when $\; f(x) = f(y) \ne \nil \;\iff\; x = y \;$ holds for all $x, y \in A$.

\paragraph{Set data types.}

For a mutable set variable $S$, we will consider to be implemented using a mixed array/linked list structure, where an array entry $S(v)$ either contains a value $\nil$ when $v \notin S$, or a node with links to at most two other nodes, representing arbitrarily designated ``previous'' and ``next'' elements of the set (using $\nil$ if one or both of these are not defined) when $v \in S$.
A specially designated master node is kept to point to the ``first'' element of the set, or $\nil$ if the set is empty: newly added elements of the set may be inserted to the beginning of this list of nodes.
This will allow for constant-time elementhood testing, element addition and removal, and comparison with the empty set, and iteration across the set can be done in $O(\card{S})$ steps; it also provides a total ordering of $S$ if required.

\subsection{Eliminating geometries with too many edges}
\label{sec:eliminatingAbundant}

As we proved in Theorem~\ref{thm:extremalUpperBound}, if $G$ has more than $kn - \binom{k+1}{2}$ edges, then $(G,I,O)$ cannot have a flow.
Then, as a preliminary step, we may reject any geometry which has more than this number of edges, so we may assume that $m \in O(kn)$for the purposes of run-time analysis.
(As we proved in Lemma~\ref{lemma:constrMaximal}, the bound of $m \le kn - \binom{k+1}{2}$ is tight for geometries with flows.)

\subsection{Finding a potential flow-function $f$ when $\card I = \card O$}

In the case where $\card I = \card O$, the uniqueness result of Theorem~\ref{thm:uniqueFamilyPaths} allows us to reduce finding a causal path cover to finding an arbitrary maximum-size vertex-disjoint family of $I$\thru$O$ paths.
This problem may be reduced to an instance of network flow, by finding a maximum integer flow in a digraph $N$ with unit edge capacities, constructed from $G$ so that edge-disjoint paths in $N$ correspond to vertex-disjoint paths in $G$~\cite{CH91}.
Here, we describe a different algorithm, based on the proof of the lemma in~\cite{BGH2001} (reproduced here with minor changes in order to facilitate description of the algorithm) used to prove Menger's Theorem :
\begin{lemma}[\protect{\cite[Theorem 3]{BGH2001}}]
	\label{thm:BGH2001}
	Let $G$ be a digraph, $I, O \subset V(G)$ such that $I$ cannot be separated from $O$ by a set of $\kappa$ or fewer vertices, for some $\kappa \in \N$.
	If $\cW \subset G$ is a subgraph consisting of $\kappa$ vertex-disjoint $I$\thru$O$ paths in $G$, there is a subgraph $\cU \subset G$ consisting of $\kappa+1$ disjoint $I$\thru$O$ paths, such that the terminal points of the paths of $\cU$ contain those of $\cW$.
\end{lemma}
\begin{proof}
	We will induct on supersets of $O$ contained in the vertex-set $V(G)$.
	If $O = V(G)$, then $I \subset O$, in which case $\cU = G[I]$ contains at least $\kappa+1$ (trivial) paths from $I$ to $O$.
	Otherwise, we suppose that the Lemma holds for all proper supersets $O' \prsupset O$.

	Let $R = O \inter V(\cW)$: as $\card R = \kappa$\,, removing $R$ from $G$ does not suffice to separate $I$ from $O$, in which case the graph $G \setminus R$ contains a path $P$ from $I$ to $O$.
	If $P$ is disjoint from $\cW$, we may let $\cU = \cW \union P$.
	Otherwise, let $x$ be the final vertex on $P$ which intersects $\cW$, and let $W$ be the path of $\cW$ containing $x$.

	Let $P_1$ be the subpath of $P$ from $x$ to $O$, $W'$ be the subpath of $W$ from $I$ to $x$, and $P_2$ be the subpath of $W$ from $x$ to $O$.
	We may augment the set $O$ by defining $O' = O \union V(P_1) \union V(P_2)$, and correspondingly reduce the path $W$ by defining $\cW' = (\cW \setminus W) \union W'$\,; then $\cW'$ consists of $\kappa$ paths from $I$ to $O'$.
	As $O \prsubset O' \subset V(G)$, by hypothesis there exists a subgraph $\cU' \subset G$ consisting of $\kappa + 1$ vertex-disjoint paths, such that the terminal points of $\cU'$ contain those of $\cW'$.

	Let $\omega \in O'$ be the terminal point in $\cU'$ which is not covered by $\cW'$, and let $z$ be the endpoint of $W$ in $O$.
	If $\omega$ is a vertex of $O \setminus \ens{z}$, then $P_2$ extends a path of $\cU'$ from $x$ to $z$; then $\cU = \cU' \union P_2$ covers $\omega$, as well as all of the vertices of $O$ covered by $\cW$.
	Otherwise, we have $\omega \in V(P_1 \union P_2)$ and $\omega \ne x$, and there is an $\omega$\thru$O$ subpath $Q \subset P_j$ (for one of $j = 1$ or $j = 2$).
	Either $P_1$ or $P_2$ will extend a path of $\cU'$ from $x$ to $O$, and $Q$ will extend a path of $\cU'$ from $\omega$ to $O$.
	Then, $\cU = \cU' \union P_{3-j} \union Q$ covers all of the vertices of $O$ covered by $\cW$, as well as the endpoint of $P$ in $O$.
\end{proof}

The following few sections will concern a translation of this proof to a depth-first search along alternating walks.

\subsubsection{Translation to depth-first search}

\begin{figure}[p]
	\vspace{-2.5ex}
	\begin{center}
	\includegraphics[height=0.82\textheight]{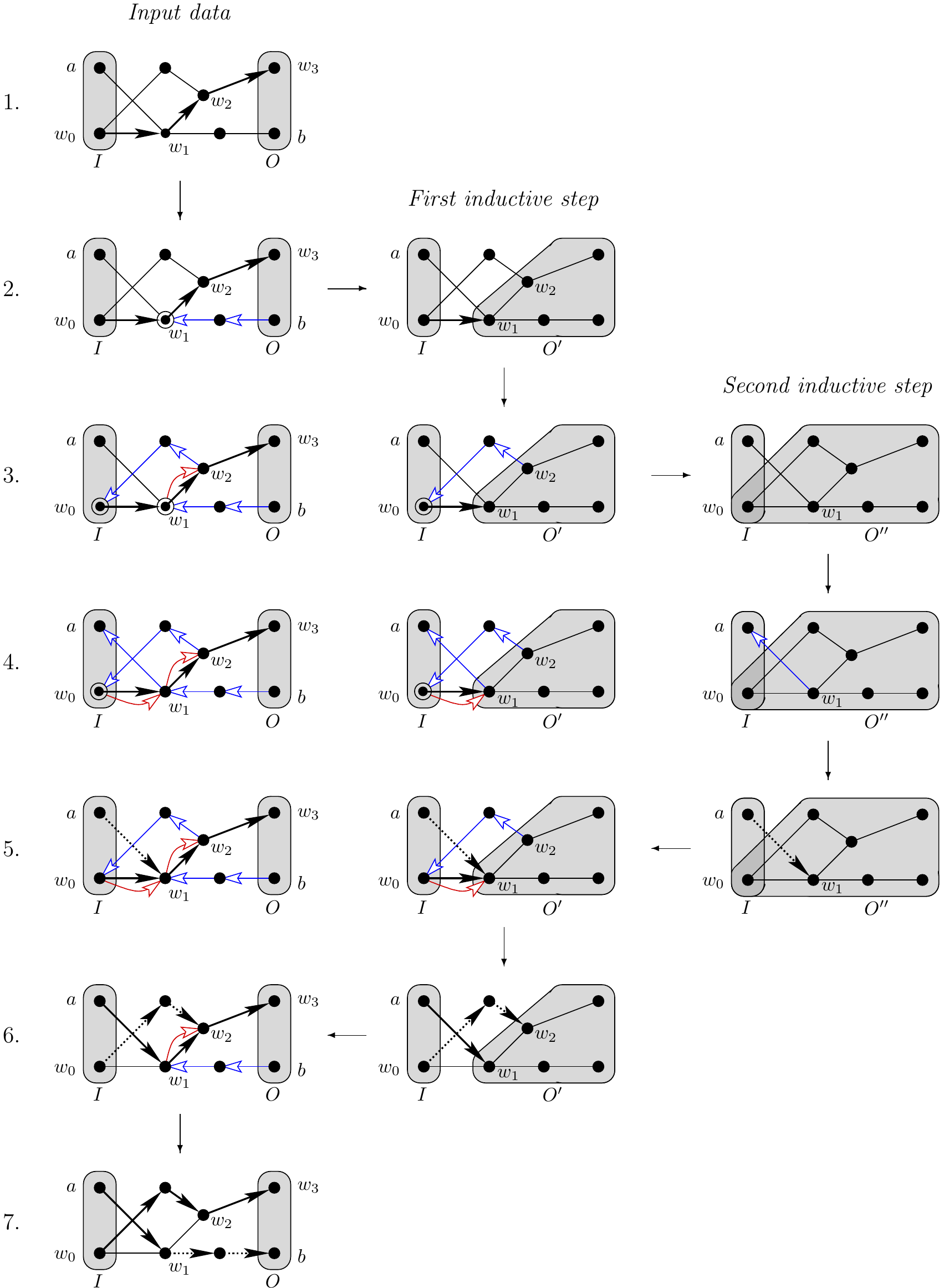}
	\end{center}
	\vspace{-2ex}
	\caption[Illustration of a depth-first search along alternating walks, based on the proof of Theorem~\ref{thm:BGH2001}.]{\label{fig:entryPoint}\colouronline
		Illustration of a depth-first search along alternating walks, based on the proof of Theorem~\ref{thm:BGH2001}.
		Each column corresponds to a successive reduction of the problem by augmenting the set $O$, as in the inductive step: arrows between the diagrams represent the problem reduction suggested by the proof.
		Steps~1\thru3 consist of a search for a path $P$, which stops when some $a \in I$ or a pre-existing path $W$ is encountered. 
		This search in the reduced problems corresponds to a search along alternating walks (indicated here with hollow arrows), illustrated in the left-most column.
		A solution for a reduced problem is extended in the parent problem, corresponding to backtracking along the walk: this amounts to ``subtracting'' the alternating walk from the existing family of paths, as described in the paragraph leading up to the equation~\eqref{eqn:adjDigraphForSubtraction}.
	}
\end{figure}

The connection between the induction proof above of Theorem~\ref{thm:BGH2001} and walks which alternate with respect to $\cW$ can be made as follows:
\begin{itemize}
\item
	If $P$ is disjoint from $\cW$, it (trivially) alternates with respect to $\cW$.

\item
	If $P$ intersects $\cW$ at a vertex not in $O$, but we have $\omega \in O \setminus \ens{z}$, this corresponds to a case where an $I$\thru$O$ path is found which is totally independent (\ie\ disjoint) from $P$, which may be found \eg~after a search from the terminal point of $P$ has been exhausted.
	Similarly, if $\omega \in V(P_1) \setminus \ens{x}$, this corresponds to a case where an $I$\thru$O$ path is found which intersects $P_1$ at some internal vertex, which may be found \eg~after a search through $x$ has been exhausted.

\item
	If $P_1$ intersects $\cW$ and $\omega \in V(P_2) \setminus \ens{x}$, this corresponds to a case where $\omega$ is an intermediary 
	point of an alternating walk which backtracks (by at least one edge) along $W$ from $\omega$ to $x$, and then progresses along $P_1$ to meet $O$.
	In this case, the induction hypothesis essentially reduces the problem to finding a vertex $\omega$ along $P_2$ such that there is an alternating walk with respect to $\cW'$ from $I$ to $\omega$.
\end{itemize}
In any case, the alternating walk may be used to construct the new family of paths $\cU$ incrementally, by including and excluding path segments as in the induction steps of the proof of Theorem~\ref{thm:BGH2001} above.
This process is illustrated in Figure~\ref{fig:entryPoint}.

The fact that this walk should be an alternating walk is not the main concern, but rather that $\cU$ can be obtained at all by searching along a walk with simple structure.
We will proceed by describing the reductions that will make it easy to prove the correctness of a depth-first search.

We may refine the proof of Theorem~\ref{thm:BGH2001} by extending the set $O \mapsto O'$ regardless of whether $P$ intersects $\cW$.
In particular: the question of whether there is an $I$\thru$O$ path $P = p_0 \cdots p_{\ell - 1} p_\ell$, which is either disjoint from $\cW$ or such that $p_j \notin V(\cW)$ for some range of consecutive vertices $r < j \le \ell$, can be reduced to whether there is a similar $I$\thru$O'$ path $P' = p_0 \cdots p_{\ell-1}$, for $O' = O \union \ens{p_{\ell-1}}$.
Similarly, the question of whether there is an $I$\thru$O$ path $P = p_0 \cdots x p_\ell$ such that $x \in V(\cW)$ can be reduced (as it is in the induction step of Theorem~\ref{thm:BGH2001}) to the question of whether there exists a path $P'$ from $I$ to some vertex of the set $O' = O \union \ens{x, x', x'', \ldots}$, where $x x' x'' \cdots$ is the tail of the path $W$ in $\cW$ which covers $x$; we also require that $P'$ does not end at $x$ itself, as this is covered by the truncated path $W' \subset W$ ending at $x$.
The depth-first search may then be described in terms of the valid ways in which we may perform an augmentation $O\supp{r} \mapsto O\supp{r+1}$, in order to reduce the search for a path $P\supp{r}$ from $I$ to $O\supp{r}$ to a search for a path $P\supp{r+1}$ from $I$ to $O\supp{r+1}$.
\label{discn:reductionRefinement}%
To re-iterate, the valid extensions would be:
\begin{enumerate}
\item
	For any $v \in O\supp{r}$, extend to $O\supp{r+1} = O\supp{r} \union \ens{w}$ for any $w \sim v$ such that $w \notin V(\cW\supp{r})$ and $w \notin O\supp{r}$.
	We then attempt to find $\cU\supp{r+1}$ which covers $w$, and the terminal points of $\cW\supp{r+1} = \cW\supp{r}$.

\item
	For any $v \in O\supp{r}$, extend to $O\supp{r+1} = O\supp{r} \union \ens{x, f(x), f^2(x), \ldots}$, for any $x \in V(\cW\supp{r}) \setminus O\supp{r}$, where $f$ is the ``successor'' function mapping each vertex covered by $\cW$ to the one which follows it (orienting the paths towards $O$).
	If $W$ is the path of $\cW\supp{r}$ on which $x$ lies, let $W$ be the segment of that path from $I$ to $x$; we then attempt to find $\cU\supp{r+1}$ which covers the terminal points of $\cW\supp{r+1} = \cW\supp{r} \symdiff \ens{W, W'}$, and one of the vertices $f^\ell(x)$ for $\ell \ge 1$.
\end{enumerate}
The path families $\cW\supp{r}$ we can keep track of by simply ``ignoring'' any arcs of $\cW$ contained within $O\supp{r}$, as such edges are guaranteed not to be in $\cW\supp{r}$ by construction.
The extended set $O\supp{r}$ will be the original set of output vertices $O$ together with the vertices traversed by the depth-first search to that point.


We will not keep track of of $O\supp{r}$ explicitly, but instead by ``marking'' most of the vertices traversed in the walk (which we may also use to avoid non-terminating recursive loops).
The exceptions, those vertices which are traversed but not marked, are what we will call \emph{entry points}.
Let us call a vertex $x$ an \emph{entry point} of the search into a path of $\cW$ if the edge $v x$ is not covered by $\cW$, where $v$ is the previous vertex in the search.
(Examples where a depth-first search encounters entry points are illustrated in Figure~\ref{fig:entryPoint}, where entry points are circled at each stage of the algorithm where they have been traversed only once: see \eg\ the vertex $w_1$ in the left column, in steps $2$\thru$3$.)
When such a vertex is discovered in the $r\th$ problem reduction for some $r \ge 0$, the set $O\supp{r}$ is augmented to a set $O\supp{r+1} = O\supp{r} \union \ens{x, f(x), \ldots}$; however, $x$ is then an element of $O\supp{r+1}$ covered by $W\supp{r+1}$ (specifically, by the path $W$ truncated after $x$), and is not a valid vertex from which to search for an augmented family of paths $\cU\supp{r+1}$.
Nevertheless, a later problem reduction may truncate the path $W'$ on which $x$ lies, in which case it becomes uncovered in that problem reduction, and a valid vertex from which to later perform a depth-first search --- in particular, it may be necessary to traverse $x$ a second time.
(Figure~\ref{fig:entryPoint} illustrates an example where visiting an entry point twice, $w_1$ in that case, is necessary to find the augmented family of paths.)
Because of this special behavior of entry points, we do not mark them when they are first traversed, in order to allow a second traversal if necessary.
However, we may still identify vertices of $O\supp{r}$ as vertices which are in $O$, vertices which are marked, or vertices covered by $\cW$ which immediately precede marked vertices (the latter case representing entry points).

\label{discn:depthFirst}
To implement a depth-first search, we will be interested in describing the problem reductions described above in terms of local transitions in the graph $G$.
We may define
\begin{subequations}
\begin{align}
		f(v)
	\;\;=&\;\;
		\begin{cases}
			\nil\,,	&	v \in O \text{~or $v$ not covered by $\cW$}	\\
			w\,,		&	v \arc w \text{~an arc of $\cF$}
		\end{cases}	\;,
	\\[1ex]
		g(w)
	\;\;=&\;\;
		\begin{cases}
			\nil\,,	&	w \in I \text{~or $v$ not covered by $\cW$}	\\
			v\,,		&	v \arc w \text{~an arc of $\cF$}
		\end{cases}.
\end{align}
\end{subequations}
In a slight abuse of the original terminology, we call $f$ the \emph{successor function} of $\cW$ (which will be a family of vertex-disjoint paths, albeit perhaps not a path cover); we similarly call $g$ the \emph{predecessor function} of $\cW$.
A transition which will be permitted for the depth-first search, in the $r\th$ problem reduction, will be one from a vertex $v \in O\supp{r}$ not covered by $\cW\supp{r}$ to a vertex $w$ satisfying one of the following conditions:
\begin{romanum}
\item
	$v \sim w$, $w$ is not covered by $\cW$, and $w \notin O\supp{r}$: this corresponds to the first reduction in the discussion above.
	The latter two conditions are equivalent to $w$ not covered by $\cW\supp{r}$, and $w$ not a marked vertex.

\item
	$w = f(x) \ne \nil$ for some $x \sim v$ on some path of $\cW$ with $x \notin O\supp{r}$, and where $v x$ is not covered by a path of $\cW\supp{r}$: this corresponds to the first two vertices in the second reduction in the discussion above.
	The conditions that $x$ be covered by $\cW\supp{r}$ and $x \notin O\supp{r}$ is equivalent to $w = f(x)$ being well defined (as we already require), and neither $x$ nor $w$ being marked.

\item
	$w = f(v) \ne \nil$, and $w$ on some path of $\cW\supp{r}$ --- in particular, where either $w \notin O\supp{r}$, or $w \in O\supp{r}$ is the vertex at the end of some path $W \in \cW\supp{r}$: this corresponds to all but the first two vertices in the second reduction of the discussion above.
	In the latter case, $w$ was an entry point of some ancestral problem reduction, in which case $w$ is unmarked, but $f(w)$ is well defined and marked.
	Then, this case is equivalent to $w = f(v) \ne \nil$ and $w$ not a marked vertex.

	In order to perform the second reduction properly, \ie\ all at once as described, this transition must be performed preferentially when it is valid; the reduction $O\supp{r} \mapsto* O\supp{r+1}$ is only complete when the endpoint of the path $W \in \cW\supp{r}$ has been reached, and other transitions explored from that vertex.
\end{romanum}
We mark a vertex when it plays the role of $w$ in such a local transition (in particular, leaving any entry point $x$ initially unmarked).
In each case, we need only consider the neighbors $z \sim v$ which have not yet been marked, such that either $z \notin O$ or $z = f(v)$; in the case where $z$ is an entry point, we further require that $w = f(z)$ has not been marked.
Then, in the depth-first search, it will suffice at each step to explore the unmarked vertices $z$ such that either $z = f(v) \ne \nil$, or such that $z \sim v$ and $z \notin O$.

Algorithms~\ref{alg:iterFindPaths} and~\ref{alg:augmentPaths} together describe an iterative procedure to build a maximum-size family of vertex-disjoint paths, based on this reduction to a depth-first search; where sets of disjoint paths $\cW$ are represented by successor/predecessor functions and set consisting of those vertices not covered by $\cW$.
In the following pages, we use the reduction just described to prove their correctness.

\subsubsection{Overview of Algorithm~\ref{alg:iterFindPaths}}

\begin{algorithm}[t]

	\PROCEDURE \ObtainMaxFamilyPaths(G,I,O):

	\Input{%
		$G$, a graph with subsets $I, O$.
	}

	\Output{%
		successor/predecessor functions $f,g: V(G) \partialto V(G)$ for a maximum-size family $\cP$ of vertex-disjoint $I$\thru$O$ paths
	}

	\BEGIN
		\LET $f,g: V(G) \partialto V(G)$\;
		\LET $\marked: V(G) \to \N$\;
		\LET $\iter \gets 0$\;
		\FORALL $v \in V(G)$ \DO																\nllabel{line:initPathsLoop}
			$\marked(v) \gets 0$\;
			$f(v) \gets \nil$;~
			$g(v) \gets \nil$\;
		\ENDFOR																						\nllabel{line:initPathsLoopEnd}
		\REPEAT																						\nllabel{line:buildPathsLoop}
			$\INCR \iter$\;
			\LET $\foundpath \gets \false$\;
			\FORALL $v \in O$ \SUCHTHAT $v \notin I$ \ANDALSO $g(v) = \nil$ \DO	\nllabel{line:iterSearchLoop}
				$\AlternPathSearch(v)$\;
				\oneline
				\IF $\foundpath$ \THEN \ESCAPE\;
			\ENDFOR																					\nllabel{line:iterSearchLoopEnd}
		\UNTIL {$\foundpath = \false$}														\nllabel{line:buildPathsLoopEnd}
		\RETURN $(f,g)$
	\END

	\caption[An iterative algorithm to obtain a maximum-size family of vertex-disjoint $I$\thru$O$ paths.]{\label{alg:iterFindPaths}%
		An iterative algorithm to obtain a maximum-size family of vertex-disjoint $I$\thru$O$ paths.
		The subroutine $\AlternPathSearch$ is described in Algorithm~\ref{alg:augmentPaths} on page~\pageref{alg:augmentPaths}.}
\end{algorithm}

The procedure $\ObtainMaxFamilyPaths$ described in Algorithm~\ref{alg:iterFindPaths} above uses a subroutine $\AlternPathSearch$ (described in Algorithm~\ref{alg:augmentPaths}, on page~\pageref{alg:augmentPaths}) to obtain larger and larger families of vertex disjoint paths, until a maximum size family is found.
Before turning to the description of the more detailed subroutine $\AlternPathSearch$, we will describe Algorithm~\ref{alg:iterFindPaths} as a whole.

As noted before, we will represent a family of vertex-disjoint paths $\cW$ using partial functions $f$ and $g$, which will store the successor and predecessor functions of $\cW$.
In the loop on lines~\ref{line:initPathsLoop}\thru\ref{line:initPathsLoopEnd}, the partial functions are initialized to represent a family of paths consisting only of trivial paths $v \in I \inter O$ (without loss of generality, we may assume that $\cW$ contains the trivial paths on whatever elements there may be in $I \inter O$): we will consider a vertex $v$ to be covered by $\cW$ if and only if at least one of $v \in I \inter O$, $f(v) \ne \nil$, or $g(v) \ne \nil$ holds.
Also defined and initialized are an array $\marked$ and a counter $\iter$: as we construct larger families of paths, $\iter$ will represent the iteration of the construction, and $\marked$ will keep track of the most recent iteration that each vertex was marked in the depth-first search along alternating walks.
These variables will all be used in the subroutine $\AlternPathSearch$ as global variables, all of which (except for $\iter$) may be modified.

The iterative component of $\ObtainMaxFamilyPaths$ increments $\iter$, and defines a flag $\foundpath$ which will indicate whether the pre-existing family of paths $\cW$ represented by $(f,g)$ has been successfully augmented; this is also a global parameter which may be modified by $\AlternPathSearch$.
The loop on lines~\ref{line:iterSearchLoop}\thru\ref{line:iterSearchLoopEnd} iterates through all of the elements of $O$ which are not covered by $\cW$.
We will maintain the constraint that $f(v) = \nil$ for all $v \in O$: then, $v \in O$ is covered by $\cW$ if and only if $g(v) \ne \nil$ or $g(v) \in I$.

We will later show that so long as $\foundpath = \false$, the elements of $O$ which have been marked are those for which $\AlternPathSearch$ has been invoked, but has failed to find a path $P$ similar to that in the proof of Theorem~\ref{thm:BGH2001}.
We will also show that, once an augmenting walk is found, $\AlternPathSearch$ updates $(f,g)$ to represent a collection of vertex-disjoint paths $\cU$ which covers the terminal points of $\cW$ and one additional vertex in $O$, and that $\foundpath$ is set to $\true$\,; we may then progress to the next iteration of the construction.
If $\foundpath$ retains the value $\false$, the variables $(f,g)$ will remain unchanged since the last iteration.
Finally, if all uncovered elements of $O$ are visited in an iteration without the value of $\foundpath$ changing, the pre-existing path family $\cW$ is of maximum size, and the variables $(f,g)$ representing it are returned.

Having described the data provided to, and the requirements made of, the subroutine $\AlternPathSearch$, we will turn to the description of it in Algorithm~\ref{alg:augmentPaths}.

\subsubsection{Analysis of Algorithm~\ref{alg:augmentPaths}}

\begin{algorithm}[p]

	\SUBROUTINE \AlternPathSearch(v):

	\Data {%
		$G$, a graph with subsets $I, O$;
		\\
		$f$ and $g$, successor/predecessor \fns\ for a family $\cW$ of disjoint $I$\thru$O$ paths;
		\\
		$\iter$, the current iteration number of path-family augmentations;
		\\
		$\foundpath$, a flag for indicating if an augmenting path was found.
	}

	\BEGIN
		$\marked(v) \gets \iter$\;

		\medskip
		\IF $v \in I$ \THEN {\nllabel{line:trivialPath}%
			$\foundpath \gets \true$\;
			\RETURN}{}
		\algline

		\medskip
		\LET $w \gets f(v)$\;																		\nllabel{line:testPathGlob-1}
		\IF $w \ne \nil$ \ANDALSO $\marked(w) < \iter$ \THEN {							\nllabel{line:testPathGlob-2}
				$\AlternPathSearch(w)$\;
				\IF \foundpath \THEN {\nllabel{line:testRemoveVertexPath}
					$f(v) \gets \nil$;~																
					$g(v) \gets \nil$\;																\nllabel{line:removeVertexPath}
					\RETURN
			}{}
		}

		\medskip

		\FOREACH $w \sim v$ \SUCHTHAT $w \notin O$ \ANDALSO $\marked(w) < \iter$ \DO						\nllabel{line:searchLoop}
			\algline
			\IF $f(w) = \nil$ \THEN {\nllabel{line:simpleRedux}%
				$\AlternPathSearch(w)$\;
				\IF \foundpath \THEN {%
					$f(w) \gets v$;~
					$g(v) \gets w$\;
					\RETURN
				}{}
			}
			\ELSE {\nllabel{line:entryPoint}%
				\algline
				\LET $x \gets w$;~
				$w \gets f(x)$\;
				\IF $\marked(w) < \iter$ \THEN {%
					$\AlternPathSearch(w)$\;
					\IF \foundpath \THEN {%
						$f(x) \gets v$;~															
						$g(v) \gets x$\;
						\RETURN
					}{}
				}{}
			}
		\ENDFOR																							\nllabel{line:searchLoopEnd}

	\medskip
	\RETURN
	\END

	\caption[A depth-first search subroutine to find an alternating walk from $I$ to $O$.]{\label{alg:augmentPaths}
		A depth-first search subroutine to find an alternating walk (with respect to a family of disjoint paths $\cW$) from $I$ to $O$, traversed in reverse.}
\end{algorithm}

The subroutine $\AlternPathSearch$ is described in Algorithm~\ref{alg:augmentPaths} (page~\pageref{alg:augmentPaths}).
%
%
%
We will analyze it in terms of the problem reductions which we derived on page~\pageref{discn:depthFirst} from the proof of Theorem~\ref{thm:BGH2001}.
In particular, we will consider an invocation of $\AlternPathSearch$ on a vertex $v$, representing the $r\th$ problem reduction (or part of it) for some $r \ge 0$.
(Note that even if $v \in O$ is not covered by $\cW$, we may have $r > 0$, if $v$ is not the first such vertex for which $\AlternPathSearch$ has been invoked in a given iteration.)

Recall that a vertex $z$ is marked iff $\marked(z) = \iter$: we will assume as in the analysis on page~\pageref{discn:depthFirst} that $O\supp{r}$ consists of vertices $z$ which are marked, vertices $x$ for which $f(x)$ is non-$\nil$ and marked, and the elements of $O$.
A truncated family of paths $\cW\supp{r}$ then corresponds to those vertices covered by $\cW$ which are not marked (and the edges covered by $\cW$ incident to those vertices).
Equivalently, we will say that the variables $(f,g)$ represent a family of vertex disjoint paths $\cW\supp{r}$ from $I$ to $O\supp{r}$ if the following hold:
\begin{romanum}
\item
	the vertices covered by $\cW\supp{r}$ are those which satisfy at least one of
	$v \in I \inter O\supp{r}$\,,
	$\big[ g(v) \ne \nil$ and $g(v) \notin O\supp{r} \big]$\,, or 
	$\big[ f(v) \ne \nil$ and $v \notin O\supp{r} \big]$\,;
	\vspace{0.5ex}

\item
	$\cW$ covers every edge $vw$ for which $v \notin O\supp{r}$, $w = f(v)$, and $v = g(w)$.
\end{romanum}
(The same variables $(f,g)$ may represent multiple families of paths from $I$ to $O\supp{r}$, corresponding to different levels $r$ of problem reduction.)

The first action performed by $\AlternPathSearch$ is to set $\marked(v) \gets \iter$, corresponding to the inclusion of $v$ into $O\supp{r}$ (and if $g(v) \ne \nil$, the inclusion of that vertex into $O\supp{r}$ as well).
If it happens that $v \in I$, then the trivial path on $v$ is a path from $I$ to $O\supp{r}$ which is disjoint from the paths of $\cW\supp{r}$\,; we may then define $\cU\supp{r} = \cW\supp{r} \union \ens{v}$, which covers $v$ and the terminal points of $\cW\supp{r}$, and which will already be represented by $(f,g)$.
We then set $\foundpath \gets \true$, and return to the previous recursive call.

If $v \notin I$, we then test on lines~\ref{line:testPathGlob-1}\thru\ref{line:testPathGlob-2} whether if $v$ is on a path and has an unmarked successor $w = f(v)$.
If so, we are in the midst of a problem reduction of the form $O\supp{r-1} \mapsto* O\supp{r-1} \union \ens{x, f(x), f^2(x), \ldots}$ for some preceding vertex $x$ on the same path as $v$.
To complete the reduction, we must accumulate any unmarked vertices $w, f(w), \ldots$ that there may be, and perform the depth-first search through them as we mark them for inclusion.
To do this, we recur the depth-first search at $w = f(v)$.

We will consider the effects of the conditional block starting on line~\ref{line:testRemoveVertexPath} further on: in any case, we regard all the operations from line~\ref{line:testRemoveVertexPath} onwards as being part of an $r\th$-level problem reduction for some $r \ge 0$, and the state of $\marked$ at this point to define $O\supp{r}$, any further changes made to $\marked$ by further recursive calls notwithstanding.

We require that at line~\ref{line:testRemoveVertexPath}, the variables $(f,g)$ represent a family of vertex disjoint paths $\cW\supp{r}$, and that $v \in O\supp{r}$ is not covered by $\cW\supp{r}$.
(In particular, this will be true when $v \in O$ is uncovered by $\cW$ and is the only vertex in $V(G)$ which is marked, \ie\ for $r = 0$).
We will then prove by induction that, after an invocation of $\AlternPathSearch(v)$ for $v$ not covered by $\cW\supp{r}$, the following hold if $\foundpath$ evaluates to $\true$:
\begin{romanum}
\item
	The variables $(f,g)$ represent a family of vertex-disjoint paths $\cU\supp{r}$ from $I$ to $O\supp{r}$, which covers the terminal points of $\cW\supp{r}$ and one other vertex;

\item
	If $v$ is not covered by $\cW$, then the additional vertex covered by $\cU\supp{r}$ is $v$, and $f(v)$ will remain unchanged;

\item
	The variables $f$ and $g$ may only change for vertices $z \in V(G) \setminus O\supp{r-1}$ (or for $z \in V(G)$ when $r=0$); 
	and they will remain unchanged for any vertex for which a recursive call $\AlternPathSearch(z)$ returns with $\foundpath = \false$.
\end{romanum}
Note that if $(f,g)$ represents a family of vertex disjoint paths $\cW\supp{r}$ which does not cover $v \in I$ at line~\ref{line:trivialPath}, the above conditions are satisfied; this forms the base case of the induction.

As we noted on page~\pageref{discn:depthFirst}, for variables $(f,g)$ representing a family of paths $\cW\supp{r}$ from $I$ to $O\supp{r}$, and for a vertex $v \in O\supp{r}$ which is not covered by $\cW\supp{r}$, the possible problem reductions can be obtained by considering the vertices which are iterated through on line~\ref{line:searchLoop} (of which there are at most $\deg(v)$ in number, and which can easily be iterated over provided an adjacency list representation of $G$).
We will now consider the reductions performed in the loop on lines~\ref{line:searchLoop}\thru\ref{line:searchLoopEnd}, together with the conditional block at line~\ref{line:testRemoveVertexPath}:
\begin{itemize}
\item
	For a vertex $w$ for which $f(w) = \nil$ (on line~\ref{line:simpleRedux}): we may attempt a problem reduction by defining $O\supp{r+1} = O\supp{r} \union \ens{w}$: to do this, we recur the depth-first search at $w$.
	If $\foundpath = \true$ after this reduction attempt, $(f,g)$ represents a new family of paths $\cU\supp{r+1}$ from $I$ to $O\supp{r+1}$ covering $w$ and the terminal points of $\cW\supp{r+1}$.
	Let $W' \in \cU\supp{r+1}$ be the path covering $w$\,: we may obtain a path $W$ covering $v$ by extending $W'$ by the edge $wv$.
	As $v$ is not covered by $\cW\supp{r+1}$, it is not covered by $\cU\supp{r+1}$, and so $\cU\supp{r} = \cU\supp{r+1} \symdiff \ens{W', W}$ is a vertex-disjoint family of paths covering $v$ and the terminal points of $\cW\supp{r}$.
	To represent $\cU\supp{r}$, we set $f(w) \gets v$ and $g(v) \gets w$.
	The only vertices for which $(f,g)$ are changed are then $v$ and those changed by the recursive call, which consist of $w$ and
	some vertices $z \in V(G) \setminus O\supp{r+1} \subset V(G) \setminus O\supp{r-1}$.

\item
	For a vertex $w = f(x)$ (on line~\ref{line:testRemoveVertexPath}): as noted before, if $w$ is well-defined, then we must perform a recursive call on $w$ to complete a reduction of the form $O\supp{r+1} = O\supp{r} \union \ens{x,f(x),\ldots}$ for some $x$ preceding $v$ and $w$ on a common path $P \in \cW\supp{r}$.
	In particular, such an $x$ is by definition an entry point, and as we will see in the next case, $\AlternPathSearch$ is not invoked on vertices which occur as entry points; then $v$ is distinct from $x$, and is not the end point of any path of $\cW\supp{r+1}$.

	If $\foundpath = \true$ after this reduction attempt, $(f,g)$ represents a new family of paths $\cU\supp{r+1}$ from $I$ to $O\supp{r+1}$ covering the terminal points of $\cW\supp{r+1}$, and one more vertex $\omega$ on the segment of $P$ starting at $w$ (in particular, we have $\omega \ne v$).
	Let $W' \in \cU\supp{r+1}$ be the path ending at $\omega$, and let $P'$, $P''$, and $P'''$ be the segment of $W$ ending at $x$, the segment of $W$ strictly bounded between $x$ and $\omega$ (and not including those endpoints\footnote{%
	Note that the path $P''$ may be an ``empty path'', consisting of zero vertices.}),
	and the segment of $W$ starting at $\omega$ respectively.
	By definition, $\omega$ is the first vertex from the end of $W$ for which the recursive call to $\AlternPathSearch$ returned with $\foundpath = \true$\,; in particular, for every vertex $z \in V(P''' \setminus \omega)$, the recursive call returned with $\foundpath = \false$.
	Then, the status of such vertices $z$ remain unchanged with respect to $(f,g,S)$\,: for each such vertex, we have $f(g(z)) = z$, and $g(f(z)) = z$ if $f(z) \ne \nil$.
	As well, by the analysis for the preceding case, the value of $g(\omega)$ will have changed to the vertex $\tilde{w}$ which precedes $\omega$ on $W'$, and $f(\omega)$ will remain unchanged.

	The reduction described in the proof of Theorem~\ref{thm:BGH2001} involves constructing a family of paths $\cU\supp{r}$ by extending the path $W'$ from $I$ to $O\supp{r+1}$ to a path $W = W' P'''$ from $I$ to $O\supp{r}$, and extending the path $P'$ from $I$ to $O\supp{r+1}$ by a single edge (as $x$ must in this case be adjacent to a vertex $\tilde{v} \in O\supp{r}$ which is not covered by $\cW\supp{r}$, from which $x$ was reached in the depth-first search).
	In particular, the vertices of $P''$ are not covered by $\cU\supp{r+1}$, and are not elements of $O\supp{r}$.
	In order to represent $\cU\supp{r}$, it is then necessary to remove the vertices $z \in V(P'')$ from the domains of $f$ and $g$.
	Therefore, we set $f(v) \gets \nil$ and $g(v) \gets \nil$\,; by induction on the length of $P''$, this will then be performed for all vertices in $V(P'')$.
	The only vertices for which $(f,g)$ have changed are then $\omega$, the vertices of $P''$ from $v$ onwards, and some vertices $z \in V(G) \setminus O\supp{r} \subset V(G) \setminus O\supp{r-1}$.

\item
	For a vertex $w$ for which $f(x) \ne \nil$ (on line~\ref{line:entryPoint}): because of the recursive call in the conditional block on line~\ref{line:testRemoveVertexPath}, the vertex $f(v)$ has already been marked, if it is defined; because $w$ is unmarked, we know that $w \ne f(v)$.
	Then either $w$ is an entry point, or $w = g(v)$.
	We relabel the possible entry point $w$ as $x$, overwrite $w \gets f(x)$, and test whether $w = f(x)$ has been marked.
	If not, then we have $x \ne g(v)$, as $f(g(v)) = v$ is a marked vertex; in this case, $x$ is indeed an entry point.
	We may then attempt a problem reduction by defining $O\supp{r+1} = O\supp{r} \union \ens{x, f(x), f^2(x), \ldots}$: to do this, we recur the depth-first search at $w$.
	(Marking $w$ will implicitly include both $x,w \in O\supp{r+1}$\,; the test at line~\ref{line:testRemoveVertexPath} ensures that all further vertices along the same path, up to its endpoint in $O\supp{r}$, will be immediately included as well.)

	If $\foundpath = \true$ after this reduction attempt, $(f,g)$ represents a new family of paths $\cU\supp{r+1}$ from $I$ to $O\supp{r+1}$ covering the endpoints of $\cW\supp{r+1}$, which include $x$, and one more vertex $f^\ell(x)$ for some $\ell \ge 1$.
	By the analysis of the preceding case, the variables $(f,g)$ also represent a family of paths $\cU'$ from $I$ to $O\supp{r} \union \ens{x}$, covering the terminal points of $\cW\supp{r}$ as well as the additional vertex $x$.
	Let $W' \in \cU'$ be the path ending at $x$\,: we may obtain a path $W$ covering $v$ by extending $W'$ by the edge $xv$.
	As $v$ is not covered by either $\cU\supp{r+1}$ or $\cW\supp{r}$ (and in particular not by the path $P \in \cW\supp{r}$ which covers $x$), $\cU\supp{r} = \cU' \symdiff \ens{W, W'}$ is a vertex-disjoint collection of paths covering $v$ and the terminal points of $\cW\supp{r}$.
	To represent $\cU\supp{r}$, we set $f(x) \gets v$ and $g(v) \gets x$.
	The only vertices for which $(f,g)$ are changed are then $v$, the vertices $O\supp{r+1} \setminus O\supp{r} = \ens{x, f(x), f^2(x), \ldots}$ for which the recursive call to $\AlternPathSearch$ returned with $\foundpath = \true$, and some vertices $z \in V(G) \setminus O\supp{r+1} \subset V(G) \setminus O\supp{r-1}$.
\end{itemize}
It is easy to verify that if the preconditions are satisfied that $(f,g)$ represent a family of vertex disjoint paths $\cW\supp{r}$ from $I$ to $O\supp{r}$, and that $v \in O\supp{r}$ is not the endpoint of such a path, that they are also satisfied for each recursive call.
Then, by induction, if initially the variables $(f,g)$ represents a family $\cW\supp{r}$ of vertex disjoint $I$\thru$O\supp{r}$ paths, $v \in O$ is not covered by $\cW\supp{r}$, and $\AlternPathSearch(v)$ returns with $\foundpath = \true$, then afterwards $(f,g)$ represents a family $\cU$ of vertex-disjoint $I$\thru$O\supp{r}$ paths covering $v$ and the terminal points of $\cU$.
If such a collection of paths $\cU$ exists, then it is the only family of vertex disjoint $I$\thru$O$ paths larger than $\cW$ in the a graph $G \setminus S$ where we define
\begin{align}
	S \;=\; \ens{z \in O\supp{r} \,\Big|\, \Big(\, g(z) = \nil \text{~or~} g(z) \in O\supp{r} \,\Big) \text{~and~} z \ne x}	\;,
\end{align}
\ie\ where we remove every element of $O\supp{r}$ not covered by $\cW$ \emph{except} for $x$.
Because Algorithm~\ref{alg:augmentPaths} reduces this problem to finding an appropriate family of paths ending in at least one of its neighbors, Theorem~\ref{thm:BGH2001} guarantees that $\AlternPathSearch(v)$ will terminate with $\foundpath = \true$ in the case that such a family of paths $\cU$ does exist.

\subsubsection{Analysis of Algorithm~\ref{alg:iterFindPaths}}

Because a call to $\AlternPathSearch(v)$ returns with $\foundpath = \true$ if and only if either $v \in I$ or a recursive call returns with $\foundpath = \true$, and because $(f,g)$ are only changed when such a recursive call returns with $\foundpath = \true$, we know that they will be unchanged if $\AlternPathSearch(v)$ returns with $\foundpath = \false$.
If this occurs for an invocation $\AlternPathSearch(v)$ where initially no vertices were marked, the loop on lines~\ref{line:iterSearchLoop}\thru\ref{line:iterSearchLoopEnd} of Algorithm~\ref{alg:iterFindPaths} will select another element $v' \in O$ such that $g(v') = \nil$, and invoke $\AlternPathSearch(v')$.
In this case, the data $(f,g)$ represent $\cW$ as a family of paths from $I$ to $O'$, for some superset $O' \supset O$ such that, for all $w \in O' \setminus O$, $w$ is not itself the endpoint of any family of $I$\thru$O'$ vertex-disjoint paths $\cU'$ which also covers the terminal points of $\cW$.
Then, any attempt to reduce the problem to finding such a family of paths for $w \in O' \setminus O$ will fail, and is therefore unnecessary.
Therefore, by induction, subsequent invocations of $\AlternPathSearch(v')$ in the loop on  lines~\ref{line:iterSearchLoop}\thru\ref{line:iterSearchLoopEnd} of Algorithm~\ref{alg:iterFindPaths} will return with $\foundpath = \true$ if and only if there exists a family $\cU'$ of vertex disjoint $I$\thru$O$ paths which covers $v'$ as well as the terminal points of $\cW$; and $(f,g)$ represents this family of paths.

In each repetition of this loop, $\AlternPathSearch$ is invoked at most once on each vertex $v \in V(G)$ in the depth-first search.
Because the assignments and conditional tests for each vertex can be performed in time $O(1)$, the time consumed by each invocation (not including the time consumed by recursive invocations) is dominated by the time required to iterate through the neighbors of $v$ on lines~\ref{line:searchLoop}\thru\ref{line:searchLoopEnd} of Algorithm~\ref{alg:augmentPaths}, of which there are $\deg(v)$.
Summing over all vertices, the time required to perform one iteration of the ``while'' loop on lines~\ref{line:buildPathsLoop}\thru\ref{line:buildPathsLoopEnd} of Algorithm~\ref{alg:iterFindPaths} is then
\begin{gather}
		O\paren{\sum_{v \in V(G)} \deg(v)	\;}
	\;\;=\;\;
		O\Big(2 \card{E(G)}\Big)
	\;\;=\;\;
		O(m)	\;.
\end{gather}

Because there are at most $k$ vertices in $O$, there can be at most $k$ vertex disjoint paths from $I$ to $O$\,; then, at most $k$ iterations of the ``while'' loop on lines~\ref{line:buildPathsLoop}\thru\ref{line:buildPathsLoopEnd} of Algorithm~\ref{alg:iterFindPaths} can terminate with $\foundpath = \true$; then this loop will terminate in time $O(km)$.
By induction, regardless of the value of $\foundpath$, the variables $(f,g)$ represent a family of vertex disjoint paths $\cW$ from $I$ to $O$ after each such iteration; and by the analysis above, we have $\foundpath = \false$ only if there does not exist a family of vertex disjoint $I$\thru$O$ paths in $G$ which is larger than $\cW$.
The run-time of Algorithm~\ref{alg:iterFindPaths} is dominated by the ``while'' loop, as the initialization loop can be performed in time $O(n)$.
Therefore:

\begin{lemma}
	\label{lemma:obtainMaxFamilyPaths}
	Let $(G,I,O)$ be a geometry with $\card{E(G)} = m$ and $\card{O} = k$.
	Then the procedure $\ObtainMaxFamilyPaths(G,I,O)$ terminates in time $O(km)$, and returns data $(f,g)$ such that there exists a maximum-size family $\cW$ of vertex disjoint $I$\thru$O$ paths with the following properties:
	\begin{romanum}
	\item
		$\dom(f) = V(\cW) \setminus O$, and $\dom(g) = V(\cW) \setminus I$\,;

	\item
		$vw \in E(G)$ is covered by $\cW$ if and only if $v \in \dom(f)$, $w \in \dom(g)$, $w = f(v)$, and $v = g(w)$.
	\end{romanum}
\end{lemma}

\paragraph{Remark on this algorithm and its run-time.}

It should be noted that the algorithms presented above are essentially an adaptation of the Ford-Fulkerson algorithm for Max-Flow,\footnote{%
For more information about the Max-Flow problem and the Ford-Fulkerson algorithm to solve it, the interested reader may consult any standard text on polynomial time graph algorithms, \eg~\cite{CLRS}.
The term ``flow'' in this context is not directly related to the definition used throughout this chapter.}
for a flow network with multiple indistinguishable sources and sinks, and unit edge capacities.
For such a flow-network, a solution to Max Flow is a maximum-size collection of \emph{edge}-disjoint $I$\thru$O$ paths: the adaptation made here is the special treatment of entry points treated as an added special case, to ensure \emph{vertex}-disjointness of the families of paths.
Thus, the run-time of $O(km)$ should be unsurprising, as this is also an upper found for the Ford-Fulkerson algorithm for a flow-network whose maximum integer flow has value $k$.

\subsubsection{Determining if we have found a successor function}

The final step towards finding a potential flow-function (\ie\ a successor function in the sense of Definition~\ref{def:successorFn}, to which we will again restrict ourselves to for the remainder of the chapter) is to check if the family of paths $\cW$ described by the variables $(f,g)$ constructed by $\ObtainMaxFamilyPaths(G,I,O)$ is in fact a path cover.
We may do this by simply verifying that
\begin{gather}
	\dom(f) \union \dom(g) \union (I \inter O) = V(G)	\,,
\end{gather}
that is, that there are no vertices $v \in I\comp \union O\comp$ for which $f(v) = g(v) = \nil$, as the set of vertices which would lie outside of both the domain and the image of a successor function is precisely $I \inter O$.
We can test whether there are any such vertices $v$ in time $O(n)$.

By Theorem~\ref{thm:uniqueFamilyPaths}, if $(G,I,O)$ is a geometry where $\card{I} = \card{O}$ and which has a causal path cover $\cC$, then that path cover is the unique maximum family of vertex-disjoint $I$\thru$O$ paths.
Then, if the family of paths $\cW$ represented by the variables $(f,g)$ is not a path cover, the geometry $(G,I,O)$ does not have a flow.
We may then reject any geometries for which this occurs.

\subsection{Constructing the natural pre-order for a successor function}

As we remarked in Section~\ref{sec:influencing}, the problem of determining if a successor function is a flow function can be reduced to the problem of constructing the natural pre-order $\preceq$ for $f$.
The natural pre-order is the transitive and reflexive closure of the influence relation $\inflRel$ for $f$ defined by Definition~\ref{def:inflRel}: thus, we may reduce finding $\preceq$ to the Transitive Closure problem.
In this section, we will describe algorithms for doing this which take advantage of the addition structure provided by the presence of the successor function $f: O\comp \to I\comp$ (and its inverse function $g: I\comp \to O\comp$) based on techniques presented in Nuutila~\cite{Nuutila95}.

\subsubsection{Adapting Tarjan's Algorithm to find the natural pre-order}

Tarjan's algorithm~\cite{Tarjan72, CLRS} is a depth-first search algorithm for determining the strongly connected components of a digraph $D$, which are the equivalence classes of vertices which are mutually reachable from each other by directed walks.
A straightforward adaptation of Tarjan's algorithm due to~\cite{SS-S88} can be used to solve the Transitive Closure problem by storing the set of vertices which can be reached from each vertex (which we refer to as the \emph{descendants} of a vertex), thus determining the transitive closure of the digraph $D$.
A clear presentation of such an algorithm can be found in~\cite[page~49]{Nuutila95}.

In order to efficiently store the sets of descendants of a given vertex, we may make use of a \emph{chain decomposition} of the digraph $D$, a technique proposed in~\cite{Simon88}.
A chain decomposition of a digraph $D$ is a decomposition of $V(D)$ into a family of vertex-disjoint directed paths $\cP$ in the digraph $D$.
Then, for each vertex $v$ and each path $P \in \cP$, we may store the minimal vertex $w \in V(P)$ such that $w$ is a descendant of $v$.
If it is easy to determine when one vertex within a path $P$ is a descendant of another, we may then use this to obtain an efficient method to make reachability comparisons between arbitrary vertices.

We may adopt these techniques to efficiently obtain a representation of the natural pre-order $\preceq$ of the successor function $f$.
Described in terms of directed graphs, the relevant problem is to find the (reflexive and) transitive closure of the influence digraph $\sI_f$ (Definition~\ref{def:inflDigraph}, page~\pageref{def:inflDigraph}).
Note that for each $v \in O\comp$ and $w = f(v)$, the arc $v \arc w$ is an arc of $\sI_f$\,: then, the path cover $\cP_f$ is a chain decomposition of $\sI_f$.
Thus, if functions $f$ and $g$ have been computed, we have a ready-made chain decomposition of $\sI_f$ which we may exploit.
As well, if $\sI_f$ contains directed circuits (\ie\ if it has strongly connected components consisting of more than one vertex), then the pre-order $\preceq$ is not antisymmetric: rather than storing strongly connected components, as in Tarjan's algorithm, we may simply determine whether we have found a directed circuit in $\sI_f$ and terminate once one is found.

The way in which the natural pre-order $\preceq$ will be constructed is as follows.
We will construct and array $\Path: V(G) \to \N$ which assigns a ``path label'' to each vertex which is constant along paths, and distinct between distinct paths, of $\cP_f$.
For each vertex $v \in V(G)$, we will also store a distance $\Dist(v)$ which measures the distance of each vertex from the initial point of the path on which it lies: more precisely, to the largest integer $\ell \in \N$ such that $g^\ell(v) \ne \nil$.
Thus, for vertices $v,w$ on a common path, we have $v \preceq w$ if and only if $\Dist(v) \le \Dist(w)$.
We will call a pair of such arrays $(\Dist,\Path)$ a \emph{path-labelling} of the orbits of $f$.

To exploit the chain decomposition to efficiently store the partial order $\preceq$\,, we will construct an array of lower bounds for each vertex, which we will use to store the ``ancestors'' rather than the descendants of a vertex (\ie\ for a vertex $w$, the vertices $v$ from which $w$ can be reached by directed walks).
The \emph{infimum} $\inf_P(w)$ of a vertex $w$ in a path $P \in \cP_f$ will simply be the maximal vertex $v$ in $\preceq$ such that $v \in V(P)$ and $v \preceq w$.
Rather than explicitly storing the vertices $\inf_P(w)$, we may instead store the distance of $\inf_P(w)$ from the initial point on its path.
Then, if we label the paths of $\cP_f$ arbitrarily by integers $p \in \ens{1, \ldots, k}$,\footnote{%
Recall that each path of $\cP_f$ ends in $O$ and each vertex of $O$ ends such a path: thus $\card{\cP_f} = \card{O} = k$.}
we may record the infima of vertices $w \in V(G)$ in an array $\Inf: V(G) \x \ens{1, \ldots, k} \to \Z$ such that $\Inf(w,p)$ is the maximum distance of a vertex $v$ such that $\Path(v) = p$ from the initial point of its path, such that $v \preceq w$.
(If there is no such vertex $v$, we may set $\Inf(w,p)$ to some negative integer.)
This suggests the following definition:

\begin{definition}
	For a geometry $(G,I,O)$ with a successor function $f: O\comp \to I\comp$, a \emph{system of infima} for a pre-order $\preceq$ on $V(G)$ is a triple $(\Dist, \Path, \Inf)$ of arrays $\Path, \Dist: V(G) \to \N$ and $\Inf: V(G) \x \N \to \Z$ such that 
	\begin{gather}
		\label{eqn:infimaCriterion}
			v \preceq w
		\;\;\iff\;\;
			\Dist(v) \le \Inf(w, \Path(v))
	\end{gather}
	holds for all $v, w \in V(G)$.
\end{definition}

Algorithms~\ref{alg:findInfima} and~\ref{alg:testNaturalPreorder} together describe an iterative procedure for constructing a system of infima for $\preceq$ as described above, and in doing so, test whether the natural pre-order $\preceq$ is antisymmetric.
In the following pages, we will prove their correctness.

\subsubsection{Analysis of Algorithm~\ref{alg:findInfima}}

Algorithm~\ref{alg:findInfima} describes a recursive subroutine $\FindInfima$, with implicit inputs defined by an embedding procedure shown in Algorithm~\ref{alg:testNaturalPreorder}~(page~\pageref{alg:testNaturalPreorder}).
Before describing the more general algorithm there, we will first describe behavior of $\FindInfima$.

\begin{algorithm}[t]

	\SUBROUTINE \FindInfima(w):

		\Data{%
			$(G,I,O)$, a geometry with $\card{O} = k$
		\\
			$f: V(G) \partialto V(G)$, a successor function for $(G,I,O)$
		\\
			$g: V(G) \partialto V(G)$, the inverse function of $f$
		\\
			an array $\status: V(G) \to \ens{\untouched, \incomplete, \complete}$
		\\
			$\Inf: V(G) \x \ens{1,\ldots,k} \to \Z$, an array to store path-infima for vertices
		}

	\BEGIN
		\algline

		\oneline
		\IF $\status(w) = \incomplete$ \THEN $\foundcircuit = \true$\;					\nllabel{line:testCircuit}
		\oneline
		\IF $\status(w) \ne \untouched$ \THEN \RETURN\;										\nllabel{line:testStatus}
		$\status(w) \gets \incomplete$\;															\nllabel{line:setIncomplete}

		\medskip

		\LET $v \in V(G)$\;
		\FOREACH $z \sim w$ \SUCHTHAT $z \ne f(w)$ \ANDALSO $\big(z = g(w) ~\ORELSE~ g(z) \ne \nil\big)$ \DO		\nllabel{line:predLoop}
			\short
			\IF $z = g(w)$ \THEN {\nllabel{line:directPred}%
				$v \gets z$
			}
			\ELSE {\nllabel{line:indirectPred}
				$v \gets g(z)$
			}	

			\smallskip
			$\FindInfima(v)$\;																								\nllabel{line:recursiveFindInfima}
			\oneline
			\IF $\foundcircuit$ \THEN \RETURN\;																			\nllabel{line:recursiveTestCircuit}

			\smallskip
			\short 
			{\FORALL $p \in \ens{1,\ldots,k}$ \DO																		\nllabel{line:augmentInfLoop}
				\oneline
				\IF $\Inf(w,p) < \Inf(v,p)$ \THEN $\Inf(w,p) \gets \Inf(v,p)$\;
			\ENDFOR}

			\smallskip
	\ENDFOR																																	\nllabel{line:predLoopEnd}

		\medskip
		$\status(w) \gets \complete$\;
		\RETURN
	\END

	\caption[An algorithm to obtain the ``ancestors'' of a vertex $w$.]{\label{alg:findInfima}%
		An algorithm to obtain the ``ancestors'' of a vertex $w$ (and by extension of all vertices $v \prec w$), and to determine in the process whether $\prec$ is antisymmetric.}
\end{algorithm}

Suppose that $(\Dist,\Path)$ is a path-labelling of the orbits of $f$.
We will say that $w$ is \emph{untouched} with respect to a function $\Inf: V(G) \x \ens{1,\ldots,k} \to \N$ if we have
$\status(w) = \untouched$ and 
\begin{gather}
	\Inf(w,p) = \begin{cases}
						\Dist(w)	\,,	&	\text{if $p = \Path(w)$}	\\
						-1	\,,			&	\text{otherwise}
	            \end{cases}
\end{gather}
for all $p \in \ens{1,\ldots,k}$\,; and that $w$ is \emph{complete} with respect to $\Inf$ if for all $v \preceq w$, the following holds:
\begin{romanum}
\item
	$\status(v) = \complete$\,,

\item
	$\Inf(v,p) \ge -1$ for all $p \in \ens{1,\ldots,k}$\,, and
	
\item
	$u \preceq v \;\iff\; \Dist(u) \le \Inf(v, \Path(u))$ for all $u \in V(G)$.
\end{romanum}
For any vertex $w$ which is complete with respect to $\Inf$, we also have $v$ complete with respect to $\Inf$ whenever $v \preceq w$: this represents a subset of $(v,w) \in V(G) \x V(G)$ where the relation $v \preceq w$ is faithfully represented by a comparison as in~\ref{eqn:infimaCriterion}. 
We will be interested in situations where $\FindInfima$ is called for an untouched vertex, conditioned on subsets of $V(G)$ being either untouched or complete with respect to $\Inf$.

Note that the instructions of the loop on lines~\ref{line:predLoop}\thru\ref{line:predLoopEnd} are executed for precisely those vertices $z = f(v) \ne w$ and $z = v = g(w)$, where $v \inflRel w$; and on lines~\ref{line:directPred} and~\ref{line:indirectPred}, we assign to the \emph{variable} $v$ that corresponding vertex $v \inflRel w$.
Then, the iteration over $z$ on line~\ref{line:predLoop} essentially stands in for an iteration over all vertices $v \inflRel w$.
We will use this fact without further comment throughout our analysis of $\FindInfima$.

Next, note that in any invocation of $\FindInfima(w)$ terminates with $\foundcircuit = \false$ only if either $\foundcircuit = \false$ and $\status(w) = \complete$ before the invocation, or if the program control proceeds beyond for loop on lines~\ref{line:predLoop}\thru\ref{line:predLoopEnd}.
The value of $\Inf(w,p)$ can only change for some value of $p \in \ens{1, \ldots, k}$ in the latter case, in which case the value of $\status(w)$ changes from $\untouched$ to $\complete$ in the course of the invocation.
Furthermore, this will only happen if every recursive call to $\FindInfima(v)$ made also returns with $\foundcircuit = \false$, in which case the same is true of all $v \inflRel w$.
Thus, an invocation of $\FindInfima(w)$ terminates with $\foundcircuit = \false$ only if:
\begin{romanum}
\item
	$\foundcircuit = \false$ before the invocation;

\item
	for any $v \in V(G)$ the value of $\Inf(v,p)$ is only changed for some $p \in \ens{1, \ldots, k}$ if $\status(v)$ is also changed; and

\item
	$\status$ is changed by the invocation only in that $\status(v)$ is set to $\complete$ for some set of vertices $v \preceq w$, including $w$ in particular.
\end{romanum}
Using this, we will consider certain conditions under which an invocation $\FindInfima(w)$ will return with $\foundcircuit = \true$.

\begin{definition}
	We call a set of vertices $S$ \emph{noxious} if, when $\status(v) = \untouched$ or $\status(v) = \complete$ for every vertex $v \in V(G) \setminus S$, an invocation of $\FindInfima(s)$ will terminate with $\foundcircuit = \true$ for any $s \in S$.
\end{definition}

\begin{observation*}
	Suppose that $\status(v) = \incomplete \iff v \in S$ for some set $S \subset V(G)$.
	Then $S$ is a noxious set, as an invocation of $\FindInfima(v)$ for any $v \in S$ will terminate with $\foundcircuit = \true$.
\end{observation*}

The concept of a vertex in a noxious set is meant as a generalization of the case above, of a vertex $v \in V(G)$ for which $\status(v) = \incomplete$.
We will apply this concept ultimately to demonstrate simple conditions on $\status$ under which a circuit in $\sI_f$ will cause an invocation of $\FindInfima$ to return with $\foundcircuit = \true$.
The main tool in this approach is the following Lemma:

\begin{lemma}
	\label{lemma:noxiousChain}
	Let $S \subset V(G)$ be a noxious set defined solely in terms of the properties of vertices with respect to adjacency in $G$, and the arrays $\status$, $f$, and $g$.
	Suppose that $w \in V(G)$ is such that for all $v \preceq w$, either $\status(v) = \untouched$, $\status(v) = \complete$, or $v \in S$; and that there is some $s \in S$ such that $s \preceq w$.
	Then an invocation of $\FindInfima(w)$ will terminate with $\foundcircuit = \true$.
\end{lemma}
\begin{proof}
	For any such $w$, let $S'$ be the subset of vertices $s \in S$ such that there is a chain
	\begin{gather}
		s \;=\; u_0 \inflRel u_1 \inflRel \cdots \inflRel u_d = w
	\end{gather}
	for some $d \ge 0$, such that $u_j \notin S$ for all $j > 0$; and let $\ell(w)$ be the length of the longest such chain.
	(We may extend this to vertices $w$ for which there are no vertices $s \preceq w$, by adopting the convention $\ell(w) = \infty \ge N$ for any $N \in \N$.)
	If $\ell(w) = 0$, we have $w \in S$, in which case an invocation of $\FindInfima(w)$ will terminate with $\foundcircuit = \true$ by definition.
	Otherwise, suppose the claim holds for all such vertices $v \in V(G)$ for which $\ell(v) \le L$, and suppose that $\ell(w) = L+1$.
	Let $N$ be the set of vertices $u \inflRel w$ such that $s \preceq u$ for some $s \in S'$, and consider the operations performed in an invocation of $\FindInfima(w)$ under the stated initial conditions, and with $\foundcircuit = \false$.

	If the invocation $\FindInfima(w)$ terminates without setting $v$ to an element of $N$ in the loop from lines~\ref{line:predLoop}\thru\ref{line:predLoopEnd}, this is because the condition on line~\ref{line:recursiveTestCircuit} was met, in which case the invocation terminates with $\foundcircuit = \true$.
	Otherwise, all of the preceding recursive invocations $\FindInfima(v)$ for $v \inflRel w$ terminated with $\status(v) = \complete$ and $\foundcircuit = \false$, in which case the value of $\status(u)$ is unchanged for any vertex $u$ such that $u \not\preceq v$ for all of the preceding values of $v$.
	Also, by the induction hypothesis, because these preceding recursive invocations $\FindInfima(v)$ did not terminate with $\foundcircuit = \true$, we have $\ell(v) > L$.

	Let $u_0 \inflRel w$ be the first vertex of $S'$ to occur as the value of $v$ in the loop on lines~\ref{line:predLoop}\thru\ref{line:predLoopEnd} of the invocation $\FindInfima(w)$.
	Then, any chain $s = u_d \inflRel u_{d-1} \inflRel \cdots \inflRel u_1 \inflRel u_0$, such that $s \in S$ and $u_j \notin S$ for $j < d$, has length $d \le L$ by the definition of $\ell(w)$.
	Because $\ell(v) > L$ for any value of $v \inflRel w$ from an earlier iteration, there cannot be a chain $s \inflRel v_{d-1} \inflRel \cdots \inflRel v_1 \inflRel v$, where $d \le \ell$, for any such $v$; and because there is no such chain for $d > \ell$ by the definition of $\ell(w)$, there cannot be a chain $u_j \inflRel v_{N-1} \inflRel \cdots \inflRel v_1 \inflRel v$ for any length $N$, and for any $0 \le j \le d$.

	Thus, for all $0 \le j \le d$ and for all $v \inflRel w$ from earlier iterations of the for loop on lines~\ref{line:predLoop}\thru\ref{line:predLoopEnd}, we have $u_h \not\prec v$.
	In particular, values of $\status(u_j)$ are unchanged from their initial values before the invocation of $\FindInfima(w)$, and so for each $0 \le j \le d$, $u_j \in S$ before the invocation of $\FindInfima(u_0)$ if and only if $u_j \in S$ before the invocation of $\FindInfima(w)$.
	Therefore, just prior to the invocation of $\FindInfima(u_0)$, there is a chain $s \inflRel u_{d-1} \inflRel \cdots \inflRel u_1 \inflRel u_0$ for $d \le L$, such that $s \in S$ and $u_j \notin S$ for $j < d$.
	By the induction hypothesis, the invocation of $\FindInfima(u_0)$ terminates with $\foundcircuit = \true$: then the invocation of $\FindInfima(w)$ also terminates with $\foundcircuit = \true$.
	By induction on $\ell(w)$, the claim then holds.
\end{proof}

\begin{definition}
	A vertex $w \in V(G)$ is \emph{circuitous} if there is a vertex $v \prec w$ for which $w \prec v$\,, \ie\ if there is a circuit in $\sI_f$ containing $w$.
\end{definition}
The order $\preceq$ is a partial order if and only if there are no such vertices.
We would like to describe simple conditions under which such a vertex causes an invocation of $\FindInfima$ to return with $\foundcircuit = \true$.
We will show:

\begin{lemma}
	\label{lemma:circuitNoxious}
	Suppose that every vertex $v \in V(G)$ is either untouched or complete with respect to $\Inf$, and in particular that every circuitous vertex is untouched with respect to $\Inf$\,.
	Then the set of circuitous vertices is noxious.
\end{lemma}
\begin{proof}
	Let $w$ be a circuitous vertex, and let $N$ be the set of vertices $v \inflRel w$ such that $w \prec v$\,.
	Because $\status(w) = \untouched$ initially, it follows that any vertex $u \in V(G)$ for which $w \preceq u \preceq w$ holds is not complete with respect to $\Inf$\,; therefore each such $u$ is untouched with respect to $\Inf$.
	In particular, for each vertex $u_j$ in any chain $w \inflRel u_{d-1} \inflRel \cdots \inflRel u_0 \inflRel w$, we have $\status(u_j) = \untouched$.

	Consider an invocation of $\FindInfima(w)$.
	On line~\ref{line:setIncomplete}, we set $\status(w) \gets \incomplete$\,.
	Then, for all $u \in N$, there is a chain $w \inflRel u_{d-1} \inflRel \cdots \inflRel u_0 = u$ such that $\status(w) = \incomplete$ and $\status(u_j) = \untouched$ for $0 \le j < d$.
	If the invocation $\FindInfima(w)$ terminates without setting $v$ to an element of $S$ in the loop from lines~\ref{line:predLoop}\thru\ref{line:predLoopEnd}, this is because the condition on line~\ref{line:recursiveTestCircuit} was met, in which case the invocation terminates with $\foundcircuit = \true$.
	Otherwise, let $u_0$ be the first vertex in $S$ which is assigned to $v$ in the loop on lines~\ref{line:predLoop}\thru\ref{line:predLoopEnd}.
	By the very fact that $v \notin N$ in the preceding iterations, we do not have $w \preceq v$ for those iterations, and in particular there are not any vertices $u'$ such that $w \preceq u' \preceq v$\,; then the values of $\status(u')$ are unchanged throughout the recursive invocation $\FindInfima(v)$ for these preceding iterations.
	In particular, for all vertices $u' \preceq u_0$, we have $\status(u') = \untouched$, $\status(u') = \complete$, or $\status(u') = \incomplete$.
	As remarked above, any set of vertices $S$ consisting of vertices $u'$ for which $\status(u') = \incomplete$ is noxious.
	Then, by Lemma~\ref{lemma:noxiousChain}, an invocation of $\FindInfima(u_0)$ will return with $\foundcircuit = \true$: so the invocation of $\FindInfima(w)$ will terminate with $\foundcircuit = \true$ as well.
	Thus, under these conditions, the set of circuitous vertices is noxious.
\end{proof}

Lemmas~\ref{lemma:noxiousChain} and~\ref{lemma:circuitNoxious} provide a sufficient condition for an invocation $\FindInfima(w)$ to return with $\foundcircuit = \true$.
The following will provide a useful partial converse:

\begin{lemma}
	\label{lemma:inductComplete}
	Suppose that, for some vertex $w \in V(G)$ every vertex $v \preceq w$ is either untouched or complete with respect to $\Inf$.
	If there are no circuitous vertices $s \preceq w$ and $\foundcircuit = \false$, an invocation $\FindInfima(w)$ will return with $\foundcircuit = \false$, and only change the values of $\status$ and $\Inf$ to make $w$ complete with respect to $\Inf$.
\end{lemma}
\begin{proof}
	If $w$ is complete with respect to $\Inf$, then $\FindInfima$ terminates immediately with no changes to any variables; otherwise, we may assume that $w$ is untouched with respect to $\Inf$.
	We will prove the claim by induction on the \emph{depth} of $w$, which we define as the length of the longest chain $v_0 \inflRel v_1 \inflRel \cdots \inflRel v_d = w$ in $V(G)$: if there are no circuitous vertices $s \preceq w$, $d$ is guaranteed to be bounded above by $n = \card{V(G)}$.

	Consider the base case $d = 0$, 
	The conditions on lines~\ref{line:testCircuit} and~\ref{line:testStatus} are not fulfilled as $\status(w) = \untouched$, so the first operation performed is to set $\status(w) \gets \incomplete$\,.
	No vertex $z$ satisfies the conditions of line~\ref{line:predLoop}, by assumption of the minimality of $w$, and so the instructions within that for loop are never executed; we then set $\status(w) \gets \complete$, and return.
	As well, because $\foundcircuit$ is left unaltered, we still have $\foundcircuit = \false$.
	Then, the only change performed by $\FindInfima(w)$ is to set $\status(w) \gets \complete$\,; but because $v \preceq w \iff v = w$ in this case, and because we had
	\begin{gather}
			\Inf(w,p)
		\;\;=\;\;
			\begin{cases}[r]
				0\,,	&	\text{for $p = \Path(w)$}	\\
				-1\,,	&	\text{otherwise}
			\end{cases}
	\end{gather}
	prior to the invocation, this is sufficient to cause $w$ to be complete with respect to $\Inf$.

	Suppose then that the claim holds for all vertices for with some depth $d \in \N$, and that $w$ has depth $d+1$.
	The conditions on lines~\ref{line:testCircuit} and~\ref{line:testStatus} are not fulfilled, as $\status(w) = \untouched$; we then set $\status(w) \gets \incomplete$.
	In the loop on lines~\ref{line:predLoop}\thru\ref{line:predLoopEnd}, we iterate over all $v \inflRel w$\,: by assumption, every such $v$ is either untouched or complete with respect to $\Inf$, as are all vertices $u \preceq v$ for such $v$.
	It is clear that $\FindInfima(v)$ simply returns without changing $\Inf$ or $\status$ when initially we have $\status(v) = \complete$; then by induction, after each recursive call to $\FindInfima(v)$ on line~\ref{line:recursiveFindInfima}, each vertex $v$ is complete with respect to $\Inf$ and $\foundcircuit = \false$.
	The condition on line~\ref{line:recursiveTestCircuit} is then unfulfilled, so we execute the for loop on line~\ref{line:augmentInfLoop} and proceed to the next vertex $z$.

	Initially, for every $p \in \ens{1,\ldots,k} \setminus \ens{\Path(w)}$, we have $\Inf(w,p) = -1$: then, for such $p$, the initial value of $\Inf(w,p)$ is bounded above by the values of $\Inf(v,p)$ for each $v \inflRel w$ after the recursive calls $\FindInfima(v)$.
	Then by induction, it is easy to show that after the loop on lines~\ref{line:predLoop}\thru\ref{line:predLoopEnd},
	\begin{align}
			\Inf(w,p)
		\;\;=&\;\;
			\max_{v \:\!\inflRel\:\! w} \;\,	\Inf(v,p)	\;,
		\quad
			\text{for $p \ne \Path(w)$}.	
	\end{align}
	For any $u \in V(G)$ with $\Path(u) = p \ne \Path(w)$, we have in particular $u \ne w$, and so $u \preceq w$ if and only if $u \preceq v$ for some $v \inflRel w$; and this holds if and only if $\Dist(u) \le \Inf(v, \Path(u)) \le \Inf(w, \Path(u))$ for that same $v$ after the loop on lines~\ref{line:predLoop}\thru\ref{line:predLoopEnd}, by the induction hypothesis.
	Finally, because there are no circuitous vertices $u \preceq w$, we have in particular $w \not\prec v$ for all $v \inflRel w$\,; then $\Inf(v, \Path(w)) < \Dist(w)$ for all $v \inflRel w$, in which case we have $\Inf(w, \Path(w)) = \Dist(w)$ after line~\ref{line:predLoopEnd}.
	Therefore, we also have $u \preceq w \,\iff\, \Dist(u) \le \Inf(w, \Path(u))$ when $\Path(u) = \Path(w)$\,: thus, the claim holds for $w$ as well.
	By induction, the Lemma holds.
\end{proof}

Lemmas~\ref{lemma:noxiousChain},~\ref{lemma:circuitNoxious}, and~\ref{lemma:inductComplete} will form the basis of our analysis of Algorithm~\ref{alg:testNaturalPreorder} in the next section.

\subsubsection{Analysis of Algorithm~\ref{alg:testNaturalPreorder}}

In the previous section, we described the behavior of the subroutine $\FindInfima$ under conditions of the vertices being either untouched or complete with respect to $\Inf$.
The role of the procedure $\TestNaturalPreorder$ in Algorithm~\ref{alg:testNaturalPreorder} (page~\pageref{alg:testNaturalPreorder}) is to establish a path-labelling $(\Dist,\Path)$ of the orbits of $f$, initialize arrays $\status$ and $\Inf$ so that every vertex is untouched with respect to $\Inf$, and then repeatedly invoke $\FindInfima$ in an attempt to change the status of every vertex to that of being complete with respect to $\Inf$. 
In doing so, we will either determine that $\sI_f$ has circuits (and so $\preceq$ is not a partial order), or assign values of $\Inf(w,p)$ for all $w \in V(G)$ and $p \in \ens{1,\ldots,k}$ so that $(\Dist,\Path,\Inf)$ forms a system of infima for $\preceq$\,.

\begin{algorithm}[p]

	\PROCEDURE \TestNaturalPreorder(G,I,O,f,g):

		\Input{%
			$(G,I,O)$, a geometry with $\card{O} = k$
		\\
			$f: V(G) \partialto V(G)$, a successor function for $(G,I,O)$
		\\
			$g: V(G) \partialto V(G)$, the inverse function of $f$
		}
		\Output{%
			Either a triple $(\Dist,\Path,\Inf)$ representing the natural pre-order for $f$ (in the case that $f$ is a flow function), or $\nil$ (in the case that $f$ is not a flow function).
		}

	\BEGIN
		\LET $\status: V(G) \to \ens{\untouched, \incomplete, \complete}$\;
		\LET $\Dist: V(G) \to \N$\;
		\LET $\Path: V(G) \to \N$\;
		\LET $\Inf: V(G) \x \ens{1,\ldots,k} \to \N$\;

		\medskip
		\LET $\nextpath \gets 1$\;
		\FORALL $w \in O$ \DO																							\nllabel{line:initPathLoop}
			$\InitializePath(w,\nextpath)$\;
			$\INCR \nextpath$\;
		\ENDFOR																												\nllabel{line:initPathLoopEnd}

		\medskip
		\LET $\foundcircuit \gets \false$\;
		\FORALL $w \in O$ \DO																					\nllabel{line:findInfimaLoop}
			$\FindInfima(w)$\;																					\nllabel{line:findInfima}
			\oneline
			\IF $\foundcircuit$ \THEN \RETURN \nil\;
		\ENDFOR																										\nllabel{line:findInfimaLoopEnd}

		\RETURN $(\Dist,\Path, \Inf)$\;
	\END

	\bigskip

	\SUBROUTINE \InitializePath(w,P):

	\BEGIN
		\oneline
		\IF $g(w) = \nil$ \THEN $\Dist(w) \gets 0$\;
		\ELSE {
			$\InitializePath(g(w))$\;
			$\Dist(w) \gets \Dist(g(w)) + 1$\;
		}
		$\Path(w) \gets P$\;

		\medskip
		\FORALL $p \in \ens{1,\ldots,k}$ \DO																		\nllabel{line:infimaInitLoop}
			\short
			\IF $p = P$ \THEN { $\Inf(w,p) \gets \Dist(w)$} \ELSE { $\Inf(w,p) \gets -1$ }
		\ENDFOR																												\nllabel{line:infimaInitLoopEnd}
	
		$\status(w) \gets \untouched$\;

		\medskip

		\RETURN
	\END

	\caption[An algorithm to test whether or not the natural pre-order of a successor function $f$ for a geometry $(G,I,O)$ is a partial order.]{\label{alg:testNaturalPreorder}%
		An algorithm to test whether or not the natural pre-order of a successor function $f$ for a geometry $(G,I,O)$ is a partial order.
		The subroutine $\FindInfima$ is described in Algorithm~\ref{alg:findInfima} on page~\pageref{alg:findInfima}.}
\end{algorithm}

To establish a path-labelling $(\Dist,\Path)$ for the orbits of $f$, we use a subroutine $\InitializePath$.
It is straightforward to verify that $\InitializePath(w,P)$ sets $\Path(v) = P$ for $v = w$ and for all vertices $v$ which precede $w$ on the same path, and correctly assigns their distance along the path from the first vertex of the path.
As well, for all such vertices $v$, it initializes $\status(v)$ and $\Inf(v,p)$ for each $p \in \ens{1,\ldots,k}$ so that $v$ is untouched with respect to $\Inf$.
In doing so, it is easy to see that it does not affect the status of any other vertex, or the labelling of any other paths.
$\TestNaturalPreorder$ calls $\InitializePath(v,p)$ for each vertex $w \in O$,\footnote{%
For an arbitrary path cover $\cC$ for a geometry $(G,I,O)$, the initial point of some paths of $\cC$ may not be a vertex in $I$.
While we are primarily interested in the case where $\card{I} = \card{O}$, in which case there is a one-to-one correspondence between paths and elements of $I$, we may make the algorithm more general by labelling paths from their endpoint.
In doing so, this algorithm may also be used to construct the natural pre-order for $f$ in cases where $\card{I} < \card{O}$.}
setting the path-label $p = \nextpath$ of each vertex in $O$ to a distinct integer $1 \le p \le k$.
This initializes every vertex to be untouched with respect to $\Inf$, and establishes a path-labelling $(\Dist,\Path)$.

Next, we set $\foundcircuit \gets \false$, and we invoke $\FindInfima(w)$ for each vertex $w \in O$, aborting in the case that we discover $\foundcircuit = \true$.
We return the triple $(\Dist, \Path, \Inf)$ if the loop on lines~\ref{line:findInfimaLoop}\thru\ref{line:findInfimaLoopEnd} terminates without ending the procedure call. 

Consider a loop precondition on the values of $\foundcircuit$, $\status$, and $\Inf$ where, prior to some iteration of the loop on lines~\ref{line:findInfimaLoop}\thru\ref{line:findInfimaLoopEnd}, we have
\begin{romanum}
\item
	$\foundcircuit = \false$,

\item
	every vertex is either untouched or complete with respect to $\Inf$, and

\item
	any circuitous vertices in particular are untouched with respect to $\Inf$.
\end{romanum}
By Lemma~\ref{lemma:circuitNoxious}, the set of circuitous vertices is noxious under these conditions; then, by Lemma~\ref{lemma:noxiousChain}, $\FindInfima(w)$ will return with $\foundcircuit = \true$ in the case that there exists some circuitous vertex $v \preceq w$.
If such a circuitous vertex $v$ exists, $\TestNaturalPreorder$ terminates with a return value of $\nil$.
Otherwise, by Lemma~\ref{lemma:inductComplete}, $\TestNaturalPreorder$ terminates with $\foundcircuit = \false$ and $w$ complete with respect to $\Inf$.
In particular, $\status$ and $\Inf$ in such a way as to make $w$ complete with respect to $\Inf$, in which case it is still the case that every vertex $v \in V(G)$ is untouched or complete with respect to $\Inf$.

By induction on the iteration over $w \in O$ on line~\ref{line:findInfimaLoop}, we may then prove that an invocation of $\TestNaturalPreorder(G,I,O,f,g)$ returns $\nil$ if the natural pre-order $\preceq$ is not antisymmetric, and returns with a triple $(\Dist, \Path, \Inf)$ forming a system of infima for $\preceq$ in the case that $\preceq$ is antisymmetric.
If there are any circuitous vertices $v$, eventually there will be an invocation $\FindInfima(w)$ for which $v \preceq w$\,; because the loop precondition above is maintained until such a vertex $w$ is found, it follows that $\TestNaturalPreorder(G,I,O,f,g)$ will terminate with return value $\nil$.
Otherwise, it will eventually terminate with return value $(\Dist,\Path,\Inf)$ such that $(\Dist,\Path)$ is a path-labelling of the orbits of $f$, and where every vertex $w \in O$ is complete with respect to $\Inf$\,.
Because every $v \in V(G)$ is bounded above by some $w \in O$, it follows that $\status(v) = \complete$\,, $\Inf(v,p) \ge -1$ for all $p \in \ens{1,\ldots,k}$\,, and
\begin{gather}
	\label{eqn:infimaCriterionRevisited}
	u \preceq v \;\iff\; \Dist(u) \le \Inf(v, \Path(u))	\qquad \text{for all $u,v \in V(G)$}.
\end{gather}
Thus, the triple $(\Dist,\Path,\Inf)$ forms a system of infima for $\preceq$.

In $\TestNaturalPreorder(G,I,O,f,g)$, the subroutine $\InitializePath(w,P)$ is run once for each vertex along each path, and so is invoked $n = \card{V(G)}$ times; and in each invocation, the run-time is dominated by the loop on lines~\ref{line:infimaInitLoop}\thru\ref{line:infimaInitLoopEnd}, which performs $k$ iterations of a constant-time assignment.
Thus, the loop on lines~\ref{line:initPathLoop}\thru\ref{line:initPathLoopEnd} runs in time $O(kn)$.
However, we will show that this time requirement is swamped by cumulative run-time of the invocations of $\FindInfima$.

For vertices $v \in V(G)$ in general, the subroutine $\FindInfima(v)$ is invoked on line~\ref{line:findInfima} for $w \in O$, and line~\ref{line:recursiveFindInfima} of Algorithm~\ref{alg:findInfima} for $v \in V(G)$ such that $v \inflRel w$ for some $w \in V(G)$.
It is easy to show that every $w \in O$ is maximal in $\preceq$, in which case it is invoked only once for such vertices; for any other vertex $v$, $\FindInfima(v)$ is invoked once for each $w \in V(G)$ such that $v \inflRel w$; because this holds precisely for the set of vertices $w = f(v)$ or $w \sim z = f(v)$, this is bounded above by the degree of $f(v)$.
The first time that $\FindInfima(v)$ is invoked for a particular vertex $v$, $v$ is untouched with respect to $\Inf$; for any subsequent invocation made, $v$ is complete with respect to $\Inf$, and the invocation terminates in time $O(1)$.
Then, the time requirements of all invocations of $\FindInfima(v)$ for a particular vertex $v$ (but not including the time consumed by recursive invocations for vertices $u \inflRel v$) is bounded above by the time spent in the first invocation of $\FindInfima(v)$, plus $\deg(f(v))$ for all subsequent invocations combined. 
The time required by the first invocation is dominated by by the loop at line~\ref{line:augmentInfLoop}, inside the loop from lines~\ref{line:predLoop}\thru~\ref{line:predLoopEnd} (of Algorithm~\ref{alg:findInfima}).
The inner loop performs $k$ iterations of condition tests and assignments requiring $O(1)$ time, which is repeated for each $u \inflRel v$ by the outer loop: as the number of such vertices $u$ is bounded above by $\deg(v)$, the outer loop then requires time $O(k \deg(v))$.

By a temporary abuse of notation, let $\deg(f(v)) = 0$ for vertices $v \in O$\,.
The total run-time of all invocations of $\FindInfima(v)$ is then $O(k \deg(v) + \deg(f(v)))$ for any $v \in V(G)$\,: summing over all vertices $v$, the cumulative run-time of the invocations of $\FindInfima$ is then
\begin{align}
		O\paren{ \sum_{v \in V(G)} \Big[ k \deg(v) + \deg(f(v)) \Big] }
	\;=&\;\;
		O\paren{ \sum_{v \in V(G)} k \deg(v) \;\;+ \sum_{v \in V(G)} \deg(f(v)) }
	\notag\\[1ex]=&\;\;
		O\Big(2 k \abs{E(G)} \,+\, 2 \abs{E(G)}\Big)
	\;=\;
		O(km)\,.
\end{align}
We have therefore shown:
\begin{lemma}
	\label{lemma:testNaturalPreorder}
	Let $(G,I,O)$ be a geometry with $\card{E(G)} = m$ and $\card{O} = k$, $f : V(G) \partialto V(G)$ be a successor function for $(G,I,O)$, and let $g: V(G) \partialto V(G)$ be the inverse of $f$.
	Then the procedure $\TestNaturalPreorder(G,I,O,f,g)$ terminates in time $O(km)$, and returns either
	\begin{itemize}
	\item
		$\nil$, if the natural pre-order $\preceq$ of $f$ is not antisymmetric; or

	\item
		a system of infima $(\Dist,\Path,\Inf)$ for $\preceq$, where in particular $(\Dist, \Path)$ is a path-labelling of the orbits of $f$.
	\end{itemize}
\end{lemma}
Restricting to those geometries satisfying the extremal bound of Theorem~\ref{thm:extremalUpperBound}, this provides a proof of Theorem~\ref{thm:findPreorder} on page~\pageref{thm:findPreorder}.

\subsection{The complete flow-finding algorithm}

As a summary of the results of the previous sections, we now describe an algorithm for discovering whether a geometry $(G,I,O)$ has a flow $(f,\preceq)$ in the case that $\card{I} = \card{O}$, and to construct one if so.

\begin{algorithm}[t]

\PROCEDURE \FindFlow(G,I,O):

\Input{%
		$G$, a graph;
	\\
		$I,O \subset V(G)$, sets of vertices of equal size.
}

\Output{%
	Either $\nil$, if $(G,I,O)$ does not have a flow; or a successor function $f: V(G) \partialto* V(G)$ for $(G,I,O)$, its inverse function $g: V(G) \partialto* V(G)$, and a system of infima $\partialOrd = (\Dist,\Path,\Inf)$ for the natural pre-order $\preceq$ for $f$, in the case where $(G,I,O)$ has a flow.
}

\BEGIN
	
	\LET $n \gets \card{V(G)}$\;
	\LET $m \gets \card{E(G)}$\;
	\LET $k \gets \card{O}$\;

	\oneline
	\IF $m > kn - \binom{k+1}{2}$ \THEN \RETURN \nil\;												\nllabel{line:extremalTest}

	\medskip
	\LET $(f,g) \gets \ObtainMaxFamilyPaths(G,I,O)$\;												\nllabel{line:constructFamilyPaths}
	\FORALL $v \in V(G)$ \DO																				\nllabel{line:testPathCoverLoop}
		\oneline
		\IF $v \notin I \inter O$ \ANDALSO $f(v) = g(v) = \nil$ \THEN \RETURN \nil\;	
	\ENDFOR
	
	\medskip
	\LET $\partialOrd \gets \TestNaturalPreorder(G,I,O,f,g)$\;									\nllabel{line:constructPreorder}
	\short
	\IF $\partialOrd = \nil$ \THEN {\nllabel{line:testPartialOrder}%
		\RETURN \nil
	}
	\ELSE { \RETURN $(f,g,\partialOrd)$ }

\END

	\caption[A complete algorithm for determining if a geometry $(G,I,O)$ has a flow in the case that $\card{I} = \card{O}$.]{\label{alg:completeFlowAlg}%
		A complete algorithm for determining if a geometry $(G,I,O)$ has a flow in the case that $\card{I} = \card{O}$, using the procedures defined in Algorithms~\ref{alg:iterFindPaths} and~\ref{alg:testNaturalPreorder} (pages~\pageref{alg:iterFindPaths} and~\pageref{alg:testNaturalPreorder} respectively).}

\end{algorithm}

On line~\ref{line:extremalTest}, we eliminate any geometry which violates the extremal bound given by Theorem~\ref{thm:extremalUpperBound}, thus restricting to geometries for which $m < kn$.
The invocation of $\ObtainMaxFamilyPaths(G,I,O)$ constructs mutually inverse partial functions on $V(G)$ whose orbits describe a maximum-size family of vertex disjoint paths, by Lemma~\ref{lemma:obtainMaxFamilyPaths}; the loop starting at line~\ref{line:testPathCoverLoop} then determines whether or not this family is a path cover, by seeing if at least one of $f$ and $g$ are defined for all vertices not in $I \inter O$.
If there are vertices $v \notin \dom(f) \union \dom(g) \union (I \inter O)$, then the family of paths obtained is in particular not a causal path cover: by Theorems~\ref{thm:uniqueFamilyPaths} and~\ref{thm:graphth-charn}, $(G,I,O)$ then has no flow.
Otherwise, $f$ is a successor function for $(G,I,O)$, and we may test whether or not its natural pre-order $\preceq$ is a partial order by invoking $\TestNaturalPreorder(G,I,O,f,g)$ on line~\ref{line:constructPreorder}.
By Lemma~\ref{lemma:testNaturalPreorder}, if $\partialOrd = \nil$ on line~\ref{line:testPartialOrder}, then $\preceq$ is not a partial order, and subsequently $\cP_f$ is not a causal path cover; again by Theorems~\ref{thm:uniqueFamilyPaths} and~\ref{thm:graphth-charn}, $(G,I,O)$ then has no flow.
Otherwise, $\partialOrd$ contains a system of infima for $\preceq$, and we return $(f,g,\partialOrd)$.

The run time for $\FindFlow$ is dominated by those of $\ObtainMaxFamilyPaths$ and $\TestNaturalPreorder$, which are both $O(km)$.
Due to the restriction imposed at line~\ref{line:extremalTest}, we then have the following:
\begin{theorem}
	\label{thm:findFlow}
	For a geometry $(G,I,O)$ with $\card{V(G)} = n$ and $\card{I} = \card{O} = k$, the procedure $\FindFlow(G,I,O)$ terminates in time $O(k^2 n)$, returns $\nil$ if $(G,I,O)$ has no flow, and a triple $(f,g,\partialOrd)$ otherwise; where $f: V(G) \partialto V(G)$ is a successor function for $(G,I,O)$, $g: V(G) \partialto V(G)$ is the inverse of $f$, and $\partialOrd$ is a system of infima for the natural pre-order $\preceq$ of $f$.
	In particular, if $(G,I,O)$ has a flow, then $(f, \preceq)$ is a flow for $(G,I,O)$.
\end{theorem}

\section{A semantic map for the DKP representation}
\label{sec:semanticDKP}

The purpose of this Section is to show how to obtain a right-inverse to the DKP representation map $\cR$, from unitary circuits to one-way measurement based algorithms.
We show how this may be done for a one-way measurement patterns with an underlying geometry $(G,I,O)$, and whose operations satisfy certain constraints related to a flow.

\subsection{Influencing walks and measurement adaptation}
\label{sec:dependencies}

For a geometry $(G,I,O)$ with a flow $(f, \preceq)$, the partial order $\preceq$ is meant to describe an order of measurements which allow a unitary embedding to be performed.
However, it represents a \emph{sufficient} condition, not a necessary one: and in particular, as shown in~\cite{DK05u}, it is a sufficient condition under the constraints that a measurement of a qubit $v \in O\comp$ affects the following operations:
\begin{romanum}
\item
	it induces a sign dependency on the measurement of its successor $f(v)$, and a $\pi$-dependency on the measurements of its other neighbors $w \sim f(v)$, and affects no other measurements; and

\item
	it induces an $X$ correction on $f(v)$ if this is an element of $O$, and a $Z$ correction on any neighbor $w \sim f(v)$ which are elements of $O$, and induces no other corrections.
\end{romanum}
These constraints essentially arise from the simplified DKP construction, as described in Section~\ref{sec:flowDefn}.

As we noted in Section~\ref{sec:DKPstandardization}, we may eliminate $\pi$-dependencies in a one-way measurement pattern based computation by performing signal shifting: this may result in more qubits depending on a given measurement result, but relaxes the measurement dependencies to allow transformations to be done with less depth complexity (neglecting the cost of computing boolean expressions for adapting the measurements).
As well, measurements along Pauli axes are not affected by any previous measurements (although their results may be affected, in the counterfactual sense described in Section~\ref{sec:flowDefn}).
We will see that, as a result, the appropriate tool for considering the measurement order for standardized one-way measurement patterns is not the partial order $\preceq$, but rather a special class of influencing walks.\footnote{%
The importance of influencing walks for depth-complexity in one-way measurement based computation was first realized by Broadbent and Kashefi, who first coined the term \emph{influencing path} while working on the results of~\cite{BK2007}.
The recognition of influencing paths as objects of interest helped to inspire simplifications in the work which I've presented in this Chapter form their original forms, and were incorporated in the published version of these results in~\cite{Beaudrap08}.
We will consider here a different approach to the analysis of depth of one-way measurement based procedures in terms of influencing walks than the work in~\cite{BK2007}.}

For a unitary circuit $C$, we may perform the DKP construction by first producing a pattern $\tilde P$ via the \emph{simplified} DKP construction.
As we remarked in Section~\ref{sec:flowDefn}, we have $v \preceq w$ for two distinct qubits involved in $\tilde P$ if the measurement of $v$ induces $X$ or $Z$ corrections onto $w$, which are absorbed into the measurement basis for $w$ if $w \in O\comp$.
These corrections are induced if and only if either $w = f(v)$, in which case the induced correction is $\Xcorr{}{\s[v]}$, or $w \sim f(v)$, in which case the induced correction is $\Zcorr{}{\s[v]}$.
We will refer to these dependencies as \emph{type \iti} and \emph{type \itii} dependencies.
Note that the relationships of qubits with dependencies of type~\iti\ or type~\itii\ are slight generalizations\footnote{%
A segment of type~\itii\ with respect to the path cover $\cP_f$ is a path of the form $v \arc z \arc w$, where the edge $zw$ is not covered by a path of $\cP_f$\,; measurement dependencies of type~\itii, by contrast, include the case $w = f(f(v))$.}
of the relationships occurring in segments of type~\iti\ and type~\itii\ of influencing walks; what is significant about these types is in particular whether the dependency arises from an $X$ or a $Z$ correction induced by the measurement of a qubit.

Consider the further operations required to standardize $\tilde P$, into a standardized pattern:
\begin{enumerate}
\item
	As we noted in Section~\ref{sec:DKPstandardization}, measurements along a Pauli axis can be performed independently of any previous measurement; then, for qubits $v \in O\comp$ and $w = f(v) \in O\comp$, we may eliminate any dependency of type~\iti\ of $w$ on $v$ if the former is to be measured with an angle $\alpha \in \frac{\pi}{2} \Z$.

\item
	For a measurement along the $Y$ axis in particular (\ie\ for measurement angles $\alpha \in \pi \Z + \pion 2$), the way in which type~\iti\ dependencies are eliminated are to replace sign-dependencies with $\pi$-dependencies, which is the effect of a dependency of type~\itii.
	Thus, when $w = f(v)$ for some $v \in O\comp$ and $w \in O\comp$ measured with an odd multiple of $\pion 2$, we may replace the type~\iti\ dependency of $w$ on $v$ with a dependency of type~\itii.

\item
	Whenever a qubit $w \in O\comp$ has a type~\itii\ dependency on a qubit $v$, the effect of the measurement result of $v$ is not to change the basis of measurement for $w$, but rather the significance of the measurement result of $w$, which is formalized in the process of signal shifting.
	Specifically, for any qubit $w$ which has a type~\itii\ dependency on another qubit $v$, we may eliminate the dependency of $w$ on $v$ if we make every qubit $x$ with a type~\iti\ dependency on $w$ also depend on $v$, and similarly for qubits $y$ with a type~\itii\ dependency on $w$.
\end{enumerate}
Recursively applying the last rule, we may describe walks in $G$ similar to influencing walks, of the kind described below:
\begin{definition}
	For a geometry $(G,I,O)$, a successor function $f: O\comp \to I\comp$, and a vector $\vec a = (\alpha_v)_{v \in O\comp}$ of default measurement angles for each qubit, a \emph{segment of an adapting walk} for $f$ is any directed path of one of the following three forms:
	\begin{romanum}
	\item
		$v \arc w$, where $w = f(v)$, and $w$ is not to be measured with an angle $\alpha \in \frac{\pi}{2} \Z$\,;

	\item
		$v \arc z \arc w$, where $v \in O\comp$, $z = f(v)$, and $w \sim z$\,;

	\item
		$v \arc w$, where $w = f(v)$, and $\alpha_w \in \pi\Z + \pion 2$\,.
	\end{romanum}
	An \emph{adapting walk} for $f$ is any directed walk $W$ in $G$ which may be formed concatenation of such segments, subject to the constraint that a type~\iti\ segment may only occur at the end of the walk, and that either $W$ ends with a segment of type~\iti\ or at a vertex in $O$.
\end{definition}
Note that an adapting segment for $f$ of type~\itiii\ is also a segment of type~\iti\ of an influencing walk of $\cP_f$\,.
Thus, all adapting walks for $f$ are also influencing walks of $\cP_f$.

We will show that adapting walks describe the propagation of measurement or correction dependency on previous measurement results for one-way measurement based patterns arising from the DKP construction, with the type of adapting walk determining the nature of the dependency, and that these dependencies may cancel when an even number of them converge on a given qubit:

\begin{theorem}
	\label{thm:dependenciesDKP}
	Let $P = \cR(C)$ be the image of a unitary circuit $C$ in the DKP construction $\cR$\,, where $C$ is given as a stable index tensor expression.
	Let $(G,I,O)$ be the geometry consisting of the graph $G$ of entangling operations on qubits of $P$ (which are indexed by wire-segments of $C$), the set $I$ of non-prepared qubits, and the set $O$ of non-measured qubits, and let $\vec a = (\alpha_v)_{v \in O\comp}$ be the vector of default measurement angles for each qubit.
	If $f: O\comp \to I\comp$ is the function mapping each deprecated index in $C$ to the corresponding advanced index, then $f$ is a flow function for $(G,I,O)$; and for two distinct qubits $v,w \in V(G)$, the following holds:
	\begin{romanum}
	\item
		there is a measurement on $w \in O\comp$ where the sign of the measurement angle depends on $\s[v]$, or a potential $X$ correction on $w \in O$ depending on $\s[v]$, if and only if there are an odd number of adapting walks for $f$ from $v$ to $w$ which end in a segment of type~\iti;

	\item
		there is a potential $Z$ correction on $w \in O$ depending on $\s[v]$ if and only if there are an odd number of adapting walks for $f$ from $v$ to $w$ which end in a segment of type~\itii.
	\end{romanum}
\end{theorem}
\begin{proof}
	Consider the one-way measurement pattern $\tilde{P}$ obtained from $C$ by the \emph{simplified} DKP construction.
	From the discussion in Section~\ref{sec:flowDefn}, the function $f$ is a flow-function by construction.
	We may obtain the one-way measurement pattern  $P$ by standardizing the procedure $\tilde P$\,.

	Consider how the process of standardization propagates the dependencies of qubits $w$ on a particular qubit $v$.
	Let us say that a qubit $w$ has an \emph{$X$-dependency} on a qubit $v$ in a measurement procedure if either the measurement of $w \in O\comp$ has a sign dependency on $\s[v]$, or if $w \in O$ may be subject to an $X$ correction depending on the value of $\s[v]$; and similarly, a qubit $w$ has a \emph{$Z$-dependency} on a qubit $v$ in a measurement procedure if either the measurement of $w$ has a $\pi$-dependency on $\s[v]$, or if $w$ may be subject to a $Z$ correction depending on the value of $\s[v]$.
	If a qubit $w$ has a $Z$-dependency on another qubit $v$, performing signal shifting on the measurement of $w$ will cause any qubits with an $X$-dependency on $w$ to
	\begin{inlinum}
	\item
		accrue an $X$-dependency on $v$, if it had none previously, or

	\item
		lose its $X$-dependency on $v$, if it already had such a dependency, due to cancellation modulo $2$\,;
	\end{inlinum}
	and similarly for $Z$-dependencies.

	We may then describe how standardization transformations cause dependencies to propagate, in terms of ``locally describable'' dependencies.
	Starting from the measurement procedure $\tilde P$, the only $X$-dependencies which exist are of vertices $f(v)$ on their respective qubits $v \in O\comp$\,; then we may reduce the problem to tracking $Z$ dependencies.
	Because the $Z$-dependencies of a qubit $w$ may always be removed unless $w \in O$, we may model this propagation in the form of a ``binary walk'' of a measurement result $\s[v]$, starting from $v$ and ending at vertices which have unremovable $Z$-dependencies.
	We do this as follows:
	\begin{enumerate}
	\item
		For a qubit $w \in O\comp$ and $z \sim f(w)$ distinct from $w$, signal shifting on the measurement at $w$ will propagate the $Z$-dependencies of $w$ to $z$ (possibly cancelling modulo $2$ as described above).

	\item
		For a qubit $w \in O\comp$ and a qubit $z = f(w)$ which is to be measured with an angle $\alpha \in \pi\Z + \pion 2$, the $X$-dependency of $z$ on $w$ is equivalent to a $Z$-dependency; signal shifting on the measurement at $w$ will then propagate the $Z$-dependencies of $w$ to $z$ (possibly cancelling modulo $2$ as described above).
	\end{enumerate}
	Then, we may describe the propagation of the effects of a measurement result $\s[v]$ by an $V(G) \x V(G)$ matrix $T$ over $\Z$, given by
	\begin{align}
		\label{eqn:shiftPropagator}
			T_{wv}
		\;\;=\;\;
			\begin{cases}
				1\,,	&	\text{if $v \in O\comp$, $w \sim f(v)$, and $w \ne v$}	\\[1ex]
				1\,,	&	\begin{minipage}[t]{15em}
								if $v \in O\comp$, $w = f(v)$, and $w$ is to be measured with angle $\alpha \in \pi\Z + \pion 2$
								\\[1ex]
				    	 	\end{minipage}	\\[-2ex]
				0\,,	&	\text{otherwise}
			\end{cases}\;.
	\end{align}
	Note that $T_{wv} = 1$ only if $v$ and $w$ are connected by a segment of an influencing walk for $\cP_f$\,, and in particular, an adapting segment for $f$ of type~\itii\ or~\itiii.
	Then, $\unit_v \cdot (T^\ell\, \unit_w) \ne 0$ only if $v$ and $w$ are connected by an influencing walk for $\cP_f$ consisting of $\ell$ segments.
	Because influencing walks of $\cP_f$ are of bounded length (in particular because there are no vicious circuits), there exists an $\ell$ for which $T^\ell = 0$\,.
	In particular, there are no eigenvectors of $T$ with eigenvalue $1$: then, the matrix $\idop_n - T$ has trivial kernel, and is therefore invertible (and in particular, invertible modulo $2$).

	The matrix $T$ represents the propagation of a $Z$-measurement dependency due to signal shifting or simplification of Pauli measurements, but neglects $X$-dependencies of qubits not to be measured along a Pauli axis, as well as the persistent $Z$-dependencies of elements of $O$, which cannot be signal-shifted --- these are dependencies which cannot be eliminated by pattern transformation.
	If after some number of standardization operations, a qubit $z$ has an unremovable dependency on a qubit $w$ with a (removable) $Z$-dependency on some qubits, the dependencies of $w$ are accumulated into the dependencies of $z$ by a signal shifting on the measurement at $w$.
	We describe this accumulation separately for $X$- and $Z$-dependencies.
	\begin{romanum}
	\item
		Consider another $V(G) \x V(G)$ matrix $F$ over $\Z$, given by
		\begin{align}
		\label{eqn:flowPropagator}
				F_{wv}
			\;\;=\;\;
				\begin{cases}
					1\,,	&	\begin{minipage}[t]{17em}
									if $v \in O\comp$, $w = f(v)$, and $w$ is not to be measured with an angle $\alpha \in \pi\Z + \pion 2$
									\\[1ex]
					    	 	\end{minipage}	\\[-2ex]
					0\,,	&	\text{otherwise}
				\end{cases}\;.
		\end{align}
		Then $F$ represents the effect of a signal shift on the $X$-dependencies of vertices $f(v)$ on their predecessors $v \in O\comp$.
		For qubits of the form $w = f(z)$ for $z \in O\comp$ arbitrary, the contribution of a possible dependency of $w$ on a qubit $v$ arising from the $\ell\th$ signal shift for $\ell \ge 0$ is then given by
		\begin{gather}
				\unit_w \cdot \paren{ F T^\ell \,\unit_v };
		\end{gather}
		and the cumulative dependency over all signal shifting operations is then given by
		\begin{align}
				\unit_w \cdot \paren{\sum_{\ell \in \N} F T^\ell\, \unit_v}
			\;\;=&\;\;
				\unit_w \cdot F \paren{\,\sum_{\ell \in \N} \, T^\ell } \unit_v
			\notag\\=&\;\;
				\unit_w \cdot F \paren{\idop_n - T}\inv\, \unit_v	\,,
		\end{align}
		as $\idop - T$ is non-singular.
		If this evaluates to zero modulo $2$, there is no $X$-dependency of $w$ on $v$ in the pattern $P$.\\[-2ex]
		
	\item
		For $w \in O$, the $Z$-dependency of $w$ on an arbitrary qubit $v$ can be computed as the cumulated contributions from each possible number $\ell$ of signal shifts, which for each $\ell \ge 1$ is given by
		\begin{gather}
				\unit_w \cdot (T^\ell \,\unit_v);
		\end{gather}
		then, similarly to the previous case, the cumulative dependency over all signal shifting operations is then given by
		\begin{align}
				\unit_w \cdot \paren{\sum_{\ell \in \N} T^\ell\, \unit_v}
			\;\;=&\;\;
				\unit_w \cdot \paren{\idop_n - T}\inv\, \unit_v	\,,
		\end{align}
		as $\idop - T$ is non-singular.
		Again, if this evaluates to zero modulo $2$, there is no $Z$-dependency of $w$ on $v$ in the pattern $P$.
	\end{romanum}
	After an ``infinite'' number of such dependency propagations performed on the initial procedure $\tilde{P}$ (equivalent to some finite number of shifts, with subsequent propagations having no effect), we obtain the measurement procedure $P$.

	Note that the coefficients of $F$ corresponds to adapting segments for $f$ of type~\iti, and those of $T$ correspond to adapting segments of type~\itii\ and~\itiii.
	In the procedure $P$, possible dependencies of the signs of measurement angles for a qubit $w$ depending on $\s[v]$, of potential $X$ corrections on a qubit $w$ depending on $\s[v]$, exist then if there are an odd number of walks from $v$ to $w$ consisting of some number of type~\itii\ and~\itiii\ adapting segments followed by a type~\iti\ segment; and possible $Z$ corrections on $w$ depending on $\s[v]$ exist if there are an odd number of walks from $v$ to $w \in O$ consisting entirely of type~\itii\ and~\itiii\ adapting segments.
	This proves the Theorem.
\end{proof}

\begin{algorithm}[t]

\PROCEDURE \TestDependencies(P, f) :

\Input{%
	$P$, a standardized one-way measurement pattern;
	\\
	$f$, a flow-function for the geometry $(G,I,O)$ underlying $P$
}

\Output{%
	A boolean result indicating if $P$ is consistent with the constraints of Theorem~\ref{thm:dependenciesDKP}.
}

\BEGIN
	\algline
	
	\LET $T \gets $ [\,matrix over $\Z_2$ depending on $f$ described in Eqn~\eqref{eqn:shiftPropagator}\,]\;
	\LET $F \gets $ [\,matrix over $\Z_2$ depending on $f$ described in Eqn~\eqref{eqn:flowPropagator}\,]\;
	\LET $D \gets (\idop_n - T)\inv$\;
	\LET $E \gets F D$\;
	\FOREACH measurement or correction operation $\Op{w}{B}$ in $P$ \DO
		\algline
		\LET $\xType \gets ((\Op{}{} = \Meas{}{}) ~\ORELSE~ (\Op{}{} = \Xcorr{}{}))$\;
		\LET $\zType \gets (\Op{}{} = \Zcorr{}{})$\;
		\LET $S \gets \ens{v \in V(G) \,\big|\, \text{$\s[v]$ occurs in $B$}}$\;		\nllabel{line:scanDependencies}

		\medskip
		{\FOREACH $v \in V(G)$ \DO
			\IF $v \in S$ \THEN {
				\short
				\IF $(\xType ~\ANDALSO~ E_{wv} = 0) ~\ORELSE~ (\zType ~\ANDALSO~ D_{wv} = 0)$ \THEN {\RETURN \false}
				\relax
			}
			\ELSE {
 				\short
				\IF $(\xType ~\ANDALSO~ E_{wv} = 1) ~\ORELSE~ (\zType ~\ANDALSO~ D_{wv} = 1)$ \THEN {\RETURN \false}
				\relax
			}
		\ENDFOR}
	\ENDFOR

	\medskip
	\RETURN \true\;
\END

	\caption[An algorithm to test whether the dependencies of a standardized one-way measurement pattern based measurement procedure is consistent with those produced by the DKP construction.]{\label{alg:testDependencies}%
		An algorithm to test whether the dependencies of a standardized one-way measurement pattern based measurement procedure is consistent with those produced by the DKP construction (see Theorem~\ref{thm:dependenciesDKP}).
		}

\end{algorithm}

\label{discn:testDependencies}
Algorithm~\ref{alg:testDependencies} describes an algorithm which, provided a one-way measurement pattern $P$ in standard form (and without shift operators) and a flow $f$ for the geometry underlying $P$, tests if the dependencies of $P$ are consistent with the flow-function those described in Theorem~\ref{thm:dependenciesDKP}.
Let $N$ be the length of the measurement pattern $P$, and let $n = \card{V(G)}$, $m = \card{E(G)}$, and $k = \card{O}$.
There will then be exactly $m$ entangling operations, $n -k$ measurement operations, and $k$ measurement operations in $P$, where the latter two classes of operations may in principle depend on arbitrary subsets of $V(G)$; then, the length of the pattern is $N \in O(m + n^2) = O(n^2)$.
The time required to construct the matrices $T$ and $F$ is $O(n^2)$, and the time to compute $D$ and $E$ (using \eg\ the algorithm of~\cite{PMH08} to perform matrix inversion modulo $2$) is $O(n^3 / \log n)$.
The set $S$ constructed on line~\ref{line:scanDependencies} may be constructed in time $O(n)$\,; then, the run-time of Algorithm~\ref{alg:testDependencies} is dominated by the nested for loops and repeated construction of $S$, both of which requires $O(Nn) \subset O(n^3)$ time in total to execute.

The characterization of dependencies between qubits in a pattern $P \in \img(\cR)$ is the penultimate step towards obtaining a semantic map for $\cR$.
Together with an algorithm to find flows, Algorithm~\ref{alg:testDependencies} provides an efficient algorithm for testing membership in $\img(\cR)$\,: however, the proof of this claim will await the construction of a complete semantic map for $\cR$.

\subsection{Star decomposition of geometries with flows}
\label{sec:starDecomposition}

For a geometry $(G,I,O)$ with a flow-function $f$, the influence relation $\inflRel$ for $f$ suggests a decomposition of $(G,I,O)$ into smaller geometries similar to the star decomposition of circuits described in Section~\ref{sec:flowDefn}.
For each vertex $v \in V(G)$, define the vertex-set
\begin{align}
		V_{v \star}
	\;=&\;
		\ens{w \in V(G) \,\Big|\, w = v \text{~or~} \big[v \inflRel w \text{~and~} w \ne f(f(v)) \big]} \,,
\end{align}
and from these, define the geometry
\begin{subequations}
\label{eqn:starGeometry}
\begin{gather}
			G_{v \star}
		\;\;=\;\;
			G[V_{v \star}]	\,-\, \ens{xw \,\big|\, x,w \ne f(v)}\,,
		\\[1.5ex]
			I_{v \star}
		\;\;=\;\;
				V_{v \star} \setminus \ens{f(v)}	\,,
		\\[1ex]
			O_{v \star}
		\;\;=\;\;
				V_{v \star} \setminus \ens{v}	\,.
\end{gather}
\end{subequations}
The graph $G_{v\star}$ consists of a star graph consisting of a single edge $v w$ for $w = f(v)$, and some additional edges $wx$ for $w = f(v)$ and $v \inflRel x$.
We will call the geometry $(G_{v \star}\,, I_{v \star}\,, O_{v \star})$ a \emph{star geometry}\footnote{%
Star geometries as we have defined them are similar to, but distinct from, the geometries underlying \emph{star patterns} in~\cite{DK06}.
The decomposition result of Lemma~\ref{lemma:starDecomp} was motivated by the discussion of star geometries in that article.}
with \emph{root} $v$, or with \emph{center} $f(v)$.

\begin{lemma}
	\label{lemma:starDecomp}
	Let $(G,I,O)$ be a geometry with flow-function $f$.
	Then there is a decomposition of $(G,I,O)$ into star geometries on each of the vertices $v \in O\comp$, together with the geometry $\big(G\big[\,\tilde{I}\,\big], I, \tilde{I}\,\big)$, where $\tilde{I} = V(G) \setminus \img(f)$.
\end{lemma}
\begin{proof}
	We proceed by induction on the size of $\img(f)$.
	If $\img(f) = \vide$, then $\dom(f) = \vide$ as well, in which case $I \subset V(G) = O$; we then have $(G,I,O) = \big(G\big[\,\tilde{I}\,\big], I, \tilde{I}\,\big)$.
	Otherwise, suppose that the claim holds for geometries $(G,I,O)$ with flow-functions $f$ in which $\card{\img(f)} \le N$ for some $N \in \N$, and that $\card{\img(f)} = N+1$.
	Because $f$ is a flow function, $\sI_f$ is acyclic; then, there is a vertex $v \in V(G)$ such that $v \inflRel w$ only for $w$ maximal in the natural pre-order $\preceq$.
	Let $\cG_{v\star} = (G_{v\star}\,, I_{v\star}\,, O_{v\star})$, and $\cG' = (G', I, O')$, where
	\begin{gather}
		\label{eqn:geometryReduce}
			G'
		\;=\;
			G \setminus f(v)	\,,\quad\text{and}
		\qquad	
			O'
		\;=\;
			O \symdiff \ens{v, f(v)}	\,.
	\end{gather}
	Note that $O_{v\star} \subset O$ by the choice of $v$, and $I_{v\star} = O_{v \star} \symdiff \ens{v, f(v)} \subset O \symdiff \ens{v, f(v)}$\,.
	Because $V(G_{v\star}) = I_{v\star} \union \ens{f(v)}$, we then have $V(G') \inter V(G_{v\star}) \subset I_{v\star} \inter O'$\,: then the composition $(\bar{G}, \bar{I}, \bar{O})  \,=\, \cG_{v\star} \circ \cG'$ is well-defined, and we have
	\begin{subequations}
	\begin{align}
			\bar{G}
		\;\;=&\;\;
			G_{v\star} \union G' 
		\notag\\=&\;\;
			\ens{xw \in G \,\big|\, w = f(v)} \union \Big[ G \setminus f(v) \Big] 
		\;\;=\;\;
			G	\,;
		\\[2ex]
			\bar{I}
		\;\;=&\;\;
			\big( I_{v\star} \setminus O' \big) \union I
		\;\;=\;\;
			\vide \union I
		\;\;=\;\;
			I	\,;
		\\[2ex]
			\bar{O}
		\;\;=&\;\;
			O_{v\star} \union \big( O' \setminus I_{v\star} \big)
		\;\;=\;\;
			O_{v\star} \union \big( O \setminus O_{v\star} \big)
		\;\;=\;\;
			O	\,.
	\end{align}
	\end{subequations}
	Then, we have $(G,I,O) = \cG_{v\star} \circ \cG'$\,.
	The restriction $f'$ of the flow-function $f$ to $\cG'$ does not include $f(v)$, and therefore has cardinality at most $N$\,; then, $\cG'$ has a decomposition into star geometries and the geometry $(G' \setminus \img(f'), I, V(G') \setminus \img(f'))$.
	Finally, note that $V(G) \setminus \img(f') = V(G) \setminus \img(f) = \tilde{I}$, by the definition of $f'$\,: the lemma then holds by induction.
\end{proof}

\begin{figure}[p]
	\begin{center}
		\includegraphics{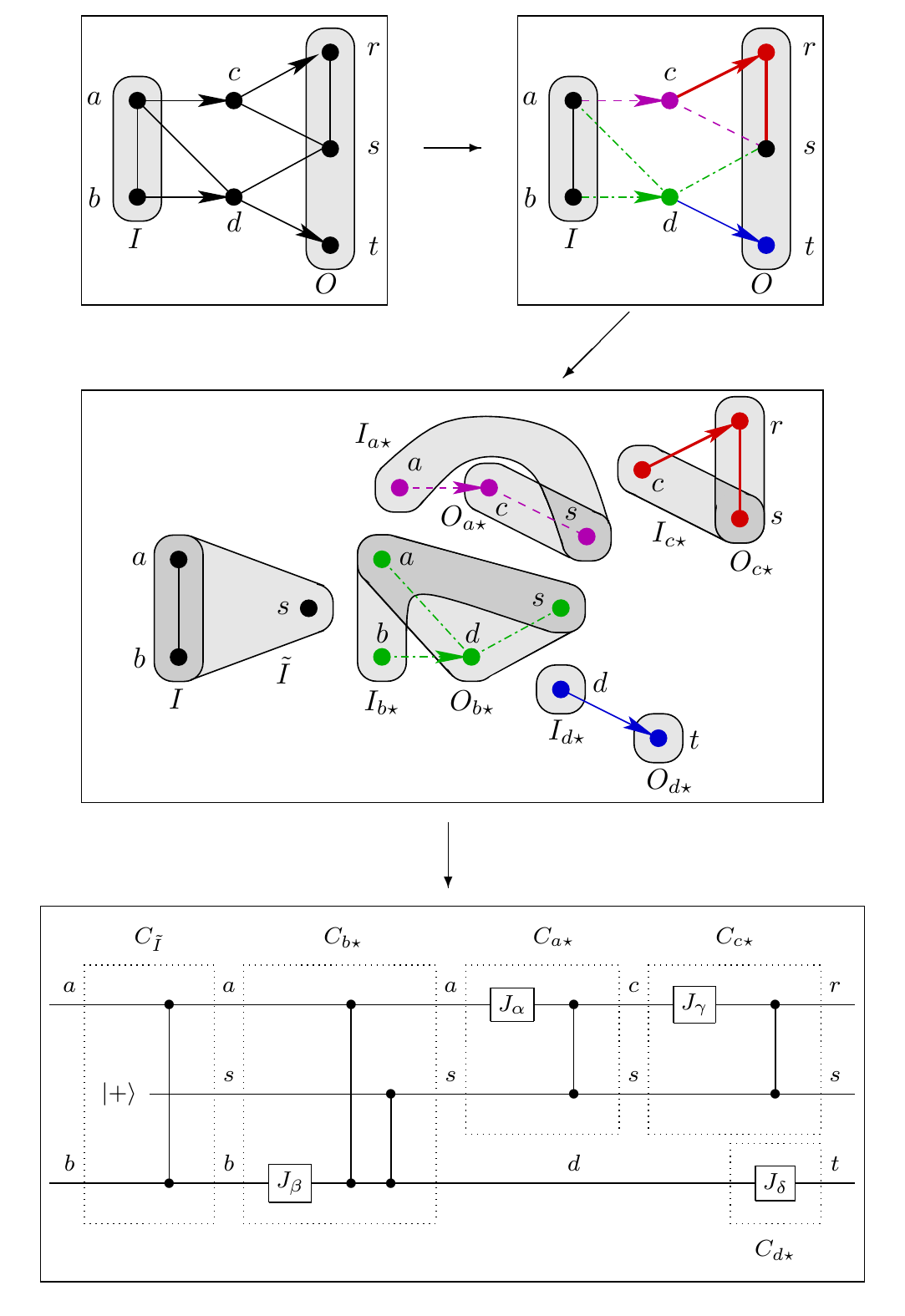}
	\end{center}
	\caption[Illustration of a decomposition of a geometry with a flow-function into star geometries.]{\label{fig:starDecomp}\colouronline
		Illustration of a decomposition of a geometry $(G,I,O)$ with a flow-function $f$ (denoted by arrows in the upper three panels) into star geometries for each $v \in O\comp$, and a reduced geometry $\big(G\big[\,\tilde I\,\big], I, \tilde I\big)$ for $\tilde{I} = V(G) \setminus \img(f)$.
		The middle panel illustrates each star geometry separately, located next to the geometries with which it may be composed to form the original geometry $(G,I,O)$; vertices are repeated for each geometry in which they occur.
		The bottom panel illustrates the unitary circuits corresponding to these star geometries (including a labelling of wire-segments, \ie\ of stable tensor indices), and the complete circuit obtained from composing them.}
\end{figure}
Figure~\ref{fig:starDecomp} illustrates a decomposition of a geometry with a flow into star geometries in the manner described above.

For a geometry $(G,I,O)$ with a given flow-function $f$, we will say that a star geometry $(G_{v\star}\,, I_{v\star}\,, O_{v\star})$ is \emph{maximal} in $(G,I,O)$ if, as in the proof of Lemma~\ref{lemma:starDecomp} above, every $w \in V(G)$ such that $v \inflRel w$ is maximal in $\preceq$\,; then, we may decompose $(G,I,O)$ by extracting maximal star geometries.
The maximal elements of $V(G)$ in the natural pre-order are exactly the elements of $O$ (as $z \preceq f(z)$ for any $z \in O\comp$).
Thus, the root $v$ of a maximal star geometry in $(G,I,O)$ is one such that all neighbors $z \sim f(v)$ are elements of $O$ or equal to $v$ itself.
There always exists such a maximal star for a geometry with a flow: an independent observation of a similar nature has since given rise to an improved algorithm for finding flows, which we describe in Section~\ref{sec:improvedFlowAlgs}.

\label{discn:maximalStarVertex}
Algorithm~\ref{alg:maximalStarVertex} describes an procedure to obtain a vertex $w = f(v) \in V(G)$ which is the center of a maximal star geometry in some geometry $(G[V'], V' \setminus W', O')$, using the fact that $w$ is the center of such a star geometry if and only if all neighbors of $w$ in the graph $G[V']$ are either equal to $v$ or are elements of $O'$.
Having found a vertex at the center of a maximal geometry of $(G,I,O)$, we may obtain the decomposition $(G,I,O) = \cG_{v\star} \circ \cG'$ by constructing these geometries according to~\eqref{eqn:starGeometry} and~\eqref{eqn:geometryReduce}, respectively.
The run-time of Algorithm~\ref{alg:maximalStarVertex} is taken up by constant-time comparisons for up to all of the neighbors of every element of $O'$, and so requires time
\begin{align}
		O\paren{\sum_{w \in O'} \deg(w)}	
	\;\;\subset\;\;
		O\paren{\sum_{w \in V(G)} \deg(w)}
	\;\;=&\;\;\;
		O(2\card{E(G)})
	\notag\\=&\;\;
		O(m)
	\;\;\subset\;\;
		O(k n)	\;,
\end{align}
by the extremal bound of Theorem~\ref{thm:extremalUpperBound}.

\begin{algorithm}[t]
	\PROCEDURE \MaximalStarVertex(G, V', W', O', f):

	\Input{%
		$G$, a graph underlying a geometry $(G,I,O)$;
	\\
		$V', W', O'$, sets describing a sub-geometry $(G[V'], V' \setminus W',O')$;
	\\
		$f: O\comp \to I\comp$, a flow-function function for $(G,I,O)$
	}

	\Output{%
		A vertex $w \in W'$ such that every neighbor $v$ of $w$ such that $v \in V'$ either satisfies $v = g(w)$ or $v \in O'$
	}

	\BEGIN
		\algline
		\FOREACH $w \in W'$ \DO
			\LET $\maximalStar \gets \true$\;
			{\FORALL $v \sim w$ \DO
				\IF $v \in V' \setminus O'$ \ANDALSO $w \ne f(v)$ \THEN {%
					$\maximalStar \gets \false$\;
					\ESCAPE\;
				}\relax
			\ENDFOR}

			\smallskip
 			\oneline
			\IF $\maximalStar$ \THEN \RETURN $w$\;
		\ENDFOR
	\END

	\caption{\label{alg:maximalStarVertex}%
		An algorithm to obtain a vertex $v$ which is the center of a maximal star geometry of a specified geometry.}

\end{algorithm}

\subsection{A semantic map from star decompositions}

Star geometries $(G_{v\star}\,, I_{v\star}\,, O_{v\star})$ correspond to simple unitary circuits in the following manner.
We may construct a circuit $C_{v\star}$ with an input wire for each qubit $a \in I_{v \star}$\,, and which performs the transformation
\begin{gather}
		C_{v\star}
	\;\;=\;\;
		\paren{\prod_{\substack{w \in O_{v\star} \\[0.25ex] w \sim f(v)}} \cZ \pseu[w, f(v)] } J(\alpha) \pseu[:f(v)/v]
\end{gather}
expressed in stable index notation (pages~\pageref{sec:nonstandardRepnsEnd}\thru\pageref{sec:nonstandardRepnsEnd}); we map the labels of the vertices in the geometries to the tensor index names in the expansion, and the angle $\alpha$ is a parameter which we may relate to the measurement angle for the qubit $v$.
Similarly, we may define a circuit $C_{\tilde I}$\ which corresponds to the geometry $(G[\,\tilde{I}\,], I, \tilde{I})$\,:
\begin{gather}
		C_{\tilde I}
	\;\;=\;\;
		\paren{\prod_{\substack{vw \in E(G[\,\tilde I\,])}} \cZ \pseu[v,w] } \paren{\prod_{v \in \tilde I \setminus I} \ket{+} \pseu[:v/] }	\;.
\end{gather}
We will call such circuits the \emph{star circuits} corresponding to these geometries; these are the same circuits described in Section~\ref{sec:flowDefn} in the star decomposition of circuits.
Examples of such circuits for a particular geometry $(G,I,O)$ are illustrated in the bottom panel of Figure~\ref{fig:starDecomp}, in which they are composed together to form a larger circuit.
In that example, we may recover the geometry in the upper left-hand panel by composing
\begin{align}
	\begin{split}
		(G,I,O)
	\;=\;
		(G_{c\star}\,, I_{c\star}\,, O_{c\star})
		\circ
		(G_{d\star}\,, I_{d\star}\,,& O_{d\star})
		\circ
		(G_{a\star}\,, I_{a\star}\,, O_{a\star}) \circ
	\\&
		(G_{b\star}\,, I_{b\star}\,, O_{b\star})
		\circ
		\big(G\big[\,\tilde{I}\,\big], I, \tilde{I}\big)	\,;
	\end{split}
\end{align}
and similarly, the larger circuit $C$ in Figure~\ref{fig:starDecomp} is obtained by composing
\begin{align}
		C
	\;=\;
		C_{c\star} \circ C_{d\star} \circ C_{a\star} \circ C_{b\star} \circ C_{\tilde I}	\;.
\end{align}
Algorithm~\ref{alg:semanticDKP} describes a procedure $\SemanticDKP(G,I,O,P)$ which provides a semantic map $\cS$ for the DKP representation map $\cR$.
\begin{algorithm}[p]

\PROCEDURE \SemanticDKP(G,I,O,P) :

\Input{%
		$P$, a standardized one-way measurement pattern, with an equal number of input and output qubits;
	\\
		$(G,I,O)$, the geometry underlying $P$ (in particular, $\card I = \card O$).
}

\Output{%
	Either \nil, if $(G,I,O)$ has no flow or the dependencies of the operations of $P$ are not consistent with the constraints of Theorem~\ref{thm:dependenciesDKP}; or a unitary circuit $C$ performing the same transformation as $P$, otherwise.}

\BEGIN
	\LET $\flow \gets \FindFlow(G,I,O)$\;
	\short
	\IF $\flow = \nil$ \THEN {%
		\RETURN \nil																									\nllabel{line:flowFail}
	}
	\ELSE {
		\LET $(f,g,\partialOrd) \gets \flow$
	}

	\medskip

	\LET $\theta: V(G) \to \R$\vspace{-1ex}\;
	\oneline
	\FOREACH measurement operation $\Meas{v}{\alpha;B}$ in $P$ \DO
		$\theta(v) \,\gets\, -\alpha$\;
	\ENDFOR
	\oneline
	\IF $\TestDependencies(P,f) = \false$ \THEN \RETURN \nil\;										\nllabel{line:dependenciesFail}

	\medskip

	\LET $O' \gets O$\;																							\nllabel{line:initGeometry}
	\LET $V' \gets V(G)$\;
	\LET $W' \gets V(G) \setminus I$\;																		\nllabel{line:initGeometryEnd}
	\LET $C 	\gets $ [\,the empty circuit \,]\;
	\medskip

	\algline
	\WHILE $W' \ne \vide$ \DO																					\nllabel{line:circuitConstrLoop}
		\LET $w \gets \MaximalStarVertex(G,V',W',O',f)$\vspace{0.1ex}\;
		\LET $v \gets g(w)$\vspace{0.3ex}\;
		\LET $C' \gets  J(\theta(v)) \pseu[:w/v]$\,\;
		\oneline
		\FORALL $z \sim w$ \SUCHTHAT $z \ne v$ \DO
			$C' \gets \cZ \pseu[w,z] \circ C'$\,\;												\nllabel{line:czNeighbors-1}
		\ENDFOR
		\medskip
		$C \gets C \circ C'$\,\;
		$V' \gets V' \setminus \ens{w}$\;																	\nllabel{line:reduceVprime}
		$W' \gets W' \setminus \ens{w}$\;																	\nllabel{line:reduceWprime}
		$O' \gets \big(O' \setminus \ens{w}\big) \union \ens{v}$\; 
			
	\ENDWHILE																										\nllabel{line:circuitConstrLoopEnd}
	\medskip

	\LET $G' \gets G[V']$\;
	\FORALL $v \in V'$ \DO																						\nllabel{line:circuitInputLoop}
		\oneline
		{\FORALL $w \in V'$ \SUCHTHAT $vw \in E(G')$ \DO
			\vspace{0.3ex}
			$C \gets C \circ \cZ \pseu[v,w]$\,\vspace{-0.2ex}\;								\nllabel{line:czNeighbors-2}
		\ENDFOR}
		$V' \gets V' \setminus \ens{v}$\,;
	\ENDFOR																									\nllabel{line:circuitInputLoopEnd}

	\medskip
	\RETURN $C$\;																								\nllabel{line:constructedCircuit}
\END

	\caption[An algorithm for the semantic map corresponding to the DKP representation map.]{\label{alg:semanticDKP}%
		An algorithm for the semantic map $\cS$ corresponding to the DKP representation map $\cR$, in the case of one-way measurement patterns with an equal number of input and output qubits, using the procedures defined in Algorithms~\ref{alg:completeFlowAlg}, \ref{alg:testDependencies}, and~\ref{alg:maximalStarVertex} (pages~\pageref{alg:completeFlowAlg}, \pageref{alg:testDependencies}, and~\pageref{alg:maximalStarVertex} respectively).}
\end{algorithm}%
We will show:
\begin{lemma}
	\label{lemma:semanticDKP}
	Let $P$ be a standardized one-way measurement pattern with underlying geometry $(G,I,O)$, where $\card{I} = \card{O}$.
	Then $\SemanticDKP(G,I,O,P)$ returns a unitary circuit $C$ such that $\cR(C) \cong P$ if $P$ is congruent to some element of $\img(\cR)$, and returns $\nil$ otherwise.
\end{lemma}
\begin{proof}
As noted before, because we can perform the DKP construction by first applying the \emph{simplified} DKP construction, we know that $(G,I,O)$ has a flow if $P \in \img(\cR)$\,.
Similarly, if $P \in \img(\cR)$, then the dependencies of $P$ will be consistent with the constraints imposed by Theorem~\ref{thm:dependenciesDKP} for some flow function.
In the case where $\card I = \card O$, there is at most one flow-function; so if $P \in \img(\cR)$, its dependencies will in particular be consistent with the constraints of Theorem~\ref{thm:dependenciesDKP} for any flow-function discovered for $(G,I,O)$ by $\FindFlow$.
Neither of these are affected by commuting independent operations of $P$ past one another.
Therefore, $\SemanticDKP(G,I,O,P)$ returns $\nil$ at lines~\ref{line:flowFail} or~\ref{line:dependenciesFail} of Algorithm~\ref{alg:semanticDKP}, $P$ is not equivalent to any $P' \in \img(\cR)$.

On lines~\ref{line:initGeometry}\thru\ref{line:initGeometryEnd}, we initialize vertex sets $V'$, $W'$, and $O'$ which will represent various sub-geometries of $(G,I,O)$ of the form
\begin{gather}
	(G[V'], V' \setminus W', O')	\,.
\end{gather}
(More precisely, $W'$ represents $\img(f')$, for successive restrictions $f'$ of the flow-function $f$ to $V'$\,; in the case $\card I = \card O$, this is equivalent to $V' \setminus I$.)
We also initialize an empty circuit $C$ to which we pre-compose other circuits in the loops on lines~\ref{line:circuitConstrLoop}\thru\ref{line:circuitConstrLoopEnd} and lines~\ref{line:circuitInputLoop}\thru\ref{line:circuitInputLoopEnd}, represented in stable index notation; we will sometimes refer to the tensor indices as ``wire segments'' (see the discussion on page~\pageref{discn:stableIndexPathIntegral}) for the sake of clarity.
We will consider the circuits which we compose to $C$ in reverse order, considering sub-circuits proceeding from the inputs to the outputs of the final circuit returned on line~\ref{line:constructedCircuit}.

\begin{itemize}
\item
	In the loop on lines~\ref{line:circuitInputLoop}\thru\ref{line:circuitInputLoopEnd}, we compose to $C$ one controlled-$Z$ operation between wire segments labelled by the vertices contained in $V'$ at that stage of the algorithm, for each edge in the graph $G' = G[V']$.
	This loop is only entered once $W' = \vide$\,; as we only transform $W'$ by removing one vertex $w$ at a time on line~\ref{line:reduceWprime} (which we also remove from $V'$ on line~\ref{line:reduceVprime}), the value of $V'$ just before line~\ref{line:circuitInputLoop} is $V' = V(G) \setminus \img(f)$.
	Thus, the cumulative effect of lines~\ref{line:circuitInputLoop}\thru\ref{line:circuitInputLoopEnd} is to compose $C_{\tilde I}$ to the beginning of $C$, for $\tilde{I} = V(G) \setminus \img(f)$.

\item
	Using the procedure $\MaximalStarVertex$ defined in Algorithm~\ref{alg:maximalStarVertex}, each iteration of the loop on lines~\ref{line:circuitConstrLoop}\thru\ref{line:circuitConstrLoopEnd} produces a vertex $w$ which is the center of a maximal star geometry of $(G[V'], V' \setminus W', O')$.
	For each such $w$, we obtain $v = g(w)$, and construct the star circuit $C'$ corresponding to the geometry $(G_{v\star}\,, I_{v\star}\,, O_{v\star})$, which we append to the input of $C$.
	We then reduce the geometry $(G[V'], V' \setminus W', O')$ by setting $V' \gets V' \setminus \ens{w}$, and $O' \gets O' \symdiff \ens{v, w}$, corresponding to the reduction described in~\eqref{eqn:geometryReduce}; we also reduce $W' \gets W' \setminus \ens{w}$, corresponding to a further restriction of $f$ to the new value of $V'$.

	By induction on $\img(f)$, as in the proof of Lemma~\ref{lemma:starDecomp}, this sequence of star geometries leads to a decomposition of the original geometry $(G,I,O)$ into star geometries, leaving behind the geometry $(G[\tilde I], I, \tilde I)$\,.
	For each such star geometry, we obtain the star circuit and compose it to the input end of $C$\,; then, prior to line~\ref{line:circuitInputLoop}, $C$ consists of a composition of the star circuits corresponding to the geometries in a star decomposition of $(G,I,O)$.
\end{itemize}
Thus, the value of $C$ on line~\ref{line:constructedCircuit} is a composition of the star circuits corresponding to the decomposition of $(G,I,O)$ into star geometries, where in particular for each gate $J(\theta(v))\pseu[:w/v]$\,, we have $\theta(v) = -\alpha_v$, for $\alpha_v$ the angle of the measurement on $v$ in the case where all of the preceding measurements $\Meas{z}{}$ produced the result $\s[z] = 0$.

By construction, the output is a circuit $C$ constructed with the elementary gate set $\Jacz(\A)$ for some $\A \subset \R$.
Such a circuit is in the domain of $\cR$\,; consider then the value of $\bar{P} = \cR(C)$ for such a circuit $C$, and let $(\bar G, \bar I, \bar O)$ be the geometry underlying $\bar P$.
\begin{itemize}
\item
	The DKP construction preserves the existing wire segment labels of $C$ as the labels of vertices in $\bar G$, $\bar I$, and $\bar O$, for the geometry $(\bar G, \bar I, \bar O)$ underlying the measurement pattern $\bar P$.
	As the labels of the wire-segments in $C$ are precisely the vertex-labels of $G$, with the non-advanced indices arising from $I$ and the non-deprecated indices arising from $O$, we then have $V(\bar G) = V(G)$, $\bar I = I$, and $\bar O = O$.

\item
	Because the flow function $f$ for $(G,I,O)$ is used to define the correspondence between deprecated and advanced indices in $C$, and (as we noted in Section~\ref{sec:flowDefn}) this correspondence is a flow-function for $(\bar G, \bar I, \bar O)$ arising from $C$ by the DKP construction, the same function $f$ is then a flow-function for $(\bar G, \bar I, \bar O)$ as well.
	In particular, for each pair of deprecated/advanced indices $v, w$ in $C$, there is an edge $vw \in E(\bar G)$.

\item
	For every operation $\cZ \pseu[v,w]$ in $C$, there will be an edge $vw \in E(\bar G)$ which is not a flow-edge for the function $f$.
	Because each such controlled-$Z$ operation can be attributed to a star circuit for a unique star geometry in a decomposition of $(G,I,O)$ --- as well as for the flow edges arising from the $J(\theta)$ operations --- we then have $E(\bar G) = E(G)$, so that $\bar G = G$.

\item
	Because the angle $\theta$ of the $J(\theta)$ gate between wire segments $v$ and $f(v)$ is the negative of the default measurement angle of $v$ in the pattern $P$ (by the construction of $\theta$) and the negative of the default measurement angle of $v$ in the pattern $\bar P$ (by the construction of $\bar P$), each qubit $v \in O\comp$ has the same default measurement angle in $\bar P$ as in $P$.

\item
	By Theorem~\ref{thm:dependenciesDKP}, the measurement dependencies in $\bar P$ are described by the matrices $F(\idop_n - T)\inv$ and $(\idop_n - T)\inv$, for matrices $T$ and $F$ given by~\eqref{eqn:shiftPropagator} and~\eqref{eqn:flowPropagator} respectively, depending on the flow-function $f$.
	However, $P$ satisfies the same constraints by the test on lines~\ref{line:dependenciesFail}.
	It follows that the measurement operations on $v$ in both $P$ and $\bar P$ are the same for $v \in O\comp$, and the correction operations are similarly identical for $v \in O$, including dependencies on other measurement results.
\end{itemize}
Thus, $P$ and $\bar P$ contain the same operations, albeit possibly in a different order.
Finally, because $P$ is well-formed by assumption, and $\bar P$ is so by construction, the only possible difference in the ordering of corresponding operations between $P$ and $\bar P$ is in the commuting of independent terms by the type composition constraints of Lemma~\ref{lemma:typeCompose}.
Thus, $P$ is congruent to $\bar P$\,; and in particular, $\SemanticDKP(G,I,O,P)$ returns $\nil$ only if $P$ is not congruent to an element of $\img(\cR)$.
\end{proof}

The above Lemma then allows us to show:

\begin{theorem}
	Let $\tilde \cR$ be the restriction of the DKP construction map $\cR$ to to unitary circuits without terminal measurements or trace-out, and with the same number of output wires as input wires; and let $\cS$ be the map taking one-way measurement pattern $P$ to the circuit $\SemanticDKP(G,I,O,P)$, where $(G,I,O)$ is the geometry underlying $P$.
	Then $\cS$ is the semantic map for $\tilde\cR$\;.
\end{theorem}
\begin{proof}
	Taken as a partial function on its input data --- and neglecting the trivial detail that its domain is on tuples $(G,I,O,P)$, rather than just on patterns $P$ --- the domain of $\SemanticDKP$ is the set of well-formed measurement procedures which are congruent to elements of $\img(\tilde\cR)$, which in particular contains $\img(\tilde\cR)$; and each element of $\img(\tilde\cR)$ will itself be mapped by $\tilde\cR \circ \cS$ to a congruent one-way measurement pattern.
	Thus, $\SemanticDKP$ satisfies property~\ref{condition:retraction} on page~\pageref{condition:retraction}.

	For a circuit $C$ for a unitary bijection composed from the gate set $\Jacz(\A)$ for some $\A \subset \R$, let $\tilde P = \tilde\cR(C)$ and $\tilde C = \cS(\tilde P)$\,, and let $(G,I,O)$ be the geometry underlying $\tilde P$; then, consider the correspondence between the elements of $C$, $\tilde P$, and $\tilde C$.
	There is by construction a bijective correspondence between the wire segments of $C$ to the vertices of $V(G)$, and from $V(G)$ to the wire segments of $\tilde C$\,.
	By construction, this bijective correspondence is an isomorphism of stable index expansions:
	\begin{romanum}
	\item
		Successive wire-segments in $C$ correspond to paths $v, f(v), f^2(v)$ in a path-cover $\cP_f$ in $G$, which in turn correspond to successive path segments in $\tilde C$\,.

	\item
		Wire-segments in $C$ which interact via a $\cZ$ gate in $C$ correspond to neighbors $v \sim w$ in $G$ such that $v \ne f(w)$ and $w \ne f(v)$\,, which in turn give rise (via the contribution of some star geometry in the decomposition of $(G,I,O)$, as remarked in the analysis of the preceding section) to wire segments in $\tilde C$ which interact via a $\cZ$ gate.
		As there are no other operations in $\tilde C$ than this, and as every operation in $C$ gives rise to an edge in $G$, the interaction graphs of $C$ and $\tilde C$ are the same.
		Because $C$ and $\tilde C$ are both defined on the same gate-set, which in particular is parsimonious (in which any multi-qubit gate preserves standard-basis states), there is a homomorphism of stable-index expressions from $C$ to $\tilde C$ (where in particular the index labels are identical).

	\item
		The wire segments in $C$ as in $\tilde C$ are separated by gates of the form $J(\theta)$\,; and by construction, in both $C$ and $\tilde C$, the angles $\theta$ of the $J(\theta)$ gates are the negative of the default measurement angle of the qubit $v$ in $\tilde P$, where $v$ is the label of the deprecated index for the $J(\theta)$ gate for $C$ and $\tilde C$ alike.
		As there is only one variety of two-qubit gate, the homomorphism between $C$ and $\tilde C$ is then an isomorphism.
	\end{romanum}
	Thus, $C$ is isomorphic to $\tilde C$ as a quantum circuit, and differ only in the ordering of commuting gates.
	Thus, the map $\cS$ is a semantic map for $\tilde\cR$ in the sense defined on page~\pageref{discn:representationMap}.
\end{proof}

\subsubsection{Remarks and run-time analysis}

The run-time for $\SemanticDKP(G,I,O,P)$ can be accounted for as follows, for $\card{V(G)} = n$, $\card{I} = \card{O} = k$, and $\card{P} = N$:
\begin{itemize}
\item
	As we showed on pages~\pageref{thm:findFlow} and~\pageref{discn:testDependencies}, $\FindFlow(G,I,O)$ requires time $O(k^2 n)$ and $\TestDependencies(P,f)$ requires time $O(Nn) \subset O(n^3)$.

\item
	For each iteration of the loop on lines~\ref{line:circuitConstrLoop}\thru\ref{line:circuitConstrLoopEnd}, we require time $O(kn)$ to run $\MaximalStarVertex(G,V',W',O',f)$, by the discussion on page~\pageref{discn:maximalStarVertex}, and time $O(\deg(v))$ (for a different $v \in O\comp$ in each iteration) for the loop on line~\ref{line:czNeighbors-1}; all other operations require constant time.
	The cumulative run time of the loop is then
	\begin{align}
			O\paren{\sum_{v \in V(G) \setminus \img(f)} \Big[ kn + \deg(v) \Big]}
		\;\;\subset&\;\;
			O\paren{\sum_{v \in V(G)} kn \quad \!\!+ \sum_{v \in V(G)} \deg(v) }
		\notag\\[1ex]=&\;\;
			O(kn^2 + 2 \card{E(G)})
		\notag\\[1ex]=&\;\;
			O(kn^2 + 2m)
		\;\;=\;\;
			O(kn^2)	\,,
	\end{align}
	by the extremal result of Theorem~\ref{thm:extremalUpperBound}.

\item
	For each iteration of the loop on lines~\ref{line:circuitInputLoop}\thru\ref{line:circuitInputLoopEnd}, we perform at most $\deg(v)$ iterations of the loop on line~\ref{line:czNeighbors-2}, plus one constant-time set removal.
	As noted above, just prior to line~\ref{line:circuitInputLoop}, we have $V' = I$\,.
	The cumulative run time of this loop is then
	\begin{align}
			O\paren{\sum_{v \in I} \deg(v)}	
		\;\;\subset\;\;
			O\paren{\sum_{v \in V(G)} \deg(v)}
		\;\;=&\;\;\;
			O(2\card{E(G)})
		\notag\\=&\;\;
			O(m)
		\;\;\subset\;\;
			O(k n)	\;,
	\end{align}
	again by the extremal bound of Theorem~\ref{thm:extremalUpperBound}.
\end{itemize}
The time required by $\SemanticDKP(G,I,O,P)$ is then dominated by the time required by $\TestDependencies(P,f)$, which is $O(n^3)$.

\section{Further developments}
\label{sec:flowsFurtherResults}

The Algorithm~\ref{alg:completeFlowAlg} of Section~\ref{sec:flowAlgs} is (a revision of) the first efficient algorithm to determine whether a broad class of geometries have a flow.
Since the presentation of the original version of that algorithm in~\cite{Beaudrap06}, improved algorithms --- and extensions to the concept of a flow itself --- have been introduced.
As well, flows provide us additional tools to compare the DKP and RBB constructions of one-way measurement patterns.
In this penultimate section of Chapter~\ref{Semantics for unitary one-way patterns}, we present an overview of these results to put the earlier results of this chapter into context.

\subsection{Flows in relation to more general measurement-based constructions}
\label{sec:extendedConstructions}

In Sections~\ref{sec:RBB} and~\ref{sec:otherConstrs}, we presented the simplified RBB construction representing the building blocks of the patterns in~\cite{RBB03}, as well as two other patterns from the more elaborate constructions which are not obviously reproducible by the DKP construction.
Using the graph-theoretic characterization of flows in this chapter, we can show that they cannot be produced by the DKP construction.

\subsubsection{Qubit reversal pattern}

The pattern described in Section~\ref{sec:qubitReversal} for reversing the order of a sequence of ``logical qubits'' in the cluster-state model can be shown not to have a flow, using the characterization of Section~\ref{sec:graphth-charn}, because the underlying geometry does not admit a causal path cover.
We may show this by sketching various constructions of families of vertex-disjoint $I$\thru$O$ paths, in the case of $\card{I} = \card{O} = k$, as follows.

In the following, we label vertices $(a,b)$ by rows and columns; the $j\th$ input vertex is in the row $2(j-1)$, and the zeroth column.
\begin{itemize}
\item
	Suppose the path $P_k$ starting on the row $2(k-1)$ leaves that row, so that it covers the vertex $(2k-2,\ell)$ but not the edge between $(2k-2,\ell)$ and $(2k-2, \ell+1)$.
	Either it returns to the row $2(k-1)$ to cover the vertex $(2k-2,\ell+1)$, or it does not cover $(2k-2,\ell+1)$.
	In the former case, consider the function $\tilde f$ mapping each vertex covered by $P_k$ to its successor in $P_k$; then we have $(2k-2,\ell+1) = f^h(2k-2,\ell)$ for some $h \in \N$.
	Then, the path $(2k-2,\ell) \arc f(2k-2,\ell) \arc \cdots \arc f^{h}(2k-2,\ell) \arc (2k-2,\ell)$ is a vicious circuit, in which case any collection of paths including $P_k$ is not a causal path cover.
	In the latter case where $P_k$ does not cover $(2k-2, \ell+1)$, it then also separates that vertex from any other $I$\thru$O$ path, as a path starting in any other row must intersect $P_k$ to reach $(2k-2,\ell+1)$.
	These two cases are illustrated in Figure~\ref{fig:swapRBBpaths}.
	\begin{figure}[t]
		\begin{center}
			~\hfill
			\includegraphics[width=58mm]{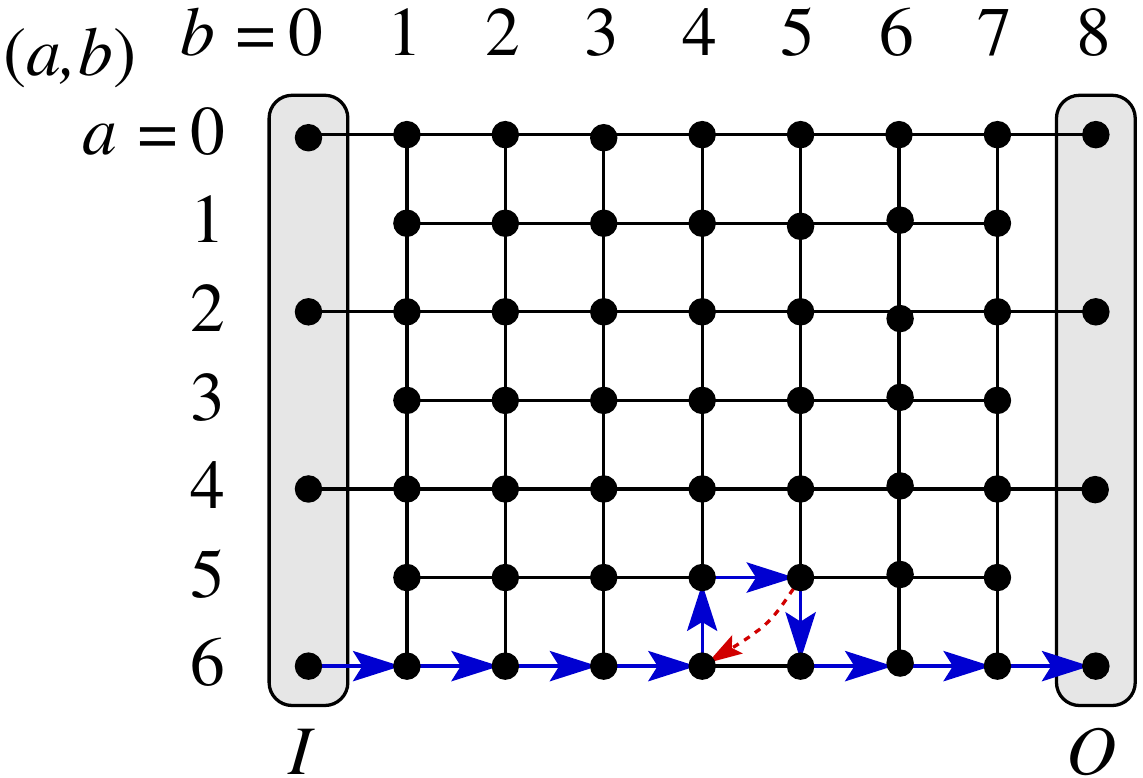}
			\hfill
			\includegraphics[width=58mm]{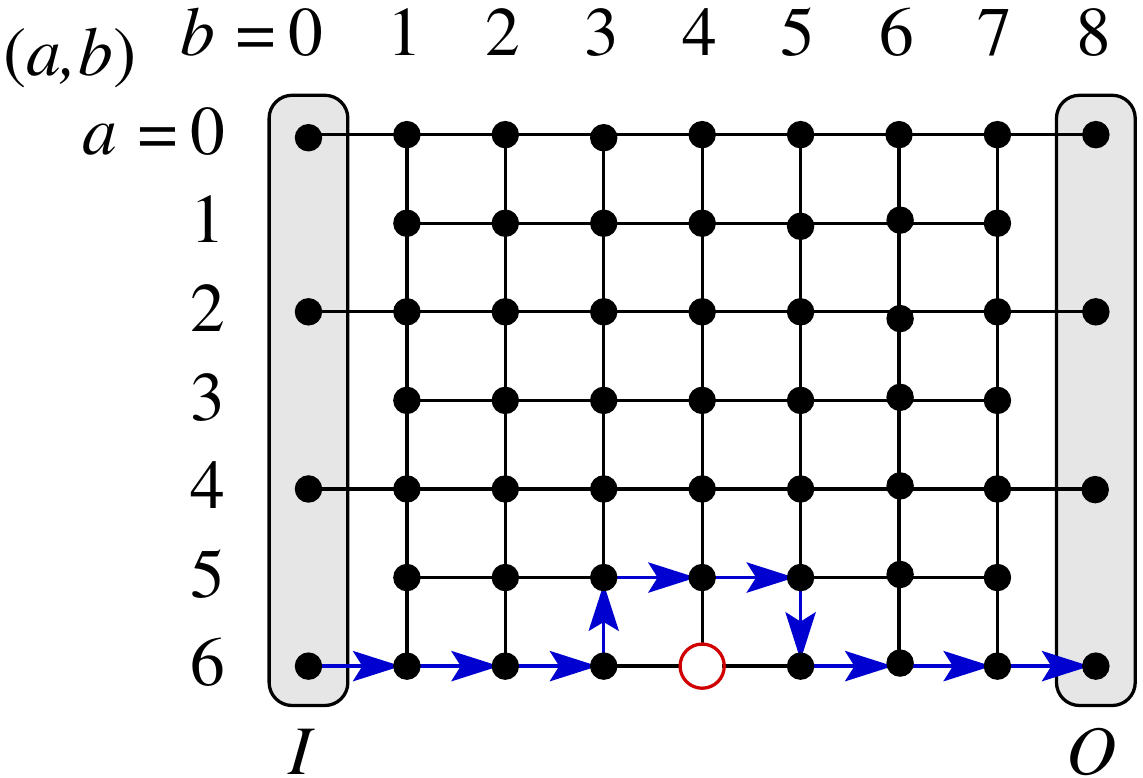}
			\hfill~\hfill~
		\end{center}
		\caption[Examples of an $I$\protect\thru$O$ path in the geometry for the qubit-reversal pattern illustrated in Figure~\protect{\ref{fig:swapRBB}}.]{\label{fig:swapRBBpaths}\colouronline
			Examples of an $I$\thru$O$ path in the geometry for the qubit-reversal pattern illustrated in Figure~\ref{fig:swapRBB}.
			Arrows on the grid lines in these diagrams represent orientations of edges of the geometry corresponding to a single $I$\thru$O$ path $P_k$.
			In the example on the left, a path leaves the row $2(k-1)$ and returns to cover all of the vertices in that row; this induces a vicious circuit (emphasized by the dotted arrow, representing an arc from the influencing digraph induced by an arc of $P_k$ and an edge not covered by $P_k$).
			In the example on the right, a path leaves the row $2(k-1)$ and returns, but fails to cover all of the vertices in that row, leaving a vertex uncovered (shown as a large hollow circle).}
	\end{figure}	

	Thus, if a path starting in the row $2(k-1)\st$ does not remain constrained to that row, any family of paths containing that path is not a causal path cover; and a similar analysis also holds for the path beginning in the first row.
	(This suffices then to show that the geometry does not have a flow \eg\ for $k = 2$.)

\item
	Suppose the path starting in the row $2(k-1) = 2k-2$ remains in that row.
	Then, the only path in a collection of vertex-disjoint $I$\thru$O$ paths that can cover vertices in the row $2k-3$ is the path $P_{k-1}$ starting in the row $2k-4$ (as the odd numbered rows do not have input vertices, and the path $P_{k-1}$ will separate any other path from the row $2k - 3$).
	In order to cover all of the vertices in the row $2k-3$, the path $P_{k-1}$ must cover the first vertex $(2k-3,1)$ in that row; and once in that row, by a similar analysis as applied for the row $2(k-1)$ above, it must remain in that row until it reaches the final vertex $(2k-3,2k-1)$ in that row.
	In order for the output vertex in row $2k-4$ to be covered, the path $P_{k-1}$ must then terminate there, determining that path.

	By induction, we can show that the paths in every lower-numbered row must follow a staircase pattern to higher-numbered rows if those higher-numbered rows are to be completely covered.
	(Such a family of paths is illustrated in Figure~\ref{fig:swapRBBstaircase}).
	\begin{figure}[t]
		\begin{center}
			\includegraphics[width=58mm]{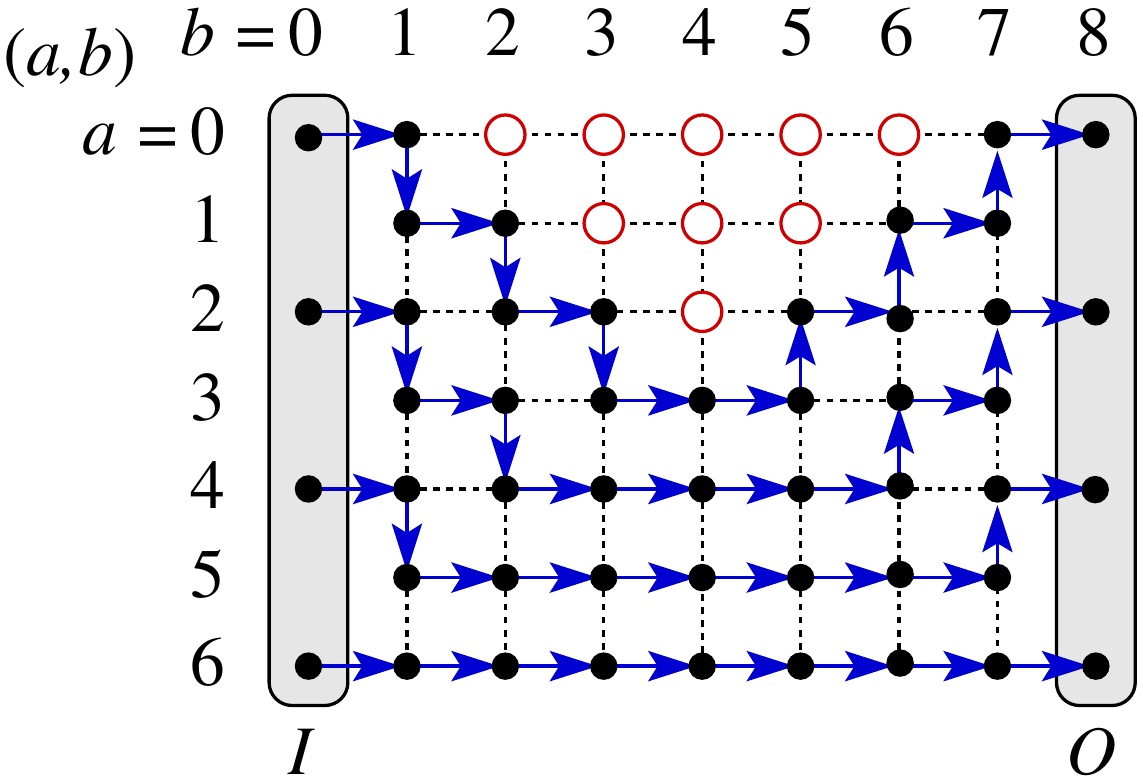}
		\end{center}
		\caption[Illustration of a family of $I$\protect\thru$O$ paths in the geometry for the qubit-reversal pattern illustrated in Figure~\protect{\ref{fig:swapRBB}}.]{\label{fig:swapRBBstaircase}\colouronline
			Illustration of a family of $I$\thru$O$ paths in the the geometry for the qubit-reversal pattern illustrated in Figure~\ref{fig:swapRBB}.
			Arrows on the grid lines in these diagrams represent orientations of edges of the geometry corresponding to a family of vertex-disjoint $I$\thru$O$ paths; the other edges in the grid have been made dashed to emphasize the flow edges.
			This family of paths is constructed to cover all of the vertices in the higher-numbered rows; as a result, vertices in the lower-numbered rows are necessarily left uncovered (represented here by large hollow circles).}
	\end{figure}
	In particular, the path $P_1$ will then leave the first row, so that by the previous case the resulting family of paths will not be a path cover.
\end{itemize}
Thus, the geometry of the qubit reversal pattern of~\cite{RBB03} does not have a causal path cover, by Theorem~\ref{thm:graphth-charn}, does not have a flow.
As a result, it cannot be produced by the DKP construction.

\begin{figure}[p]
	\begin{center}
			\includegraphics{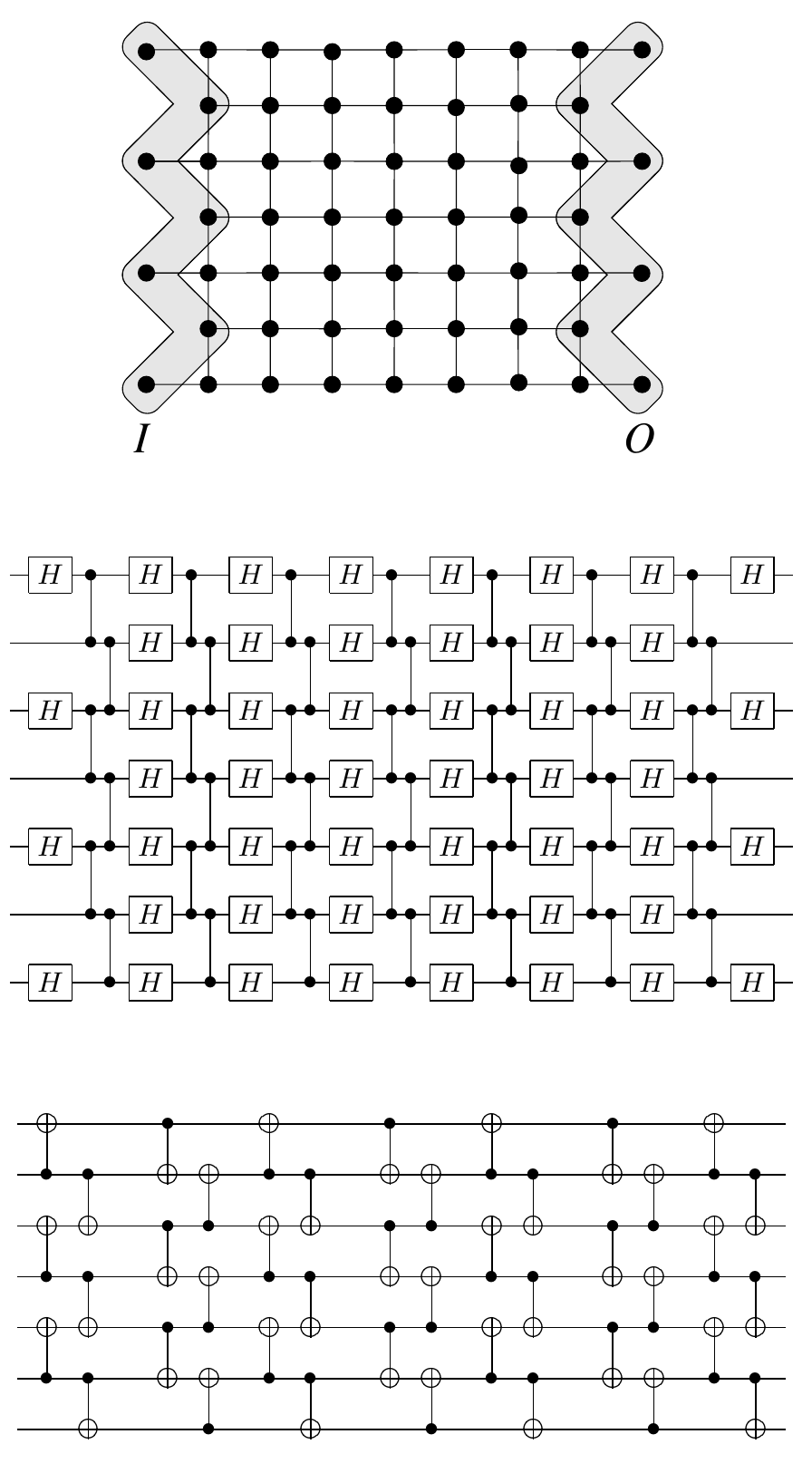}
	\end{center}
	\caption[Illustration of the geometry for the qubit-reversal pattern illustrated in Figure~\protect{\ref{fig:swapRBB}} with augmented input and output subsystems, and corresponding unitary circuits.]{\label{fig:swapRBBredux}%
		Illustration of the geometry for the qubit-reversal pattern illustrated in Figure~\ref{fig:swapRBB} with augmented input and output subsystems, and corresponding unitary circuits.
		The circuit diagram in the middle represents the circuit produced by the procedure $\SemanticDKP$, assuming that each default measurement is zero, and that the dependencies are those consistent with the DKP construction, given a flow function mapping each vertex in the geometry to the one immediately to the right.
		The circuit diagram on the bottom represents an equivalent circuit resulting from gathering the Hadamard operations of the circuit above.}
\end{figure}

Despite this, we may gain some information about the qubit-reversal pattern using the smenatic map for the DKP construction by an appropriate extension of the input and output systems, as exhibited in~\cite{CLM2005}.
It is easy to see that if we include the left-most element of each row into $I$, and the right-most into $O$, the resulting altered geometry has a flow with a successor function which maps each vertex to the one immediately to the right.
Chains in the natural pre-order $\preceq$ have monotonically increasing column numbers: and therefore $\preceq$ is antisymmetric.
Given that the default measurement angle for each qubit is zero, and assuming that the dependencies of the measurement pattern are consistent with the those of the DKP construction given the flow function described, we may then obtain a measurement pattern which is in the range of the DKP construction.
The geometry for this measurement pattern, and the corresponding unitary circuits in the $\Jacz(\A)$ model and a related unitary circuit involving $\cX$ gates, are illustrated in Figure~\ref{fig:swapRBBredux}.
The final circuit illustrated there can easily be described as a reversible classical computation: for seven input bits $(v_0, \ldots, v_6)$, the computation performed by that circuit is
\begin{align}
	\lefteqn{(v_0,\; v_1,\; v_2,\; v_3,\; v_4,\; v_5,\; v_6)}\qquad\qquad&\notag\\
	\mapsto&\quad
	(v_6,\; v_4 \oplus v_5 \oplus v_6,\; v_4,\; v_2 \oplus v_3 \oplus v_4,\; v_2,\; v_0 \oplus v_1 \oplus v_2,\; v_0)	\;.
%
\end{align}
If we apply this circuit instead to qubits, with the odd-indexed qubits initialized to the $\ket{\splus}$ state (\ie\ a uniform superposition of the bit-values described above), we the result is that the odd-indexed qubits remain in the $\ket{\splus}$ state, and the even-indexed qubits are reversed, as required.

We may then understand the qubit-reversal pattern in~\cite{RBB03} as corresponding to a circuit which prepares auxiliary qubits in the $\ket{\splus}$ state, and subsequently removes them by an $\axisX$ measurement.
The fact that we are able to obtain the circuit in this case is possible because the highly structured geometry makes it easier to guess which qubit measurements correspond to terminal measurements in a unitary circuit.
Without exploiting this structure, however, there are no known approaches to \emph{algorithmically} provide semantics for such a measurement pattern in terms of unitary circuits.

\subsubsection{Concise pattern for $Z \ox \cdots \ox Z$ Hamiltonians}

As we noted in Section~\ref{sec:zzMany}, the pattern $\mathfrak {Zz}_{u,v}$ on two qubits $u$ and $v$ of the simplified RBB construction (Section~\ref{sec:RBB}) is a special case of the family of patterns $\mathfrak {Zz\cdots z}^\theta_{v_1, \ldots, v_k}$ on many qubits.
The geometry for that pattern consists of the $k$ qubits $I = O = \ens{v_1, \ldots, v_k}$, and a single auxiliary qubit $a \notin I, O$.
For this geometry, there is exactly one injective function $f: O\comp \to I\comp$\,, which is the identity function on $\ens{a}$\,, which does not satisfy the constraint that $a \sim f(a)$.
(For the sake of uniformity, we may wish to restrict ourselves to patterns performing $\XY$ plane measurements only, in which case we would consider the pattern for this CPTP map given by~\eqref{eqn:zzManyXY}; however, the identity function on $\ens{a_1, a_2}$ suffers from the same problem as that on $\ens{a}$ for the pattern~\eqref{eqn:zzMany}, and the natural pre-order for the involution on $\ens{a_1, a_2}$ is not antisymmetric.)
Thus, this pattern cannot be produced via the DKP construction.

The $\YZ$-plane measurement performed by the pattern $\mathfrak {Zz\cdots z}^\theta$, with respect to the $\ket{\spm \sfy\sfz_{\theta}}$ basis defined in~\eqref{eqn:defYZtheta}, is equivalent to a measurement of the observable
\begin{gather}
	\label{eqn:observableYZ}
		\sO^{\YZ,\theta}
	\;\;=\;\;
		\ket{\splus\sfy\sfz_{\theta}}\bra{\splus\sfy\sfz_{2\theta}}
		\;\;-\;\;
		\ket{\sminus\sfy\sfz_{\theta}}\bra{\sminus\sfy\sfz_{2\theta}}
	\;\;=\;\;
		\cos({\theta}) Z + \sin({\theta}) Y	\;;
\end{gather}
this is a Hermitian operator which anticommutes with the Pauli $X$ operator for any $\theta$.
We can use this to obtain an extended sense of a flow which arises from the simplified RBB construction in the same way that flows arise from the DKP construction.

\subsection{Flows as a certificate of unitarity, and extended flows}
\label{sec:eflow}

In this section, we will consider the significance of flows not only as a structure which may underly one-way patterns, but also as a certificate which makes it possible to verify that a one-way pattern performs a unitary transformation.
In doing so, we will arrive naturally at an extension of the definition of flows which describes structures underlying the simplified RBB construction, and more generally constructions of one-way patterns from $\mathfrak J^\alpha$ and $\mathfrak{Zz\cdots z}^\theta$ patterns.

Recall that for the \emph{simplified} DKP construction, for each qubit $v \in O\comp$, the dependencies which arise are an $X$-dependency on $f(v)$, and a $Z$-dependency on every qubit adjacent to $f(v)$ except for $v$ itself; this may be considered to arise from a correction operation
\begin{gather}
		B_v
	\;\;=\;\;
		X_{f(v)} \paren{ \prod_{\substack{w \sim f(v) \\ w \ne v}} Z_w }
\end{gather}
which is absorbed into the measurements on those respective qubits, to be applied in the case that $\s[v] = 1$.
This unitary operation closely corresponds to the generator $K_{f(v)}$ of the open graphical encoding performed by the preparation map and entanglement graph at the beginning of the measurement pattern,
\begin{gather}
		K_{f(v)}
	\;\;=\;\;
		X_{f(v)} \paren{ \prod_{w \sim f(v)} Z_w }	\;:
\end{gather}
the state of the system just prior to the measurement phase in the simplified DKP construction is stabilized by the group $\gen{K_w}_{w \in I\comp}$ of such operators.
Flows may then be understood as a way of certifying that a measurement pattern performs a unitary transformation by the stabilizer formalism.
We may observe that the analysis of measurements by a single-qubit observable $P_v$ in the stabilizer formalism (as presented in Section~\ref{sec:stabilizerFormalism}) does not at any point require that $P_v$ be a Pauli operator, so long as it either commutes or anticommutes with every generator of the subgroup of the Pauli group which stabilizes the state of the system.
Note that, just as we have described above for $\YZ$-plane measurements, $\XY$-plane measurements can be described by measurements with respect to observables of the form
\begin{gather}
	\label{eqn:observableXY}
		\sO^{\XY,\theta}
	\;\;=\;\;
		\ket{\psub\theta}\bra{\psub\theta}
		\;\;-\;\;
		\ket{\msub\theta}\bra{\msub\theta}
	\;\;=\;\;
		\cos(\theta) X + \sin(\theta) Y	\;,
\end{gather}
and that this is an observable which anticommutes with the Pauli $Z$ operator.
Then, for a flow function $f: O\comp \to I\comp$ for a geometry $(G,I,O)$, an observable $\sO^{\XY,\theta}_v$ anticommutes with $K_{f(v)}$; and $\preceq$ describes an order in which stabilizer generators may be ``removed'' so that the entire computation may be treated via the stabilizer formalism.

We may then show that if $(f,\preceq)$ is a flow for a geometry $(G,I,O)$ that $\preceq$ is a partial specification of a total ordering of the qubits $v \in O\comp$, and $f$ a map associating each qubit $v \in O\comp$ to a generator $K_w$ for $w \in I\comp$, such that the measurement of each such qubit $v$ can be certified to perform an isometry by the stabilizer formalism via successive removals of the generators $K_{f(v)}$ --- provided that the corresponding corrections are performed in response to each measurement, and that the measurements are performed in an ordering extending $\preceq$.
In order to do this, we require that $\sO^{\XY, \theta}_v$ commutes for any $v \in O\comp$ with any remaining generator $K_w$\,; for this to hold for arbitrary $\theta$, this entails that such a generator $K_w$ does not act on $v$ --- which can be guaranteed by requiring that any such $K_w$ in the original stabilizer group be eliminated prior to the measurement of $v$.
This can be done by requiring the measurement a qubit $z$ such that $w = f(z)$ prior to measuring $v$.
Thus, it is sufficient for all $z$ to that $z \preceq f(z)$ and that $v \sim f(z) \,\implies\, z \preceq v$\,, which the flow conditions guarantee. 
We can then interpret flows as a way of connecting the DKP construction to the stabilizer formalism, by identifying the generators $\ens{K_w}_{w \in I\comp}$ in some order as the generators by which each measurement may be treated.

The converse statement --- that any ordering the individual generators $\ens{K_w}_{w \in I\comp}$ as a means of treating measurement-based computation by the stabilizer formalism corresponds to a flow --- is not true, however; and the $\mathfrak{Zz\cdots z}^\theta$ pattern is a counterexample.
We may define an \emph{extended flow}\footnote{%
	This extended sense of flow was described in~\cite{DK05u} for the special case of $\axisY$ measurements, but does not seem to have been seriously examined, having been superceded by the concept of generalized flows (or \emph{glows})~\cite{BKMP07} which we will discuss in Section~\ref{sec:gflows}.} 
(or \emph{eflow}) as follows:
\begin{definition}
	\label{def:eflow}
	For a tuple $(G,I,O,T)$ consisting of a geometry $(G,I,O)$ and a set $T \subset O\comp \inter I\comp$ of qubits to be measured in the $\YZ$-plane (all other qubits assumed to be measured in the $\XY$-plane), an \emph{extended flow} is an ordered pair $(f, \preceq)$ consisting of a function $f: O\comp \to I\comp$, and a partial order $\preceq$ on $V(G)$, such that the conditions
	\begin{subequations}
	\begin{gather}
			\label{eflow:a}		f(v) \sim v	\;, \;\;\text{for $v \in T\comp$}\,;
		\\
			\label{eflow:b}		f(v) = v	\;,\;\;\text{for $v \in T$}\,;
		\\
			\label{eflow:c}		v \preceq f(v)\;;\;\;\text{and}
		\\
			\label{eflow:d}		w \sim f(v)	\;\implies\; v \preceq w
	\end{gather}
	\end{subequations}
	hold for all $v \in O\comp$ and $w \in V(G)$.
\end{definition}
The special case of qubits indicated as being measured in the $\YZ$-plane applies in the case of $\mathfrak {Zz\cdots z}^\theta$ as defined in~\eqref{eqn:zzMany}: the identity function on $\ens{a}$ is an ``eflow function'', with equality being an ``eflow order''.
This corresponds to an identification of the operator $K_a$ as a generator of the open graphical encoding with which $\sO^{\YZ,\theta}$ anticommutes for any $\theta$.
This indicates that the unitarity of the pattern $\mathfrak {Zz\cdots z}^\theta$ may also be certified by the stabilizer formalism.

In the same way as for flows, an eflow for a tuple $(G,I,O,T)$ certifies that any pattern on with the geometry $(G,I,O)$ and performing the appropriate types of measurement will perform a unitary transformation, provided also that it performs the appropriate corrections (or measurement adaptations) in response to the measurements.
Again as for flows, eflows do this by attributing generators $\ens{K_w}_{w \in I\comp}$ to qubit $v \in O\comp$ to be measured.
To show this, it suffices to show that the observables for each measurement commutes with the generators which remain at the point that the measurement is performed, except for a single one with which it anticommutes.
The argument is similar to the one for flows: it is necessary and sufficient for $K_{f(v)}$ to be the only generator which acts on $v$ at the time of the measurement of $v$\,; which holds if and only if $v$ is measured no earlier than any qubit $z$ for which $v \sim f(z)$ or $v = f(z)$\,, which are guaranteed for eflows as for flows.

Any attribution of generators $K_{f(v)}$ for $f(v) \in I\comp$ (as opposed to nontrivial products of such operators) to qubits $v \in O\comp$ must satisfy $f(v) = v$ or $f(v) \sim v$ in order for $K_{f(v)}$ to act on $v$\,; and in order to anticommute with the appropriate measurement observable describing the measurement on the qubit $v$ for \emph{arbitrary} default measurement angles, we require precisely the eflow conditions to hold.
Thus, eflows characterize those geometries, together with specifications of measurement planes, which may underlie measurement patterns which can be certified to perform unitary transformations by the stabilizer formalism.

The simplified RBB construction give rise to tuples $(G,I,O,T)$ with eflows just as the DKP construction gives rise to geometries with flows: the $Z$ corrections in the pattern $\mathfrak{Zz}$ of~\eqref{eqn:zz} used for two-qubit interactions commute with all of the entangler maps of subsequent measurement patterns which are composed with it, and induce $\pi$-dependencies on the qubits they act upon; the same informal meaning of influencing the measurement result applies as that described in Section~\ref{sec:flowDefn}.
After Pauli simplifications and signal shifting, we may arrive at a similar characterization of the dependencies arising in the simplified RBB construction as that described in Theorem~\ref{thm:dependenciesDKP} for the DKP construction: the same matrices $T$ and $F$ describe the propagation of $X$- and $Z$-dependencies in that case.
As well, the analysis of star decompositions may similarly be generalized: we may augment the definition of a star geometry to include those, like that of the pattern $\mathfrak {Zz\cdots z}^\theta$, for which $I = O$ and for which $I\comp$ is a singleton set consisting of a qubit to be measured in the $\YZ$-plane; the decomposition of geometries $(G,I,O)$ into maximal geometries then generalizes straightforwardly.
Then, provided an algorithm to efficiently find eflows, we may similarly arrive at a semantic map for the simplified RBB construction --- or for any augmentation of the DKP construction by including patterns of the form $\mathfrak{Zz\cdots z}^\theta$ --- just as we have above for the DKP construction.

\subsection{A faster and more general algorithm for finding flows}
\label{sec:improvedFlowAlgs}

The algorithm $\FindFlow$ for finding flows presented in Section~\ref{sec:flowAlgs} runs in time $O(k^2 n)$, where $k = \card{I} = \card{O}$ and $n = \card{V(G)}$\,; in particular, it is constrained to the case where the input and output subsystems are of the same size.
In~\cite{MP08}, Mhalla and Perdrix present an algorithm which is both more general than $\FindFlow$, in that it can also determine the presence of a flow for geometries such that $\card{I} < \card{O}$,\footnote{%
	Note that a geometry with $\card{I} > \card{O}$ cannot have a flow, as the set of input vertices must be a subset of the initial vertices of a path cover, and the set of output points must coincide with the set of final vertices of a path cover.}
and which runs in time $O(kn)$.

The speed-up associated with the algorithm of~\cite{MP08} essentially arises out of the observation that the flow function $f$ and partial order $\preceq$ can be found simultaneously, by considering the length of the longest chains of each qubit from the output set, and constructing ``layer sets'' consisting of vertices of similar depth.
The main analytical tool for the algorithm is that of a ``maximally delayed'' flow:
\begin{definition}[{\cite[Definitions~4~\&~5]{MP08}}]
	\label{def:delayLayers}
	For a geometry $(G,I,O)$ and a flow $(f,\le)$, let $V^<_0(G)$ be the set of vertices of $V(G)$ which are maximal in $\le$, and let $V^<_{j+1}(G)$ be the maximal set of vertices of the set
	\begin{gather}
				W^<_j
			\;\;=\;\;
				V(G) \;\setminus\; \paren{\bigcup_{\ell = 0}^j \; V^<_j}
	\end{gather}
	in the order $\preceq$, for any $j > 0$.
	A flow $(f,\preceq)$ is \emph{more delayed} than another causal order $(f',\le)$ if we have $\card{W^\prec_j} \le \card{W^<_j}$ for all $j \in \N$\,, where this inequality is strict for at least one such $j$\,; and $(f,\preceq)$ is \emph{maximally delayed} if no flow for $(G,I,O)$ is more delayed.
\end{definition}
Mhalla an Perdrix show that a maximally delayed flow is also a flow of minimum depth, \ie\ that the causal order of such a flow has the same depth as the natural pre-order for the given flow-function $f$.
For a maximally delayed flow $(f,\preceq)$, let $\ell: V(G) \to \N$ be the function mapping each vertex $v \in V(G)$ to the integer $j$ such that $v \in V^\prec_j$.
Then, the partial order $\le$ characterized by $v \le w \iff [v = w \text{\;or\;} \ell(v) > \ell(w)]$ also yields a minimum-depth flow by construction.
To find a flow of minimum depth, it then suffices to find the sets $V^\prec_j$ of a maximally delayed flow $(f,\preceq)$.

Mhalla and Perdrix show for a maximally delayed flow $(f,\preceq)$ for $(G,I,O)$ that $V^\prec_0 = O$, and $V^\prec_1$ is the set of elements of $W^\prec_0$ such that, for some $w \in O$, $v$ is the only neighbor of $w$ which is contained in $W^\prec_0$.
By induction (by augmenting the set $O$ to include successive ``layers'' $V^\prec_j$), one may show that we then have
\begin{gather}
		V^\prec_{j+1}
	\;\;=\;\;
		\ens{v \in W^\prec_j \;\Big|\; \exists w \in V(G) \setminus W^\prec_j\,:\, \ngbh[G]{w} \inter W^\prec_j = \ens{v}}	\;,
\end{gather}
where again $\ngbh[G]{w}$ is the set of vertices adjacent to $w$ in $G$.
By a depth-first search from $O$, we may find in a finite number of iterations of $j$ those qubits $w \in V(G) \setminus W^\prec_j$ which have such a neighbor $v$, and accumulate them into new layers $V^\prec_{j+1}$.

This decomposition of the geometry $(G,I,O)$ into layers $V^\prec_j$ corresponds strongly to the star decomposition of the geometry $(G,I,O)$ into maximal star geometries; if (as in~\cite{MP08}) we accumulate sets $V^\prec_j$ to produce successively larger output subsystems $O\supp{j} = V(G) \setminus W^\prec_j$, the set $V^\prec_{j+1}$ as described above are precisely those vertices which are the root of a maximal geometry.\footnote{%
	While the algorithm of~\cite{MP08} predates the analysis of star geometries presented in this thesis, the two results were developed independently.}
Thus, in addition to finding flows more efficiently, the algorithm of~\cite{MP08} may also be used to unify the stages of finding a flow for the geometry underlying a pattern $P$, and obtaining the corresponding unitary circuit for $P$, in a semantic map for the DKP construction when $\card{I} = \card{O}$.\footnote{%
	In the case that $\card{I} < \card{O}$, there are additional difficulties which arise due to the fact that there need not be a unique flow-function --- which implies that there is not a unique set of $X$- and $Z$-dependencies which must arise for measurement patterns with such geometries.
	We discuss these issues in Section~\ref{sec:flowsReview}.}

\subsection{An efficient algorithm for finding extended flows}
\label{sec:eflowAlgs}

We may easily extend the analysis and the algorithm of~\cite{MP08} for flows to also efficiently find eflows.
For an eflow $(f,\preceq)$ for a tuple $(G,I,O,T)$\,, define the sets $V^\prec_j$ and $W^\prec_j$ as in Definition~\ref{def:delayLayers}.
Then for each $j \in \N$, $V^\prec_{j+1}$ consists of those vertices $v \in W^\prec_j$ such that, for any $w \in V(G)$ such that $w = f(v)$ or $w \sim f(v)$, we have $w = v$ or $w \in V(G) \setminus W^\prec_j$.
For $v \notin T$\,, this implies that (as in the analysis of~\cite{MP08}) the only neighbor of $f(v)$ which lies in $W^\prec_j$ is $v$ itself; and for $v \in T$, this implies that $v$ has no neighbors in $W^\prec_j$.
Algorithm~1 of~\cite{MP08} can then be easily extended to test for the latter condition, yielding a procedure to construct an eflow.
We will now describe such an algorithm.
\begin{algorithm}[t]
	\SUBROUTINE \AddToLayer(v,w) :
	
	\Data{%
		$\ell$, an array $V(G) \to \N$ for storing layers of vertices of a graph $G$
	\\
		$W$, the set of vertices for which $\ell$ is not yet well-defined
	\\
		$\layer$, the value of the current layer
	\\
		$\emptyLayer$, a flag indicating if the present layer is empty
	\\
		$D$, a directed graph such that vertices with non-zero in-degree are in $W$
	\\
		$T$, a set of vertices representing qubits measured in the $\YZ$-plane
	\\
		$f$, an array $V(G) \partialto V(G)$ for storing ``successors'' of vertices
	\\
		$S'$, a set of vertices to examine as possible ``successors'' in the next layer
	\\
		$v \in V(G)$, a vertex to add to the current layer
	\\
		$w \in V(G)$, a vertex to be designated the ``successor'' of $v$
	}

	\Effect {%
		Sets $f(v) \gets w$, $\ell(v) \gets \layer$, removes all of the arcs in $D$ into $v$, and checks the neighbors of $v$ in $D$ for terminal vertices in $T$ to include into $S'$ (for inclusion into the next layer)
	}

	\BEGIN
		$\emptyLayer \gets \false$\;
		$f(v) \gets w$~;
		$\ell(v) \gets \layer$\;
		$W \gets W \setminus \ens{v}$\;
		\FORALL $z \in \neighbors(D,v)$ \DO
			$\neighbors(D,z) \gets \neighbors(D,z) \setminus \ens{v}$\;
			\oneline
			\IF $z \in T$ \ANDALSO $\neighbors(D,z) = \vide$ \THEN $S' \gets S' \union \ens{z}$\;
		\ENDFOR
	\END
	\caption[A subroutine to attribute a vertex to the current layer, for use in an iterative procedure for finding the layer sets of a maximally-delayed eflow.]{\label{alg:addToLayer}%
		A subroutine to attribute a vertex to the current layer, for use in an iterative procedure for finding the layer sets of a maximally-delayed eflow.
		This subroutine is used in particular to maintain the arcs of a digraph $D$ (a ``sub-diagraph'' of a graph $G$, interpreted as a symmetric digraph) from which we remove any arc ending in a vertex with a well-defined layer.
	}
\end{algorithm}

Algorithm~1 of~\cite{MP08} operates on the basis of maintaining a set of ``correcting vertices'' which, for each new layer, may potentially be the successor of some vertex in the new layer.
We take a similar approach, but must somehow accomodate the vertices of $T$, whose neighbors we must check \emph{before} including them in a layer.
In order to ensure that this modification does not substantially increase the runtime of the flow-finding algorithm of~\cite{MP08}, it will prove helpful to maintain a digraph whose arcs point in either direction along each edge of $G$, and from which we delete any arc ending at a vertex with a defined layer.
As a result, we may easily determine whether a vertex with a defined layer has a single neighbor which does not yet have a defined layer, and also easily determine whether a vertex without a defined layer has any neighbors without a defined layer.
Algorithm~\ref{alg:addToLayer} defines a subroutine $\AddToLayer$ to perform the necessary edge deletions whenever we define the layer of a vertex, as well as several other operations useful to an iterative procedure for building an eflow by layer-sets, by operating on several global variables.

Using the subroutine $\AddToLayer$, we may easily define a procedure similar to Algorithm~1 of~\cite{MP08} to build an eflow by layer sets.
\begin{algorithm}[p]

	\PROCEDURE \FindEflow(G,I,O,T):

	\Input{%
		$G$, a graph with subsets $I, O, T$, where $T \subset V(G) \setminus O$.
	}

	\Output{%
		Either $\nil$, if $(G,I,O,T)$ does not have an eflow; or an ordered pair $(f,\ell)$ consisting of a partial function $f: V(G) \partialto* V(G)$ and a function $\ell: V(G) \to* \N$ such that $(f,\preceq)$ is an eflow, for a partial order $\preceq$ given by $v \preceq w \;\iff\; [\, v = w \text{~or~} \ell(v) > \ell(w) \,]$.
	}

	\BEGIN
		\LET $f: V(G) \partialto V(G)$\;
		\LET $\ell: V(G) \to \N$\;
		\LET $W, S \gets \vide$\;
		\LET $D \gets \text{[\,completely disconnected digraph on $V(G)$\,]}$\;
		\algline
		\FORALL $v \in V(G)$ \DO																						\nllabel{line:eflowInitLoop}
			$\neighbors(D, v) \gets \neighbors(G, v) \setminus O$\;
			\IF $v \in O$ \THEN { 
				$\ell(v) \gets 0$~;
				$f(v) \gets \nil$\;
				\oneline
				\IF $v \notin I$ \THEN $S \gets S \union \ens{v}$\; }
 			\ELSE {
				$W \gets W \union \ens{v}$\;
			}
		\ENDFOR																												\nllabel{line:eflowInitEnd}
		\medskip
%
%
		\LET $\layer \gets 0$\;
		\LET $\emptyLayer \gets \false$\;
		\REPEAT																											\nllabel{line:layerLoop}
			$\emptyLayer \gets \true$~;
			$\INCR \layer$\;
			\LET $S' \gets \vide$\;

			\FORALL $w \in S$ \DO {
				\IF $\neighbors(D,w) = \ens{v}$ \SUCHTHAT $v \notin T$ \THEN {							\nllabel{line:checkSuccessor}
					$S' \gets S' \union \ens{v}$\;																\nllabel{line:newSuccessor}
					$\AddToLayer(v,w)$\;
				} \ELSEIF $w \in T$ \THEN {																		\nllabel{line:checkYZmeas}
					$\AddToLayer(w,w)$\;
				} \ELSE {
					$S' \gets \ens{w}$\;																				\nllabel{line:propSuccessor}
				}
			} \ENDFOR
		\smallskip
		$S \gets S'$\;																									\nllabel{line:successorCopy}
	\UNTIL { $\emptyLayer$ }																						\nllabel{line:layerLoopEnd}

	\medskip
		
	\short
	\IF $W = \vide$
		\THEN {\RETURN $(f,\ell)$ }
		\ELSE { \RETURN \nil }
	\END

	\caption[An iterative algorithm to obtain an eflow.]{\label{alg:findEflow}%
		An iterative algorithm to obtain an eflow, using the maximally-delayed layers technique of~\cite{MP08} and the subroutine $\AddToLayer$.}
\end{algorithm}
We begin by creating the digraph $D$, copying the neighborhood relations of $G$, but removing any arcs ending in $O$ (which is necessarily the zeroeth layer).
Any output vertex which is in the range of a ``successor'' function $f: O\comp \to I\comp$ is a potential successor: we define a set of candidate successors from the elements of $O \setminus I$.
Vertices without a defined layer (initially the elements of $O\comp$) are included in the set $W$.
For each layer, we iterate through the set of potential successors, which initially includes no elements of $T$; we maintain the invariant that $v \in T$ becomes a candidate successor only if it has out-degree zero (\ie\ all of its neighbors are in some defined layer).
For any viable successor $w$ --- one satisfying the eflow constraints that $w \in T$ and $w \prec u$ implies that $u \in V^\prec_j$ for some $j < \layer$, or that $w = f(v)$ for $v$ the unique vertex not in $V^|prec_j$ for some $j < \layer$ --- we assign $w$ as the successor of the appropriate vertex $v$, and define the layer of that vertex $v$.
If this vertex $v$ is different from $w$, it then becomes a potential successor for the next layer; and any neighbors of $v$ which are in $T$, and which (subsequent to the removal of any arcs ending in $v$) have out-degree zero, also become viable successors for the next layer.
Otherwise, if $w$ was not a viable successor for this layer, we carry it forward for the next layer.
We assign the list of new candidate successors to $S$ before the end of the iteration; and we repeat until we cannot assign any new vertices to a layer.

If we cannot assign any vertices to new layers, this is either because we have exhausted the vertices (in which case $W = \vide$), or because there are no vertices which are reachable from the existing layers which have viable successors.
In the latter case, there is no eflow; and in the former, we return the eflow ``sucessor'' function and the layer function.

\subsubsection{Analysis of Algorithm~\ref{alg:findEflow}}
\label{sec:findEflowAnalysis}

In the subroutine $\AddToLayer(v,w)$, the run-time is dominated by the loop over the neighbors $z$ of $v$\,, which requires time at most $O(\deg_G(v))$\,.
This subroutine is called at most once for every vertex $w \in V(G)$\,: then, the cumulated run-time of $\AddToLayer(v,w)$ over the execution of $\FindEflow$ is $O\big(\sum\limits_v \deg_G(v) \big) = O(m)$, where $m = \card{E(G)}$.

Let $n = \card{V(G)}$ and $k = \card{O}$.
Initially, there are $k = \card{O}$ elements of $S$, none of which are elements of $T$\,; and as every element $w \in S \setminus T$ corresponds to a unique element of $S' \setminus T$ (either itself, or the vertex $v$ for which $w$ is the successor), there always remain at most $k$ elements of $S \setminus T$ at each iteration of the loop on lines~\ref{line:layerLoop}\thru\ref{line:layerLoopEnd}.
The comparisons on lines~\ref{line:checkSuccessor} and~\ref{line:checkYZmeas} can both be performed in constant time using the set data structures described at the beginning of Section~\ref{sec:flowAlgs}, as can the set-inclusions on lines~\ref{line:newSuccessor} and~\ref{line:propSuccessor}.
The assignment on line~\ref{line:successorCopy} can be performed simply by re-attributing the element list of $S'$ to $S$, as we will immediately re-assign $S'$ to the empty set in any subsequent iteration.
There are at most $n$ iterations of the loop on lines~\ref{line:layerLoop}\thru\ref{line:layerLoopEnd}, as it only repeats as long as at least one vertex has been assigned to a new layer.
Then, setting aside the work performed for vertices $w \in S \inter T$ in the loop on lines~\ref{line:layerLoop}\thru\ref{line:layerLoopEnd} and the work performed by $\AddToLayer$, the work performed by that loop is $O(kn)$.

As elements of $T$ are only included into $S'$ when they are guaranteed to be attributed to the next layer, the total work performed by that loop (again aside from work performed by $\AddToLayer$) for elements of $T$ is then $O(n)$.
The run-time of $\FindFlow(G,I,O,T)$ is then $O(kn + m)$, dominated by the work performed by $\AddToLayer(v,w)$ for each $w \in V(G)$, and by the iteration of the loop on lines~\ref{line:layerLoop}\thru\ref{line:layerLoopEnd} for at most $n$ layers.

This is precisely the run-time described for Algorithm~1, except that Mhalla and Perdrix also use the extremal result of Theorem~\ref{thm:extremalUpperBound} presented in~\cite{BP08} to simplify this to $O(kn)$.
It is an open question whether there is a similar extremal result for eflows.

As we remarked following Definition~\ref{def:eflow}, an efficient algorithm for eflows allows us to obtain a semantic map for the simplified RBB construction using similar techniques as we have presented for the DKP construction throughout the Chapter, as well as for any extension of the DKP or simplified RBB constructions which makes use of the more general $\mathfrak{Zz\cdots z}^\theta$ pattern.

\subsection{Generalized flows}
\label{sec:gflows}

In Definition~\ref{def:eflow}, we have described one sense in which flows may be generalized, by observing that the role of flows is essentially to permit the stabilizer formalism to certify that the measurement pattern performs a unitary operation by attributing some generator $K_w$ to each qubit $v \in O\comp$ to be measured.
A more general approach than this was defined in~\cite{BKMP07} in \emph{generalized flows} (or \emph{gflows}), which extends this both to arbitrary generators of the initial stabilizer group $\gen{K_w}_{w \in I\comp}$ (\ie\ to non-trivial products of the operators $K_w$) and to multiple planes of measurement.

Analogously to \XY-plane measurements and \YZ-plane measurements as we have described above, we may define \emph{\XZ-plane measurements} to be measurements with respect to a basis of the form
\begin{gather}
		\ket{\spm\sfx\sfz_\theta}
	\;\;=\;\;
		\sfrac{1}{\sqrt 2} \ket{\splus \sfy}
	\;\pm\;
		\sfrac{\e^{i\theta}}{\sqrt 2} \ket{\sminus \sfy}	\;.
\end{gather}
(For the duration of this section, we will write $\ket{\spm \sfx\sfy_\theta} = \ket{\pmsub\theta}$ for the sake of uniformity.)
Generalized flows encode a set of conditions imposed on measurements in the \XY, \YZ, and \XZ\ planes so that, as we described in Section~\ref{sec:extendedConstructions}, each observable to be measured anticommutes with exactly one of the remaining generators of the initial stabilizer group, commuting with the rest.
Let $\lambda: O\comp \to* \ens{\XY, \YZ, \XZ}$ be a function which designates, for each qubit to be measured, the plane in which it is to be measured.
The generalized flow conditions are as follows:
\begin{definition}[(paraphrased from~\cite{BKMP07})]
	\label{def:gflow}
	Let $(G,I,O)$ be a geometry and $\lambda$ be a function defined on $O\comp$ indicating measurement planes as above; and let $\cP$ be the power-set function, and $\odd(S)$ be the set of vertices in $G$ which are adjacent to an odd number of elements of $S$.
	Then a \emph{gflow} for $(G,I,O,\lambda)$ is a tuple $(g,\preceq)$ consisting of a function $g: O\comp \to \cP(I\comp)$ and a partial order $\preceq$ which satisfy the following conditions:
	\begin{subequations}
	\begin{align}
			w \in g(v) \;\implies\;&	v \preceq w ,
		\\
			w \in \odd(g(v)) \;\implies\;& v \preceq w ,
		\\
			\lambda(v) = \XY \;\implies\;& v \in \odd(g(v)) \setminus g(v)	, 
		\\
			\lambda(v) = \YZ \;\implies\;& v \in g(v) \setminus \odd(g(v)) ,
		\\
			\lambda(v) = \XZ \;\implies\;& v \in g(v) \inter \odd(g(v)) .
	\end{align}
	\end{subequations}
\end{definition}
The conditions above are sufficient conditions for there to exist a measurement pattern, with underlying geometry $(G,I,O)$ and performing measurements in the planes specified by $\lambda$, which performs a unitary for all specific choices of measurement angles. 
We may show this as follows.

%

We represent measurements in the \XY, \YZ, and \XZ\ planes by observables $\sO^{\XY,\theta}$, $\sO^{\YZ,\theta}$, and $\sO^{\XZ,\theta}$ respectively, which (like Pauli operators) have eigenvalues $\pm 1$, with corresponding eigenvectors $\ket{\spm\sfx\sfy_\theta}$, $\ket{\spm\sfy\sfz_\theta}$, and $\ket{\spm\sfx\sfz_\theta}$ respectively:
\begin{subequations}
\label{eqn:pauliPlaneObs}
\begin{gather}
		\sO^{\XY,\theta}
	\;\;=\;\;
		\ket{\splus\sfx\sfy_\theta}\bra{\splus\sfx\sfy_\theta}
		\;\;-\;\;
		\ket{\sminus\sfx\sfy_\theta}\bra{\sminus\sfx\sfy_\theta}
 	\;\;=\;\;
		\cos(\theta) X + \sin(\theta) Y	\;;
	\\[1ex]
		\sO^{\YZ,\theta}
	\;\;=\;\;
		\ket{\splus\sfy\sfz_\theta}\bra{\splus\sfy\sfz_\theta}
		\;\;-\;\;
		\ket{\sminus\sfy\sfz_\theta}\bra{\sminus\sfy\sfz_\theta}
	\;\;=\;\;
		\cos(\theta) Z + \sin(\theta) Y	\;,
	\\[1ex]
		\sO^{\XZ,\theta}
	\;\;=\;\;
		\ket{\splus\sfx\sfz_\theta}\bra{\splus\sfx\sfz_\theta}
		\;\;-\;\;
		\ket{\sminus\sfx\sfz_\theta}\bra{\sminus\sfx\sfz_\theta}
	\;\;=\;\;
		\cos(\theta) X + \sin(\theta) Z	\;.
\end{gather}
\end{subequations}
After applying the preparation and entangling operations $\Ent{G} \New{I\comp}{}$ in a one-way measure\-ment pattern, the state-space is stabilized by the group $\sS_{G,I} \,=\, \gen{K_v}_{v \in I\comp}$\,.
As we noted in Section~\ref{sec:extendedConstructions}, to describe how measurement of these observables transform the space stabilized by $\sS_{G,I}$, the analysis of measurements by a single-qubit observable $P_v$ in the stabilizer formalism requires only that each measurement observable either commutes or anticommutes with every generator of the subgroup of the Pauli group which stabilizes the state of the system.
If in particular, for all $1 \le t \le \card{O\comp}$, the state of the system after the $j\th$ measurement is stabilized by a group $\sS\supp{t}$\,, such that the $t+1\st$ measurement operator anticommutes with at least one element of $\sS\supp{t}$, the evolution of the system at each stage will be unitary, and the state post-measurement (selecting for the $+1$ measurement result) will be stabilized by a subgroup $\sS\supp{t+1}$ of the group $\sS\supp{t}$.

Suppose we require that the system at all times be in a joint $+1$-eigenstate of some subgroup of $\sS$ after each measurement.
It then suffices to find a generating set $\ens{S_1, \ldots, S_{\card I\comp}}$ for $\sS_{G,I}$ such that the state of the system after the $t\th$ measurement (and a classically controlled correction, if necessary) is stabilized by $\gen{\,S_j\,}_{j > t}$\,.
In particular, we require $S_t$ to be the generator which anticommutes with the observable of the $t\th$ measurement (\ie\ with 
$\sO_v^{\XY,\theta}$, $\sO_v^{\YZ,\theta}$, or $\sO_v^{\XZ,\theta}$ as appropriate), and for every $S_j$ to commute with that observable for $j > t$, \emph{regardless} of the value of $\theta$.
For a measurement on a qubit $v$, this imposes the following constraints:
\begin{itemize}
\item
	for an $\XY$-plane measurement, we require that $S_k$ acts on $v$ with a $Z$ operation, and $S_j$ acts on $v$ with the identity operation for $j > k$;

\item
	for a $\YZ$-plane measurement, we require that $S_k$ acts on $v$ with an $X$ operation, and $S_j$ acts on $v$ with the identity operation for $j > k$;

\item
	for an $\XZ$-plane measurement, we require that $S_k$ acts on $v$ with an $Y$ operation, and $S_j$ acts on $v$ with the identity operation for $j > k$.
\end{itemize}
In particular, we require that the stabilizer groups $\gen{S_j}_{j > k}$ for each $1 \le k  \le \card{O\comp}$ do not act on the first $k$ qubits which were measured.

From these constraints, we can re-derive the gflow conditions, as follows.
Consider a fixed choice of generating set $\ens{S_j}_{j = 1}^{\card{I\comp}}$ for which the conditions above may be satisfied, for \emph{some} ordering of the generators: each such operator consists of a product of operators $K_v$ for $v \in I\comp$\,.
Define the function $\gamma: \ens{1,\ldots,\card{O\comp}} \to \cP(I\comp)$ such that
\begin{align}
	\label{eqn:presentGenerators}
		S_j
	\;\;=\;\;
		\prod_{v \in \gamma(j)} K_v
	\;\;=&\;\;
		\prod_{v \in \gamma(j)} \paren{ X_v \prod_{w \sim v} Z_w }
	\notag\\\propto&\;\;
		\sqparen{\prod_{v \in \gamma(j)} \!\! X_v \;} \sqparen{\prod_{w \in \odd(\gamma(j))} \!\! Z_w \;} .
\end{align}
Let $\sigma: O\comp \to \N$ describe an arbitrary sequence in which the qubits of $O\comp$ are measured; let $\preceq$ describe the coarsest partial order (relative to the choice of generators $\ens{S_j}_{j = 1}^{\card{I\comp}}$) consistent with every such ordering of $O\comp$\,; and let $g = \gamma \circ \sigma$.
In order for the commutation/anticommutation properties above to hold, we then require the following for the qubit $v$ which is the $k\th$ qubit measured:
\begin{romanum}
\item
	As $S_k$ does not act on any qubits yet to be measured, any qubit $w \in \gamma(k)$ must be measured later than $v$\,, and similarly for any qubit $w \in \odd(\gamma(k))$.
	Therefore, if either $w \in g(v)$ or $w \in \odd(g(v))$, we require $v \preceq w$.

\item
	If $v$ is to be measured in the \XY\ plane, $S_k$ must act on $v$ with a $Z$ operation, in order for the observable $\sO_\theta^{\XY}$ to anticommute with $S_k$; this holds if and only if $v \in \odd(g(v))$ and $v \notin g(v)$.

\item
	If $v$ is to be measured in the \YZ\ plane, $S_k$ must act on $v$ with an $X$ operation, in order for the observable $\sO_\theta^{\YZ}$ to anticommute with $S_k$; this holds if and only if $v \in g(v)$ and $v \notin \odd(g(v))$.

\item
	If $v$ is to be measured in the \XZ\ plane, $S_k$ must act on $v$ with an $Y$ operation, in order for the observable $\sO_\theta^{\XZ}$ to anticommute with $S_k$; this holds if and only if $v \in g(v)$ and $v \in \odd(g(v))$.
\end{romanum}
Then, the conditions for applying the extended stabilizer formalism independently of the measurement angles are precisely those of of Definition~\ref{def:gflow}.

As noted above, if each measurement results in a collapse to their respective $+1$-eigenvalues, the resulting transformation of the state-space is an isometry; and for any $t$ such that the $t\th$ measurement results in the $-1$-eigenvalue of the measurement observable, we may apply $S_t$ as a correction operation to steer it into the post-measurement state corresponding to the $+1$ measurement result.

Therefore, the existence of a gflow for $(G,I,O,\lambda)$ acts as a certificate that a measurement pattern which performs the appropriate corrections (or measurement adaptations) performs a unitary transformation, via the stabilizer formalism.
This fully generalizes the extent to which the stabilizer formalism (relaxed to admit non-Pauli observables) can be used to define certificates for unitarity without requiring precise information about individual measurement observables.\label{discn:gflowsSpecialCases}
In particular, eflows are a special case for gflows, where we require that $g(v)$ is a singleton set for all $v \in O\comp$\,; and flows are a further restriction where we require $v \ne f(v)$ for all $v \in O\comp$.

It is not known whether gflows correspond to any construction of interest for one-way measurement patterns (in the sense that flows correspond to the DKP construction and eflows to the simplified RBB construction, as described in Sections~\ref{sec:flowDefn} and~\ref{sec:extendedConstructions} respectively); however, as an extension of flows, generalized flows seem to extend the semantic map $\SemanticDKP$ to accommodate measurement patterns which may be translated into circuits using closed time-like curves~\cite{Kashefi07priv} (a notion introduced into quantum information theory by Deutsch~\cite{Deutsch91}).\footnote{%
	In general, the availability of closed time-like curves are a powerful computational resource for both classical and quantum computation~\cite{AW08}; however, the extension of $\SemanticDKP$ to patterns with gflows suggested by Kashefi does not seem to produce arbitrary circuits which use closed time-like loops (\cf\ the discussion in the introduction to Chapter~\ref{Measurement Pattern Interpolation} on postselection).}
By virtue of having an eflow, the pattern $\mathfrak{Zz\cdots z}$ of~\eqref{eqn:zzMany} has a gflow.
However, for any measurement pattern whose corresponding tuple $(G,I,O,\lambda)$ has a gflow, every measurement will produce independent and maximally random measurement results; therefore, the qubit-reversal pattern of Section~\ref{sec:qubitReversal} (which, as we noted on page~\pageref{discn:qubitReversalCertainty}, yields some measurement results with certainty conditioned on prior results) does not have a gflow.

In~\cite{MP08}, Mhalla and Perdrix describe an $O(n^4)$ algorithm (where $n = \card{V(G)}$) for determining whether a geometry $(G,I,O)$ has a gflow, conditioned on every qubit being measured in the $\XY$-plane; it is not known whether there exists an efficient algorithm for finding a gflow under more general conditions, although this seems likely.

\subsection{Ignoring measurement dependencies when inferring unitary circuits}

In the preceding discussion of flows, eflows, and gflows as certificates of unitarity in terms of the stabilizer formalism, we have quickly brushed over the issues of the appropriate measurement adaptations and corrections depending on measurement results.
The stabilizer formalism itself furnishes a choice of immediate corrections which suffice to guarantee unitarity, which can then be absorbed into measurements and further refined by Pauli simplifications and signal shifting; however, in situations where there exist more than one ``flow-like'' certificate (which may occur for flows when $\card{I} < \card{O}$, or for gflows), it may be difficult to verify whether a particular measurement pattern meets conditions which are sufficient to show that it performs a unitary transformation.
However, if the geometry underlying a particular measurement pattern (together with additional information such as the planes of measurement for each qubit) admits a flow-like certificate, we may if we wish consider the scheme of dependencies which arise from that certificate in order to ``force'' the pattern to represent of a unitary transformation, changing the measurement/correction dependencies as required.
Because the dependencies serve only to simulate the selection of a particular post-measurement result, the particular dependency scheme does not matter (except possibly for complexity reasons): then one one arising from an arbitrary certificate of this kind will suffice.
We will consider a potential application of this approach in Chapter~\ref{Measurement Pattern Interpolation}.

If we are willing to ignore any existing dependencies in a measurement pattern and simply impose the dependencies arising from some flow-like certificate, we may streamline any algorithm for semantic maps corresponding to constructions of one-way measurement patterns by omitting the step for checking the dependencies, which \eg\ dominates the run-time of Algorithm~\ref{alg:semanticDKP}.

\section{Review and open problems}
\label{sec:flowsReview}

In this chapter, we have reviewed the concept of flows presented in~\cite{DK06}, identifying it as arising out of combinatorial constraints on expressions for unitary circuits.
We have presented a characterization of flows in terms of path covers, and presented uniqueness and extremal results, which contributed to the discovery of the first efficient algorithm for finding flows for geometries of one-way measurement patterns, and also help to bound the run-time of the best existing algorithms for finding flows.
We used these results to describe a function $\cS$ which provides semantics for mapping measurement patterns congruent to those produced by the DKP construction, in terms of unitary circuit models $\Jacz(\A)$ for $\A \subset \R$ defined in Section~\ref{sec:unitaryCircuits}, when those patterns have input and output subsystems of the same size.

The final constraint above in the semantic map $\cS$, that $\card{I} = \card{O}$, does not merely arise out of the corresponding constraint for the procedure $\FindFlow$ of Algorithm~\ref{alg:completeFlowAlg}.
In order to verify that the measurement and correction dependencies for a pattern suffice for the pattern to perform a unitary, the condition $\card{I} = \card{O}$ provides the guarantee (by Theorem~\ref{thm:uniqueFamilyPaths}) that there is at most one flow-function, and therefore at most one set of dependencies which are possible after Pauli simplifications and signal shifting.
For a pattern where $\card{I} < \card{O}$ whose geometry admits more than one flow-function, it becomes necessary to verify that there exists a flow function which gives rise to the dependencies observed in the pattern.
It is possible that the dependency information in the pattern may prove useful for discovering whether such a flow-function exists, but no such algorithm is known.

As we have observed at many points throughout this Chapter, with regards to uniqueness, extremal results, and decompositions in terms of star geometries, the presence of a flow gives rise to many structural constraints, some of which limit the possible complexity of the measurement pattern.
It seems plausible that these may lead to interesting bounds on the size and depth complexity of measurement patterns arising from the DKP construction, or on the time required to compute the dependency matrices described in Theorem~\ref{thm:dependenciesDKP}.


	\chapter[Measurement Pattern Interpolation]{
		Measurement Pattern Interpolation (from Quadratic Form Expansions)}


Can measurement-based quantum computation provide new idioms for designing quantum algorithms?
That is: in addition to being universal for quantum computation by \eg\ representing unitary circuits via the DKP or RBB constructions, can it also provide analytical tools which suggest how currently \emph{unsolved} problems might be solved on a quantum computer?
There are currently no obvious high-level concepts of the sort provided by the circuit model (such as eigenvalue estimation or amplitude amplification) or quantum walks (such as deflection by engineered energy barriers) which may lead to interesting quantum algorithms; however, the ability to describe novel solution techniques is part of what distinguishes a potentially useful model of computation from a model which is ``merely'' a potentially useful scheme for physical implementation.

The largest obstacle to high-level descriptions of algorithms in the measurement based model --- without depending on semantics being provided by another model, \eg\ a unitary circuit model --- seems to be in the details of the classical feed-forward which specifies how measurements are to be adapted based on previous measurement results.
To perform a unitary transformation with a measurement-based algorithm requires in practise a case analysis at nearly every measurement, which is a distraction from any potential ``uniform'' description of the computation.
Therefore, a likely prerequisite for developing intuitions about how to solve problems in measurement-based models is to abstract away this detail of measurement-based computation.

By definition, in a measurement-based procedure which performs a unitary transformation, the final quantum state is independent of the measurement results, even if the operations performed to obtain that state were not.
This suggests an approach where we omit all details of the measurement results or the probabilities of them occurring, except for some particular ``distinguished'' sequence of measurement results which can occur with non-zero probability.
For instance, we may consider the measurements performed by the ``default'' measurement angles (with respect to which all of the measurement adaptations are performed): if the measurement results $\s[v] = 0$ occur for each $v \in O\comp$ with non-zero probability, the adaptation in a procedure which performs a unitary serves only to select for the same transformation as would occur for that branch.
In this case, a measurement-based procedure for a unitary transformation may be regarded as \emph{simulating postselection} of this ``distinguished'' sequence of results, which is feasible due to the special choice of initial state and measurements involved.\footnote{%
	Aaronson~\cite{Aaronson05} shows that allowing postselection even of single-qubit measurement results, together with a set of (approximately) universal unitary operations and preparation of the $\ket{0}$ state, leads to circuit models which can efficiently solve problems in \PP: this is taken as a sign of the unreasonability of models allowing postselection.
	In contrast, the way in which measurements are simulated in the stabilizer formalism (and in the context of the flow/eflow/gflow techniques outlined in Section~\ref{sec:flowsFurtherResults}) allows the evolution of the state-space to be efficiently modelled due to the existence of a highly structured system of correlations of the states being measured, and the relationships between the measurement observables and those correlations.}

If we could actually postselect the distinguished measurement results, then we would not need to perform any adaptation of the measurement procedure, and dispense with classical feed-forward.
This would leave us with a set of data from which one could plausibly form an analytical intuition:
\begin{enumerate}
\item
	a set of qubits $V$\,, with an input subsystem $I \subset V$ supporting an arbitrary input state, with the others prepared initially in the $\ket{\splus}$ state\,;

\item
	a graph $G$ on $V$, describing a set of entangling operations performed between pairs of qubits;

\item
	a \emph{set} (without any ordering) of measurements to perform on the qubits of $V$; and

\item
	the subset of qubits $O \subset V$ on which no measurements are to be performed, in order to leave a residual quantum state as output.
\end{enumerate}
Such data represents a procedure to solve a problem using a \emph{postselection-based} (rather than measurement-based) model of quantum computing. 

Supposing that one can form intuitions about quantum computation in these terms, and even derive tuples $(G,I,O,M)$ consisting of  a geometry $(G,I,O)$ and a set of default measurements $M$ encoding a solution to a computational problem, it would then remain to produce from this a description of a realizable algorithm.
In this case, it is sufficient to specify an ordering of the measurements in the postselection-based ``algorithm'', as well as a specification of how those measurements are to be adapted and any final corrections to be applied in case the distinguished measurement results do not occur.
If this is possible, the postselection-based procedure is meaningfully an algorithm, which may be extended in principle to a realizable procedure to perform the operation with bounded error.

In the case where every measurement is performed in the \XY\ plane, the basis of each measurement is of the form $\ket{\pmsub\alpha} \propto \ket{0} \pm \e^{i\alpha} \ket{1}$ for some angle $\alpha$, as defined in~\eqref{eqn:pmsubAlpha}.
Suppose that we are given a geometry $(G,I,O)$, and a vector $\vec{a} = (\alpha_v)_{v \in O\comp}$ of the default measurement angles for each $v \in O\comp$, which together define a postselection-based transformation $\Upphi: \D(I) \to \D(O)$ in the case where all of the measurements on the qubits $v$ yield $\s[v] = 0$.
Such a postselection-based procedure will also be a special case of some measurement-based procedure, with different computational branches for each sequence of measurement results: we refer to the branch which performs the desired postselection procedure, given by the measurement sequence $\big(\s[v]\big)_{v \in O\comp} \equiv 0$, as the \emph{positive branch}.

Under these constraints, we may formalize the problem of mapping postselection-based procedures to measurement-based algorithms as follows:
\begin{problem}[Measurement Pattern Interpolation (\MPI)]
	\label{problem:MPI}
	Let $(G,I,O,\vec{a})$ be a partial specification of a measurement pattern $\mathfrak P$ in the $\OneWay(\A)$ model for some $\A \subset \R$, where $\mathfrak P$ has underlying geometry $(G,I,O)$ and default measurement angles $\vec{a} = (\alpha_v)_{v \in O\comp} \in \A^{O\comp}$\,.
	Given that $\mathfrak P$ performs a unitary transformation $U: \cH[I] \to \cH[O]$ conditioned on its' positive branch, determine a measurement pattern $\overline{\mathfrak P}$ with underlying geometry $(G,I,O)$ which performs $U$ unconditionally, if such a pattern exists.
\end{problem}
(This problem may readily be extended beyond measurements performed in the \XY-plane, to which $\OneWay(\A)$ is restricted; however, we will be primarily concerned with the formulation presented above.)

In this chapter, I will describe solutions for special cases of the Measurement Pattern Interpolation problem, which revolve around generalized flow conditions on the geometry $(G,I,O)$, or on the special case of ``Pauli measurements'', \ie\ when we have $\ket{\spm_{\alpha_v}} = \ket{\spm \sfx}$ or $\ket{\spm_{\alpha_v}} = \ket{\spm \sfy}$ for each $v \in O\comp$.

In the absence of ``algorithmic insights'' provided by measurement-based computation --- and setting aside theoretical motivations, such as obtaining a more robust understanding of the scope of those measurement-based procedures which perform unitary transformations --- solving \MPI\ is only of practical importance if there exists a source of ``problem instances'' $(G,I,O,\vec a)$ for which we wish to obtain measurement patterns. 
In this chapter, I also present \emph{quadratic form expansions}, which is a format in which unitaries may be expressed concisely.
These are plausibly a ``natural'' source of instances of the Measurement Pattern Interpolation problem, in that these expressions (and similar ones) have arisen multiple times in the course of research in quantum information independently of the formulation of \MPI.

\paragraph{Previous appearances of this work.}

Some of the work in this chapter are developments original to this thesis, including the substitution described in~\eqref{eqn:diagonalExpansions}, and the solved case of the \MPI\ for geometries derived from eflows described in Section~\ref{sec:synthEflows}.
The rest of the work of this chapter is the result of a collaboration with Elham Kashefi, Vincent Danos, and Martin Roetteler, and first appeared in~\cite{BDKR08}.

\section{Quadratic Form Expansions}

\subsection{Notation and definitions}

We first present a notational device which will prove most useful:
\begin{notation}
	For an index set $V$, and for a vector $\vec{x} \in \ens{0,1}^V$, and for a subset $S \subset V$, the vector $\vec{x}_S$ denotes the restriction of $\vec{x}$, considered as a function $\vec x: V \to \ens{0,1}$, to the set $S$.
	In particular, for an operator $\ket{\vec x} \in \cH^{\ox V}$ given by
	\begin{gather}
			\ket{\vec x}
		\;\;=\;\;
			\bigotimes_{v \in V} \;\ket{x_v}\,,
	\end{gather}
	we define the corresponding standard-basis vector $\ket{\vec x_S} \in \cH^{\ox S}$ by
	\begin{gather}
			\ket{\vec x_S}
		\;\;=\;\;
			\bigotimes_{v \in S} \;\ket{x_v}\,.
	\end{gather}
\end{notation}
\begin{definition}
	\label{def:quadFormExpan}
	Let $V$ be a set of $n$ indices, and $I, O \subset V$ be (possibly intersecting) subsets.
	Then a \emph{quadratic form expansion} for an operator $M: \cH^{\ox I} \to \cH^{\ox O}$ is an expression of the form
	\begin{align}
		\label{eqn:quadFormExpansion}
			M
		\;=\;\;
			C\! \sum_{\vec{x} \in \ens{0,1}^V} \!\e^{i Q(\vec{x})} \;\ket{\vec{x}_O}\bra{\vec{x}_I}	\;,
	\end{align}
	where $Q$ is a quadratic form\footnote{%
	A \emph{quadratic form} is a polynomial where every term has degree $2$.}
	in $\R^V$, and $C$ is a complex scalar.
\end{definition}
For an operator expressed in matrix form, a quadratic form expansion provides an expression for each matrix coefficient as a scalar multiple of a sum of complex numbers with modulus $1$ (or \emph{complex units}), where the scalar factor and the arguments of these complex units are given by a uniform formula.
As such, it is essentially an interpretation of an operator as an abstract ``sum over paths'', where each path has the same amplitude, but a different phase: each such ``path'' is given by a particular string $\vec x$, and the coefficient for a transition $\vec{a} \mapsto \vec{b}$ is given by summing over all paths satisfying the constraints $\vec x_I = \vec a$ and $\vec x_O = \vec b$.

\subsection{Related work in the literature}
\label{sec:relatedWork}

Similar expressions to those in~\eqref{eqn:quadFormExpansion} have occurred several times previously in quantum information or quantum mechanics literature (albeit without the terminology or notation presented here).
We will provide a brief overview of some examples.

\subsubsection{Stabilizer codes and Clifford group operations}

Perhaps the earliest presentation of quadratic form expansions is by Schlingemann and Werner~\cite{SW2001}, where they are used to describe a special class of stabilizer codes (but using significantly different notation, and where the expansions are extended to $\vec x \in \Z_d^V$ for $d \ge 2$ rather than only for $\vec x \in \ens{0,1}^V$).
Their analysis corresponds in the boolean case to quadratic form expansions where $Q$ is obtained from the adjacency matrix of a graph, and where $I, O$ form a partition of the index set $V$.
This treatment was extended to describe generalized stabilizer codes and Clifford group operations over $\U(d)$ in Schlingemann~\cite{Schl04}.
The link between the work of Schlingemann and Werner (as well as their notations), and quadratic form expansions as they are defined and treated in this thesis, is perhaps best hilighted by the result of~\cite{GKR2002}, which proved that all stabilizer codes are locally equivalent to some graphical code.

Dehaene and de Moor~\cite{DM03} also present an analysis of Clifford group operations over $\cH$, in which they present an operator formula for a generic Clifford group operation which may be obtained from a description of how it transforms the Pauli group by conjugation.
We will describe this matrix formula, and specify how it may be computed, in Section~\ref{sec:synthClifford}.

\subsubsection{The quantum Fourier transform}

One concrete example of a unitary matrix which is naturally expressed as a quadratic form expansion is the quantum Fourier transform over the cyclic group $\Z_{2^k}$, which we may write
\begin{align}
		\cF_{2^k}
	\;\;=&\;\;
		\sfrac{1}{\sqrt{2^k}} \, \sum_{x,y \in \Z_{2^k}} \e^{2\pi i x y/2^k} \ket{y}\bra{x}
	\notag\\[2ex]\cong&\;\;
		\sfrac{1}{\sqrt{2^k}}
		\sum_{\substack{\vec{x} \in \ens{0,1}^k \\ \vec{y} \in \ens{0,1}^k}}
		\exp\!\paren{\sfrac{i\pi}{2^{k-1}}
			\sqparen{ \sum\limits_{\phantom{h}\!\!j=0}^{k-1} 2^h x_j }
			\sqparen{ \sum\limits_{\phantom{j}\!\!h=0}^{k-1} 2^j y_h }}
			\ket{\vec y} \bra{\vec x},
\end{align}
identifying the integers $x,y \in \Z_{2^k}$ with their usual representations as binary strings; this is a quadratic form expansion with $V$ partitioned into disjoint index sets associated with the bits of $\vec{x}$ and $\vec{y}$, and a quadratic form $Q: \R^{2k} \to \R$ given by
\begin{align}
	Q(\vec x, \vec y) \;\;=\;\; \sum_{h = 0}^{k-1} \; \sum_{j = 0}^{k-h-1} \;\sfrac{2^{(h+j)}}{2^{k-1}} \;\pi \,x_j \,y_h \;.
\end{align}
On page~\pageref{sec:fourierTransformEg}, we will show how this structure can be exploited to reproduce the linear-nearest neighbor implementation of $\cF_{2^k}$ described by Fowler, Devitt, and Hollenberg in~\cite{FDH04} by essentially automatic techniques.

\paragraph{Remark.} The operation $\cF_{2^k}$ is an instance of a generalized Clifford group operation over $\SU(2^k)$\,; however, the description of $\cF_{2^k}$ above differs from the treatment of \eg~\cite{Schl04} in that we expand it here in terms of operations on qubits, rather than leaving it as an atomic operation on $\cH[2^k]$.

\subsubsection{Sums over paths}
\label{sec:sumsOverPaths}

A similar representation of operators as a sum over paths as in~\eqref{eqn:quadFormExpansion} was used by Dawson \etal\ in~\cite{DHHMNO04} to provide a succinct proof of $\BQP \subset \PP$, using cubic polynomials over $\ens{0,1}$ rather than quadratic polynomials; however, comments made in Section~VI of their work anticipate quadratic form expansions with discretized coefficients, and imply a method (which we will describe in Section~\ref{sec:quadFormExpansionsFromCircuits}) of obtaining these expansions from unitary circuits using the gate-set $\Phases(\frac{\pi}{4}\Z)$ defined on page~\pageref{eqn:elemGateSets}.

The expressions considered in~\cite{DHHMNO04} are explicitly described as a sum over computational paths which arise in the evolution of a quantum state through a unitary circuit: the evolution of the system in each path is described by a time-ordered sequence of standard basis states given by boolean strings, where these states differ in at most one bit position for consecutive time-steps.
The ``paths'' in this case are walks in the boolean cube $\ens{0,1}^k$ (for a circuit acting on $k \le n$ qubits) between boolean strings which differ in zero or one bit position (\ie\ including self-loops).
The bit whose value may change at each step is the same across all paths; these walks can then be described by the initial position, followed by one bit for each edge traversed, which may be described by a boolean function $\ens{0,1}^V$ describing the history of the computation in that path.
Summing over all computational paths gives rise to the unitary transition matrix.

The analogy of such matrix formulae to a sum-over-paths, suggested both in~\cite{DHHMNO04} and in this thesis, is prompted in part by the formal similarity between \eg\ quadratic form expansions and \emph{path integrals}.
Introduced by Feynman~\cite{Feynman48}, the path integral formalism is an alternative approach to describing the evolution of quantum states by associating amplitudes to trajectories of the system (\eg\ through space-time or phase space), and describing the probabilities of events upon measurement using the cumulated amplitude over all paths starting from the initial state and ending at a given configuration.
This ``summation over paths'' can be arrived at by taking integrals over all possible configurations at intermediate times (taking the limit as the intervals between these intermediate times tend to zero), which amounts to integrating over a collection of ``virtual events'' for these intermediate times.
This leads to a formulation where the transition amplitude between the initial and final configurations depends on some quantity with the units of energy (usually the \emph{action} of the system) depending on physical parameters associated with these virtual events, which can then be attributed to trajectories in space.
Further details can be found in~\cite{Feynman48, FH65, Schulman81}.
In a quadratic form expansion, the indices of the vector $\vec x$ which are neither in $I$ or $O$ may be considered to correspond to these virtual events, and the quadratic form associated with the ``computational paths'' of the computation seems to correspond to the action (which in most applications of path integration determines the phase of each path).
This suggests that a quadratic form expansion, or similar matrix expression, may be a means of discretizing a path integral in a way which may be amenable to study for quantum algorithms.

There are some technical discrepancies on a formal level, most notably that the phase of a system with multi-body interactions generally is not a quadratic polynomial in the dynamical variables of the system;\label{discn:higherDegree} or we may wish to evaluate path integrals where different paths have amplitudes of different weights (\eg\ with lower amplitudes away from the stationary points of the action, due to cancellation of contributions of similar paths) in an attempt to obtain an approximation which is easier to estimate.
We will show in in Section~\ref{sec:higherDegree} how matrix expressions similar to quadratic form expansions, with \emph{unequally weighted} paths or where the action is of degree higher than $2$, may be transformed into quadratic form expansions.
The simplicity of these transformations suggest that quadratic form expansions are not unduly restrictive from the point of view of representing the discretization of a path integral.
Thus, an examination of when we can synthesize quantum algorithms from quadratic form expansions would plausibly lead to automatic techniques for efficiently estimating path integrals on a quantum computer.

\subsubsection{Remarks on quadratic form expansions as a representation of operators}

Given the possibility of quadratic form expansions arising generically from path-integral descriptions of physical systems, as well as the fact that these and similar matrix formulae have repeatedly arisen in the literature, I conjecture that quadratic form expansions can be expected to arise naturally as mathematical expressions of interest for operators in quantum information theory.
Based on this, I will suppose for the purposes of this thesis that a quadratic form expansion is a reasonable format in which a unitary operator $U$ may be specified, in the context of the problem of obtaining a succinct decomposition of a CPTP map performing $U$ in some standard model of quantum computing (\eg\ a unitary circuit model or measurement based model).

Setting aside the special case of the quantum Fourier transform $\cF_{2^k}$, we will see in Section~\ref{sec:synthClifford} how such an expression of a unitary $U$ may be obtained (without first having a unitary circuit or measurement-based algorithm for $U$) in the case of Clifford group operations.
While this represents a limited class of operations, it nevertheless represents a generic family of operators $U$ for which quadratic form expansions may be readily obtained.

\subsection{Constructing quadratic form expansions}

As noted above, we will be primarily interested in the problem of transforming a quadratic form expansion into a quantum circuit or measurement-based procedure.
However, to illustrate the generality of quadratic form expansions, and how they relate to unitary circuit models and the measurement based model of quantum computing, we will show how to perform the reverse transformation, from unitary circuits and measurement-based procedures to quadratic form expansions.

\subsubsection{Construction as a sum over computational paths in circuits}
\label{sec:quadFormExpansionsFromCircuits}

As we remarked in Section~\ref{sec:sumsOverPaths}, Dawson~\etal\ describe a method for obtaining quadratic form expansions from unitary circuits in~\cite[Section~VI]{DHHMNO04}; the particular set of unitary transformations they use is the set $\ens{H, R_\sfz(\pion 4), \cX}$.
The construction is closely related to the stable index circuit notation presented in Chapter~\ref{Fundamentals of quantum computation} (on page~\pageref{def:stableIndex}: we will sketch this construction using the somewhat more general parameterized gate set $\ens{H, Z^t, \cZ^t}$, where
\begin{align}
		Z^t
	\;=&\;
		\left[\begin{matrix} \;1 & 0 \\[0.5ex] \;0 & \;\e^{i\pi t} \end{matrix}\,\right]
	\;\propto\;
		R_\sfz(\pi t),
	&
		\cZ^t
	\;=&\;
		\left[\begin{matrix} \;1 & \;0 & \;0 & \!0 \\[0.5ex] \;0 & \;1 & \;0 & \!0 \\[0.5ex]
									 \;0 & \;0 & \;1 & \!0 \\[0.5ex] \;0 & \;0 & \;0 & \e^{i\pi t} \end{matrix}\,\right],
\end{align}
and where $t$ is allowed to range over some subset of $(-1,\, 1\;\!]$.

Consider a quantum circuit $C$ implementing $U$ exactly, using the operations $H$, $\cZ^t$, and $Z^t$.
As in a stable index representation of $C$, we assign indices (or \emph{path labels}, to use the terminology of~\cite{DHHMNO04}) to the segments of each wire in the circuit, with each segment bounded by an input terminal, and output terminal or a Hadamard operation.
In particular:
\begin{enumerate}
\item
	We enumerate the ``input'' segments of the circuit from $1$ to $k$, and for each wire $1 \le j \le k$ introduce a path label $x_j$ corresponding to an input bit $x_j \in \ens{0,1}$; and we set $I = \ens{1,\ldots,k}$.

\item
	Reading the circuit from the inputs to the outputs, we introduce new path variables $x_j$ for $j > k$ for each wire segment ``introduced'' by the presence of a Hadamard operation, corresponding to the ``random'' bit which is introduced by applying the Hadamard to the ``previous value'' of the bit.
\end{enumerate}
We may then describe computational paths described by the circuit by a walk in the boolean cube $\ens{0,1}^k$, with one bit possibly being flipped each time a Hadamard gate is performed.

If the total number of path labels introduced is $n$, a single computational path can therefore be described by setting all of the the path variables $x_1 \cdots x_n$ collectively to some particular binary string in $\ens{0,1}^n$.
For a path which is given by a string $\vec x \in \ens{0,1}^n$, the phase of that path (which governs how it interferes with other paths in the unitary evolution described by $C$) is given by a function $\varphi(\vec x)$ which may be iteratively constructed from the gates of $C$ as follows:
\begin{romanum}
\item
	For every Hadamard gate between two wire segments labelled with path variables $x_h$ and $x_j$, we include a term $x_h x_j$\,.

\item
	For every $\cZ^t$ operation between two wires, with a path variable $x_h$ labelling the segment of one wire and $x_j$ labelling the segment of the other wire in which the $\cZ^t$ operation is performed, we add a term $t\, x_h x_j$\,.

\item
	For every $Z^t$ operation on a wire segment labelled with a path variable $x_j$\,, we add a term $t \, x_j^2$.
	(As the path variable $x_j$ ranges over $\ens{0,1}$, the extra power of $2$ has no effect.)
\end{romanum}
We may collect $\cZ^t$ and $Z^t$ terms so that there is at most one $\cZ^t$ gate between segments labelled with $x_h$ and $x_j$ (for $h \ne j$) for some value of $t$, and at most one $Z^t$ gate on a segment labelled with $x_h$, for any $h,j \in \ens{1,\ldots,n}$.
As there is at most one Hadamard operation between two wire segments by construction of the circuit, this implies that the construction of $\varphi$ will include at most one $x_h x_j$ term for each $\ens{h,j} \subset \ens{1,\ldots,n}$.
The phase of a given path, described by a bit-string $\vec{x} \in \ens{0,1}^n$, is then given by $(-1)^{\varphi(\vec x)} = \e^{i\pi\varphi(\vec x)}$.
Each path also has an associated amplitude of $2^{-r/2}$, where $r = n - k$ is the number of Hadamard gates in the circuit.\footnote{%
Although it is quite reasonable to consider $\varphi$ to be simply a polynomial over $\R$, in terms of the descriptions used in Section~VI of~\cite{DHHMNO04}, one may consider $\varphi$ to be a polynomial over the ring $\R / 2\Z$.
If we restrict to $t \in \frac{\pi}{4} \Z$, we may simplify this to the finite ring $\Z_8$ by multiplying all of the coefficients by $4$, and using it to describe powers of $\sqrt{i}$ rather than of $-1$ (but still labelling each path by a vector $\vec x \in \ens{0,1}^n \subset \Z_8^n$).}

Let $O$ be the set of indices $j$ such that some wire is labelled by the path-variable $x_j$ at its' output end.
Then, the initial points of computational paths are described by bit-vectors $\vec{a} \in \ens{0,1}^I$\,, and the
terminal points of paths are described by $\vec{b} \in \ens{0,1}^O$\,. The coefficients
${U_{\vec a}}^{\vec b}$ can then be given as the sum of the contributions of all paths beginning at $\vec{x}_I = \vec{a}$ and ending
at $\vec{x}_O = \vec b$\,:
\begin{align}
		{U_{\vec a}}^{\vec b}
	\;\;=\;\;
		\bra{\vec b} U \ket{\vec a}
	\;\;=&\;\;
		\frac{1}{\sqrt {2^r}}
		\sum_{\substack{\vec{x} \in \ens{0,1}^n \\[0.5ex] \vec{x}_I = \vec{a} \\ \vec{x}_O = \vec{b}}} \e^{i \pi \varphi(\vec x)}	\;.
\end{align}
As $\varphi(\vec{x})$ is a quadratic form over $\vec x$ by construction, this is an expression of the coefficients of $U$ as a quadratic form expansion.

Any stable index tensor expression for $U$ (using any set of elementary gates) can be interpreted as a sum-over-paths formulation of $U$, identifying paths with a setting of all of the indices in the expression.
However, this will not in general yield a \emph{quadratic form expansion}, as the modulus of the amplitude of each path need not be uniform, as in a quadratic form expansion: we then cannot express $U$ a uniformly normalized sum.
As well, gates with generic three-body or higher-degree interactions will give rise to cubic terms in the phase function $\varphi$.
Thus, as a construction for quadratic form expansions, the construction outlined above cannot be extended to arbitrary gate sets (nor even arbitrary ``parsimonious'' gate sets, in the sense of Definition~\ref{def:parsimonious}), but apparently only to gate sets closely related to $\Phases(\A)$.

\subsubsection{Inductive construction from circuit decompositions}

An equivalent, and perhaps more illustrative, construction to the above can be made by supposing that one has a unitary circuit $C$ decomposed into unitaries, each of which has a known quadratic form expansion.
In such a case, we can compose the quadratic form expansions by matrix multiplication.

Suppose we are given a stable index tensor representation of a unitary $U$ in terms of any set of unitary transformations, provided that every unitary transformation $W$ in the decomposition has a known quadratic form expansion,
\begin{gather}
		W
	\;\;=\;\;
		C_W\!\!
		\sum_{\vec x \in \ens{0,1}^{V_{\smaller W}}} \e^{i Q_{\smaller W}(\vec x)} \ket{\vec x_{O_{\smaller W}}}\bra{\vec x_{I_{\smaller W}}}	,
\end{gather}
for some constant $C_W \in \C$, quadratic form $Q_W$, and sets $V_W, I_W, O_W$ depending on $W$. 
Relabelling the elements of the particular sets $V_W$, $I_W$, and $O_W$ in this expansion does not substantially change the operator, except in changing the labels of the space on which it operates.
Consider an instance $W\supp{j}$ of the unitary $W$ in the stable index tensor expression for $U$, with input indices $I\supp{j}$ and output indices $O\supp{j}$\,: we may then make a copy of the above expansion by bijectively mapping $I_W \to I\supp{j}$ and $O_W \mapsto O\supp{j}$, and defining an extended index set $V_W \mapsto V\supp{j}$ of indices such that, for any other set $V\supp{h}$ corresponding to any other unitary in the circuit, we have $v \in V\supp{j} \inter V\supp{h}$ only if $v \in I\supp{j}$ or $v \in O\supp{j}$ (in which case we also have $v \in O\supp{h}$ or $v \in I\supp{h}$ respectively).

Having obtained a quadratic form expansion for \emph{each} unitary in the decomposition of $U$, we may then apply the following Lemma, whose proof is trivial:
\begin{lemma}
	\label{lemma:quadFormExpanComposition}
	Let $W_1$, $W_2$ be operators given by quadratic form expansions of the form
	\begin{align}
			W_j
		\;=&\;\;
			C_j \!\! \sum_{\vec x \in \ens{0,1}^{V_j}} \e^{i Q_j(\vec x)} \; \ket{\vec x_{O_j}} \bra{\vec x_{I_j}}	\;.
	\end{align}
	In the following, $C = C_1 C_2$\,, and sums are over $\ens{0,1}^{V_1 \union V_2}$.
	\begin{romanum}
	\item
		If $V_1 \inter V_2 = I_2 = O_1$\,, then $U_2 U_1 \,=\,
			C\, \sum\limits_{\vec x} \; \e^{\,i Q_1(\vec x) \,+\, i Q_2(\vec x)} \, \ket{\vec x_{O_2}} \bra{\vec x_{I_1}} .
		$\vspace{1ex}
	\item
		If $V_1$ and $V_2$ are disjoint, then $\smash{U_1 \ox U_2 \,=\,
			C\, \sum\limits_{\vec x} \; \e^{\, i Q_1(\vec x) \,+\, i Q_2(\vec x)} \, \ket{\vec x_{O}} \bra{\vec x_{I}} ,
		}$
		where $I = I_1 \union I_2$ and $O = O_1 \union O_2$\,.
	\end{romanum}
\end{lemma}
By construction, the quadratic form expansions derived for each of the unitaries in the decomposition of $U$ satisfy the above properties, and so we may compose them to obtain a quadratic form expansion for the whole.

For the gates $H$, $Z^t$, and $\cZ^t$ in particular, described in the sum-over-paths construction, we have the following expansions:
\begin{subequations}
\begin{align}
		H
	\;=&\;\;
		\sfrac{1}{\sqrt 2}\!
		\sum_{\vec x \in \ens{0,1}^2}
			\e^{i\pi x_1 x_2}
			\ket{x_2}\bra{x_1},
			\quad
	&
		V =& \ens{1,2},& I =& \ens{1},& O =& \ens{2};
	\\[1.5ex]
		Z^t
	\;=&
		\sum_{\vec x \in \ens{0,1}^1}
			\e^{i\pi t x_{\:\!\!1}^{\,2}}
			\ket{x_1}\bra{x_1},
			\quad
	&
		V =& \ens{1},& I =& \ens{1},& O=& \ens{1};
	\\[1.5ex]
		\cZ^t
	\;=&
		\sum_{\vec x \in \ens{0,1}^2}
			\e^{i\pi t x_1 x_2}
			\ket{x_1 x_2}\bra{x_1 x_2},
			\quad
	&
		V =& \ens{1,2},& I =& \ens{1,2},& O=& \ens{1,2}.
\end{align}
\end{subequations}
None of these unitaries involve ``internal'' indices in the sense described above, and so a stable index tensor expression for a unitary $U$ in terms of these unitaries may be easily transformed into a quadratic form expansion for $U$.
(The quadratic form expansions constructed with this set of unitaries will be equivalent to those obtained from the sum-over-paths construction of the previous section.)

\paragraph{Remark.}
We may also decompose unitaries $U$ as a product of \emph{non-unitary} operators with known phase-map decompositions as well.
For instance, we may define the matrix
\begin{align}
	\label{eqn:nonunitary}
		L
	\;\;=&\;\;
		\e^{-i \pi/6} \left[ \begin{matrix}
                        \;1 & \e^{2 \pi i/3} \\ \;1 & -1
                                        \end{matrix} \,\right]
	\;\;=\;\;
 		\e^{-i \pi/6} \!\!\!
 		\sum_{x_1,x_2 \in \ens{0,1}}\!
 			\e^{i\pi(2 x_1 \,+\, x_1 x_2)/3}
 			\ket{x_2}\bra{x_1}	\;;
\end{align}
it is easy to verify that $L$ is self-inverse, in which case the above expansion gives rise to another quadratic form expansion decomposition for $\idop_2$\,:
\begin{gather}
		\idop_2
	\;\;=\;\;
 		\e^{-i \pi/3} \!
 		\sum_{\vec{x} \in \ens{0,1}^3}\!
 			\e^{i\pi \!\!\:\big[(2 x_1^{\,2} \,+\, x_1 x_2) \,+\, (2x_2^{\, 2} + x_2 x_3)\big]\!\!\:/3}
 			\ket{x_3}\bra{x_1}	\;.
\end{gather}

\subsubsection{Construction from one-way measurement patterns}

The qubit-reversal pattern and the many-qubit measurement pattern $\mathfrak{Zz\cdots z}^\theta$, which we have considered in Sections~\ref{sec:qubitReversal} and~\ref{sec:zzMany}, do not have any decomposition into smaller elements due to (currently known) generic constructions from unitary circuits into the one-way model.
This motivates considering quadratic form expansions in terms of measurement-based quantum computing, independently of unitary circuit models.

Suppose we have a measurement-based procedure with underlying geometry $(G,I,O)$ which performs a unitary transformation $U: \cH[I] \to \cH[O]$, and that there is a sequence of measurement results which occurs with non-zero probability for which there are no final corrections to be performed on the qubits in $O$.
Let $(\alpha_v)_{v \in O\comp}$ be the choice of measurement angles for this branch of the computation: we then designate this branch of the computation as the positive branch.
Then we may express $U$ (modulo a scalar factor due to the probability of postselecting the positive branch) in the form
\begin{align}
		U
	\;\propto&\;\;
		\sqparen{\bigotimes_{v \in O\comp} \bra{+_{\ssstyle \alpha_v}}_v}
		\sqparen{\prod_{uv \in E(G)} \!\!\cZ_{u,v}\;}
		\sqparen{\bigotimes_{v \in I\comp} \ket{+}_v},
\end{align}
which represents the operations of preparing the qubits of $I\comp$, performing the entangling operations, and postselecting the states $\ket{\splus_{\alpha_v}}$ for $v \in O\comp$ which occur in the positive branch.
Note that we can factor the postselection operators $\bra{\splus_\alpha}$ into a $Z$-rotation and postselection of the state $\ket{\splus}$,
\begin{gather}
		\bra{+_{\alpha_v}}
	\;\;=\;\;
		\bra{+} R_z(-\alpha_v)	;
\end{gather}
factoring these rotations into the middle, we may collect them together with the $\cZ$ operations corresponding to the edges of the entanglement graph $G$ into a single diagonal unitary operation arising from the exponential of a diagonal Hermitian matrix,
\begin{subequations}
\begin{gather}
		\sqparen{\bigotimes_{v \in O\comp} R_\sfz(-\alpha_v)}
		\sqparen{\prod_{uv \in E(G)} \cZ_{u,v}}
	\;\;=\;\;
		\e^{i J},	\qquad\text{where}
	\\[2ex]
	\label{eqn:phaseMapHamiltonian}
		J
	\;\;\;=
		\sum_{\substack{\ens{u,v} \subset V(G) \\ u \ne v}} \!\!\pi \bigg[ \ket{1}\bra{1}_u \ox \ket{1}\bra{1}_v \bigg]
		\;\;-\;\;
		\sum_{v \in O\comp} \alpha_v \ket{1}\bra{1}_v	\;.
\end{gather}
\end{subequations}
Define a quadratic form $Q: \R^{V(G)} \to \R$ by 
\begin{align}
		Q(\vec x)
	\;\;\;\;=\;
		\sum_{\substack{\ens{u,v} \subset V(G) \\ u \ne v}} \!\!\pi x_u x_v
		\;\;-\;\;
		\sum_{v \in O\comp} \alpha_v x_v^2 \;,
\end{align}
and note that, for any $S \subset V(G)$, we can expand
\begin{align}
		\bigotimes_{v \in S\comp} \ket{\splus}_v
	\;\;\;\propto\;
		\sum_{\vec x' \in \ens{0,1}^{S\comp}} \idop_2^{\,\ox S} \ox \ket{\vec x'}
	\;\;\;\;=&\;
		\sum_{\vec x \in \ens{0,1}^{V(G)}} \Big[ \ket{\vec x_S}\bra{\vec x_S} \Big] \ox \ket{\vec x_{S\comp}}
	\notag\\[0.5ex]=&\;
		\sum_{\vec x \in \ens{0,1}^{V(G)}} \ket{\vec x}\bra{\vec x_S} 	\;.
\end{align}
Then, we may obtain
\begin{align}
		U
	\;\;\propto&\;\;
		\sqparen{\bigotimes_{v \in O\comp} \bra{+_{\ssstyle \alpha_v}}_v}
		\e^{i J}
		\sqparen{\bigotimes_{v \in I\comp} \ket{+}_v}
	\notag\\[1ex]\propto&\;\;
		\sqparen{\sum_{\vec y \in \ens{0,1}^{V(G)}} \!\! \ket{\vec y_O}\bra{\vec y}}
		\e^{i J}\!
		\sqparen{\sum_{\vec x \in \ens{0,1}^{V(G)}} \!\! \ket{\vec x}\bra{\vec x_I}}
	\notag\\[1ex]=&\;\;\;
		\sum_{\vec x, \vec y \in \ens{0,1}^{V(G)}} \!\! \ket{\vec y_O} \bigg[ \e^{i \, \bra{\vec y}\!\!\; J \ket{\vec x}} \bigg] \bra{\vec x_I}
	\;\;\;\;=
		\sum_{\vec x \in \ens{0,1}^{V(G)}} \!\! \e^{iQ(\vec x)} \; \ket{\vec x_O}\bra{\vec x_I},
\end{align}
which is a quadratic form expansion for $U$.

\subsection{Interpretation as a description of a postselection procedure}
\label{sec:postselectionProcedure}

In the context of \MPI\ (page~\pageref{problem:MPI}), the construction above of quadratic form expansions from measurement-based computations hints at an approach for doing the reverse: to turn a quadratic form expansion into a measurement-based procedure, and more precisely into an instance of the \MPI.
Consider a unitary $U$ given by a quadratic form expansion as in~\eqref{eqn:quadFormExpansion}, where the quadratic form $Q$ is given by
\begin{align}
	\label{eqn:quadForm}
		Q(\vec x)
	\;\;=&\;
	\sum_{\ens{u,v} \subset V} \theta_{uv} x_u x_v
\end{align}
for any $\vec x \in \ens{0,1}^V$, for some real parameters $\ens{\theta_{uv}}_{u,v \in V}$\,; and where the sum includes singleton sets corresponding to the case $u = v$.
We may express $Q(\vec x)$ as an inner product $\bra{\vec x} \mspace{-3mu} J \mspace{-3mu} \ket{\vec x}$\,, where $J$ is a $2$-local diagonal operator similar in construction to that of~\eqref{eqn:phaseMapHamiltonian}:
\begin{subequations}
\begin{gather}
		J
	\;\;=\;\;
		J_O \,+\, J_1 \,+\, J_2	\;,\qquad\text{where}
	\\[2ex]
	\begin{aligned}
		J_O
	\;\;=&\;\;
		\sum_{w \in O} \theta_{ww} \ket{1}\bra{1}_w	\;,
	&\qquad
		J_1
	\;\;=&\;\;
		\sum_{v \in O\comp} \theta_{vv} \ket{1}\bra{1}_v	\;,
	\end{aligned}
	\\[1ex]
	\!\!\!\!\!
		J_2
	\;=
		\sum_{\substack{\ens{u,v} \subset V \\ u \ne v}} \!\!\theta_{uv} \bigg[\ket{1}\bra{1}_u \ox \ket{1}\bra{1}_v \bigg].		
\end{gather}
\end{subequations}
Then we may decompose $U$ as
\begin{align}
	\label{eqn:phaseMapDecomp}
		U
	\;\;\propto\;
		\sum_{\vec{x} \in \ens{0,1}^V} \!\!\ket{\vec{x}_O} \bra{\vec x}\;\! \e^{i J} \ket{\vec x} \bra{\vec{x}_I}\,
	\;\;\propto&\;\;
			\sqparen{\bigotimes_{w \in O\comp} \bra{\splus}}
			\e^{i J_O} \e^{i J_1} \e^{i J_2}\!\!\:
			\sqparen{\bigotimes_{v \in I\comp} \ket{\splus}}	\;.
\end{align}
Note that $\e^{iJ_1}$ consists simply of single-qubit $Z$ rotations applied to the elements of $O\comp$,
\begin{align}
		\e^{i J_1}
	\;\;=\;\;
		\prod_{v \in O\comp} \e^{i \theta_{vv} \ket{1}\bra{1}_v}
	\;\propto&\;
			\bigotimes_{v \in O\comp} R_\sfz(\theta_{vv})\;,
\end{align}
and similarly for $\e^{iJ_O}$; and we may express $\e^{iJ_2}$ by a product of (potentially fractional) powers of controlled-$Z$ on distinct pairs of qubits:
\begin{align}
		\e^{i J_2}
	\;\;=\;\;
		\prod_{\substack{\ens{u,v} \subset V \\ u \ne v}} \e^{i \theta_{uv} \ket{1}\bra{1}_u \ox \, \ket{1}\bra{1}_v}
	\;\;=&\;\;
		\prod_{\substack{\ens{u,v} \subset V \\ u \ne v}} {\cZ_{u,v}}\big.^{\!\!\!\!\!\!\theta_{uv}/\pi}	\;.
\end{align}
Let $\alpha_v = -\theta_{vv}$ for each $v \in O\comp$\,: then, commuting $\e^{iJ_O}$ leftwards and (similarly) absorbing $\e^{iJ_1}$ into the projection operations, we obtain
\begin{align}
		U
	\;\;\propto\;
			\e^{i J_O}
			\sqparen{\bigotimes_{w \in O\comp} \bra{+_{\alpha_v} }}
			\sqparen{\prod_{\substack{\ens{u,v} \subset V \\ u \ne v}} {\cZ_{u,v}}\big.^{\!\!\!\!\!\!\theta_{uv}/\pi}}
			\sqparen{\bigotimes_{v \in I\comp} \ket{\splus}}	\;,
\end{align}
where we have applied the relation $\ket{\splus_\alpha} \propto R_\sfz(\alpha) \ket{\splus}$.
This is an expression of $U$ as the preparation of some number of $\ket{+}$ states, followed by a diagonal unitary operator consisting of two-qubit (fractional) controlled-$Z$ operations, followed by post-selection of the states $\ket{\splus_{\alpha_v}}$ for $v \in O\comp$ (which succeeds with non-zero probability, provided the initial scalar factor of the quadratic form expansion was nonzero), and finally applying some single-qubit $Z$ rotations on the remaining qubits.

If $\theta_{uv} \in \ens{0,\pi}$ for all distinct $u,v \in V$, and $\theta_{vv} = 0$ for $v \in O$, the above describes precisely the action of the positive branch of a measurement-based computation in which the default measurement angles on the qubits $v \in O\comp$ are given by $-\theta_{vv}$, and where we condition on all measurements yielding the result $\s[v] = 0$.
Thus, such a quadratic form expansion encodes an instance of \MPI.

\begin{figure}[tp]
	\begin{center}
		\includegraphics{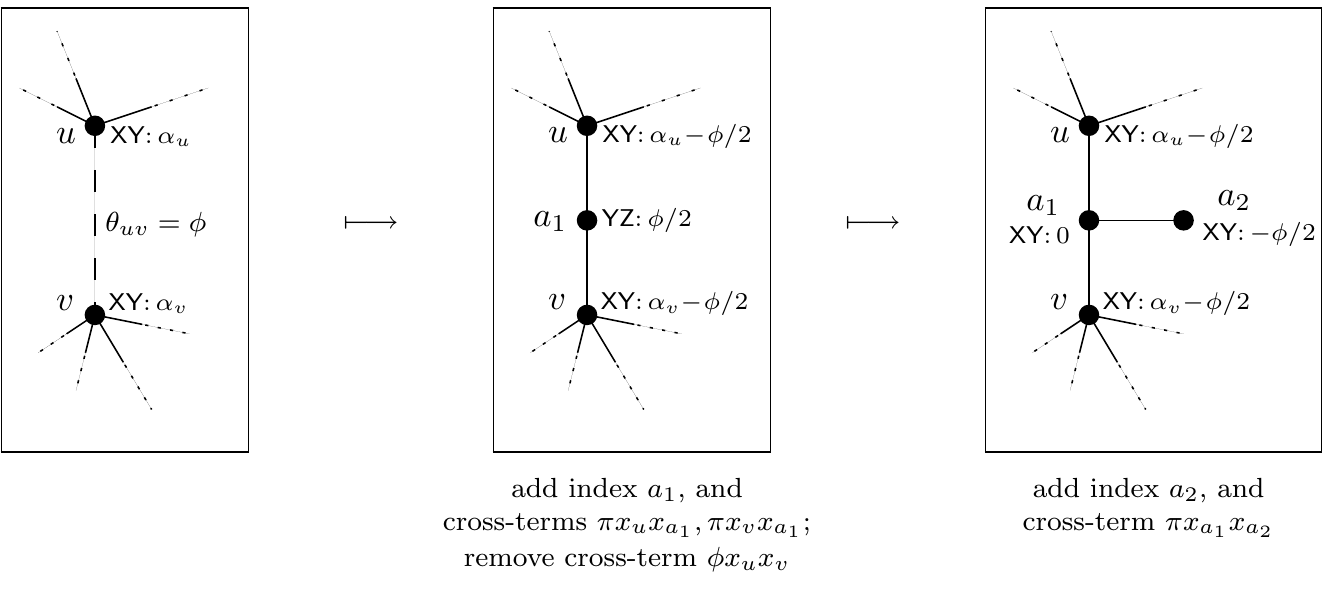}
	\end{center}
	\caption[A transformation of quadratic form expansions to eliminate cross-term angles $\theta_{uv} \notin \ens{0,\pi}$.]{%
		\label{fig:diagonalYZ}%
		A transformation of quadratic form expansions to eliminate cross-term angles $\theta_{uv} \notin \ens{0,\pi}$ which would correspond to fractional entangling operations, illustrated here in terms of geometries and measurement operations of a postselection procedure.
		This transformation essentially consists of applying the decomposition of $\cZ$ into $\mathfrak{Zz}$ and single-qubit rotations described in~\eqref{eqn:clusterEntanglerSubst}, extended to make use of the description of $\mathfrak{Zz\cdots z}^{\phi/2}$ in terms of \XY-plane measurements for more general $\phi$, which is illustrated in Figure~\ref{fig:zzMany}.
		The coefficients of the quadratic form expansion in the left and right panels are given by the corresponding measurement angles in the \XY\ plane; the center panel does not directly correspond to any quadratic form expansion, as it involves a measurement outside of the \XY\ plane.}
\end{figure}

Furthermore, we may transform any quadratic form expansion into such a form.
For quadratic form expansions with $\theta_{vv} \ne 0$ for $v \in O$, we may obtain a modified geometry by composing to such output qubits a quadratic form expansion for the identity operation,
\begin{align}
	\label{eqn:idopQuadFormExpan}
		\idop_2
	\;\;=\;\;
		\sum_{x_v, x_{v''}} \delta_{x_v,x_{v''}} \ket{x_{v''}} \bra{x_v}
	\;\;=\;\; 
		\sfrac{1}{2} \sum_{x_v, x_{v'}, x_{v''}} (-1)^{x_v x_{v'} + x_{v'} x_{v''}} \ket{x_{v''}} \bra{x_v}
\end{align}
which corresponds to the pattern $\mathfrak{Id}^2$ of~\eqref{eqn:identityPatterns}; this yields a quadratic form expansion with output indices $O'$ such that $v \notin O'$, but where $v'' \in O'$ and $\theta_{v'' v''} = 0$.
To eliminate cross-term coefficients which are not integer multiples of $\pi$, we may use the equality
\begin{gather}
\label{eqn:diagonalExpansions}
		\e^{i \phi (x_u \oplus x_v)}
	\;\;=\;\;
		\sfrac{1}{2}\! \sum_{\vec a \in \ens{0,1}^2} \e^{i\big[\pi(x_u a_1 + x_v a_1 + a_1 a_2) \,+\, \phi a_2^{\, 2}\big]}
\end{gather}
where $x_u \oplus x_v = x_u + x_v - 2x_u x_v$ is the sum of $x_u$ and $x_v$ modulo $2$; we may eliminate factors of $\e^{i \theta_{uv} x_u x_v}$ such that $\theta_{uv} \notin \pi \Z$ for distinct $u,v$ 
by substituting the equality above.
Using this substitution amounts to decomposing the operator $\cZ^t$ using the construction for $\mathfrak{Zz\cdots z}^\theta$ illustrated in Figure~\ref{fig:zzMany}, as Figure~\ref{fig:diagonalYZ} illustrates.

In this way, we may translate any quadratic form expansion into a postselection procedure which uses only the entangling and measurement operations available to the $\OneWay(\A)$ model, where $\A$ is the set of angles which occur in the quadratic form.
In particular, for a quadratic form expansion as above where $\theta_{uv} \in \ens{0,\pi}$ for distinct $u,v$ and $\theta_{vv} = 0$ for $v \in O$, we may obtain a tuple $(G,I,O,\vec a)$ describing an instance of \MPI\ by defining $G$ to be the graph whose vertices are the indices of $V$ and whose edges are those pairs $uv$ such that $\theta_{uv} \ne 0$ and by defining $\vec{a} = (-\theta_{vv})_{v \in O\comp}$\,.
Thus, any such tuple for which the \MPI\ has an efficient solution corresponds to a quadratic form expansion which we may translate to a measurement-based procedure performing a unitary transformation.

\section{Solved cases of Measurement Pattern Interpolation}

At present, there is no general solution to \MPI, which seems to be a difficult problem.
Current approaches towards a solution involve analyzing those measurements which can be treated essentially by the stabilizer formalism, where (as in Sections~\ref{sec:extendedConstructions} and~\ref{sec:gflows}) we may relax the requirement that the measurement observables be Pauli operators.
These lead to the following solved cases:
\begin{enumerate}
\item
	Tuples $(G,I,O,\vec a)$ such that, for the constant function $\lambda: O\comp \to \ens{\XY}$, the tuple $(G,I,O,\lambda)$ has a gflow (Definition~\ref{def:gflow});

\item
	Tuples $(G,I,O,\vec a)$ such that, by applying certain simple graph transformations on $G$ to obtain a smaller graph $\tilde G$ and a vertex set $T \subset V(\tilde G)$, we obtain a tuple $(\tilde G,I,O,T)$ having an eflow (Definition~\ref{def:eflow});

\item
	Tuples $(G,I,O,\vec a)$ such that $\alpha_v \in \frac{\pi}{2}\Z$ for every angle $\vec a = (\alpha_v)_{v \in O\comp}$ (\ie\ where each measurement is performed with respect to either the $\ket{\spm \sfx}$ basis or $\ket{\spm \sfy}$ basis).
\end{enumerate}
In this section, we describe these cases and illustrate how \MPI\ may be solved for each.

As we described in the preceding section, any quadratic form expansion can be transformed into an instance of \MPI\ by first transforming it into another quadratic form expansion whose cross-terms $\theta_{uv}$ are integer multiples of $\pi$.
However, it will in some cases be convenient to synthesize implementations for unitaries directly from a quadratic form expansion having cross-terms which are fractional multiple of $\pi$, corresponding to postselecting in an extended measurement-based model which may use fractional powers of $\cZ$.
This motivates the corresponding extension of the notion of a ``geometry'', which we define here for quadratic form expansions:
\begin{definition}
	\label{def:quadraticFormGeometry}
	For a quadratic form expansion	
	\begin{align}
		\label{eqn:quadFormExpansionAgain}
		C\!\! \sum_{\vec{x} \in \ens{0,1}^V} \e^{i Q(\vec{x})} \;\ket{\vec{x}_O}\bra{\vec{x}_I}
		&&
		\text{where}\quad
		Q(\vec x)
	\;=&\;
		\sum_{\substack{\ens{u,v} \subset V}} \theta_{uv} x_u x_v \;\,,
	\end{align}
	the \emph{geometry induced by the quadratic form} is a triple $(G,I,O)$\,, where $G$ is a \emph{weighted} graph
	with vertex-set $V$, edge-set $\ens{uv \mid u \ne v ~\text{and}~ \theta_{uv} \ne 0}$\,, and edge-weights
	$W_G(uv) = \frac{\theta_{uv}\!}{\pi}$.
	Without loss of generality, we may require $W_G(uv) \in (-1,\, 1\,]$ for all $uv$; we may then refer to edges $uv$ being of \emph{unit weight} if $W_G(uv) = 1$, and \emph{fractional weight} if $\theta_{uv} \notin \ens{0,1}$.
\end{definition}
Geometries with only edges of unit weight correspond to geometries as described in Definition~\ref{def:geometry} in Section~\ref{sec:geometriesOneWay}; we will refer to such geometries as \emph{unit weight geometries}.
Geometries having one or more fractional edge-weights we refer to as \emph{fractional weight geometries}.
We will note in a given context whether or not we are including geometries with fractional weight edges in the context of any given instance of \MPI; however, unless otherwise noted, a reference to a ``geometry'' should be understood as referring to the unit weight geometries which arise in the one-way measurement model.

\subsection{Geometries with gflows}
\label{sec:interpolateGflows}

The approach taken for \MPI\ based on gflows is to ignore all angle information, and consider those tuples $(G,I,O,\vec a)$ which represent postselection-schemes which may be interpolated regardless of the value of $\vec a$.
By doing so, the objective is to consider postselection schemes which \emph{generically} perform unitary transformations, in the following sense:
\begin{problem}[Generic Measurement Pattern Interpolation (\GMPI)]
	\label{problem:GMPI}
	Given an input geometry $(G,I,O)$, determine if there exists a family of measurement patterns $\mathfrak P(\vec{a})$ in the model $\OneWay(\R)$ having the underlying geometry $(G,I,O)$ and default measurement angles $\vec{a} \in \R^{O\comp}$, such that every pattern $\mathfrak P(\vec a)$ performs a unitary embedding.
\end{problem}
For a tuple $(G,I,O,\vec a)$, if the geometry $(G,I,O)$ is a ``\emph{yes}'' instance of \GMPI, we can then solve \MPI\ by constructing the particular pattern $\mathfrak P(\vec a)$.

The genericity of the measurement angles corresponds to the constraint that each measurement may be performed with respect to an arbitrary observable which anticommutes with $Z$.
The approach taken by analysis with gflows, as we described in Section~\ref{sec:gflows}, is to determine whether this suffices to ensure unitarity of the post-selection procedure via the stabilizer formalism.
In the case that the stabilizer formalism does guarantee the unitarity of the transformation, we also obtain a correction strategy which provides an interpolation of the measurement procedure to all possible measurement sequences.
Recall the definition of a gflow:
\begin{definition}
	Let $(G,I,O)$ be a geometry and $\lambda$ be a function defined on $O\comp$ indicating measurement planes as above; and let $\cP$ be the power-set function, and $\odd(S)$ be the set of vertices in $G$ which are adjacent to an odd number of elements of $S$.
	Then a \emph{gflow} for $(G,I,O,\lambda)$ is a tuple $(g,\preceq)$ consisting of a function $g: O\comp \to \cP(I\comp)$ and a partial order $\preceq$ which satisfy the following conditions:
	\begin{subequations}
	\begin{align}
			w \in g(v) \;\implies\;&	v \preceq w ,
		\\
			w \in \odd(g(v)) \;\implies\;& v \preceq w ,
		\\
			\lambda(v) = \XY \;\implies\;& v \in \odd(g(v)) \setminus g(v)	, 
		\\
			\lambda(v) = \YZ \;\implies\;& v \in g(v) \setminus \odd(g(v)) ,
		\\
			\lambda(v) = \XZ \;\implies\;& v \in g(v) \inter \odd(g(v)) .
	\end{align}
	\end{subequations}
\end{definition}
As we saw in Section~\ref{sec:gflows}, the function $g: O\comp \to \cP(I\comp)$ essentially encodes a list of generators $\ens{S_j}_{j = 1}^{n-k}$ for the state-space prepared by the map $\Ent{G} \circ \New{I\comp}{}$\,.
Provided that the qubits $v \in O\comp$ are measured in an order extending $\preceq$, the set $g(v)$ describes the set of qubits acted on with the $X$ operations of some stabilizer generator; qubits $w \in \odd(g(v))$ are acted on with $Z$ operations, and qubits in $g(v) \inter \odd(g(v))$ are acted on by both (or equivalently, by a $Y \propto X Z$ operation).
The generator $S_j$ described then anticommutes with the measurement observable $\sO_v^{\lambda(v),\alpha_v}$ defined in~\eqref{eqn:pauliPlaneObs}, for $\alpha_v$ arbitrary.
The post-measurement state corresponding to the result $\s[v] = 0$ can then be selected by performing $S_j$ as a post-measurement correction, which we may re-express as an adaptation the measurement observables for the qubits on which it acts.

\subsubsection{Geometries of unit weight with \XY-plane gflows}

In the case where $\lambda$ is a constant function $\lambda: O\comp \to \ens{\XY}$ and where $\card{V(G)} = n$, Mhalla and Perdrix~\cite{MP08} present an efficient algorithm to find a gflow $(g,\preceq)$, where $\preceq$ is defined in terms of a layer function $\ell: V(G) \to \N$ such that $v \preceq w \iff [\,v = w \text{~or~} \ell(v) > \ell(w)\,]$.
This leads immediately to a solution to \MPI\ in the case of unit weight geometries $(G,I,O)$ when the geometry has a gflow, as follows.
We may obtain a linear order $\le$ extending $\preceq$ by constructing level sets $\sV_j$ for each $j \in \N$\,, inserting each vertex $v \in V(G)$ into the set $\sV_{\ell(v)}$\,.\label{discn:linearExtension}
Using the representation of sets described in Section~\ref{sec:flowAlgs}, the sets $\sV_j$ are themselves linearly ordered by construction, leading to a linear ordering of $V(G)$ by merging these linear orders in such a way that any $v \in \sV_{j+1}$ precedes any $w \in \sV_j$ for each $j$.
We may then simulate the postselection of the residual state corresponding to the result $\s[v] = 0$ for each qubit $v \in O\comp$, by composing patterns of the form
\begin{gather}
	\sqparen{\prod_{u \in g(v)} \!\! \Xcorr{u}{\s[v]} \;} \sqparen{\prod_{\substack{w \in \odd(g(v)) \\ w \ne v}} \!\! \Zcorr{w}{\s[v]} \;} \!\Meas{v}{\alpha_v}
\end{gather}
in the ordering given by $\le$.
(The operations used here are those of the $\OneWay(\R)$ model defined by Definition~\ref{def:oneWayModel}.
The restriction of $w \ne v$ in the right-most product is due to the fact that $v \in \odd(g(v))$ has been discarded by the measurement.)
Then, the following CPTP map performs a unitary embedding for any choice of angles $\vec a = (\alpha_v)_{v \in O\comp}$\,:
\begin{align}
	\label{eqn:gflowPattern}
	\paren{
		\ordprod[\ge]_{v \in O\comp}
			\sqparen{\prod_{u \in g(v)} \!\! \Xcorr{u}{\s[v]} \;} \sqparen{\prod_{\substack{w \in \odd(g(v)) \\ w \ne v}} \!\! \Zcorr{w}{\s[v]} \;} \!\Meas{v}{\alpha_v} }
	\paren{
		\prod_{uv \in E(G)} \Ent{u,v} }
	\paren{
		\prod_{v \in I\comp} \New{v}{\sfx} },
\end{align}
where the left-most product is ordered left-to-right according to the order $\ge$.
By the analysis of Section~\ref{sec:postselectionProcedure}, if $(G,I,O, \vec a)$ is an instance of the \MPI\ derived from a quadratic form expansion for an operator $U$, and $(G,I,O)$ is a unit weight geometry, the unitary operation which is performed is $U$.

\paragraph{Run-time analysis.}

Let $n = \card{V(G)}$, $m = \card{E(G)}$, and $k = \card{O} \ge \card{I}$.
Determining whether a geometry has a gflow for the special case where all measurements are performed in the \XY-plane, and constructing the gflow $(g,\preceq)$ if so, can be done in time $O(n^4)$ by Algorithm~2 of~\cite{MP08}.
Having a gflow $(g, \preceq)$, the measurement procedure in~\eqref{eqn:gflowPattern} is well-defined and requires $O(n^3)$ time to construct, dominated by the product over $v \in O\comp$ of the correction operations for $u \in g(v)$ and $w \in \odd(g(v))$, where the latter require time $O(m) \subset O(n^2)$ to compute.
Bringing this measurement procedure into a standard form by the standardization procedure of Section~\ref{sec:DKPstandardization} also requires time $O(n^3)$\,; the resulting measurement pattern contains $n-k$ measurement operations and up to $k$ correction operations, each of which involves $O(n)$ measurement dependencies.
We may therefore conclude:

\begin{theorem}
	\label{thm:synthGflowsExact}
	For a unitary embedding $U$ given as a quadratic form expansion, with unit weight geometry $(G,I,O)$ such that $\card{V(G)} = n$, there is an $O(n^4)$ algorithm which either determines that $(G,I,O)$ has no gflow, or constructs pattern in the $\OneWay(\R)$ model whose default measurement angles are given by the square terms of the quadratic form and consisting of $O(n^2)$ operations and which performs the transformation $U$.
\end{theorem}

As we note in Theorem~\ref{thm:synthGflowsExact}, the one-way measurement pattern constructed by this technique is in the model $\OneWay(\A)$, where $\A \subset \R$ contains the coefficients for the square terms of the quadratic form.
We may strengthen this to obtain an algorithm producing one-way patterns in a fixed-precision model of measurement-based quantum computing using any set $\A$ sufficient for generating single-qubit unitaries, such as $\A = \frac{\pi}{4}\Z$.
A measurement in the \XY-plane with respect to the basis $\ket{\pmsub\alpha}$ can be transformed into a \axisZ-axis measurement by a $J(-\alpha)$ rotation~\eqref{eqn:basicGates}: more generally, we have
\begin{gather}
	\Meas{v}{\alpha;\beta}	\;\;=\;\;	\Meas{v}{\sfz} \; \sfJ_v^{(- 1)^{1 + \beta} \cdot \alpha}	\;.
\end{gather}
By the universality of the $\Jacz(\frac{\pi}{4}\Z)$ gate set defined on page~\pageref{eqn:elemGateSets}, and by the Solovay-Kitaev Theorem, we may approximate the unitary operator $J(\alpha) \in \U(2)$ to any precision $\epsilon > 0$ by some product of $N$ operations of the form $J(\theta_j)$ for $\theta_j \in \frac{\pi}{4}\Z$\,, where $N \in \polylog(1/\epsilon)$:
\begin{gather}
	\label{eqn:approxJ}
	\norm{J(\alpha) \Big.\;-\;\Big. J(\theta_{N-1}) \cdots J(\theta_1) J(\theta_0)}_\infty <\; \sfrac{\epsilon}{2}	\;.
\end{gather}
This approximation can be found in time $\polylog(1/\epsilon)$ time by the Solovay-Kitaev theorem.
In order to represent the conditional change of sign, note that
\begin{subequations}
\begin{gather}
		J(-\alpha)
	\;\;=\;\;
		H R_z(-\alpha)
	\;\;=\;\;
		H X R_z(\alpha) X
	\;\;=\;\;
		Z J(\alpha) X \;.
\end{gather}
\end{subequations}
Then, we may approximate the effect of a measurement $\Meas{v}{\alpha}$ for arbitrary $\alpha \in \R$ by a (non-standardized) measurement pattern of the form
\begin{align}
		\Meas{v}{\alpha;\beta}
	\;\;\cong\;&\;\;
		\Meas{v'}{\sfz}
		\sfJ^{(-1)^{1+\beta}\cdot\alpha}_{v/v'}
	\notag\\\approx_\epsilon&\;\;
		\Meas{v'}{\sfz}
		\;\Zcorr{v'}{1 + \beta}
		\;\mathfrak J^{\theta_{N-1}}_{v'/a_{N-1}}
		\;\cdots\,
		\;\mathfrak J^{\theta_2}_{a_2/a_1}
		\;\mathfrak J^{\theta_1}_{a_1/v}
		\;\Xcorr{v}{1 + \beta}
	\notag\\=\;&\;\;
		\Meas{v'}{\sfz}
		\;\mathfrak J^{\theta_{N-1}}_{v'/a_{N-1}}
		\;\cdots\,
		\;\mathfrak J^{\theta_2}_{a_2/a_1}
		\;\mathfrak J^{\theta_1}_{a_1/v}
		\;\Xcorr{v}{1 + \beta}	\;,
\end{align}
using the $\mathfrak{J}^\theta_{u/v}$ patterns defined in~\eqref{eqn:elemOpsDKP}; we introduce the auxiliary qubits $a_1, \ldots, a_{N-1}$ in order to approximate $\mathfrak J^\alpha_{v/v'}$ to precision $\epsilon$ using the approximation of~\eqref{eqn:approxJ}, and substituting the measurement result $\s[v']$ throughout the rest of the measurement pattern wherever $\s[v]$ occurs.

To obtain an approximation of the CPTP map $\rho \mapsto U \rho U\herm$ with precision $\epsilon$, it suffices to approximate each of the $n - k$ measurement operations with precision $\frac{\epsilon}{n - k}$ by the development of~\eqref{eqn:errorAdditiveCPTP}.
We then obtain the following corollary:
\begin{corollary}
	\label{cor:synthGflowsApprox}
	For a unitary embedding $U$ given as a quadratic form expansion, with unit weight geometry $(G,I,O)$ such that $\card{V(G)} = n$, for each $\epsilon > 0$ and $\delta > 0$ there exists an algorithm running in time $O(n^4 + n \log(n/\epsilon)^{3 + \delta})$ which either determines that $(G,I,O)$ has no gflow, or constructs a pattern in the $\OneWay(\frac{\pi}{4}\Z)$ model consisting of $O(n^2 + n \log(n/\epsilon)^{3 + \delta})$ operations and which approximates the CPTP map $\rho \mapsto U \rho U\herm$ to precision $\epsilon$.
\end{corollary}

\subsubsection{Geometries of fractional weight with flows}

Geometries with fractional edges correspond, as we remarked earlier, to extended one-way measurement procedures with include entangling operations of the form $\Ent[t]{u,v}(\rho) \;=\; \cZ^t\, \rho \;\cZ^{-t}$.
The state space of the system is then not necessarily a stabilizer code, in which case the approach given above for gflows then do not directly apply.

As we noted in Section~\ref{sec:postselectionProcedure}, we may attempt to treat such quadratic form expansions by applying the substitutions of~\eqref{eqn:diagonalExpansions} in order to obtain an equivalent expansion which only involved cross-term angles in $\ens{0,\pi}$.
However, this approach still fails to produce a problem instance approachable by the technique of gflows, as it produces a quadratic form with an extra auxiliary index $a_1$ which has no square term.
It is easy to see that the absence of an $a_1^{\, 2}$ term with a generic coefficient in this expansion is crucial, by considering the corresponding formula where the quadratic form includes a $\theta a_1^{\, 2}$ term:\label{discn:squareTermConstrained}
\begin{align}
		\sfrac{1}{2}\! \sum_{\vec a \in \ens{0,1}^2} &\e^{i\big[\pi(x_u a_1 + x_v a_1 + a_1 a_2) \,+\, \theta a_1^{\, 2} \,+\, \phi a_2^{\, 2}\big]}
	\notag\\[1ex]=&\;\;
		\sfrac{1}{2} \sqparen{
				\e^{i [ 0 ]}
			\;+\;
				\e^{i [\pi (x_u + x_v) \,+\, \theta]}
			\;+\;
				\e^{i [\phi]}
			\;+\;
				\e^{i [\pi (x_u + x_v + 1) \,+\, \theta \,+\, \phi]}
		}
	\notag\\[2ex]=&\;\;
		\sfrac{\e^{i\phi/2}}{2} \sqparen{
				\paren{
					\e^{-i \phi/2} \,+\, \e^{i\phi/2}
				}
			\;+\;
				\e^{i\theta} (-1)^{x_u \oplus x_v} \paren{
					\e^{-i \phi/2} \,-\, \e^{i \phi/2}
				}
		}
	\notag\\[2ex]=&\;\;
		\e^{i\phi/2} \sqparen{
				\cos(\tfrac{\phi}{2})
			\;-\;
				i \, \e^{i\theta} (-1)^{x_u \oplus x_v} \sin(\tfrac{\phi}{2})
		}	\;.
\end{align}
This expression is not a complex unit for all values of $x_u, x_v \in \ens{0,1}$ and angles $\phi \in (-\pi, \pi]$; in particular, for $\phi = \pion 2$\,, we may see that
\begin{gather}
	\label{eqn:nonuniformity}
		\sfrac{1}{2}\! \sum_{\vec a \in \ens{0,1}^2} \e^{i\big[\pi(x_u a_1 + x_v a_1 + a_1 a_2) \,+\, \frac{\pi}{2} a_1^{\, 2} \,+\, \frac{\pi}{2} a_2^{\, 2}\big]}
	\;\;=\;\;
		\begin{cases}
			1 + i	\,,	&	\text{if $x_u \oplus x_v = 0$}	\\[1ex]
			0\,,					&	\text{if $x_u \oplus x_v = 1$}	
		\end{cases}	\;.
\end{gather}
This is a very different expression from the expression $\e^{i \phi (x_u \oplus x_v)}$ in the case $\theta = 0$\,: thus, even if the result is a unitary in the latter case, it is by no means clear that quadratic form expansions with the same geometry should be ``generically'' unitary, regardless of the coefficients of their square terms.
(As an extreme example, where $u,v \in I \inter O$, the resulting operation would be singular for $\theta = \phi = \pion 2$, even though the operator may be unitary for $\theta = 0$.)
However, this genericity is necessary for the geometry to be a ``yes'' instance of the \GMPI.

We can attempt to resolve this by noting that the substitutions in~\eqref{eqn:diagonalExpansions} are a result of an extended measurement-based procedure which allows arbitrary measurements in the \YZ\ plane: measurements on the two auxiliary qubits $a_1$ and $a_2$ of~\eqref{eqn:diagonalExpansions}, where $a_1$ is measured with an \axisX-measurement and $a_2$ with an arbitrary \XY-plane measurement, are equivalent to an arbitrary \YZ-plane measurement on a single auxiliary qubit, as illustrated in the middle panel of Figure~\ref{fig:diagonalYZ}.\label{discn:simulateYZmeas}
In particular, as we noted in Section~\ref{sec:extendedConstructions}, this construction has an eflow, which is a special case of a gflow where $g(v)$ are singleton sets for each $v \in O\comp$.
We may then treat such quadratic form expansions by performing a transformation of the underlying \emph{geometry} $(G,I,O)$, to yield a unit-weight geometry $(G',I,O)$ with additional information indicating sets of vertices $T$ to be measured in the \YZ\ plane, and try to find an eflow for $(G',I,O,T)$.
We will use a similar technique in Section~\ref{sec:synthEflows} to synthesize one-way measurement patterns for certain cases involving auxiliary vertices of degree $1$.

For a unitary $U$ which is specified by a quadratic form expansion, whose geometry has fractional weight edges but which can be transformed into an equivalent unit-weight geometry $(G',I,O)$ and set $T$ such that $(G',I,O,T)$ has an eflow $(f,\preceq)$, we may apply the same techniques as outlined above for gflows to efficiently arrive at an implementation (or approximation) of $U$.
In this case, the vertices of $T$ can be identified with the edges of fractional weight in $G$; conversely, any qubit $v \in V(G) \setminus O$ lies outside of $T$, and so the edges $v \,f(v)$ in $G$ are of unit weight.
By the comments at the end of Section~\ref{sec:extendedConstructions}, the star geometry rooted at vertices $v \in T$ correspond to operations $\cZ^t$ in a unitary circuit via the simplified RBB construction, with the edges of the form $v \, f(v)$ representing $J(\theta)$ gates as in the analysis of flows in terms of the DKP construction.

There is, however, a simpler approach to synthesizing an implementation for $U$ by observing that the above also essentially describes an extension of the semantic map $\SemanticDKP$ constructed in Section~\ref{sec:semanticDKP}, where we ignore the absence of information about the adaptation of measurements, and allow non-flow edges to have fractional weight to represent fractional powers of $\cZ$ in the resulting unitary circuit.
First, if necessary, we transform the quadratic form expansion into a form where each $v \in O$ has $\theta_{vv} = 0$, using the technique described on page~\pageref{eqn:idopQuadFormExpan}: it is easy to verify that the flow function can be extended to the new vertex set arising from this transformation, so we may without loss of generality suppose that the input geometry has $\theta_{vv} = 0$ for $v \in O$.\label{discn:synthFlows}
Then, we may straightforwardly interpret the quadratic form expansion as a unitary circuit, by noting that it exactly represents the matrix formula produced by making explicit the summations in a stable index tensor product:
\begin{romanum}
\item
	the edges $v w$ for $w = f(v)$ correspond to a pair of deprecated/advanced tensor indices in an operation $J(\theta_{vv})\pseu[:w/v]$ operation;

\item
	any non-flow edge $vw$ corresponds to a pair of stable tensor indices in a $\cZ\pseu[v,w]$ operation.
\end{romanum}
In the above, we still obtain a unitary circuit by replacing non-flow edges with edges of fractional weight; these correspond to $\cZ^t\pseu[v,w]$ operations for fractional powers $t$.
This motivates an extension of the definition of flows to fractional-weight geometries:
\begin{definition}
	\label{def:flowFractional}
	Suppose $(G,I,O)$ is a fractional-weight geometry.
	A flow for $(G,I,O)$ is then an ordered pair $(f,\preceq)$ satisfying properties \eqref{flow:a}\thru\eqref{flow:c}, such that for all $ab \in E(G)$ with $W_G(ab) < 1$, we also have $f(a) \ne b$ and $f(b) \ne a$ (\ie, all flow-edges in $G$ have unit weight).
\end{definition}
If a fractional-weight geometry $(G,I,O)$ has a flow, and if the square terms for $v \in O$ are all zero, we may synthesize a circuit from the quadratic form expansion for $U$ using the description above.

We may verify the correctness of this circuit construction by inducting on the number of edges of fractional weight.
For the base case, if $(G,I,O)$ has a known flow $(f,\preceq)$ and no fractional-weight edges, we may synthesize a circuit for $U$ using the algorithm $\SemanticDKP$ (using the flow $(f,\preceq)$ rather than searching for one using the procedure $\FindFlow$).
The resulting circuit has a gate $J(\theta_{vv})\pseu[:x_{f(v)}/x_v]$ for each $v \in O\comp$ and a gate $\cZ\pseu[x_v, x_w]$ for each non-flow edge $vw \in E(G)$\,, which can be expressed as
\begin{gather}
	\label{eqn:unitWeightFlowCircuit}
		C
	\;\;=\;\;
		\Bigg(\prod_{v \in O\comp} J(\theta_{vv})\pseu[:x_{f(v)}/x_v]\Bigg)
		\Bigg(\prod_{\substack{vw \in E(G) \\ v \ne f(w) \\ w \ne f(v)}} \!\!\cZ\,\pseu[x_v , x_w ]\Bigg)
\end{gather}
in stable index notation.
We may then induct for geometries with fractional edge-weights if we can show we can decompose the geometry into ones with fewer fractional edge-weights.

\begin{figure}[t]
	\begin{center}
		\includegraphics{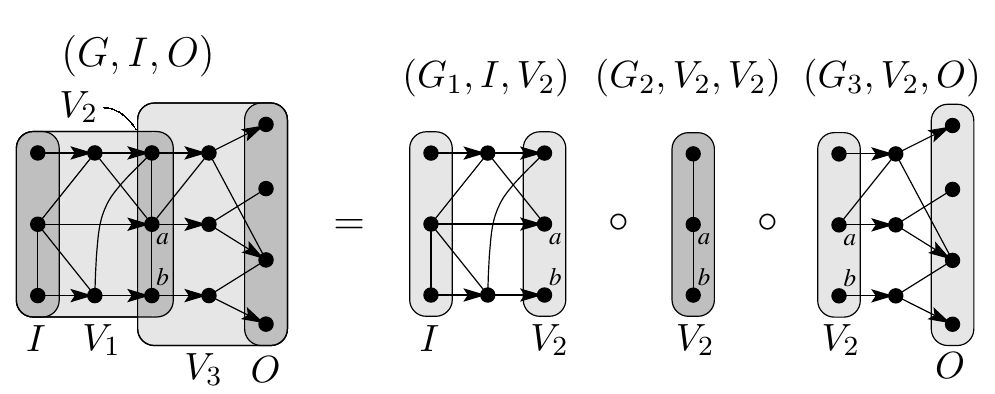}
		\caption[Illustration of the decomposion of a quadratic form expansion about an edge, expressed in terms of geometries.]{\label{fig:decompose}%
		Illustration of the decomposion of a quadratic form expansion about an edge $ab$, expressed in terms of geometries. 
		(Again, the composition of geometries $(G_3, V_2, O) \circ (G_2, V_2, V_2) \circ (G_1, I, V_2)$ is illustrated in the opposite order in the diagram above.)
		$V_2$ is a set of maximal vertices under the constraint of being bounded from above, by the vertices $a$ and $b$, in a partial order $\preceq$ associated with a flow for the fractional-edge geometry $(G,I,O)$. 
		Arrows represent the action of the corresponding flow function $f$.
		}
	\end{center}
\end{figure}

For an arbitrary fractional edge $ab \in E(G)$ and each each $z \in O$, we may define $m(ab,z)$ to be the maximal vertex $v \in V(G)$ in the ordering $\preceq$ subject to $z$ being in the orbit of $v$ under $f$ (that is, $z = f^\ell(v)$ for some $\ell \ge 0$), such that at least one of $v \preceq a$ or $v \preceq b$ holds.
For a set $S \subset V(G)$, let $G[S]$ represent the subgraph of $G$ induced by $S$ (\ie\ by deleting all vertices in $G$ not in $S$). Then, define the following subgraphs of $G$, and corresponding geometries:
\begin{itemize}
\item
	Let $V_2$ be the set of vertices $m(ab,z)$ for each $z \in O\comp$: it is easy to show that $a,b \in V_2$. Let $G_2 = G[V_2]$,
	and let $\cG_2 = (G_2, V_2, V_2)$.

\item
	Let $V_1$ be the set of vertices $u \in V(G)$ such that $u \preceq v$ for some $v \in V_2$; let $G_1 = G[V_1] \setminus \ens{uv \,\big|\, u,v \in V_2}$\,; and let $\cG_1 = (G_1, I, V_2)$.

\item
	Let $V_3$ be the set of vertices $u \in V(G)$ such that $u \succeq v$ for some $v \in V_2$; let $G_3 = G[V_3] \setminus \ens{uv \,\big|\, u,v \in V_2}$; and let $\cG_3 = (G_3, V_2, O)$.
\end{itemize}
This decomposes the geometry $(G,I,O)$ into three (possibly still fractional-weight) geometries with flows, as illustrated in Figure~\ref{fig:decompose}.

Let $Q_1$ be a quadratic form on $\ens{0,1}^{V_1}$ consisting of the terms $x_u x_v$ of $Q$ for $u \in V_1$ or $v \in V_1$,
but not both; $Q_2$ be a quadratic form on $\ens{0,1}^{V_2}$ consisting of the terms $x_u x_v$ of $Q$ for \emph{distinct}
$u, v \in V_2$; and similarly let $Q_3$ be defined on $\ens{0,1}^{V_3}$, and consist of the remaining terms of $Q$.
Then $Q_1$, $Q_2$, and $Q_3$ define quadratic form expansions for some operations $U_1$, $U_2$, and $U_3$ (respectively)
with geometries $\cG_1$, $\cG_2$, and $\cG_3$ (respectively).
\begin{itemize}
\item
	$U_2$ in particular will be a product of operations $\cZ^{W_G(uv)}$ for distinct $u,v \in V_2$\,, as it is a quadratic form expansion whose input and output indices coincide. Then $U_2$ can be represented as a circuit with a wire for each $u \in V_2$, with fractional controlled-$Z$ gates $\cZ^{W_G(uv)}$ for each edge $uv \in E(G)$.
\item
	Both $\cG_1$ and $\cG_3$ have flows, and fewer fractional edges than $(G,I,O)$.
	By induction, $U_1$ and $U_3$ are also unitary embeddings, and have stable index expressions including terms $J(\theta_{vv}) \pseu[:f(x_v)/x_v]$ for each $v \in V(G_j) \setminus O_j$ and $\cZ^{W_G(uv)} \pseu[x_u \,, x_v]$ for each non-flow edge $uv \in E(G_j)$, for $j \in \ens{1,3}$.
\end{itemize}
Because $Q_1(\vec{x}_{\text{\raisebox{-0.3ex}{$V_1$}}}) + Q_2(\vec{x}_{\text{\raisebox{-0.3ex}{$V_2$}}}) + Q_3(\vec{x}_{\text{\raisebox{-0.3ex}{$V_3$}}}) = Q(\vec{x})$ for all $\vec{x} \in \ens{0,1}^{V}$ by construction, the composite operation $U_3 U_2 U_1$ can differ from $U$ by at most a scalar factor by Lemma~\ref{lemma:quadFormExpanComposition}.
We then obtain a composite circuit
\begin{gather}
		C
	\;\;=\;\;
		\Bigg(\prod_{v \in O\comp} J(\theta_{vv})\pseu[:x_{f(v)}/x_v]\Bigg)
		\Bigg(\prod_{\substack{vw \in E(G) \\ v \ne f(w) \\ w \ne f(v)}} \!\!\cZ^{W_G(vw)}\pseu[x_v , x_w ]\Bigg)
\end{gather}
performing the operation $U$, generalizing the construction given for unit-weight geometries described in~\eqref{eqn:unitWeightFlowCircuit}.
A direct way to verify that this circuit is correct is to expand the tensor product in terms of the coefficients of each operator $J(\theta)$ and $\cZ^t$\,:
\begin{subequations}
\begin{align}
	\label{eqn:coeffsJalpha}
		J(\theta_{vv})\pseu[:x_w/x_v]
	\;\;=&\;\;
		\sfrac{1}{\sqrt 2} \, \e^{i (-\theta_{vv}/2 \,+\, \theta_{vv} x_v \,+\, \pi x_v x_w)}
	\notag\\[0.5ex]\propto&\;\;
		\sfrac{1}{\sqrt 2} \, \e^{i (\theta_{vv} x_v^2 \,+\, \pi x_v x_w)}	\;,
	\\[1.5ex]
		\cZ\pseu[x_v, x_w]
	\;\;=&\;\;
		(-1)^{x_v x_w}	\;;
\end{align}
\end{subequations}
by summing over indices which are both advanced and deprecated, we may then verify that the coefficients of the unitary operation performed by $C$ are the same as those of $U$ (up to a product of the scalar factors $\e^{-i\theta_{vv}/2}$ in the proportionality of~\eqref{eqn:coeffsJalpha} above).

\paragraph{Run-time analysis.}

Let $n = \card{V(G)}$, $m = \card{E(G)}$, and $k = \card{O}$.
Determining whether a geometry has a flow, and constructing the flow $(f,\preceq)$ if so, can be done in time $O(kn)$ by Algorithm~1 of~\cite{MP08}.
For each edge $uv$, we may check whether one of $W_G(uv) = 1$ or $\big[ u \ne f(v)$ and $v \ne f(u) \big]$ holds: if all edges satisfy this constraint, the unitary circuit described above is well-defined, and may be constructed in time $O(m)$ by iterating through the edges of $V(G)$.
By the extremal result of Theorem~\ref{thm:extremalUpperBound}, any geometry with a flow has $m \le kn - \binom{k+1}{2}$: by eliminating geometries with more edges than this, we may reduce the total running time of this algorithm to $O(kn)$.

In the case $\card{I} = \card{O}$, a flow function $f$ is unique if it exists by Theorems~\ref{thm:graphth-charn} and~\ref{thm:uniqueFamilyPaths} (pages~\pageref{thm:graphth-charn} and~\pageref{thm:uniqueFamilyPaths}, respectively); so in this case, if $(G,I,O)$ has a flow when we neglect the edge-weights of $G$, but there is an edge $v \, f(v)$ of fractional weight, there is no flow for $(G,I,O)$ considered as a fractional-weight geometry.
Finally, note that there is an operation in the circuit for each edge in the geometry: if the original quadratic form expansion had non-zero coefficients $\theta_{vv}$ for $v \in O$, the \emph{expanded} geometry $(G',I,O')$ obtained by applying the technique on page~\pageref{eqn:idopQuadFormExpan} will have at most $m + 2k \le kn - \binom{k+1}{2} + 2k \le kn + 1$ edges (where $m = \card{E(G)}$ is the number of edges of the \emph{original} geometry), by the upper bound of Theorem~\ref{thm:extremalUpperBound}.
Therefore, we have the following result:
\begin{theorem}
	\label{thm:synthFlowsExact}
	For a transformation $U \in \U(2^k)$ given as a quadratic form expansion with geometry $(G,I,O)$ such that $\card{V(G)} = n$ and $\card{E(G)} = m$, there is an $O(kn)$ algorithm which either determines that $(G,I,O)$ has no flow, or constructs a stable index expression representing a unitary circuit consisting of at most $m + 2k \le kn + 1$ operations which performs $U$.
\end{theorem}

In order to obtain a circuit using a finite elementary gate set, we may again approximate each gate $J(\theta_{vv})$ and each gate $\cZ^t$ to precision $\frac{\epsilon}{m}$ by a product of $N$ gates in the $\Jacz(\frac{\pi}{4})$ model, where $N \in \polylog(m/\epsilon)$ for arbitrary $\epsilon > 0$, by the Solovay-Kitaev Theorem; and by the analysis of~\eqref{eqn:errorAdditiveCPTP}, this suffices to approximate $U$ to precision $\epsilon$.
We then obtain:\label{discn:synthFlowsEnd}
\begin{corollary}
	\label{cor:synthFlowsApprox}
	For a transformation $U \in \U(2^k)$ given as a quadratic form expansion with geometry $(G,I,O)$ such that $\card{V(G)} = n$ and $\card{E(G)} = m$, for each $\epsilon > 0$ and $\delta > 0$ there exists an algorithm running in time $O(kn \log(kn/\epsilon)^{3 + \delta})$ which either determines that $(G,I,O)$ has no flow, or constructs a stable index expression representing a unitary circuit in the $\Jacz(\frac{\pi}{4}\Z)$ model, which consists of $O(kn \log(kn/\epsilon)^{3 + \delta})$ operations and approximates $U$ to precision $\epsilon$.
\end{corollary}

\subsubsection{The quantum Fourier transform as an example}
\label{sec:fourierTransformEg}

For the most part, quadratic form expansions have occurred in the literature as analytical tools rather than as the preferred means of representing unitary transformations of interest.
While there is a possible source in path integrals, as we remarked on page~\pageref{sec:sumsOverPaths}, it is not yet known whether expressions resulting from discretizing path integrals would yield tuples $(G,I,O,\lambda)$ with generalized flows.
At the same time, the study of flows and gflows has thus far been essentially restricted to their potential as tools to analyze procedures in the one-way model, in an attempt to characterize procedures which perform unitaries under certain constraints (as in the \GMPI) or provide automatically verifiable semantics for measurement-based computations (as in Chapter~\ref{Semantics for unitary one-way patterns}).
Therefore, there are few known ``natural'' examples of a unitary whose definition lends itself to the techniques of this section.

\begin{figure}[tb]
	\begin{center}
		\includegraphics{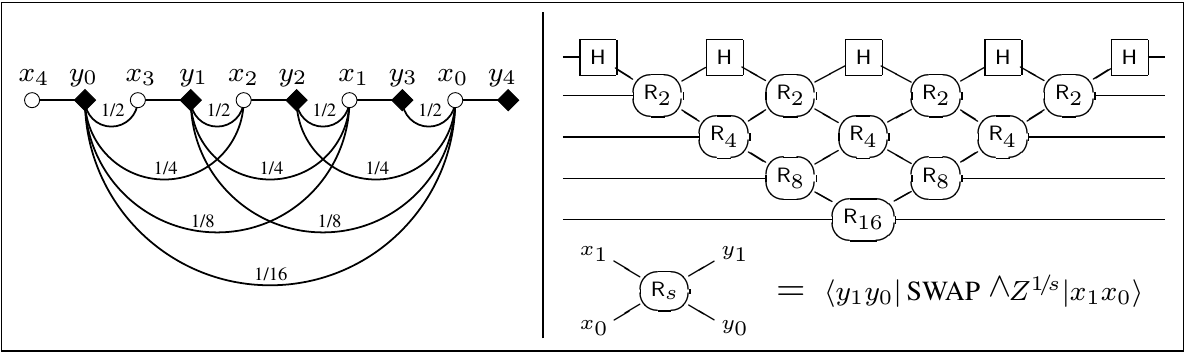}
	\end{center}
	\caption[The geometry for the quadratic form expansion of the QFT for $\Z_{32}$, and the corresponding circuit.]{\label{fig:qftSynth}
		The geometry for the quadratic form expansion of the QFT for $\Z_{32}$, and the corresponding circuit due to~\cite{FDH04}.
		In the geometry (on the left), input vertices are labelled by circles, output vertices by lozenges, and fractional edges are labelled with their edge-weights.}
\end{figure}

However, as we remarked in Section~\ref{sec:relatedWork}, there is one very prominent example of a unitary transformation which is naturally defined as a quadratic form expansion: the quantum Fourier transform $\cF_{2^k}$ over $\Z_{2^k}$\,.
Recall that $\cF_{2^k}$ has an expansion with the quadratic form
\begin{align}
	\label{eqn:qftGeom}
	Q(\vec x, \vec y) \;\;=\;\; \sum_{h = 0}^{k-1} \; \sum_{j = 0}^{k-h-1} \;\sfrac{2^{(h+j)}}{2^{k-1}} \;\pi \,x_j \,y_h \;,
\end{align}
where the vectors $\vec x$ and $\vec y$ correspond to the qubits of the input system $I$ and the output system $O$, respectively.
The (fractional-edge) geometry $\cG$ arising from this quadratic form expansion is illustrated\footnote{%
The problem of finding suitable graphical presentations of graphs is an open research problem in computational geometry; the presentation of the geometry in Figure~\ref{fig:qftSynth} was chosen by hand to best exhibit the symmetries of the geometry.}
in Figure~\ref{fig:qftSynth} for the case $n = 5$.

We may show that the fractional-weight geometry $\cG$ has a flow without having to perform a flow-finding algorithm, as follows.
Because the only edges which have unit weight in $\cG$ are those of the form $x_j y_{\text{\raisebox{-0.5ex}{$k \minus j \minus 1$}}}$ for each $0 \le j < k$, these must be the flow edges if $\cG$ has a flow.
To show that the function $f(x_j) = y_{\text{\raisebox{-0.5ex}{$k \minus j \minus 1$}}}$ is a flow-function, consider the influence relation $\inflRel$ induced by $f$ (Definition~\ref{def:inflRel}, page~\pageref{def:inflRel}).
By construction, the graph of the geometry $\cG$ is a bipartite graph, with partitions $\ens{x_j}_{j = 0}^{k-1}$ and $\ens{y_h}_{h = 0}^{k-1}$\,; then, for each vertex $x_j$, aside from $f(x_j)$, the only vertices $w$ such that $x_j \inflRel w$ are the other vertices $x_\ell$ adjacent to $y_{\text{\raisebox{-0.5ex}{$k \minus j \minus 1$}}}$.
The set of vertices adjacent to a given vertex $y_h$ are the vertices $x_\ell$ for $0 \le \ell \le k - h - 1$\,, by~\eqref{eqn:qftGeom}.
Thus, the vertices $x_\ell$ such that $x_j \inflRel x_\ell$ satisfy $\ell \ne j$ and $0 \le \ell < k - (k-j-1) - 1$, which holds if and only if $0 \le \ell < j$.
Then, every walk in the influence digraph $\sI_f$ (Definition~\ref{def:inflDigraph}, page~\pageref{def:inflDigraph}) either ends at a vertex $y_h$ for some $h$, or visits a sequence of vertices $x_{s_1} x_{s_2} \cdots$ for some monotonically decreasing sequence $\ens{s_j}_{j \ge 1}$ of indices.
In particular, $\sI_f$ contains no circuits; thus, by Lemma~\ref{lemma:causalOrderFlow} (page~\pageref{lemma:causalOrderFlow}), the geometry $\cG$ has a fractional-edge flow.

Illustrated in Figure~\ref{fig:qftSynth} alongside the presentation of $\cG$ is a representation of the circuit that may be constructed from $\cG$ together with the flow-function $f$.
Note that the static index tensor representation of unitary circuits in the $\Jacz(\R)$ model does not impose any \emph{a priori} restrictions on the layout or interactions of the qubits (in particular, it does not explicitly represent {\small SWAP} gates) --- partly because the gate-set $\Jacz(\R)$ itself places no restrictions on interactions between qubits (and doesn't contain {\small SWAP} gates).
However, by
\begin{alphanum}
\item
	representing the input and output wires in the conventional way, with vertical position increasing in order of the significance of each bit;

\item
	identifying the edges of unit weight in the geometry with $H = J(0)$ operations\footnote{%
	The presence of Hadamards in the circuit is due to the lack of any square terms in the quadratic form, corresponding to measurement angles of $0$ in the post-selection description of the unitary operation; this in turn leads to $J(0)$ gates in the constructed circuit.}
	in the unitary circuit, and arranging the other elements of the circuit diagram about those gates; and

\item
	allowing the circuit wires to cross each other, and run in directions other than horizontally (in particular, producing a crossing for each fractional edge illustrated),
\end{alphanum}
we are led to a presentation of this circuit as above, where we have collected the wire-crossings and fractional $\cZ$ operations into a family of compound gates as illustrated.

The circuit illustrated in Figure~\ref{fig:qftSynth} was first described by Fowler, Devitt, and Hollenberg in~\cite{FDH04}, and is presented in a similar graphical format in~\cite{RBB03}: therefore, the above presentation is neither new nor an independent discovery.
This example is meant only to illustrate simple heuristics which, together with the techniques of the preceding sections, can at least reproduce what is known for the quantum Fourier transform, and in a model which places spatial restrictions on multi-qubit operations (unlike the $\Phases(\A)$ or $\Jacz(\A)$ models).

\subsection{Geometries with simple modifications admitting eflows}
\label{sec:synthEflows}

As we noted on pages~\pageref{discn:squareTermConstrained}\thru\pageref{discn:simulateYZmeas}, there are instances of quadratic form expansions which represent unitary operations, and where this unitarity may depend crucially on the value of some of the coefficients $\theta_{vv}$ for $v \in O\comp$\,.
However, such quadratic form expansions can nevertheless be related to generalized flows, if we can identify it with a postselection procedure in which constrained postselections in the \XY\ plane can be regarded as allowing the postselection of states in some other plane (such as the \YZ\ plane) to be simulated.

In a one-way measurement pattern, a qubit $v \in O\comp \inter I\comp$ with degree $1$ may represent (considered as a simulation of some element of a unitary circuit) either the introduction of an auxiliary qubit prepared in the $\ket{\splus}$ state which undergoes further transformations, or a terminal measurement on some qubit which has undergone some transformations prior to measurement.
In the latter case, the preceding transformations may be interpreted as a change of reference frame prior to measurement, in order to simulate measurement in some other plane by an \XY-plane measurement.
In particular, to simulate a \YZ-plane measurement with an \XY-plane measurement, it suffices (as we noted on page~\pageref{eqn:simulateYZwXY}) to perform a Hadamard transformation prior to the \XY-plane measurement.

It is easy to verify that the only single-qubit rotations of the form $J(\theta)$ which do this are for $\theta \in \pi \Z$\,: measuring an observable $\sO^{\XY,\alpha} = \cos(\alpha) X + \sin(\alpha) Y$ as defined in~\eqref{eqn:pauliPlaneObs} after performing a $J(\theta) = H R_\sfz(\theta)$ rotation is equivalent to measuring the observable
\begin{gather}
	\label{eqn:changeFrameXY}
	J(\theta)\herm \; \sO^{\XY,\alpha} \, J(\theta)
	\;\;=\;\;
	\cos(\alpha) \, Z \;\;-\;\; \sin(\alpha) \, R_\sfz(\theta)\herm Y R_\sfz(\theta)	\;,
\end{gather}
which is of the form $\sO^{\YZ,\alpha'}$ only for $\theta$ a multiple of $\pi$.
Considering a geometry of unit weight (possibly obtained by performing the substitution described in~\eqref{eqn:diagonalExpansions}), any vertex $w \in V(G)$ of degree $1$ which is adjacent to a vertex $v \in V(G)$ with $\theta_{vv} \in \ens{0,\pi}$ may then be considered to be expressing a postselection procedure which selects for some state $\ket{\splus \sfy\sfz_\alpha}_v$ as defined in~\eqref{eqn:defYZtheta}.
Furthermore, from the fact that
\begin{subequations}
\label{eqn:telescopingJ}
\begin{align}
		J(\alpha) J(0) J(\beta) J(\gamma) \;\;=&\;\; H R_\sfz(\alpha) R_\sfz(\beta) J(\gamma) \;\;=\;\; J(\alpha + \beta) J(\gamma)	\;,
	\\
		J(\alpha) J(\pi) J(\beta) J(\gamma) \;\;=&\;\; H R_\sfz(\alpha) X R_\sfz(\beta) H R_\sfz(\gamma) \;\;=\;\; J(\alpha - \beta) J(\gamma + \pi)	\;,
\end{align}
\end{subequations}
we may extend this to arbitrary paths of odd length: if $G$ contains a path of the form $v_0 v_1 v_2 \cdots v_{N\minus1} v_N$ for $N$ odd such that $v_N$ has degree $1$, $v_j$ has degree $2$ for $1 \le j < N$, and $v_j \in \pi\Z$ for $j$ even, the corresponding postselection procedure may be regarded as simulating the postselection of some state in the \YZ\ plane state on the qubit $v_0$.
More generally, we may consider \emph{arbitrary} products of operators $J(\theta)$ which map measurement observables of interest (\eg, observables of the form $\sO^{\YZ,\alpha}$ or $\sO^{\XZ,\alpha}$) to \XY-plane observables.

By detecting when chains of vertices in a geometry represent postselections in either the \YZ\ or \XZ\ plane, we can obtain smaller geometries and a measurement-plane function $\lambda$ which may represent ``\emph{yes}'' instances of \GMPI\ (where we extend the domain of that problem in the natural way to include measurement operations in each of the \XY, \YZ, and \XZ\ planes).
Thus, for any algorithm for finding gflows for non-constant maps $\lambda$, this yields another approach to synthesizing measurement-based patterns from quadratic form expansions, by contracting paths which may be regarded as representing the simulation of a postselection of observables in the \YZ, or \XZ\ planes.

Currently, the only efficient algorithm for obtaining gflows for $(G,I,O,\lambda)$ for $\lambda$ not a constant function is Algorithm~\ref{alg:findEflow}, which determines whether or not a tuple $(G,I,O,T)$ has an eflow.
As we noted on page~\pageref{discn:gflowsSpecialCases}, this is equivalent to determining whether the tuple $(G,I,O,\lambda)$ has a gflow, for $\lambda$ given by
\begin{gather}
		\lambda(v)
	\;\;=\;\;
		\begin{cases}
			\YZ	\,,	&	\text{if $v \in T$}	\\
			\XY	\,,	&	\text{otherwise}
		\end{cases}	\;,
\end{gather}
and where we require that $g(v)$ be a singleton set for each $v \in O\comp$.
We may then attempt to map a quadratic form expansion provided as input to a unit-weight geometry $(G,I,O)$ and a set of vertices $T \subset V(G)$ such that $(G,I,O,T)$ has an eflow.

\subsubsection{Treating fractional-weight geometries}

As we have noted before, by applying the substitution of~\eqref{eqn:diagonalExpansions}, we may transform any geometry with fractional edges, corresponding to coefficients $\theta_{uv} \notin \pi \Z$ into a geometry with unit edge-weights.
As noted in Figure~\ref{fig:diagonalYZ}, this will yield a certain number of pairs of qubits which represent postselections in the \YZ\ plane: this should be immediately apparent from the fact that the substitution amounts to a decomposition of $\cZ^t$ into a pattern of the form $\mathfrak{Zz\cdots z}^{\pi t/2}$\,, which itself has an eflow.

In the case where the original geometry has a flow (in the sense defined for fractional-edge geometries in Definition~\ref{def:flowFractional}), rather than attempting to obtain a unit-weight geometry with an eflow, it may prove more convenient (\eg\ if the objective is to obtain a circuit representation of a unitary) to follow the analysis of pages~\pageref{discn:synthFlows}\thru\pageref{discn:synthFlowsEnd}.

\subsubsection{Transformations of quadratic form expansions to find eflows}

In order to consider whether a unit-weight quadratic form expansion can be mapped to a one-way measurement pattern using eflows, we may first simplify the expansion by taking advantage of identities of the same form as~\eqref{eqn:telescopingJ} above.
Consider once more a quadratic form expansion
\begin{align}
	\label{eqn:quadFormExpansion-3}
	C\!\! \sum_{\vec{x} \in \ens{0,1}^V} \e^{i Q(\vec{x})} \;\ket{\vec{x}_O}\bra{\vec{x}_I}
	&&
	\text{where}\quad
	Q(\vec x)
\;=&\;
	\sum_{\substack{\ens{u,v} \subset V}} \theta_{uv} x_u x_v \;\,,
\end{align}
and suppose there are indices $a,b,c,v \in V \setminus (I \union O)$ such that $Q(\vec x)$ contains the terms
\begin{gather}
	\begin{split}
		&\pi x_a x_b + \pi x_b x_c + \pi x_c x_v	\\
		&\quad + \theta_{aa} x_a^2 + \theta_{bb} x_b^2 + \theta_{cc} x_c^2 + \theta_{vv} x_v^2	\;,
	\end{split}
\end{gather}
and no other terms in $x_b$, $x_c$, or $x_v$ (though it may contain many other terms in $x_a$).
In the geometry induced by this quadratic form expansion, $v$ has degree $1$, and $b$ and $c$ have degree $2$.
If $\theta_{cc} \in \pi\Z$, we may then apply one of the following equalities:
\begin{subequations}
\label{eqn:telescopingQuadFormExpan}
\begin{align}
		\sum_{b,c,v} &
			\exp\paren{i\sqparen{\pi x_a x_b + \pi x_b x_c + \pi x_c x_v + \theta_{aa} x_a^2 + \theta_{bb} x_b^2 + 0 \!\cdot\! x_c^2 + \theta_{vv} x_v^2}}
	\notag\\[0.5ex]=&\;\;\;
		\sum_{b,v} \paren{
			\e^{i\sqparen{\pi x_a x_b + \theta_{aa} x_a^2 + \theta_{bb} x_b^2 + \theta_{vv} x_v^2}}
			\;\;+\;\;
			(-1)^{x_b \oplus x_v}
			\e^{i\sqparen{\pi x_a x_b + \theta_{aa} x_a^2 + \theta_{bb} x_b^2 + \theta_{vv} x_v^2}}}
	\notag\\[0.5ex]=&\;\;\;
		\sum_{b,v} \; 2 \, \delta_{b,v} \,
			\e^{i\sqparen{\pi x_a x_b + \theta_{aa} x_a^2 + \theta_{bb} x_b^2 + \theta_{vv} x_v^2}}
	\notag\\[0.5ex]=&\;\;\;
		2 \sum_{v} \e^{i\sqparen{\pi x_a x_v + \theta_{aa} x_a^2 + (\theta_{bb} + \theta_{vv}) x_v^2}}	\;;
	\\[-4.3em]\notag
	\intertext{}
		\sum_{b,c,v} &
			\exp\paren{i\sqparen{\pi x_a x_b + \pi x_b x_c + \pi x_c x_v + \theta_{aa} x_a^2 + \theta_{bb} x_b^2 + \pi \!\cdot\! x_c^2 + \theta_{vv} x_v^2}}
	\notag\\[0.5ex]=&\;\;\;
		\sum_{b,v} \paren{
			\e^{i\sqparen{\pi x_a x_b + \theta_{aa} x_a^2 + \theta_{bb} x_b^2 + \theta_{vv} x_v^2}}
			\;\;-\;\;
			(-1)^{x_b \oplus x_v}
			\e^{i\sqparen{\pi x_a x_b + \theta_{aa} x_a^2 + \theta_{bb} x_b^2 + \theta_{vv} x_v^2}}}
	\notag\\[0.5ex]=&\;\;\;
		\sum_{b,v} \; 2 \, \delta_{b,(1 - v)} \,
			\e^{i\sqparen{\pi x_a x_b + \theta_{aa} x_a^2 + \theta_{bb} x_b^2 + \theta_{vv} x_v^2}}
	\notag\\[0.5ex]=&\;\;
		2 \sum_{v} \e^{i\sqparen{\pi x_a (1 - x_v) + \theta_{aa} x_a^2 + \theta_{bb}(1 - 2x_v + x_v^2) + \theta_{vv} x_v^2}}
	\notag\\[0.5ex]=&\;\;
		2 \sum_{v} \e^{i\sqparen{\pi x_a x_v + (\pi + \theta_{aa}) x_a^2 + (-\theta_{bb} + \theta_{vv}) x_v^2}}	\;.
\end{align}
\end{subequations}
This yields a quadratic form expansion with two fewer indices, and where $v$ is still an index of degree $1$.
We may repeat this substitution as long as $v$ remains adjacent to a vertex $c'$ of degree $2$, which is itself adjacent to a vertex of degree $2$ and such that $\theta_{c'c'} \in \pi\Z$.
Any such chain will of vertices in the quadratic form will then be reduced to a vertex $v$ which is adjacent to some qubit $a$ which either has $\theta_{aa} \notin \pi\Z$, or which is involved in more than two cross-terms of the original quadratic form.

Performing the substitutions above can only help to obtain a tuple $(G',I,O,T)$ which can be analyzed using eflows.
By removing additional indices, we reduce the complexity of the quadratic form expansion, and also (in the corresponding instance of \MPI) reduce the number of qubits whose particular measurement angles may affect whether the transformation performed by the positive branch is an isometry.
For a given vertex $v$ of degree $1$ in the geometry induced by the quadratic form expansion, provided both the quadratic form and also a representation of the geometry in terms of adjacency lists, determining whether one of the substitutions of~\eqref{eqn:telescopingQuadFormExpan} can be applied can be determined in constant time.
Simplifying the quadratic form expansion provided as input as much as possible, and updating the coefficients of the square terms, is therefore a relatively inexpensive procedure.
We may assume for the remainder of this section that the geometry $(G,I,O)$ induced by the quadratic form provided as input has unit weight, and that for any vertex $v \in V(G) \setminus (I \union O)$ with degree $1$ that is adjacent to a vertex $a \in O\comp$, either the degree of $a$ is at least $3$ or $\theta_{aa} \notin \pi\Z$.

Provided a quadratic form expansion whose geometry $(G,I,O)$ satisfies the above constraints, define $\vec a = (\alpha_v)_{v \in O\comp}$ given by $\alpha_v = -\theta_{vv}$.
Consider the set $S$ of vertices $v$ of degree $1$: for such vertices $v$ and for the unique neighbor $a \sim v$, as we noted in the analysis of~\eqref{eqn:changeFrameXY}, the postselection of the states $\ket{\psub{\alpha_a}}$ and $\ket{\psub{\alpha_v}}$ is equivalent to post-selecting some state $\ket{\splus\sfy\sfz_{\alpha'}}$ if and only if $\alpha_a \in \pi\Z$\,.
Let $S' \subset S$ be the subset of those degree $1$ vertices whose unique neighbors $a$ have $\theta_{aa} \in \pi\Z$\,: then, using the relations
\begin{subequations}
\begin{align}
		\bra{\psub{\alpha_v}}_v \bra{\splus}_a \cZ_{av} \ket{\splus}_v
	\;\;\propto&\;\;\;
		\bra{0}_v \bra{\splus}_a  \ket{\splus}_v
	\;+\;
		\e^{-i\alpha_v} \bra{1}_v \bra{\sminus}_a \ket{\splus}_v
	\notag\\=&\;\;\;
		\sfrac{1}{\sqrt 2} \bra{\splus \sfx}_a \;\;+\;\;\; \sfrac{\e^{-i\alpha_v}\!}{\sqrt 2} \bra{\sminus \sfx}
	\notag\\=&\;\;\;
		\bra{\splus\sfy\sfz_{(\minus \alpha_v)}}_a	\;,
	\\[2ex]
		\bra{\psub{\alpha_v}}_v \bra{\sminus}_a \cZ_{av} \ket{\splus}_v
	\;\;\propto&\;\;\;
		\bra{0}_v \bra{\sminus}_a  \ket{\splus}_v
	\;+\;
		\e^{-i\alpha_v} \bra{1}_v \bra{\splus}_a \ket{\splus}_v
	\notag\\\propto&\;\;\;
		\sfrac{1}{\sqrt 2} \bra{\splus \sfx}_a \;\;+\;\;\; \sfrac{\e^{i\alpha_v}\!}{\sqrt 2} \bra{\sminus \sfx}
	\notag\\=&\;\;\;
		\bra{\splus\sfy\sfz_{\alpha_v}}_a	\;,
\end{align}
\end{subequations}
we may infer that the instance of the \MPI\ represented by the tuple $(G,I,O,\vec a)$ is equivalent to an instance $(G', I, O, T, \vec a')$, defined as follows:
\begin{romanum}
\item
	$G' = G \setminus S'$\,;

\item
	$T$ consists of the neighbors $a$ of the elements of $v \in S'$ in the graph $G$, which are to be measured in the \YZ\ plane; and

\item
	$\vec a' = (\alpha'_v)_{v \in V(G') \setminus O}$\,, where $\alpha'_a = -\e^{i \alpha_a} \alpha_v$ for vertices $a \in T$ which are adjacent in $G$ to some vertex $v \in S'$, and $\alpha'_w = \alpha_w$ otherwise.
\end{romanum}
(Again, this is an instance of an \emph{extension} of \MPI, in which measurements in the \YZ\ plane are permitted.)
We may then test whether the triple $(G',I,O,T)$ has an eflow using Algorithm~\ref{alg:findEflow}.

As in the case of gflows described on page~\pageref{eqn:gflowPattern}, an eflow leads immediately to a solution to (the original version of) \MPI\ for the unit weight geometries $(G,I,O)$ which we have assumed we are given as input.
Algorithm~\ref{alg:findEflow} represents the partial order $\preceq$ of an eflow $(f,\preceq)$ on $(G', I, O, T)$ in terms of a layer function $\ell: V(G') \to \N$ such that $v \preceq w \iff [\,v = w \text{~or~} \ell(v) > \ell(w)\,]$.
We may obtain a linear order $\le$ extending $\preceq$ by constructing level sets $\sV_j$ for each $j \in \N$\,, inserting each vertex $v \in V(G')$ into the set $\sV_{\ell(v)}$\,.
By the remarks on page~\pageref{discn:linearExtension}, this allows us to obtain a linear ordering of $V(G)$ extending $\preceq$.
For qubits $v \in V(G') \setminus T$, we may select for the result $\s[v] = 0$ using the pattern
\begin{subequations}
\begin{align}
	\label{eqn:eflowPatternXYmeas}
		\mathfrak {Ps}_v
	\;\;=\;\;
		\Xcorr{f(v)}{\s[v]} \sqparen{\prod_{\substack{w \sim f(v) \\ w \ne v}} \!\! \Zcorr{w}{\s[v]} \;} \!\Meas{v}{\alpha_v} 	\;;
\end{align}
for qubits $a \in T$, using the approach described in the analysis of~\eqref{eqn:zzManyXY}, we may select for the result $\s[a] = 0$ of a \YZ-plane measurement in the $\ket{\spm\sfy\sfz_{\alpha'_a}}$ basis in the extended instance of \GMPI, by performing the pattern
\begin{gather}
	\label{eqn:eflowPatternYZmeas}
		\mathfrak {Ps}_a
	\;\;=\;\;
		\sqparen{\prod_{\substack{u \sim a \\ u \ne v}} \Zcorr{u}{\s[v]}}
		\Meas{v}{\alpha_v;\s[a]}
		\Meas{a}{\,\alpha_a}	\;,
\end{gather}
\end{subequations}
where $v \in V(G) \setminus V(G')$ is the corresponding vertex of degree $1$ adjacent to $a$.
Then, the following CPTP map performs a unitary embedding for any choice of angles $\vec a = (\alpha_v)_{v \in O\comp}$ such that $\alpha_a \in \pi\Z$ for $a \in T$\,:
\begin{align}
	\label{eqn:eflowPattern}
	\paren{
		\ordprod[\mspace{-30mu}\ge\mspace{15mu}]_{v \in V(G') \setminus O}
			\mathfrak{Ps}_v }
	\paren{
		\prod_{uv \in E(G)} \Ent{u,v} }
	\paren{
		\prod_{v \in I\comp} \New{v}{\sfx} },
\end{align}
where the left-most product is ordered left-to-right according to the order $\ge$.
By the analysis of Section~\ref{sec:postselectionProcedure}, if $(G,I,O, \vec a)$ is an instance of the \MPI\ derived from a quadratic form expansion for an operator $U$, the unitary operation which is performed is $U$.

\paragraph{Run-time analysis.}

Let $n = \card{V(G)}$, $m = \card{E(G)}$, and $k = \card{O} \ge \card{I}$.
Determining whether a geometry has an eflow can be done by Algorithm~\ref{alg:findEflow} in time $O(kn + m)$ by the analysis on page~\pageref{sec:findEflowAnalysis}.
Having an eflow $(f, \preceq)$, the measurement procedure in~\eqref{eqn:eflowPattern} is well-defined and requires $O(m)$ time to construct, dominated by the product over $v \in V(G') \setminus O$ of the patterns $\mathfrak {Ps}_v$, each of which have complexity at most $O(\deg_G(v))$. 
Bringing this measurement procedure into a standard form by the standardization procedure of Section~\ref{sec:DKPstandardization} (also requires time $O(m)$\,, by a similar analysis as applies for patterns constructed by the DKP construction.
The resulting measurement pattern contains $n-k$ measurement operations and up to $k$ correction operations, each of which involves $O(n)$ measurement dependencies.
Finally, each substitution in~\eqref{eqn:diagonalExpansions} and~\eqref{eqn:telescopingQuadFormExpan} requires $O(1)$ time to perform provided a graph-based representation of the geometry underlying the original quadratic form expansion; the former expansion can be applied at most $m$ times (once for each edge), and the latter at most $m + n$ times (one for each vertex in the geometry, after the addition of vertices due to fractional edges) after spending $O(n)$ time finding all vertices of degree $1$.
We may therefore conclude:

\begin{theorem}
	\label{thm:synthEflowsExact}
	For a unitary embedding $U$ given as a quadratic form expansion such that $\card{V(G)} = n$, $\card{E(G)} = m$, and $\card{O} = k$, there is an $O(kn + m)$ algorithm which determines whether or not the quadratic form expansion cannot be transformed into a set of default measurement angles $\vec a'$ and a tuple $(G',I,O,T)$ which has an eflow.
	If so, the algorithm also constructs a pattern in the $\OneWay(\R)$ model consisting of $O(kn + m)$ operations which performs the transformation $U$.
\end{theorem}

As for the analysis in terms of gflows on page~\pageref{thm:synthGflowsExact}, we may strengthen this result to obtain an algorithm producing one-way patterns in a model with a finite gate set such as $\A = \frac{\pi}{4}\Z$\,: given that $m \ge n - k$ for a geometry with an eflow (from which we may infer that there are $O(m)$ measurement operations in the exact measurement pattern we must be approximated), using the same techniques as in the case for gflows, we then obtain the following corollary:

\begin{corollary}
	\label{cor:synthEflowsApprox}
	For a unitary embedding $U$ given as a quadratic form expansion such that $\card{V(G)} = n$, $\card{E(G)} = m$, and $\card{O} = k$, for each $\epsilon > 0$ and $\delta > 0$ there exists an algorithm running in time $O(kn + m \log(m / \epsilon)^{3+\delta})$ algorithm which determines whether or not the quadratic form expansion cannot be transformed into a set of default measurement angles $\vec a'$ and a tuple $(G',I,O,T)$ which has an eflow.
	If so, the algorithm also constructs a pattern in the $\OneWay(\frac{\pi}{4}\Z)$ model consisting of $O(kn + m \log(m/\epsilon)^{3 + \delta})$ operations which approximates the CPTP map $\rho \mapsto U \rho U\herm$ to precision $\epsilon$.
\end{corollary}

\subsection{Synthesizing measurement patterns for the Clifford group}
\label{sec:synthClifford}

If a quadratic form expansion has a geometry whose edges all have unit weight, and its' other coefficients are multiples of $\pion 2$, then it corresponds to the positive branch of a one-way measurement pattern which measures only $X$ or $Y$ observables.
A measurement pattern of this sort, if it performs a unitary operation, performs a Clifford group operation, as we may show (using the stabilizer formalism) that it transforms the Pauli group to itself.

An algorithm of Aaronson and Gottesman~\cite{AG04} can produce a Clifford circuit of size $O(n^2 / \log(n))$ in classical deterministic time $O(n^3 / \log(n))$ for a Clifford group operation $U$ acting on $n$ qubits, from a description of how $U$ transforms Pauli operators by conjugation.\label{discn:AG04}
By converting the circuit into a measurement-based algorithm, and performing the graph transformations of~\cite{HEB04} to remove auxiliary qubits, we may obtain a pattern of at most $3n$ qubits\footnote{%
In~\cite{BB06}, Clifford operations on $n$ qubits are described as having minimal patterns for are described as requiring at most $2n$ qubits; however, this only holds up to local Clifford operations on the output qubits.
These local Clifford operations may be used to adapt subsequent measurements, if the output state is to be operated on by more measurement based procedures, or simply measured to obtain a classical result; however, in the context of a decomposition of unitary transformations in \eg\ the $\OneWay(\frac{\pi}{2}\Z)$ model, these local Clifford operations should be explicitly described.}
in time $O(n^4 / \log(n))$.
Building on the results of~\cite{DM03}, we show how to classically compute such a minimal pattern in time $O(n^3 / \log(n))$ by solving \MPI\ for a quadratic form expansion for $U$.

\subsubsection{Obtaining a quadratic form expansion}

We now outline the relevant results of~\cite{DM03}.
Define the following notation for bit-flip and phase-flip operators on a qubit $t$ out of a collection $\ens{1,\ldots,n}$:
\begin{align}
		P_t
	\;\;=&\;\;
		X_t	\,,
	&
		P_{n+t}
	\;\;=&\;\;
		Z_t	\,.
\end{align}
Let $\diag(M) \in \Z_2^{\, m}$ represent the vector of the diagonal elements of any square boolean matrix $M$; and let $\vec{d}(M) = \diag\left(M\trans\! \left[\begin{smallmatrix} 0 & \idop_n\! \\ 0 & 0 \end{smallmatrix} \right] M\right) \in \Z_2^{\, 2n}$
for a $2n \x 2n$ matrix $M$ over $\Z_2$. Then, we may represent an $n$ qubit Clifford operation $U$ by a $2n \x 2n$ boolean matrix $C$ and a vector $\vec{h} \in \ens{0,1}^{2n}$, whose coefficients are jointly given by
\begin{align}
	\label{eqn:LeuvenTableau}
		U P_t U\herm
	\;\;=&\quad
		i\big.^{d_t(C)} \big(-1\big)^{h_t}
		\bigotimes_{j = 1}^n \sqparen{ Z_j^{\, C_{(n+j)t}} X_j^{\, C_{jt}} }
\end{align}
for each $1 \le t \le 2n$.~ (Note that the factor of $i^{d_t(C)}$ is only necessary to ensure that the image of $P_t$ is Hermitian, and does not serve as a constraint on the value of $C$.)
We will call an ordered pair $(C,\vec h)$ a \emph{Leuven tableau}\footnote{%
This terminology is motivated by the similarity between the block matrix $\big[\;\! C\trans \, \vec{h} \;\!\big]$ and \emph{destabilizer tableaux}, as defined in~\cite{AG04}.}
for a Clifford group element $U$ if it satisfies~\eqref{eqn:LeuvenTableau}.

Provided a Leuven tableau $(C,\vec h)$ for a Clifford group operation $U$, \cite{DM03}~provides a matrix formula for $U$ which we may obtain for $U$.
We outline the derivation of this formula as follows.
Decompose $C$ as a block matrix $C = \left[ \begin{smallmatrix} \,E\, & \,F\, \\ \,G\, & \,H\, \end{smallmatrix} \right]$ with $n \x n$ blocks, and then find invertible matrices $\tilde{R}_1, \tilde{R}_2$ over $\Z_2$ such that $\tilde{R}_1\inv G \tilde{R}_2 = \left[ \begin{smallmatrix} \,0\; & \!0\; \\ \,0\; & \idop_r \end{smallmatrix}\right]$ for some $r < n$~ (using \eg\ the decomposition algorithm of~\cite{PMH08} to obtain $\tilde{R}_1$ and $\tilde{R}_2$ in terms of $O(n^2/\log(n))$ elementary row operations).
Then, define the matrices
\begin{align}
		\left[ \begin{matrix}
			\,\tilde{E}_{11}\, & \,\tilde{E}_{12}\, \\[1ex]
			\,\tilde{E}_{21}\, & \,\tilde{E}_{22}\,
		\end{matrix} \right]
	\,=&\;\;
		\tilde{R}_1\trans E \tilde{R}_2	\,,
	&
		R_1
	\,=&\,\,
		\tilde{R}_1	\;,
	&
		R_2
	\;=&\,\,
		\left[\begin{matrix}
			\,\tilde{E}_{11}\inv\, & 0 \\[1ex]
			\;0\; & \;\idop_r\;
		\end{matrix}\right]\trans\!\!
		\tilde{R}_2\trans	\;,
\end{align}
where $\tilde{E}_{11}$ is taken to be a block of size $(n-r) \x (n-r)$. We may then obtain the block matrices
\begin{align}
		\left[\begin{smallmatrix}
			\idop_{n \minus r} & 		E_{12}		& F_{11} & F_{12} \\[1ex]
					E_{21}		& 		E_{22} 		& F_{21} & F_{22} \\[1ex]
					0				&			0			& H_{11} & H_{12}	\\[1ex]
					0				&	\idop_r	& H_{21} & H_{22}
				 \end{smallmatrix}\right]
	\;\;=&\;\;
		\left[\begin{smallmatrix} R_1\trans & 0 \\ 0 & R_1\inv \end{smallmatrix}\right]
		C
		\left[\begin{smallmatrix} R_2\trans & 0 \\ 0 & R_2\inv \end{smallmatrix}\right]	\,,
\end{align}
and use these to construct the $n \x n$ boolean matrices
\begin{align}
		M_{br}
	\;&=\;\;
		\left[\begin{matrix} \;F_{11} + E_{12} H_{21}\; & \;E_{12}\; \\[1ex] \;E_{12}\trans\; & \;E_{22}\; \end{matrix}\right]\,,
	&
		M_{bc}
	\;=&\;\;
		\left[\begin{matrix} \;0\; & \;H_{21}\trans\; \\[1ex] \;H_{21}\; & \;H_{22}\; \end{matrix}\right]\,.
\end{align}
Next, define 
\begin{align}
	\begin{split}
			\vec{d}_{br}
		&=\;
			\diag(M_{br}) \,,
		\\
			L_{br}
		\;&=\;
			\Lower\!\big(M_{br} + \vec{d}_{br} \vec{d}_{br}\trans\big) \,,
	\end{split}
	&
	\begin{split}
			\vec{d}_{bc}
		&=\;
			\diag(M_{bc}) \,,
		\\
			L_{bc}
		\;&=\;
			\Lower\!\big(M_{bc} + \vec{d}_{bc} \vec{d}_{bc}\trans\big) \,,
	\end{split}
\end{align}
where $\Lower(M)$ is the strictly lower-triangular part of a square matrix $M$ (with all other coefficients set to $0$).
Finally, define $\Pi_r = \left[\begin{smallmatrix} 0 & 0 \\ 0 & \idop_r\! \end{smallmatrix}\right]$ and $\Pi_r\dual
= \idop_n - \Pi_r$ for the sake of brevity, and let\footnote{%
These formulae for $\vec{t}$ and $\vec{h}_{bc}$ may be obtained by repeated application of Theorem~2 of~\cite{DM03} to the data specified by Theorem~6.}
\begin{subequations}
\begin{align}
		\vec{t}
	\;\;=&\;\;
		\left[\begin{matrix} \,\idop_n \,&\, 0 \;\end{matrix}\right] \vec{h}
		\;+\;	\diag\Big( \sqparen{ R_2\inv \Pi_r  } \!L_{br} \sqparen{ R_2\inv \Pi_r  }\trans \Big)		\;,
	\\[0.5ex]
	\begin{split}
		\vec{h}_{bc}
	\;\;=&\;\;
		 \left[\begin{matrix}\; 0 \,&\, R_2\intrans \,\end{matrix}\right] \vec{h}
		\;+\;	R_2\intrans \diag\Big(
					R_2\trans \Big[ L_{bc} \,+\, \Pi_r M_{bc} 
	\\[-0.5ex]&\mspace{100mu}
					\,+\, \paren{\Pi_r\dual + \Pi_r M_{bc}} L_{br} \paren{\Pi_r\dual +  M_{bc} \Pi_r} \Big] R_2
				\Big)	\;.
	\end{split}
\end{align}
\end{subequations}
Then Theorem~6 of~\cite{DM03} states that the unitary operation $U$ for the Clifford operation characterized by $(C, \vec{h})$ is given by the matrix formula
\begin{align}
	\label{eqn:cliffordFormula}
	\begin{split}
		U
	\,=\:\!
		\sfrac{1}{\sqrt{2^r}}\!\!\!
		\sum_{\substack{\vec{x}_b \in \ens{0,1}^{n \minus r} \\ \vec{x}_c, \vec{x}_r \in \ens{0,1}^r}}\!
		\bigg[\,
			(&-1)^{\paren{
									\vec{x}_{br}\trans L_{br} \vec{x}_{br} \,+\, 
									\vec{x}_r \trans \vec{x}_c \,+\,
									\vec{x}_{bc}\trans L_{bc} \vec{x}_{bc}	\,+\,
									\vec{h}_{bc}\trans \vec{x}_{bc} 
									}} \;\x
	\\[-4ex]&
			(-i)^{\paren{\vec{d}_{br}\trans \vec{x}_{br} \,+\, \vec{d}_{bc}\trans\vec{x}_{bc}}}	\;
			\ket{R_1\big. \vec{x}_{br}}\bra{R_2\inv \vec{x}_{bc} + \vec{t}}
		\,\bigg]	\,,
	\end{split}
\end{align}
where $\vec{x}_{br} = \left[\begin{smallmatrix} \vec{x}_b \\ \vec{x}_r \end{smallmatrix}\right]$ and
$\vec{x}_{bc} = \left[\begin{smallmatrix} \vec{x}_b \\ \vec{x}_c \end{smallmatrix}\right]$ are $n$ bit boolean
vectors.

The formula in~\eqref{eqn:cliffordFormula} shows strong similarities to a quadratic form expansion.
In particular, consider disjoint sets of indices $V_b$, $V_r$, and $V_c$, with $\card{V_b} = n-r$ and $\card{V_r} = \card{V_c} = r$.
Let $V = V_b \union V_c \union V_r$, $I = V_b \union V_c$, and $O = V_b \union V_r$, and define the following notation for $\vec{x} \in \ens{0,1}^V$\,:
\begin{align}
		\vec{x}_I
	\;=&\;
		\left[\begin{matrix} \vec{x}_b \\ \vec{x}_c \end{matrix}\right]
	\;=\;
		\left[\begin{matrix} \vec{x}_{\text{\raisebox{-0.3ex}{$V_b$}}} \\ \vec{x}_{\text{\raisebox{-0.3ex}{$V_c$}}} \end{matrix}\right] \in \ens{0,1}^{I}	\;,
	&
		\vec{x}_O
	\;=&\;
		\left[\begin{matrix} \vec{x}_b \\ \vec{x}_r \end{matrix}\right]
	\;=\;
		\left[\begin{matrix} \vec{x}_{\text{\raisebox{-0.3ex}{$V_b$}}} \\ \vec{x}_{\text{\raisebox{-0.3ex}{$V_r$}}} \end{matrix}\right] \in \ens{0,1}^{O}	\;,
\end{align}
\begin{align}
	\begin{split}
		Q(\vec x)
	\;=&\;\;
		\pi\Big( \vec{x}_O\trans L_{br} \vec{x}_O \,+\,
							\vec{x}_O\trans \Pi_r \vec{x}_I \,+\, \vec{x}_I\trans L_{bc} \vec{x}_I \,+\,
							\vec{x}_I\trans \vec{h}_{bc} \vec{h}_{bc}\trans \vec{x}_I	\Big)
	\\&\mspace{170mu}-\;
		\sfrac{\pi}{2} \Big(	\vec{x}_O\trans \vec{d}_{br}\vec{d}_{br}\trans \vec{x}_O \,+\,
									\vec{x}_I\trans \vec{d}_{bc}\vec{d}_{bc}\trans \vec{x}_I \Big)	\;.
	\end{split}
\end{align}
Then,~\eqref{eqn:cliffordFormula} is equivalent to
\begin{align}
	\label{eqn:cliffordAlmostQuadFormExpan}
		U
	\;\;=\;\;
		\sfrac{1}{\sqrt{2^r}}
		\sum_{\vec{x} \in \ens{0,1}^V}
			\e^{i Q(\vec x)} \ket{R_1\big. \vec{x}_O}\bra{R_2\inv \vec{x}_I + \vec{t}}	\;,
\end{align}
\begin{figure}[tb]
	\begin{center}
		\includegraphics{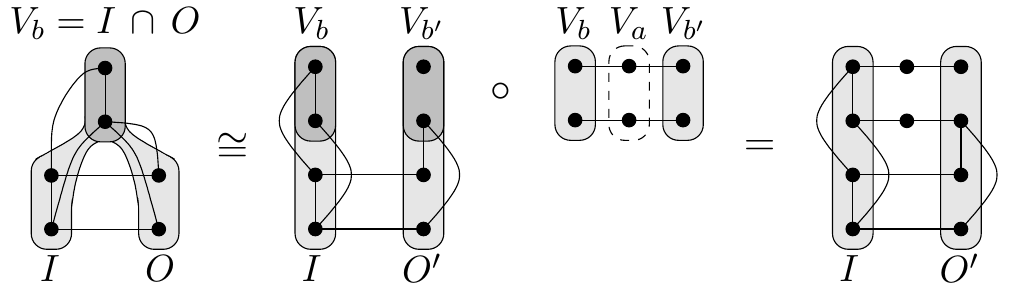}
		\caption[Illustration of an equivalence of two quadratic form expansions, via their geometries.]{\label{fig:idxSplit}%
		Illustration of an equivalence of two quadratic form expansions, via their geometries.
		On the left is a geometry whose inputs and output intersect; on the right is a geometry from an equivalent quadratic form expansion, constructed so that the input and output indices are disjoint.} 
	\end{center}
\end{figure}%
which is essentially a quadratic form expansion sandwiched between two networks of controlled-not and $X$ gates.
To obtain a simple quadratic form expansion, we would like to perform a change of variables on $\vec{x}_I$ and $\vec{x}_O$; but this cannot be done as $I$ and $O$ intersect at $V_b$, and the changes of variables do not necessarily respect the partitioning of $I$ and $O$ with respect to this intersection.
However, we may add auxiliary variables in order to produce an expansion with disjoint input and output indices.
Recall the expansion of the identity of~\eqref{eqn:idopQuadFormExpan},
\begin{align}
		\idop_2
	\;\;=\;\;
		\sum_{\vec{x} \in \ens{0,1}^2} \delta_{x_1,x_2} \ket{x_2} \bra{x_1}
	\;\;=\;\; 
		\sfrac{1}{2} \sum_{\vec{x} \in \ens{0,1}^3} (-1)^{x_1 x_3 + x_2 x_3} \ket{x_2} \bra{x_1}	\;,
\end{align}
where $\delta_{x,y}$ is the Kronecker delta.
Let $V_a$ and $V_{b'}$ be disjoint copies of $V_b$, and set $V' = V \union V_a \union V_{b'}$ and $O' = V_{b'} \union V_r$.
Writing $\vec{x}_a$ and $\vec{x}_{b'}$ for the restriction of $\vec{x} \in \ens{0,1}^{V'}$ to $V_a$ and $V_{b'}$, we then define
\begin{align}
		\vec{x}_I
	\;=&\;
		\left[\begin{matrix} \vec{x}_b \\ \vec{x}_c \end{matrix}\right] \in \ens{0,1}^{\, I}	\;,
	&
		\vec{x}_{O'}
	\;=&\;
		\left[\begin{matrix} \vec{x}_{b'} \\ \vec{x}_r \end{matrix}\right] \in \ens{0,1}^{\, O'}	\;,
\end{align}
\vspace{-1ex}
\begin{align}
	\mspace{-10mu}
		Q'(\vec{x}_I\,,\, \vec{x}_a\,,\, \vec{x}_{O'})
	\;=&\;\;
		\pi\Big(	\vec{x}_{O'}\trans L_{br} \vec{x}_{O'} \,+\,
							\vec{x}_{O'}\trans \Pi_r \vec{x}_I \,+\, \vec{x}_I\trans L_{bc} \vec{x}_I \,+\,
							\vec{h}_{bc}\trans \vec{x}_I	\Big)
	\notag\\&\mspace{10mu}
		\;+\;	\pi \vec{x}_I\trans \!\left[\begin{smallmatrix} \idop_{n \minus r} \\[0.5ex] 0 \end{smallmatrix}\right]\! \vec{x}_a
		\;+\;	\pi \vec{x}_{O'}\trans \!\left[\begin{smallmatrix} \idop_{n \minus r} \\[0.5ex] 0 \end{smallmatrix}\right]\! \vec{x}_a
		\;-\;	\sfrac{\pi}{2} \Big(	\vec{d}_{br}\trans \vec{x}_{O'} \,+\,	\vec{d}_{bc}\trans \vec{x}_I \Big)	\;.
\end{align}
Note that the difference between the expressions for $Q'$ and $Q$ is essentially that all instances of $\vec{x}_O$ have been replaced with $\vec{x}_{O'}$ (which is independent from $\vec{x}_I$), and the presence of the terms involving $\vec{x}_a$.
We therefore have
\begin{align}
	\label{eqn:quadFormTransform}
		\sum_{\vec{x} \in \ens{0,1}^V}
			\e^{i Q(\vec x)} &\ket{R_1\big. \vec{x}_O}\bra{R_2\inv \vec{x}_I + \vec{t}}
	\notag\\=&\;\;\;
		\sum_{\vec{x}_I,\, \vec{x}_{O'}}
			\delta_{\vec x_b, \vec x_{b'}} \, \e^{i Q'(\vec x_I,\, \vec{0},\, \vec x_{O'})} \ket{R_1\big. \vec{x}_{O'}}\bra{R_2\inv \vec{x}_I + \vec{t}}
	\notag\\=&\;\;\;	
		\sfrac{1}{2^{n-r}} \sum_{\vec{x} \in \ens{0,1}^{V'}}
			\e^{i Q'(\vec{x}_I,\, \vec{x}_a,\, \vec{x}_{O'})} \ket{R_1\big. \vec{x}_{O'}}\bra{R_2\inv \vec{x}_I + \vec{t}} \,.
\end{align}
This transformation is illustrated in Figure~\ref{fig:idxSplit} as a transformation of the geometries underlying $Q$ and $Q'$\,.
Substituting the final expression of~\eqref{eqn:quadFormTransform} into~\eqref{eqn:cliffordAlmostQuadFormExpan} and performing the appropriate change of variables, we have
\begin{align}
	\label{eqn:cliffordQuadFormExpan}
		U
	\;\;=\;\;
		\sfrac{\sqrt{2^r}}{2^n}
		\sum_{\vec{x} \in \ens{0,1}^{V'}}
			\e^{i Q'(R_2(\vec{x}_I + \vec{t}),\, \vec{x}_a,\, R_1^{\minus 1} \vec{x}_{O'})} \ket{\vec{x}_{O'}}\bra{\vec{x}_I}	\;:
\end{align}
expanding the expression in the exponent, it is easy to see that this is a quadratic form expansion for $U$.

\subsubsection{Interpolating the measurement pattern}

Note that the quadratic form of the expansion in~\eqref{eqn:cliffordQuadFormExpan} has only angles $\theta_{uv}$ which are multiples of $\pion 2$, with $\theta_{uv} \in \ens{0, \pi}$ for $u \ne v$.
This then represents the positive branch of a one-way measurement pattern using only \axisX- and \axisY-axis measurements, and having only $n-r$ auxiliary vertices.
From the analysis of the stabilizer formalism in Section~\ref{sec:stabilizerFormalism}, the fact that the result of the post-selection procedure is in particular a \emph{unitary} transformation implies that each individual post-selection induces a unitary transformation; then, we can then obtain a correction procedure to obtain a complete measurement pattern for $U$ via the stabilizer formalism.

First, we may produce a complete measurement procedure in which we perform correction operations after each measurement.
After applying the preparation and entangling operations $\Ent{G} \New{I\comp}{}$ before the measurements, the state-space is stabilized by the group $\sS_{G,I} \,=\, \gen{K_v}_{v \in I\comp}$\,.
Because post-selecting the measurement results $\s[v]$ for $v \in O\comp$ performs an isometry of the encoded space, there must be a list of generators $(S_v)_{v \in O\comp}$ for $\sS_{G,I}$ such that, for some ordering $v_1, v_2, \ldots$ of the qubits of $V(G)$,
\begin{romanum}
\item
	$S_{v_j}$ anticommutes with the measurement observable for the measurement on $v_j$\,, where we define
	\begin{gather}
			\sO_{v_j}
		\;\;=\;\;\
			\begin{cases}
				X	\,,	&	\text{if $\theta_{v_j v_j} \in \pi\Z$}	\\
				Y	\,,	&	\text{if $\theta_{v_j v_j} \in \pi \Z + \pion 2$}
			\end{cases}\;;
	\end{gather}

\item
	$S_{v_j}$ commutes with the measurement observable $\sO_{v_h}$ for any $h > j$\,.

\end{romanum}
We may produce such a list of generators for any given enumeration $(v_j)_{j = 1}^n$ of the elements of $V(G)$ in polynomial time, essentially by row-reduction.

\begin{figure}[p]
	\begin{center}
		\includegraphics[width=120mm]{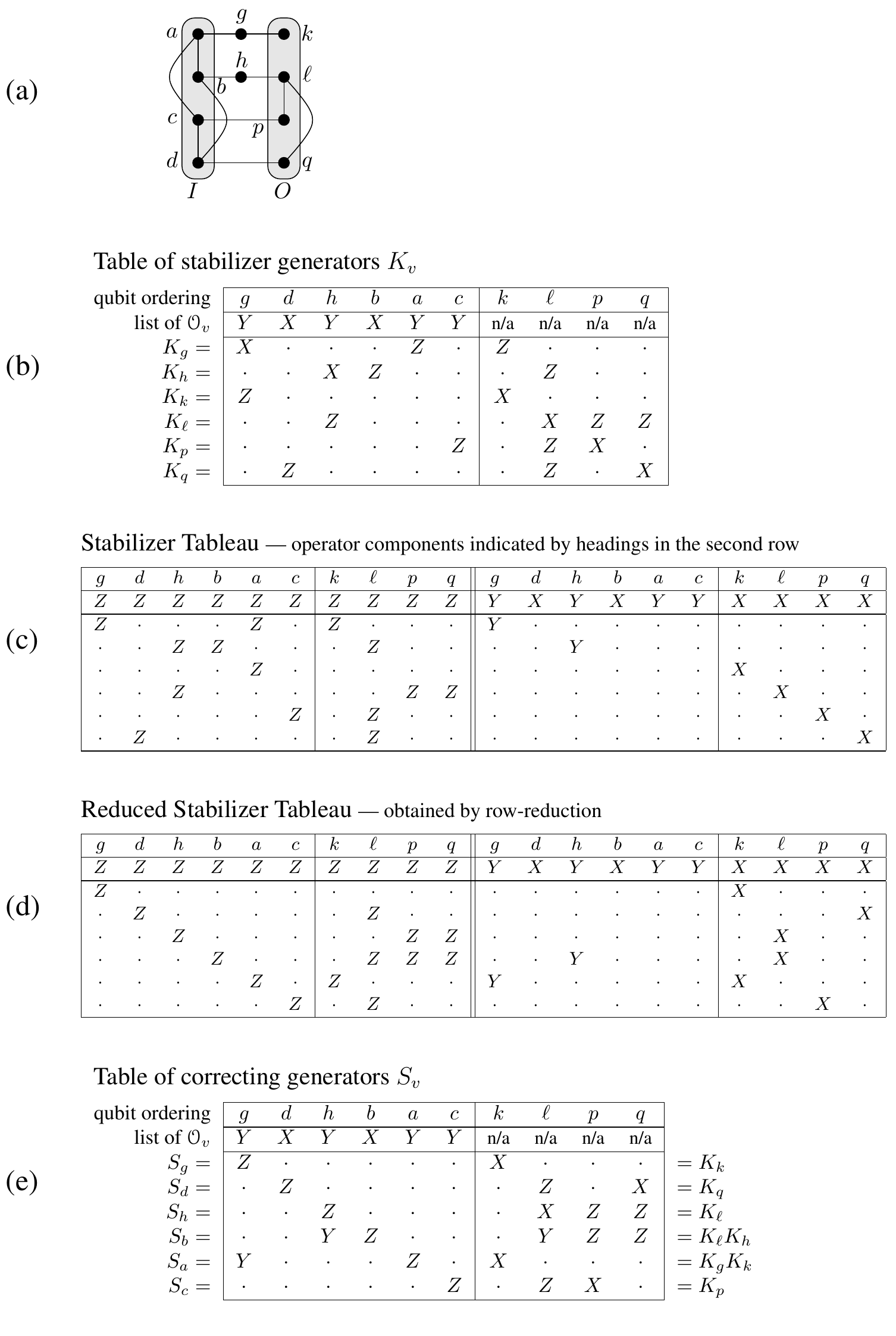}
		\vspace{-1ex}
	\end{center}
	\caption[Illustration of an algorithm for obtaining the correction operations for a postselection-based procedure for a unitary $U$ using measurement observables $X$ and $Y$.]{\label{fig:stabReduce}%
		Illustration of an algorithm for obtaining the correction operations for a postselection-based procedure for a unitary $U$ using measurement observables $X$ and $Y$.
		\begin{inlinum}
		\item The input geometry.
		\item	The stabilizer generators $K_v$ arising from the geometry, as well as an arbitrary measurement order of the qubits of $O\comp$ and a specification of the measurement observables $\sO_v$\,; ``dots'' represent identity operations $\idop_2$.
		\item The same generators presented in a ``stabilizer tableau'', in which each generator is decomposed into a $Z$ component (which anticommutes with every measurement observable) and a non-$Z$ component (chosen to commute with the observable $\sO_v$ for each qubit $v \in O\comp$).
		Note that this tableau is essentially an $(n - \card{I}) \x 2n$ boolean matrix.
		\item The reduced row-echelon form arising from the previous table.
		For each $j \ge 1$, the $j\th$ row represents an operator which anticommutes with the $j\th$ measurement observable, and commutes with every other measurement observable.
		\item A more explicit representation of the preceding table.
		\end{inlinum}
		}
\end{figure}

We arrange the generators $K_v$ for $v \in O\comp$ in rows in arbitrary order, with the columns denoting the qubits on which each operator acts.
We then decompose each operator $K_v$ into a product of operators $\sO_w$ for $w \in O\comp$ and $X_w$ for $w \in O$ (which we call the \emph{commuting} part of $K_v$) and a product of $Z_w$ operators for $w \in V(G)$ (which we call the \emph{anticommuting} part of $K_v$).
It is obvious that the anticommuting part of each operator $K_v$ will anticommute with any $\sO_w$ for a qubit $w \in O\comp$ on which it acts, and similarly the commuting part of $K_w$ commutes with every observable $\sO_w$.
We may use these components to form a stabilizer tableau, as illustrated in Figure~\ref{fig:stabReduce}, with the ``anticommuting'' $Z$ part of the operator in the leading columns, and the ``commuting'' part in the trailing columns.
Row transformations of this matrix then produce different generators of $\sS_{G,I}$, where the commutation relations with the observables $\sO_w$ for $w \in O\comp$ can be easily read from the $Z$ component part of the table.
By putting this table into reduced row-echelon form, we may then obtain a collection of generators $\ens{S_v}_{v \in O\comp}$ for $\sS_{G,I}$ which not only satisfy the criteria above, but such that $S_v$ anticommutes with $\sO_w$ only for $v = w$\,.
As each correction operation only acts on the qubit $v \in O\comp$ whose measurement it depends upon (and which we are considering to be measured destructively), these generators $S_v$ also represent the final correction operations for the measurement pattern: at the end of the measurement sequence, it suffices to perform the correction $S_v^{\s[v]}$ for each $v \in O\comp$ to obtain a unitary transformation.

\paragraph{Run-time analysis.}

For a Clifford operation $U \in \U(2^n)$ given as a Leuven tableau $(C, \vec h)$, the quadratic form of~\eqref{eqn:cliffordAlmostQuadFormExpan} can be found in time $O(n^3 / \log( n))$, which is dominated by the time required to compute $R_1$ and $R_2$, using the techniques of~\cite{PMH08}.\footnote{%
The algorithm of~\cite{PMH08} obtains an LU decomposition for matrix over $\Z_2$\,: while the result is described as applying to invertible matrices, it in fact can be applied to arbitrary boolean matrices.}
From this, the quadratic form expansion of~\eqref{eqn:cliffordQuadFormExpan} can be obtained by a constant number of matrix multiplications, and by inverting $R_1$ and $R_2$ (which may be computed alongside $R_1$ and $R_2$ themselves).
The boolean row-reductions required to compute the correcting generators $S_v$ for $v \in O\comp$ also requires time $O(n^3 / \log(n))$ by using a modified version of the algorithm of~\cite{PMH08} to do simple row-reduction.\footnote{%
It is sufficient to modify the algorithm of~\cite{PMH08} to eliminate ones above the diagonal as well as below, and to forego the column-reduction phase (consisting of the transposition and all operations afterwards).}
The resulting measurement pattern contains at most $3n$ qubits, and therefore requires $O(n^2)$ entangling operations, $O(n)$ measurement operations, and $O(n)$ correction operations each of complexity at most $O(n)$, yielding a total size of $O(n^2)$.
Therefore, we have:

\begin{theorem}
	\label{thm:synthClifford}
	For an $n$-qubit Clifford group operation $U$ given in the form of a Leuven tableau, there is an $O(n^3 / \log(n))$ algorithm which produces a pattern of size $O(n^2)$ in the $\OneWay(\frac{\pi}{2})$ model which performs $U$.
\end{theorem}

\subsubsection{Minimality of patterns synthesized from quadratic form expansions}

Hein, Eisert, and Briegel present techniques in~\cite{HEB04} for the elimination of Pauli observable measurements in graph states by
\begin{inlinum}
\item
	transformations of the underlying geometry, and

\item
	local unitary transformations of the qubits, which amount to changes in the observable to be performed for some of the measurements. 
\end{inlinum}
Paraphrasing~\cite{HEB04} and extending it to open graphical encodings, these transformations are as follows:
\begin{lemma}[{\cite[Proposition~1]{HEB04}}]
	\label{lemma:HEB04}
	Let $G$ be a graph: throughout the following, let $\sim$ denote adjacency in $G$.
	Define a unitary embedding
	\begin{gather}
		\cE_{G,I} = \paren{\prod_{u \sim v} \cZ_{uv} } \paren{ \bigotimes_{v \in I\comp} \ket{\splus}_v}
	\end{gather}
	performing an open graphical encoding.
	For a vertex $u \in V(G)$, let $G \ast u$ represent the \emph{local complementation} of $G$ about $u$, which is the graph on $V(G)$ where
	\begin{gather}
			vw \in E(G \ast u)
		\;\;\iff\;\;
			\begin{cases}
				vw \in E(G)		\;,	&	\text{if $v \not\sim u$ or $w \not\sim u$}	\\
				vw \notin E(G)	\;,	&	\text{if $v \sim u$ and $w \sim u$}
			\end{cases}	.
	\end{gather}
	If an $X_a$, $Y_a$, or $Z_a$ observable measurement is performed on some qubit $a \in I\comp$ for a joint state of the form $\cE_{G,I} \ket{\psi}$ for any $\ket{\psi} \in \cH[I]$, then the resulting pure state, conditioned on the outcome $\pm 1$, are as follows:
	\begin{subequations}
	\begin{romanum}
	\item
		 for a $Z_a$ observable measurement, the post-measurement state is
		\begin{gather}
			\ket{\spm \sfz}_a \ox U_{a,\pm\sfz} \, \cE_{G'_\sfz,I} \ket{\psi}
		\end{gather}
		where $G'_\sfz = G \setminus a$, $U_{a,\splus\sfz} = \idop_{V(G)}$, and $U_{a,\sminus\sfz} = \prod\limits_{v \sim a} Z_v$\;;

	\item
		for a $Y_a$ observable measurement, the post-measurement state is
		\begin{gather}
			\ket{\spm \sfy}_a \ox U_{a,\pm\sfy} \, \cE_{G'_\sfy,I} \ket{\psi}
		\end{gather}
		where $G'_\sfy = (G \ast a) \setminus a$, and $U_{a,\spm\sfy} = \prod\limits_{v \sim a} R_\sfz(\pm\pion 2)_v$\;;

	\item
		for an $X_a$ observable measurement, for an arbitrarily chosen $b \in I\comp$ adjacent to $a$, the post-measurement state is
		\begin{gather}
			\ket{\spm \sfx}_a \ox U_{a,\pm\sfx} \, \cE_{G'_\sfx,I} \ket{\psi}
		\end{gather}
		where $G'_\sfx = (G \ast a \ast b \ast a) \setminus a$, and where we define
		\begin{align}
				U_{a,\splus\sfx}
			\;=&\;\;
				R_\sfy(\mpion 2)_b \prod_{\substack{v \sim a \\ v \not\sim b_0 \\ v \ne b_0}} Z_v	\;;
			&
				U_{a,\sminus\sfx}
			\;=&\;\;
				R_\sfy(\pion 2)_b \prod_{\substack{v \sim b_0 \\ v \not\sim a \\ v \ne a}} Z_v	\;.
		\end{align}
	\end{romanum}
	\end{subequations}
\end{lemma}
It can be easily verified that~\cite[Proposition~1]{HEB04} can be extended as above, as the analysis of that article depends only on the presence of the stabilizer generators $K_v$ for $v \in I\comp$.\footnote{%
The input subsystem $I$ does not contribute any generators $K_v$ for $v \in I$, precisely because the input subsystem is not in general stabilized by any non-trivial Pauli operators.
Despite this, local complementations still affect adjacency relations which are entirely restricted to the input subsystem.
This may be verified by noting that an open graphical encoding can be characterized by $2^{\card I}$ stabilizer groups in parallel, with each group describing a scenario where each qubit $v \in I$ is stabilized by $X_v$ or $Z_v$ prior to encoding.
Collectively, these groups characterize any unitary transformations performed on the state, by linearity.
For each such group, any qubit $v \in I$ which is initially stabilized by $X_v$ does contribute a generator $K_v$ to the stabilizer group in that case; the generators of such groups are transformed in the same way as in graph states.}
The connection with local complementations of graphs (which is not made explicit in the presentation of Proposition~1) can be verified by application of Proposition~8 in the same article.

The results of~\cite{HEB04} would seem to be pertinent for the measurement patterns produced from quadratic form expansions, as described by the constructions above.
This suggests that we may be able to apply them to obtain measurement patterns involving fewer qubits than the ones produced by our construction.
Despite this, we may show that the measurement patterns produced by the techniques of this section cannot be reduced using Lemma~\ref{lemma:HEB04}.

Let $A$ denote the set of auxiliary qubits, corresponding to the indices of the vector $\vec{x}_a$ adjoined in~\eqref{eqn:quadFormTransform}.
In the resulting measurement pattern, these are all to be measured with the observable $X$, as there is no square term in the quadratic form $Q'$ for any of these indices.
As well, they are adjacent only to the input/output vertices (corresponding to the indices of $\vec{x}_I$ and $\vec{x}_O$).
These properties still hold in the final quadratic form expansions after the final change of variables, which do not act on the indices of $A$.
To eliminate a vertex $v \in A$ using Lemma~\ref{lemma:HEB04} in the open graphical encoding of the resulting one-way pattern, we must identify an output qubit $b' \in O$ adjacent to $v$, and transform the graph $G$ of the underlying geometry by to the graph $G'_{\sfx} \,=\, (G \ast v \ast b' \ast v) \setminus a$.
This would result in a geometry where $b'$ has the former neighbors of $v$ in $G$ (excepting itself): in particular, it will only be adjacent to one other qubit $b \in I$.
As well, a $R_\sfy(\pm\pion 2)$ rotation must be applied to $b$ after the entangling procedure.
This rotation is essentially unavoidable, in the following sense:
\begin{itemize}
\item
	Because $b' \in O$, it will not be measured, and so the $R_\sfy(\pm\pion 2)$ operation cannot be absorbed into a measurement.

\item
	Because $b'$ is not adjacent to any elements of $V(G'_{\sfx}) \setminus I$, the only stabilizer generator of the output space of the encoding $\cE_{G'_\sfx, I}$ is $K'_{b'} = X_{b'} Z_b$, which does not commute with $R_\sfy(\pm\pion 2)$.
	Then, the $R_\sfy(\pm\pion 2)$ is not equivalent (in the open graphical encoding) to any operator which does not act on $b'$.
\end{itemize}
Then, to represent the result of the transformation described by Lemma~\ref{lemma:HEB04} on a qubit $v \in I\comp$ which is to be measured, we must perform a local Clifford operation outside of the Pauli group to one of the qubits of the resulting encoding.
In the $\OneWay(\frac{\pi}{2}\Z)$ model, this local unitary will be represented by a pattern involving at least one additional qubit; thus, the number of qubits in the total pattern are not decreased.

Given a Clifford operation $U$ with a one-way measurement pattern synthesized from a quadratic form expansion, the techniques of~\cite{HEB04} may be applied if $U$ is to be composed with some other one-way pattern, or if the output subsystem is to be measured to produce a final classical result; however, as a representation of a unitary transformation, the above analysis shows that the number of qubits in the measurement pattern cannot be reduced using those techniques.

\subsubsection{Comparison with the algorithm of Aaronson and Gottesman}

As we noted on page~\pageref{discn:AG04}, an approximately optimal Clifford circuit for a Clifford group operation (consisting of $O(n^2 / \log(n))$ gates) can be found in time $O(n^3 / \log(n))$.
This may be done from a Leuven tableau by transforming it into a destabilizer tableau (another representation of transformations of Pauli operators by $U$ which differs trivially from a Leuven tableau), and then applying the algorithm of~\cite{AG04}.
To obtain a measurement pattern from such a circuit by composing the patterns for each gate, opportunistically removing vertices using the result of Lemma~\ref{lemma:HEB04} (with each removal taking $O(n^2)$ time), leads instead to an $O(n^4 / \log(n))$ time algorithm.
Thus, by Theorem~\ref{thm:synthClifford}, we may synthesize a minimal one-way measurement pattern more efficiently via quadratic form expansions than from the circuits produced by the algorithm of~\cite{AG04}.

The upper bound in Theorem~\ref{thm:synthClifford} on the complexity of the measurement pattern in $O(n^2)$, in contrast with the complexity of $O(n^2/\log(n))$ achieved by the circuits of~\cite{AG04}.
It is an open question whether the complexity of the patterns produced by the algorithm outlined above can be bounded more strongly than $O(n^2)$.
It is not obvious, however, that the number of edges in the entanglement graph or (nearly equivalently) the complexity of the final correction operations should saturate theupper bound of $O(n^2)$, or that the one-way measurement model itself imposes enough structural constraints to cause an asymptotically significant inflation in the optimal representation of a unitary transformation.
I therefore conjecture that the complexity of the measurement patterns synthesized by the algorithm above can also be improved to $O(n^2 / \log(n))$, as in the optimal bound for Clifford circuits.

\section{Further Remarks}

In this Chapter, we have focused largely on how we can transform or construct quadratic form expansions in such a way as to obtain an instance of \MPI, or a related problem such as \GMPI, which we are able to efficiently solve.
Before closing, we will remark on some of the ways in which the work in this Chapter may be extended.

\subsection{Reductions from more complex operator formulae}
\label{sec:higherDegree}

As we remarked on page~\pageref{discn:higherDegree}, path integrals are a potential ``natural'' source of quadratic form expansions, for which we may wish to obtain one-way measurement patterns.
In the case of path integrals, the most natural analogue of the quadratic form is often the integral of the action along a trajectory through space; however, these trajectories usually manifest as a polynomial of degree higher than $2$ for non-trivial systems.
Direct attempts to discretize a path integral may then lead to formulae similar to quadratic form expansions, except where the quadratic form is substituted by a higher degree polynomial.
As well, natural approaches to efficiently estimating the path integral of a physical system may involve attributing different weights to the discretized paths, corresponding to different amounts of constructive/destructive interference of the amplitudes of similar paths which differ from the trajectories of minimum action.
We outline here how variations on quadratic form expansions, involving higher degree polynomials and amplitudes which are non-uniform across the path variables $\vec{x} \in \ens{0,1}^n$, may be reduced to quadratic form expansions.

\subsubsection{Higher degree polynomials governing phases}

In order to treat matrix formulae with higher degree polynomials governing the phase contributions of the terms in the sum, consider a generalization of the substitution described by~\eqref{eqn:diagonalExpansions} to more than two bits.
Considering the construction of the pattern $\mathfrak{Zz\cdots z}^\theta$ described in~\eqref{eqn:zzManyXY}, it is possible to show that
\begin{gather}
	\label{eqn:diagonalExpansionBig}
		\e^{i\phi(x_1 \oplus \cdots \oplus x_d)}
	\;\;=\;\;
		\sfrac{1}{2}\! \sum_{\vec a \in \ens{0,1}^2} \e^{i\big[\pi(x_1 a_1 + \cdots + x_d a_1 + a_1 a_2) \,+\, \phi a_2^{\, 2}\big]}
\end{gather}
for any $d \ge 1$.
We may easily prove by induction that
\begin{gather}
		x_1 \oplus \cdots \oplus x_d
	\;\;=\;\;
		\sum_{\substack{S \subset \ens{1,\ldots,d} \\ S \ne \vide}} \; \prod_{j \in S} (-2)^{\card S - 1} x_j	\;;
\end{gather}
equivalently, we have
\begin{gather}
		\prod_{j = 1}^d x_j
	\;\;=\;\;
		\paren{\sfrac{-1}{2}}^{d-1}
		\sqparen{
			-\paren{x_1 \oplus \cdots \oplus x_d}
			\;\;\;+\!\!
			\sum_{\substack{S \prsubset \ens{1,\ldots,d} \\ S \ne \vide}} \prod_{j \in S} (-2)^{\card S - 1} x_j	\;
		}\;,
\end{gather}
which we can apply recursively to products of the bits $x_1$ through $x_d$ to obtain the formula
\begin{gather}
		\prod_{j = 1}^d x_j
	\;\;=\;\;
		-\paren{\sfrac{-1}{2}}^{d-1}
		\sum_{\substack{S' \subset \ens{1,\ldots,d} \\ S' \ne \vide}} \paren{\,\bigoplus_{j \in S} \; x_j}	\;.
\end{gather}
Thus a monomial of degree $d$ may be expanded into a sum of $2^d - 1$ parities; we may then use the substitution of~\eqref{eqn:diagonalExpansionBig} to express any factor of the form $\e^{i \theta x_1 \cdots x_d}$ in terms of a sum of $2^{d+1} - 2$ auxiliary indices, with two such indices for each sum modulo $2$ to be computed.
If the degree of the terms of the action of a physical system is bounded by some constant, this then leads to a constant factor overhead in the size of a quadratic form expansion, in order to represent a similar expression whose ``paths'' have phases given by higher degree polynomials.

As a result of the analysis above, the degree of the action in the dynamical variables of a physical system need not be an overriding concern in the formulation of a quadratic form expansion, provided that the degree is bounded above by a small constant.
A comparison between this and techniques for ``decoupling'' higher degree terms in path integrals in the physics literature (such as the Hubbard--Stratonivich transformation: see~\cite{Auerbach94,Gubernatis85} for expositions) may provide insights into a high-level physical interpretation of this substitution.

\subsubsection{Unequal amplitude contributions across paths}

Quadratic form expansions are a normalized sum of contributions of operators $\ket{\vec x_O}\bra{\vec x_I}$ for $\vec x \in \ens{0,1}^n$\,, each of which has one singular value of $1$ and the rest equal to zero.
That is to say, as a sum over paths, each path is given equal weight; only the phases differ.
However, it is natural to consider operator sums where terms may have different weights for different boolean strings $\vec x$\,.
One reason may be that, in order to discretize a path integral, contributions of many continuous paths are attributed to each discrete path string $\vec x$\,, and these will in general result in contributions with different amplitudes.
Therefore, matrix expressions in which each coefficient is a sum of complex numbers with different moduli (instead of complex units), normalized over all matrix coefficients, may be of interest.

In case the amplitudes of the paths depend uniformly on the value of a single bit $x_v$ in the path string $\vec{x}$ (\ie\ if all paths with $x_v = 0$ have the same weight, and similarly for $x_v = 1$), we may represent this bias in the amplitudes simply by means of a choice of measurement basis for $v$.
Considering the relationship between quadratic form expansions and instances of \MPI\ shown in Section~\ref{sec:postselectionProcedure}, we may attribute the uniform modulus of the paths in that case to the fact that every measurement is performed with a state which is a uniform-weight superposition of the standard basis: the projection of the states $\ket{0}$ and $\ket{1}$ onto $\ket{\psub\alpha}$ for any $\alpha$ has the same modulus.
In order to achieve a systematic bias towards either $x_v = 0$ or $x_v = 1$, it suffices to consider a measurement of $v$ where the preferred measurement result is similarly biased towards the corresponding standard basis state.

More generally, we may express the bias in the amplitudes of the paths in terms of a boolean polynomial on $\vec{x}$\,.
To achieve the necessary biases, we may consider substitutions of the sort shown in~\eqref{eqn:nonuniformity}.
The non-uniformity of the sum which presented problems for an analysis in terms of \GMPI\ in Section~\ref{sec:interpolateGflows}, here allows us to express amplitudes for paths in terms of the parity of two path variables.
By using parities of boolean expressions, we may produce biases depending on the value variable $\vec x$ consisting of a monomial of arbitrary degree, using the techniques of the preceding page.
By summing over such monomial variations, we may obtain arbitrary dependencies of path amplitudes on the path variables.
Again, if the dependency of the amplitudes on the path variables is of bounded degree, this yields a constant factor increase in the complexity of the resulting matrix expression, and ultimately yields a sum with uniform weights across all terms.

\subsection{Techniques for simulating postselection via error correction}

The results of this chapter have revolved about rather special conditions on the structure of a quadratic form expansion, under which the simulation of postselection is essentially possible generically due to restrictions on the graphical encoding performed by the preparation and entangling phases (in the case of ``\emph{yes}'' instances of different versions of the \GMPI) or of the default measurement angles (in the case of Clifford group unitaries).
In order to achieve a deeper understanding of how postselection-based procedures may be extended to measurement-based patterns, it would be useful to arrive at characterization of sequences of measurement observables on the one hand, and graphical encodings on the other, such that postselection of the $+1$ result of each measurement in sequence performs an isometry.

In~\cite{MCSB98}, necessary and conditions are laid out for conditions under which a CPTP map (such as a measurement operation) performs a reversible operation on a given subspace of the possible states of a physical system.
Considering the special case of single-qubit measurements on open graphical encodings, and supplemented by an explicit account of the evolution of the state-space under such measurements, may plausibly lead to a fuller understanding of \MPI.

\section{Review and open Problems}

In this chapter, we have considered the possibility of developing new techniques for developing quantum algorithms via the one-way measurement model by presenting techniques for transforming quadratic form expansions, and for synthesizing realizable procedures for unitary transformations by considering solvable instances of \MPI.
We have also presented quadratic form expansions themselves as interesting expressions in their own right, predicting that they are likely to recur as an important form in which unitary operations may be expressed, and suggesting in particular that they may be obtained by discretizing path integral formulations of propagators for physical systems.

The major open problem of this chapter which may be clearly addressed is whether a meaningful link between path integrals and quadratic form expansions is possible, and whether the \MPI\ could be solved for typical instances which would be generated in this way.
Affirmative answers for both would represent the most likely way in which the results of this chapter (or extensions of them for more general instances of the \MPI) could see concrete application.

From a purely theoretical point of view, a subject which has not been addressed in this chapter is the question of when a quadratic form expansion represents a unitary transformation.
Also of potential interest is what transformation is performed on the input subsystem by the postselection procedure described by a quadratic form expansion, when that operator is not a unitary transformation.
Answers to these questions may also shed light on the one-way measurement based model, as a potentially useful model of computation.

Finally, note that the construction of the one-way measurement patterns produced for Clifford group operations in Section~\ref{sec:synthClifford} does not hint at any obvious circuit structure.
In particular, while the depth of the measurement pattern is constant, an asymptotically optimal circuit for a Clifford group operation has size $O(n^2/\log(n))$, and therefore depth similar to $n/\log(n)$.
Is it possible to arrive at a semantic map for Clifford patterns in some sense similar to that which was done for the DKP and simplified RBB constructions in Chapter~\ref{Semantics for unitary one-way patterns}?
	\chapter[Further research directions]{
		Further research directions ()}


This thesis has been concerned with the one-way measurement model as a means of representing unitary transformations.
This has largely revolved around \emph{flows} on one hand, as a combinatorial structure arising out of a convenient representation of unitary circuits with certain ``simple'' gate sets, and generalizations of flows, which arise from links between flows and the techniques of the stabilizer formalism.
From this analysis, we have obtained efficient algorithms for recognizing certain one-way measurement patterns which perform unitary transformations, and translating them into unitary circuit models.
These techniques then proved useful for describing how we might be able to automatically obtain realizable procedures for unitary transformations which are specified by a certain type of operator formula which has recurred in the literature, and which are similar in form to a path integral formulation of propagators for a physical system.

The most natural (if perhaps ambitious) research direction from a mathematical standpoint is to characterize those one-way measurement patterns which perform unitary transformations.
This problem seems to demand an extension of the techniques of the stabilizer formalism.
The most likely necessary ingredient is one which is contained in the analysis of certain measurement patterns in~\cite{RBB03}: given an open graphical encoding stabilized by some subset of the Pauli group, we may obtain Hamiltonians which have a non-trivial kernel, in which the state of the system lies.
This allows us to consider the continuous group of unitary operators which also stabilize the state, which may be derived from continuous-time evolution of these Hamiltonians.
Judiciously chosen operators of this form can be used to transform measurement observables into Pauli operators whose measurement can be treated by the stabilizer formalism.
Connecting these rotation operations to the conditions of~\cite{MCSB98} for reversibility of quantum operations may yield an analysis of sequences of measurements, for which a characterization may be obtained.
Solving this problem will require a characterization of how stabilizer codes transform after successive information-preserving single-qubit measurements; this amounts to finding a more general class of codes than stabilizer codes, which is closed under such operations.

After the problem of characterizing measurement patterns performing unitary transformations, the next natural problem is to recognize instances of \MPI\ which may be ``interpolated'' to satisfy such a characterization (where we extend \MPI\ to admit arbitrary ``default measurement observables'' in place of the vector of default measurement angles).
A successful analysis of this problem would represent essentially complete low-level knowledge of the one-way measurement model.

While the above problems may have interesting practical dividends, the immediate question from the point of view of application is where one may expect to obtain instances of \MPI, in order to have something to apply these techniques to.
I have tried to motivate quadratic form expansions as a recurring form of operator expression in the quantum computing literature, but this does not describe a programme for actually obtaining such expressions.
The natural question is then whether or not it is possible to transform path integrals for physical systems of interest into quadratic form expansions, for which we may solve \MPI\ and thereby obtain efficient algorithms for simulating those systems.
At the same time, an exploration of a possible connection between quadratic form expansions and path integrals may itself yield interesting theoretical insights into the one-way measurement model, or into quantum computation in general.


\renewcommand{\bibname}{References}
\addcontentsline{toc}{chapter}{\textbf{\bibname}}

\bibliographystyle{plain}

\def\articleIndex{$\langle\textrm{\itshape article index}\rangle$}
\newcommand\BibliographyPreamble{%
	Many of the articles below are freely avaliable on the Cornell E-print Archive: such articles are indicated by the notation [\myurlstyle{arXiv:\articleIndex}] in their bibliographical entries.
	For such articles, an abstract and links to Postscript/PDF versions of the article can be found at the URL  [\myurlstyle{www.arXiv.org/abs/\articleIndex}] corresponding to the article's index.
	
	\ifElectronicVersion\ifFinalVersion
	After each bibliographical citation, the list of numbers indicate the page numbers on which the article is cited.
	\fi\fi
}

\bibliography{\BibliographyFile}

\nocite{*}

\end{document}